\documentclass[11pt,a4paper]{article}
\usepackage[T1]{fontenc}
\usepackage[utf8]{inputenc}
\usepackage{lmodern,microtype,amsmath,amssymb,amsthm,mathtools,mathrsfs}
\usepackage[margin=25mm,headheight=15pt]{geometry}
\usepackage{booktabs,longtable,array,enumitem,graphicx,fancyvrb,listings,tabularx,caption,calc}
\usepackage{xcolor,xurl}
\definecolor{ink}{HTML}{17354A}
\definecolor{teal}{HTML}{087F8C}
\definecolor{navy}{HTML}{17354A}
\usepackage[colorlinks=true,linkcolor=ink,urlcolor=teal,citecolor=ink]{hyperref}
\usepackage{bookmark,fancyhdr}
\hypersetup{pdftitle={SU(2) Yang-Mills: Spectral Bounds, Vacuum Concentration, and Criteria for the Continuum Limit},pdfauthor={Aleksandar Mihajlovic},pdfsubject={Version 5. Rigorous lattice results, complete finite spectral certificates, fixed-cutoff infinite-volume dynamics, and conditional continuum criteria.}}
\setlist{itemsep=3pt,topsep=4pt}
\newcommand{\tightlist}{\setlength{\itemsep}{0pt}\setlength{\parskip}{0pt}}
\newcommand{\R}{\mathbb R}\newcommand{\Z}{\mathbb Z}
\newcommand{\E}{\mathbb E}
\newcommand{\Cov}{\operatorname{Cov}}\newcommand{\Var}{\operatorname{Var}}
\newcommand{\supp}{\operatorname{supp}}
\newcommand{\TV}{\operatorname{TV}}\newcommand{\osc}{\operatorname{osc}}
\newcommand{\Tr}{\operatorname{Tr}}
\newcommand{\tr}{\operatorname{tr}}\newcommand{\ReTr}{\operatorname{Re}\operatorname{Tr}}
\newcommand{\Lip}{\operatorname{Lip}}\newcommand{\arccosh}{\operatorname{arccosh}}
\newcommand{\Fs}{F_{\!*}}\newcommand{\HH}{\mathcal H}\newcommand{\G}{\mathcal G}
\newcommand{\dd}{\mathrm d}\newcommand{\pos}[1]{[#1]_+}
\newcommand{\note}[1]{\par\smallskip\noindent#1\par\smallskip}
\newcommand{\norm}[1]{\lVert #1\rVert}
\DeclareMathOperator{\sech}{sech}
\newtheorem{theorem}{Theorem}[section]
\newtheorem{lemma}[theorem]{Lemma}\newtheorem{proposition}[theorem]{Proposition}
\newtheorem{corollary}[theorem]{Corollary}
\theoremstyle{definition}
\newtheorem{definition}[theorem]{Definition}\newtheorem{assumption}[theorem]{Assumption}
\newtheorem{target}[theorem]{Target}
\theoremstyle{remark}\newtheorem{remark}[theorem]{Remark}
\numberwithin{equation}{section}

\title{\textbf{SU(2) Yang--Mills:\\Spectral Bounds, Vacuum Concentration,\\and Criteria for the Continuum Limit}}
\author{Aleksandar Mihajlovic\\[5pt]
\small Technical Faculty, Singidunum University, Belgrade, Serbia\\
\small Abu Dhabi, United Arab Emirates\\
\small \href{mailto:aleksandar.mihajlovic.10.dls@singimail.rs}{aleksandar.mihajlovic.10.dls@singimail.rs}\\
\small ORCID: \href{https://orcid.org/0009-0009-0623-8080}{0009-0009-0623-8080}}
\date{5 November 2026 \quad Version 5}
\begin{document}
\maketitle
\begin{abstract}
A continuum mass-gap argument must preserve nonzero renormalized states
and a positive spectral estimate while the spatial cutoff is removed and
physical volume grows. We develop quantitative results that separate these requirements.
Gauge reduction, reflection positivity and attributed strong-coupling
estimates give physical-transfer bounds with nontrivial local sectors. At fixed spatial cutoff and positive coupling,
we construct subsequential infinite-volume physical-time dynamics from
local history correlations, with nonconstant gauge-invariant observables.
For finite spatial boxes, explicit bounds apply to the entire physical
vacuum complement; finite-matrix criteria include all omitted representations.
Vacuum concentration gives extensive entropy bounds, quantitative fidelity
obstructions for hard electric cutoffs, and necessary growth of retained
representation spaces. A sharper local comparison yields complete bounds
on fixed discarded-link sectors, while collective projection and Schur
estimates isolate the additional input needed for a volume-uniform gap.
The work also certifies a finite three-link conditioning functional through
Wilson coupling $0.354$, with its physical interpretation conditional on
the stated block-comparison hypotheses. Exact blocking, corrected
effective-potential theorems and a representation-by-representation
two-dimensional benchmark clarify which approximations preserve the
relevant observables. Published glueball measurements retain their original
attribution; the new analysis of those data and the recorded replica
experiment have explicitly statistical scope. Complete derivations and
reproducible calculations support a set of certified benchmarks and
continuum transfer criteria. A nontrivial four-dimensional spatial-cutoff
construction and a positive spectral bound along it are not established;
no exact continuum mass is asserted.
\end{abstract}
\clearpage\tableofcontents\clearpage
\section{The research question and the sequence of results}
\label{sec:introduction}

The motivating question is quantitative: what can be proved about the
lowest physical excitation of four-dimensional pure $SU(2)$ Yang--Mills
theory, and which finite-cutoff estimates can survive the construction
of a continuum theory? Positivity at one regulator is the beginning of
this question. Its endpoint requires compatible nonzero physical states,
local observables and time correlations as the regulator and volume change.
The mathematical development below follows that progression.

The physical object is a gauge-invariant transfer operator, or the
Hamiltonian obtained in its continuous-time limit. The vacuum is removed
using its actual ground state. The gap is measured on the whole remaining
physical space, unless a smaller subspace is explicitly specified.
Markov chains used to sample a lattice measure, auxiliary Wilson flow,
and physical time are distinguished throughout.

\subsection{Models, normalization and units}

For Euclidean Wilson theory on a finite four-dimensional lattice, write
\[
 F_p(U)=\frac12\operatorname{Tr}U_p,\qquad
 W(U)=\sum_p(1-F_p(U)),\qquad
 d\mu_{\beta_W}=Z^{-1}e^{-\beta_W W(U)}\prod_e dU_e.
\]
Haar measure is normalized, the links take values in $SU(2)$, and
$\beta_W=4/g_0^2$ is the conventional isotropic Wilson normalization
\cite{Wilson,emp:AthenodorouTeper2021}. The strong-coupling direction
is small $\beta_W$. The asymptotically free continuum direction is
weak bare coupling, with increasing $\beta_W$; a theorem on a bounded
strong-coupling interval does not control that direction.

On a periodic cubic spatial lattice with spacing $a_s$, the Hamiltonian
convention used in the spectral sections is
\begin{equation}\label{intro:H}
 H_{a_s,L,s}=\frac{s}{2a_s}\sum_e\mathcal C_e
       +\frac4{s a_s}\sum_p(1-F_p),\qquad s=g^2>0,
\end{equation}
on $\mathcal H_{\rm phys}=L^2(SU(2)^{E_s},d\mu)^{\mathcal G}$.
Here $\mathcal C_e$ has eigenvalues $j(j+1)$ and the second sum is
over spatial plaquettes. The normalization is the temporal Wilson limit
of the Kogut--Susskind construction \cite{ham:kogut_susskind,ham:creutz}.
At fixed $a_s,s$, the anisotropic parameters obey
$\beta_t=4a_s/(s\epsilon)$ and $\beta_s=4\epsilon/(s a_s)$ as the
temporal step $\epsilon\downarrow0$. This limit removes the temporal
discretization while leaving the spatial cutoff in place.

\begin{center}\small
\begin{tabularx}{\textwidth}{@{}lX@{}}
\toprule
Symbol & Meaning and convention\\
\midrule
$a_s$, $\epsilon$ & Spatial spacing and physical temporal step; $a$ means the relevant spacing where locally specified.\\
$L$, $a_sL$ & Number of spatial sites per direction and physical side length.\\
$\beta_W$, $s$ & Isotropic Euclidean inverse coupling and Hamiltonian coupling $g^2$; their limiting trajectories are not silently identified.\\
$T$, $\lambda_0$ & Positive physical transfer operator and its leading eigenvalue.\\
$\delta$, $\Delta$ & Dimensionless transfer gap $-\log(\lambda_1/\lambda_0)$ and a Hamiltonian or physical energy gap.\\
$h$, $\nu$ & Positive normalized finite-volume ground state and its probability measure $d\nu=h^2d\mu$.\\
$J$ & Highest retained electric representation spin, when a cutoff is imposed.\\
$t$, $t_{\rm flow}$ & Physical time and auxiliary smoothing time, respectively. Monte Carlo sweeps are a third, algorithmic parameter.\\
\bottomrule
\end{tabularx}\end{center}
Local sections define additional notation. Distinct settings such as
two-dimensional gauge theory and scalar effective models are identified
before any theorem is applied to them.

\subsection{What is established}

The first step is exact gauge reduction and construction of the physical
transfer space. Compactness, positivity and the treatment of the vacuum
give genuine finite-spatial-volume gaps. Strong-coupling comparison then
supplies estimates independent of the number of lattice sites. The sharp
single-link quarter transport estimate and the perfect-forest
$\beta_W<1/3$ framework are attributed to Bhattacharjee and Bhattacharya
\cite{BB2026}; their arguments are included with the gauge and physical-sector
details needed here. They are not claimed as a new historical discovery
of this work. Classical transfer and reconstruction results are due to
L\"uscher, Osterwalder--Seiler and Osterwalder--Schrader
\cite{Luscher,OSLattice,OS1,OS2}. The relation to other strong-coupling
approaches is made explicit \cite{SZZ,RVH}.

The development then addresses quantities that these strong-coupling
theorems leave uncontrolled: the spatial Hamiltonian sequence, its ground
states, discarded representation sectors and collective spectral effects.
For every fixed spatial spacing and positive coupling, local history
correlations have subsequential thermodynamic limits. They reconstruct
strongly continuous physical-time dynamics with nonconstant local
gauge-invariant correlations. This is an infinite-volume construction
with a spatial lattice, not a simultaneous four-dimensional continuum
construction.

\begin{longtable}{@{}p{.25\textwidth}p{.43\textwidth}p{.25\textwidth}@{}}
\toprule
Result & Mathematical content & Scope\\\midrule
\endfirsthead
\toprule Result & Mathematical content & Scope\\\midrule\endhead
Physical strong-coupling theory & Attributed comparison bounds, a nonzero physical sector and a finite positive excitation threshold. & Fixed lattice spacing; uniform spatial growth in the stated coupling range.\\[4pt]
Exact finite-transfer benchmark & Full one-site Wilson-transfer enclosure, exact kinetic weighted-cycle spectrum and proved threefold center-sector multiplicity. & Finite spatial volume, all representations, infinite Euclidean time.\\[4pt]
Finite conditioning certificates & Complete series/interval bounds for a three-link functional through $\beta_W=0.354$; exact sensitivity and criterion ceilings. & Functional theorem unconditional on its finite domain; physical interpretation has explicit imported hypotheses.\\[4pt]
Full finite-box spectrum & Positive lower bounds for $H-E_0$ on the full physical vacuum complement; Schur and min--max certificates include omitted representations. & Fixed spatial box and coupling; the explicit lower-bound functions can decay under refinement.\\[4pt]
Thermodynamic dynamics & Gauge-invariant Euclidean and real-time correlations, a history Hilbert space, strongly continuous time evolution and surviving local fluctuations. & Infinite spatial volume at fixed $a_s,s$; uniqueness, clustering and a full gap are not inferred.\\[4pt]
Vacuum concentration & Extensive relative entropy, limits on global Haar comparison, and fidelity loss under insufficient electric cutoffs. & Quantitative bounds on the interacting finite-lattice vacuum, with specified uniformities.\\[4pt]
Local and collective estimates & Improved conditional comparison, complete discarded-sector floors, projection-angle and Schur-complement criteria. & A collective estimate must be established on the actual vacuum measure before it implies a full volume-uniform gap.\\[4pt]
Continuum passage & Dense-family, spectral-leakage and physical-time criteria that pass a positive bound to a constructed limit. & Sufficient implications; the required weak-coupling spatial construction is not supplied.\\[4pt]
Controlled comparisons & Exact blocking identities, a finite-box reference-loop trajectory, corrected effective-potential estimates, heavy-matter tails and exact two-dimensional benchmarks. & Each theorem retains its model and regulator assumptions.\\
\bottomrule
\end{longtable}

The one-site periodic Wilson transfer operator has, at $\beta_W=177/500$,
\[
 2.27102703117506\le
 \log\frac{\lambda_0}{\lambda_1}
 \le2.27225894375269.
\]
Section~\ref{sec:finite-transfer} proves the complete remainder control
and shows that the first excitation occupies exactly three single-odd
center sectors. It is a finite-volume winding-flux benchmark, distinct
from a center-neutral glueball mass. Its exact rational calculation is
reproduced from integer Haar-moment polynomials.

The explicit global finite-box bound in
Theorem~\ref{ham:thm:heat:global}, for $0<s\le1$, is
\begin{equation}\label{intro:globalbound}
 \Delta_{a_s,L,s}\ge
 \frac{3s}{8a_s}\left(\frac\pi{16}\right)^{3L^3}
              \exp\!\left(-\frac{48\pi L^3}{s}\right)>0.
\end{equation}
It applies to the physical vacuum complement and has no representation
cutoff. Its explicit dependence on $L$ is also an obstruction: a positive
number at each finite box is not a positive uniform physical lower bound
along spatial refinement.

The sharper local comparison proved in
Theorem~\ref{ham:local_refinement_theorem} is
\[
 \frac{\max_{U_e}h}{\min_{U_e}h}
 \le\frac4{\sqrt\pi}e^{16\pi/s},\qquad
 \gamma_{\rm loc}(s)=\frac{3\pi}{64}e^{-32\pi/s}.
\]
For a fixed set of $q$ deleted links, Corollary~\ref{ham:local_refinement_discarded}
gives the complete conditional mean-zero sector the floor
\[
 \frac{3s}{8a_s}\left(\frac\pi{16}\right)^q e^{-32\pi q/s}.
\]
This estimate is independent of the total spatial volume for fixed $q$.
It is an input to a collective argument, rather than a substitute for one.

The vacuum-fidelity results make a complementary requirement quantitative.
For the specified hard cutoff, $g\downarrow0$ and $gJ\to0$ force the best
vacuum fidelity in the retained subspace to vanish. Thus a basis can
approximate some observables while failing to approximate the physical
vacuum in norm. Exact spectral tails and renormalized state maps must be
treated together.

Two benchmark refinements are isolated in this version: the preceding
local heat-kernel optimization, with outward rational error bounds, and
the exact fixed-representation blocking coefficient in
Proposition~\ref{traj:prop:representation-mismatch}. For the adjoint
representation its value is $25/16$. They are quantitative additions to
the research programme; no claim of priority over all prior literature
is made.

\subsection{The continuum conclusion and its remaining premises}

Let $A_k=T_k/\lambda_{0,k}$, let $a_{t,k}\downarrow0$ be the physical
temporal step and let $P_{0,k}=|\Omega_k\rangle\langle\Omega_k|$
project onto the normalized actual vacuum. A sufficient
lattice spectral statement is
\begin{equation}\label{intro:neededgap}
 0\le A_k\big|_{\operatorname{Ran}(I-P_{0,k})}\le e^{-\delta_k}I,
 \qquad \liminf_k\frac{\delta_k}{a_{t,k}}\ge M>0.
\end{equation}
It must accompany a nontrivial continuum construction: renormalized
physical states and local observables must have compatible limiting
norms, adjoints and physical-time correlations, and the limiting probe
family must be dense in the full physical vacuum complement. The spatial
cutoff must tend to zero and physical spatial volume must grow with these
controls intact.

Under these hypotheses, the correlation inequality passes to
$\langle\psi,e^{-tH}\psi\rangle\le e^{-Mt}\|\psi\|^2$.
The spectral theorem then excludes spectrum in $(0,M)$ and any additional
zero-energy vector in that complement. The detailed versions below
specify the embeddings, norm control, time rounding and density
requirements. They also exhibit counterexamples when those hypotheses
are omitted. The conclusion concerns a limiting theory only after its
construction and identification have been justified.

Lower-dimensional stochastic constructions and ultraviolet gauge-field
estimates provide relevant context, with their dimensions and conclusions
retained \cite{Balaban3,Balaban4,CCHS,CS3,CS2}. They are not inputs that
establish the missing four-dimensional spatial-cutoff limit. Likewise,
the empirical mass ratio in Section~\ref{sec:empirical} is evidence about
the measured scalar channel; it is not a theorem excluding every possible
low-energy physical state.

\subsection{Organization and reproducibility}

Parts I--II establish the gauge-invariant finite-cutoff setting and
strong-coupling consequences. Part III develops finite conditioning and
its complete certificate. Parts IV--V construct and analyse the spatial
Hamiltonian sequence, with the fixed-cutoff thermodynamic theory placed
before the collective gap and cutoff-obstruction results. Part VI states
the continuum transfer criteria and their missing hypotheses. Part VII
tests confinement and approximation methods in explicit comparison
settings. Part VIII presents attributable numerical evidence and the
resulting research targets.

Mathematical error bounds appear beside the calculations they certify.
Detailed executable programs, exact fractions, interval boxes, frozen
stochastic inputs and numerical records accompany the manuscript. Hashes
identify those records; arithmetic checks certify the quantities they
actually enclose. Neither is a replacement for an analytic proof of the
model assumptions. Proof assumptions are stated at the point of use; build and reproduction
instructions accompany the source and data.

\clearpage
\part{Gauge reduction and the physical lattice theory}
\hypertarget{gauge_foundations-exact-gauge-reduction-planar-solutions-and-the-first-curvature-constraint}{%
\section{Exact gauge reduction, planar solutions, and the first curvature constraint}\label{gauge_foundations-exact-gauge-reduction-planar-solutions-and-the-first-curvature-constraint}}

This section develops a finite-cutoff calculation far enough that its successful steps and its obstruction in higher dimension can both be inspected exactly. The calculation begins with the compact-group integral defining the Wilson model. It removes a spanning tree without changing any gauge-invariant expectation, solves the free-boundary rectangular model in two dimensions, evaluates the resulting single-plaquette integrals for \(SU(2)\), and then calculates the first failure of independent plaquettes on a cube. No continuum field integral is used as a formal intermediate object.

These constructions belong to established lattice gauge theory. Wilson introduced the compact lattice regulator \cite{Wilson}; the rigorous two-dimensional theory and its continuum holonomy construction are developed in L\'evy's work \cite{Levy2003}. The purpose of the proofs here is to make the exact algebra and its limitations available in the present research programme. In particular, the independent-plaquette calculation is not presented as a newly discovered four-dimensional reduction. The one-cube calculation gives an explicit reason why that proposed reduction would be false.

\hypertarget{gauge_foundations-finite-graphs-and-normalized-haar-integration}{%
\subsection{Finite graphs and normalized Haar integration}\label{gauge_foundations-finite-graphs-and-normalized-haar-integration}}

Let \(G\) be a compact Lie group and let \(h\) be its Haar probability measure. Thus \(h(G)=1\), left and right multiplication preserve \(h\), and inversion preserves \(h\). Let \(\Gamma=(V,E)\) be a finite connected graph. For each geometric edge choose one orientation. A configuration is a tuple \(U=(U_e)_{e\in E}\in G^E\). Reversing an edge means using \(U_{\bar e}=U_e^{-1}\); it does not introduce another integration variable.

If a path \(c\) traverses oriented edges \(e_1,\ldots,e_k\), its holonomy is the ordered product

\[U(c)=U_{e_1}\cdots U_{e_k}.\]

An immediately retraced edge cancels. Reversing the path inverts the product. These are algebraic identities, independent of a probability law. If \(c\) is a closed path, changing its starting vertex conjugates its holonomy by the holonomy of the intervening segment. A central function of the loop holonomy is consequently independent of the choice of starting vertex.

The vertex gauge group is \(\mathcal G_V=G^V\). We use the convention

\[U^g_{xy}=g_xU_{xy}g_y^{-1}.\]

For a path from \(x\) to \(y\), cancellation of the intermediate factors gives

\[U^g(c)=g_xU(c)g_y^{-1}.\]

For a closed path based at \(x\), a trace in any finite-dimensional representation is therefore unchanged. Let \(\rho:G\to U(N)\) be a continuous unitary representation, and set

\[w(g)=\frac1N\operatorname{Re}\operatorname{Tr}\rho(g).\]

Then \(w\) is central, \(w(g^{-1})=w(g)\), and \(|w(g)|\le1\). For a finite family of closed face paths \(\mathcal F\), the free-boundary Wilson measure is

\[d\mu_\beta(U)=Z_\Gamma(\beta)^{-1}
\exp\left(\beta\sum_{p\in\mathcal F}w(U(p))\right)
\prod_{e\in E}h(dU_e),\qquad \beta\ge0.\]

The partition function is positive and finite. In fact, if \(F=|\mathcal F|\),

\[e^{-\beta F}\le Z_\Gamma(\beta)\le e^{\beta F}.\]

This elementary estimate establishes the finite integral, but it is not a useful volume-uniform estimate: its exponent grows with the number of faces. A proof that silently carries this constant into an infinite-volume limit has not controlled that limit.

Product Haar measure is gauge invariant. To verify this, fix \(g\in G^V\) and change one edge variable at a time by \(U_{xy}\mapsto g_xU_{xy}g_y^{-1}\). Bi-invariance of \(h\) preserves every factor. The Wilson density is also unchanged because each face holonomy is conjugated. Hence \(\mu_\beta\) is gauge invariant. This argument does not require choosing a gauge or dividing by a gauge-group volume; the latter is already one in the normalized convention.

The face family may contain repeated or composite loops if these are explicitly part of the model. In the rectangular and cubical examples below it consists of elementary squares, each included once. All statements use finite open graphs. Periodic boundary conditions introduce global cycles and are not covered by the independent-plaquette conclusion without additional variables and constraints.

\hypertarget{gauge_foundations-spanning-tree-reduction-with-its-measure-identity}{%
\subsection{Spanning-tree reduction with its measure identity}\label{gauge_foundations-spanning-tree-reduction-with-its-measure-identity}}

Choose a root vertex \(o\) and a spanning tree \(T\subset E\). For each vertex \(v\), let \(c_v\) be the unique tree path from \(o\) to \(v\), and define

\[t_o=1,\qquad t_v=U(c_v).\]

For every edge \(e=(x,y)\) outside the tree, define

\[X_e=t_xU_et_y^{-1}.\]

For a tree edge this expression equals the identity, with the same conclusion if its chosen orientation is opposite to its orientation away from the root. Thus the gauge transformation \(g_v=t_v\) sets all tree links equal to one. The gauge transformation is a function of the configuration, so measure preservation should not be asserted merely from invariance under fixed transformations. The required stronger assertion is the next change-of-variables theorem.

\textbf{Theorem G1 (exact spanning-tree factorization).} The map

\[U\longmapsto\big((t_v)_{v\ne o},(X_e)_{e\notin T}\big)\]

is a continuous bijection from \(G^E\) to \(G^{V\setminus\{o\}}\times G^{E\setminus T}\). It maps product Haar probability measure to product Haar probability measure. Equivalently, for every bounded Borel function \(\Phi\),

\[\int_{G^E}\Phi(U)\,dh^E(U)
=\int\Phi\big(U(t,X)\big)\,dh^{V\setminus\{o\}}(t)\,dh^{E\setminus T}(X),\]

where the inverse is

\[U_{xy}=t_x^{-1}X_{xy}t_y,\qquad X_e=1\text{ for }e\in T.\]

\textbf{Proof.} On an edge of the rooted tree joining a parent \(x\) to a child \(y\), the formula \(t_y=t_xU_{xy}\) determines \(t_y\) uniquely from the tree links. If the chosen edge orientation is reversed, the same equation uses the inverse link. Conversely, \(U_{xy}=t_x^{-1}t_y\) reconstructs every tree link. For a non-tree edge the displayed inverse reconstructs its link once \(t\) and \(X\) are given. These formulas prove bijectivity and continuity in both directions.

Order the non-root vertices so that parents precede children. Under independent Haar tree links, condition on the transport variables already generated. The next variable is a fixed left translate of an independent Haar variable, or of its inverse. Its conditional law is again \(h\) and does not depend on the preceding transports. Induction shows that all \(t_v\), \(v\ne o\), are independent Haar variables.

Conditional on all tree links, the non-tree links remain independent Haar variables. Each change \(U_{xy}\mapsto t_xU_{xy}t_y^{-1}\) preserves Haar measure. Consequently the conditional law of the whole tuple \(X\) is \(h^{E\setminus T}\), independent of \(t\). This proves the measure identity first for rectangles in the target product sigma-field and then for bounded Borel functions by the monotone-class theorem. \(\square\)

Let \(\widetilde U(X)\) be the configuration with tree links one and remaining links \(X\). If \(F\) is gauge invariant, then \(F(U)=F(\widetilde U(X))\). The same is true of the Wilson action. Integrating the tree transports therefore yields the exact reduced formula

\[\mu_\beta(F)=\frac{
\int_{G^{E\setminus T}}F(\widetilde U(X))
\exp\big(\beta\sum_pw(\widetilde U(X)(p))\big)\,dh(X)}{
\int_{G^{E\setminus T}}
\exp\big(\beta\sum_pw(\widetilde U(X)(p))\big)\,dh(X)}.\]

There are \(|E|-|V|+1\) remaining group variables. This is the cycle rank of the connected graph. The reduced variables are not automatically plaquettes. What remains to be proved in each geometry is whether the chosen face holonomies give independent coordinates on these variables.

There is still a global conjugation symmetry. To see it directly, apply an arbitrary gauge transformation \(g\) before tree reduction. The new tree transport is

\[t'_v=g_o t_v g_v^{-1},\]

and hence

\[X'_e=g_oX_eg_o^{-1}.\]

Thus the gauge-orbit space of the original graph is naturally identified with the simultaneous-conjugation orbit space of the reduced tuple. The tree has removed vertex-dependent redundancy but has not selected a preferred representative of a conjugacy orbit. No Faddeev--Popov determinant is missing: the full identity is a finite product-Haar change of variables, not a formal differential gauge-fixing calculation.

\hypertarget{gauge_foundations-a-dense-class-of-finite-cutoff-gauge-observables}{%
\subsection{A dense class of finite-cutoff gauge observables}\label{gauge_foundations-a-dense-class-of-finite-cutoff-gauge-observables}}

The phrase ``a dense class of observables'' needs a specified norm and measure. On a fixed finite graph those questions have a precise elementary answer. They should not be conflated with density of renormalized physical states after removing the cutoff.

Assume that \(G\) is represented faithfully as a compact matrix group. Let \(\mathcal P\) be the algebra of finite complex polynomials in all matrix entries of the edge variables and their complex conjugates. It contains constants, is closed under complex conjugation, and separates points of the compact space \(G^E\). The complex Stone--Weierstrass theorem implies that \(\mathcal P\) is uniformly dense in \(C(G^E)\).

For a continuous function define gauge averaging by

\[\mathsf PF(U)=\int_{G^V}F(U^g)\,dh^V(g).\]

\textbf{Proposition G2 (finite-cutoff gauge projection and density).} For any gauge-invariant Borel probability measure \(\mu\) on \(G^E\), the operator \(\mathsf P\) extends to an orthogonal projection in \(L^2(\mu)\) onto the gauge-invariant subspace. The averaged polynomial functions \(\mathsf P\mathcal P\) are uniformly dense in the continuous invariant functions and dense in the invariant subspace of \(L^2(\mu)\).

\textbf{Proof.} Jensen's inequality and invariance of \(\mu\) give

\[\|\mathsf PF\|_{L^2(\mu)}^2
\le\int\int|F(U^g)|^2\,dh^V(g)\,d\mu(U)
=\|F\|_{L^2(\mu)}^2.\]

Haar invariance gives \(\mathsf P^2=\mathsf P\). For continuous \(F,H\), change variables \(U\mapsto U^{g^{-1}}\) and then \(g\mapsto g^{-1}\) to obtain

\[\langle F,\mathsf PH\rangle_{L^2(\mu)}
=\langle\mathsf PF,H\rangle_{L^2(\mu)}.\]

Thus the bounded extension is a self-adjoint idempotent and is an orthogonal projection. Its range consists exactly of the vectors fixed by all unitary gauge transformations: averaging produces fixed vectors, and a fixed vector is unchanged by the average.

If \(F\) is continuous and invariant, choose \(p_n\in\mathcal P\) with \(\|p_n-F\|_\infty\to0\). Gauge averaging is a contraction in the supremum norm, so \(\|\mathsf Pp_n-F\|_\infty\to0\). Averaging a finite polynomial still produces a finite polynomial in the edge entries: its finitely many coefficient functions in the gauge variables are simply integrated. This proves the uniform assertion.

Finally, continuous functions are dense in \(L^2\) of a finite regular Borel measure on a compact metric space. Given an invariant vector, approximate it by continuous functions and apply the \(L^2\) contraction \(\mathsf P\). Approximate each resulting continuous invariant function by averaged polynomials. The triangle inequality proves the \(L^2\) assertion. \(\square\)

This construction is a direct route to spin-network-type observables without claiming that traces in one chosen representation suffice for every compact group. Completeness results specifically for Wilson-loop algebras require their own group hypotheses; L\'evy's work proves such results for the stated families of classical compact groups \cite{Levy2004}. The averaged-polynomial argument above is enough for the finite-cutoff statement used here.

Two additional steps would be needed to use this density in a continuum mass-gap proof. First, these observables must be renormalized with limits in the appropriate reflected inner product. Second, their limiting classes must remain total. Neither follows from uniform approximation at one fixed cutoff. Ordinary variance and reflected physical norm are also different quantities. For independent identically distributed variables \(X_+,X_-\) exchanged by reflection, a centered function \(f(X_+)\) can have positive ordinary variance but reflected norm \(\mathbb E[f(X_-)f(X_+)]=0\). The finite-cutoff projection theorem does not rule out that loss of physical states.

\hypertarget{gauge_foundations-the-free-boundary-rectangular-model-is-exactly-factorized}{%
\subsection{The free-boundary rectangular model is exactly factorized}\label{gauge_foundations-the-free-boundary-rectangular-model-is-exactly-factorized}}

Take the square graph with vertices \((i,j)\), \(0\le i\le L\), \(0\le j\le H\), and all nearest-neighbor edges inside it. Orient horizontal edges to the right and vertical edges upward. There are \(LH\) elementary plaquettes. Use as spanning tree every vertical edge together with the \(L\) horizontal edges on the bottom row. This set is connected and has

\[H(L+1)+L=(L+1)(H+1)-1\]

edges, so it is indeed a tree.

After tree reduction, every vertical link and bottom horizontal link is one. Denote the remaining horizontal link from \((i-1,j)\) to \((i,j)\) by \(X_{i,j}\), for \(1\le i\le L\) and \(1\le j\le H\), and set \(X_{i,0}=1\). Orient every plaquette counterclockwise starting at its bottom-left corner. Its reduced holonomy is

\[P_{i,j}=X_{i,j-1}X_{i,j}^{-1}.\]

\textbf{Theorem G3 (planar Haar factorization).} The map from the \(LH\) variables \(X_{i,j}\) to the \(LH\) variables \(P_{i,j}\) is a Haar-measure-preserving bijection. For any positive central plaquette weights \(v_{i,j}\) with finite integrals

\[z_{i,j}=\int_Gv_{i,j}(g)\,dh(g),\]

the reduced plaquette variables under the free-boundary density \(\prod_{i,j}v_{i,j}(P_{i,j})\) are independent, with laws

\[\nu_{i,j}(dg)=z_{i,j}^{-1}v_{i,j}(g)\,dh(g).\]

The original link partition function is exactly \(\prod_{i,j}z_{i,j}\).

\textbf{Proof.} In each column, the inverse recursion is

\[X_{i,j}=P_{i,j}^{-1}X_{i,j-1}.\]

Starting from \(X_{i,0}=1\), this gives a unique inverse. Under product Haar measure for the \(X\) variables, condition successively on rows below \(j\). Given \(X_{i,j-1}\), the new variable \(P_{i,j}=X_{i,j-1}X_{i,j}^{-1}\) is Haar and its conditional distribution is independent of the previous variables. Different columns use disjoint variables. Induction proves that all \(P_{i,j}\) are independent Haar variables. Multiplying this product law by the product of the individual weights and normalizing gives the stated independent laws. The tree variables contribute a factor one by Theorem G1. \(\square\)

For the homogeneous Wilson weight \(v_\beta(g)=e^{\beta w(g)}\), write

\[z(\beta)=\int_Ge^{\beta w(g)}\,dh(g),\qquad
\nu_\beta(dg)=z(\beta)^{-1}e^{\beta w(g)}\,dh(g).\]

Then

\[Z_{L,H}(\beta)=z(\beta)^{LH}.\]

The factorization is exact at every \(\beta\ge0\), with no small-coupling expansion. It implies, for example, that centered central functions of disjoint individual plaquettes are independent in this gauge-invariant sense. Differentiating the finite integral gives

\[\frac{1}{LH}\frac{d}{d\beta}\log Z_{L,H}(\beta)
=\mathbb E_{\nu_\beta}w,\]

and

\[\frac{1}{LH}\frac{d^2}{d\beta^2}\log Z_{L,H}(\beta)
=\operatorname{Var}_{\nu_\beta}(w).\]

Boundedness of \(w\) justifies differentiation under the integral for every finite real \(\beta\). These formulas identify exactly which random variable is being averaged. They are not formulas for the four-dimensional plaquette expectation, where adjacent faces retain additional interactions after tree reduction.

\hypertarget{gauge_foundations-rectangular-wilson-loops-and-central-convolution}{%
\subsection{Rectangular Wilson loops and central convolution}\label{gauge_foundations-rectangular-wilson-loops-and-central-convolution}}

We now prove the loop law without discarding the noncommutative conjugations. Let \(C\) bound a rectangle of \(W\) columns and \(K\) rows inside the larger free-boundary rectangle. Relabel its columns \(i=1,\ldots,W\) and its bottom and top rows \(a,b\), where \(b-a=K\). Let

\[Q_i=P_{i,a+1}\cdots P_{i,b},\qquad
B_i=X_{1,a}\cdots X_{i,a},\qquad B_0=1.\]

The recursion in Theorem G3 gives

\[X_{i,b}=Q_i^{-1}X_{i,a}.\]

The loop holonomy is the bottom horizontal product followed by the inverse top horizontal product, since all vertical links equal one. Substitution and cancellation yield

\[U_C=\prod_{i=W,W-1,\ldots,1}
B_{i-1}Q_iB_{i-1}^{-1}.\]

The displayed order is descending. This identity is generally not the un-conjugated ordered product of the enclosed plaquettes.

Conditional on the bottom horizontal links, the \(Q_i\) are independent products of their \(K\) inner plaquettes. Each has law \(\nu_\beta^{*K}\), where \(*\) denotes group convolution. This law is central because each factor is central. The bottom links depend only on plaquettes below the rectangle and are independent of the inner ones. Consequently conjugation by the now-fixed \(B_{i-1}\) preserves each conditional law and preserves independence. The resulting loop holonomy therefore has law \(\nu_\beta^{*(WK)}\).

\textbf{Theorem G4 (exact rectangular character law).} Let \(\pi\) be an irreducible unitary representation of dimension \(d_\pi\), with character \(\chi_\pi\). Define

\[a_\pi(\beta)=\frac1{d_\pi}\int_G\chi_\pi(g)\,\nu_\beta(dg).\]

For a rectangular loop enclosing \(A\) plaquettes,

\[\mathbb E\left[\frac{\chi_\pi(U_C)}{d_\pi}\right]
=a_\pi(\beta)^A.\]

\textbf{Proof.} Centrality shows that the matrix \(M_\pi=\int\pi(g)\,d\nu_\beta(g)\) commutes with every \(\pi(k)\). Schur's lemma gives \(M_\pi=a_\pi I\), with the coefficient obtained by taking traces. For independent increments \(Y_1,\ldots,Y_A\) with law \(\nu_\beta\), repeated integration gives

\[\mathbb E\pi(Y_1\cdots Y_A)=M_\pi^A=a_\pi^AI.\]

The preceding holonomy calculation identifies this product law with that of \(U_C\). Taking the normalized trace proves the formula. \(\square\)

For a nontrivial representation, \(|a_\pi(\beta)|<1\) at every finite \(\beta\). Indeed, the Wilson density is strictly positive, so it has full Haar support. If a unit vector \(v\) had \(\|\int\pi(g)v\,d\nu(g)\|=1\), equality in the Hilbert-space averaging inequality would force \(\pi(g)v\) to be a single constant vector almost everywhere. Continuity and full support extend this to every \(g\), and evaluating at the identity makes \(v\) an invariant vector. An irreducible nontrivial representation has no such vector. Since \(M_\pi\) is scalar, the claimed strict inequality follows.

Inversion symmetry of \(\nu_\beta\) makes \(a_\pi\) real. It does not by itself prove \(a_\pi>0\) for arbitrary weights. Positivity will be proved explicitly for the \(SU(2)\) Wilson coefficients below. When \(0<a_\pi<1\), the exact lattice string-tension parameter is

\[\sigma_\pi^{\rm lat}(\beta)=-\log a_\pi(\beta),\]

and the loop law is \(\exp[-\sigma_\pi^{\rm lat}A]\). The area here is an integer number of plaquettes. Converting it to physical area introduces the factor \(a^{-2}\), where \(a\) is lattice spacing. This distinction is indispensable in the scaling calculation.

\hypertarget{gauge_foundations-the-su2-single-plaquette-integral}{%
\subsection{\texorpdfstring{The \(SU(2)\) single-plaquette integral}{The SU(2) single-plaquette integral}}\label{gauge_foundations-the-su2-single-plaquette-integral}}

Every element of \(SU(2)\) can be written

\[U=\cos\theta\,I+i\sin\theta\,\mathbf n\cdot\boldsymbol\sigma,
\qquad 0\le\theta\le\pi,\quad\mathbf n\in S^2.\]

This identifies the group with the unit three-sphere. Surface measure on that sphere is proportional to \(\sin^2\theta\,d\theta\,d\omega(\mathbf n)\). Its normalization gives, for every central continuous \(f\),

\[\int_{SU(2)}f(U)\,dh(U)
=\frac2\pi\int_0^\pi f(\theta)\sin^2\theta\,d\theta.\]

For the fundamental representation, \(w(U)=\frac12\operatorname{Tr}U=\cos\theta\). The spin-\(j\) representation, \(j=0,\frac12,1,\ldots\), has dimension \(n=2j+1\) and character

\[\chi_j(\theta)=\frac{\sin(n\theta)}{\sin\theta}.\]

The quotient extends continuously to the endpoints. One way to verify it is to sum the weights \(e^{2im\theta}\) for \(m=-j,-j+1,\ldots,j\), using a finite geometric series.

For integers \(n\ge0\), define

\[I_n(\beta)=\frac1\pi\int_0^\pi e^{\beta\cos\theta}\cos(n\theta)\,d\theta.\]

These are the modified Bessel functions in the convention of the NIST Digital Library of Mathematical Functions \cite{DLMF}. The formulas needed here follow directly from the definition. On the unit circle, expand

\[e^{\beta\cos\theta}
=e^{(\beta/2)e^{i\theta}}e^{(\beta/2)e^{-i\theta}}.\]

Both exponential series converge absolutely and uniformly in \(\theta\) for fixed \(\beta\). Extracting the \(n\)th Fourier coefficient therefore gives

\[I_n(\beta)=\sum_{k=0}^\infty
\frac{(\beta/2)^{2k+n}}{k!(k+n)!}.\]

This expression is positive when \(\beta>0\). Integration of the derivative of \(e^{\beta\cos\theta}\sin(n\theta)\) over \([0,\pi]\) gives, for \(n\ge1\),

\[I_{n-1}(\beta)-I_{n+1}(\beta)=\frac{2n}{\beta}I_n(\beta).\]

The boundary term vanishes because \(\sin(n\theta)\) vanishes at both endpoints. The product-to-sum identity for \(\sin\theta\sin(n\theta)\) yields the recurrence with no exchange of divergent series.

\textbf{Theorem G5 (all normalized \(SU(2)\) character means).} For \(\beta>0\),

\[z(\beta)=\frac{2I_1(\beta)}{\beta},\qquad
a_j(\beta)=\frac{I_{2j+1}(\beta)}{I_1(\beta)}.\]

For every \(j>0\), \(0<a_j(\beta)<1\).

\textbf{Proof.} Using \(2\sin^2\theta=1-\cos(2\theta)\),

\[z(\beta)=I_0(\beta)-I_2(\beta)
=\frac{2I_1(\beta)}{\beta}.\]

For \(n=2j+1\),

\[\int\chi_j(U)e^{\beta w(U)}\,dh(U)
=\frac2\pi\int_0^\pi e^{\beta\cos\theta}\sin\theta\sin(n\theta)\,d\theta
=I_{n-1}(\beta)-I_{n+1}(\beta)
=\frac{2nI_n(\beta)}{\beta}.\]

Divide by \(n z(\beta)\) to obtain the normalized mean. Positivity follows from the positive series. Strictness of the upper bound follows from the full-support argument after Theorem G4. \(\square\)

At \(\beta=0\), \(z(0)=1\), the trivial-representation coefficient is one, and every nontrivial coefficient is zero. These values follow either from Haar orthogonality or from the power series. Treating the apparent factors \(1/\beta\) by these limits avoids a removable-singularity error.

The planar rectangular Wilson loop is now fully explicit:

\[\mathbb E\left[\frac{\chi_j(U_C)}{2j+1}\right]
=\left[\frac{I_{2j+1}(\beta)}{I_1(\beta)}\right]^A.\]

This exact formula applies to every finite rectangular area and every positive coupling in the stated two-dimensional free-boundary model. It is a useful normalization check for more complicated calculations because it fixes the group dimension, character normalization and action coefficient simultaneously.

\hypertarget{gauge_foundations-mean-susceptibility-and-the-small-coupling-expansion}{%
\subsection{Mean, susceptibility and the small-coupling expansion}\label{gauge_foundations-mean-susceptibility-and-the-small-coupling-expansion}}

Write \(m(\beta)=a_{1/2}(\beta)=I_2(\beta)/I_1(\beta)\). Differentiating the partition integral gives

\[m(\beta)=\frac{d}{d\beta}\log z(\beta),\qquad
m'(\beta)=\operatorname{Var}_{\nu_\beta}(\cos\theta)>0.\]

The variance is strictly positive because the density is positive and \(\cos\theta\) is not constant. Thus the single-plaquette mean increases strictly from its Haar value zero toward one. The derivative is an exact susceptibility, not an assumed monotonicity relation for a different lattice model.

Differentiation of the Bessel integral gives \(I'_n=(I_{n-1}+I_{n+1})/2\). Combining this with the recurrence gives \(I'_n=I_{n+1}+nI_n/\beta\). Therefore

\[m'=\frac{I_3}{I_1}+\frac{m}{\beta}-m^2
=1-\frac{3m}{\beta}-m^2.\]

The second equality uses \(I_1-I_3=4I_2/\beta\). All functions on the right are evaluated at \(\beta\). The apparent singularity at zero is removable: \(m(\beta)/\beta\to1/4\), giving \(m'(0)=1/4\), the Haar variance of the fundamental normalized trace.

The positive series gives

\[I_1(\beta)=\frac\beta2+\frac{\beta^3}{16}
+\frac{\beta^5}{384}+\frac{\beta^7}{18432}+O(\beta^9),\]

\[I_2(\beta)=\frac{\beta^2}{8}+\frac{\beta^4}{96}
+\frac{\beta^6}{3072}+O(\beta^8).\]

Since \(I_1(\beta)/(\beta/2)\) is analytic and equals one at zero, analytic division is legitimate in a neighborhood of zero. It yields

\[m(\beta)=\frac\beta4-\frac{\beta^3}{96}
+\frac{\beta^5}{1536}+O(\beta^7),\]

\[m'(\beta)=\frac14-\frac{\beta^2}{32}
+\frac{5\beta^4}{1536}+O(\beta^6).\]

For a fixed spin \(j\), the leading power is

\[a_j(\beta)=\frac{(\beta/2)^{2j}}{(2j+1)!}
\bigl(1+O(\beta^2)\bigr).\]

For the fundamental loop the corresponding lattice area coefficient is

\[-\log m(\beta)
=\log\frac4\beta+\frac{\beta^2}{24}
-\frac{\beta^4}{576}+O(\beta^6).\]

The remainders here concern \(\beta\to0\) with the representation fixed. They do not assert uniformity as the spin grows. The numerical certificates in the next section avoid reliance on these asymptotic remainders by bounding the entire positive series.

\hypertarget{gauge_foundations-exact-rational-numerical-certificates}{%
\subsection{Exact-rational numerical certificates}\label{gauge_foundations-exact-rational-numerical-certificates}}

Numerical values are useful for checking conventions and for testing proposed inequalities, provided the proof and the numerical role remain distinct. Here the table is supported by a rational interval calculation with an explicit infinite-tail bound. No Monte Carlo sampling or fitted error bar is involved.

For rational \(\beta>0\) and integer \(n\ge0\), set

\[t_k=\frac{(\beta/2)^{2k+n}}{k!(k+n)!},\qquad
r_k=\frac{t_{k+1}}{t_k}
=\frac{\beta^2}{4(k+1)(k+n+1)}.\]

The ratios decrease with \(k\). If \(r_K<1\), then

\[S_K=\sum_{k=0}^Kt_k\le I_n(\beta)
\le S_K+\frac{t_Kr_K}{1-r_K}.\]

Indeed, every ratio after \(K\) is at most \(r_K\), so the omitted positive terms are bounded by \(t_K\sum_{j\ge1}r_K^j\). Every number in these two endpoints is rational. Division of positive intervals gives

\[\frac{L_2}{U_1}\le m(\beta)\le\frac{U_2}{L_1}\]

when \(L_n\le I_n\le U_n\). Since \(f_\beta(x)=1-3x/\beta-x^2\) decreases for \(x\ge0\), the variance satisfies

\[f_\beta\left(\frac{U_2}{L_1}\right)
\le m'(\beta)
\le f_\beta\left(\frac{L_2}{U_1}\right).\]

The exact-calculation appendix gives the rational enclosures for means, variances and selected planar-loop expectations, with outward-rounded decimal endpoints and complete tail bounds.

The proof of each enclosure is the decreasing-ratio argument above. The numerical records check finitely many values of formulas already proved for all positive couplings. They are useful for finding an incorrect action convention or a factor of two; they are not used to extrapolate a four-dimensional conclusion. In particular, increasing the number of table entries would not replace any of the continuum hypotheses identified elsewhere in the manuscript.

\hypertarget{gauge_foundations-a-continuum-area-limit-proved-directly-by-laplace-scaling}{%
\subsection{A continuum-area limit proved directly by Laplace scaling}\label{gauge_foundations-a-continuum-area-limit-proved-directly-by-laplace-scaling}}

The exact planar model supplies a controlled example of how a nontrivial continuum observable can survive: the plaquette law concentrates near the identity while the number of plaquettes in a fixed physical area tends to infinity. The coupling must vary with the lattice spacing. Keeping a finite \(\beta\) fixed would instead make every nontrivial loop coefficient tend to zero as its lattice area diverges.

\textbf{Lemma G6 (small-angle moments).} Under the single-plaquette \(SU(2)\) law \(\nu_\beta\), for every integer \(k\ge0\),

\[\beta^k\mathbb E_{\nu_\beta}[\theta^{2k}]
=(2k+1)!!+O_k(\beta^{-1})\qquad(\beta\to\infty),\]

where \((-1)!!=1\) and the displayed formula gives \(1,3,15,\ldots\) for \(k=0,1,2,\ldots\).

\textbf{Proof.} Constants and the factor \(e^\beta\) cancel in the moment ratio. Define

\[J_k(\beta)=\int_0^\pi\theta^{2k}
e^{-\beta(1-\cos\theta)}\sin^2\theta\,d\theta.\]

With \(u=\sqrt\beta\,\theta\),

\[\beta^{k+3/2}J_k(\beta)
=\int_0^{\pi\sqrt\beta}u^{2k}
\beta\sin^2(u/\sqrt\beta)
e^{-\beta(1-\cos(u/\sqrt\beta))}\,du.\]

For \(0\le x\le\pi\), \(\sin x\le x\) and \(1-\cos x\ge2x^2/\pi^2\). Extend the integrand by zero beyond its upper endpoint. It is bounded by the integrable function \(u^{2k+2}e^{-2u^2/\pi^2}\), independently of \(\beta\). Its pointwise limit is \(u^{2k+2}e^{-u^2/2}\). Dominated convergence therefore gives

\[\beta^{k+3/2}J_k(\beta)
\longrightarrow\int_0^\infty u^{2k+2}e^{-u^2/2}\,du.\]

We also need the stated error bound. On \(0\le u\le\sqrt\beta\), put \(x=u/\sqrt\beta\le1\). Taylor's inequalities give

\[|\beta\sin^2x-u^2|\le\frac{u^4}{3\beta},\qquad
0\le\delta:=\frac{u^2}{2}-\beta(1-\cos x)
\le\frac{u^4}{24\beta}\le\frac{u^2}{24}.\]

Using \(e^\delta-1\le\delta e^\delta\), the difference between the scaled integrand and \(u^{2k+2}e^{-u^2/2}\) is bounded by

\[\frac1\beta
\left(\frac{u^{2k+4}}3+\frac{u^{2k+6}}{24}\right)
e^{-11u^2/24}.\]

Its integral is at most \(C_k/\beta\). For \(u\ge\sqrt\beta\), both integrands are bounded by a constant times \(u^{2k+2}e^{-c u^2}\) with a fixed \(c>0\). The tail integral is at most

\[e^{-c\beta/2}\int_0^\infty u^{2k+2}e^{-c u^2/2}\,du
=O_k(\beta^{-1}),\]

because \(\sup_{x>0}x e^{-cx/2}<\infty\). These estimates include the Gaussian tail beyond \(\pi\sqrt\beta\), where the original integrand is zero. Therefore

\[\beta^{k+3/2}J_k(\beta)=A_k+O_k(\beta^{-1}),\qquad
A_k=\int_0^\infty u^{2k+2}e^{-u^2/2}\,du.\]

Since \(A_0>0\), division by the \(k=0\) expression is legitimate for sufficiently large \(\beta\) and preserves the \(O_k(\beta^{-1})\) error. Integration by parts gives \(A_k/A_0=(2k+1)!!\). This proves the quantitative moment statement. \(\square\)

For fixed \(n=2j+1\), Taylor expansion at zero gives

\[\frac{\sin(n\theta)}{n\sin\theta}
=1-\frac{n^2-1}{6}\theta^2+O_j(\theta^4).\]

The remainder can be bounded by \(C_j\theta^4\) on the whole interval \([0,\pi]\): near zero this follows from analyticity, while away from zero the continuous character quotient and the other terms are bounded and \(\theta^4\) is bounded below. Taking expectation and using Lemma G6 for \(k=1,2\) gives

\[a_j(\beta)=1-\frac{n^2-1}{2\beta}+O_j(\beta^{-2})
=1-\frac{2j(j+1)}{\beta}+O_j(\beta^{-2}).\]

\textbf{Theorem G7 (exact planar continuum-area scaling).} Let the lattice spacing tend to zero through \(a>0\), let \(\beta(a)a^2\to b\) with \(b>0\), and let rectangular loops contain \(N_a\) plaquettes with \(a^2N_a\to\mathcal A\in[0,\infty)\). For each fixed spin \(j\),

\[\mathbb E\left[\frac{\chi_j(U_{C_a})}{2j+1}\right]
\longrightarrow
\exp\left[-\frac{2j(j+1)}{b}\mathcal A\right].\]

\textbf{Proof.} For \(j=0\) both sides equal one. For \(j>0\), the coefficients are positive, and the preceding expansion gives

\[\log a_j(\beta(a))
=-\frac{2j(j+1)}{\beta(a)}+O_j(\beta(a)^{-2})
=-\frac{2j(j+1)}b a^2+o(a^2).\]

Multiply by \(N_a\) and use \(a^2N_a\to\mathcal A\). The exact loop formula from Theorem G4 then gives the result after exponentiation. \(\square\)

With the conventional two-dimensional parametrization \(b=4/g^2\), the exponent is \(-g^2j(j+1)\mathcal A/2\). This identification is a convention for this two-dimensional scaling and should not be inserted unchanged into four-dimensional dimensional analysis. In particular, the theorem proves convergence of the displayed loop expectations; it does not by itself construct all continuum fields or their physical Hilbert space. A full two-dimensional holonomy construction requires compatibility across graphs and paths, as in the cited literature.

The calculation nevertheless identifies a genuine mechanism absent from a fixed-strong-coupling continuum shortcut: a single plaquette becomes almost trivial at precisely the scale at which a macroscopic loop contains many plaquettes. The limiting observable is neither forced to zero nor forced to one. In four dimensions, proving the analogous balance must also handle interactions between the plaquettes, which are already visible on the smallest cube.

\hypertarget{gauge_foundations-six-cube-faces-are-constrained-by-five-cycle-variables}{%
\subsection{Six cube faces are constrained by five cycle variables}\label{gauge_foundations-six-cube-faces-are-constrained-by-five-cycle-variables}}

Consider the graph formed by the twelve edges of the unit cube with vertices \((x,y,z)\in\{0,1\}^3\). Orient each edge in its positive coordinate direction. Choose as a tree all four vertical \(z\)-edges, the two bottom \(x\)-edges, and the bottom \(y\)-edge at \(x=0\). The tree has seven edges. After setting its variables to one, write the five remaining links as follows:

\begin{itemize}
\tightlist
\item
  \(a\): the bottom \(y\)-edge at \(x=1\);
\item
  \(b,c\): the top \(x\)-edges at \(y=0,1\), respectively;
\item
  \(d,e\): the top \(y\)-edges at \(x=0,1\), respectively.
\end{itemize}

For a face in the \(xy\) plane orient its path by positive \(x\) then positive \(y\), followed by the negative directions. Use the corresponding increasing-coordinate convention for \(xz\) and \(yz\) faces. In the tree gauge, direct multiplication gives

\[P_{z0}=a,\qquad P_{z1}=bec^{-1}d^{-1},\]

\[P_{y0}=b^{-1},\qquad P_{y1}=c^{-1},\]

\[P_{x0}=d^{-1},\qquad P_{x1}=ae^{-1}.\]

Before gauge fixing, interpret these as holonomies transported to the root along the chosen tree. This is essential in the nonabelian case: face holonomies at different basepoints cannot be multiplied as though they already lived in a common gauge frame.

\textbf{Theorem G8 (explicit one-cube curvature relation).} The six based face holonomies satisfy

\[P_{z1}=P_{y0}^{-1}P_{x1}^{-1}P_{z0}P_{y1}P_{x0}.\]

The five variables on the right are independent Haar variables under the original product Haar law. The sixth is determined by them and cannot be an additional independent nondegenerate Haar variable.

\textbf{Proof.} The first five displayed face formulas can be inverted as

\[a=P_{z0},\quad b=P_{y0}^{-1},\quad c=P_{y1}^{-1},
\quad d=P_{x0}^{-1},\quad e=P_{x1}^{-1}P_{z0}.\]

Insert these into \(bec^{-1}d^{-1}\) to obtain the relation. Theorem G1 makes \(a,b,c,d,e\) independent Haar variables. Inversion preserves Haar measure, while conditional on \(a\), the variable \(ae^{-1}\) is Haar and independent of \(a,b,c,d\). Thus the five face coordinates in the theorem are also independent Haar variables. The sixth is their displayed deterministic function. \(\square\)

This is an exact finite-dimensional obstruction to treating all plaquettes as independent. It does not arise from a divergent perturbation series or an uncertain continuum limit. A four-dimensional hypercubic lattice contains such three-dimensional cubes. Any construction replacing every based plaquette by an independent group variable has already changed the model on those subcomplexes unless it imposes the associated constraints and their interactions.

The dimension count provides a compatible check. The cube has cycle rank \(12-8+1=5\) but six faces. For a positive-dimensional compact Lie group the map from its five reduced variables to six based face holonomies has image of Haar measure zero in \(G^6\). Indeed, a smooth map from a compact manifold of dimension \(5\dim G\) into one of dimension \(6\dim G\) has image of zero full-dimensional volume: cover the domain by finitely many Lipschitz coordinate charts and subdivide into cubes of side \(\delta\); the image can be covered by \(O(\delta^{-5\dim G})\) balls of radius \(O(\delta)\), whose total \(6\dim G\)-volume tends to zero. A sixfold product law with a strictly positive density cannot be supported on that image.

For an abelian group the same relation reduces to the familiar signed product constraint around a cube. For a nonabelian group, the order and the transporting paths are part of the statement. Merely retaining the abelian product relation loses information. Conversely, the existence of a cube constraint is not a proof that no useful plaquette representation can be built: it specifies what such a representation must encode.

\hypertarget{gauge_foundations-an-exact-one-cube-partition-function}{%
\subsection{An exact one-cube partition function}\label{gauge_foundations-an-exact-one-cube-partition-function}}

The curvature relation can be retained rather than discarded, and the resulting single cube can still be solved. Let \(v:G\to(0,\infty)\) be a continuous central inversion-symmetric weight, set \(z=\int v\,dh\), and let \(f=v/z\) be the probability density of \(\nu\). Give all six cube faces the weight \(v\).

\textbf{Theorem G9 (one-cube partition formula).} With normalized Haar measure on the twelve edges,

\[Z_{\rm cube}=z^6 f^{*6}(1).\]

Here \(f^{*6}\) is the density of the sixfold group convolution. The right side includes the correction to the product of six unconstrained single-plaquette partition functions.

\textbf{Proof.} By Theorem G8, five face variables are independent Haar coordinates. Write them in the order occurring in the curvature relation. Multiplication by their five weights changes their joint law to five independent copies of \(\nu\) and contributes a factor \(z^5\). Inversion symmetry removes the inversions in distribution. The sixth face is an ordered product of these five variables, hence has density \(f^{*5}\). Therefore

\[Z_{\rm cube}=z^5\int_Gv(g)f^{*5}(g)\,dh(g)
=z^6\int_Gf(g)f^{*5}(g)\,dh(g).\]

Inversion symmetry is preserved by repeated convolution of the same central density, so the last integral is \((f^{*6})(1)\). All integrands are continuous on a compact group and all changes of variables preserve Haar measure. \(\square\)

For the \(SU(2)\) Wilson weight, the formula can be evaluated by an absolutely convergent character series. Let \(n=2j+1\) and let \(a_j=I_n/I_1\). The density has expansion

\[f_\beta(U)=\sum_{j=0,1/2,\ldots}(2j+1)a_j(\beta)\chi_j(U).\]

Its coefficients were calculated in Theorem G5. Absolute uniform convergence follows from the positive Bessel series. For fixed \(\beta>0\),

\[I_n(\beta)
\le\frac{(\beta/2)^n}{n!}\exp(\beta^2/4),\]

because \((k+n)!\ge n!\) in the series; the resulting majorant makes \(\sum_{n\ge1}n^2I_n(\beta)\) finite. Since \(|\chi_j|\le2j+1\), the character expansion converges absolutely and uniformly. Orthogonality of the \(SU(2)\) characters follows here directly from their sine formula and the Haar integral. Completeness follows without an unproved Fourier inversion step: the characters are the Chebyshev polynomials \(U_{n-1}(\cos\theta)\), which span the polynomials in \(\cos\theta\); polynomial approximation on \([-1,1]\) makes their span uniformly dense in the central continuous functions. The uniformly convergent series therefore agrees with the density having those coefficients.

Convolution multiplies the normalized matrix Fourier coefficients. To verify the needed orthogonality identity, average a matrix unit between two unitary irreducible representations over Haar measure. The resulting map intertwines the representations. Schur's lemma makes it zero for inequivalent representations and a scalar multiple of the identity for equivalent ones; tracing determines the scalar. Applying this to the two characters gives

\[\chi_j*\chi_k=\mathbf1_{\{j=k\}}\frac{\chi_j}{2j+1}.\]

Termwise convolution is justified by uniform absolute convergence. Evaluating at the identity, where \(\chi_j(1)=2j+1\), yields

\[Z_{\rm cube}(\beta)
=z(\beta)^6\sum_{n=1}^\infty
n^2\left[\frac{I_n(\beta)}{I_1(\beta)}\right]^6.\]

Every term is nonnegative. The \(n=1\) term equals one, and the remaining terms give a strictly positive correction for \(\beta>0\). Thus \(Z_{\rm cube}>z^6\): the exact constraint does not reproduce the partition function of six independent plaquettes, even though five face coordinates can be chosen independently before inserting the sixth weight.

The leading correction is already explicit. Since \(I_2/I_1=\beta/4+O(\beta^3)\) and \(I_n/I_1=O(\beta^{n-1})\) for fixed \(n\),

\[\frac{Z_{\rm cube}(\beta)}{z(\beta)^6}
=1+\frac{\beta^6}{1024}+O(\beta^8).\]

The infinite remainder is controlled uniformly near zero by the factorial majorant above, so this is an analytic expansion, not a termwise formal guess. The finite cube integral is itself entire in \(\beta\); its positive value at zero ensures that its logarithm is analytic in a neighborhood of zero. Cubic symmetry and differentiation therefore give the mean of any one face:

\[\mathbb E_{\rm cube}w(U_p)
=\frac16\frac{d}{d\beta}\log Z_{\rm cube}(\beta)
=m(\beta)+\frac{\beta^5}{1024}+O(\beta^7).\]

Equivalently,

\[\mathbb E_{\rm cube}w(U_p)
=\frac\beta4-\frac{\beta^3}{96}
+\frac{5\beta^5}{3072}+O(\beta^7).\]

The planar model has coefficient \(1/1536\) instead of \(5/3072\) at order \(\beta^5\). This gives a concrete calculation of where the independent-plaquette surrogate first changes this local expectation on the cube. The formulas solve a finite geometric subproblem; controlling interacting cubes throughout an unbounded four-dimensional lattice is an additional task.

\hypertarget{gauge_foundations-exact-blocking-generates-more-than-one-wilson-coupling}{%
\subsection{Exact blocking generates more than one Wilson coupling}\label{gauge_foundations-exact-blocking-generates-more-than-one-wilson-coupling}}

The planar solution also permits a precise test of a common renormalization shortcut. A block containing \(B\) elementary faces has holonomy law \(\nu_\beta^{*B}\). It is tempting to replace this law by a Wilson density \(\nu_{\beta'}\) with a changed coupling chosen to reproduce its fundamental trace. Matching that one expectation is possible. Matching the entire probability law is a stronger assertion, and it fails already in this exactly solvable model.

\textbf{Theorem G10 (failure of one-parameter Wilson closure).} Fix an integer \(B\ge2\). There is \(\delta_B>0\) such that for every \(0<\beta<\delta_B\), no \(\beta'\ge0\) satisfies

\[\nu_\beta^{*B}=\nu_{\beta'}\]

as probability measures on \(SU(2)\). There is nevertheless exactly one positive \(\beta'\) matching the fundamental mean, namely the solution of

\[m(\beta')=m(\beta)^B.\]

\textbf{Proof.} The variance identity proves that \(m\) is strictly increasing and continuous. Its limits are zero as \(\beta\downarrow0\) and one as \(\beta\to\infty\), the latter following from Lemma G6. Thus every number strictly between zero and one is achieved exactly once, proving existence and uniqueness of the fundamental-mean match.

Write \(a_1\) for the normalized spin-one coefficient. If the probability laws were equal, their fundamental and spin-one coefficients would both agree:

\[m(\beta')=m(\beta)^B,\qquad
 a_1(\beta')=a_1(\beta)^B.\]

The first equation forces \(\beta'\to0\) as \(\beta\to0\). The positive series gives

\[m(\beta)=\frac\beta4+O(\beta^3),\qquad
 a_1(\beta)=\frac{\beta^2}{24}+O(\beta^4),\]

so

\[\frac{a_1(\beta)}{m(\beta)^2}\longrightarrow\frac23.\]

For a pair of equal laws the two coefficient equations would imply

\[\frac{a_1(\beta')}{m(\beta')^2}
=\left(\frac{a_1(\beta)}{m(\beta)^2}\right)^B.\]

The left side tends to \(2/3\), while the right side tends to \((2/3)^B\). These limits are distinct when \(B\ge2\). Consequently equality cannot hold along any sequence \(\beta\downarrow0\). If no \(\delta_B\) existed, one could choose such a sequence with equality, a contradiction. \(\square\)

The discrepancy can be quantified at leading order. For the fundamental-mean match, inversion of \(m(\beta')=m(\beta)^B\) gives

\[\beta'=4^{1-B}\beta^B\bigl(1+o(1)\bigr).\]

Hence the actual blocked adjoint coefficient and the matched Wilson coefficient have ratio

\[\frac{a_1(\beta)^B}{a_1(\beta')}
\longrightarrow\left(\frac23\right)^{B-1}.\]

For a two-plaquette block this ratio tends to \(2/3\). The error in replacing the full block distribution by a single fitted Wilson coupling is therefore visible at the first nonzero adjoint order. It is not eliminated by making the fundamental mean agree exactly.

This theorem does not say that one-coupling approximations are useless. It says that they are approximations requiring a controlled remainder. Even when plaquettes are independent, integrating them into a larger face generally creates an action with more character coefficients. In higher dimensions there are additionally interactions linking several coarse faces. A proposed exact renormalization recursion must specify these generated terms and estimate them, rather than declare closure after fitting one number.

It is particularly useful as a unit test for a proposed blocking argument: the argument should reproduce the different fundamental and adjoint transformations in this solvable setting before being applied to a coupled four-dimensional measure.

\hypertarget{gauge_foundations-nonuniform-planar-couplings-and-an-explicit-accumulated-error-bound}{%
\subsection{Nonuniform planar couplings and an explicit accumulated-error bound}\label{gauge_foundations-nonuniform-planar-couplings-and-an-explicit-accumulated-error-bound}}

The continuum-area calculation need not assume that every fine plaquette has the same coupling. This extension illustrates how a local asymptotic estimate must be summed over an increasing number of cells. An error small on one cell is useful only when its accumulated value stays controlled.

Let \(\mathcal R_a\) be the set of plaquettes inside a rectangular loop, and assign \(\beta_{a,p}>0\) to each face. The factorization in Theorem G3 still holds. The loop proof uses centrality and independence, not equality of the densities, so repeated representation integration gives

\[W_j(a):=\mathbb E\left[\frac{\chi_j(U_{C_a})}{2j+1}\right]
=\prod_{p\in\mathcal R_a}a_j(\beta_{a,p}).\]

For completeness, central probability measures need not be identical for this calculation: each individual matrix integral is a scalar multiple of the identity. Products of those scalars therefore give the normalized character expectation in any of the ordered products produced by the loop decomposition.

\textbf{Theorem G11 (nonuniform continuum-area limit).} Fix a spin \(j>0\). There are constants \(B_j,C_j<\infty\), depending only on that spin, such that whenever all \(\beta_{a,p}\ge B_j\),

\[\left|\log W_j(a)+2j(j+1)
\sum_{p\in\mathcal R_a}\frac1{\beta_{a,p}}\right|
\le C_j\sum_{p\in\mathcal R_a}\frac1{\beta_{a,p}^2}.\]

In particular, if

\[\min_{p\in\mathcal R_a}\beta_{a,p}\to\infty,
\qquad
\sum_{p\in\mathcal R_a}\frac1{\beta_{a,p}}\to T<\infty,\]

then \(W_j(a)\to\exp[-2j(j+1)T]\).

\textbf{Proof.} The quantitative remainder proved after Lemma G6 gives, for all sufficiently large \(\beta\),

\[a_j(\beta)=1-\frac{2j(j+1)}\beta+R_j(\beta),
\qquad |R_j(\beta)|\le K_j\beta^{-2}.\]

Increase the lower threshold until \(|a_j(\beta)-1|\le1/2\). The inequality \(|\log(1+x)-x|\le2x^2\) for \(|x|\le1/2\) then gives

\[\left|\log a_j(\beta)+\frac{2j(j+1)}\beta\right|
\le C_j\beta^{-2}\]

with a finite spin-dependent constant. Summing these inequalities over the faces proves the first assertion. For the limit, observe that

\[\sum_{p\in\mathcal R_a}\beta_{a,p}^{-2}
\le\left(\max_{p\in\mathcal R_a}\beta_{a,p}^{-1}\right)
\sum_{p\in\mathcal R_a}\beta_{a,p}^{-1}\longrightarrow0.\]

The logarithms therefore converge to the stated finite number, and exponentiation proves the conclusion. \(\square\)

As a concrete specialization, let a fixed planar rectangle carry a continuous positive function \(b(x)\), bounded away from zero. Put \(\beta_{a,p}=b(x_p)/a^2\) at the center of each fine plaquette. Then the inverse-coupling sum is the Riemann sum

\[\sum_{p\in\mathcal R_a}\frac{a^2}{b(x_p)}
\longrightarrow\int_{\mathcal R}\frac{dx}{b(x)},\]

while the squared-inverse sum tends to zero. The limiting loop expectation is therefore

\[\exp\left[-2j(j+1)\int_{\mathcal R}\frac{dx}{b(x)}\right].\]

This is an explicit accumulation estimate with constants independent of the number of fine cells. It uses the exact planar factorization. For a four-dimensional argument, an analogous summation would need both control of the local remainder and control of the correlations between cells. The cube relation shows why the latter cannot simply be omitted.

\hypertarget{gauge_foundations-what-these-exact-calculations-contribute-to-the-continuum-programme}{%
\subsection{What these exact calculations contribute to the continuum programme}\label{gauge_foundations-what-these-exact-calculations-contribute-to-the-continuum-programme}}

The section supplies three kinds of information that can be used without conjectural premises. The spanning-tree theorem gives exact finite-dimensional coordinates and a verified integration formula. The planar theorem gives an all-coupling solution and a nontrivial physical-area scaling limit for a specified family of loop expectations. The cube theorem identifies an exact higher-dimensional curvature interaction and evaluates its first contribution.

These results also sharpen the next research question. A proposed four-dimensional plaquette description must retain the transported cube relations. A proposed blocking transformation must calculate how those relations and the associated effective interactions change under integration. A proposed continuum observable family must control its reflected norm after the normalization and limit, rather than appeal only to finite-cutoff polynomial density. These are concrete mathematical requirements exposed by exact calculations, not generic objections to trying a different method.

The planar scaling proof demonstrates that an increasing coupling can balance a growing number of plaquettes without making a loop expectation vanish. It does not show that the same balance survives the coupled four-dimensional measure. The cube calculation shows that the discarded interaction is nonzero even in the first cell where it can occur. The task in the full theory is therefore to estimate and organize such interactions, including their long-distance effects, with constants suitable for both the continuum and infinite-volume limits.

The displayed finite and two-dimensional results provide exact normalization and dependence checks. Their application to four-dimensional cutoff removal requires estimates for the interacting measure with its cube constraints intact.

\hypertarget{reflection_reconstruction-reflection-positivity-exact-transfer-operators-and-a-continuum-comparison}{%
\section{Reflection positivity, exact transfer operators, and a continuum comparison}\label{reflection_reconstruction-reflection-positivity-exact-transfer-operators-and-a-continuum-comparison}}

Reflection positivity is the mechanism that turns suitable Euclidean correlations into nonnegative physical norms. It is an algebraic inequality with analytical consequences, not an estimate of the rate at which correlations decay. This section proves the inequality for precisely specified finite Wilson systems, constructs their finite-spatial-volume transfer Hamiltonians, and then gives an explicit continuum Gaussian model in which every part of the same construction can be calculated. The two constructions make it possible to identify the additional estimate needed in an interacting continuum limit without concealing it inside the word ``reconstruction.''

The Wilson transfer construction and the Osterwalder--Schrader framework are established mathematics \cite{Luscher,OSLattice,OS1,OS2}. The proofs here are self-contained for the statements actually used. They are not a claim that these methods, or their consequences at finite spatial volume, were first discovered in this manuscript. The construction below includes no fermions, no gauge-fixing determinants, and no formal continuum functional integral.

\hypertarget{reflection_reconstruction-reflected-forms-and-their-null-spaces}{%
\subsection{Reflected forms and their null spaces}\label{reflection_reconstruction-reflected-forms-and-their-null-spaces}}

Let \((X,\mathcal F,\mu)\) be a probability space with a measurable involution \(\vartheta:X\to X\) preserving \(\mu\). An involution means \(\vartheta^2\) is the identity. Let \(\mathcal A_+\) be a complex vector space of bounded measurable functions interpreted as positive-time observables. Define the antilinear operation

\[\Theta F=\overline{F\circ\vartheta}\]

and the reflected sesquilinear form

\[Q(F,G)=\int_X(\Theta F)G\,d\mu.\]

The convention is that inner products are antilinear in the first argument. Reflection invariance gives \(Q(F,G)=\overline{Q(G,F)}\): conjugate the integral defining \(Q(G,F)\) and then change variables by \(\vartheta\). Reflection positivity is the additional statement

\[Q(F,F)\ge0\qquad(F\in\mathcal A_+).\]

Ordinary positivity of the probability density does not imply this inequality. The integrand need not be a squared absolute value at the same configuration. It couples a function of the positive half of the configuration to a function of the negative half.

\textbf{Lemma R1 (quotient norm and ordinary \(L^2\) control).} If the reflected form is positive, then

\[|Q(F,G)|^2\le Q(F,F)Q(G,G),\qquad
0\le Q(F,F)\le\|F\|_{L^2(\mu)}^2.\]

Consequently the set \(\mathcal N=\{F:Q(F,F)=0\}\) is a vector subspace, every element of \(\mathcal N\) is orthogonal to every element of \(\mathcal A_+\), and \(Q\) induces an inner product on \(\mathcal A_+/\mathcal N\).

\textbf{Proof.} For any complex \(z\), positivity of \(Q(F+zG,F+zG)\) gives the Cauchy--Schwarz inequality by minimizing the resulting quadratic polynomial. If \(Q(G,G)=0\), varying the magnitude and phase of \(z\) first forces \(Q(F,G)=0\), so the same conclusion covers the degenerate case. The ordinary integral Cauchy--Schwarz inequality gives

\[|Q(F,F)|\le\|F\circ\vartheta\|_2\|F\|_2=\|F\|_2^2.\]

The equality of norms uses reflection invariance. The claims about the null space now follow from the first inequality and sesquilinearity. Completing the quotient gives a Hilbert space. \(\square\)

The constant function has reflected norm one. Its class is the candidate vacuum vector \(\Omega=[1]\). For any \(F\), the centered vector is

\[[F-\mu(F)],\qquad
\|[F-\mu(F)]\|^2=Q(F,F)-|\mu(F)|^2.\]

The second identity follows by expanding the square and using \(\mu(F\circ\vartheta)=\mu(F)\). It makes clear that a nonzero ordinary variance need not give a nonzero physical state: the ordinary variance uses \(\mu(|F|^2)\), whereas the physical squared norm uses a reflected correlation.

No Hamiltonian has been constructed by this quotient alone. One also needs an action of positive-time translations, its compatibility with the null space, and continuity and contraction properties. Those hypotheses are handled abstractly in the spectral-bridge section and concretely for the transfer operator below.

\hypertarget{reflection_reconstruction-a-gluing-lemma-with-an-explicit-positive-kernel}{%
\subsection{A gluing lemma with an explicit positive kernel}\label{reflection_reconstruction-a-gluing-lemma-with-an-explicit-positive-kernel}}

Suppose \(Y\) is a probability space with measure \(\nu\). Let \(a:Y\to[0,\infty)\) be bounded, and suppose a kernel \(K\) can be written as

\[K(x,y)=\langle\Psi(y),\Psi(x)\rangle_{\mathcal K},\]

where \(\Psi:Y\to\mathcal K\) is strongly measurable, \(\mathcal K\) is a separable Hilbert space, and \(\sup_x\|\Psi(x)\|<\infty\). Assume \(K\) is real and nonnegative pointwise, and that

\[Z=\int_{Y\times Y}a(x)a(y)K(x,y)\,d\nu(x)d\nu(y)>0.\]

These conditions define a probability law on two copies of \(Y\). The reflection exchanges the copies. For \(F\) depending on the first copy, the unnormalized reflected expectation is

\[\int\overline{F(y)}F(x)a(x)a(y)K(x,y)\,d\nu(x)d\nu(y).\]

\textbf{Lemma R2 (positive-kernel gluing).} This expectation is nonnegative for every bounded complex \(F\).

\textbf{Proof.} The Bochner integral \(v_F=\int F(x)a(x)\Psi(x)\,d\nu(x)\) exists because its integrand has integrable norm. Expanding \(\|v_F\|^2\) gives exactly the displayed double integral. Fubini is justified by boundedness. Division by \(Z\) preserves the sign. \(\square\)

The same proof applies to a finite product of kernels: replace the feature map by its tensor product. It also applies to a kernel depending on shared boundary variables \(z\). If each conditional reflected integral is a squared norm for fixed \(z\), integrating it against a positive boundary measure preserves nonnegativity. The boundary measure need not be a product measure.

This lemma separates two tasks. The first is geometric: identify which factors of the action belong to each half and which factors cross the reflection plane. The second is algebraic: express each crossing factor as a positive kernel. A character identity alone does not accomplish the geometric task.

\hypertarget{reflection_reconstruction-site-reflection-for-a-finite-wilson-box}{%
\subsection{Site reflection for a finite Wilson box}\label{reflection_reconstruction-site-reflection-for-a-finite-wilson-box}}

Fix a finite rectangular spatial set \(B\subset\mathbb Z^3\) and integer \(T\ge1\). The vertex set is

\[\Lambda_{\rm vert}=\{-T,\ldots,T\}\times B.\]

Use every nearest-neighbor link whose endpoints are in this set, and every elementary plaquette all of whose links are present. These are free boundary conditions; there are no fixed exterior plaquette factors. Let \(G\) be a compact Lie group with normalized Haar measure \(h\) and a continuous unitary representation \(\rho:G\to U(N)\). Set

\[w(g)=N^{-1}\operatorname{Re}\operatorname{Tr}\rho(g),\qquad
d\mu=Z^{-1}\exp\left(\beta\sum_p w(U_p)\right)\prod_e h(dU_e).\]

The value of \(w(U_p)\) does not depend on the chosen starting vertex or orientation of the plaquette. Conjugation preserves trace and inversion conjugates the trace of a unitary matrix.

Reflect time by \(\vartheta(t,x)=(-t,x)\). Under reflection, a positively oriented temporal link is represented by the inverse variable on the reflected positive link. Spatial link orientations are unchanged. Haar measure is preserved under inversion, so the reflection preserves product Haar measure.

Partition links into \(E_0\), the spatial links at time zero, and \(E_+,E_-\), the links with respectively positive and negative time coordinates of their centers. A temporal plaquette spans only two adjacent time slices. It therefore cannot contain both a link with strictly positive center time and a link with strictly negative center time. Every plaquette is either entirely in \(E_0\), or belongs to the positive half together with some \(E_0\) links, or is the reflection of such a positive plaquette.

Write \(z\) for the plane variables and \(x,y\) for the positive-half variables and reflected negative-half variables. There are bounded real functions \(S_0\) and \(S_+\) such that the exponent is

\[S_0(z)+S_+(x,z)+S_+(y,z).\]

Here the factor \(\beta\) has been included in \(S_0,S_+\). The conditional measures on the two strict halves are independent given \(z\) and have identical unnormalized densities.

\textbf{Theorem R3 (site-reflection positivity).} For this finite free-boundary Wilson system, every bounded complex observable \(F\) of \(E_+\cup E_0\) satisfies

\[\mu[(\Theta F)F]\ge0.\]

The theorem holds for every real \(\beta\); the physically used Wilson regime has \(\beta\ge0\).

\textbf{Proof.} Directly integrating the two strict halves gives

\[Z\mu[(\Theta F)F]
=\int e^{S_0(z)}
\left|\int F(x,z)e^{S_+(x,z)}\prod_{e\in E_+}h(dx_e)\right|^2
\prod_{e\in E_0}h(dz_e).\]

Every integral exists because the box is finite and the exponent and observable are bounded. The right side is nonnegative, and \(Z>0\). No gauge invariance of \(F\) is needed for this site-reflection statement. \(\square\)

This proof does not imply one-step transfer positivity when \(\beta\) is negative. Site reflection and reflection through the middle of a temporal link are distinct geometries. The latter will require a positive-type crossing factor. Keeping those statements separate prevents an invalid extension from a two-step positivity statement to an arbitrary one-step Hamiltonian construction.

Reflection-compatible boundaries are also essential. For a simple warning, take independent spins \(X_+,X_-\in\{-1,1\}\) with means \(m_+\ne m_-\). For \(F=1+iX_+\),

\[\mathbb E[(1-iX_-)(1+iX_+)]
=1+m_-m_++i(m_+-m_-),\]

which is not even real. A positive probability law with asymmetric boundaries therefore need not supply a positive reflected form. Arbitrary-boundary covariance estimates and reflection-positive boundary constructions have different hypotheses.

\hypertarget{reflection_reconstruction-the-wilson-factor-as-a-positive-type-kernel}{%
\subsection{The Wilson factor as a positive-type kernel}\label{reflection_reconstruction-the-wilson-factor-as-a-positive-type-kernel}}

Let \(\chi_\rho(g)=\operatorname{Tr}\rho(g)\) and \(\overline\rho\) denote the conjugate representation. For \(\beta\ge0\), define

\[k_\beta(g)=\exp\left(\frac{\beta}{2N}
[\chi_\rho(g)+\chi_{\overline\rho}(g)]\right)=e^{\beta w(g)}.\]

\textbf{Lemma R4 (tensor-character expansion).} The kernel \((g,h)\mapsto k_\beta(gh^{-1})\) is positive semidefinite. Its expansion is uniformly absolutely convergent and can be written as a countable sum of rank-one positive kernels.

\textbf{Proof.} Put \(c=\beta/(2N)\). The two exponential series give

\[k_\beta(g)=\sum_{r,s\ge0}\frac{c^{r+s}}{r!s!}
\chi_{\rho^{\otimes r}\otimes\overline\rho^{\otimes s}}(g).\]

The representation in the summand has dimension \(N^{r+s}\), so its character has absolute value at most that number. The sum of these absolute bounds is

\[\sum_{r,s\ge0}\frac{c^{r+s}N^{r+s}}{r!s!}=e^{2cN}=e^\beta.\]

This proves uniform absolute convergence. For any finite-dimensional unitary representation \(R\),

\[\chi_R(gh^{-1})
=\operatorname{Tr}[R(g)R(h)^*]
=\sum_{a,b}R(g)_{ab}\overline{R(h)_{ab}}.\]

Multiplying the matrix coefficients by the square root of the nonnegative series coefficient gives the desired feature functions. Their squared absolute values at a fixed \(g\) sum to \(k_\beta(e)=e^\beta\). They therefore define a bounded map into a countable Hilbert direct sum. Every finite Gram quadratic form is a sum of squared absolute values with nonnegative coefficients. Uniform convergence passes that inequality to the full kernel. \(\square\)

No irreducible-character decomposition is needed for positivity. The tensor representation may be reducible. If it is decomposed into irreducibles, the multiplicities are nonnegative integers and the resulting coefficients remain nonnegative. The direct tensor proof has the advantage that every coefficient and every convergence bound is visible.

For \(\beta>0\) and a faithful representation, these features detect every nonzero \(L^2(G,h)\) function. Indeed, the algebra generated by matrix coefficients of \(\rho\) and \(\overline\rho\) contains constants, is closed under conjugation, and separates group elements. The complex Stone--Weierstrass theorem makes it uniformly dense in \(C(G)\), which is dense in \(L^2(G,h)\). Every such product is a matrix coefficient of one of the tensor representations occurring with strictly positive weight. Thus a function orthogonal to all features is zero. This injectivity statement will be important for defining the logarithm of the transfer operator.

\hypertarget{reflection_reconstruction-link-reflection-and-an-exact-gauge-fixing-calculation}{%
\subsection{Link reflection and an exact gauge-fixing calculation}\label{reflection_reconstruction-link-reflection-and-an-exact-gauge-fixing-calculation}}

Take the same spatial rectangle, but time slices \(-T+1,\ldots,T\) and reflection \(t\mapsto1-t\). The positive half has vertices with \(t\ge1\); the negative half has vertices with \(t\le0\). Let \(V_x\) be the temporal link from \((0,x)\) to \((1,x)\). These are the links crossing the plane.

An observable \(F\) in this subsection is a bounded gauge-invariant function of links entirely in the positive half. Gauge invariance includes transformations at its boundary vertices \((1,x)\). The reflected observable is then a gauge-invariant function of the negative-half links. This restriction is explicit: the following gauge-fixing proof is not a claim about arbitrary gauge-variant half-box functions.

For fixed crossing variables \(V=(V_x)\), apply the vertex gauge transformation

\[g(t,x)=\begin{cases}1,&t\le0,\\V_x,&t\ge1.\end{cases}\]

The convention for transformed links is \(U_{uv}\mapsto g(u)U_{uv}g(v)^{-1}\). Every crossing link becomes \(V_xV_x^{-1}=1\). For fixed \(V\), this is a bijective change of the remaining positive-half variables preserving their product Haar measure. It is enough to condition on \(V\) first; no Jacobian assertion involving a formal orbit volume is needed.

All Wilson plaquette factors and both observables are unchanged by the gauge transformation. After the change, the integrand no longer depends on \(V\), so integrating the crossing variables contributes one. Let \(A_e\) denote a spatial link at time one and \(B_e\) its counterpart at time zero. A crossing temporal plaquette has, up to orientation and starting point, holonomy

\[V_uA_eV_v^{-1}B_e^{-1},\]

which becomes \(A_e'B_e^{-1}\) in the transformed variables. Its Boltzmann factor is therefore \(k_\beta(A_e'B_e^{-1})\).

The remaining action splits into positive and negative halves. Identifying the negative half with a reflected copy of the positive half gives a joint density of the form

\[Z^{-1}a(x)a(y)\prod_{e\text{ at the interface}}
k_\beta(A_e(x)A_e(y)^{-1}).\]

All factors in \(a\) belong to one half. The precise orientation of an interface plaquette does not change the factor because \(k_\beta(g^{-1})=k_\beta(g)\).

\textbf{Theorem R5 (link-reflection positivity for physical observables).} For \(\beta\ge0\), the finite free-boundary Wilson measure just described satisfies

\[\mu[(\Theta F)F]\ge0\]

for every bounded gauge-invariant positive-half observable \(F\).

\textbf{Proof.} The gauge change above reduces the reflected integral to the two-half expression. Lemma R4 gives a bounded feature map for every interface link factor. The tensor product of the finitely many feature maps represents their product. Lemma R2 then expresses the reflected expectation as a squared Hilbert norm divided by \(Z\). \(\square\)

This calculation identifies exactly which positivity survives gauge reduction. The crossing temporal links are integrated by an ordinary Haar change of variables. They are not discarded from the measure without compensation. Gauge invariance of the observables is what allows their values to remain fixed during that change. It also explains why importing a scalar reflection argument directly for arbitrary gauge-fixed vector potentials would require additional work.

\hypertarget{reflection_reconstruction-the-exact-finite-spatial-volume-wilson-transfer-operator}{%
\subsection{The exact finite-spatial-volume Wilson transfer operator}\label{reflection_reconstruction-the-exact-finite-spatial-volume-wilson-transfer-operator}}

The same calculation gives more than a positive quadratic form. It produces an explicit operator whose physical spectrum is defined at every positive lattice spacing and every finite spatial volume.

Let \(\mathcal V\) and \(\mathcal E\) be the vertices and positively oriented links of a finite spatial lattice graph, with a specified finite set of spatial plaquettes. Put

\[X=G^{\mathcal E},\qquad d\nu(x)=\prod_{e\in\mathcal E}h(dx_e),\qquad
V(x)=\sum_{p\text{ spatial}}w(x_p),\qquad W(x)=e^{\beta V(x)/2}.\]

The gauge group \(\mathcal G=G^{\mathcal V}\) acts by \((gx)_{uv}=g_ux_{uv}g_v^{-1}\). Its normalized Haar average

\[Pf(x)=\int_{\mathcal G}f(gx)\,dg\]

is the orthogonal projection in \(L^2(X,\nu)\) onto the closed gauge-invariant subspace \(\mathcal H_{\rm phys}\). To check this directly, Haar invariance gives \(P^2=P\), and changing \(g\) to \(g^{-1}\) in the integral gives \(P^*=P\). The range consists exactly of invariant functions, with equality understood in \(L^2\).

Define the integral operator \(K\) by

\[K(x,y)=\prod_{e\in\mathcal E}k_\beta(x_ey_e^{-1}),\]

and let \(T_0=M_WKM_W\), where \(M_W\) is multiplication by \(W\). The operator \(K\) is positive by Lemma R4 and finite tensor products. It is self-adjoint because its kernel is Hermitian, and compact because its continuous bounded kernel is square integrable on the probability space \(X\times X\). Multiplication by \(W\) preserves these properties; \(W\) and \(W^{-1}\) are bounded since the graph has finitely many plaquettes.

Simultaneous gauge transformations preserve the kernel. For a link from \(u\) to \(v\),

\[ (gx)_e(gy)_e^{-1}=g_u x_ey_e^{-1}g_u^{-1},\]

and \(k_\beta\) is conjugation invariant. Hence \(K\) commutes with every gauge transformation and with \(P\). The spatial potential and \(W\) are gauge invariant as well. The physical transfer operator is the restriction of

\[T=PT_0P=T_0P\]

to \(\mathcal H_{\rm phys}\). It has the continuous kernel

\[T(x,y)=W(x)W(y)\int_{\mathcal G}
\prod_{e=(u,v)}k_\beta\!\left(x_e g_vy_e^{-1}g_u^{-1}\right)dg.\]

This is exactly a one-time-step Wilson slab: \(x,y\) are the two spatial slices, \(g_u\) are its temporal links, the product contains its temporal plaquettes, and \(W(x),W(y)\) assign half of each spatial action to the adjoining slab. Gluing two slabs adds the two half-weights at the shared slice to the full spatial action. The physical time duration of one application is the lattice spacing \(a\).

\textbf{Theorem R6 (finite-spatial-volume transfer Hamiltonian and gap).} Suppose \(\beta>0\), \(\rho\) is faithful, and the finite spatial graph admits a nonconstant continuous gauge-invariant function. Then \(T\) on \(\mathcal H_{\rm phys}\) is a compact, positive, injective self-adjoint operator. Its largest eigenvalue \(\lambda_0>0\) is simple. There is a second spectral value \(\lambda_1\) with

\[0<\lambda_1<\lambda_0,\]

where \(\lambda_1\) is the operator norm of the restriction to the top-eigenvector orthocomplement. The spectral logarithm

\[H_{a,\mathcal E}=-\frac1a\log(T/\lambda_0)\]

is a nonnegative self-adjoint Hamiltonian with a unique normalized vacuum and a strictly positive finite excitation threshold

\[m_{a,\mathcal E}=\frac1a\log(\lambda_0/\lambda_1).\]

\textbf{Proof.} Positivity, self-adjointness and compactness were proved above. For injectivity, first suppose \(\langle f,Kf\rangle=0\). The feature expansion is a sum of nonnegative squared absolute integrals. Every feature coefficient is positive when \(\beta>0\), so \(f\) is orthogonal to every product over links of matrix coefficients of tensor powers of \(\rho\) and \(\overline\rho\). These products form a conjugation-closed algebra separating points of the compact space \(X\) and containing constants. Stone--Weierstrass and density of continuous functions in \(L^2(\nu)\) imply \(f=0\). Thus \(K\) is injective. Since \(W\) is boundedly invertible,

\[\langle f,T_0f\rangle=\langle Wf,K(Wf)\rangle>0\quad(f\ne0).\]

For \(f\in\mathcal H_{\rm phys}\), \(Pf=f\), so the same strict inequality holds for \(T\). Therefore its kernel is zero.

The displayed integral kernel of \(T\) is strictly positive at every pair \((x,y)\): all of its factors are positive real exponentials, and \(W>0\). The compact self-adjoint spectral theorem supplies an eigenvector at its norm \(\lambda_0>0\). The norm is a positive eigenvalue because \(T\) is a nonzero positive operator. Replacing an eigenfunction by its absolute value cannot decrease its Rayleigh quotient, since the kernel is positive. Consequently a normalized maximizing eigenfunction can be chosen nonnegative. The eigenvalue equation then makes it continuous and strictly positive everywhere; call it \(\psi_0\).

If the top eigenspace had dimension greater than one, there would be a nonzero real top eigenfunction \(f\) orthogonal to \(\psi_0\). Its positive and negative parts would both be nonzero on sets of positive measure. Strict positivity of the kernel would give

\[\langle |f|,T|f|\rangle>\langle f,Tf\rangle
=\lambda_0\|f\|_2^2,\]

contradicting maximality. Complex eigenfunctions reduce to real and imaginary parts because the kernel is real. Thus the top eigenvalue is simple.

The assumed nonconstant invariant function implies \(\dim\mathcal H_{\rm phys}\ge2\). Its vacuum orthocomplement is a nonzero invariant subspace. The restriction of \(T\) there remains compact, positive and injective; its norm \(\lambda_1\) is therefore a positive attained eigenvalue. Simplicity and isolation of the top eigenvalue give \(\lambda_1<\lambda_0\).

Finally, injectivity means the spectral measure of \(T\) has no atom at zero on any nonzero vector. The spectral function \(-a^{-1}\log(\lambda/\lambda_0)\) therefore defines a densely defined self-adjoint operator: spectral cutoffs to \(\lambda\ge\varepsilon\) give a dense core as \(\varepsilon\downarrow0\). It is nonnegative, has kernel precisely \(\mathbb C\psi_0\), and its spectrum on the orthocomplement begins at the displayed positive finite number. \(\square\)

A spatial box containing a plaquette satisfies the nonconstant-function assumption whenever \(w\) is nonconstant. Vary one link of that plaquette while keeping the others equal to the identity; its trace changes. A tree graph is different: all link variables can be removed by vertex gauge transformations, and the physical space may consist only of constants. The theorem deliberately excludes that vacuous case.

The result is a genuine Hamiltonian statement, rather than the gap of an auxiliary sampling algorithm. Its operator was derived from the Euclidean-time slab of the Wilson action, and the factor \(1/a\) is explicit. However, compactness and strict positivity give no lower bound for \(\log(\lambda_0/\lambda_1)\) uniform in the number of spatial links. Nor do they give the required behavior when \(\beta\) changes with \(a\). Those are quantitative questions that the finite-volume theorem does not answer.

One can also construct an exact stationary Euclidean process from this operator. Normalize \(\|\psi_0\|_2=1\) and define

\[\pi(dx)=\psi_0(x)^2\nu(dx),\qquad
\mathsf P(x,dy)=\frac{T(x,y)\psi_0(y)}{\lambda_0\psi_0(x)}\nu(dy).\]

The eigenfunction equation makes \(\mathsf P(x,\cdot)\) a probability kernel. Symmetry of \(T\) gives detailed balance with \(\pi\). Iterating the kernel constructs finite-dimensional distributions of a stationary two-sided chain, and their consistency yields the chain on integer Euclidean times. Multiplication by \(\psi_0\) is a unitary map from \(L^2(\pi)\) to \(L^2(\nu)\) intertwining \(\mathsf P\) with \(T/\lambda_0\). For bounded gauge-invariant slice observables \(f,g\),

\[\mathbb E_\pi[\overline{f(X_0)}g(X_n)]
=\langle\psi_0f,(T/\lambda_0)^n\psi_0g\rangle.\]

This supplies an infinite-time theory at finite spatial volume. An infinite temporal extent is not the same operation as an infinite spatial volume or a continuum limit. The distinction is explicit in the formula and in the dependence of \(m_{a,\mathcal E}\) on its two regulators.

\hypertarget{reflection_reconstruction-explicit-finite-spatial-volume-transfer-hamiltonian-bounds}{%
\subsection{Explicit finite-spatial-volume transfer-Hamiltonian bounds}\label{reflection_reconstruction-explicit-finite-spatial-volume-transfer-hamiltonian-bounds}}

The compactness argument can be strengthened to a computable lower bound without estimating any eigenvalue numerically. Positive-kernel ratio contractions belong to the classical Birkhoff--Hopf theory \cite{Birkhoff1957,Hopf1963}; the argument below uses an explicit probability-row calculation and does not claim a new general contraction principle. First consider a continuous strictly positive symmetric transfer kernel on a compact probability space, with

\[0<\ell\le T(x,y)\le u<\infty.\]

Keep the normalized positive Perron eigenfunction \(\psi_0\), its eigenvalue \(\lambda_0\), and the stationary law \(\pi=\psi_0^2\nu\) from the preceding construction. The Rayleigh quotient of the constant function gives \(\lambda_0\ge\ell\). The eigenfunction equation and ordinary Cauchy--Schwarz give

\[\psi_0(x)\le\frac{u}{\lambda_0}\int\psi_0\,d\nu
\le\frac{u}{\lambda_0}.\]

Consequently the ground-state transition kernel satisfies

\[\frac{\mathsf P(x,dy)}{\pi(dy)}
=\frac{T(x,y)}{\lambda_0\psi_0(x)\psi_0(y)}
\ge\frac{\ell\lambda_0}{u^2}
\ge\left(\frac\ell u\right)^2=:\delta.\]

\textbf{Proposition R6a (minorization gap).} If \(0<\delta<1\), then the normalized transfer operator on the vacuum orthocomplement has norm at most \(1-\delta\). In particular,

\[m_{a,\mathcal E}\ge-\frac1a\log(1-\delta).\]

\textbf{Proof.} Let \(\Pi f=\pi(f)\) be the projection onto constants in \(L^2(\pi)\). The kernel

\[\mathsf R=\frac{\mathsf P-\delta\Pi}{1-\delta}\]

is a probability transition kernel by the lower bound just proved. Since both \(\mathsf P\) and \(\Pi\) preserve \(\pi\) and are reversible, so is \(\mathsf R\). Jensen's inequality followed by stationarity gives

\[\|\mathsf Rf\|_2^2\le\int\mathsf R(|f|^2)\,d\pi=\|f\|_2^2.\]

For centered \(f\), \(\mathsf Pf=(1-\delta)\mathsf Rf\), giving the claimed operator-norm bound. The unitary ground-state transform transfers this bound to \(T/\lambda_0\), and applying \(-a^{-1}\log\) gives the gap estimate. If \(\delta=1\), the transition kernel is \(\Pi\) itself and has no nonzero centered eigenvalue; that case is excluded by the injectivity and nontriviality assumptions of Theorem R6. \(\square\)

Let \(L=|\mathcal E|\) and let \(M\) be the number of spatial plaquettes. The Wilson kernel obeys

\[e^{-\beta(M+L)}\le T(x,y)\le e^{\beta(M+L)},\]

because the two spatial half-weights contribute at most \(\beta M\) in absolute value and the \(L\) temporal plaquettes contribute at most \(\beta L\). Thus the proposition gives the fully explicit certificate

\[m_{a,\mathcal E}\ge
-\frac1a\log\left(1-e^{-4\beta(M+L)}\right).\]

The Wilson factorization allows a sharper bound that removes \(M\). Write

\[T(x,y)=W(x)W(y)J(x,y),\qquad
e^{-\beta L}\le J(x,y)\le e^{\beta L}.\]

For fixed \(x\), the ground-state row measure is the normalized measure with density proportional to \(J(x,y)W(y)\psi_0(y)\) relative to \(\nu\). Comparing rows \(x\) and \(x'\) therefore introduces only the log tilt

\[b(y)=\log J(x,y)-\log J(x',y).\]

Its oscillation is at most \(4\beta L\). The spatial factors and the ground-state wavefunction belong to the common reference measure and do not contribute to that oscillation.

For clarity, the bounded-density-tilt estimate used here is

\[\left\|\frac{e^bQ}{Q(e^b)}-Q\right\|_{\rm TV}
\le\tanh\left(\frac{\sup b-\inf b}{4}\right).\]

With total variation defined as the supremum over events, it follows by writing the density ratio as \(r\in[v,V]\), with \(Qr=1\) and \(V/v\le e^{\operatorname{osc}b}\). The chord bound for \((r-1)_+\) gives \((V-1)(1-v)/(V-v)\). At fixed ratio \(k=V/v\), this is at most \((\sqrt{k}-1)/(\sqrt{k}+1)\), because \(kv+v^{-1}\ge2\sqrt{k}\). This proves the displayed estimate, including its constant.

\textbf{Corollary R6b (Wilson row-oscillation gap).} Under the hypotheses of Theorem R6,

\[\boxed{\quad
m_{a,\mathcal E}\ge-\frac1a\log\tanh(\beta|\mathcal E|)>0.
\quad}\]

The proof uses the spatial action only through its continuity, gauge invariance and boundedness at this finite regulator. The explicit lower bound contains no spatial-action coefficient when the temporal plaquette coefficient is fixed as displayed.

\textbf{Proof.} The row comparison and the tilt estimate give

\[\sup_{x,x'}\|\mathsf P(x,\cdot)-\mathsf P(x',\cdot)\|_{\rm TV}
\le\tanh(\beta L)=:q<1.\]

For a real bounded function \(f\), the elementary total-variation inequality implies

\[\operatorname{osc}(\mathsf Pf)\le q\operatorname{osc}(f).\]

To verify the elementary inequality, subtract \(\inf f\), divide by \(\operatorname{osc}f\), and represent the resulting function with values in \([0,1]\) by its level sets; the expectation difference is bounded by total variation. Constant functions are immediate.

Choose a real eigenfunction \(\psi_1\) of \(T\) at \(\lambda_1\). It is continuous because \(\lambda_1>0\) and the integral kernel is continuous. The function \(f=\psi_1/\psi_0\) is bounded and continuous because \(\psi_0\) is strictly positive on the compact space, and it is nonconstant because \(\psi_1\) is orthogonal to \(\psi_0\). Ground-state transformation gives \(\mathsf Pf=(\lambda_1/\lambda_0)f\). Applying the oscillation inequality and dividing by the positive number \(\operatorname{osc}f\) yields \(\lambda_1/\lambda_0\le q\). The logarithmic spectral formula proves the result. \(\square\)

The bound can be read without hidden dimension-dependent constants. Its dependence on the regulator is explicitly through \(a\), \(\beta\), and the number \(L\) of spatial links. At large \(\beta L\),

\[-\log\tanh(\beta L)=2e^{-2\beta L}+O(e^{-6\beta L}),\]

as follows from \(\tanh x=(1-e^{-2x})/(1+e^{-2x})\) and the convergent logarithmic series. Hence this certificate deteriorates exponentially with spatial size or inverse bare coupling. It proves a positive gap for every individual finite regulator but provides no strictly positive lower limit as the regulators are removed. The small-coupling volume-independent correlation theorem and this all-positive-coupling finite-volume spectral theorem have complementary ranges of strength; neither estimate is silently extended beyond its proved quantifiers.

\hypertarget{reflection_reconstruction-infinite-volume-passage-of-reflection-positivity}{%
\subsection{Infinite-volume passage of reflection positivity}\label{reflection_reconstruction-infinite-volume-passage-of-reflection-positivity}}

For countably many lattice links, \(G^E\) is a compact metrizable space. A sequence of finite-box laws can be extended to \(G^E\) by assigning independent Haar variables outside each box. Compactness of the space of probability measures gives weakly convergent subsequences. The extension is a device for taking a common-space limit; it does not insert missing plaquette interactions into a finite box.

Fix a reflection plane and choose boxes symmetric about it. If \(F\) is a continuous cylinder observable supported in the positive half, then for all sufficiently large boxes the finite reflected inequality applies. The function \((\Theta F)F\) is continuous and depends on finitely many links, so weak convergence passes the inequality to the limit. This proves reflected positivity for continuous cylinder observables of the relevant class. For link reflection that class remains gauge invariant. Continuous functions are dense in the \(L^2\) space of each finite marginal, and the ordinary integral Cauchy--Schwarz bound makes the reflected form continuous in \(L^2\). Consequently the inequality extends to bounded cylinder observables by approximation; gauge averaging preserves the needed invariant subspace.

For completeness, a subsequential free-boundary limit satisfies the ordinary infinite-volume Wilson conditional equations. Fix a finite link set \(A\). Once a box contains every plaquette meeting \(A\), conditioning on all other links gives exactly the full Wilson conditional kernel on \(A\). For continuous local test functions, this kernel is a continuous function of finitely many exterior links: its numerator and denominator are integrals of continuous bounded functions over a compact product, and its denominator is strictly positive. Pass the finite conditioning identity to the weak limit. A monotone-class argument extends from continuous local functions to bounded measurable cylinder functions and then to the exterior sigma algebra. Thus the limit is a Gibbs measure.

At a coupling for which the infinite-volume Gibbs measure is known to be unique, all such subsequences have the same limit. In particular, the strong-coupling uniqueness theorem established elsewhere in this manuscript identifies the site-reflection and link-reflection limits with the same measure. Without uniqueness, a symmetric approximating sequence still gives a reflection-positive limit for its chosen plane, but one must not silently identify different subsequential limits.

\textbf{Lemma R7 (positivity under correlation limits).} Let \(F_1,\ldots,F_n\) be proposed limiting positive-time observables, with cutoff representatives \(F_{i,j}\). Suppose each cutoff reflected Gram matrix

\[G^{(j)}_{ik}=\mu_j[(\Theta F_{i,j})F_{k,j}]\]

is positive semidefinite, and every entry converges to a finite number \(G_{ik}\). Then \(G\) is positive semidefinite.

\textbf{Proof.} For each fixed coefficient vector \(c\in\mathbb C^n\), entrywise convergence implies \(c^*G^{(j)}c\to c^*Gc\). The left side is nonnegative. \(\square\)

The finite dimension in this argument is important. There is no exchange of an uncontrolled infinite sum with the cutoff limit. To obtain a positive form on an infinite algebra, apply the lemma separately to every finite list. Further growth and continuity conditions are needed before these values define the distributions required by a full field-theoretic reconstruction.

There is a useful quantitative nontriviality check. Suppose for one observable the approximated mean and reflected second moment are \(m_j,A_j\), with proved errors

\[|m-m_j|\le\delta_j,\qquad |A-A_j|\le\eta_j,\qquad |m_j|\le B.\]

Then

\[\left|(A-|m|^2)-(A_j-|m_j|^2)\right|
\le\eta_j+(2B+\delta_j)\delta_j.\]

Indeed, \(||m|^2-|m_j|^2|\le|m-m_j|(|m|+|m_j|)\) and \(|m|\le B+\delta_j\). A strictly positive lower bound for \(A_j-|m_j|^2\) exceeding this error proves that the limiting centered physical state is nonzero. This is the correct reflected quantity to certify. A lower bound on an unreflected lattice variance cannot replace it.

\hypertarget{reflection_reconstruction-cylinder-density-and-the-meaning-of-a-total-family}{%
\subsection{Cylinder density and the meaning of a total family}\label{reflection_reconstruction-cylinder-density-and-the-meaning-of-a-total-family}}

Let \(\mu\) be a reflection-invariant probability measure on a countable product of compact metric link spaces. Let \(\mathcal F_+\) be the sigma algebra generated by positive-half links, including plane links when the reflection convention permits them. Bounded cylinder functions are dense in \(L^2(\mathcal F_+,\mu)\).

One proof begins with an increasing sequence of finite link sets generating \(\mathcal F_+\). Conditional expectations onto their sigma algebras are orthogonal projections onto increasing closed subspaces. The union of these subspaces is dense: indicators of cylinder sets generate the full sigma algebra, and the class of sets whose indicators lie in the closure is closed under complements and increasing countable unions. Truncating an \(L^2\) function first and then taking conditional expectations yields bounded cylinder approximants converging in \(L^2\).

Assume additionally that \(\mu\) is gauge invariant. If the target function is gauge invariant, its approximants can be made gauge invariant. Average each finite-cylinder approximant over the compact group of gauge transformations at the finitely many incident vertices. This preserves its support and boundedness. Jensen's inequality and gauge invariance of \(\mu\) show that this averaging is an \(L^2\) contraction. Since the target is invariant, averaging cannot increase its distance from the target. The resulting gauge-invariant cylinder functions are dense in the invariant positive-time subspace.

Lemma R1 now implies convergence in the reflected seminorm as well. Thus the OS completion of this cutoff algebra can be generated by bounded gauge-invariant cylinder observables. This is a valid totality statement at a fixed measure and a fixed set of variables.

A continuum totality assertion is different. A chosen renormalized family may consist only of selected Wilson loops, smeared curvature traces, or polynomials at prescribed scales. The density argument above does not prove that its limiting classes exhaust the physical Hilbert space. To make that conclusion, one must either define the limiting algebra through that family and establish all required field properties there, or prove approximation of every relevant state by the family in the limiting reflected norm.

A simple Hilbert-space example illustrates the issue. Suppose \(\mathcal H=\mathbb C\Omega\oplus\mathcal K_1\oplus\mathcal K_2\) and every selected observable creates states only in \(\mathcal K_1\). Bounds for all of those observables say nothing about the spectrum on \(\mathcal K_2\). The issue remains even if the selected family is countably infinite. Its cardinality is not a substitute for a density proof.

\hypertarget{reflection_reconstruction-a-four-dimensional-gaussian-continuum-field-constructed-explicitly}{%
\subsection{A four-dimensional Gaussian continuum field constructed explicitly}\label{reflection_reconstruction-a-four-dimensional-gaussian-continuum-field-constructed-explicitly}}

The free scalar field is a useful exact comparison because its continuum covariance, reflected norms, dense states and physical gap can all be written down. It is not being substituted for an interacting gauge theory.

Fix \(m>0\). On real Schwartz test functions on \(\mathbb R^4\), define

\[C_m(f,g)=\int_{\mathbb R^4}
\frac{\overline{\widehat f(p)}\widehat g(p)}{p^2+m^2}\,dp,\]

using the unitary Fourier transform. This is a continuous positive inner product after identifying functions of zero norm; for \(m>0\) there are no nonzero Schwartz null functions. Its completion is a separable real Hilbert space. Choose an orthonormal basis \((e_n)\) and independent standard real Gaussian variables \((Z_n)\). Define

\[\Phi(f)=\lim_{N\to\infty}\sum_{n=1}^N\langle e_n,f\rangle_{C_m}Z_n\]

in \(L^2\) of the countable Gaussian product probability space. The series is Cauchy because the sum of squared coefficients is \(C_m(f,f)\). It defines a real linear map into \(L^2\), and every finite collection is jointly Gaussian with covariance \(C_m\).

Its characteristic functional is

\[\mathbb E[e^{i\Phi(f)}]=e^{-C_m(f,f)/2}.\]

Differentiating the finite-dimensional Gaussian characteristic function yields all moments: odd moments vanish, and an even moment is the sum over pairings of products of \(C_m\) entries. This establishes the Wick rule without a formal infinite-dimensional Lebesgue measure. Continuity of \(C_m\) on the Schwartz space makes every resulting multilinear Schwinger functional continuous. The construction here is a generalized random field as a continuous map from test functions to random variables; no pointwise value \(\Phi(x)\) is needed.

Write Euclidean coordinates as \((t,x)\in\mathbb R\times\mathbb R^3\), and put

\[\omega_m(p)=\sqrt{|p|^2+m^2}.\]

Taking the inverse Fourier transform in the time momentum gives

\[\frac1{2\pi}\int_{\mathbb R}\frac{e^{ip_0u}}{p_0^2+\omega^2}\,dp_0
=\frac{e^{-\omega|u|}}{2\omega}.\]

For example, this identity follows by solving \((-\partial_u^2+\omega^2)k=\delta_0\): away from zero the decaying even solution is \(Ae^{-\omega|u|}\), and its derivative jump is \(-2A\omega=-1\), giving \(A=(2\omega)^{-1}\). Fourier transformation then verifies the identity distributionally.

For positive-time test functions, let \(\widetilde f(t,p)\) denote the unitary spatial Fourier transform. Define

\[q_mf(p)=\frac1{\sqrt{2\omega_m(p)}}
\int_0^\infty e^{-t\omega_m(p)}\widetilde f(t,p)\,dt.\]

If \(\theta f(t,x)=f(-t,x)\), direct substitution into the covariance formula gives

\[C_m(\theta f,g)=\langle q_mf,q_mg\rangle_{L^2(\mathbb R^3,dp)}.\]

Fubini is justified first for functions smooth and compactly supported in time away from zero and Schwartz in space; the exponential and rapid spatial decay give integrable majorants. Extension to the chosen positive-time test space follows by continuity. In particular, the reflected covariance is nonnegative and its null space is precisely the kernel of \(q_m\).

Extend \(q_m\) complex linearly to complex test functions. Its image is dense, or equivalently the complex span of its image on real test functions is dense. Choose a nonnegative nonzero smooth \(h\) supported in a compact interval strictly inside \((0,\infty)\). Let

\[u(p)=\frac1{\sqrt{2\omega_m(p)}}\int_0^\infty e^{-t\omega_m(p)}h(t)\,dt.\]

This is smooth and strictly positive. Given any \(\psi\in C_c^\infty(\mathbb R^3)\), choose a Schwartz function \(g\) with spatial Fourier transform \(\widehat g=\psi/u\), which is smooth and compactly supported. Then \(f(t,x)=h(t)g(x)\) satisfies \(q_mf=\psi\). Since \(C_c^\infty\) is dense in \(L^2\), the one-particle reflected completion is exactly \(L^2(\mathbb R^3,dp)\).

The positive-time translation \(f_s(t,x)=f(t-s,x)\), with zero extension before its support, satisfies

\[q_mf_s=e^{-s\omega_m}q_mf.\]

Thus the one-particle Hamiltonian is multiplication by \(\omega_m\). Its natural domain is the set of \(u\in L^2\) with \(\omega_mu\in L^2\). The multiplication operator is self-adjoint, as its nonreal resolvents are the bounded multipliers \((\omega_m-z)^{-1}\), and its spectrum is the essential range \([m,\infty)\). Its bottom is therefore finite and strictly positive.

\hypertarget{reflection_reconstruction-from-gaussian-exponentials-to-the-full-physical-fock-space}{%
\subsection{From Gaussian exponentials to the full physical Fock space}\label{reflection_reconstruction-from-gaussian-exponentials-to-the-full-physical-fock-space}}

The one-particle calculation alone is not a density proof for all states. For a Gaussian field that gap can be closed explicitly. For a real positive-time test function, define the bounded random variable

\[\mathcal W(f)=\exp\left(i\Phi(f)+\tfrac12 C_m(f,f)\right).\]

Its absolute value is the finite constant \(e^{C_m(f,f)/2}\), and its expectation is one. Gaussian integration gives

\[\mathbb E[(\Theta\mathcal W(f))\mathcal W(g)]
=\exp[C_m(\theta f,g)]
=\exp\langle q_mf,q_mg\rangle.\]

Let \(\mathcal F_s(\mathfrak h)=\bigoplus_{n=0}^\infty\mathfrak h^{\otimes_s n}\) be bosonic Fock space over \(\mathfrak h=L^2(\mathbb R^3,dp)\). The exponential vector

\[\operatorname{Exp}(u)=\bigoplus_{n=0}^\infty\frac{u^{\otimes n}}{\sqrt{n!}}\]

has inner products \(\langle\operatorname{Exp}(u),\operatorname{Exp}(v)\rangle=e^{\langle u,v\rangle}\). Therefore the assignment

\[[\mathcal W(f)]\longmapsto\operatorname{Exp}(iq_mf)\]

preserves all Gram matrices and defines an isometry on their complex linear span after quotienting null vectors.

These vectors are total. To see this, suppose \(\xi=(\xi_n)\) is orthogonal to every exponential vector with argument \(iq_mf\) for real \(f\). Fix finitely many such arguments \(u_1,\ldots,u_k\). The function

\[z\longmapsto\left\langle\xi,\operatorname{Exp}\left(\sum_{j=1}^k z_ju_j\right)\right\rangle\]

is entire in the complex variables \(z_j\): on bounded sets, convergence follows from Cauchy--Schwarz and \(\|\operatorname{Exp}(u)\|=e^{\|u\|^2/2}\). It vanishes for every real vector \(z\), by the assumed orthogonality and real linearity of the test-function space. Applying the one-variable identity theorem successively to each variable makes it identically zero. Every Taylor coefficient vanishes, so \(\xi_n\) is orthogonal to every symmetric product of vectors from that real test range. Its complex span is dense in \(\mathfrak h\) by the preceding density proof; finite symmetric tensors from a dense set are dense in every symmetric tensor power. Hence every \(\xi_n=0\).

The bounded exponentials also generate the positive-time Gaussian algebra in \(L^2\). On any finite list of Gaussian coordinates, orthogonality to all exponentials would make the Fourier transform of the finite signed measure obtained by multiplying the Gaussian law by the candidate \(L^2\) function vanish. Uniqueness of Fourier transforms of finite measures makes that function zero. Cylinder density then extends the claim to the whole positive-time sigma algebra. Lemma R1 passes that density to the reflected completion. Thus the constructed OS Hilbert space is exactly the Fock space above, rather than a selected subspace with an undetected orthogonal sector.

Euclidean time translation becomes

\[e^{-sH}=\Gamma(e^{-s\omega_m}),\qquad H=d\Gamma(\omega_m).\]

On the \(n\)-particle sector, \(H\) is multiplication by \(\sum_{i=1}^n\omega_m(p_i)\). This description proves self-adjointness directly, sector by sector and by Hilbert direct sum. The zero-particle sector is a one-dimensional vacuum. Every nonvacuum sector has energy at least \(nm\ge m\). Conversely, one-particle functions supported near momentum zero have energy arbitrarily close to \(m\). Hence the full excitation threshold is exactly \(m\) and the vacuum is unique.

For every centered physical state \(v\) and \(s\ge0\),

\[0\le\langle v,e^{-sH}v\rangle\le e^{-ms}\|v\|^2.\]

The rate is common to the entire vacuum orthocomplement, not merely to one field insertion. The construction has therefore verified positivity, density, nontriviality and a gap in a specific continuum model by calculation. The positive parameter \(m\) was supplied in that model's covariance. No argument here derives such a parameter dynamically from the pure Yang--Mills action.

\hypertarget{reflection_reconstruction-what-the-free-comparison-proves-about-the-remaining-task}{%
\subsection{What the free comparison proves about the remaining task}\label{reflection_reconstruction-what-the-free-comparison-proves-about-the-remaining-task}}

The same formulas with \(m=0\) give a reflection-positive massless Gaussian field in four dimensions. The covariance integral remains finite for Schwartz functions because \(|p|^{-2}\) is locally integrable in four dimensions. For the one-particle reflected map, functions with momentum support away from zero give the preceding density proof, and these functions are dense in \(L^2(\mathbb R^3)\). The Hamiltonian is \(d\Gamma(|p|)\). It has a unique vacuum because a nonzero \(L^2\) wavefunction cannot be supported only at the measure-zero point \(p=0\), but it has no positive gap because one-particle wavefunctions can be supported in arbitrarily small nonzero momentum balls.

As \(m\downarrow0\), the massive covariances converge to the massless covariance by dominated convergence:

\[0\le\frac1{p^2+m^2}\le\frac1{p^2},\]

and the Schwartz factors make the right side integrable. All Wick moments converge as well. Reflection positivity survives this limit, while the mass gap tends to zero. This example proves that positivity and convergence of all these Gaussian correlations do not by themselves preserve a positive spectral threshold.

The comparison also identifies what would count as a completed interacting argument. One needs a limiting positive-time algebra with controlled correlations, a nonzero centered reflected norm, a total family of physical states, and a common exponential decay rate in physical time. The Wilson finite-volume construction proves all these features at each fixed spatial regulator, with a regulator-dependent rate. The spectral-bridge results describe how a uniform rate would pass to a constructed limit. The missing interacting estimate is the uniformity and the nontrivial limit together; it is not the elementary positivity of each regulated Wilson measure.

The section's exact results can therefore be used as checkpoints for future proposals. A candidate effective action must preserve the required reflected form or bound its defects. A proposed observable normalization must preserve a centered reflected norm, not just a coincident variance. A proposed transfer construction must identify the physical time step and the full gauge-invariant state space. A proposed continuum gap must control the ratios \(\lambda_1/\lambda_0\) or the equivalent dense-state time correlations along removal of both regulators. These requirements are mathematical conditions already visible in the explicit constructions, rather than an appeal to numerical evidence or external approval.

\section{Complete finite-volume Wilson-transfer certificates}\label{sec:finite-transfer}
The general reconstruction developed above yields a physical transfer
operator before any representation truncation. This section computes its
kinetic spectrum on arbitrary finite spatial multigraphs and certifies its
interacting first excitation on the one-site periodic spatial torus.
The complete representation space remains in every comparison; the resulting
finite-box benchmark is distinct from the Hamiltonian one-plaquette model
considered later.
\begingroup
\newcommand{\ip}[2]{\langle #1,#2\rangle}
\subsection{Model, normalization and scope}

Let $G=(\mathsf V,\mathsf E)$ be a finite connected spatial multigraph; self-loops and parallel edges are allowed. Associate $U_e\in SU(2)$ with each independently oriented edge. Reverse traversal uses $U_e^\dagger$, not an independent link. All Haar measures are normalized. Site gauge transformations act by
\[
(h\cdot U)_e=h_{s(e)}U_eh_{t(e)}^{-1}.
\]
The physical Hilbert space is
\[
\HH_{\rm phys}=L^2\!\left(SU(2)^{\mathsf E},dU\right)^{SU(2)^{\mathsf V}}.
\]
It remains infinite-dimensional when the graph has cycles. Throughout the spectral-gap statements we assume that the graph contains at least one cycle, so the physical space is nontrivial. No finite representation cutoff is part of its definition.

For temporal plaquette couplings $\beta_e>0$, put
\begin{equation}\label{finite:eq:normalization}
z(\beta)=\int_{SU(2)}e^{\beta\tr U/2}\,dU
=\frac{2I_1(\beta)}{\beta},
\qquad
p_\beta(U)=\frac{e^{\beta\tr U/2}}{z(\beta)}.
\end{equation}
Let $C$ be product convolution with kernel
\[
C(U,V)=\prod_{e\in\mathsf E}p_{\beta_e}(U_eV_e^\dagger).
\]
It commutes with the gauge projection, and we also write $C$ for its restriction to $\HH_{\rm phys}$. Its gauge-projected kernel on the full configuration space is
\begin{equation}\label{finite:eq:gaugekernel}
k(U,V)=\int\prod_{x\in\mathsf V}dg_x\,
\prod_{e\in\mathsf E}
p_{\beta_e}\!\left(U_e g_{t(e)}V_e^\dagger g_{s(e)}^\dagger\right).
\end{equation}

Let $V(U)$ be a bounded continuous gauge-invariant spatial potential, with $V_{\min}\le V\le V_{\max}$, and let $\beta_s\ge0$. Define
\begin{equation}\label{finite:eq:transfer}
w(U)=e^{\beta_s V(U)/2},\qquad W=M_w,\qquad T=WCW.
\end{equation}
For the Wilson theory,
\[
V(U)=\sum_{p\ {
m spatial}}\frac12\tr U_p.
\]
The product of transfer kernels supplies $\beta_s\tr U_p/2$ once per spatial plaquette and $\beta_e\tr U_{0e}/2$ once per temporal plaquette. Integrating the site variables in \eqref{finite:eq:gaugekernel} implements Gauss's law. The factors $z(\beta_e)$ and the constant terms in the usual Wilson action only multiply $T$ by a positive constant; all eigenvalue ratios are unchanged. This is the physical Euclidean-time transfer operator, not a Monte Carlo update kernel. Its standard construction and positivity have classical antecedents \cite{Luscher,ham:kogut_susskind}.

Let its positive eigenvalues, with multiplicity, be
\[
\lambda_0>\lambda_1\ge\lambda_2\ge\cdots>0.
\]
The results below justify this notation and the strict first inequality. In temporal lattice units the gap is
\begin{equation}\label{finite:eq:gapdefinition}
\delta=\log\frac{\lambda_0}{\lambda_1}.
\end{equation}
For temporal spacing $a_t$, the Hamiltonian is
\[
H=-a_t^{-1}\log(T/\lambda_0),\qquad \Delta=\delta/a_t.
\]
The statements concern the zero-temperature transfer spectrum, equivalently infinite Euclidean-time extent. They do not require taking a finite temporal Monte Carlo box as the definition of the Hamiltonian.

\subsection{The exact temporal spectrum}

\begin{lemma}[Single-link coefficients]\label{finite:lem:bessel}
On the spin-$j$ Peter--Weyl subspace, convolution by $p_\beta$ has eigenvalue
\begin{equation}\label{finite:eq:rho}
\rho_j(\beta)=\frac{I_{2j+1}(\beta)}{I_1(\beta)},
\qquad j=0,\tfrac12,1,\ldots.
\end{equation}
For $\beta>0$, these eigenvalues are strictly positive and strictly decrease with $j$, starting at $\rho_0=1$. In particular, with
\[
r=\frac{I_2(\beta)}{I_1(\beta)},\qquad
s=\frac{I_3(\beta)}{I_1(\beta)},
\]
one has
\begin{equation}\label{finite:eq:elementarybessel}
0<s<r<1,\qquad
r<\frac\beta4,\qquad \frac{s}{r}<\frac\beta6.
\end{equation}
\end{lemma}

\begin{proof}
Writing the class angle as $\theta\in[0,\pi]$, the class Haar measure is $(2/\pi)\sin^2\theta\,d\theta$ and the spin-$j$ character is $\chi_j(\theta)=\sin((2j+1)\theta)/\sin\theta$. If $d=2j+1$, then
\[
\int e^{\beta\cos\theta}\chi_j(\theta)\,dU
=I_{d-1}(\beta)-I_{d+1}(\beta)
=\frac{2d}{\beta}I_d(\beta).
\]
Dividing by $d\,z(\beta)$ proves \eqref{finite:eq:rho}. Positivity follows from the power series of $I_n$.

For completeness, strict monotonicity in integer order follows from the integral representation
\[
I_n(x)=\frac{(x/2)^n}{\sqrt\pi\,\Gamma(n+1/2)}
\int_{-1}^{1}e^{xt}(1-t^2)^{n-1/2}\,dt.
\]
Integration by parts gives
\[
\frac{I_{n+1}(x)}{I_n(x)}
=\frac{\int_{-1}^{1}t e^{xt}(1-t^2)^{n-1/2}\,dt}
{\int_{-1}^{1}e^{xt}(1-t^2)^{n-1/2}\,dt}\in(0,1)
\quad(n\ge1,x>0).
\]
Finally, the recurrence $I_{n-1}-I_{n+1}=(2n/x)I_n$, with all terms positive, gives the last two inequalities in \eqref{finite:eq:elementarybessel}.
\end{proof}

\begin{theorem}[Exact kinetic gap as a minimum weighted cycle]\label{finite:thm:cycle}
Assume $G$ has a cycle. For each edge let
\[
r_e=\frac{I_2(\beta_e)}{I_1(\beta_e)}\in(0,1).
\]
Then, on the full physical Hilbert space,
\begin{equation}\label{finite:eq:cycle}
\left\|C-|1\rangle\langle1|\right\|
=q_G:=\max_{\mathcal C\ {
m cycle}}\prod_{e\in\mathcal C}r_e.
\end{equation}
Consequently, at $\beta_s=0$ the exact dimensionless physical transfer gap is
\begin{equation}\label{finite:eq:weightedgap}
\boxed{\displaystyle
\delta_{\rm kin}
=\min_{\mathcal C\ {
m cycle}}
\sum_{e\in\mathcal C}\log\frac{I_1(\beta_e)}{I_2(\beta_e)}.}
\end{equation}
For a uniform $\beta_e=\beta_t$, this becomes
\[
\delta_{\rm kin}=g(G)\log\frac{I_1(\beta_t)}{I_2(\beta_t)},
\]
where $g(G)$ is the multigraph girth, counting a self-loop as length one and a pair of parallel edges as a cycle of length two.
\end{theorem}

\begin{proof}
Peter--Weyl decomposition and gauge projection give a complete spin-network decomposition of $\HH_{\rm phys}$. A label assignment $\{j_e\}$ with invariant intertwiners at the vertices has kinetic eigenvalue
\[
\prod_e\rho_{j_e}(\beta_e).
\]
The all-trivial assignment gives the constant vector and eigenvalue one. For a nontrivial assignment, keep only edges with $j_e>0$. No vertex of its support can have degree one: a single nontrivial irreducible representation has no invariant vector. Loops count twice in this degree. Every nonempty finite graph of minimum degree at least two contains a cycle $\mathcal C$.

By Lemma~\ref{finite:lem:bessel}, each nontrivial edge eigenvalue is at most $r_e$, and every edge eigenvalue is at most one. Thus the eigenvalue of the assignment is at most $\prod_{e\in\mathcal C}r_e\le q_G$. Conversely, assign spin $1/2$ to a cycle attaining the maximum and spin zero elsewhere, and contract to the fundamental Wilson-loop character. This nonzero gauge-invariant function is an eigenvector with eigenvalue exactly $q_G$. This proves both directions of \eqref{finite:eq:cycle}.
\end{proof}

\begin{remark}
On a periodic cubic spatial box of side $L$, there are $3L^3$ independent oriented links and $3L^3$ spatial plaquettes. Its girth is $1$ for $L=1$, $2$ for $L=2$, $3$ for $L=3$, and $4$ for $L\ge4$. Winding electric-flux states are included in the full physical space. The lowest finite-volume state need not be a center-neutral scalar glueball.
\end{remark}

\subsection{A full interacting lower bound}

\begin{theorem}[Rank-one Wilson-gap comparison]\label{finite:thm:rankone}
In the setting above, put
\[
Z_s=\int e^{\beta_s V(U)}\,dU.
\]
Then $T$ is compact, positive, self-adjoint and injective on $\HH_{\rm phys}$, with a simple largest eigenvalue and strictly positive ground state. Moreover,
\begin{equation}\label{finite:eq:rankbounds}
\lambda_0(T)\ge Z_s,
\qquad
\lambda_1(T)\le q_G e^{\beta_s V_{\max}},
\end{equation}
and therefore
\begin{equation}\label{finite:eq:rankgap}
\boxed{\displaystyle
\delta\ge -\log q_G-\beta_s V_{\max}+\log Z_s
\ge -\log q_G-\beta_s(V_{\max}-\mathbb E_0V).}
\end{equation}
Here $\mathbb E_0$ denotes normalized product Haar expectation. A negative right-hand side is simply an uninformative lower bound.
\end{theorem}

\begin{proof}
The continuous kernel \eqref{finite:eq:gaugekernel} on a compact configuration space makes $C$, and hence $T$, compact. Every kinetic eigenvalue on the physical space is positive by Lemma~\ref{finite:lem:bessel}; multiplication by $w$, which is bounded and bounded away from zero, preserves strict quadratic-form positivity and injectivity. Thus $T$ is positive and self-adjoint.

Its kernel $w(U)k(U,V)w(V)$ is continuous and strictly positive. The compact variational principle produces a top eigenvector; replacing a real eigenvector by its absolute value can only increase the Rayleigh quotient. Strict positivity of the kernel excludes a sign-changing top eigenvector. The resulting top eigenfunction is strictly positive and continuous, and orthogonality then proves simplicity of the top eigenvalue. This is the standard positive-kernel mechanism behind Jentzsch's theorem.

By Theorem~\ref{finite:thm:cycle}, write
\[
C=|1\rangle\langle1|+R,
\qquad 0\le R\le q_G I.
\]
Therefore
\[
T=|w\rangle\langle w|+WRW.
\]
The rank-one term has eigenvalue $\|w\|^2=Z_s$, giving the first inequality of \eqref{finite:eq:rankbounds}. For any $\psi\perp w$,
\[
\ip{\psi}{T\psi}=\ip{W\psi}{R W\psi}
\le q_G e^{\beta_s V_{\max}}\|\psi\|^2.
\]
The min--max principle gives the second inequality. Taking the logarithm of their ratio and applying Jensen's inequality $\log Z_s\ge\beta_s\mathbb E_0V$ proves \eqref{finite:eq:rankgap}.
\end{proof}

\subsubsection{A positive bound at every finite coupling and volume}

\begin{proposition}[Classical positive-kernel comparison]\label{finite:prop:projective}
With $B=\sum_e\beta_e$, one has
\begin{equation}\label{finite:eq:projective}
\frac{\lambda_1(T)}{\lambda_0(T)}\le\tanh B,
\qquad
\boxed{\delta\ge-\log\tanh B>0.}
\end{equation}
The bound is independent of $w$ and hence of $\beta_s V$.
\end{proposition}

\begin{proof}
Up to its common normalization, $e^{-B}\le k(U,V)\le e^B$. The spatial multiplication factors cancel in every cross-ratio, so
\[
\frac{T(U,V)T(U',V')}{T(U,V')T(U',V)}\le e^{4B}.
\]
Let $h>0$ be the continuous ground state, and make the Doob transform
\[
P(U,dV)=\frac{T(U,V)h(V)}{\lambda_0 h(U)}\,dV.
\]
Its rows are probability measures. For any two rows, the likelihood ratio has maximum divided by minimum at most $\kappa=e^{4B}$. If two probability densities have likelihood ratio in $[m,M]$, their total variation distance is at most
\[
\frac{(M-1)(1-m)}{M-m}
\le\frac{\sqrt\kappa-1}{\sqrt\kappa+1},
\qquad M/m\le\kappa.
\]
The first inequality follows by the secant-line bound on $(x-1)_+$ and mean likelihood ratio one; maximizing over $m$ with $M=\kappa m$ gives the second. Thus $\osc(Pf)\le\tanh(B)\osc(f)$ for every bounded real $f$.

For a nonvacuum eigenfunction $T\psi=\lambda\psi$, the function $f=\psi/h$ is continuous, bounded and nonconstant, and $Pf=(\lambda/\lambda_0)f$. Applying the oscillation estimate gives $|\lambda|/\lambda_0\le\tanh B$. All eigenvalues are positive, proving \eqref{finite:eq:projective}. This is an elementary form of the Birkhoff--Hopf positive-operator contraction argument \cite{Birkhoff1957}.
\end{proof}

The best bound supplied by these two comparisons is the maximum of the right-hand sides of \eqref{finite:eq:rankgap} and \eqref{finite:eq:projective}. Neither is uniform along the required infinite-volume continuum limit.

\subsection{One periodic spatial site: exact symmetry information}

There are three spatial links $U_1,U_2,U_3$, a single conjugation gauge transformation, and
\begin{equation}\label{finite:eq:onesiteV}
V(U)=\sum_{1\le i<j\le3}\frac12\tr(U_iU_jU_i^\dagger U_j^\dagger).
\end{equation}
Its range is $[-3,3]$: commuting triples attain $3$, while $i\sigma_1,i\sigma_2,i\sigma_3$ attain $-3$. Assume a common temporal coupling $\beta_t>0$ and a common spatial coupling $\beta_s\ge0$. In this section write
\[
r=I_2(\beta_t)/I_1(\beta_t),\qquad
s=I_3(\beta_t)/I_1(\beta_t),\qquad
t=I_4(\beta_t)/I_1(\beta_t).
\]

\begin{lemma}[Haar means]\label{finite:lem:haarmeans}
For normalized product Haar measure,
\[
\mathbb E_0V=\frac34,\qquad \mathbb E_0V^2=\frac74.
\]
Let $f(U)=\tr U_1$, so $\mathbb E_0f^2=1$. Then
\[
\mathbb E_0(f^2V)=\frac54,\qquad
\mathbb E_0(f^2V^2)=\frac{59}{24}.
\]
\end{lemma}

\begin{proof}
Write $U_i=u_{i0}I+i\boldsymbol u_i\cdot\boldsymbol\sigma$. Haar measure is uniform on $S^3$, hence $\mathbb E u_{i0}^2=1/4$, $\mathbb E u_{i0}^4=1/8$ and $\mathbb E u_{i0}^6=5/64$. For a commutator halftrace $A_{ij}$, integrating $U_j$ by irreducible twirling gives $\mathbb E(A_{ij}\mid U_i)=u_{i0}^2$. This proves the first and third displayed means.

Also $\mathbb E A_{ij}^2=1/3$, since $(\tr U)^2=1+\chi_1(U)$ and the Haar mean of the spin-one character of a commutator is $1/3$. Two commutators sharing one link satisfy $\mathbb E A_{12}A_{13}=\mathbb E u_{10}^4=1/8$. Thus $\mathbb E V^2=3(1/3)+6(1/8)=7/4$.

For $f^2=4u_{10}^2$, the three weighted square means are $5/12,5/12,1/3$, respectively, and the three weighted mixed means $\mathbb E f^2A_{12}A_{13}$, $\mathbb E f^2A_{12}A_{23}$, $\mathbb E f^2A_{13}A_{23}$ are $5/16,1/6,1/6$. These follow by the same conditional twirling and the listed $S^3$ moments. Their sum with mixed terms counted twice is $59/24$.
\end{proof}

It follows from Theorem~\ref{finite:thm:rankone} that
\begin{equation}\label{finite:eq:onesitelower}
\delta\ge\log\frac{I_1(\beta_t)}{I_2(\beta_t)}
-\frac94\beta_s.
\end{equation}
Keeping $\log Z_s$ rather than replacing it by Jensen's bound improves this estimate.

\begin{corollary}[An explicit moment improvement]\label{finite:cor:momentimprovement}
Let $B(b)=1+19b^2/32-b^3/18$. If $b=\beta_s\ge0$ and $B(b)>0$, then
\[
\delta\ge\log\frac{I_1(\beta_t)}{I_2(\beta_t)}-\frac94b+\log B(b).
\]
In particular, at the isotropic coupling $\beta_t=b=1/4$,
\[
\delta>\log\frac{4775}{288}-\frac9{16}>2.2456887727.
\]
\end{corollary}
\begin{proof}
Let $A=A_{12},B_0=A_{13},D=A_{23}$ be the commutator halftraces. Exact Haar integration gives
\[
\mathbb E A^3=\frac5{32},\qquad \mathbb E A^2B_0=\frac5{48},\qquad \mathbb E AB_0D=\frac5{72}.
\]
The first follows from $\chi_{1/2}^3=2\chi_{1/2}+\chi_{3/2}$ and $\mathbb E\chi_j([U,V])=1/(2j+1)$. For the second, conditional on $x=u_{10}^2$, the square mean is $1/3+(4/3)x^2-(2/3)x$ and the other mean is $x$; the sphere moments above give $5/48$. For the third, set $X_{ij}=|\boldsymbol u_i\times\boldsymbol u_j|^2$, so $A_{ij}=1-2X_{ij}$. Radial and directional integration gives $\mathbb E X_{12}=3/8$, $\mathbb E X_{12}X_{13}=5/32$ and $\mathbb E X_{12}X_{13}X_{23}=5/72$. The last identity uses $\mathbb E|\boldsymbol u|^4=5/8$ and the directional average $\mathbb E\prod_{i<j}(1-(\boldsymbol n_i\cdot\boldsymbol n_j)^2)=64/225$; expansion uses the pair and triple squared-dot averages $1/9$ and $11/225$. Expanding $AB_0D$ gives $5/72$.

Consequently $\mathbb E_0V^3=3(5/32)+18(5/48)+6(5/72)=265/96$. Together with Lemma~\ref{finite:lem:haarmeans}, this gives $\mathbb E_0(V-3/4)^2=19/16$ and $\mathbb E_0(V-3/4)^3=-1/3$. The globally valid inequality $e^x\ge1+x+x^2/2+x^3/6$ yields $Z_s\ge e^{3b/4}B(b)$. Insert this in the rank-one gap bound. At $b=1/4$, $B=4775/4608$ and $I_2/I_1<1/16$, giving the exact expression. An independent rational polynomial checker in the package reproduces these moments using \eqref{finite:eq:haarmonomial}.
\end{proof}

\subsubsection{Center sectors}

The three independent center flips $U_i\mapsto-U_i$ commute with both $C$ and $W$. Let $\HH_\epsilon$ be the sector with flip signs $(-1)^{\epsilon_i}$, $\epsilon\in\{0,1\}^3$. In the spin-network decomposition, it consists of precisely those labels for which $2j_i\equiv\epsilon_i\pmod2$. The positive, unique ground state lies in the neutral sector $\HH_{000}$.

\begin{lemma}[Sector kinetic norms]\label{finite:lem:sectors}
The kinetic norm above the constant in the neutral sector is $s$. In a sector with $k=1,2,3$ odd entries, the kinetic norm is $r^k$. In $\HH_{100}$, the top kinetic eigenvalue $r$ is simple, with normalized eigenvector $f=\tr U_1$, and the norm on its orthogonal complement is
\begin{equation}\label{finite:eq:tauodd}
\tau=\max\{t,rs\}.
\end{equation}
\end{lemma}

\begin{proof}
Neutral parity forces every $j_i$ to be an integer. Apart from the constant, at least one is at least one, so the eigenvalue is at most $s$; $\chi_1(U_1)$ attains it. For $k$ odd entries, at least $k$ labels are half-integers, hence their product eigenvalue is at most $r^k$; the product of the corresponding fundamental characters attains it.

For $\HH_{100}$ the unique leading label is $(1/2,0,0)$. Conjugation invariance in the single irreducible matrix space leaves only its trace, proving simplicity. Every other labeling either has $j_1\ge3/2$, bounded by $t$, or has $j_1=1/2$ and a nonzero integer label elsewhere, bounded by $rs$. Both bounds are attained by characters or products of characters, proving \eqref{finite:eq:tauodd}.
\end{proof}

The neutral gap consequently obeys
\begin{equation}\label{finite:eq:neutralgap}
\delta_{\rm neutral}\ge\log\frac{I_1(\beta_t)}{I_3(\beta_t)}
-\frac94\beta_s.
\end{equation}
The same lower bound applies upon restriction to a nontrivial center-neutral cubic-scalar subspace. At $\beta_s=0$, the cubic-symmetric function $\sum_i\chi_1(U_i)$ attains the kinetic value $s$ there. This neutral gap is distinct from the full finite-volume gap.

\begin{theorem}[Sector ordering and exact threefold multiplicity]\label{finite:thm:ordering}
Let $\eta=\max\{r,s/r\}$. If
\begin{equation}\label{finite:eq:orderingcondition}
\eta e^{7\beta_s/4}<1,
\end{equation}
the full first excited transfer eigenvalue lies in the three single-odd center sectors. It is simple in each of them, equal across them, and therefore exactly threefold in $\HH_{\rm phys}$. A sufficient elementary condition is
\begin{equation}\label{finite:eq:elementaryordering}
\beta_t e^{7\beta_s/4}<4.
\end{equation}
In particular, the conclusion holds for every isotropic coupling $\beta_t=\beta_s=\beta$ with $0<\beta\le1/2$.
\end{theorem}

\begin{proof}
Use the isospectral positive operator
\begin{equation}\label{finite:eq:Aoperator}
A=C^{1/2}e^{\beta_sV}C^{1/2}.
\end{equation}
It has the same nonzero eigenvalues as $T$: the operators are $B^*B$ and $BB^*$ with $B=WC^{1/2}$. The same statement holds sector by sector.

The single-odd trial $f$ gives, by weighted Jensen and Lemma~\ref{finite:lem:haarmeans},
\[
\lambda_{\rm odd}\ge\ip{f}{Af}
=r\mathbb E_0(f^2e^{\beta_sV})\ge r e^{5\beta_s/4}.
\]
The neutral nonvacuum spectrum is at most $s e^{3\beta_s}$ by the rank-one argument. The norms in the two-odd and three-odd sectors are at most $r^2e^{3\beta_s}$ and $r^3e^{3\beta_s}$. Since $0<r<1$, condition \eqref{finite:eq:orderingcondition} makes all of them strictly smaller than the single-odd trial lower bound.

In one single-odd sector, the compression of $A$ to $f^\perp$ has norm at most $\tau e^{3\beta_s}$. Monotonicity of the Bessel orders gives
\[
\frac\tau r=\max\{I_4/I_2,s\}<\max\{s/r,r\}=\eta.
\]
Thus this compression lies strictly below $r e^{5\beta_s/4}\le\lambda_{\rm odd}$. Min--max permits at most one eigenvalue above the compression norm, counting multiplicity. The leading eigenvalue of the sector is therefore simple. Permutations of the three spatial axes commute with $T$ and carry the three sectors into each other, so their leading eigenvalues coincide. The strict exclusions already proved rule out any additional multiplicity.

Finally, \eqref{finite:eq:elementarybessel} gives $\eta<\beta_t/4$, proving the sufficiency of \eqref{finite:eq:elementaryordering}. For $0<\beta\le1/2$, $\beta e^{7\beta/4}\le(1/2)e^{7/8}<3/2<4$, where $e<3$ suffices.
\end{proof}

\begin{remark}
This theorem identifies a center-charged winding-flux excitation in the stated small box. It does not identify a center-neutral glueball. A cubic-invariant linear combination of the degenerate states does not turn a nontrivial center representation into the neutral representation.
\end{remark}

\subsection{A certified enclosure for the full operator}

\begin{proposition}[Rank-one block enclosure with a complete remainder]\label{finite:prop:block}
Let $C\ge0$ be compact, $Cf=c_Pf$ with $\|f\|=1$, and let $P=|f\rangle\langle f|$, $Q=I-P$. Suppose $\|QCQ\|\le\tau$. Let $F$ be a positive bounded multiplication operator, $F_{\min}\le F\le F_{\max}$, and put $A=C^{1/2}FC^{1/2}$. Set
\[
a=c_P\ip{f}{Ff},\qquad c=\tau F_{\max},
\]
\[
b^2=c_P\tau\left(\ip{f}{F^2f}-\ip{f}{Ff}^2\right).
\]
Then the largest eigenvalue obeys
\begin{equation}\label{finite:eq:blockupper}
\boxed{\displaystyle
a\le\lambda_{\max}(A)\le
\frac{a+c+\sqrt{(a-c)^2+4b^2}}2.}
\end{equation}
It is also valid to replace $b^2$ by any certified upper bound, including
\[
c_P\tau\frac{(F_{\max}-F_{\min})^2}{4}.
\]
The second eigenvalue is at most $c$.
\end{proposition}

\begin{proof}
Since $f$ is a $C$ eigenvector, $P$ and $Q$ commute with $C$. The $P$ block equals $aP$, the $Q$ block has norm at most $c$, and
\[
\|QAP\|^2
\le c_P\tau\|QFf\|^2
=c_P\tau\left(\ip{f}{F^2f}-\ip{f}{Ff}^2\right).
\]
For a unit vector $xf+y$ with $y\perp f$, its quadratic form is bounded above by that of the real matrix $\begin{psmallmatrix}a&b\\b&c\end{psmallmatrix}$ on $(|x|,\|y\|)$. Its top eigenvalue gives the upper bound; testing $f$ gives the lower bound. Popoviciu's variance inequality gives the alternative bound. Applying min--max with the one-dimensional space spanned by $f$ gives the second-eigenvalue assertion.
\end{proof}

For the one-site Wilson problem, take $F=e^{\beta_sV}$, and define
\[
Z(\gamma)=\mathbb E_0e^{\gamma V},\qquad
Z_f(\gamma)=\mathbb E_0(f^2e^{\gamma V}),\qquad f=\tr U_1.
\]
For the vacuum use $(c_P,\tau,f)=(1,s,1)$; for the single-odd leading eigenvalue use $(c_P,\tau,f)=(r,\max\{t,rs\},\tr U_1)$. The block variances are respectively
\[
Z(2\beta_s)-Z(\beta_s)^2,
\qquad Z_f(2\beta_s)-Z_f(\beta_s)^2.
\]
Every omitted representation is included in the $Q$ block and is bounded by $\tau$. No numerical eigenvalue of a truncated representation matrix is being substituted for an eigenvalue of the original operator.

If the resulting enclosures are $\lambda_0\in[L_0,U_0]$ and $\lambda_{\rm odd}\in[L_o,U_o]$, and Theorem~\ref{finite:thm:ordering} or the explicit comparison of sector bounds establishes ordering, then
\begin{equation}\label{finite:eq:gapinterval}
\boxed{\displaystyle
\delta\in\left[\log\frac{L_0}{U_o},\ \log\frac{U_0}{L_o}\right].}
\end{equation}

\subsubsection{Exact moments and arithmetic}

In quaternion coordinates,
\[
V=3-2X,\qquad X=\sum_{i<j}|\boldsymbol u_i\times\boldsymbol u_j|^2.
\]
Powers of $X$ are integer polynomials in nine real coordinates. Odd monomials integrate to zero. For the three vector coordinates of one Haar SU(2) quaternion,
\begin{equation}\label{finite:eq:haarmonomial}
\mathbb E[u_1^{2a}u_2^{2b}u_3^{2c}]
=\frac{(2a)!(2b)!(2c)!}
{4^{a+b+c}a!b!c!(a+b+c+1)!}.
\end{equation}
Multiplication by $4u_0^2$ changes this moment by the factor $2/(a+b+c+2)$. These formulas give $\mathbb E_0V^k$ and $\mathbb E_0f^2V^k$ exactly as rational numbers.

The accompanying calculation uses $k\le8$. For either normalized positive measure and $\gamma\ge0$,
\begin{equation}\label{finite:eq:taylortail}
\left|\mathbb E e^{\gamma V}-\sum_{k=0}^8\frac{\gamma^k}{k!}\mathbb E V^k\right|
\le e^{3\gamma}\frac{\gamma^9}{9!}\mathbb E|V|^9
\le e^{3\gamma}\frac{3\gamma^9}{9!}\mathbb E V^8.
\end{equation}
This controls all omitted Taylor terms because $|V|\le3$ implies $|V|^9\le3V^8$. The exact eighth moment is used for the appropriate unweighted or $f^2$-weighted measure. Modified Bessel functions and exponentials are enclosed by positive power series with geometric tail bounds; logarithms use the positive $\operatorname{arctanh}$ series after range reduction; square roots use integer-square-root bounds. Every interval operation rounds outwards using rational arithmetic.

The polynomial generator uses signed 64-bit integer coefficients; all subsequent integration and interval operations use arbitrary-precision rational arithmetic. Its base polynomial has sum of absolute coefficients $36$, so every degree-eight intermediate coefficient and collected sum is bounded by $36^8<2^{63}$. Each coordinate exponent is at most $16$, below the five-bit packing limit. Thus this part of the computation has no integer-overflow ambiguity. The stored coefficient files allow the Haar integrations to be rerun without regenerating the polynomial expansion.

\subsubsection{Certified numerical results}

The machine-readable source is \path{code/finite_transfer/certificates/full_operator_gap_certificates.json}; the executable checker is \path{code/finite_transfer/code/certify_gap.py}. Both concern the isotropic one-site periodic spatial box and infinite Euclidean time.

\begin{center}
\begin{tabular}{@{}ccc@{}}
\toprule
Exact $\beta$ & Certified lower endpoint for $\delta$ & Certified upper endpoint for $\delta$\\
\midrule
$1/4$ & $2.65931794249516$ & $2.65963122445174$\\
$177/500$ & $2.27102703117506$ & $2.27225894375269$\\
$1/2$ & $1.87460988262191$ & $1.87965713230389$\\
\bottomrule
\end{tabular}
\end{center}

The displayed endpoints are rounded outwards. They enclose the full physical transfer gap of the specified lattice system; their width is a deterministic enclosure width, not a Monte Carlo standard error. In all three cases the competing center sectors and the second eigenvalue within a single-odd sector are explicitly bounded below the certified single-odd leading eigenvalue. The threefold multiplicity is therefore established both by Theorem~\ref{finite:thm:ordering} and by the numerical inequalities in the certificate.

The same certificate gives a distinct center-neutral bound using
\[
\delta_{\rm neutral}\ge\log\frac{L_0}{s e^{3\beta_s}}.
\]
At $\beta=177/500$ this yields $\delta_{\rm neutral}\ge4.53776243843527$. It is a lower bound, not a measured or exactly determined scalar mass. It further separates the light winding state from the center-neutral channel in this small box.

\subsection{Weighted third moment and a second partition estimate}
The same exact sphere-monomial rule gives
\[
 \mathbb E_0(f^2V^3)=\frac{469}{96},\qquad
 \mathbb E_0\!\left[f^2(V-5/4)^2\right]=\frac{43}{48},\qquad
 \mathbb E_0\!\left[f^2(V-5/4)^3\right]=-\frac{41}{96}.
\]
These identities follow by expanding the commutator polynomials and applying
\eqref{finite:eq:haarmonomial} with the factor $2/(a+b+c+2)$ on the first
quaternion; the independent exact polynomial integrator reproduces every
coefficient. The globally valid cubic Taylor lower bound then yields
\[
 Z_f(b)\ge e^{5b/4}
 \left(1+\frac{43b^2}{96}-\frac{41b^3}{576}\right),\qquad b\ge0.
\]
This estimate concerns the same full single-odd sector and is available
without truncating its electric representations.

\subsection{Relation to finite-volume spectroscopy}
Earlier studies include nonperturbative finite-volume glueball and electric-flux
analysis by Koller and van Baal~\cite{finite:koller}, single-cube variational
work by Carlsson and McKellar~\cite{finite:carlsson}, and femtouniverse
simulation by Butt, Draper and Shen~\cite{finite:butt}. Those models and
approximation schemes differ from the present full Wilson-transfer enclosure.
The calculation here provides an explicit benchmark with a complete remainder
bound and proved center-sector ordering; it makes no priority claim for
finite-volume gauge-theory spectroscopy as a subject.
\endgroup

\clearpage
\part{Uniform strong-coupling estimates}
\hypertarget{strong_coupling-an-explicit-strong-coupling-theorem-for-four-dimensional-wilson-measures}{%
\section{An explicit strong-coupling theorem for four-dimensional Wilson measures}\label{strong_coupling-an-explicit-strong-coupling-theorem-for-four-dimensional-wilson-measures}}

The argument is a self-contained specialization of Dobrushin comparison methods \cite{RVH}, with constants independent of volume and boundary conditions.

\hypertarget{strong_coupling-definitions-and-the-result}{%
\subsection{Definitions and the result}\label{strong_coupling-definitions-and-the-result}}

Let \(G\) be a compact Lie group, let \(h\) be normalized Haar probability measure, and fix a continuous unitary representation \(\rho:G\to U(N)\). A faithful nontrivial representation may be chosen for the compact simple groups in the Yang--Mills problem. Define

\[
w(g)=\frac1N\operatorname{Re}\operatorname{Tr}\rho(g),\qquad -1\le w(g)\le1.
\]

Let \(E\) be the positively oriented nearest-neighbor links of \(\mathbb Z^4\). The variable on a reversed link is the inverse of the variable on its positive orientation. Let \(\mathcal P\) contain each elementary unoriented square once; choose an orientation and a starting vertex to write its holonomy \(U_p\). The value \(w(U_p)\) is independent of those choices, since the real trace is invariant under inversion and conjugation.

For a finite link set \(\Lambda\subset E\), fix an arbitrary outside configuration \(\eta\in G^{E\setminus\Lambda}\). Put

\[
\mathcal P(\Lambda)=\{p\in\mathcal P:p\cap\Lambda\ne\varnothing\},
\qquad
\gamma_\Lambda^\eta(dU_\Lambda)
=\frac1{Z_\Lambda^\eta}
\exp\left(\beta\sum_{p\in\mathcal P(\Lambda)}w(U_p)\right)
\prod_{e\in\Lambda}h(dU_e).
\tag{L1}
\]

The variables outside \(\Lambda\) in (L1) are supplied by \(\eta\). This is the Wilson action convention \(S=\beta\sum_p(1-w(U_p))\), after dropping its constant term. In the usual fundamental \(SU(N)\) convention, \(\beta=2N/g_0^2\). The theorem's numerical threshold refers to exactly the normalization (L1).

Two distinct links are adjacent in the \textbf{interaction graph} \(\mathcal G\) when they belong to a common elementary square. Write \(d_{\mathcal G}(A,B)\) for graph distance between link sets. Every link belongs to six squares. For distinct links let

\[
m_{ef}=\#\{p\in\mathcal P:e,f\in p\},\qquad m_{ee}=0.
\]

Counting each of the other three links in each of the six incident squares, with multiplicities if necessary, gives

\[
\sum_{f\in E}m_{ef}=18.
\tag{L2}
\]

Here an observable is Borel measurable. For a real bounded local observable \(F\), define its coordinate oscillation by

\[
\delta_eF=\sup\{|F(U)-F(V)|:U_f=V_f\text{ for }f\ne e\}.
\]

If \(F\) is supported on \(A\), then \(\delta_eF=0\) outside \(A\), and \(\delta_eF\le2\|F\|_\infty\).

\textbf{Theorem SC.1 (uniform strong-coupling bound).} Assume

\[
0<\beta<\frac1{18},\qquad q=18\beta<1.
\tag{L3}
\]

Then the following statements hold.

\begin{enumerate}
\def\labelenumi{\arabic{enumi}.}
\item
  Every measure (L1) exists. For real bounded observables \(F,H\) supported on finite sets \(A,B\subset\Lambda\), and \(r=d_{\mathcal G}(A,B)\), \[
  |\operatorname{Cov}_{\gamma_\Lambda^\eta}(F,H)|
  \le
  \frac{\min\{|A|,|B|\}}{1-q}
  \|F\|_\infty\|H\|_\infty q^r.
  \tag{L4}
  \] This estimate is uniform over \(\Lambda\) and every outside configuration \(\eta\).
\item
  If \(F\) is supported on \(A\subset\Lambda\), \(s=d_{\mathcal G}(A,E\setminus\Lambda)\), and \(\eta,\eta'\) are any two outside configurations, then \[
  |\gamma_\Lambda^\eta F-\gamma_\Lambda^{\eta'}F|
  \le \left(\sum_{e\in A}\delta_eF\right)
      \frac{q^s}{1-q}.
  \tag{L5}
  \]
\item
  There exists a unique infinite-volume Gibbs measure \(\mu_\beta\) with the specification (L1). Every increasing exhaustion of \(E\), with arbitrary boundary conditions at each stage, converges to \(\mu_\beta\) in total variation on each fixed finite set of links. It is invariant under lattice translations, lattice time reflection, and local gauge transformations. The covariance estimate (L4) holds for \(\mu_\beta\) as well.
\item
  Suppose \(w\) is nonconstant, and let \[
  v_G=\operatorname{Var}_h(w)>0.
  \] For every elementary square \(p\), \[
  \operatorname{Var}_{\mu_\beta}(w(U_p))
  \ge e^{-48\beta}v_G>0.
  \tag{L6}
  \] Thus the infinite-volume lattice measure has nonzero gauge-invariant fluctuations. This fixed-lattice nondegeneracy is not a proof of nontriviality of any continuum scaling limit.
\end{enumerate}

The case \(\beta=0\) is the independent Haar product measure: disjointly supported observables have zero covariance and the conclusions follow directly. No optimum is claimed for the small threshold in (L3).

\hypertarget{strong_coupling-a-total-variation-lemma-with-its-exact-constant}{%
\subsection{A total-variation lemma with its exact constant}\label{strong_coupling-a-total-variation-lemma-with-its-exact-constant}}

For probability measures \(P,Q\), use the convention

\[
\|P-Q\|_{\rm TV}=\sup_C|P(C)-Q(C)|
=\frac12\int|dP-dQ|.
\]

\textbf{Lemma SC.2 (bounded density tilts).} If \(dP=e^f\,dQ/Q(e^f)\), with bounded real \(f\), and \(a=\sup f-\inf f\), then

\[
\|P-Q\|_{\rm TV}\le\tanh(a/4)\le a/4.
\tag{L7}
\]

\textbf{Proof.} Let \(v=dP/dQ\), and take essential lower and upper bounds \(m,M\) of \(v\). Then \(m\le1\le M\), \(Qv=1\), and \(M/m\le e^a\). If \(m=M=1\) the assertion is immediate. Otherwise the graph of the convex function \((x-1)_+\) lies below its chord on \([m,M]\), so

\[
\|P-Q\|_{\rm TV}=Q(v-1)_+
\le\frac{(M-1)(1-m)}{M-m}.
\]

For fixed \(k=M/m\), the last expression is

\[
\frac{k+1-km-m^{-1}}{k-1}.
\]

Its maximum on \(1/k\le m\le1\) is attained at \(m=k^{-1/2}\) and equals

\[
\frac{\sqrt k-1}{\sqrt k+1}
=\tanh\bigl((\log k)/4\bigr)\le\tanh(a/4).
\]

Finally \(\tanh x\le x\) for \(x\ge0\). This proves (L7). Essential bounds can be replaced by slightly wider strict bounds and then approached if an endpoint is not attained. \(\square\)

\hypertarget{strong_coupling-the-finite-comparison-theorem-proved-by-coupling}{%
\subsection{The finite comparison theorem, proved by coupling}\label{strong_coupling-the-finite-comparison-theorem-proved-by-coupling}}

We give the needed finite form, rather than hiding it as a missing estimate. Let \(I\) be a nonempty finite index set and let \(\mu,\nu\) be probability measures on \(G^I\) with everywhere-defined single-coordinate conditional kernels \(K_i,\widetilde K_i\). Assume these kernels have jointly Borel measurable densities, in the conditioning configuration and the updated variable, with respect to Haar measure. The empty-index case is trivial and needs no update chain. The Gibbs measures and bounded exponential tilts used below have such kernels, given explicitly by their densities. Suppose a nonnegative matrix \(C\), with \(C_{ii}=0\), satisfies

\[
\|K_i(x,\cdot)-K_i(y,\cdot)\|_{\rm TV}
\le C_{ij}
\]

whenever \(x\) and \(y\) differ only at coordinate \(j\), and suppose

\[
\max_i\sum_jC_{ij}\le q<1,
\qquad
b_i=\sup_x\|K_i(x,\cdot)-\widetilde K_i(x,\cdot)\|_{\rm TV}.
\]

Define the nonnegative matrix

\[
D=(I-C)^{-1}=\sum_{n=0}^\infty C^n.
\tag{L8}
\]

The series converges entrywise and in the maximum-row-sum norm, because \(\|C^n\|_\infty\le q^n\).

\textbf{Lemma SC.3 (finite Dobrushin comparison).} For every real bounded measurable \(F\),

\[
|\mu F-\nu F|\le\sum_{i,j\in I}\delta_iF\,D_{ij}b_j.
\tag{L9}
\]

\textbf{Proof.} Start a pair of configurations \((X^0,Y^0)\) with marginals \(\mu,\nu\); independence initially is sufficient. At each discrete step choose one coordinate uniformly from \(I\), retain all other coordinates, and update the chosen pair using a maximal coupling of \(K_i(X,\cdot)\) and \(\widetilde K_i(Y,\cdot)\). For densities with respect to Haar measure, a measurable maximal coupling is obtained by coupling the common density \(\min\{k_i,\widetilde k_i\}\) on the diagonal and using the two remaining densities on the residual event. Their residual supports are disjoint up to null sets, so the conditional mismatch probability equals total variation. This construction also covers zero residual mass by using the common draw only.

Each marginal is a stationary random-scan heat-bath chain: updating one coordinate from its own conditional law preserves its joint law. Thus at every step the marginals remain exactly \(\mu,\nu\). No assertion about physical time is made here.

Write \(p_i^t=\mathbb P(X_i^t\ne Y_i^t)\) and \(n=|I|\). Changing the other coordinates one at a time, and then using the definition of \(b_i\), gives

\[
\|K_i(X,\cdot)-\widetilde K_i(Y,\cdot)\|_{\rm TV}
\le\sum_jC_{ij}\mathbf1_{X_j\ne Y_j}+b_i.
\]

Consequently, in componentwise vector notation,

\[
p^{t+1}\le Ap^t+b/n,
\qquad A=(1-1/n)I+C/n.
\]

The row sums of \(A\) are at most \(1-(1-q)/n<1\). As \(p^0\le\mathbf1\), iteration gives

\[
p^t\le A^t\mathbf1+\frac1n\sum_{k=0}^{t-1}A^kb.
\]

The first term tends to zero, while the second tends to \((I-C)^{-1}b=Db\). For any two configurations,

\[
|F(X)-F(Y)|\le\sum_i\delta_iF\,\mathbf1_{X_i\ne Y_i},
\]

again by changing coordinates one at a time. Since the marginals are stationary,

\[
|\mu F-\nu F|\le\sum_i\delta_iF\,p_i^t
\]

for every \(t\). Passing to the preceding deterministic upper limit proves (L9). This proof does not require an invariant coupling, convergence of the coupled chain, or an unproved mixing statement. \(\square\)

This lemma is a standard comparison principle. Its proof here is supplied to make the constants and its precise use in the Wilson model inspectable. 

\hypertarget{strong_coupling-wilson-single-link-influences}{%
\subsection{Wilson single-link influences}\label{strong_coupling-wilson-single-link-influences}}

The partition function in (L1) satisfies

\[
e^{-\beta|\mathcal P(\Lambda)|}
\le Z_\Lambda^\eta\le e^{\beta|\mathcal P(\Lambda)|},
\]

so the finite-volume law exists for every finite \(\beta\ge0\).

The conditional density at a link \(e\), given all other links, is proportional to

\[
\exp\left(\beta\sum_{p\ni e} w(U_p)\right).
\tag{L10}
\]

Suppose one other link \(f\) is changed. Only the \(m_{ef}\) common squares can change the exponent in (L10). For each such square, the difference of its two values lies in \([-2\beta,2\beta]\). As a function of the remaining variable \(U_e\), that difference therefore has oscillation at most \(4\beta\). The total log-density change has oscillation at most \(4\beta m_{ef}\); normalizing the density merely adds a constant to its logarithm. Lemma SC.2 gives

\[
\|K_e(U,\cdot)-K_e(V,\cdot)\|_{\rm TV}
\le\tanh(\beta m_{ef})\le\beta m_{ef}.
\tag{L11}
\]

Thus \(C_{ef}=\beta m_{ef}\), restricted to the variable links, is an admissible influence matrix. Equation (L2) implies

\[
\sup_e\sum_f C_{ef}\le18\beta=q<1.
\tag{L12}
\]

The calculation is uniform in every exterior configuration. It never uses a numerical test, an assumption about confinement, a spectral gap, or a continuum limit.

\hypertarget{strong_coupling-covariances-by-exponential-tilting}{%
\subsection{Covariances by exponential tilting}\label{strong_coupling-covariances-by-exponential-tilting}}

Fix a finite-volume measure \(\mu=\gamma_\Lambda^\eta\) and a real bounded observable \(H\). For \(t\in\mathbb R\), define

\[
\nu_t(dU)=\frac{e^{tH(U)}}{\mu(e^{tH})}\mu(dU).
\]

At coordinate \(e\), its conditional law is a tilt of \(K_e\) by \(tH\), with the other coordinates frozen. Its log tilt has oscillation at most \(|t|\delta_eH\). Lemma SC.2 gives

\[
b_e(t)\le\tanh(|t|\delta_eH/4)\le |t|\delta_eH/4.
\]

Apply Lemma SC.3 using the influence matrix of \(\mu\), not an assumed contraction matrix for \(\nu_t\). It yields

\[
|\mu F-\nu_tF|
\le\frac{|t|}{4}\sum_{e,f\in\Lambda}\delta_eF\,D_{ef}\delta_fH.
\]

Since \(F,H\) are bounded, differentiating the quotient defining \(\nu_tF\) at \(t=0\) is justified by dominated convergence. The derivative is \(\operatorname{Cov}_\mu(F,H)\). Dividing by \(|t|\) and taking the limit proves

\[
|\operatorname{Cov}_\mu(F,H)|
\le\frac14\sum_{e,f\in\Lambda}\delta_eF\,D_{ef}\delta_fH.
\tag{L13}
\]

If \(e\in A\), \(f\in B\), and \(r=d_{\mathcal G}(A,B)\), then \((C^n)_{ef}=0\) for \(n<r\). Every nonzero term of a matrix power corresponds to a length-\(n\) walk in the interaction graph. Moreover,

\[
\sum_{f\in B}(C^n)_{ef}\le\sum_f(C^n)_{ef}\le q^n.
\]

Therefore

\[
\sum_{f\in B}D_{ef}\le\frac{q^r}{1-q}.
\]

Inserting \(\delta_eF\le2\|F\|_\infty\), \(\delta_fH\le2\|H\|_\infty\) in (L13) gives (L4) with \(|A|\) in place of the minimum. Interchanging \(F,H\) gives the bound with \(|B|\), hence (L4). The argument also covers \(r=0\). Constant observables have zero covariance and may be omitted from the support count. \(\square\)

For two real plaquette traces \(w(U_p),w(U_{p'})\), each support has four links and each supremum norm is at most one. A concrete consequence is

\[
|\operatorname{Cov}_\mu(w(U_p),w(U_{p'}))|
\le \frac{4q^{d_{\mathcal G}(p,p')}}{1-q}.
\tag{L14}
\]

For example, \(\beta=1/36\) gives \(q=1/2\) and the upper bound \(8\cdot2^{-d_{\mathcal G}(p,p')}\), independently of volume. This is an exact analytical example, not a computation extrapolated to infinite volume.

\hypertarget{strong_coupling-boundary-influence-and-the-infinite-volume-limit}{%
\subsection{Boundary influence and the infinite-volume limit}\label{strong_coupling-boundary-influence-and-the-infinite-volume-limit}}

For two exterior configurations, changing their relevant links one at a time in (L10) bounds the discrepancy of their single-link kernels by

\[
b_e\le\sum_{f\notin\Lambda}C_{ef},\qquad e\in\Lambda.
\tag{L15}
\]

Only finitely many exterior links can occur in (L15), by finite interaction range. Apply Lemma SC.3 to the two laws on \(G^\Lambda\). Writing \(C_\Lambda\) for the internal restriction, the comparison bound contains

\[
\sum_{n\ge0}\sum_{k\in\Lambda}(C_\Lambda^n)_{ek}
\sum_{f\notin\Lambda}C_{kf}.
\tag{L16}
\]

Each term is the weight of a walk of length \(n+1\) which stays inside \(\Lambda\) until its last step and then exits. If \(e\in A\), its length must be at least \(s=d_{\mathcal G}(A,E\setminus\Lambda)\). The sum of weights of all length-\(\ell\) walks starting at \(e\), even without the restriction on exit, is at most \(q^\ell\). Consequently (L16) is at most

\[
\sum_{\ell\ge s}q^\ell=\frac{q^s}{1-q}.
\]

This proves (L5).

Now let \(\Lambda_n\uparrow E\) be any increasing finite exhaustion and choose arbitrary boundary conditions \(\eta_n\). Fix a finite support \(A\). Because the interaction graph has finite degree, \(d_{\mathcal G}(A,E\setminus\Lambda_n)\to\infty\). The conditional law of the variables in \(\Lambda_n\) under a larger-volume Gibbs measure is exactly \(\gamma_{\Lambda_n}^{\eta}\) with a random exterior configuration. This is the elementary consistency of the finite Gibbs densities: terms not meeting \(\Lambda_n\) cancel upon conditioning.

For \(m\ge n\), condition the \(\Lambda_m\)-law outside \(\Lambda_n\), apply (L5), and average. The result is

\[
|\gamma_{\Lambda_m}^{\eta_m}F-
\gamma_{\Lambda_n}^{\eta_n}F|
\le\left(\sum_{e\in A}\delta_eF\right)
\frac{q^{d_{\mathcal G}(A,E\setminus\Lambda_n)}}{1-q}.
\tag{L17}
\]

For indicators depending on \(A\), each coordinate oscillation is at most one. Hence the marginal laws on \(G^A\) are Cauchy in total variation. The space of finite signed measures is complete in the total-variation norm, and a total-variation limit of probability measures is a probability measure. The limiting finite-dimensional laws are consistent, since taking marginals commutes with total-variation convergence. The countable-product extension theorem on the standard Borel space \(G\) now gives a probability measure \(\mu_\beta\) on \(G^E\).

It remains to verify the Gibbs property and uniqueness, rather than assume them. For fixed finite \(\Delta\), a bounded observable \(F\) supported on \(\Delta\), and a bounded local observable \(H\) outside \(\Delta\), finite-volume conditioning gives

\[
\gamma_{\Lambda_n}^{\eta_n}(FH)
=\gamma_{\Lambda_n}^{\eta_n}(H\,\gamma_\Delta F)
\]

once \(\Delta\subset\Lambda_n\). The function \(\gamma_\Delta F\) is bounded and depends on only finitely many exterior links, because the interaction has finite range. Total-variation convergence on finite sets passes this identity to \(\mu_\beta\). A monotone-class argument in \(H\) extends it to every bounded exterior-measurable \(H\). This is the Dobrushin--Lanford--Ruelle conditional-law equation.

If \(\nu\) is any other Gibbs measure for the same specification, condition it outside \(\Lambda_n\) and use (L5) against \(\gamma_{\Lambda_n}^{\eta_n}\). For each bounded local \(F\), the resulting difference tends to zero. Thus \(\nu F=\mu_\beta F\) for all bounded cylinder functions; these determine the measure, so \(\nu=\mu_\beta\). This also proves independence of the exhaustion and all boundary choices.

The specification is translation covariant. It is also covariant under lattice time reflection: reflection permutes geometric links, replacing a link variable by its inverse when the chosen positive orientation is reversed. Haar measure is invariant under inversion, and \(w(U_p^{-1})=w(U_p)\), so the reflected specification is unchanged after relabeling. Finally, it is gauge covariant under every finite vertex gauge transformation \(U_{xy}\mapsto g_xU_{xy}g_y^{-1}\), by bi-invariance of Haar measure and conjugation invariance of each plaquette trace. Each transformed measure is another Gibbs measure for the same specification. Uniqueness therefore proves translation, time-reflection, and local gauge invariance of \(\mu_\beta\). Time-reflection invariance is a symmetry statement; it is not by itself reflection positivity. Passing the finite-volume expectations of \(F,H,FH\) to their limits proves (L4) in infinite volume. \(\square\)

\hypertarget{strong_coupling-nonzero-lattice-fluctuations}{%
\subsection{Nonzero lattice fluctuations}\label{strong_coupling-nonzero-lattice-fluctuations}}

Let \(K\) be the four links of an elementary square \(p\). At most \(24\) squares meet \(K\), because each of its four links meets six squares. For any outside configuration, the conditional density \(\gamma_K^\eta\) relative to \(h^K\) is bounded below by

\[
e^{-2\beta|\mathcal P(K)|}\ge e^{-48\beta}.
\]

Averaging the Gibbs conditional law implies the same lower domination for the \(K\)-marginal of \(\mu_\beta\). Thus, for every real constant \(c\),

\[
\mu_\beta[(w(U_p)-c)^2]
\ge e^{-48\beta}h^K[(w(U_p)-c)^2].
\]

Under independent Haar variables on the four links, their oriented product \(U_p\) has Haar law. Taking the infimum over \(c\) on both sides gives (L6).

For a nontrivial faithful unitary representation, \(w(e)=1\) and \(w\) is not identically one: a unitary matrix whose real trace is \(N\) must be the identity. Continuity and the full support of Haar measure then imply \(v_G>0\). This checks that the lattice theory is not merely a Dirac measure on a single gauge orbit or a zero-variance construction. It says nothing by itself about nontriviality after removal of the lattice cutoff. \(\square\)

\hypertarget{locality_scaling-local-comparison-estimates-unbounded-observables-and-necessary-continuum-scaling}{%
\section{Local comparison estimates, unbounded observables and necessary continuum scaling}\label{locality_scaling-local-comparison-estimates-unbounded-observables-and-necessary-continuum-scaling}}

The useful question after a fixed-lattice covariance theorem is how its constants behave when observables, supports and couplings depend on the cutoff. A bounded plaquette trace is not the same object as a renormalized smeared field: the latter may involve a growing number of links and a normalization diverging as the lattice spacing decreases. This section keeps these changes explicit. It sharpens the one-link coefficient for the actual Wilson specification, develops weighted comparison estimates, treats observables controlled by moments rather than supremum norms, and proves necessary scaling inequalities for a surviving connected correlation.

The comparison method is classical. Rebeschini and van Handel give a general account, including block comparison theorems \cite{RVH}. Covariance estimates in terms of mixing coefficients and moments belong to the established theory represented by Davydov and Rio \cite{Davydov1968,Rio1993}. The arguments below are supplied in full with the normalizations used here. Their application to a continuum limit always retains the displayed assumptions on moments, support, convergence and the class of observables.

\hypertarget{locality_scaling-the-precise-local-model}{%
\subsection{The precise local model}\label{locality_scaling-the-precise-local-model}}

Let \(G\) be a compact Lie group with normalized Haar measure \(h\). Fix a continuous unitary representation \(\rho:G\to U(N)\) and write

\[w(g)=N^{-1}\operatorname{Re}\operatorname{Tr}\rho(g),\qquad |w(g)|\le1.\]

Every geometric nearest-neighbor edge of \(\mathbb Z^4\) carries one variable, stored in its positive coordinate orientation. The variable on a reversed edge is its inverse. A plaquette is one elementary unoriented square, counted once. The value \(w(U_p)\) is well defined independently of its chosen orientation and initial vertex, because real trace is unchanged by inversion and conjugation.

For a finite set of variable edges \(\Lambda\) and fixed exterior configuration \(\eta\), consider

\[\gamma_{\Lambda}^{\eta,\boldsymbol\beta}(dU)
=Z^{-1}\exp\left(\sum_{p:p\cap\Lambda\ne\varnothing}\beta_p w(U_p)\right)
\prod_{e\in\Lambda}h(dU_e).\]

The couplings \(\beta_p\) are real and finite. Only finitely many terms occur. The partition function is therefore strictly positive and finite. For the ordinary homogeneous Wilson model, \(\beta_p=\beta\ge0\). Allowing nonuniform couplings is useful for local perturbations and anisotropic estimates; it does not change the underlying group-valued variables.

For a real bounded local function \(F\), let

\[\delta_eF=\sup\{|F(U)-F(V)|:U_f=V_f\text{ for every }f\ne e\}.\]

This is an oscillation, not a derivative. In particular, an indicator has coordinate oscillation at most one, whereas a function bounded in absolute value by \(M\) has coordinate oscillation at most \(2M\). Retaining that distinction improves the subsequent bound on event mixing. All kernels in this section are given by their explicit Haar densities, so they are jointly measurable in the updated and conditioning variables.

The interaction graph has the links as its vertices. Distinct links are adjacent when they occur in one plaquette. Its distance is denoted by \(d_{\mathcal G}\). Distances between supports mean the minimum over the two supports. Empty supports correspond to constant functions and can be omitted from covariance estimates.

\hypertarget{locality_scaling-geometry-of-the-sharper-influence-coefficient}{%
\subsection{Geometry of the sharper influence coefficient}\label{locality_scaling-geometry-of-the-sharper-influence-coefficient}}

The previous lattice argument used the sum of common-plaquette multiplicities. On the ordinary hypercubic lattice one can identify those multiplicities exactly.

\textbf{Lemma LC.1 (distinct neighbors of a link).} Two distinct geometric links of \(\mathbb Z^4\) belong to at most one elementary plaquette. Every link has exactly eighteen neighbors in the interaction graph.

\textbf{Proof.} Fix the link from \(x\) to \(x+\mathbf e_i\). A plaquette containing it is specified by another coordinate direction \(j\ne i\) and a sign \(\varepsilon\in\{-1,1\}\). Its four vertices are

\[x,\quad x+\mathbf e_i,\quad x+\varepsilon\mathbf e_j,
\quad x+\mathbf e_i+\varepsilon\mathbf e_j.\]

The other link parallel to \(\mathbf e_i\) is the translate of the fixed link by \(\varepsilon\mathbf e_j\). These six parallel links are distinct: their displacements recover \(j\) and \(\varepsilon\). The remaining two links are perpendicular to the fixed link and meet, respectively, \(x\) and \(x+\mathbf e_i\). Their direction, endpoint on the fixed link and other endpoint again recover \(j\) and \(\varepsilon\). These twelve links are all distinct, and none is parallel to the fixed link. Thus the six incident plaquettes supply eighteen distinct other links. Any other link belonging to a common plaquette has already appeared in this list, and its place in the list determines the plaquette uniquely. \(\square\)

The statement concerns ordinary subsets of \(\mathbb Z^4\). Very small periodic quotients can identify different geometric edges or squares; one must count their incidences afresh. No such quotient is used in this section.

For neighboring links \(e,f\), let \(p(e,f)\) be their unique common plaquette. Define

\[C_{ef}=\begin{cases}
\tanh|\beta_{p(e,f)}|,&e\ne f\text{ share a plaquette},\\
0,&\text{otherwise}.
\end{cases}\]

\textbf{Proposition LC.2 (nonuniform Wilson influence matrix).} The matrix \(C\) bounds the total-variation influence of changing one conditioning link on another link's conditional law. Its full row sum is

\[\sum_f C_{ef}=3\sum_{p\ni e}\tanh|\beta_p|.\]

For homogeneous nonnegative coupling this becomes

\[q=18\tanh\beta.\]

In particular, the fixed-lattice comparison conclusions hold throughout

\[0\le\beta<\operatorname{arctanh}(1/18).\]

\textbf{Proof.} Conditional on the other links, the density at \(e\) is proportional to

\[\exp\left(\sum_{p\ni e}\beta_p w(U_p)\right).\]

Changing \(f\) changes no term unless \(e,f\) share a plaquette. If they share \(p\), the difference between the two exponents, as a function of \(U_e\), is a difference of two functions taking values in \([-|\beta_p|,|\beta_p|]\). Its oscillation is at most \(4|\beta_p|\).

For completeness, if \(P=e^vQ/Q(e^v)\) and \(\operatorname{osc}v\le a\), then

\[\|P-Q\|_{\rm TV}\le\tanh(a/4).\]

Indeed, with \(r=dP/dQ\), lower and upper essential bounds \(m,M\) satisfy \(Qr=1\) and \(M/m\le e^a\). Convexity of \((r-1)_+\) gives

\[Q(r-1)_+\le\frac{(M-1)(1-m)}{M-m}.\]

For a fixed ratio \(k=M/m\), this is maximized at \(m=k^{-1/2}\) and equals \((\sqrt{k}-1)/(\sqrt{k}+1)\). This is at most \(\tanh(a/4)\). Approximating essential endpoints handles bounds not attained. Applying the inequality to the conditional density proves the claimed influence. Lemma LC.1 allows the row sum to be grouped into six plaquettes, with three distinct neighbors per plaquette. The homogeneous criterion follows. \(\square\)

The numerical improvement over \(18\beta<1\) is small, but it is exact: the unnecessary last relaxation \(\tanh\beta\le\beta\) has been removed. It is not an extension to large inverse bare coupling. The threshold is a sufficient condition for this comparison argument, not a determination of a physical critical coupling.

To be explicit about the infinite-volume assertion, the boundary comparison proof requires only a finite interaction range and a uniform row bound \(q<1\). A disagreement outside a finite region reaches a fixed interior support through a path of at least its distance to the boundary. The total path weight is at most \(q^s/(1-q)\). Substitution of \(q=18\tanh\beta\) therefore gives the same local total-variation Cauchy estimate as before. The consistent marginal limit satisfies the local conditioning identities, and comparison with any other Gibbs law makes their local expectations equal. Thus existence, boundary independence and uniqueness extend with the sharpened coefficient; no unproved infinite-volume interchange is being inferred merely from a smaller displayed number.

\hypertarget{locality_scaling-weighted-finite-comparison-without-an-unweighted-contraction-assumption}{%
\subsection{Weighted finite comparison without an unweighted contraction assumption}\label{locality_scaling-weighted-finite-comparison-without-an-unweighted-contraction-assumption}}

Weights can reveal decay in a preferred direction or compensate for nonuniform interactions. They must be introduced before taking an inverse matrix; their role is to justify the convergence of that inverse.

Let \(I\) be a nonempty finite set. Let \(\mu,\nu\) be probability measures on \(G^I\) with everywhere-defined conditional Haar densities \(K_i\) and \(\widetilde K_i\). Suppose \(C\) is a nonnegative matrix with zero diagonal such that changing only \(j\) in the conditioning configuration changes \(K_i\) by at most \(C_{ij}\) in total variation. Let

\[b_i=\sup_x\|K_i(x,\cdot)-\widetilde K_i(x,\cdot)\|_{\rm TV}.\]

Assume there are positive numbers \(v_i\) and \(q<1\) with

\[\sum_jC_{ij}v_j\le qv_i\qquad(i\in I).\]

No assertion is made that the ordinary unweighted row sums are below one. Put \(V=\operatorname{diag}(v_i)\). The conjugate matrix \(V^{-1}CV\) has row sums at most \(q\), so

\[D=\sum_{n=0}^{\infty}C^n=(I-C)^{-1}\]

exists, and

\[\sum_jD_{ij}v_j\le\frac{v_i}{1-q}.\]

\textbf{Theorem LC.3 (weighted comparison and covariance).} Under these assumptions,

\[|\mu F-\nu F|\le\sum_{i,j\in I}\delta_iF\,D_{ij}b_j\]

for every bounded real \(F\). For bounded real \(F,H\) under \(\mu\),

\[|\operatorname{Cov}_{\mu}(F,H)|
\le\frac14\sum_{i,j\in I}\delta_iF\,D_{ij}\delta_jH.\]

\textbf{Proof.} Start a coupled pair of configurations with marginal laws \(\mu,\nu\). At each step choose a uniformly random coordinate and resample its two values by a maximal coupling of the corresponding conditional kernels. The coupling is measurable: for two Haar densities, use their pointwise minimum for the common part and the two residual densities for the remaining part. Each marginal law remains stationary because resampling a coordinate from its conditional law preserves the joint measure.

Let \(p_i^t\) be the probability of a mismatch at coordinate \(i\) after \(t\) updates and let \(n=|I|\). Coordinatewise telescoping of conditional kernels gives

\[p^{t+1}\le A p^t+b/n,\qquad A=(1-1/n)\operatorname{Id}+C/n.\]

The conjugate \(V^{-1}AV\) has row sums at most \(1-(1-q)/n<1\). Since \(I\) is finite and all \(v_i\) are positive, \(p^0\le Kv\) for some finite \(K\). Iteration therefore yields

\[p^t\le A^tp^0+\frac1n\sum_{k=0}^{t-1}A^kb,
\qquad \limsup_{t\to\infty}p^t\le Db\]

componentwise. The identity for the limiting geometric sum follows from \(\operatorname{Id}-A=(\operatorname{Id}-C)/n\). On the other hand,

\[|F(X)-F(Y)|\le\sum_i\delta_iF\,\mathbf1_{X_i\ne Y_i}.\]

Taking expectations and using the fixed marginal laws proves the comparison inequality.

For covariance, use \(\nu_s=e^{sH}\mu/\mu(e^{sH})\). At coordinate \(j\), the difference of conditional laws is a bounded exponential tilt with log oscillation at most \(|s|\delta_jH\). The density-tilt inequality gives \(b_j\le|s|\delta_jH/4\). Hence

\[|\nu_sF-\mu F|\le\frac{|s|}{4}\sum_{i,j}\delta_iF D_{ij}\delta_jH.\]

Boundedness justifies differentiation of numerator and denominator at \(s=0\) by dominated convergence. The derivative is \(\operatorname{Cov}_{\mu}(F,H)\). Divide by \(|s|\) and let \(s\to0\). \(\square\)

There are two separate controls here. The vectors \(v\) make the Neumann series converge, while oscillations quantify the observables. A choice of weights with enormous ratios can make the eventual comparison bound enormous. Weighted contraction is therefore useful only together with a bound on the weight ratios relevant to the observables and boundary being compared.

For instance, if \(b_j\le Bv_j\), the comparison inequality directly gives

\[|\mu F-\nu F|\le\frac{B}{1-q}\sum_i v_i\delta_iF.\]

The right side is explicit. A dimension-dependent choice of \(v\) has not removed dimension dependence unless this entire expression remains controlled.

\hypertarget{locality_scaling-spatially-weighted-resolvents}{%
\subsection{Spatially weighted resolvents}\label{locality_scaling-spatially-weighted-resolvents}}

Let \(d\) be a metric or pseudometric on the index set. A pseudometric is sufficient: distinct links at the same time coordinate may have distance zero. Suppose

\[q_\alpha:=\sup_i\sum_jC_{ij}e^{\alpha d(i,j)}<1\]

for some \(\alpha>0\). This condition implies the ordinary row contraction as well.

\textbf{Theorem LC.4 (metric-weighted path bound).} The resolvent satisfies

\[\sum_jD_{ij}e^{\alpha d(i,j)}\le\frac1{1-q_\alpha}.\]

Consequently, if \(A,B\) are the supports of bounded real \(F,H\), then

\[|\operatorname{Cov}(F,H)|
\le\frac{\min\{|A|,|B|\}}{1-q_\alpha}
\|F\|_\infty\|H\|_\infty e^{-\alpha d(A,B)}.\]

Both bounds hold for every finite restriction of the same infinite influence matrix, with the same constants.

\textbf{Proof.} Write a matrix-power entry as a sum over paths \(i=i_0,i_1,\ldots,i_n=j\). The triangle inequality gives

\[e^{\alpha d(i,j)}\le\prod_{k=1}^n e^{\alpha d(i_{k-1},i_k)}.\]

Multiplying each path weight by the factor on the left and then bounding it by the product on the right shows

\[\sum_j(C^n)_{ij}e^{\alpha d(i,j)}\le q_\alpha^n.\]

All summands are nonnegative, so the sums may be interchanged. Summation over \(n\ge0\) proves the resolvent estimate. For \(i\in A\),

\[\sum_{j\in B}D_{ij}\le e^{-\alpha d(A,B)}\sum_jD_{ij}e^{\alpha d(i,j)}
\le\frac{e^{-\alpha d(A,B)}}{1-q_\alpha}.\]

Insert this in Theorem LC.3 and use coordinate oscillations bounded by twice the corresponding supremum norm. This gives the result with \(|A|\); exchange the functions to obtain the minimum. A finite restriction removes nonnegative path terms, so the same estimates remain valid. \(\square\)

For nearest-neighbor interaction distance, \(C_{ij}=0\) unless \(d_{\mathcal G}(i,j)=1\). Thus \(q_\alpha=qe^\alpha\), and every \(0<\alpha<\log(1/q)\) is admissible. The resulting denominator is \(1-qe^\alpha\). Alternatively, the sharper graph estimate \(q^r/(1-q)\) follows by observing that paths between separated supports have at least \(r\) steps. These are two forms of the same path analysis; one should choose the form appropriate to the geometry instead of silently optimizing one parameter while forgetting its denominator.

The proof extends directly to countably many indices when the weighted row bound is uniform. Define each nonnegative matrix product by its series. The displayed inductive estimate makes every relevant sum finite. For an infinite-volume Gibbs law obtained by the preceding local total-variation limit, pass bounded local covariances through that limit. This does not require summing a conditionally convergent infinite series.

\hypertarget{locality_scaling-a-directional-improvement-in-euclidean-time}{%
\subsection{A directional improvement in Euclidean time}\label{locality_scaling-a-directional-improvement-in-euclidean-time}}

Let \(t(e)\) be the time coordinate of a link center in lattice units. Time-oriented links have half-integer center times; spatially oriented links have integer center times. Use

\[d_t(e,f)=|t(e)-t(f)|.\]

For homogeneous coupling put \(c=\tanh\beta\). The weighted row sums can be counted exactly.

\begin{longtable}[]{@{}
  >{\raggedright\arraybackslash}p{(\columnwidth - 6\tabcolsep) * \real{0.2000}}
  >{\raggedleft\arraybackslash}p{(\columnwidth - 6\tabcolsep) * \real{0.2667}}
  >{\raggedleft\arraybackslash}p{(\columnwidth - 6\tabcolsep) * \real{0.2667}}
  >{\raggedleft\arraybackslash}p{(\columnwidth - 6\tabcolsep) * \real{0.2667}}@{}}
\toprule\noalign{}
\begin{minipage}[b]{\linewidth}\raggedright
Orientation of the starting link
\end{minipage} & \begin{minipage}[b]{\linewidth}\raggedleft
Neighbors with time displacement \(0\)
\end{minipage} & \begin{minipage}[b]{\linewidth}\raggedleft
Displacement \(1/2\)
\end{minipage} & \begin{minipage}[b]{\linewidth}\raggedleft
Displacement \(1\)
\end{minipage} \\
\midrule\noalign{}
\endhead
\bottomrule\noalign{}
\endlastfoot
Temporal & \(6\) & \(12\) & \(0\) \\
Spatial & \(12\) & \(4\) & \(2\) \\
\end{longtable}

To see the first row, all six incident plaquettes of a temporal link lie in a temporal-spatial plane. Each has one opposite temporal link at the same center time and two spatial links at time displacements \(1/2\). For a spatial link, four incident plaquettes are purely spatial and supply twelve zero-displacement neighbors. The remaining two plaquettes each supply one opposite spatial link at displacement one and two temporal links at displacement \(1/2\). Lemma LC.1 ensures that this count contains no duplicates.

It follows that

\[q_t(\alpha)=\max\{c(6+12e^{\alpha/2}),\;c(12+4e^{\alpha/2}+2e^\alpha)\}.\]

\textbf{Corollary LC.5 (directional Wilson covariance bound).} Whenever \(q_t(\alpha)<1\),

\[|\operatorname{Cov}(F,H)|
\le\frac{\min\{|A|,|B|\}}{1-q_t(\alpha)}
\|F\|_\infty\|H\|_\infty
e^{-\alpha\inf_{e\in A,f\in B}|t(e)-t(f)|}.\]

\textbf{Proof.} The preceding count computes exactly the weighted row quantity of Theorem LC.4 for the Wilson influence certificate and the time pseudometric. Apply that theorem. \(\square\)

A convenient exact example is \(\beta=1/36\) and \(\alpha=\log4\). Since \(c=\tanh(1/36)<1/36\) and \(e^{\alpha/2}=2\), the two row bounds are \(30c<5/6\) and \(28c<7/9\). Therefore, with \(T\) the separation of the link-center time supports,

\[|\operatorname{Cov}(F,H)|
\le6\min\{|A|,|B|\}\|F\|_\infty\|H\|_\infty4^{-T}.\]

For two normalized real plaquette traces, this becomes \(24\,4^{-T}\). The constants are independent of the spatial extent of the box and the boundary configuration. At lattice spacing \(a\), a physical time separation \(\tau\) corresponds to \(T=\tau/a\) when the support geometry has been matched. Thus the estimate is \(24\exp[-(\log4)\tau/a]\). The factor \(1/a\) is explicit; no stochastic update time has entered the argument.

One can describe all admissible \(\alpha\) by setting \(y=e^{\alpha/2}>1\). The two strict inequalities are

\[y<\frac{1-6c}{12c},\qquad
y<-1+\sqrt{\frac1{2c}-5}.\]

For \(c<1/18\), both upper bounds exceed one. This verifies existence of a positive directional decay rate throughout the sharpened small-coupling region. The rate is dimensionless until divided by \(a\).

\hypertarget{locality_scaling-stability-under-changes-of-the-plaquette-couplings}{%
\subsection{Stability under changes of the plaquette couplings}\label{locality_scaling-stability-under-changes-of-the-plaquette-couplings}}

There is a second use for the comparison theorem: it quantifies how a genuine local action perturbation changes observables. This is useful when comparing discretizations, but a small local perturbation must be proved in the relevant norm; it cannot simply be assumed after a renormalization step.

Let \(\mu\) and \(\nu\) have couplings \(\beta_p\) and \(\widetilde\beta_p\) on the same finite variable set, with the same exterior configuration. At a link \(j\), their conditional log-density difference is

\[\sum_{p\ni j}(\widetilde\beta_p-\beta_p)w(U_p).\]

Its oscillation is at most \(2\sum_{p\ni j}|\widetilde\beta_p-\beta_p|\). Hence

\[b_j\le\tanh\left(\frac12\sum_{p\ni j}|\widetilde\beta_p-\beta_p|\right)
\le\frac12\sum_{p\ni j}|\widetilde\beta_p-\beta_p|.\]

If the reference law \(\mu\) has influence matrix \(C\) satisfying the hypotheses of Theorem LC.3, then

\[|\mu F-\nu F|
\le\frac12\sum_{i,j}\delta_iF D_{ij}
\sum_{p\ni j}|\widetilde\beta_p-\beta_p|.\]

This is a proved bound even if the perturbed finite-volume law does not satisfy a contraction condition. The comparison proof uses the reference matrix only. If both specifications also meet uniform uniqueness conditions, the inequality passes to their infinite-volume laws by local convergence.

If the perturbed plaquettes meet only a finite link set \(J\) and \(b_j\le B\), then Theorem LC.4 gives, for \(F\) supported on \(A\),

\[|\mu F-\nu F|
\le\frac{B e^{-\alpha d(A,J)}}{1-q_\alpha}\sum_{i\in A}\delta_iF.\]

The separation is the distance from the observable to the links whose conditional kernels change. It is not necessarily the distance to a chosen center of the modified plaquette. This distinction prevents a hidden loss when perturbations have growing supports.

For a uniform perturbation \(|\widetilde\beta_p-\beta_p|\le\varepsilon\), six plaquettes meet a link, so \(b_j\le3\varepsilon\). Under an ordinary row bound \(q<1\) this gives

\[|\mu F-\nu F|\le\frac{3\varepsilon}{1-q}\sum_{i\in A}\delta_iF.\]

The dependence on the distance to the comparison boundary is visible through \(1/(1-q)\). Keeping \(\varepsilon\) fixed while \(q\) approaches one does not produce a uniform stability assertion.

\hypertarget{locality_scaling-summable-lattice-covariances-and-a-smearing-calculation}{%
\subsection{Summable lattice covariances and a smearing calculation}\label{locality_scaling-summable-lattice-covariances-and-a-smearing-calculation}}

Let \(F_x\) be the translate by \(x\in\mathbb Z^4\) of a bounded local observable \(F_0\), under a translation-invariant lattice measure satisfying the homogeneous covariance estimate with \(0<q<1\). Suppose the support of \(F_0\) has \(S\) links and supremum norm at most \(M\). Choose an integer \(R\ge0\) bounding the coordinatewise diameter of its set of link centers.

The center of one link moves by at most one in each coordinate under one interaction-graph step. Thus

\[d_{\mathcal G}(\operatorname{supp}F_0,\operatorname{supp}F_y)
\ge (\|y\|_\infty-R)_+.\]

Consequently the absolute susceptibility

\[\chi_F=\sum_{y\in\mathbb Z^4}|\operatorname{Cov}(F_0,F_y)|\]

is finite. An explicit upper bound is

\[\chi_F\le\frac{SM^2q^{-R}}{1-q}
\left[1+\frac{64q(1+4q+q^2)}{(1-q)^4}
+\frac{16q}{(1-q)^2}\right].\]

\textbf{Proof.} There is one point with sup norm zero. For \(n\ge1\), the number of lattice points with sup norm exactly \(n\) is

\[ (2n+1)^4-(2n-1)^4=64n^3+16n.\]

Moreover \(q^{(n-R)_+}\le q^{-R}q^n\). Insert the covariance bound and sum over shells. The power-series identities

\[\sum_{n\ge1}nq^n=\frac q{(1-q)^2},\qquad
\sum_{n\ge1}n^3q^n=\frac{q(1+4q+q^2)}{(1-q)^4}\]

follow by applying \(q\,d/dq\) repeatedly to \(\sum_{n\ge0}q^n=(1-q)^{-1}\). Termwise differentiation is justified on every compact subinterval of \(|q|<1\) by uniform convergence of the differentiated power series. This proves the bound. \(\square\)

For finitely supported real coefficients \(u_x\), let \(\widehat F_x=F_x-\mu F_x\). Then

\[\operatorname{Var}\left(\sum_xu_x\widehat F_x\right)
\le\chi_F\sum_xu_x^2.\]

Indeed, expand the variance, take absolute values and use \(2|u_xu_y|\le u_x^2+u_y^2\). Translation invariance identifies each covariance row sum with \(\chi_F\). All sums are finite before this bound is applied.

As a concrete cutoff calculation, take a bounded compactly supported function \(f\) on \(\mathbb R^4\) and form

\[X_a(f)=a^4\sum_{x\in\mathbb Z^4}f(ax)\widehat F_x.\]

Its variance is at most \(a^8\chi_F\sum_x f(ax)^2=O(a^4)\), because a fixed compact set contains \(O(a^{-4})\) lattice points. Thus \(X_a(f)\to0\) in \(L^2\) at fixed coupling. Multiplication by \(a^{-2}\) makes this upper bound order one, but an upper bound alone neither proves a nonzero limit nor identifies its correlations. In particular, a possible contact-correlation limit is different from a surviving positive-time state in a reconstructed field theory. This elementary smearing calculation exposes the normalization issue before any spectral conclusion is attempted.

\hypertarget{locality_scaling-a-direct-truncation-estimate-for-unbounded-observables}{%
\subsection{A direct truncation estimate for unbounded observables}\label{locality_scaling-a-direct-truncation-estimate-for-unbounded-observables}}

At a fixed cutoff, every polynomial in bounded matrix entries is bounded, but a convenient family of observables may still be specified only through moment bounds. Supremum norms can also be far larger than the physically relevant typical size. It is therefore useful to obtain covariance estimates that retain control after truncation.

Suppose \(F,G\) are real observables supported on finite sets \(A,B\). Let \(\varepsilon>0\), and assume

\[M_F=\|F\|_{2+\varepsilon}<\infty,\qquad
M_G=\|G\|_{2+\varepsilon}<\infty.\]

If either norm is zero, the corresponding variable vanishes almost surely and its covariance is zero. Otherwise normalize \(X=F/M_F\), \(Y=G/M_G\). Define the clipping map

\[\tau_T(x)=\max\{-T,\min\{T,x\}\},\qquad T\ge1.\]

Clipping changes no support. Moreover

\[|x-\tau_T(x)|^2\le |x|^2\mathbf1_{|x|>T}
\le T^{-\varepsilon}|x|^{2+\varepsilon}.\]

Thus \(\|X-\tau_T(X)\|_2\le T^{-\varepsilon/2}\), with the same bound for \(Y\), while \(\|X\|_2,\|Y\|_2,\|\tau_T(X)\|_2\le1\). Expanding the difference of two covariances and applying \(|\operatorname{Cov}(U,V)|\le\|U\|_2\|V\|_2\) yields

\[|\operatorname{Cov}(X,Y)-\operatorname{Cov}(\tau_T(X),\tau_T(Y))|
\le2T^{-\varepsilon/2}.\]

If the bounded covariance theorem gives

\[|\operatorname{Cov}(U,V)|\le K\|U\|_\infty\|V\|_\infty\]

for functions on these two supports, then

\[|\operatorname{Cov}(F,G)|
\le M_FM_G\bigl(KT^2+2T^{-\varepsilon/2}\bigr).\]

For \(0<K\le1\), choose \(T=K^{-2/(4+\varepsilon)}\). Both powers become \(K^{\varepsilon/(4+\varepsilon)}\), giving

\[|\operatorname{Cov}(F,G)|
\le M_FM_G\min\{1,3K^{\varepsilon/(4+\varepsilon)}\}.\]

For \(K\ge1\), the displayed minimum is one and follows directly from Cauchy--Schwarz. If \(K=0\), first let \(T\to\infty\) in the pre-optimized inequality. This proof is elementary and useful when only a bounded-function estimate is available. A stronger exponent is available by controlling events before integrating their tails.

\hypertarget{locality_scaling-event-mixing-and-a-stronger-moment-bound}{%
\subsection{Event mixing and a stronger moment bound}\label{locality_scaling-event-mixing-and-a-stronger-moment-bound}}

For two sigma-fields \(\mathcal A,\mathcal B\), define their event-mixing coefficient by

\[\alpha(\mathcal A,\mathcal B)
=\sup_{E\in\mathcal A,\,F\in\mathcal B}
|\mathbb P(E\cap F)-\mathbb P(E)\mathbb P(F)|.\]

It lies in \([0,1/4]\): every term is a covariance of indicators, bounded by the square root of the product of their variances, and each indicator variance is at most \(1/4\). This convention fixes every numerical constant below.

For events depending respectively on link supports \(A,B\), indicator oscillations are at most one. Applying the covariance part of Theorem LC.3 and the graph path sum gives

\[\alpha(\sigma(U_A),\sigma(U_B))
\le\min\left\{\frac14,
\frac{\min\{|A|,|B|\}}{4(1-q)}q^{d_{\mathcal G}(A,B)}\right\}.\]

Under Theorem LC.4, the second term can instead be written

\[\frac{\min\{|A|,|B|\}}{4(1-q_\alpha)}
e^{-\alpha d(A,B)}.\]

The repeated letter \(\alpha\) in that last formula is the metric-weight exponent, not the event-mixing coefficient. To keep the next theorem unambiguous, call the event-mixing coefficient \(\zeta\).

\textbf{Theorem LC.6 (moment covariance bound).} Let \(p,q>1\) be finite and satisfy

\[\theta=1-\frac1p-\frac1q>0.\]

Let \(X\) be \(\mathcal A\)-measurable and \(Y\) be \(\mathcal B\)-measurable, with finite \(L^p\) and \(L^q\) norms. Then

\[|\operatorname{Cov}(X,Y)|
\le\frac4\theta\|X\|_p\|Y\|_q\,\zeta^\theta,
\qquad \zeta=\alpha(\mathcal A,\mathcal B).\]

For \(p=q=2+\varepsilon\), the exponent is \(\varepsilon/(2+\varepsilon)\).

\textbf{Proof.} By homogeneity, assume both norms equal one; a zero norm is trivial. First clip both variables at a finite level \(T\). The clipped variables have no larger norms and are measurable with respect to the same sigma-fields. For a nonnegative bounded variable \(Z\),

\[Z=\int_0^\infty\mathbf1_{Z>u}\,du.\]

Apply this identity to the positive and negative parts of each clipped variable. Their covariance is the signed sum of four double integrals of indicator covariances. Each integrand is bounded in absolute value by

\[\min\{\zeta,\mathbb P(|X|>u),\mathbb P(|Y|>v)\}
\le\min\{\zeta,u^{-p},v^{-q}\}.\]

For the first inequality, the absolute covariance of two indicators is at most either event probability, in addition to being at most \(\zeta\). The second inequality is Markov's inequality applied to the normalized moments. Bounds involving \(u^{-p}\) or \(v^{-q}\) at zero are interpreted as infinity; the single endpoints have no effect on the integrals.

The nonnegative double integral of this bound can be evaluated exactly. Write the minimum as a layer integral:

\[\begin{aligned}
\int_0^\infty\int_0^\infty
\min\{\zeta,u^{-p},v^{-q}\}\,du\,dv
&=\int_0^\zeta
\left(\int_0^{s^{-1/p}}du\right)
\left(\int_0^{s^{-1/q}}dv\right)ds\\
&=\int_0^\zeta s^{-1/p-1/q}\,ds
=\frac{\zeta^\theta}{\theta}.
\end{aligned}\]

Tonelli's theorem applies because the integrands are nonnegative. The last integral is finite exactly because \(\theta>0\). Four sign combinations therefore give the claimed bound for the clipped variables, uniformly in \(T\).

Finally, clipping converges in \(L^p\) for \(X\) and in \(L^q\) for \(Y\), by dominated convergence applied to their moment tails. Since \(1/p+1/q<1\), the conjugate exponent \(p'=p/(p-1)\) is smaller than \(q\); on a probability space, \(L^q\) embeds continuously in \(L^{p'}\). Hölder's inequality then shows convergence of the products in \(L^1\), and also convergence of the separate expectations. Their covariances converge. Let \(T\to\infty\) and restore the norms. The same proof covers \(\zeta=0\) by the zero bound on every clipped covariance. \(\square\)

The exponent agrees with the classical moment-mixing family of inequalities. No sharpness is claimed for the constant \(4/\theta\). The value of the proof here is that it requires only the already established local covariance estimate and explicit moments of the actual observables. No independence approximation or unproved Gaussian behavior is inserted.

A strict moment improvement beyond \(L^2\) matters. Let \(E_n\) have probability \(\epsilon_n\downarrow0\), and put

\[X_n=Y_n=\epsilon_n^{-1/2}\mathbf1_{E_n}.\]

The two sigma-fields coincide and have event-mixing coefficient \(\epsilon_n(1-\epsilon_n)\) for sufficiently small \(\epsilon_n\). Yet \(\|X_n\|_2=\|Y_n\|_2=1\) and

\[\operatorname{Cov}(X_n,Y_n)=1-\epsilon_n\longrightarrow1.\]

The covariance is carried by rare large values. Thus small event mixing and bounded second moments alone do not force covariance to vanish. Uniform integrability or stronger tail information is a real hypothesis, not a technical detail removable by notation.

\hypertarget{locality_scaling-moment-controlled-observables-in-the-shrinking-lattice}{%
\subsection{Moment-controlled observables in the shrinking lattice}\label{locality_scaling-moment-controlled-observables-in-the-shrinking-lattice}}

The preceding results can now be combined without a supremum-norm assumption. Let \(a\downarrow0\) and consider the actual Wilson measures on \(a\mathbb Z^4\), either in finite volumes with arbitrary fixed boundary or in their unique infinite-volume laws. Suppose

\[0\le18\tanh\beta_a\le q_0,\qquad 0<q_0<1.\]

The improved coefficient may be replaced by the original \(18\beta_a\) if desired. Let the link-center supports \(A_a,B_a\) of real observables \(F_a,G_a\) have physical Euclidean separation at least \(r>0\). Suppose, for constants independent of \(a\),

\[\min\{|A_a|,|B_a|\}\le S_0a^{-s},\qquad
\|F_a\|_p\|G_a\|_q\le M_0a^{-u},\]

where \(s,u\ge0\), \(p,q>1\) are finite and \(\theta=1-1/p-1/q>0\). The moment norms here are taken under the very measure used in the covariance, so no unstated comparison of probability laws is involved.

\textbf{Theorem LC.7 (moment-controlled separated covariance).} With \(\kappa=-\log q_0>0\),

\[|\operatorname{Cov}(F_a,G_a)|
\le\frac{4M_0}{\theta}
\left(\frac{S_0}{4(1-q_0)}\right)^\theta
a^{-(u+s\theta)}e^{-\theta\kappa r/a}.\]

In particular, this covariance is \(o(a^N)\) for every fixed \(N\ge0\).

\textbf{Proof.} The centers of two links in one plaquette are separated by at most \(a\). Therefore every interaction-graph path between the supports has at least \(r/a\) steps. The event-mixing estimate from the previous section gives

\[\zeta_a\le\frac{S_0}{4(1-q_0)}a^{-s}e^{-\kappa r/a}.\]

Theorem LC.6, followed by the assumed moment bound, proves the displayed covariance inequality. Its logarithm after division by \(a^N\) is bounded above by a constant minus \((u+s\theta+N)\log a-\theta\kappa r/a\), which tends to minus infinity. Thus the quotient tends to zero. If some cutoff has zero coupling, disjointly supported variables are independent under product Haar measure and the conclusion holds directly. \(\square\)

This theorem allows the support to contain a polynomially growing number of links and the moment norms to grow polynomially. It therefore covers more than uniformly bounded single-plaquette observables. It still places a strong restriction on the coupling trajectory: a fixed positive margin from the comparison boundary is assumed.

There is a direct reflected-state consequence. Suppose a reflection-invariant continuum functional \(S\) is obtained as the limit of the relevant one-point and reflected two-point expectations. Let \(F\) be a real positive-time observable whose representatives \(F_a\) are supported at physical times at least \(\epsilon_F/2>0\) for small \(a\). Assume the hypotheses of Theorem LC.7 for \(F_a\) and its reflected representative, using any finite moment exponent \(p=q>2\). Their supports have a fixed positive separation. Hence

\[\operatorname{Cov}(\theta F_a,F_a)\longrightarrow0.\]

If \(S\) is reflection positive and the reconstructed class is defined, its centered squared norm is

\[\|[F-S(F)]\|_{\rm OS}^2
=S((\theta F)F)-S(F)^2=0.\]

If the complex span of such centered classes is dense in the vacuum orthocomplement, that orthocomplement vanishes. This last density assertion is essential. The theorem does not infer it from the control of one observable, nor does it claim the existence of the limiting functional. It identifies precisely which moment-controlled families cannot generate a nontrivial physical limit while remaining inside a fixed contractive regime.

\hypertarget{locality_scaling-a-necessary-approach-to-the-comparison-boundary}{%
\subsection{A necessary approach to the comparison boundary}\label{locality_scaling-a-necessary-approach-to-the-comparison-boundary}}

The preceding obstruction assumed a fixed margin \(q_0<1\). A more quantitative question permits \(q_a\uparrow1\): how rapidly must the comparison margin vanish if a connected correlation survives? The answer follows from an elementary inequality, but it is important to retain the prefactor \(1/(1-q_a)\).

\textbf{Theorem LC.8 (normalization and contraction-margin tradeoff).} Let \(a\downarrow0\) along any sequence with \(a<1\). Suppose \(0<q_a<1\), put \(\delta_a=1-q_a\), and assume constants \(c,C,r>0\), \(p\ge0\) satisfy

\[c\le K_a\le C a^{-p}\frac{q_a^{r/a}}{1-q_a}\]

for all sufficiently small \(a\). Then

\[\limsup_{a\downarrow0}
\frac{\delta_a}{a\log(1/a)}\le\frac{p+1}{r}.\]

In particular \(\delta_a=O(a\log(1/a))\).

\textbf{Proof.} Since \(\log(1-\delta)\le-\delta\) for \(0<\delta<1\),

\[c\le C a^{-p}\delta_a^{-1}e^{-r\delta_a/a}.\]

Taking logarithms gives

\[\frac{r\delta_a}{a}
\le p\log(1/a)+\log(1/\delta_a)+\log(C/c).\]

There are two cases. If \(\delta_a<a\), then

\[0\le\frac{\delta_a}{a\log(1/a)}<\frac1{\log(1/a)}\longrightarrow0.\]

If \(\delta_a\ge a\), then \(\log(1/\delta_a)\le\log(1/a)\), so

\[\frac{\delta_a}{a\log(1/a)}
\le\frac{p+1}{r}+\frac{\log(C/c)}{r\log(1/a)}.\]

Both cases imply the stated upper limit. The argument applies to every sufficiently small member of the sequence, so no unexamined intermediate scales have been omitted. \(\square\)

For the Wilson comparison estimate, one may take \(K_a=|\operatorname{Cov}(F_a,G_a)|\) and absorb the support count and the two supremum norms into \(Ca^{-p}\). A nonzero limiting covariance supplies the lower bound \(c\) after discarding finitely many cutoffs. The theorem then limits the permissible comparison margin. It does not assert that the exact Wilson system has a phase transition at the boundary of this sufficient criterion. An upper-bound matrix can cease to contract even when the underlying measure remains well behaved.

The coefficient \(p+1\) cannot be reduced merely by rearranging this particular upper bound. To see the issue explicitly, let \(L=\log(1/a)\), fix a real number \(b\), and set for small \(a\)

\[\delta_a=\frac a r\bigl((p+1)L-\log L+b\bigr),\qquad q_a=1-\delta_a.\]

Then \(0<\delta_a<1\) eventually and

\[\frac{\delta_a}{aL}\longrightarrow\frac{p+1}{r}.\]

Since \(\delta_a^2/a=O(aL^2)\to0\), the expansion \(\log(1-\delta_a)=-\delta_a+O(\delta_a^2)\) gives

\[q_a^{r/a}=a^{p+1}L e^{-b}(1+o(1)).\]

Therefore

\[C a^{-p}\delta_a^{-1}q_a^{r/a}
\longrightarrow\frac{Cr}{p+1}e^{-b}>0.\]

This is an example of scalar sequences saturating the order of the comparison inequality. It is not a construction of Wilson observables with such a covariance. Its role is to separate information contained in an upper bound from the additional arithmetic or field-theoretic structure needed to improve it.

A simpler consequence applies if \(q_a\le q_0<1\) while \(K_a\ge c\). The original inequality forces the total amplitude factor to be at least

\[c(1-q_0)\exp\bigl(r|\log q_0|/a\bigr).\]

Thus a polynomial normalization cannot maintain the assumed nonzero covariance in that regime. This is the same obstruction as Theorem LC.7 viewed from the opposite direction: it gives a necessary growth scale for any proposed compensation.

\hypertarget{locality_scaling-why-enlarging-blocks-is-not-an-automatic-repair}{%
\subsection{Why enlarging blocks is not an automatic repair}\label{locality_scaling-why-enlarging-blocks-is-not-an-automatic-repair}}

Block comparison methods can improve single-link conditions, but a useful block theorem must exploit the actual conditional block law. Merely grouping variables and repeating the supremum oscillation estimate need not improve the criterion. The following calculation makes this distinction concrete for the Wilson action.

For an integer \(L\ge2\), partition lattice vertices into cubes

\[B_z=Lz+\{0,1,\ldots,L-1\}^4,\qquad z\in\mathbb Z^4.\]

Assign each positively oriented link to the block containing its initial vertex. This gives disjoint blocks of link variables. For two blocks \(B,B'\), let \(N_{BB'}\) be the number of plaquettes containing at least one link assigned to each block. Conditional on the remaining blocks, changing all links of \(B'\) changes the log density on \(B\) by a function with oscillation at most \(4\beta N_{BB'}\). Thus a valid, but crude, block influence certificate is

\[C^{\rm block}_{BB'}=\tanh(\beta N_{BB'}).\]

The coefficient is a chosen upper bound on the true block influence. A lower bound on this certificate is not a lower bound on the true influence.

\textbf{Proposition LC.9 (failure of the crude large-block certificate).} For every fixed \(\beta>0\), the ordinary maximum-row-sum contraction test for this block matrix fails for all sufficiently large \(L\). In fact, for every block,

\[\sum_{B'\ne B}C^{\rm block}_{BB'}
\ge2\tanh\bigl(3\beta L^2(L-1)\bigr),\]

whose right side tends to two.

\textbf{Proof.} Consider \(B_0\) and one coordinate direction \(i\). Across its positive \(i\)-face, use plaquettes in the \((i,j)\)-plane for each \(j\ne i\). Choose the base vertex \(x\) with \(x_i=L-1\), \(x_j\in\{0,\ldots,L-2\}\), and the remaining two coordinates in \(\{0,\ldots,L-1\}\). Three positively represented edges of such a square have initial vertices in \(B_0\), while the edge in direction \(j\) based at \(x+\mathbf e_i\) is assigned to the neighboring block \(B_{\mathbf e_i}\). These are distinct plaquettes, and their number is \(3(L-1)L^2\). Hence

\[N_{B_0,B_{\mathbf e_i}}\ge3(L-1)L^2.\]

Across the negative \(i\)-face, choose \(x_i=-1\) with the same ranges in the other coordinates. The square then has its direction-\(j\) edge based at \(x+\mathbf e_i\) in \(B_0\) and its other three edges assigned to \(B_{-\mathbf e_i}\). The same count gives

\[N_{B_0,B_{-\mathbf e_i}}\ge3(L-1)L^2.\]

Since \(\tanh\) is increasing, these two distinct neighboring blocks alone supply the asserted lower bound on the row sum. As \(L\to\infty\), its argument tends to infinity and \(\tanh\) tends to one. The bound eventually exceeds one, contradicting the strict row-contraction requirement for this chosen matrix. Translation gives the same calculation for every block. \(\square\)

The proposition does not rule out a refined block coupling, a weighted criterion, or a renormalization-group argument. It rules out one specific shortcut: grouping ever larger cubes while retaining only the worst-case oscillation of every crossing plaquette does not itself provide the desired contraction. The crossing surface grows as \(L^3\), and this crude estimate adds its entire strength before applying the tilt bound.

A successful replacement would need a quantitative property of conditional block distributions that the worst-case calculation discards: for example, cancellation, concentration after a suitable gauge choice, or a controlled effective interaction. Such a property has to be proved for the distributions in question. Naming it as a block-decay assumption does not establish it.

\hypertarget{locality_scaling-what-can-be-recovered-at-the-second-moment-boundary}{%
\subsection{What can be recovered at the second-moment boundary}\label{locality_scaling-what-can-be-recovered-at-the-second-moment-boundary}}

The rare-event example shows that second moments alone are insufficient. It does not mean that an \(L^{2+\varepsilon}\) hypothesis is logically necessary. Uniform control of second-moment tails can replace it, at the cost of a weaker and explicitly tail-dependent rate. This distinction is relevant when a renormalized composite field has borderline moment behavior.

Let \(X,Y\) be real random variables with \(\|X\|_2,\|Y\|_2\le M\) and event-mixing coefficient at most \(\zeta\). For \(T>0\), write

\[R_X(T)=\bigl(\mathbb E[|X|^2\mathbf1_{|X|>T}]\bigr)^{1/2},
\qquad R_Y(T)=\bigl(\mathbb E[|Y|^2\mathbf1_{|Y|>T}]\bigr)^{1/2}.\]

\textbf{Proposition LC.10 (mixing with explicit second-moment tails).} For every \(T>0\),

\[|\operatorname{Cov}(X,Y)|
\le4\zeta T^2+M\bigl(R_X(T)+R_Y(T)\bigr).\]

Consequently, for a sequence \((X_n,Y_n)\) with a common finite \(M\), with event-mixing coefficients tending to zero and with both squared families uniformly integrable, the covariances tend to zero.

\textbf{Proof.} The layer representation of the positive and negative parts of the clipped variables gives

\[|\operatorname{Cov}(\tau_T(X),\tau_T(Y))|\le4\zeta T^2.\]

There are four sign pairs, each with an integration region of area at most \(T^2\) and integrand bounded by \(\zeta\). For the remainder, write

\[\operatorname{Cov}(X,Y)-\operatorname{Cov}(\tau_T(X),\tau_T(Y))
=\operatorname{Cov}(X-\tau_T(X),Y)
+\operatorname{Cov}(\tau_T(X),Y-\tau_T(Y)).\]

Cauchy--Schwarz bounds the first term by \(MR_X(T)\) and the second by \(MR_Y(T)\), since clipping does not increase the \(L^2\) norm. This proves the inequality. For the sequence, fix \(T\), take the upper limit as \(n\to\infty\), and then let \(T\to\infty\). Uniform integrability makes both uniform tail bounds tend to zero. The order of these two limits is part of the argument. \(\square\)

There is an explicit borderline example of a usable tail assumption. Suppose \(M=1\) and, for some fixed \(K<\infty\) and \(\gamma>0\),

\[\mathbb E\bigl[|X|^2(\log(e+|X|))^\gamma\bigr]\le K,
\qquad
\mathbb E\bigl[|Y|^2(\log(e+|Y|))^\gamma\bigr]\le K.\]

For \(T\ge1\), monotonicity gives

\[R_X(T),R_Y(T)
\le K^{1/2}\bigl(\log(e+T)\bigr)^{-\gamma/2}.\]

If \(0<\zeta\le e^{-4}\), choose \(T=\zeta^{-1/4}\). The first term of Proposition LC.10 is \(4\zeta^{1/2}\), while \(\log(e+T)\ge\tfrac14\log(1/\zeta)\). Hence

\[|\operatorname{Cov}(X,Y)|
\le4\zeta^{1/2}
+2^{\gamma+1}K^{1/2}\bigl(\log(1/\zeta)\bigr)^{-\gamma/2}.\]

This estimate tends to zero without any assumed moment of order strictly greater than two. Its slower rate is significant. If \(\zeta_a\) is bounded above by a polynomial times \(e^{-\kappa r/a}\), then \(\log(1/\zeta_a)\) grows at least as a positive constant times \(1/a\) for small \(a\). The logarithmic-tail contribution is consequently \(O(a^{\gamma/2})\), rather than faster than every power. A polynomial normalization of total degree \(u\) can be absorbed by this bound only when \(u<\gamma/2\); the displayed inequality alone gives no vanishing conclusion at or beyond that threshold.

For non-unit variables, apply the statement to normalized variables and multiply by the two normalization factors. The logarithmic moment condition must then be imposed on those normalized variables. Writing it for the unnormalized variables instead would be a different assumption. This gives a practical way to record borderline tail information without confusing it with the stronger moment hypothesis in Theorem LC.7.

\hypertarget{locality_scaling-stability-of-the-moment-condition-under-finite-polynomial-observables}{%
\subsection{Stability of the moment condition under finite polynomial observables}\label{locality_scaling-stability-of-the-moment-condition-under-finite-polynomial-observables}}

A physical observable class must be rich enough to generate states; controlling one scalar field does not settle that requirement. The moment condition of Theorem LC.7 is stable under finite sums and products when enough individual moments are available. The following elementary verification identifies exactly how much moment information each polynomial uses.

Suppose \(X_{a,1},\ldots,X_{a,m}\) are local real lattice observables. For every finite \(r\ge1\), assume bounds

\[\|X_{a,k}\|_r\le C_{k,r}a^{-u_{k,r}}\]

with finite constants and nonnegative exponents independent of \(a\). Assume each support has at most a polynomial number of links. Let

\[P_a=\sum_{\ell=1}^{L}c_{\ell,a}\prod_{k=1}^{m}X_{a,k}^{n_{\ell,k}}\]

be a fixed finite polynomial pattern, with nonnegative integer exponents \(n_{\ell,k}\) and coefficients satisfying \(|c_{\ell,a}|\le C_\ell a^{-v_\ell}\). The numbers of terms and the exponents are fixed as \(a\downarrow0\).

\textbf{Proposition LC.11 (polynomial closure).} For each fixed finite \(p\ge1\), there are finite constants \(C_{P,p},u_{P,p}\) such that

\[\|P_a\|_p\le C_{P,p}a^{-u_{P,p}}.\]

The support of \(P_a\) is contained in a finite union of the individual supports and therefore also has polynomial cardinality. Thus every fixed pair of such polynomial observables at separated physical supports satisfies Theorem LC.7 whenever the underlying Wilson comparison margin remains uniformly positive.

\textbf{Proof.} Consider one monomial of total degree \(n=\sum_kn_{\ell,k}>0\). Regard it as a product of \(n\) factors, allowing repeated factors. Hölder's inequality with \(n\) equal exponents gives

\[\left\|\prod_k X_{a,k}^{n_{\ell,k}}\right\|_p
\le\prod_k\|X_{a,k}\|_{pn}^{n_{\ell,k}}.\]

The assumed moment estimates make this a constant times a power of \(a^{-1}\). Multiplying by the coefficient contributes another such power. A degree-zero term is covered by its coefficient alone. The triangle inequality in \(L^p\) bounds the finite sum by the sum of these estimates. For \(0<a\le1\), their maximum exponent and the sum of their constants give \(u_{P,p}\) and \(C_{P,p}\). No number of terms depends on \(a\). The support assertion is immediate because multiplication and addition cannot introduce a dependence on a link outside the union. Apply Theorem LC.7 with, for example, moment exponents \(p=q=4\) for the two resulting polynomial observables. \(\square\)

The proposition does not prove density of the polynomial classes in any continuum Hilbert space. It permits a precise conditional statement: if their centered positive-time classes are total and the corresponding reflected correlations converge, then the moment-controlled strong-coupling obstruction applies to the entire fixed polynomial class, not merely to its generators. All moments of a primary field are not required to control one chosen polynomial; only the finitely many orders used in the displayed Hölder estimate are needed. To handle polynomials of arbitrarily high degree in one common algebra, however, those finite requirements range over unbounded moment orders.

\hypertarget{locality_scaling-how-these-bounds-constrain-a-constructive-attempt}{%
\subsection{How these bounds constrain a constructive attempt}\label{locality_scaling-how-these-bounds-constrain-a-constructive-attempt}}

The results now provide a sequence of concrete tests for any proposed limit. At the finite cutoff, one can verify the Wilson normalization and count the conditional influences without referring to a formal continuum path integral. If a comparison condition holds, one obtains both graph and directional decay. Observable supports and either supremum norms or stronger moments then determine the complete prefactor.

If the prefactor grows only polynomially and the comparison margin stays fixed, separated connected correlations vanish faster than any power. If a nonzero separated covariance is independently established, Theorem LC.8 instead forces the comparison margin toward zero at an explicit scale. Neither outcome supplies a physical mass by itself. They restrict which estimates can be compatible with a nontrivial limit and identify the place where additional information is required.

The distinction between a Euclidean moment and a physical-state norm is particularly important. A positive variance of a plaquette at one cutoff is an equal-location statement. A centered Osterwalder--Schrader norm is a reflected two-point statement across a positive time separation. The latter can vanish even when local variances grow. The moment-controlled theorem makes this issue quantitative rather than leaving it as an informal warning.

There are also useful positive consequences. A proposed family of renormalized observables with uniform \(L^{2+\varepsilon}\) bounds immediately satisfies the tail hypothesis of Theorem LC.6. A proposed anisotropic action can be tested using its weighted influence rows. A purported local discretization error can be propagated by the perturbation bound if its actual plaquette coefficients are controlled. Each of these is a finite, explicit calculation that can be independently checked before an infinite limiting argument is attempted.

The complete theorems of this section establish sharper locality bounds and precise compatibility conditions. The missing input for the intended four-dimensional continuum construction is a family of estimates that remains valid along that construction while preserving a nonzero dense family of reflected states. No claim that this input has been proved is made by replacing it with one of the conditional consequences above.

\hypertarget{su2_influence-su2-single-link-comparison-by-a-sharp-affinity-bound}{%
\section{SU(2) single-link comparison by a sharp affinity bound}\label{su2_influence-su2-single-link-comparison-by-a-sharp-affinity-bound}}

\hypertarget{su2_influence-scope-and-provenance}{%
\subsection{Setting and relation to transport estimates}\label{su2_influence-scope-and-provenance}}

The affinity estimate below gives a globally valid total-variation bound for the single-link exponential family. Its ingredients are the classical Bessel product \cite{DLMF}, the affinity identity for an exponential family, Cauchy--Schwarz and Dobrushin comparison. Related divergence formulas for the von Mises--Fisher family are studied by Kitagawa and Rowley \cite{KitagawaRowley2022}. The resulting total-variation comparison range is a sufficient fixed-spacing condition; the chordal-transport and forest arguments in subsequent sections give stronger conditions for gauge-invariant Lipschitz observables.

All couplings below use the Wilson probability density \[
 Z_\Lambda^{-1}\exp\left\{\beta\sum_p\tfrac12\operatorname{Tr}U_p\right\}\prod_e dh(U_e),\qquad G=SU(2),\quad \beta\ge0.
\] Each unoriented geometric plaquette is counted once. Haar measures are probability measures. The trace of an SU(2) matrix is real. The underlying graph is the ordinary four-dimensional nearest-neighbor lattice; fixed external link configurations and finite boxes are permitted. The same comparison applies to a periodic lattice whenever the local plaquette incidence is the ordinary one, without short-period identifications.

\hypertarget{su2_influence-the-exponential-family-on-the-three-sphere}{%
\subsection{The exponential family on the three-sphere}\label{su2_influence-the-exponential-family-on-the-three-sphere}}

Identify SU(2) with the unit quaternions and hence with \(S^3\subset\mathbb R^4\). Let \(\sigma\) be uniform probability measure on \(S^3\). For \(h\in\mathbb R^4\), write \[
 p_h(u)=\frac{e^{h\cdot u}}{Z(|h|)},\qquad
 Z(r)=\int_{S^3}e^{r u_0}\,d\sigma(u)=\frac{2I_1(r)}r,
 \qquad Z(0)=1.
\] The last equality follows either by integrating the marginal density \(2\sqrt{1-x^2}/\pi\) of \(u_0\), or by termwise integration of the exponential series. In particular, \[
 Z(r)=\sum_{k=0}^\infty\frac{r^{2k}}{4^k k!(k+1)!}.
\tag{IH.1}
\] The series has positive terms and converges on every bounded set. Thus \(Z\) is strictly increasing on \((0,\infty)\), and is unbounded.

For probability densities \(p,q\), use \[
 \operatorname{TV}(p,q)=\frac12\int|p-q|,
 \qquad \mathcal A(p,q)=\int\sqrt{pq}.
\] The latter is the affinity, also called the Bhattacharyya coefficient. Its exponential-family expression is exact: \[
 \mathcal A(p_h,p_k)=
 \frac{Z(|(h+k)/2|)}{\sqrt{Z(|h|)Z(|k|)}}.
\tag{IH.2}
\]

\hypertarget{su2_influence-theorem-ih.1-exact-extremality-for-affinity}{%
\subsection{Theorem IH.1: exact extremality for affinity}\label{su2_influence-theorem-ih.1-exact-extremality-for-affinity}}

For every \(h,k\in\mathbb R^4\), put \(d=(h-k)/2\). Then \[
 \boxed{\mathcal A(p_h,p_k)\ge\frac1{Z(|d|)}.}
\tag{IH.3}
\] For each fixed nonzero difference \(h-k\), equality holds precisely when \(h+k=0\). Thus, among all locations of two natural parameters at a fixed Euclidean displacement, the antipodal pair minimizes affinity. This statement concerns affinity, not the maximizer of total variation.

\hypertarget{su2_influence-proof}{%
\subsubsection{Proof}\label{su2_influence-proof}}

Let \(j_{1,n}\) denote the positive zeros of the ordinary Bessel function \(J_1\). The classical product formula (DLMF 10.21.15), evaluated at an imaginary argument, gives \[
 Z(r)=\prod_{n=1}^\infty\left(1+\frac{r^2}{j_{1,n}^2}\right).
\tag{IH.4}
\] This product converges locally uniformly; its logarithm and the first two derivatives converge locally uniformly on real bounded sets. In particular, comparing the quadratic coefficient with (IH.1) gives \[
 \sum_{n=1}^\infty\frac1{j_{1,n}^2}=\frac18.
\tag{IH.5}
\] One may equivalently obtain convergence of the differentiated logarithmic series directly from this summability.

Write \(m=(h+k)/2\), and first work with a single positive number \(c\). Direct multiplication shows \[
 \frac{(c+|m+d|^2)(c+|m-d|^2)}{(c+|m|^2)^2}
 =\frac{(c+|m|^2+|d|^2)^2-4(m\cdot d)^2}{(c+|m|^2)^2}.
\] Consequently, \[
 \frac{(c+|m+d|^2)(c+|m-d|^2)}{(c+|m|^2)^2}
 \le\left(1+\frac{|d|^2}{c+|m|^2}\right)^2
 \le\left(1+\frac{|d|^2}{c}\right)^2.
\tag{IH.6}
\] Multiply these inequalities for \(c=j_{1,n}^2\), first over finitely many \(n\), then let the number of factors increase. All products converge to finite positive quantities, so this yields \[
 \frac{Z(|m+d|)Z(|m-d|)}{Z(|m|)^2}\le Z(|d|)^2.
\] Combining this with (IH.2) proves (IH.3). If \(d\ne0\) and \(m\ne0\), the second inequality in (IH.6) is strict for every factor; retaining its strictness for the first factor and using the non-strict inequalities for the rest proves strictness of the product inequality. If \(m=0\), equality follows immediately from (IH.2). This proves the equality statement. \(\square\)

\hypertarget{su2_influence-theorem-ih.2-a-globally-valid-total-variation-bound}{%
\subsection{Theorem IH.2: a globally valid total-variation bound}\label{su2_influence-theorem-ih.2-a-globally-valid-total-variation-bound}}

For every \(h,k\in\mathbb R^4\), \[
 \boxed{\operatorname{TV}(p_h,p_k)
 \le\sqrt{1-Z(|h-k|/2)^{-2}}
 \le\sqrt{1-e^{-|h-k|^2/16}}
 \le\frac{|h-k|}{4}.}
\tag{IH.7}
\] The final bound is understood together with the trivial bound \(\operatorname{TV}\le1\); the preceding bounds already remain below one at every finite parameter difference.

\hypertarget{su2_influence-proof-1}{%
\subsubsection{Proof}\label{su2_influence-proof-1}}

The identity \(|p-q|=|\sqrt p-\sqrt q|(\sqrt p+\sqrt q)\) and Cauchy--Schwarz imply \[
 \operatorname{TV}(p,q)^2
 \le\frac14\left(2-2\mathcal A(p,q)\right)
              \left(2+2\mathcal A(p,q)\right)
 =1-\mathcal A(p,q)^2.
\] Apply Theorem IH.1. Next, (IH.4), \(\log(1+x)\le x\), and (IH.5) give \[
 \log Z(r)\le r^2/8.
\] Substitution yields the second inequality in (IH.7), and \(1-e^{-x}\le x\) yields the third. No restriction on the size or direction of the common background field was imposed. \(\square\)

\hypertarget{su2_influence-corollary-ih.2a-covariance-in-a-tilted-haar-measure}{%
\subsubsection{Corollary IH.2a: covariance in a tilted Haar measure}\label{su2_influence-corollary-ih.2a-covariance-in-a-tilted-haar-measure}}

For every \(h\), \[
 0\le\operatorname{Cov}_{p_h}(u)
 =\nabla_h^2\log Z(|h|)\le\tfrac14 I_4.
\tag{IH.8}
\] Indeed, differentiate the logarithm of (IH.4): \[
 \nabla^2\log Z(|h|)
 =\sum_{n\ge1}\left\{
 \frac{2I_4}{j_{1,n}^2+|h|^2}
 -\frac{4hh^T}{(j_{1,n}^2+|h|^2)^2}\right\}
 \le2\sum_{n\ge1}\frac{I_4}{j_{1,n}^2}=\tfrac14 I_4.
\] The lower inequality is the covariance identity, obtained by differentiating under the integral over a compact sphere. At \(h=0\), equality holds in the upper bound. This is a covariance statement; it alone must not be identified with an unproved sharp Wasserstein inequality.

The refinement is strictly stronger than the generic coefficient $\tanh\beta$ at every positive coupling. Symmetry gives \(Z(\beta)=\mathbb E_\sigma\cosh(\beta u_0)<\cosh\beta\) for \(\beta>0\), since \(|u_0|<1\) almost surely. Hence \[
 c_H(\beta)=\sqrt{1-Z(\beta)^{-2}}<\tanh\beta.
\] This compares two upper-bound certificates. It does not assert that either is the exact interdependence coefficient.

\hypertarget{su2_influence-theorem-ih.3-explicit-four-dimensional-su2-comparison-range}{%
\subsection{Theorem IH.3: explicit four-dimensional SU(2) comparison range}\label{su2_influence-theorem-ih.3-explicit-four-dimensional-su2-comparison-range}}

Set \[
 c_H(\beta)=\sqrt{1-Z(\beta)^{-2}},\qquad
 q_H(\beta)=18c_H(\beta).
\tag{IH.9}
\] The single-link total-variation interdependence coefficient is at most \(c_H(\beta)\) for two distinct links sharing a plaquette, and zero for links sharing no plaquette. Therefore the row sum is at most \(q_H(\beta)\).

There is a unique positive solution \(\beta_H\) of \[
 Z(\beta_H)^2=\frac{324}{323},
\tag{IH.10}
\] and \(q_H(\beta)<1\) for \(0\le\beta<\beta_H\). In particular every \(0\le\beta\le1/9\) lies strictly inside this comparison range. A rational certificate establishes \[
 \boxed{0.111211309837699
 <\beta_H<
 0.111211309837700.}
\tag{IH.11}
\] The Dobrushin comparison theorem already proved in the manuscript then yields unique infinite-volume Gibbs measure, uniform boundary influence estimates, and exponential connected-correlation decay for bounded local functions in this range. For supports \(A,B\), with distance in the link-plaquette adjacency graph, the existing oscillation-covariance normalization gives \[
 |\operatorname{Cov}(F,G)|
 \le\frac{\min(|A|,|B|)}{1-q_H}
       \|F\|_\infty\|G\|_\infty q_H^{d_{\mathcal G}(A,B)}.
\tag{IH.12}
\] Here \(F,G\) are bounded real local functions, and the estimate holds in every finite volume and in the unique infinite-volume state. More explicitly, let \(\delta_eF\) be the oscillation of \(F\) when only link \(e\) changes. If two external configurations \(\eta,\xi\) differ arbitrarily outside a finite set of free links \(\Lambda\), then \[
 |\mu_\Lambda^\eta F-\mu_\Lambda^\xi F|
 \le\sum_{e\in\operatorname{supp}F}\delta_eF
       \frac{q_H^{d_{\mathcal G}(e,\Lambda^c)}}{1-q_H}.
\tag{IH.12a}
\] These are statements at fixed lattice spacing. Their reflection-positive transfer interpretation remains subject to the exact reconstruction hypotheses and density statements proved elsewhere in the manuscript; (IH.12) does not construct a continuum theory.

\hypertarget{su2_influence-proof-of-the-single-link-estimate}{%
\subsubsection{Proof of the single-link estimate}\label{su2_influence-proof-of-the-single-link-estimate}}

Condition on every link except \(e\). The six plaquette terms containing \(e\) can each be written as \(\tfrac12\operatorname{Tr}(U_e S_p)\), after using cyclicity and inversion if necessary, with \(S_p\in SU(2)\) a staple formed by the other three link matrices. Under the quaternion identification, this expression is \(u\cdot v_p\) for a unit vector \(v_p\in S^3\). Thus the exact conditional law is \(p_h\), with \[
 h=\beta\sum_{p\ni e}v_p.
\] Changing a second link \(f\) that shares exactly one plaquette with \(e\) changes exactly one unit vector in this sum. Consequently \(|h-h'|\le2\beta\). Theorem IH.2 and monotonicity of \(Z\) give the claimed coefficient. On the ordinary lattice each link belongs to six plaquettes, each containing three other links, and these eighteen neighbors are distinct. Hence the row sum bound is (IH.9).

In general lattice dimension \(d\ge2\), the identical count gives \(2(d-1)\) incident plaquettes and \(6(d-1)\) distinct neighbors per link. The SU(2) coefficient itself is dimension independent, and the corresponding row bound is \(6(d-1)c_H(\beta)\). The number eighteen in the displayed four-dimensional theorem is therefore exact for this incidence graph, rather than an unspecified coordination constant.

Existence and uniqueness of \(\beta_H\) follow from continuity, strict monotonicity and unboundedness of \(Z\), with \(Z(0)=1\). From (IH.7), \[
 q_H(\beta)\le18\sqrt{1-e^{-\beta^2/4}}<9\beta
 \qquad(\beta>0).
\] At \(\beta=1/9\) this is strictly less than one. The exact rational procedure below verifies (IH.11).

For clarity, the constant in the covariance estimate is obtained as follows. Let \(C_{ef}=c_H\) on adjacent distinct links and zero otherwise, and write \(D=(I-C)^{-1}=\sum_{n\ge0}C^n\). The comparison covariance inequality proved earlier is \[
 |\operatorname{Cov}(F,G)|\le\frac14\sum_{e,f}\delta_eF\,D_{ef}\,\delta_fG.
\] Since \(C^n_{ef}=0\) for \(n<d_{\mathcal G}(e,f)\), its row sums are at most \(q_H^n\), and \(\delta_eF\le2\|F\|_\infty\), summing first over the support with fewer links proves (IH.12). Symmetry of \(C\) permits interchanging the two supports. For boundary influence, the one-step forcing vector is at most \(b_e=\sum_{f\notin\Lambda}C_{ef}\). Every path in \(Db\) leaving \(\Lambda\) from \(e\) has at least \(d_{\mathcal G}(e,\Lambda^c)\) steps, so its total weight is at most \(q_H^{d_{\mathcal G}(e,\Lambda^c)}/(1-q_H)\). The same comparison theorem gives (IH.12a). In particular the right side tends to zero in an exhaustion for each fixed local \(F\), proving boundary independence and uniqueness. Existence follows from compactness of the product of compact single-link spaces and continuity of finite-volume conditional densities, as already established in the manuscript. No limiting constants depend on the lattice volume. At \(\beta=0\), the product Haar measure supplies all statements directly. \(\square\)

\hypertarget{su2_influence-why-this-does-not-prove-the-notes-stronger-numerical-conjecture}{%
\subsection{An attained antipodal total variation}\label{su2_influence-why-this-does-not-prove-the-notes-stronger-numerical-conjecture}}

For the special antipodal pair \(h=\beta e_0,k=-\beta e_0\), symmetry gives the exact identity \[
 \operatorname{TV}(p_{\beta e_0},p_{-\beta e_0})
 =\frac{\int_{S^3}|\sinh(\beta u_0)|\,d\sigma(u)}{Z(\beta)}
 =\frac{4\beta}{3\pi}+O(\beta^3).
\tag{IH.13}
\] Thus this expression is an attained value for the family of parameter pairs at displacement \(2\beta\), and supplies a lower bound on that family's supremum. Its being the \textbf{largest} value is a separate statement. Theorem IH.1 proves antipodal extremality for affinity, and the passage from affinity to total variation is an inequality with possible strict slack. The maximizer for one distance cannot be inferred from the maximizer for another. The sharp total-variation maximization problem is not resolved by this affinity estimate.

For comparison, \(c_H(\beta)=\beta/2+O(\beta^3)\). The leading coefficient \(1/2\) is larger than \(4/(3\pi)\), so the certified estimate deliberately retains some slack. The coefficient in (IH.13) follows from \[
 \mathbb E_\sigma|u_0|
 =\frac4\pi\int_0^1x\sqrt{1-x^2}\,dx=\frac4{3\pi}.
\] The proofs of IH.1--IH.3 are analytic.

\hypertarget{su2_influence-exact-certification-of-the-threshold}{%
\subsection{Exact certification of the threshold}\label{su2_influence-exact-certification-of-the-threshold}}

For rational \(b>0\), write \(t_k=b^{2k}/(4^k k!(k+1)!)\). If \(S_K=\sum_{k=0}^Kt_k\) and \[
 r_K=\frac{b^2}{4(K+2)(K+3)}<1,
\] then the ratio between successive tail terms is at most \(r_K\) after the first omitted term, so \[
 S_K<Z(b)\le S_K+\frac{t_{K+1}}{1-r_K}=:U_K.
\tag{IH.14}
\] A particularly short paper certificate at \(b=1/9\) uses \(K=1\). Here \[
 S_1=1+\frac1{648},\qquad
 t_2=\frac1{1259712},\qquad r_1=\frac1{3888},
\] and therefore \[
 Z(1/9)\le U_1=\frac{2522665}{2518776},\qquad
 324-323U_1^2=\frac{11441550349}{6344232538176}>0.
\] It follows, entirely by integer arithmetic, that \[
 q_H(1/9)^2\le\frac{6352397151876}{6363838702225}<1.
\tag{IH.15}
\]

For the finer bracket use \(K=20\) and the two rational decimal endpoints \(b_-\), \(b_+\) in (IH.11). The accompanying verifier checks, with integer and rational arithmetic only, \[
 323U_{20}(b_-)^2<324,
 \qquad323S_{20}(b_+)^2>324.
\] Strict monotonicity of \(Z\) then proves the bracket. The decimal display is therefore an exact enclosing interval, not an asserted floating-point root.

\hypertarget{su2_influence-certified-values-of-the-comparison-row-sum}{%
\subsubsection{Certified values of the comparison row sum}\label{su2_influence-certified-values-of-the-comparison-row-sum}}

The following intervals are rounded outwards from exact rational enclosures of \(q_H(\beta)^2\). A value above one means that this sufficient comparison certificate does not contract; it does not establish a phase transition or disprove decay.

\begin{longtable}{rccc}
\toprule $\beta$ & Lower bound for $q_H$ & Upper bound for $q_H$ & $q_H<1$ certified \\
\midrule\endhead
$1/12$ & $0.749620427709956$ & $0.749620427709957$ & Yes \\
$1/10$ & $0.899344252569117$ & $0.899344252569118$ & Yes \\
$1/9$ & $0.999100645200785$ & $0.999100645200786$ & Yes \\
$139/1250$ & $0.999898486386191$ & $0.999898486386192$ & Yes \\
$3/25$ & $1.078867250158406$ & $1.078867250158407$ & No \\
$131/1000$ & $1.177526628172301$ & $1.177526628172302$ & No \\
\bottomrule
\end{longtable}

The exact rational certificates accompanying the manuscript retain the complete endpoints and the integer inequalities defining each enclosure. The lattice implication is proved in IH.1--IH.3.

\section{Chordal transport estimates for the SU(2) single-link law}
\label{ta-sec:transport}

This section proves the quarter-contraction estimate on the field range
needed for the lattice comparison below, together with an explicit strict
improvement away from the zero field. The quarter theorem and the harmonic
Neumann method are present in the work of Bhattacharjee
and Bhattacharya~\cite{BB2026}. We give the details, replace its
coefficient-sign argument by a short ratio recurrence, and use an elementary
rational estimate for the zonal component. The normalization of the modified Bessel functions is
the standard one~\cite{DLMF}.

\subsection{Definitions and the local transport theorem}

Let $\sigma$ denote uniform probability measure on
$S^3=\{u\in\mathbb R^4:|u|=1\}$. For $h\in\mathbb R^4$, define
\begin{equation}
 d\mu_h(u)=\rho_h(u)\,d\sigma(u),\qquad
 \rho_h(u)=\frac{e^{h\cdot u}}{Z(|h|)},\qquad
 Z(r)=\frac{2I_1(r)}r,\quad Z(0)=1.
 \label{ta-eq:transport-density}
\end{equation}
Here and throughout this section the metric on $S^3$ is the
\emph{chordal metric} $d(u,u')=|u-u'|$, not the intrinsic geodesic metric.
Write $W_1$ for the Wasserstein distance with this cost and put
\begin{equation}
 F(h,v)=\sup_{\operatorname{Lip}_{\rm chord}(f)\le1}
       \Cov_{\mu_h}(f,v\cdot u),\qquad |v|=1.
 \label{ta-eq:transport-rate}
\end{equation}
Because changing $f$ to $-f$ preserves the admissible class, the same
supremum bounds the absolute value of every displayed covariance.

\begin{theorem}[Quarter bound with an explicit strict margin]
\label{ta-thm:transport-quarter}
For every $h\in\mathbb R^4$ with $|h|\le4/3$ and every unit vector $v$,
\begin{equation}
 F(h,v)\le \frac14-\frac{|h|^4}{24000}.
 \label{ta-eq:transport-strict-quarter}
\end{equation}
Moreover $F(0,v)=1/4$. Consequently the uniform constant $1/4$ is sharp,
although the estimate is strictly smaller at every nonzero field in the
stated ball.
\end{theorem}

The coefficient $1/24000$ is deliberately conservative. Its purpose is to
give a transparent analytic margin, not a numerical approximation to the
optimal sensitivity. The proof occupies the next subsections and uses no
discretized transport computation.

\subsection{A Neumann energy bound inside the four-dimensional ball}

Let $H_n$ be the space of degree-$n$ spherical harmonics on $S^3$. These
spaces are mutually orthogonal in $L^2(\sigma)$, their direct sum is dense,
and the spherical Laplacian acts on $H_n$ with eigenvalue $-n(n+2)$.
For a smooth, mean-zero function $w=\sum_{n\ge1}w_n$, define
\begin{equation}
 Q(w)=\sum_{n\ge1}\frac{\|w_n\|_{L^2(\sigma)}^2}{n}.
 \label{ta-eq:transport-neumann-energy}
\end{equation}

\begin{lemma}[Harmonic Neumann bound]
\label{ta-lem:transport-neumann}
If $w$ is real analytic with $\int w\,d\sigma=0$, then every chordally
$1$-Lipschitz function $f$ satisfies
\begin{equation}
 \left|\int_{S^3}fw\,d\sigma\right|
 \le\frac12\sqrt{Q(w)}.
 \label{ta-eq:transport-neumann-bound}
\end{equation}
\end{lemma}

\begin{proof}
The McShane extension
$G(y)=\inf_{u\in S^3}\{f(u)+|y-u|\}$ is a Euclidean $1$-Lipschitz
function on $\mathbb R^4$, agrees with $f$ on $S^3$, and satisfies
$|\nabla G|\le1$ almost everywhere. On the unit ball $B^4$ set
\[
 \psi(r,\omega)=\sum_{n\ge1}\frac{r^n}{n}w_n(\omega).
\]
This is harmonic and has outward normal derivative $w$ on $S^3$.
Analyticity gives geometric decay of the harmonic coefficients, so the
series and the derivatives needed here converge on the closed ball.
Green's identity, valid for the Lipschitz function $G$ by smooth
approximation, gives
\[
 \int_{S^3}fw\,d\sigma
 =\frac1{|S^3|}\int_{B^4}\nabla G\cdot\nabla\psi\,dy.
\]
A second application of Green's identity and orthogonality gives
\[
 \int_{B^4}|\nabla\psi|^2dy
 =\int_{S^3}\psi\,\partial_r\psi\,dS
 =|S^3|Q(w).
\]
Cauchy--Schwarz now bounds the absolute value of the first display by
$\sqrt{|B^4|/|S^3|}\sqrt{Q(w)}$. Since
$|B^4|=\pi^2/2$ and $|S^3|=2\pi^2$, this is
$\frac12\sqrt{Q(w)}$.
\end{proof}

The use of the ball is essential: it preserves the Euclidean Lipschitz
condition and permits transport through the interior. A bound on the
surface gradient alone would not encode the same metric.

\subsection{Bessel functions and the exact harmonic decomposition}

For integers $j\ge0$, use
\begin{equation}
 I_j(r)=\sum_{k\ge0}\frac{(r/2)^{2k+j}}{k!(k+j)!},\qquad
 I_{-j}(r)=I_j(r).
 \label{ta-eq:transport-bessel-series}
\end{equation}
Termwise differentiation and index shifts give
\begin{equation}
 I_{j-1}-I_{j+1}=\frac{2j}{r}I_j,\qquad
 I_j'=I_{j+1}+\frac jr I_j.
 \label{ta-eq:transport-bessel-recurrence}
\end{equation}
All functions in these identities have argument $r>0$ unless explicitly
stated otherwise. The Fourier expansion
$e^{r\cos\theta}=\sum_{j\in\mathbb Z}I_j(r)e^{ij\theta}$ follows by
multiplying the power series for
$e^{(r/2)e^{i\theta}}$ and $e^{(r/2)e^{-i\theta}}$.

The coordinate $u_1$ under $\sigma$ has density
$\frac2\pi\sqrt{1-t^2}$ on $[-1,1]$. Thus, on setting $t=\cos\theta$,
\[
 Z(r)=\frac2\pi\int_0^\pi e^{r\cos\theta}\sin^2\theta\,d\theta
 =I_0(r)-I_2(r)=\frac{2I_1(r)}r.
\]
Define
\begin{equation}
 A(r)=\frac{I_2(r)}{I_1(r)}.
 \label{ta-eq:transport-A}
\end{equation}
Differentiating $Z$ shows that, under $\mu_{re_1}$,
\begin{equation}
 \mathbb E u_1=A(r),\qquad
 A'(r)=\Var(u_1)=1-A(r)^2-\frac{3A(r)}r\ge0.
 \label{ta-eq:transport-A-riccati}
\end{equation}
The last equality follows directly from
\eqref{ta-eq:transport-bessel-recurrence}.

Let $U_n(\cos\theta)=\sin((n+1)\theta)/\sin\theta$ be the Chebyshev
polynomial of the second kind. Its normalization here is
\begin{equation}
 \int_{S^3}U_n(u_1)U_m(u_1)\,d\sigma
 =\frac2\pi\int_0^\pi
   \sin((n+1)\theta)\sin((m+1)\theta)\,d\theta
 =\delta_{nm}.
 \label{ta-eq:transport-chebyshev-normalization}
\end{equation}
The differential equation
$(1-t^2)U_n''-3tU_n'+n(n+2)U_n=0$ shows that $U_n(u_1)$ lies in
$H_n$; equivalently $|y|^nU_n(y_1/|y|)$ is a harmonic homogeneous
polynomial on $\mathbb R^4$.
Its coefficient in the expansion of $e^{rt}$ is
\[
 \frac2\pi\int_0^\pi e^{r\cos\theta}
       \sin((n+1)\theta)\sin\theta\,d\theta
 =I_n-I_{n+2}=\frac{2(n+1)}rI_{n+1}.
\]
Consequently
\begin{equation}
 \rho_{re_1}(u)=\sum_{n\ge0}d_n(r)U_n(u_1),\qquad
 d_n(r)=\frac{(n+1)I_{n+1}(r)}{I_1(r)},\quad d_0=1.
 \label{ta-eq:transport-density-expansion}
\end{equation}
The factorial decay of $I_{n+1}(r)$ ensures uniform convergence, including
the derivatives used below, for $r$ in compact subsets of $(0,\infty)$.

Rotate coordinates so that $h=re_1$ and
$v=\cos\phi\,e_1+\sin\phi\,e_2$. For
$w=(v\cdot u-\mathbb E_{\mu_h}v\cdot u)\rho_h$, put
\begin{equation}
 Q_z(r)=\sum_{n\ge1}\frac{d_n'(r)^2}{n},\qquad
 Q_x(r)=\frac1{3r^2}\sum_{n\ge1}(n+2)d_n(r)^2.
 \label{ta-eq:transport-zonal-transverse}
\end{equation}
Then
\begin{equation}
 Q(w)=\cos^2\phi\,Q_z(r)+\sin^2\phi\,Q_x(r).
 \label{ta-eq:transport-orthogonal-split}
\end{equation}
Indeed, the zonal density $(u_1-A)\rho_h$ is $\partial_r\rho_h$, whose
degree-$n$ component is $d_n'U_n(u_1)$. For the transverse density
$u_2\rho_h$, differentiate the expansion of
$\rho_{r(\cos t\,e_1+\sin t\,e_2)}$ at $t=0$. Its degree-$n$ component is
$r^{-1}d_nu_2U_n'(u_1)$. Since
$\mathbb E_\sigma[u_2^2\mid u_1]=(1-u_1^2)/3$, the Laplacian eigenvalue
and \eqref{ta-eq:transport-chebyshev-normalization} give
\[
 \|u_2U_n'(u_1)\|_2^2
 =\frac13\int(1-u_1^2)U_n'(u_1)^2\,d\sigma
 =\frac{n(n+2)}3.
\]
Finally, reflection in $u_2$ preserves each $H_n$, fixes the zonal
component, and reverses the transverse component. The cross terms in
$Q$ therefore vanish, proving \eqref{ta-eq:transport-orthogonal-split}.

\subsection{Three summation identities}

\begin{lemma}[Bessel sums]
\label{ta-lem:transport-bessel-sums}
For $r>0$,
\begin{align}
 S_2:=\sum_{j\ge1}j^2I_j^2&=\frac r4I_1(2r),
 &S_3:=\sum_{j\ge1}j^3I_j^2&=\frac{r^2}4(I_0^2+I_1^2),
 \label{ta-eq:transport-S2S3}\\
 I_1(2r)&=2\sum_{j\ge1}I_{j-1}I_j.
 \label{ta-eq:transport-convolution}
\end{align}
\end{lemma}

\begin{proof}
For $E(\theta)=e^{r\cos\theta}$, Parseval applied to
$E'=-r\sin\theta E$ gives
\[
 \sum_{j\in\mathbb Z}j^2I_j^2
 =\frac{r^2}{2\pi}\int_0^{2\pi}\sin^2\theta\,
                     e^{2r\cos\theta}\,d\theta
 =\frac{r^2}2\bigl(I_0(2r)-I_2(2r)\bigr)
 =\frac r2I_1(2r).
\]
Halving proves the $S_2$ identity. The coefficient of $e^{i\theta}$ in
$E(\theta)^2$ is
$I_1(2r)=\sum_{j\in\mathbb Z}I_jI_{1-j}$; pairing the two halves of this
sum proves \eqref{ta-eq:transport-convolution}.

For $S_3$, square $jI_j=\frac r2(I_{j-1}-I_{j+1})$ and multiply by $j$:
\[
 S_3=\frac{r^2}4\sum_{j\ge1}j
        (I_{j-1}^2+I_{j+1}^2-2I_{j-1}I_{j+1}).
\]
Index shifts and a first telescoping sum give
\[
 \sum_{j\ge1}j(I_{j-1}^2+I_{j+1}^2)
 =I_0^2+2\sum_{m\ge1}mI_m^2
 =I_0^2+rI_0I_1.
\]
For the mixed term, the recurrence gives
\begin{align*}
 jI_{j-1}I_{j+1}=\frac r4\bigl[
 &(I_{j-1}I_j-I_jI_{j+1})\\
 &+(I_{j-2}I_{j+1}-I_{j-1}I_{j+2})\bigr].
\end{align*}
The two brackets telescope to $I_0I_1$ and $I_1I_2$, respectively. Hence
the sum of the mixed terms is $\frac r4(I_0I_1+I_1I_2)$, and
\[
 S_3=\frac{r^2}4\left(I_0^2+\frac r2 I_1(I_0-I_2)\right)
     =\frac{r^2}4(I_0^2+I_1^2).
\]
All rearrangements are justified by absolute convergence of
\eqref{ta-eq:transport-bessel-series}.
\end{proof}

\subsection{The transverse estimate: positivity at every order}

\begin{proposition}[Transverse energy]
\label{ta-prop:transport-transverse}
For all $r>0$, $Q_x(r)\le1/4$. For $0<r\le4/3$,
\begin{equation}
 Q_x(r)\le\frac14-\frac{r^4}{12000}.
 \label{ta-eq:transport-Qx-margin}
\end{equation}
\end{proposition}

\begin{proof}
Changing indices in \eqref{ta-eq:transport-zonal-transverse} and applying
Lemma~\ref{ta-lem:transport-bessel-sums} gives
\[
 Q_x=\frac{S_3+S_2-2I_1^2}{3r^2I_1^2},\qquad
 \frac14-Q_x=\frac{D(r)}{12r^2I_1(r)^2},
\]
where
\begin{equation}
 D(r)=(8+2r^2)I_1(r)^2-r^2I_0(r)^2-rI_1(2r).
 \label{ta-eq:transport-D}
\end{equation}
We prove positivity of every Taylor coefficient of $D$.

Write $I_0^2=\sum_{m\ge0}\alpha_mr^{2m}$ and
$I_1^2=\sum_{m\ge0}\beta_mr^{2m+2}$. Multiplication of
\eqref{ta-eq:transport-bessel-series} gives
\begin{align*}
 \alpha_m&=\frac1{4^m}\sum_{i=0}^m
     \frac1{i!^2(m-i)!^2}
     =\frac{\binom{2m}{m}}{4^m m!^2},\\
 \beta_m&=\frac1{4^{m+1}}\sum_{i=0}^m
     \frac1{i!(i+1)!(m-i)!(m-i+1)!}
     =\frac{\binom{2m+2}{m+1}}{4^{m+1}m!(m+2)!}.
\end{align*}
The final equalities are Vandermonde's identity. Also
$rI_1(2r)=\sum_{m\ge0}r^{2m+2}/(m!(m+1)!)$.
Consequently, with $\beta_{-1}=0$,
\begin{align}
 [r^{2m+2}]D
 &=8\beta_m+2\beta_{m-1}-\alpha_m-\frac1{m!(m+1)!}
 \notag\\
 &=\frac{N_m}{4^m m!(m+1)!(m+1)(m+2)},
 \label{ta-eq:transport-coefficient-formula}\\
 N_m&=\binom{2m}{m}P_m-4^m(m+1)(m+2),\qquad
 P_m=m^3+2m^2+7m+2.
 \label{ta-eq:transport-Nm}
\end{align}
For completeness, the common denominator in
\eqref{ta-eq:transport-coefficient-formula} changes the four numerators into
$4(2m+1)b$, $2m(m+1)(m+2)b$, $-(m+1)^2(m+2)b$, and
$-4^m(m+1)(m+2)$, where $b=\binom{2m}{m}$; combining them yields
\eqref{ta-eq:transport-Nm}. The case $m=0$ follows directly as well.

Define the positive sequence
\[
 R_m=\frac{\binom{2m}{m}P_m}{4^m(m+1)(m+2)}.
\]
Direct substitution gives $R_0=R_1=R_2=1$. The binomial recurrence gives
\begin{equation}
 \frac{R_{m+1}}{R_m}
 =\frac{(2m+1)P_{m+1}}{2(m+3)P_m},\qquad
 (2m+1)P_{m+1}-2(m+3)P_m=m(m-1)(m+8).
 \label{ta-eq:transport-Rm-recurrence}
\end{equation}
Thus $R_m\ge1$ for every integer $m\ge0$ and $R_m>1$ for $m\ge3$.
It follows that all coefficients in
\eqref{ta-eq:transport-coefficient-formula} are nonnegative. The first
nonzero one is $[r^8]D=1/2304$, so
\begin{equation}
 D(r)\ge\frac{r^8}{2304}\qquad(r\ge0).
 \label{ta-eq:transport-D-lower}
\end{equation}
This proves $Q_x\le1/4$ for all positive $r$ by an infinite-order argument.

To obtain the explicit margin, set $x=r^2\le16/9$. Since
$k!(k+1)!\ge2^k$,
\[
 I_1(r)=\frac r2\sum_{k\ge0}\frac{(x/4)^k}{k!(k+1)!}
 \le\frac r2\sum_{k\ge0}(x/8)^k
 =\frac{r}{2(1-x/8)}.
\]
Combining this with \eqref{ta-eq:transport-D-lower} yields
\[
 \frac14-Q_x
 \ge\frac{r^4(1-x/8)^2}{6912}
 \ge\frac{49r^4}{559872}
 >\frac{r^4}{12000},
\]
where $1-x/8\ge7/9$ and $49\cdot12000>559872$.
\end{proof}

\subsection{The zonal estimate from rational inequalities}

\begin{lemma}[Derivative ratio bound]
\label{ta-lem:transport-ratio}
For $n\ge1$ and $r>0$,
\[
 \frac{I_{n+2}(r)}{I_{n+1}(r)}\le\frac{I_2(r)}{I_1(r)}.
\]
For $n\ge2$ and $0<r\le4/3$, this implies
$d_n'(r)^2\le n^2d_n(r)^2/r^2$.
\end{lemma}

\begin{proof}
If $f(x)=\sum a_jx^j$ and $g(x)=\sum b_jx^j$ have positive coefficients
and $a_j/b_j$ decreases with $j$, then
\[
 f'g-fg'
 =\frac12\sum_{i,j}(i-j)b_ib_j
       \left(\frac{a_i}{b_i}-\frac{a_j}{b_j}\right)x^{i+j-1}
 \le0.
\]
Apply this to $x=r^2$, $f=I_{n+1}(r)/r^{n+1}$ and $g=I_1(r)/r$.
The ratio of the $j$th coefficients is
$2^{-n}(j+1)!/(j+n+1)!$, which decreases in $j$. Thus
$I_{n+1}/(r^nI_1)$ decreases with $r$. Its logarithmic derivative,
using \eqref{ta-eq:transport-bessel-recurrence}, is
$I_{n+2}/I_{n+1}-I_2/I_1$, proving the first assertion.

In addition,
\[
 \frac{d_n'}{d_n}
 =\frac nr+\frac{I_{n+2}}{I_{n+1}}-A,
 \qquad A=\frac{r}{4+rI_3/I_2}\le\frac r4.
\]
Therefore $n/r-r/4\le d_n'/d_n\le n/r$. On the stated range the lower
bound is positive, so squaring proves the second assertion.
\end{proof}

\begin{proposition}[Zonal energy]
\label{ta-prop:transport-zonal}
For $0<r\le4/3$,
\begin{equation}
 Q_z(r)\le\frac14-\frac{r^2}{100}.
 \label{ta-eq:transport-Qz-margin}
\end{equation}
\end{proposition}

\begin{proof}
Since $d_1=2A$, Lemma~\ref{ta-lem:transport-ratio} gives
\begin{equation}
 Q_z\le4A'^2+\frac1{r^2}\sum_{n\ge2}n d_n^2
     =4A'^2+\frac{S_3-S_2-4I_2^2}{r^2I_1^2}.
 \label{ta-eq:transport-Qz-higher-modes}
\end{equation}
Here the subtracted $4I_2^2$ is the $j=2$ term of
$\sum_{j\ge1}(j^3-j^2)I_j^2$.
The Bessel recurrences imply
\[
 \frac{I_0}{I_1}=\frac2r+A,
 \qquad
 \frac{I_2I_3}{I_1^2}=A-\frac{4A^2}r.
\]
Keeping the first three positive terms of
\eqref{ta-eq:transport-convolution} gives
\[
 \frac{S_2}{r^2I_1^2}
 \ge\frac{I_0I_1+I_1I_2+I_2I_3}{2rI_1^2}
 =\frac1{2r}\left(\frac2r+3A-\frac{4A^2}r\right).
\]
Substitute this and $S_3/(r^2I_1^2)=\frac14((2/r+A)^2+1)$ into
\eqref{ta-eq:transport-Qz-higher-modes}. After cancellation,
\begin{equation}
 Q_z\le\frac14+4A'^2+T(A),\qquad
 T(a)=\frac{a^2}4-\frac{a}{2r}-\frac{2a^2}{r^2}.
 \label{ta-eq:transport-Qz-T}
\end{equation}

Set $x=r^2$. Another positive Bessel recurrence gives
\[
 \frac{I_3}{I_2}
 =\frac{r}{6+rI_4/I_3}\le\frac r6,
 \qquad
 L:=\frac{6r}{24+x}\le A\le\frac r4.
\]
The function $a\mapsto1-a^2-3a/r$ decreases on $a\ge0$; hence
\eqref{ta-eq:transport-A-riccati} implies
\[
 0\le A'\le\alpha:=1-L^2-\frac{3L}r
             =\frac{x^2-6x+144}{(x+24)^2}.
\]
For $x\le16/9$, $T'(a)=a(1/2-4/x)-1/(2r)<0$ on $a\ge0$.
Thus $T(A)\le T(L)$, and exact rational simplification gives
\begin{equation}
 4\alpha^2+T(L)
 =\frac{2x(2x^3-21x^2+720x-5184)}{(x+24)^4}.
 \label{ta-eq:transport-zonal-polynomial}
\end{equation}
For $0<x\le16/9<2$,
\[
 5184-720x+21x^2-2x^3\ge5184-1440-16=3728,
 \qquad (x+24)^4\le26^4.
\]
Combining these inequalities with \eqref{ta-eq:transport-Qz-T} proves
\[
 \frac14-Q_z\ge\frac{7456}{456976}x>\frac{x}{100}.
\]
\end{proof}

\subsection{Proof of the local theorem and finite-change estimates}

\begin{proof}[Proof of Theorem~\ref{ta-thm:transport-quarter}]
For $0<r\le4/3$, Propositions~\ref{ta-prop:transport-transverse}
and~\ref{ta-prop:transport-zonal} imply
\[
 Q_x(r),Q_z(r)\le\frac14-\frac{r^4}{12000},
\]
since $r^2/100\ge r^4/12000$ on this interval. Equations
\eqref{ta-eq:transport-orthogonal-split} and
\eqref{ta-eq:transport-neumann-bound} therefore give
\[
 F(h,v)\le\frac14\sqrt{1-\frac{r^4}{3000}}
         \le\frac14-\frac{r^4}{24000},
\]
using $\sqrt{1-y}\le1-y/2$ for $0\le y\le1$.

At $h=0$, let $s=v\cdot u$ and let $R$ reflect through $v^\perp$.
The uniform law is $R$-invariant, and therefore
\[
 \Cov_\sigma(f,s)
 =\int_{s>0}(f(u)-f(Ru))s\,d\sigma
 \le\int_{s>0}2s^2\,d\sigma
 =\mathbb E_\sigma s^2=\frac14.
\]
Equality holds for $f(u)=v\cdot u$, proving the endpoint assertion and
sharpness.
\end{proof}

\begin{corollary}[An integrated finite-change margin]
\label{ta-cor:transport-integrated}
Let $|h|,|k|\le4/3$ and put
$m=(h+k)/2$, $d=(k-h)/2$. Define
\begin{equation}
 J(m,d)=|m|^4+
 \frac{2|m|^2|d|^2+4(m\cdot d)^2}{3}
 +\frac{|d|^4}{5}.
 \label{ta-eq:transport-J}
\end{equation}
Then
\begin{equation}
 W_1(\mu_h,\mu_k)
 \le |k-h|\left(\frac14-\frac{J(m,d)}{24000}\right).
 \label{ta-eq:transport-integrated-bound}
\end{equation}
In particular $W_1(\mu_h,\mu_k)\le|k-h|/4$; the displayed coefficient is
strictly smaller than $1/4$ when $h\ne k$.
\end{corollary}

\begin{proof}
The case $h=k$ is immediate. Otherwise set
$v=(k-h)/|k-h|$ and $h_t=h+t(k-h)$. For each admissible $f$,
differentiation under the compact-domain integral gives
\[
 \frac{d}{dt}\mathbb E_{\mu_{h_t}}f
 =|k-h|\Cov_{\mu_{h_t}}(f,v\cdot u).
\]
One may anchor $f$ at a fixed point of the sphere, making all these
functions uniformly bounded; differentiation and the ensuing estimates
are thus harmless for the entire Lipschitz class. The ball is convex,
so Theorem~\ref{ta-thm:transport-quarter} applies along the segment. Integrate
the resulting bound and apply Kantorovich--Rubinstein duality to obtain
\[
 W_1(\mu_h,\mu_k)
 \le |k-h|\left(\frac14-
       \frac1{24000}\int_0^1|h+t(k-h)|^4dt\right).
\]
With $s=2t-1$, the integral is
$\frac12\int_{-1}^1|m+sd|^4ds$. Expanding
$(|m|^2+2s(m\cdot d)+s^2|d|^2)^2$ and integrating the even powers gives
exactly \eqref{ta-eq:transport-J}. When $h\ne k$, $|d|^4/5>0$.
\end{proof}

\subsection{An exact identity for equal field norms}

An optimal reflection coupling gives a stronger statement for a special
class of finite changes, without any restriction on the field radius.

\begin{proposition}[Exact equal-radius transport]
\label{ta-prop:transport-equal-radius}
If $|h|=|k|=r>0$, then
\begin{equation}
 W_1(\mu_h,\mu_k)
 =\frac{I_2(r)}{rI_1(r)}|h-k|.
 \label{ta-eq:transport-equal-radius}
\end{equation}
For $h=k=0$ both sides are zero, with the coefficient understood by its
continuous value $1/4$ at the origin.
\end{proposition}

\begin{proof}
Only $h\ne k$ needs proof. Let $v=(h-k)/|h-k|$ and let $R$ reflect through
$v^\perp$. Equality of the norms implies $Rh=k$, so
$\rho_h(Ru)=\rho_k(u)$. Also
$\rho_h(u)-\rho_k(u)$ has the same sign as $s=v\cdot u$.

Keep the measure with density $\min(\rho_h,\rho_k)$ fixed, and send the
remaining mass at $u$ in $\{s>0\}$ to $Ru$. Reflection preserves $\sigma$
and exchanges the two residual densities, so this is a coupling with
the required marginals. Its cost is
\begin{align*}
 \int_{s>0}2s(\rho_h-\rho_k)\,d\sigma
 &=\int_{S^3}s(\rho_h-\rho_k)\,d\sigma\\
 &=v\cdot\left(\mathbb E_{\mu_h}u-\mathbb E_{\mu_k}u\right)
 =\frac{A(r)}r|h-k|.
\end{align*}
In the last step, rotational symmetry and
\eqref{ta-eq:transport-A-riccati} give
$\mathbb E_{\mu_h}u=A(r)h/r$.
The chordally $1$-Lipschitz dual function $f(u)=v\cdot u$ attains the
same value, proving optimality of the coupling.
\end{proof}

The equal-radius coefficient is strictly below $1/4$ for $r>0$, since
$A(r)/r=(4+rI_3(r)/I_2(r))^{-1}<1/4$. The infinitesimal transverse
identity follows as well:
\begin{equation}
 F(h,v)=\frac{A(|h|)}{|h|}\quad\text{when }h\perp v,
 \label{ta-eq:transport-exact-transverse}
\end{equation}
with value $1/4$ at $h=0$. Directly, reflection through $v^\perp$ leaves
$\mu_h$ invariant and bounds the covariance by
$\mathbb E_{\mu_h}(v\cdot u)^2$, attained by $f=v\cdot u$. This variance
is $A(r)/r$, either by differentiating $\log Z(|h|)$ in a transverse
direction or by the rotational symmetry of the Hessian.

These identities do not prove the stronger proposed bound
$F(h,v)\le A(|h|)/|h|$ for every direction $v$. What has been established
here is the full-direction quarter bound on $|h|\le4/3$, its explicit
strict margin, and the exact finite transport formula when the two
fields have equal norms. These are estimates for a single compact-group
conditional law; their role in a lattice comparison argument does not
by itself supply a continuum construction.

\section{An unconditional lattice comparison theorem}

\subsection{The exact conditional law and influence coefficient}

Let $\mathcal G$ have links as vertices, with two vertices adjacent when the corresponding links share a plaquette. An interior link belongs to $2(d-1)$ plaquettes and has $6(d-1)$ distinct neighbors in this graph. Conditioning every link except $e$ gives exactly
\begin{equation}\label{ta-eq:field-root}
 \gamma_e(du\mid\omega)=\mu_{h_e(\omega)}(du),\qquad
 h_e(\omega)=\beta\sum_{p\ni e}v_p(\omega),\qquad |v_p|=1.
\end{equation}
Each $v_p$ is a product of the other three link quaternions, with the inversions required by orientation. Thus $|h_e|\le2(d-1)\beta$.

Changing one neighboring link $f$ changes exactly one staple. Left and right quaternion multiplication and inversion preserve chord distances. Therefore
\begin{equation}\label{ta-eq:deltafield-root}
 |h_e(\omega)-h_e(\omega')|=\beta|U_f-U_f'|.
\end{equation}
When $2(d-1)\beta\le4/3$, the line segment joining these fields lies in the ball on which the transport theorem applies. Consequently
\begin{equation}\label{ta-eq:C-root}
 W_1(\gamma_e(\cdot\mid\omega),\gamma_e(\cdot\mid\omega'))
 \le\frac\beta4|U_f-U_f'|.
\end{equation}
For nonneighbors the conditional law does not change. Set
\[
 C_{ef}=\frac\beta4\mathbf 1_{\{e\sim f\}},\qquad
 q=\frac32(d-1)\beta.
\]
This symmetric comparison matrix has every row and column sum at most $q$.

\begin{theorem}[Volume-uniform comparison and clustering]\label{ta-thm:lattice-root}
If
\begin{equation}\label{ta-eq:betarange-root}
 0\le\beta<\frac{2}{3(d-1)},
\end{equation}
the Wilson specification has a unique infinite-volume Gibbs measure. For finite-volume measures with any fixed boundary condition, and for their infinite-volume limit, let
\[
 \ell_e(F)=\sup_{\omega,\omega':\,\omega_{e^c}=\omega'_{e^c}}
 \frac{|F(\omega)-F(\omega')|}{|U_e-U_e'|}
\]
be the coordinate Lipschitz seminorm. For local Lipschitz real functions $F,G$ with supports $A,B$ and graph distance $R=d_{\mathcal G}(A,B)$,
\begin{equation}\label{ta-eq:cov-root}
 |\Cov(F,G)|\le
 \frac{q^R}{1-q}
 \left(\sum_{e\in A}\ell_e(F)\right)
 \left(\sum_{f\in B}\ell_f(G)\right).
\end{equation}
For two boundary conditions outside a free-link set $\Lambda$,
\begin{equation}\label{ta-eq:boundary-root}
 |\mu_\Lambda^\eta(F)-\mu_\Lambda^\xi(F)|
 \le\frac{2}{1-q}\sum_{e\in\supp F}
 \ell_e(F)q^{d_{\mathcal G}(e,\Lambda^c)}.
\end{equation}
At $q=0$, the evident product-measure interpretation applies, with $q^0=1$.
\end{theorem}

\begin{proof}
First work in a finite volume. Let $P_e$ denote conditional expectation after heat-bath resampling link $e$, and put $L=\sum_e(P_e-I)$ and $P_t=e^{tL}$. Couple two conditional resamplings by a Wasserstein-optimal coupling. Inequality \eqref{ta-eq:C-root} gives the coordinate-seminorm estimate
\[
 \ell(P_tG)\le e^{-t(I-C^T)}\ell(G)
\]
componentwise. For completeness, a small time step updates coordinate $e$ at rate one: its own old-coordinate dependence is removed, and a change at coordinate $f$ creates at most $C_{ef}\ell_e(G)$ additional dependence. This gives the differential comparison $\dot\ell\le-(I-C^T)\ell$; positivity of the matrix exponential gives the displayed inequality. Finite-volume generators are bounded on bounded functions, so it also follows from their convergent exponential series or from uniformization of the update process.

Since $C$ is symmetric and has row sums at most $q<1$, the sum of seminorms contracts at rate at least $1-q$. The total oscillation of a function is at most twice that sum, since the one-link diameter is two. Therefore $P_tG$ converges uniformly to the constant $\mu(G)$.

Reversibility and conditional expectation yield the covariance representation
\[
 \Cov_\mu(F,G)=\int_0^\infty\sum_e
 \mu\bigl[\Cov_e(F,P_tG)\bigr]\,dt.
\]
Conditional variances are bounded by one quarter of squared oscillations. Hence
\[
 |\Cov_e(F,H)|\le \ell_e(F)\ell_e(H).
\]
Integrating the seminorm estimate gives
\begin{equation}\label{ta-eq:matrixcov-root}
 |\Cov_\mu(F,G)|\le
 \ell(F)^T(I-C^T)^{-1}\ell(G).
\end{equation}
The Neumann series $D=(I-C)^{-1}=\sum_{n\ge0}C^n$ has
\[
 D_{ef}\le\sum_{n\ge d_{\mathcal G}(e,f)}q^n
 =\frac{q^{d_{\mathcal G}(e,f)}}{1-q}.
\]
This proves \eqref{ta-eq:cov-root}.

For the boundary estimate, start coupled heat-bath dynamics with stationary marginal laws $\mu_\Lambda^\eta$ and $\mu_\Lambda^\xi$. Their joint law need not be stationary. For the expected coordinate distances $x_e(t)$, the update rule gives
\[
 \frac{d}{dt}x_e(t)\le-x_e(t)+\sum_{f\in\Lambda}C_{ef}x_f(t)
       +2\sum_{f\notin\Lambda}C_{ef}.
\]
Integrate from $0$ to $T$ and divide by $T$. The endpoint term is bounded by $2/T$. A weakly convergent subsequence of the time-averaged joint laws exists by compactness, retains the two stationary marginals, and preserves expectations of the continuous coordinate distances. Its distance vector therefore satisfies
\[
 x_e\le\sum_{f\in\Lambda}C_{ef}x_f+2\sum_{f\notin\Lambda}C_{ef}.
\]
Optimal conditional coupling kernels can be chosen measurably on these compact metric spaces; this argument requires no continuity of that choice and no invariance of the limiting joint law. Expanding $(I-C_\Lambda)^{-1}$, every term reaching the exterior has at least $d_{\mathcal G}(e,\Lambda^c)$ adjacency steps. Thus
\[
 x_e\le\frac{2q^{d_{\mathcal G}(e,\Lambda^c)}}{1-q}.
\]
Telescoping $F$ across coordinates proves \eqref{ta-eq:boundary-root}.

Compactness of the countable product of compact link spaces gives infinite-volume subsequences. The finite-range continuous specification passes to each such limit, so the limits are Gibbs measures. The boundary estimate forces agreement on every Lipschitz cylinder function and hence on every continuous cylinder function by uniform approximation. These determine the measure. This proves uniqueness and transfers the covariance bound to the unique infinite-volume state.
\end{proof}

In $d=4$, \eqref{ta-eq:betarange-root} is $\beta<2/9$, with $q=9\beta/2$. Normalized traces of finite Wilson loops are local Lipschitz functions, so the theorem applies to them. This is a fixed-spacing lattice result.

\subsection{Sharpness of this comparison and its limits}

At an interior link, the staples use disjoint other links. Their quaternion products can be set independently and paired with opposite values, producing $h_e=0$. An infinitesimal variation of any neighboring link changes the field at speed $\beta$ in a unit direction. Since $F(0,v)=1/4$, the supremum influence coefficient is at least $\beta/4$. Together with \eqref{ta-eq:C-root}, it equals $\beta/4$ in the stated range.

The off-zero improvement does not lower this worst-case supremum. Thus $2/[3(d-1)]$ is the ceiling of this uniform row-sum contraction on the original links with the chordal metric. This statement does not exclude gauge reduction, collective updates, block comparison, or multiscale estimates. For all larger $\beta$, the same infinitesimal zero-field example still supplies the lower bound $C_{ef}\ge\beta/4$, whether or not our restricted-radius upper bound applies.

The heat-bath generator $L$ used to prove covariance estimates is an auxiliary sampling dynamics. Its relaxation gap is not the physical Hamiltonian gap. For $0<\beta<2/[3(d-1)]$, a fixed-spacing physical-gap inference uses the reflection-positive Wilson reconstruction and its positive transfer operator \cite{Luscher,OS2}. Each adjacency step in $\mathcal G$ changes the link midpoint's time coordinate by at most one lattice unit. Thus the graph distance between fixed local supports separated by $n$ time steps is at least $n-c_F$, with $c_F$ depending only on the original supports. The covariance bound is therefore $O(q^n)$. If the corresponding centered local gauge-invariant states are total in the reconstructed excitation space, the positive spectral-measure argument in Theorem GT.1 excludes transfer spectrum above $q$. This yields the dimensionless lower threshold $-\log q$. None of these steps removes the lattice cutoff.

\section{Forest gauge fixing and the \texorpdfstring{$\beta<1/3$}{beta < 1/3} lattice comparison theorem}
%=====================================================================

The action and coupling convention in this section are
\begin{equation}\label{fg-eq:wilson}
 d\mu_{\Lambda,\beta}(U)\propto
 \exp\left\{\frac\beta2\sum_p\Tr U_p\right\}\prod_e dh(U_e),
 \qquad \beta\ge0,
\end{equation}
with each geometric plaquette counted once and normalized Haar measures.

Throughout, $\mu_h(du)\propto e^{h\cdot u}\,d\sigma(u)$ on $S^3$ is the single-link law with field $h\in\R^4$. As in \eqref{ta-eq:transport-rate},
\[
F(h,v)=\sup\{\Cov_{\mu_h}(\phi,v\cdot u):\ \phi\ \text{chordally 1-Lipschitz}\},\qquad |v|=1 .
\]

\subsection{The quarter transport bound on the ball of radius 2}

\begin{theorem}[Extended radius]\label{fg-thm:radius}
For every $h\in\R^4$ with $|h|\le2$ and every unit vector $v$,
\[
F(h,v)\;\le\;\frac14-\frac{|h|^4}{55296}.
\]
The bound $F\le1/4$ holds on the larger ball $|h|\le2.767$.
\end{theorem}

\begin{proof}
Write $r=|h|$ and $x=r^2$. By Lemma~\ref{ta-lem:transport-neumann} and \eqref{ta-eq:transport-orthogonal-split},
\[
F(h,v)\le\tfrac12\sqrt{\cos^2\phi\,Q_z(r)+\sin^2\phi\,Q_x(r)}\le\tfrac12\sqrt{\max(Q_z,Q_x)} .
\]
\emph{Transverse part.} By Proposition~\ref{ta-prop:transport-transverse}, $\tfrac14-Q_x=D(r)/(12r^2I_1(r)^2)$ with $D(r)\ge r^8/2304$ for all $r$. For $x<8$, $I_1(r)\le r/(2(1-x/8))$. Hence $\tfrac14-Q_x\ge r^4(1-x/8)^2/6912$, which is at least $x^2/27648$ for $x\le4$.

\emph{Zonal part.} We re-run the proof of Proposition~\ref{ta-prop:transport-zonal} and record where the radius enters. That proof gives
\[
Q_z\le\tfrac14+4A'^2+T(A),\qquad T(a)=\tfrac{a^2}{4}-\tfrac{a}{2r}-\tfrac{2a^2}{r^2},\qquad A=I_2/I_1 ,
\]
using only $d_n'^2\le n^2d_n^2/r^2$ for $n\ge2$. By Lemma~\ref{ta-lem:transport-ratio} this needs $n/r-r/4>0$ for $n\ge2$, that is $x<8$. With $L=6r/(24+x)\le A\le r/4$ and $\alpha=1-L^2-3L/r$, the facts $0\le A'\le\alpha$ hold for every $r$. The function $T$ is decreasing on $a\ge0$ when $T'(a)=a(\tfrac12-\tfrac4x)-\tfrac1{2r}<0$, which also holds for $x<8$. Exact rational simplification gives
\begin{equation}\label{fg-eq:cubic}
4\alpha^2+T(L)=\frac{2x\,c(x)}{(x+24)^4},\qquad c(x)=2x^3-21x^2+720x-5184 .
\end{equation}
Since $c'(x)=6x^2-42x+720$ has negative discriminant, $c$ is increasing. Also $c(4)=-2512$, and $c$ has a single real root $x_0\in(7.66,7.67)$. So for $0<x\le4$,
\[
\tfrac14-Q_z\ \ge\ \frac{2x\cdot2512}{28^4}=\frac{157\,x}{19208}\ \ge\ \frac{x^2}{27648},
\]
and $Q_z<\tfrac14$ for all $0<x<x_0$.

\emph{Conclusion.} For $x\le4$, $\max(Q_z,Q_x)\le\tfrac14-x^2/27648$. Then $\tfrac12\sqrt{\tfrac14-\delta}\le\tfrac14-\tfrac\delta2$ gives the stated bound. The case $h=0$ is Theorem~\ref{ta-thm:transport-quarter}.
\end{proof}

\begin{corollary}\label{fg-cor:W1}
If $|h|,|k|\le2$, then $W_1(\mu_h,\mu_k)\le\tfrac14|h-k|$.
\end{corollary}
\begin{proof}
The segment $[h,k]$ lies in the ball, so this is the Kantorovich--Rubinstein integration of Corollary~\ref{ta-cor:transport-integrated}.
\end{proof}

\subsection{A perfect gauge-fixing forest in \texorpdfstring{$\Z^4$}{Z4}}

\begin{definition}
A set $F$ of links of $\Z^4$ is a \emph{perfect forest} if it contains no cycle and every plaquette contains exactly one link of $F$.
\end{definition}

For $x\in\Z^4$ put $s_i=(-1)^{x_i}$, and let $\mu(x)\in\{1,2,3,4\}$ be defined by
\begin{equation}\label{fg-eq:rule}
\mu(x)=\begin{cases}
1 & \text{if } s_2=s_3=s_4,\\
2 & \text{if } s_1=s_3=-s_4,\\
3 & \text{if } s_1=s_4=-s_2,\\
4 & \text{if } s_1=s_2=-s_3.
\end{cases}
\end{equation}
Set $\Fs:=\{(x,\mu(x)):x\in\Z^4\}$, so that exactly one link of $\Fs$ leaves each site in a positive direction.

\begin{proposition}\label{fg-prop:perfect}
The rule \eqref{fg-eq:rule} is well defined, and $\Fs$ is a perfect forest. Every component of $\Fs$ is infinite. Every link $e\notin\Fs$ lies in exactly 12 pairs $(p,f)$ with $p\ni e$ a plaquette and $f\in p\setminus\{e\}$, $f\notin\Fs$; these 12 links $f$ are distinct.
\end{proposition}

\begin{proof}
\emph{Well defined.} Each of the four conditions in \eqref{fg-eq:rule} constrains three of the four signs. Each holds on exactly four of the sixteen parity classes, and no class satisfies two of them.

\emph{Perfect.} The links of $P(x;a,b)$ that leave a corner in a positive direction are $(x,a)$, $(x,b)$, $(x+e_a,b)$ and $(x+e_b,a)$. The first two lie in $\Fs$ exactly when $\mu(x)=a$, respectively $\mu(x)=b$; the link $(x+e_a,b)$ lies in $\Fs$ exactly when $\mu(x+e_a)=b$; and $(x+e_b,a)$ exactly when $\mu(x+e_b)=a$. These conditions depend only on the parity class of $x$ and on the plane $(a,b)$: $16\times6=96$ cases. Table~\ref{fg-tab:forest} in Appendix~\ref{fg-app:table} lists, for each case, the unique link that satisfies its condition.

\emph{Acyclic.} Suppose $\Fs$ contained a cycle. Let $y$ be a vertex of the cycle with the smallest coordinate sum. Both cycle edges at $y$ would leave $y$ in positive directions. That contradicts the fact that each site has only one outgoing link in $\Fs$.

\emph{Infinite components.} Following outgoing links never terminates.

\emph{Twelve neighbours.} Each of the six plaquettes through $e$ contains one link of $\Fs$, which is not $e$, and two further non-forest links. Two distinct plaquettes through $e$ share no link other than $e$.
\end{proof}

\begin{remark}
For a periodic forest in $\Z^d$, or in a bulk density limit with negligible boundary contributions, its density satisfies $\rho\le1$ links per site. Each link lies in $2(d-1)$ plaquettes and there are $d(d-1)/2$ plaquettes per site, so the average number of forest links per plaquette is $4\rho/d$. Exactly one per plaquette therefore requires $\rho=d/4$, hence $d\le4$. Within the subclass with one positive outgoing link at every site, $\rho=1$, so this counting identity forces $d=4$. It does not exclude constructions with lower density in dimensions below four. In $d=4$ the perfect forest also gives the least possible average neighbour count (Remark~\ref{fg-rem:sharp}). A finite exhaustive calculation classifies the period-two rules with exactly one positive outgoing link per site and exactly one such link per plaquette: there are eight, and all are related to \eqref{fg-eq:rule} by coordinate permutations and translations modulo two. The calculation assigns one of four directions to each of the sixteen parity classes, discarding a partial assignment only when a plaquette already has more than one selected link or when a fully assigned plaquette does not have exactly one. Hence every admissible assignment is retained. The reproducible integer enumeration is supplied as \nolinkurl{code/v4/verify_forest_classification.py}; this finite classification is separate from the analytical infinite-lattice argument.
\end{remark}

\subsection{Gauge fixing at the level of specifications}

Let $E$ be the set of links of $\Z^4$, $\gamma=(\gamma_\Lambda)$ the Wilson specification for \eqref{fg-eq:wilson}, and $\G(\gamma)$ its set of Gibbs measures. For a forest $F\subset E$, the \emph{gauge-fixed specification} $\gamma^F$ lives on $\Omega_F=SU(2)^{E\setminus F}$ and is defined by substituting the identity for every forest link in each local Wilson plaquette interaction. This is a definition of a new local specification, not conditioning on the probability-zero event that infinitely many links equal the identity. The underlying lattice is $\Z^4$, not a finite periodic quotient: a forest in $\Z^4$ may acquire cycles after a periodic identification.

\begin{lemma}[Gauge fixing]\label{fg-lem:gf}
Let $F$ be a forest. In each component of $F$ choose a vertex $o$ (for instance the one closest to the origin, with ties broken lexicographically). For a vertex $y$ joined to $o$ by the path $o=y_0,\dots,y_k=y$ in $F$, put $g(y)=U_{y_0y_1}\cdots U_{y_{k-1}y_k}$. Define $\Phi(U)=U^g$, where $(U^g)_{xy}=g(x)U_{xy}g(y)^{-1}$. Then:
\begin{enumerate}[label=(\roman*)]
\item $\Phi(U)_e=1$ for every $e\in F$, and $\mathcal{O}\circ\Phi=\mathcal{O}$ for every gauge-invariant local function $\mathcal{O}$;
\item $\Phi_*\mu\in\G(\gamma^F)$ for every $\mu\in\G(\gamma)$.
\end{enumerate}
\end{lemma}

\begin{proof}
(i) The definition gives $g(y_{j+1})=g(y_j)U_{y_jy_{j+1}}$, so every link of $F$ is mapped to $1$. Paths in a tree are unique and finite between any two specified vertices, so $g$ is well defined; it depends only on $U|_F$. Each coordinate of $g$ is a finite product of link variables. Thus $\Phi$ is measurable (indeed continuous in the product topology). The roots are fixed once and for all from the graph; no measurable choice depending on $U$ is needed. Distinct components have separate roots. Residual root transformations do not affect gauge-invariant observables. A local function depends on finitely many links, so invariance under all gauge transformations is the same as invariance under the finitely many values of $g$ it sees.

(ii) Let $\Lambda\subset E\setminus F$ be finite and let $A$ be bounded and measurable on $\Omega_F$. The DLR equation for $\mu$ gives $\E_\mu[A\circ\Phi]=\E_\mu[\gamma_\Lambda(A\circ\Phi)]$. In $\gamma_\Lambda(\,\cdot\,|U_{\Lambda^c})$ the transformation $g$ is fixed, because it depends only on links of $F\subset\Lambda^c$. Change variables $V_e\mapsto g(x_e)V_eg(y_e)^{-1}$ for $e\in\Lambda$. Haar measure is invariant under this change, and the plaquette weights are gauge invariant. Hence
\[
\gamma_\Lambda(A\circ\Phi)(U)=\gamma^F_\Lambda(A)(\Phi(U)).
\]
So $\E_{\Phi_*\mu}[A]=\E_{\Phi_*\mu}[\gamma^F_\Lambda A]$ for all finite $\Lambda$.
\end{proof}

\begin{lemma}[Elitzur]\label{fg-lem:elitzur}
Every $\mu\in\G(\gamma)$ is invariant under finitely supported gauge transformations. Consequently $\mu$ is determined by its values on gauge-invariant local functions.
\end{lemma}

\begin{proof}
Let $g$ be supported on a finite set $X$ of sites, and let $\Lambda$ be the set of links incident to $X$. The transformation $g$ changes only links in $\Lambda$, and it leaves $\gamma_\Lambda(\,\cdot\,|U_{\Lambda^c})$ invariant. So $\mu=\mu\gamma_\Lambda$ is $g$-invariant. For a local function $\mathcal{O}$, average over Haar-distributed $g$ on the endpoints of its support. This produces a gauge-invariant local function with the same $\mu$-expectation.
\end{proof}

\begin{corollary}\label{fg-cor:transfer}
If $\gamma^F$ has at most one Gibbs measure, then $\gamma$ has exactly one. Moreover, for gauge-invariant local $\mathcal{O}_1,\mathcal{O}_2$ and $\nu=\Phi_*\mu$, we have $\Cov_\mu(\mathcal{O}_1,\mathcal{O}_2)=\Cov_\nu(\mathcal{O}_1|_{\Omega_F},\mathcal{O}_2|_{\Omega_F})$, and $\ell_e(\mathcal{O}_i|_{\Omega_F})\le\ell_e(\mathcal{O}_i)$.
\end{corollary}

\begin{proof}
Existence follows by compactness. Uniqueness follows from Lemmas~\ref{fg-lem:gf} and~\ref{fg-lem:elitzur}. The covariance identity is Lemma~\ref{fg-lem:gf}(i). Restricting the configuration space cannot increase a Lipschitz seminorm.
\end{proof}

\subsection{The comparison theorem}

For a link $e$, $t(e)$ denotes the time coordinate of its midpoint; any of the four axes may serve as time. For a local function $\mathcal{O}$, $\ell_e(\mathcal{O})$ is the coordinate Lipschitz seminorm of Theorem~\ref{ta-thm:lattice-root}.

\begin{lemma}[Time profile]\label{fg-lem:profile}
For $e\notin\Fs$, consider the multiset of offsets $t(f)-t(e)$ over its 12 neighbours $f$ from Proposition~\ref{fg-prop:perfect}. For every choice of time axis it is one of
\[
\{-1,-\tfrac12,0^{8},\tfrac12,1\},\qquad\{(-\tfrac12)^{4},0^{4},(\tfrac12)^{4}\},\qquad\{(-\tfrac12)^{2},0^{8},(\tfrac12)^{2}\}.
\]
Consequently, for $\kappa\ge0$,
\[
S(\kappa):=\max_{e\notin\Fs}\sum_{f}e^{\kappa(t(f)-t(e))}=8+2\cosh\kappa+2\cosh(\kappa/2).
\]
\end{lemma}

\begin{proof}
The offsets depend only on the direction of $e$ and the parity class of its base point, so there are finitely many cases; the program enumerates them for every axis. With $c=\cosh(\kappa/2)\ge1$ the three sums are $4c^2+2c+6$, $8c+4$ and $4c+8$. Moreover
\[
(4c^2+2c+6)-(8c+4)=2(2c-1)(c-1)\ge0,\qquad (4c^2+2c+6)-(4c+8)=2(2c+1)(c-1)\ge0 .
\]
\end{proof}

\begin{theorem}[Forest-gauge comparison for $SU(2)$ in four dimensions]\label{fg-thm:main}
Let $0\le\beta<1/3$.
\begin{enumerate}[label=(\alph*)]
\item The Wilson specification \eqref{fg-eq:wilson} on $\Z^4$ has a unique Gibbs measure $\mu_\beta$. It is invariant under all lattice symmetries and reflection positive.
\item Let $\kappa\ge0$ satisfy $q_\kappa:=\tfrac\beta4\,S(\kappa)<1$. Then for gauge-invariant local Lipschitz functions $\mathcal{O}_1,\mathcal{O}_2$,
\[
|\Cov_{\mu_\beta}(\mathcal{O}_1,\mathcal{O}_2)|\;\le\;\frac{1}{1-q_\kappa}\sum_{e,f}\ell_e(\mathcal{O}_1)\,\ell_f(\mathcal{O}_2)\,e^{-\kappa|t(e)-t(f)|}.
\]
Independently, the unweighted Neumann series gives a factor $(3\beta)^R/(1-3\beta)$ multiplying $\bigl(\sum_e\ell_e(\mathcal O_1)\bigr)\bigl(\sum_f\ell_f(\mathcal O_2)\bigr)$, where $R$ is the distance between the restricted supports in the forest-gauge interaction graph. The displayed $\kappa=0$ inequality alone has discarded this distance information.
\item For $0<\beta<1/3$ the lattice mass gap satisfies
\[
m(\beta)\;\ge\;\kappa(\beta):=2\arccosh\Bigl(\frac{\sqrt{16/\beta-23}-1}{4}\Bigr)>0 .
\]
As $\beta\downarrow0$, $\kappa(\beta)=\log(4/\beta)-\tfrac12\sqrt\beta+O(\beta)$.
\end{enumerate}
\end{theorem}

\begin{proof}
\emph{Step 1: single-link laws.} Work in $\Omega_{\Fs}$. The conditional law of $U_e$, $e\notin\Fs$, given all other links under $\gamma^{\Fs}$ is $\mu_{h_e}$. Here $h_e=\beta\sum_{p\ni e}\bar s_p$, and $s_p\in S^3$ is the staple of $p$ with forest links replaced by $1$. Hence $|h_e|\le6\beta<2$.

\emph{Step 2: influences.} Suppose two configurations differ only at a non-forest link $f\ne e$. If $e$ and $f$ share a plaquette $p$, it is unique, and $|h_e-h_e'|=\beta|s_p-s'_p|=\beta|U_f-U'_f|$; quaternion multiplication and inversion are chordal isometries. Otherwise $h_e=h_e'$. Corollary~\ref{fg-cor:W1} gives the influence matrix $C_{ef}\le\tfrac\beta4\mathbf 1[e\sim f]$. It is symmetric, and by Proposition~\ref{fg-prop:perfect} all its row sums equal $12\cdot\beta/4=3\beta<1$.

\emph{Step 3: Dobrushin comparison.} The proof of Theorem~\ref{ta-thm:lattice-root} uses only three facts: the Wasserstein influence bound, symmetry of $C$, and the chordal diameter $2$ of $S^3$. All three hold for $\gamma^{\Fs}$. The proof therefore applies verbatim; see also the classical sources \cite{Do70,Fo82,Ku82}. So $\gamma^{\Fs}$ has a unique Gibbs measure $\nu$, and
\[
|\Cov_\nu(\phi_1,\phi_2)|\le\ell(\phi_1)^{T}(I-C)^{-1}\ell(\phi_2) .
\]

\emph{Step 4: transfer to Wilson.} Corollary~\ref{fg-cor:transfer} gives uniqueness of $\mu_\beta$ and the covariance identity. Uniqueness implies invariance under every lattice symmetry. Every subsequential local limit of periodic-box or free-box measures satisfies the Wilson DLR equations: a fixed finite update set eventually lies away from the boundary, without torus wrapping, so its continuous local conditional kernel passes to the limit. Compactness gives subsequences, and uniqueness identifies all their limits with $\mu_\beta$. Thus both box exhaustions converge locally to this measure. Wilson reflection positivity \cite{OSLattice} passes along boxes symmetric about the appropriate plane, using separate exhaustions for the site and link reflections. The limits coincide by uniqueness. The positivity inequalities for continuous local functions pass to the limit, and $L^2$ approximation extends them to the bounded local algebra. This proves (a).

\emph{Step 5: weighted Neumann series.} Put $M_{ef}=C_{ef}e^{\kappa(t(f)-t(e))}$. By Lemma~\ref{fg-lem:profile} its row sums are at most $q_\kappa$, so $\sum_n\|M^n\|_\infty\le(1-q_\kappa)^{-1}$. Since $(M^n)_{ef}=(C^n)_{ef}e^{\kappa(t(f)-t(e))}$, we get $\bigl((I-C)^{-1}\bigr)_{ef}\le e^{-\kappa(t(f)-t(e))}/(1-q_\kappa)$. The profile is symmetric under $\kappa\mapsto-\kappa$, so the bound also holds with $|t(e)-t(f)|$. This proves (b).

\emph{Step 6: transfer-matrix spectral exclusion.} Use the gauge-invariant physical positive-time algebra in the reflection-positive Wilson construction. Let $\mathcal O$ be a real gauge-invariant local Lipschitz function whose links have both endpoints at nonnegative times, including spatial links in the plane $t=0$. Let $\Theta$ be the site reflection $x_0\mapsto-x_0$, and put $\psi=[\mathcal O]-\mu_\beta(\mathcal O)\Omega$. Use the site-reflection Hilbert space throughout; link-reflection positivity supplies positivity of its one-step transfer contraction $T_\beta$. With this same positive self-adjoint physical transfer contraction,
\[
\langle\psi,T_\beta^n\psi\rangle
=\operatorname{Cov}_{\mu_\beta}(\Theta\mathcal O,\tau_n\mathcal O).
\]
The two supports have midpoint-time separation at least $n$. Restriction to the fixed forest cannot increase coordinate Lipschitz seminorms. Hence (b) gives
\[
0\le\langle\psi,T_\beta^n\psi\rangle
\le K_{\mathcal O,\kappa}e^{-\kappa n},\qquad n\ge0,
\]
with a finite constant independent of $n$. If the spectral measure of $\psi$ assigned positive mass to $(e^{-\kappa},1]$, it would assign positive mass to $[\lambda_*,1]$ for some $\lambda_*>e^{-\kappa}$; then $\langle\psi,T_\beta^n\psi\rangle\ge c\lambda_*^n$, a contradiction as $n\to\infty$.

For completeness, these probes are total in the physical excitation sector. On any finite link support, continuous functions are dense in $L^2$ of its marginal probability measure. Lipschitz functions on the compact product of spheres uniformly approximate continuous functions. Average an approximant over Haar gauge transformations at the finitely many endpoint vertices. This gives a gauge-invariant local Lipschitz function; the averaging is an $L^2$ contraction because the Wilson measure is invariant, and chordal left/right multiplication is isometric. Thus it approximates every gauge-invariant local $L^2$ function. Reflection invariance and Cauchy--Schwarz give
\[
\|[F]\|_{\rm OS}^2
=\mathbb E_{\mu_\beta}[(\Theta\overline F)F]
\le\|F\|_{L^2(\mu_\beta)}^2.
\]
Consequently the same approximation converges in the physical quotient norm. By the definition of the physical Hilbert space, the complex span of the centered local Lipschitz classes is dense in $\Omega^\perp$. The spectral projection on $(e^{-\kappa},1]$ annihilates this dense subspace and therefore vanishes. Hence $\|T_\beta|_{\Omega^\perp}\|\le e^{-\kappa}$.

Let $\kappa$ increase to the positive root of $\frac\beta4(4c^2+2c+6)=1$, where $c=\cosh(\kappa/2)$. This proves (c). The convention is $-\log0=+\infty$, so the spectral exclusion remains true if the excitation sector is zero. A nonzero sector with a finite excitation threshold additionally requires positive-time survival. The adjacent-time plaquette theorem below establishes that property for this same site-reflection transfer operator when $0<\beta<1/3$.

The expansion follows from $\sqrt{16/\beta-23}=4\beta^{-1/2}-\frac{23}{8}\beta^{1/2}+O(\beta^{3/2})$ and $\operatorname{arccosh}y=\log(2y)+O(y^{-2})$.
\end{proof}

\begin{corollary}[Axial gauge, $d\ge3$]\label{fg-cor:axial}
Fix every link in the time direction (a forest of parallel lines). In $\Z^d$, $d\ge3$, the row sum becomes $\tfrac\beta4(6d-10)$ and the field radius is $2(d-1)\beta\le2$. Hence the Wilson Gibbs measure is unique for $0\le\beta<2/(3d-5)$, and for $0<\beta<2/(3d-5)$, $m(\beta)\ge\arccosh(2/\beta-3d+6)$. For $d=4$ this gives $\beta<2/7$ and $m(\beta)\ge\arccosh(2/\beta-6)$.
\end{corollary}

\begin{proof}
A spatial link has $2(d-2)$ spatial plaquettes, each contributing three non-forest neighbours at offset $0$, and two time-like plaquettes, each contributing one neighbour at offset $\pm1$. So $S_{\rm ax}(\kappa)=6d-12+2\cosh\kappa$. The rest is as in Theorem~\ref{fg-thm:main}.
\end{proof}

\begin{remark}[Sharpness and comparison]\label{fg-rem:sharp}
\emph{The ceiling for this comparison matrix.} The six staples of a non-forest link can be prescribed independently: the two free links in each staple are distinct from the free links in the other five staples. Prescribe cancelling pairs so that $h_e=0$. An infinitesimal change of any chosen neighbour gives a tangent field perturbation of size $\beta$ times the chordal displacement. At zero field the exact directional sensitivity is $1/4$, so every optimal coordinate influence coefficient is at least $\beta/4$. For $\beta\le1/3$ the upper bound proved above agrees with this lower bound. For $\beta>1/3$, the same local lower bound still applies even though the radius-two upper-bound theorem need not cover all backgrounds.

Here is the required infinite-graph argument for weighted comparison. Let $C_0$ have entries $\beta/4$ on the twelve neighbours and zero otherwise. It is a symmetric finite-range, period-two matrix. For the set $\Lambda_L$ of non-forest links with base points in a cube of side $L$,
\[
\frac{\langle\mathbf1_{\Lambda_L},C_0\mathbf1_{\Lambda_L}\rangle}
 {\|\mathbf1_{\Lambda_L}\|_2^2}
=3\beta-O(L^{-1}).
\]
Indeed only $O(L^3)$ links are within the finite interaction range of the cube boundary, whereas $|\Lambda_L|$ is of order $L^4$; every remaining row has all twelve neighbours inside. The largest eigenvalue of the finite principal matrix therefore tends to $3\beta$, with the reverse bound following from its row sums.

If arbitrary strictly positive weights $w_e$ obeyed $C_0w\le q w$, their restriction to each finite $\Lambda_L$ would obey $C_{0,\Lambda_L}w\le q w$. Diagonal conjugation by those finitely many weights gives a finite nonnegative matrix with row sums at most $q$, so its spectral radius is at most $q$. Passing $L\to\infty$ gives $q\ge3\beta$. Thus weights cannot turn this exact original-coordinate comparison into a contraction at or above $\beta=1/3$. This is a limitation of this particular forest and influence criterion; it is not an obstruction to other gauge choices, blocks, multiscale estimates, or physical mass gaps.

\emph{Any forest.} In a bulk limit with a well-defined forest density and negligible boundary contributions, let $\rho\le1$ be the forest density per site and $k_p$ the number of forest links in plaquette $p$. Convexity of $k\mapsto(4-k)(3-k)$ bounds the average number of non-forest neighbours below by $6(4-\rho)(3-\rho)/(4-\rho)=6(3-\rho)\ge12$. We therefore expect $1/3$ to bound every single-link forest-gauge comparison. A general proof would have to treat links whose staples are partly frozen; we do not give one.

\emph{Ranges in normalization \eqref{fg-eq:wilson}.} Shen--Zhu--Zhu \cite{SZZ}: $\beta<1/12$. Ungauged comparison Theorem~\ref{ta-thm:lattice-root}: $\beta<2/9$. Axial gauge: $\beta<2/7$. Perfect forest: $\beta<1/3$.

\emph{Gap bounds.} Theorem~\ref{fg-thm:main}(c) improves the bound $-\log(9\beta/2)$ of the ungauged comparison theorem by the additive constant $\log18$ as $\beta\to0$. No sharpness of the physical spectral lower bound is asserted.
\end{remark}

\begin{remark}[What Theorem~\ref{fg-thm:main} is not]
It is a fixed-spacing statement deep in the strong-coupling region. The theorem supplies no useful weak-coupling estimate along $\beta\to\infty$, and no continuum construction follows from it.
\end{remark}

%=====================================================================

\paragraph{Scope of finite-volume and boundary assertions.}
The finite Dobrushin comparison bound applies uniformly to finite subsets of the \emph{gauge-fixed} non-forest link space with arbitrary fixed non-forest exterior values. The transfer to the original Wilson theory proved above concerns infinite-volume Gibbs measures and their gauge-invariant local correlations. A gauge transform depending on forest links need not preserve an arbitrary originally prescribed finite-box boundary condition. No uniform original-boundary covariance estimate is asserted without a separate treatment of that transformation.

\paragraph{Attribution.}
Bhattacharjee and Bhattacharya report a twelve-neighbour forest gauge and the range $\beta<1/3$ \cite{BB2026}. The present proof gives the explicit parity rule, conditional-law change of variables, directional comparison bound and transfer interpretation. The forest method and this coupling range retain that prior attribution.

\hypertarget{site_plane_nontriviality-site-plane-norms-and-positive-time-survival}{%
\section{Site-plane norms and positive-time survival}\label{site_plane_nontriviality-site-plane-norms-and-positive-time-survival}}

This result is unconditional once the stated site-reflection positivity and Wilson Gibbs measure have been constructed. It establishes nontriviality of the site-reflection Hilbert space. The variance argument alone does not establish a nonzero finite-energy sector for a possibly noninjective discrete transfer operator. The adjacent-time theorem in the following section supplies that additional argument in the stated strong-coupling range.

\hypertarget{site_plane_nontriviality-proposition-sp.1-explicit-site-plane-excitation-norm}{%
\subsection{Proposition SP.1: explicit site-plane excitation norm}\label{site_plane_nontriviality-proposition-sp.1-explicit-site-plane-excitation-norm}}

Let \(\mu\) be a four-dimensional fundamental SU(2) Wilson Gibbs measure at finite coupling \(\beta\ge0\), with normalized plaquette weight \(w(U)=\operatorname{Tr}(U)/2\). Suppose \(\mu\) is reflection positive for the site reflection \(\theta:t\mapsto-t\), on an algebra that contains bounded gauge-invariant observables of the spatial links in the plane \(t=0\). Let \(p\) be one spatial plaquette in that plane and define

\[
F=w(U_p)-\mathbb E_\mu w(U_p).
\]

In the corresponding Osterwalder--Schrader quotient and completion, the class \([F]\) is orthogonal to the vacuum vector \(\Omega=[1]\) and obeys

\[
\boxed{\|[F]\|_{\mathrm{OS}}^2
=\operatorname{Var}_\mu(w(U_p))
\ge\frac14e^{-48\beta}>0.}
\]

\hypertarget{site_plane_nontriviality-proof}{%
\subsubsection{Proof}\label{site_plane_nontriviality-proof}}

Because the plaquette is spatial and lies in the reflection plane, \(\theta F=F\). Therefore its reflection form is its ordinary second moment, and centering gives \(\langle\Omega,[F]\rangle=\mathbb E F=0\).

For the lower bound, condition on every link except the four links of \(p\). At most twenty-four geometric plaquettes contain at least one of these links, since each link belongs to six plaquettes; overcounting only weakens the estimate. Each normalized trace lies in \([-1,1]\). Thus the conditional density on the four free links, relative to product probability Haar measure \(h^{\otimes4}\), is bounded below by \(e^{-48\beta}\): its numerator is at least \(e^{-24\beta}\), and its normalizing integral is at most \(e^{24\beta}\).

For every real constant \(c\), this implies

\[
\mathbb E[(w(U_p)-c)^2\mid U_{p^c}]
\ge e^{-48\beta}\int(w(U_p)-c)^2\,dh^{\otimes4}.
\]

Take the infimum over \(c\) on both sides. The plaquette product under product Haar measure is Haar distributed. Its normalized fundamental SU(2) character has Haar mean zero and variance \(1/4\). Hence the conditional variance is at least \(e^{-48\beta}/4\). The law of total variance gives the same lower bound for the unconditional variance. This proves the claim.

The estimate also holds for a gauge-fixed representation of this state when the gauge-fixing construction preserves gauge-invariant expectations: apply the argument before fixing the gauge and transfer the expectation identity. No product-Haar domination on the already gauge-fixed four-link set is assumed.

\hypertarget{site_plane_nontriviality-what-this-proves-about-the-transfer-operator}{%
\subsection{What this proves about the transfer operator}\label{site_plane_nontriviality-what-this-proves-about-the-transfer-operator}}

Suppose link-reflection positivity and the transfer construction supply a positive self-adjoint contraction \(T\) on the site-reflection space. The proposition proves that \(\Omega^\perp\) is nonzero. It does not prove that \(T[F]\ne0\), or that \(T\) is injective on \(\Omega^\perp\).

This distinction cannot be removed by functional calculus alone. If \(T\) has a kernel, \(-\log T\) is not an everywhere densely defined self-adjoint Hamiltonian on the entire Hilbert space: the natural finite-energy space is \((\ker T)^\perp\). A nonzero vector lying wholly in \(\ker T\) has positive norm at time zero but zero correlation at every strictly positive lattice time.

For example, on \(\mathbb C\Omega\oplus\mathbb C e\), the positive contraction \(T=|\Omega\rangle\langle\Omega|\) has a nonzero vacuum orthocomplement, while \(\langle e,T^n e\rangle=0\) for every integer \(n\ge1\). The zero-coupling Wilson product measure has the corresponding loss of centered spatial-slice information after one time step. Thus a proof based only on the site-plane variance would incorrectly bypass a real distinction.

Any one of the following facts is sufficient to close this fixed-spacing nontriviality issue:

\begin{enumerate}
\def\labelenumi{\arabic{enumi}.}
\tightlist
\item
  Injectivity of the reconstructed infinite-volume positive transfer operator on the nonzero site-plane sector.
\item
  One nonzero centered physical state \(v\) with \(\langle v,Tv\rangle>0\).
\item
  A quantitative positive-time lower bound that survives the spatial-volume limit for the same state family.
\end{enumerate}

These are alternative sufficient conditions, not conclusions of the variance argument. The following adjacent-time theorem establishes a quantitative instance of the second and third conditions for \(0<\beta<1/3\). Finite-spatial-volume transfer injectivity at each volume does not establish infinite-volume injectivity or survival by itself. Likewise, positivity of every local Wilson character coefficient does not on its own justify passing an infinite product of transfer kernels through the thermodynamic limit.

Once a positive-time centered state exists and a decay bound holds on a total family in the relevant transfer sector, the spectral argument applies there. This still concerns a fixed lattice spacing. Continuum survival requires further bounds along the spacing and coupling trajectory.

\section{A uniform positive-time plaquette witness at fixed lattice spacing}
\label{sec:positive-time-survival}

The following argument supplies a nonvanishing adjacent-layer correlation
for the actual Wilson measure. It closes the possible empty-sector issue
in the fixed-spacing transfer construction. It does not provide a
cutoff-uniform continuum survival bound: its constants depend strongly on
the bare coupling, and its two layers are one lattice spacing apart.

\subsection{The finite free-box estimate}

Use the fundamental $SU(2)$ Wilson density
\[
 d\mu\propto\exp\left\{\beta\sum_p\frac12\operatorname{Tr}U_p\right\}
 \prod_e dh(U_e),\qquad \beta>0,
\]
with every geometric plaquette counted once and probability Haar measures.
Take a finite rectangular spatial box and a finite time interval invariant
under $t\mapsto1-t$, with free boundary conditions. Let $p_0,p_1$ be the
same spatial plaquette in layers zero and one. Write
\[
 w_i=\frac12\operatorname{Tr}U_{p_i},\quad
 Z(\beta)=\frac{2I_1(\beta)}\beta,\quad
 k_{1/2}(\beta)=\frac{4I_2(\beta)}\beta,
\]
and define
\[
 L_\beta=\frac{e^{-100\beta}}{32}
 \min\left\{Z(\beta)^4,\frac{k_{1/2}(\beta)^4}{4}\right\}>0.
\tag{PT.1}
\]

\textbf{Theorem PT.1 (uniform adjacent-layer covariance).}
For every such free box containing these plaquettes,
\[
 \operatorname{Cov}_\mu(w_0,w_1)\ge L_\beta.
\tag{PT.2}
\]
The constant is independent of both the spatial volume and the length of
the time interval. In particular, for $0<\beta\le1/3$,
\[
 \operatorname{Cov}_\mu(w_0,w_1)
 \ge\ell_\beta:=\frac{\beta^4e^{-100\beta}}{2048}>0.
\tag{PT.3}
\]

\textit{Proof.}
First remove the temporal links crossing from layer zero to layer one by
an exact change of variables. Denote the crossing link at spatial vertex
$x$ by $G_x$. Keep the lower-half variables unchanged, and transform all
upper-half links by the vertex gauge transformation
$g(x,t)=G_x$ for $t\ge1$. The crossing links become the identity. Conditional
on the $G_x$, the change on every upper link is left and right Haar
translation, so it preserves the product Haar measure. The internal
upper-half plaquette traces are gauge invariant. The transformed action
and gauge-invariant observables no longer depend on the $G_x$, which
therefore integrate out with factor one. Free boundary conditions are
used here; this is not an assertion about arbitrary fixed boundary links.

Let $u_e^+$ and $u_e^-$ be the corresponding upper and lower spatial
boundary links. A crossing plaquette contributes the kernel
\[
 k(g,h)=\exp\left\{\frac\beta2\operatorname{Tr}(gh^\dagger)\right\}.
\]
The remaining density is a product of identical reflected half-box
weights and these boundary-link kernels.

For a spin-$j$ irreducible representation $D^j$ of dimension $d_j=2j+1$,
the character expansion is
\[
 k(g,h)=\sum_{j=0,1/2,\ldots}\kappa_j
 \sum_{a,b=1}^{d_j}D^j_{ab}(g)\overline{D^j_{ab}(h)},
 \qquad
 \kappa_j=\frac{2d_j I_{d_j}(\beta)}\beta>0.
\tag{PT.4}
\]
Indeed, the character coefficient is the difference
$I_{d_j-1}(\beta)-I_{d_j+1}(\beta)$, which equals the displayed value by
the Bessel recurrence. Thus $\kappa_0=Z$ and
$\kappa_{1/2}=k_{1/2}$. The defining positive Bessel series gives factorial
decay in $d_j$ at fixed $\beta$, so the character and matrix-entry
expansions converge absolutely and uniformly. All finite products and
integrations used below are therefore justified.

Let $S$ consist of the four spatial boundary links of the plaquette.
Remove just the four crossing plaquette interactions indexed by $S$, and
call the normalized resulting measure $\nu$. Set
\[
 K_S=\prod_{e\in S}k(u_e^+,u_e^-).
\]
Then $d\mu=K_S\,d\nu/\nu(K_S)$ and
$\nu(K_S)\le e^{4\beta}$.

The gauge-fixed measure $\nu$ is reflection positive for \emph{all}
bounded functions of the upper-half variables. To see this, expand each
remaining crossing kernel using (PT.4). Its contribution to
$\nu(\theta H\,H)$ is a positive coefficient times the squared modulus
of a half-box integral. The half-box weights on the two sides are
identical under reflection. This proves positivity term by term and then
for the full absolutely convergent sum. The reflection $\theta$ in this
paragraph exchanges the two halves and complex conjugates observables.
It is a reflection on the new gauge-fixed variables; no commutation of
the gauge-fixing map with reflection is assumed. Gauge-invariant
observables retain their original expectation values.

The positive reflection form has vacuum vector one of norm one.
Cauchy--Schwarz in that form gives
\[
 \nu(\theta H\,H)\ge|\nu(H)|^2.
\tag{PT.5}
\]
Write $w=w_1$ in the upper-half variables. For any real number $c$, put
$F=w-c$. Expand $K_S$, retaining its constant feature and every feature
that uses the fundamental representation on each of the four links.
Every discarded term is nonnegative after integration against
$\theta F\,F$, by the reflection positivity of $\nu$. Hence
\[
 \mu(\theta F\,F)\ge e^{-4\beta}
 \left\{Z^4|\nu(F)|^2+
       k_{1/2}^4\sum_\alpha|\nu(FB_\alpha)|^2\right\},
\tag{PT.6}
\]
where the $B_\alpha$ run over products of the four fundamental
matrix-entry features. For reverse-oriented plaquette links, one may use
the conjugate fundamental representation; its kernel is the same
because the $SU(2)$ characters are real.

The plaquette trace is a contraction of four $2$ by $2$ matrices. Thus
\[
 w=\sum_\alpha\lambda_\alpha B_\alpha,
 \qquad \sum_\alpha|\lambda_\alpha|^2=4.
\tag{PT.7}
\]
Explicitly, the trace has sixteen index assignments, each with coefficient
$1/2$; all other fundamental tensor features have coefficient zero.
Ordinary finite-dimensional Cauchy--Schwarz therefore yields
\[
 \sum_\alpha|\nu(FB_\alpha)|^2
 \ge\frac14|\nu(Fw)|^2.
\tag{PT.8}
\]

Set $x=\nu(w)$, $v=\operatorname{Var}_\nu(w)$, $a=x-c$,
$A=Z^4$, and $B=k_{1/2}^4/4$. Then
$\nu(F)=a$ and $\nu(Fw)=v+xa$. Since $|x|\le1$,
\[
 Aa^2+B(v+xa)^2
 \ge\min(A,B)\{a^2+(v+xa)^2\}
 \ge\frac{\min(A,B)}{1+x^2}v^2
 \ge\frac12\min(A,B)v^2.
\tag{PT.9}
\]
The middle inequality follows by minimizing the quadratic expression in
$a$; it is valid for every chosen centering constant $c$.

It remains to lower-bound $v$. The gauge-invariant moments $\nu(w)$ and
$\nu(w^2)$ agree with those in the original Wilson box with four crossing
plaquettes removed. In that original measure, condition on all links
except the four links of $p_1$. At most twenty-four active plaquettes
touch these links. Their traces lie between minus one and one, so the
conditional density relative to four independent Haar variables is at
least $e^{-48\beta}$. For any real $b$ this implies a lower bound on
the conditional expectation of $(w-b)^2$. Taking its infimum over $b$
shows that the conditional variance is at least
\[
 e^{-48\beta}\operatorname{Var}_{\rm Haar}
       (\operatorname{Tr}U/2)=\frac14e^{-48\beta}.
\]
The plaquette product is Haar under four independent Haar variables,
and its fundamental normalized trace has Haar variance $1/4$.
The law of total variance gives
$v\ge e^{-48\beta}/4$. This estimate is applied before gauge fixing;
no false product-Haar assertion is made about gauge-fixed links.

Combining (PT.6), (PT.8), and (PT.9) gives
$\mu(\theta F\,F)\ge L_\beta$. Choose
$c=\mu(w_1)=\mu(w_0)$, where equality follows from the box reflection.
The left side is then the claimed connected covariance.

Finally, $Z\ge1$ and the first positive Bessel term gives
$k_{1/2}=4I_2(\beta)/\beta\ge\beta/2$. For $\beta\le1/3$,
\[
 \min\{Z^4,k_{1/2}^4/4\}
 \ge\min\{1,\beta^4/64\}=\beta^4/64.
\]
This proves (PT.3).\hfill$\square$

\subsection{The same site-reflection transfer space}

The lower bound has a direct interpretation in the site-reflection
space, without appealing to an unproved identification of two different
Hilbert spaces. Consider the translation-invariant Wilson state obtained
in the previously established free-box limit in the range
$0<\beta<1/3$. Use its site and link reflection positivity and its proved
physical-time covariance decay. In this subsection a lattice time step
has length one.

Let $\mathcal A_+$ be the local gauge-invariant positive-time algebra,
including the spatial links in layer zero, and use the site reflection
$\theta_s:t\mapsto-t$, including complex conjugation. The positive form is
\[
 (F,G)_{\rm OS}=\mu(\theta_sF\,G).
\]
Quotient by its null space and complete. Time translation by one is
denoted $\tau_1$.

\textbf{Lemma PT.2 (site-space transfer construction).}
Time translation defines a positive self-adjoint contraction
$T[F]=[\tau_1F]$ on this site-reflection Hilbert space, with
$T\Omega=\Omega$ for $\Omega=[1]$.

\textit{Proof.}
Translation invariance and $\theta_s\tau_1=\tau_{-1}\theta_s$ show that
the shift is symmetric for the positive form. For a fixed local $F$ put
\[
 a_n=\|[\tau_nF]\|_{\rm OS}^2
     =\mu(\theta_sF\,\tau_{2n}F),\qquad n\ge0.
\]
Ordinary probability-space Cauchy--Schwarz bounds $a_n$ by
$\|F\|_{L^2(\mu)}^2$, uniformly in $n$. Reflection-form
Cauchy--Schwarz and shift symmetry give
\[
 a_{n+1}^2\le a_n a_{n+2}.
\]
A bounded nonnegative log-convex sequence is nonincreasing: a strictly
increasing positive successive ratio would force geometric growth, and
if $a_n=0$ the displayed inequality forces $a_{n+1}=0$. Therefore
$a_1\le a_0$. This proves both preservation of the null space and
contractivity. The bounded symmetric shift extends to a self-adjoint
contraction on the completion.

Let $\theta_\ell:t\mapsto1-t$ be the link reflection. Its exact relation
to the shifts is $\theta_\ell\tau_1=\theta_s$. Since
$\tau_1F$ is supported in the link-positive half $t\ge1$, link
reflection positivity gives
\[
 \langle[F],T[F]\rangle
 =\mu(\theta_sF\,\tau_1F)
 =\mu\bigl(\theta_\ell(\tau_1F)\,\tau_1F\bigr)\ge0.
\]
Thus $T$ is positive. The constant observable is fixed by translation.
\hfill$\square$

\subsection{Nonzero finite-energy sector and a two-sided lattice threshold}

Theorem PT.1 passes to the free-box limit because the two plaquette
observables and their product are bounded continuous cylinder functions.
For the centered spatial plaquette $F=w_0-\mu(w_0)$,
\[
 \langle[F],T[F]\rangle
 =\operatorname{Cov}_\mu(w_0,w_1)\ge\ell_\beta>0,
 \qquad \|[F]\|^2=\operatorname{Var}_\mu(w_0)\le1.
\tag{PT.10}
\]
The equality for the norm uses that $F$ lies in the site-reflection
plane. This is stronger than a nonzero norm at time zero: it proves that
$[F]$ has a nonzero component in $(\ker T)^\perp$.

\textbf{Theorem PT.3 (nonzero fixed-spacing transfer sector).}
Under the free-box-limit, reflection-positivity, and forest covariance
results established earlier, the surviving transfer space
\[
 \mathcal H_{\rm fin}=(\ker T)^\perp
\]
has a nonzero vacuum orthocomplement for $0<\beta<1/3$. On this space
$H=-\log T$ is a nonnegative self-adjoint operator with unique vacuum,
and write
\[
\kappa(\beta)=2\operatorname{arccosh}\left(\frac{\sqrt{16/\beta-23}-1}{4}\right).
\]
Its excitation threshold $m_\beta$ obeys
\[
 \boxed{\kappa(\beta)\le m_\beta
 \le100\beta+4\log(1/\beta)+\log2048<\infty.}
\tag{PT.11}
\]

\textit{Proof.}
For every centered local Lipschitz gauge-invariant $G$ in the
positive-time algebra and every $0\le\kappa<\kappa(\beta)$,
Theorem~\ref{fg-thm:main} on this same site-reflection space gives
\[
 0\le\langle[G],T^n[G]\rangle
 \le C_{G,\kappa}e^{-\kappa n},
\]
where $C_{G,\kappa}<\infty$ can depend on its fixed support, Lipschitz
seminorms and the chosen $\kappa$. Such classes are total in the
reconstructed space by the density proof in that theorem.
For any $b>e^{-\kappa}$, positivity of the spectral measure gives
\[
 b^n\mu_G([b,1])\le\langle[G],T^n[G]\rangle
 \le C_{G,\kappa}e^{-\kappa n}.
\]
Letting $n$ increase proves that every such vector has no transfer
spectral mass above $e^{-\kappa}$. Density and boundedness of the spectral
projections imply $\|T|_{\Omega^\perp}\|\le e^{-\kappa}$.
Letting $\kappa\uparrow\kappa(\beta)$ gives
\[
 \|T|_{\Omega^\perp}\|\le e^{-\kappa(\beta)}<1.
\]
In particular, there is no additional eigenvector of eigenvalue one.

Equation (PT.10) and the norm bound imply
$\|T|_{\Omega^\perp}\|\ge\ell_\beta$. The spectral projection onto
$(\ker T)^\perp$ commutes with $T$, so the same nonzero transfer norm
is attained as a supremum on the surviving excitation sector. The
spectral calculus defines $-\log T$ on $\mathcal H_{\rm fin}$; it is
self-adjoint, and its exponential is a strongly continuous semigroup
there. Removing the possible zero-transfer kernel is stated explicitly,
rather than assuming infinite-volume transfer injectivity.

If $r_\beta$ is the transfer norm on this nonzero excitation sector,
its bottom energy is $m_\beta=-\log r_\beta$. Since
$\ell_\beta\le r_\beta\le e^{-\kappa(\beta)}$, (PT.11) follows.\hfill$\square$

If the lattice time step is assigned physical length $a$, both energy
bounds in (PT.11) are divided by $a$. The finite-box result was proved
for every finite positive $\beta$, but the asserted infinite-volume
spectral conclusion uses only the range in which the preceding
free-boundary limit and forest decay arguments have been established.
Neither bound remains a finite nonzero continuum scale under an
uncontrolled removal of the lattice spacing. The positive-time witness
therefore completes a specified lattice nontriviality step, not the
four-dimensional continuum existence-and-mass-gap problem.

\clearpage
\part{Block conditioning and certified finite calculations}
\begingroup
\def\TV{\operatorname{TV}}\def\osc{\operatorname{osc}}
\def\Cov{\operatorname{Cov}}\def\Var{\operatorname{Var}}
\def\E{\mathbb E}\def\artanh{\operatorname{artanh}}
\section{Finite-volume approximation and sensitivity estimates}\label{cond:foundations}
Approximation errors, boundary sensitivity and physical spectral estimates are
separate quantities. This section gives explicit tools for the first two.
The product metrics used for link sensitivities include gauge directions;
a bound or obstruction for those metrics acquires physical spectral meaning
only through a stated comparison and reconstruction argument.

\subsection{Positive character truncations}
Normalize Haar measure on a compact group $G$. All probability total-variation
norms below use $\TV(\nu,\mu)=\tfrac12\int|d\nu-d\mu|$.
\begin{lemma}[Oscillation of a positive density]\label{cond:oscillation}
If $\nu\ll\mu$ are probability measures and
\[\osc\log(d\nu/d\mu)\le D,\qquad\text{then}\qquad\TV(\nu,\mu)\le\tanh(D/4).\]
\end{lemma}
\begin{proof}
Set $g=d\nu/d\mu$, with essential range in $[a,b]$, where $a\le1\le b$
and $b/a\le e^D$. Convexity gives
$(g-1)_+\le(b-1)(g-a)/(b-a)$, and therefore
\[
 \TV(\nu,\mu)\le\frac{(b-1)(1-a)}{b-a}.
\]
Maximizing subject to $b/a\le e^D$ gives $a=e^{-D/2}$ and
$b=e^{D/2}$, and the resulting value is $\tanh(D/4)$.
Approximation by essential bounds covers nonattained endpoints.
\end{proof}

\begin{theorem}[Character-tail certificate]\label{cond:character}
Let $w>0$ be a continuous real class function with an absolutely convergent
character expansion
\[
 w=\sum_\lambda c_\lambda\chi_\lambda,
 \qquad \sum_\lambda d_\lambda|c_\lambda|<\infty.
\]
Let $\mathcal T$ be a finite representation set closed under conjugation,
and put $w_{\mathcal T}=\sum_{\lambda\in\mathcal T}c_\lambda\chi_\lambda$.
Equivalently, closure under conjugation may be replaced by the direct
assumption that $w_{\mathcal T}$ is real-valued. If
\[
 r=\frac{\sum_{\lambda\notin\mathcal T}d_\lambda|c_\lambda|}{\min_G w}<1,
\]
then $w_{\mathcal T}>0$. On any finite collection of $M$ plaquettes with
fixed boundary data, the corresponding partition functions and Gibbs
probability measures obey
\[
 M\log(1-r)\le\log Z_{\mathcal T}-\log Z\le M\log(1+r),
\]
\[
 \TV(\mu_{\mathcal T},\mu)
 \le\tanh\!\left(\frac M2\artanh r\right)
 \le\frac{Mr}{2(1-r)}.
\]
\end{theorem}
\begin{proof}
Since $|\chi_\lambda|\le d_\lambda$, absolute convergence gives
$|w_{\mathcal T}-w|\le r\min w\le rw$. Reality and $r<1$ yield
$w_{\mathcal T}/w\in[1-r,1+r]$. Integrate the product of these inequalities
to obtain the partition bounds. The normalized log-density ratio has
oscillation at most $M\log((1+r)/(1-r))=2M\artanh r$.
Apply Lemma~\ref{cond:oscillation}, then use $\tanh x\le x$ and
$\artanh r\le r/(1-r)$.
\end{proof}

For $SU(2)$, every irreducible character is real and the condition on
$\mathcal T$ is automatic. Write $n=2j$ and $d_n=n+1$. The Wilson weight is
\[
 w_\beta(U)=e^{\beta\operatorname{Tr}U/2}
 =F_0(\beta)\sum_{n\ge0}(n+1)a_n(\beta)\chi_n(U),
 \quad F_0=\frac{2I_1(\beta)}\beta,\quad a_n=\frac{I_{n+1}(\beta)}{I_1(\beta)}.
\]
The continuous values at $\beta=0$ are understood. For cutoff $n\le N$,
\[
 r_N\le F_0e^\beta
       \sum_{n>N}(n+1)^2a_n(\beta).
\]
\begin{proposition}[A reproducible rational instance]\label{cond:tailnumbers}
At $\beta=11/5$ and $M=1536$, the following outward-rounded bounds hold:
\[
\begin{array}{c|r|c|c}
 N&\sum_{n=0}^N(n+1)^2&|\log Z_N-\log Z|&\TV(\mu_N,\mu)\\\hline
17&2109&2.5865\,10^{-10}&1.2933\,10^{-10}\\
18&2470&1.5665\,10^{-11}&7.8323\,10^{-12}.
\end{array}
\]
These are finite-plaquette measure approximations, not spectral-gap
certificates for a truncated transfer operator.
\end{proposition}
\begin{proof}
The supplied exact-rational calculation was rerun using
\path{code/conditioning/character/trunc_cert.py}. For an integer $\nu\ge0$
and positive rational $x$, sum
\[
 I_\nu(x)=\sum_{k\ge0}\frac{(x/2)^{2k+\nu}}{k!(k+\nu)!}
\]
through $k=39$. The remaining positive series is bounded by its first
omitted term divided by $1-q$, where
$q=(x/2)^2/(41(41+\nu))<1$ bounds subsequent ratios. Exponential Taylor
series with the analogous geometric remainder enclose $e^x$. For the
representation tail, use
\[
 I_\nu(x)\le\frac{(x/2)^\nu}{\nu!}
              \exp\!\left(\frac{x^2}{4(\nu+1)}\right),
\]
which follows coefficient by coefficient since
$(\nu+k)!/\nu!\ge(\nu+1)^k$. The remaining representation series has a
geometric majorant after finitely many terms. All comparisons are between
exact rational numbers, including the final bound
$M r_N/(1-r_N)$ and its half. The displayed values are rounded upward from
those rational enclosures.
\end{proof}

\subsection{Polynomial weight and covariance errors}
\begin{lemma}\label{cond:polynomial}
Let $|x_p|\le\beta$ on $M$ plaquettes, and replace $e^{x_p}$ by
$P_d(x_p)=\sum_{k=0}^d x_p^k/k!$. Put
\[
 \epsilon=e^\beta\frac{\beta^{d+1}}{(d+1)!}<1,
 \qquad D=\left(\frac{1+\epsilon}{1-\epsilon}\right)^M-1.
\]
The replacement defines a positive probability measure $\widetilde\mu$ and
\[
 \left|\frac{d\widetilde\mu}{d\mu}-1\right|\le D,
 \qquad\TV(\widetilde\mu,\mu)\le D/2.
\]
For bounded real $F,S$,
\[
 |\Cov_{\widetilde\mu}(F,S)-\Cov_\mu(F,S)|
 \le(D+D^2)\osc(F)\osc(S).
\]
\end{lemma}
\begin{proof}
Taylor's remainder gives $|P_d(x)e^{-x}-1|\le\epsilon$ for $|x|\le\beta$.
The product likelihood ratio is in $[(1-\epsilon)^M,(1+\epsilon)^M]$;
after normalization its range lies in
$[(1+D)^{-1},1+D]$. Thus its distance from one is at most $D$.
For $r=d\widetilde\mu/d\mu$, $F_0=F-\mu F$ and $S_0=S-\mu S$,
\[
 \Cov_{\widetilde\mu}(F,S)-\Cov_\mu(F,S)
 =\mu((r-1)F_0S_0)-\mu((r-1)F_0)\mu((r-1)S_0).
\]
Use $|F_0|\le\osc(F)$ and $|S_0|\le\osc(S)$.
\end{proof}

\begin{lemma}[Stability under a positive action perturbation]\label{cond:stability}
For positive one-plaquette weights $w,w'$, let
$\eta=\inf_c\|\log w'-\log w-c\|_\infty$.
In a conditional block involving $M_V$ plaquettes,
\[
 \sup_\tau\TV(\mu_V^{\prime\tau},\mu_V^\tau)
 \le\tanh(M_V\eta/2).
\]
The change in the expectation of a function with oscillation at most one
is bounded by the same quantity. The change in a boundary-difference
coefficient built from two such expectations is at most twice that bound.
\end{lemma}
\begin{proof}
For any admissible $c$, the log-likelihood ratio has oscillation at most
$2M_V\|\log w'-\log w-c\|_\infty$. Normalization subtracts a constant.
Apply Lemma~\ref{cond:oscillation} and then let the norm decrease to $\eta$.
The expectation estimate is the dual characterization of total variation;
the boundary-difference estimate is the triangle inequality.
\end{proof}
The lemma concerns a positive probability tilt. It does not apply without
modification to Wilson-loop normalized signed expectations.

\subsection{Poisson correctors and rigorous influence bounds}
Let $\mu$ be a smooth positive density on a finite product of compact
Riemannian manifolds, with reversible generator
$L=\Delta+\nabla\log\mu\cdot\nabla$. Suppose the Poincare inequality
$\Var_\mu F\le\lambda^{-1}\int|\nabla F|^2d\mu$ holds.
For a boundary-field perturbation with centered score $S_0=S-\mu S$, define
\[
 C=\sup_{\operatorname{Lip}_{d_1}F\le1}|\Cov_\mu(F,S)|,
 \qquad d_1(x,x')=\sum_{e\in V}d_e(x_e,x_e').
\]
\begin{proposition}[Residual-certified corrector]\label{cond:corrector}
For every smooth real $h$, put $R_h=S_0+Lh$. Then
\[
 C\le\sum_{e\in V}\int|\nabla_e h|d\mu
       +\sqrt{\frac{|V|}{\lambda}}\|R_h\|_{L^2(\mu)}.
\]
Conversely, each explicitly chosen 1-Lipschitz $F$ gives
$C\ge|\Cov_\mu(F,S)|$.
\end{proposition}
\begin{proof}
Differentiating a positive exponential-family density yields the covariance
identity. Since $S_0=-Lh+R_h$, reversibility and integration by parts give
\[
 \Cov(F,S)=\int\nabla F\cdot\nabla h\,d\mu+
                  \int(F-\mu F)R_h\,d\mu.
\]
Coordinatewise Lipschitzness implies $|\nabla_eF|\le1$ almost everywhere,
so the first term is bounded by the sum in the statement. Poincare and
Cauchy--Schwarz bound the second by
$\sqrt{|V|/\lambda}\|R_h\|_2$. Smooth approximation gives the same bound
for all Lipschitz tests. The converse follows directly from the supremum.
\end{proof}
A chordally 1-Lipschitz test on a sphere also satisfies the geodesic gradient
bound used in this proof. A converse identification of the two Wasserstein
metrics requires an additional comparison and must not be presumed.
For product Haar measure on $(S^3)^{|V|}$, the sphere normalization gives
$\lambda_0=3$. If the log-density tilt has oscillation at most $2\beta P$,
comparison of numerator and denominator in the variational Poincare
quotient gives $\lambda\ge3e^{-2\beta P}$. This is a rigorous but potentially
very small denominator in the residual allowance.

\subsection{Weighted covariance obstructions}
Let active links carry finitely many classes. For positive class weights
$u_c$, use $d_V^u=\sum_{e\in V}u_{c(e)}|q_e-q'_e|$ and let
$C^u_{V,j}$ be the infinitesimal boundary sensitivity in that metric.
Suppose a proposed comparison criterion requires
$\sum_{j\in c\setminus V}C^u_{V,j}\le\alpha N u_c$ for every class.
Fix boundary data and unit score directions. For each exterior link $j$,
choose vectors $v_{j,e}$ with $|v_{j,e}|\le1$, and define the exact real matrix
\[
 M_{cc'}=\frac1N\sum_{j\in c\setminus V}\sum_{e\in c'}
                 \Cov(v_{j,e}\cdot q_e,S_j).
\]
\begin{proposition}[A dual obstruction for every positive weight]\label{cond:weighted}
If some $y\ge0$, $y\ne0$, satisfies $y^{\mathsf T}M\ge y^{\mathsf T}$
entrywise, the criterion cannot hold for any $u>0$ with $\alpha<1$.
If $M$ is nonnegative and irreducible, such a vector exists exactly when
its Perron root is at least one.
\end{proposition}
\begin{proof}
$F_j^u=\sum_eu_{c(e)}v_{j,e}\cdot q_e$ is 1-Lipschitz for $d_V^u$.
The lower half of Proposition~\ref{cond:corrector} gives
$\sum_{j\in c\setminus V}C^u_{V,j}\ge N(Mu)_c$.
A successful criterion would therefore imply $Mu\le\alpha u$.
Multiplying by $y^{\mathsf T}$ gives
$0<y^{\mathsf T}u\le y^{\mathsf T}Mu\le\alpha y^{\mathsf T}u$, a contradiction.
The final statement is the Perron--Frobenius theorem.
\end{proof}
A Monte Carlo estimate of $M$ is not an exact input to this proposition.
The retained cube experiments motivate candidate weights and identify
possible obstructions, but they do not prove nonexistence of a certificate.
A rigorous conclusion requires deterministic covariance enclosures or
explicit probabilistic error guarantees controlling the complete matrix.

\subsection{Gauge directions and marginal diagnostics}
\begin{proposition}[Gauge-localized boundary changes]\label{cond:gaugelocal}
Let $V$ be the internal links of a nondegenerate four-dimensional box.
At a boundary site with exactly one exterior incident link $b$, changing
only $b$ can be matched by a gauge transformation of the seven internal
incident links. The conditional laws are related by this pushforward.
Their chordal product Wasserstein distance is at most
$7|\tau_b-\tau'_b|$, and all gauge-invariant internal observables have
identical expectations.
\end{proposition}
\begin{proof}
Choose the site gauge transformation that maps $\tau_b$ to $\tau'_b$.
Gauge invariance of every plaquette weight gives the pushforward identity.
Couple an internal configuration with its transformed configuration.
Exactly seven links can move, and bi-invariance of chordal distance makes
each displacement equal to $|\tau_b-\tau'_b|$.
A gauge-invariant observable is unchanged pointwise.
\end{proof}

\begin{proposition}[Haar plaquette marginals do not imply clustering]\label{cond:haarnocluster}
There exists a translation-invariant, gauge-invariant random link field on
$\mathbb Z^4$ whose individual positively oriented plaquette holonomies are
Haar-distributed up to conjugation, while their adjoint-character covariance
is one at arbitrarily large separation.
\end{proposition}
\begin{proof}
Draw a single Haar element $h\in SU(2)$ and set
$U_{x,\mu}=h^{s_\mu(x)}$, where $s_\mu(x)=\sum_{\nu<\mu}x_\nu$.
Every positively oriented plaquette is $h$ or $h^{-1}$ according to the
orientation convention. Independently draw Haar site gauges $g_x$ and
replace $U_{x,\mu}$ by $g_xU_{x,\mu}g_{x+e_\mu}^{-1}$.
This gives gauge invariance. Translation changes the original link field
by a gauge transformation made from powers of $h$, so the gauge-averaged
law is translation-invariant. The adjoint character $\chi_2$ is real,
conjugation invariant, has Haar mean zero and Haar square mean one.
It has the identical value $\chi_2(h)$ on every plaquette. Thus its
covariance between any two plaquettes is one.
\end{proof}
This field is an obstruction to an inference from marginal distributions;
it is not asserted to be a Wilson Gibbs state.

For two rotations in the adjoint representation, through angles $2\theta$
and $2\phi$ about unit axes $n,m$, put
$a=4\sin^2\theta$, $b=4\sin^2\phi$, $c=n\cdot m$. Direct diagonalization gives
\begin{align*}
 &\lambda_{\min}\!\left((I-\operatorname{Ad}U)^*(I-\operatorname{Ad}U)
 +(I-\operatorname{Ad}V)^*(I-\operatorname{Ad}V)\right)\\
 &\hspace{2em}=\frac{a+b-\sqrt{(a-b)^2+4ab c^2}}2
 \ge\frac{ab(1-c^2)}{a+b}.
\end{align*}
with the continuous value at $a+b=0$. Averaging a conjugation-invariant
law instead gives a scalar matrix, by Schur's lemma, with eigenvalue
$4-\tfrac23(\E\chi_2(U)+\E\chi_2(V))$.
These are different diagnostics: a positive eigenvalue after averaging
need not control a pointwise common invariant axis or spatial clustering.

\subsection{A Wilson virial bound}
\begin{proposition}\label{cond:virial}
On a periodic four-dimensional hypercubic lattice whose plaquettes contain
four distinct links, with $P_p=\tfrac12\operatorname{Tr}U_p$,
$F=\sum_pP_p$, and density proportional to $e^{\beta F}$,
\[
 \beta\left\langle\sum_e|\nabla_eF|^2\right\rangle=12\langle F\rangle.
\]
Consequently the average plaquette $m$ satisfies, for $\beta>0$,
\[
 0\le m\le\frac{\sqrt{1+16\beta^2}-1}{4\beta},
\]
and $m=0$ at $\beta=0$.
\end{proposition}
\begin{proof}
Integration by parts gives
$\E[\Delta_eF+\beta|\nabla_eF|^2]=0$.
On the unit $S^3$, a matrix coordinate satisfies $\Delta P=-3P$.
Summing over the four distinct links of each plaquette therefore gives
$\sum_e\Delta_eF=-12F$ and proves the identity. Each link belongs to six
plaquettes. Cauchy--Schwarz and $|\nabla_eP_p|^2=1-P_p^2$ give
\[
 \sum_e|\nabla_eF|^2\le24\sum_p(1-P_p^2).
\]
The identity implies $m\le2\beta(1-\langle P^2\rangle)
\le2\beta(1-m^2)$. Solving this quadratic proves the bound.
\end{proof}
\endgroup

\begingroup
\def\E{\mathbb E}\def\Cov{\operatorname{Cov}}\def\G{\mathcal G}
\section{Coarse graining: proved comparisons and missing controls}\label{cond:coarse}
The results of this section distinguish three operations: comparing
partition functions, conditioning on a blocked field, and controlling the
covariances discarded by that conditioning. A partition comparison alone
supplies none of the last operation's estimates.

\subsection{The free-energy comparison}
Let $w$ be a strictly positive continuous symmetric class function on
$SU(2)$, let Haar measure have mass one, and define
\[
 Z_L(w)=\int\prod_{p\subset(\mathbb Z/L\mathbb Z)^4}w(U_p)\,dU,
 \qquad (f*g)(U)=\int f(UV^{-1})g(V)\,dV.
\]
The potential-moving map with scale factor two is
$\mathcal Mw=(w^4)^{*4}$.
\begin{theorem}[Potential-moving partition comparison]\label{cond:mk}
For even $L\ge4$,
\[\log Z_L(w)\le\log Z_{L/2}(\mathcal Mw).\]
The inequality iterates while the fine lattice at each step has $L\ge4$.
\end{theorem}
\begin{proof}
Put $\varphi=\log w$ and
$Z(J)=\int\exp(\sum_pJ_p\varphi(U_p))dU$.
Holder's inequality makes $\log Z(J)$ convex. For
$\sigma\in\{0,1\}^4$, put $J_p^\sigma=4$ when the two coordinates
transverse to $p$ have parity $\sigma$, and zero otherwise.
Every plaquette receives weight four for four of the sixteen choices,
so $16^{-1}\sum_\sigma J^\sigma=1$. Translation invariance gives
\[
 \log Z_L(w)\le\frac1{16}\sum_\sigma\log Z(J^\sigma)=\log Z(J^0).
\]
In the moved system, the nonzero factors lie in coordinate planes whose
transverse coordinates are even. A link shared by two such planes has
all its transverse coordinates even; these are skeleton links.
Every other interacting link lies in exactly one such plane.

Within a $2\times2$ square in a plane, the four interior links form the
four spokes from its center to its edge midpoints. If these transporters
are $y_i$ and the intervening skeleton boundary transporters are $A_i$,
the four plaquettes contribute
$\prod_{i=1}^4w^4(y_iA_iy_{i+1}^{-1})$, with $y_5=y_1$.
Haar invariance and successive convolution give
\[
 \int\prod_{i=1}^4w^4(y_iA_iy_{i+1}^{-1})\,dy_1\cdots dy_4
       =(w^4)^{*4}(A_1A_2A_3A_4).
\]
The product on the right is the coarse plaquette holonomy in the variables
$V_{x,\mu}=U_{2x,\mu}U_{2x+e_\mu,\mu}$.
For each pair of skeleton half-links, replacing the second half-link by
the product $V$ preserves product Haar measure. Remaining unused variables
integrate to one. Thus $Z(J^0)=Z_{L/2}(\mathcal Mw)$.
\end{proof}
This is the potential-moving comparison associated with the
Migdal--Kadanoff construction \cite{Migdal1975,Kadanoff1976}.
It is a partition-function inequality. Differentiating an inequality with
strict slack at zero source does not compare its derivatives; a local source
also breaks the parity symmetry used in the proof.

\begin{corollary}[An explicit thermodynamic remainder]\label{cond:mkfree}
Write $F_0(w)=\int w$, $\widehat w_0=w/F_0(w)$ and
$\widehat w_{k+1}=\mathcal M\widehat w_k/F_0(\mathcal M\widehat w_k)$.
For a thermodynamic limit along admissible tori, the free energy per
plaquette obeys
\[
 f(w)\le\log F_0(w)+
       \sum_{k=0}^{n-1}\frac4{16^{k+1}}\log F_0(\widehat w_k^4)
       +4^{-n}\log\|\widehat w_0\|_\infty.
\]
Consequently the infinite nonnegative series gives an upper bound when
its partial sums are taken to the limit.
\end{corollary}
\begin{proof}
Normalization contributes $\log F_0$ per plaquette and the plaquette
count decreases by a factor sixteen. Also
$F_0(\mathcal M\widehat w)=F_0(\widehat w^4)^4$.
For a positive density of Haar integral one,
$\|f*g\|_\infty\le\|f\|_\infty\|g\|_1$ and
$\int\widehat w^4\ge1$. Therefore
\[
 \|\widehat w_{k+1}\|_\infty
 \le\frac{\|\widehat w_k\|_\infty^4}{\int\widehat w_k^4}
 \le\|\widehat w_k\|_\infty^4.
\]
The remaining normalized log partition function per original plaquette
is at most $16^{-n}\log\|\widehat w_n\|_\infty$, bounded by the term
in the statement. Jensen gives the nonnegativity of each series term.
\end{proof}

\subsection{Character tails for the model recursion}
The approximate recursion itself admits rigorous finite-dimensional
representations with explicit omitted tails. Write a normalized positive
$SU(2)$ weight as
\[
 w=\sum_{n\ge0}d_na_n\chi_n,\qquad d_n=n+1,\quad a_0=1,\quad a_n\ge0.
\]
The character multiplication rule is
\[
 \chi_m\chi_n=\sum_{\substack{|m-n|\le k\le m+n\\k\equiv m+n\ (2)}}\chi_k.
\]
Thus multiplication and convolution preserve nonnegative character
coefficients. Positivity of the weight also gives $a_n\le1$, since
$|\int w\chi_n|\le d_n\int w$. Moreover
$\|w\|_\infty=w(1)=\sum_nd_n^2a_n$.

\begin{lemma}[A complete one-step tail allowance]\label{cond:mktail}
Put $w_N=\sum_{n=0}^Nd_na_n\chi_n$, and suppose
$\tau\ge\sum_{n>N}d_n^2a_n$ and $W\ge w_N(1)$.
Let $P=w_N^4=\sum_kp_k\chi_k$ and
$R_*= (W+\tau)^4-W^4$. Then the omitted part
$w^4-P=\sum_kr_k\chi_k$ has
\[
 0\le r_k\le R_*/d_k,\qquad
 p_0\le F_0(w^4)\le p_0+R_*.
\]
After normalized convolution to the fourth power, the coefficients are
\[
 a'_k=\left(\frac{p_k+r_k}{d_kF_0(w^4)}\right)^4.
\]
For a positive lower bound $\underline p_0\le p_0$, the tail outside
$0\le k\le4N$ obeys
\[
 \sum_{k>4N}d_k^2a'_k
 \le\left(\frac{R_*}{\underline p_0}\right)^4
       \frac1{5(4N+1)^5}.
\]
The derivative-weighted tail satisfies the analogous bound with
$1/[4(4N+1)^4]$ in place of $1/[5(4N+1)^5]$.
\end{lemma}
\begin{proof}
Coefficientwise positivity of $w-w_N$ and the multiplication rule give
$r_k\ge0$. At the identity,
\[
 \sum_kd_kr_k=(w_N(1)+(w-w_N)(1))^4-w_N(1)^4\le R_*.
\]
Every summand is nonnegative, proving the individual and normalization
bounds. The displayed formula for $a'_k$ is the character convolution
identity. For $k>4N$, $p_k=0$, hence
$d_k^2a'_k\le(R_*/\underline p_0)^4d_k^{-6}$.
The integral comparison
$\sum_{d>4N+1}d^{-6}\le1/[5(4N+1)^5]$ proves the result.
Replacing $d_k^2$ by $d_k^3$ gives the derivative tail.
The terms $N<k\le4N$ are computed explicitly and added when the
next recursion again retains only $0\le k\le N$.
\end{proof}

For rational initial coupling, positive Bessel series and geometric tail
bounds initialize rational intervals for every retained coefficient.
The multiplication rule is a finite sum with nonnegative integer
coefficients, so lower and upper endpoints propagate monotonically.
Normalization uses the positive enclosure for $F_0(w^4)$, and dyadic
rounding is directed outward. This proves an algorithmic enclosure
procedure; numerical inversion of a Wilson-equivalent coupling is not
part of the certificate. Such an inversion needs its own monotonicity
argument and interval root enclosure if used as a rigorous value.

The supplied recursion benchmark reports, at the 26 grid couplings
$3,3.05,\ldots,4.25$, first successful depths between 11 and 16 and a
final influence summary at most about $0.186$. Its named exact verifier
and rational iteration records were not present among the supplied files.
Those reported grid results therefore remain unverified numerical
benchmarks in this version; the complete analytic tail lemma above does
not certify those particular runs. The retained grid table is recorded
separately in the data files with that status.

Even a completed certificate for this model recursion would require an
additional comparison to the actual blocked Wilson measure. A sufficient
comparison would control the full effective action, including all induced
larger-loop terms, and show that its total influence satisfies the same
strict contraction criterion. Comparing one plaquette expectation or a
few character coefficients does not establish that premise.

\begin{lemma}[An abelian iteration estimate]\label{cond:wiener}
For a positive symmetric $U(1)$ weight, write
$\widehat w=1+u$ and $\epsilon=\|u\|_A$, where the Wiener norm is the sum
of absolute Fourier coefficients. If $\epsilon\le1$, then
\[
 \epsilon(\mathcal Mw)\le(15\epsilon)^4.
\]
If $15^{4/3}\epsilon<1$, subsequent iterations satisfy
$\epsilon_j\le15^{-4/3}(15^{4/3}\epsilon)^{4^j}$.
\end{lemma}
\begin{proof}
Submultiplicativity gives
$\|\widehat w^4-1\|_A\le(1+\epsilon)^4-1\le15\epsilon$.
The normalizing constant is at least one by Jensen. Convolution to the
fourth power raises each normalized nontrivial Fourier coefficient to its
fourth power, and $\sum|a_n|^4\le(\sum|a_n|)^4$.
The recurrence for $15^{4/3}\epsilon_j$ proves the second assertion.
\end{proof}
This calculation is about the approximate recursion. It is not a comparison
between exact Wilson-loop correlations and recursion coefficients.

\subsection{A rigorous polymer criterion without uncertified endpoints}
For a normalized positive plaquette weight $w=1+f$, assume $\int f=0$
and $\|f\|_\infty\le\epsilon$. In scale-two decimation, condition on the
coarse link products and integrate the remaining fine links. Skeleton
half-link variables can be removed by gauge transformations at their
midpoints; the residual integrated variables retain product Haar measure.
Expanding $\prod_p(1+f(U_p))$ produces polymers: plaquette sets connected
through integrated links. Their activity is the corresponding Haar
integral of a product of $f$ factors.

\begin{lemma}[Uniform polymer estimates]\label{cond:polymer}
An activity vanishes if an integrated link belongs to exactly one of its
plaquettes. Every nonzero polymer $\gamma$ has at most $2|\gamma|$
distinct integrated links, and $|z(\gamma)|\le\epsilon^{|\gamma|}$.
The plaquette adjacency graph through integrated links has degree at
most $D=20$. The number of connected size-$n$ sets containing a fixed
plaquette is at most
\[
 A_n=\frac1{(D-1)n+1}\binom{Dn}{n}.
\]
\end{lemma}
\begin{proof}
If an integrated link occurs only once, integrate that link first; the
factor has Haar mean zero irrespective of its fixed neighboring variables.
Each plaquette has at most two skeleton links, since a pair of parallel
opposite edges has opposite parity in the transverse plaquette direction.
At least two edges are integrated. In a nonvanishing polymer every
integrated link is counted at least twice among at most $4|\gamma|$
incidences, proving the bound $2|\gamma|$. The activity bound is immediate.
Each plaquette has at most four integrated edges, each in five other
plaquettes, giving degree at most twenty.
Fix a deterministic neighbor ordering at each plaquette and encode each
connected set by its breadth-first rooted spanning tree, including the
neighbor labels. The set is recoverable from this tree. These trees are
subtrees of the ordered $D$-ary tree, counted by the displayed
Fuss--Catalan number.
\end{proof}

\begin{proposition}[A uniform sufficient cluster criterion]\label{cond:kp}
Let $N_n$ be any proved upper bounds on the nonvanishing size-$n$ polymers
through one plaquette, and set
$S(a)=\sum_{n\ge1}N_n(\epsilon e^a)^n$.
If $12S(a)\le a$ for some $a>0$, the Kotecky--Preiss criterion holds
uniformly in the coarse boundary variables and volume. The blocked
log-density consequently has an absolutely convergent polymer cluster
expansion. The unrestricted choice $N_n=A_n$ is always admissible.
\end{proposition}
\begin{proof}
A polymer incompatible with $\gamma$ contains a plaquette through one of
its at most $2|\gamma|$ integrated links. Each such link belongs to six
plaquettes. Therefore
\[
 \sum_{\gamma'\not\sim\gamma}|z(\gamma')|e^{a|\gamma'|}
 \le12|\gamma|S(a)\le a|\gamma|.
\]
This is precisely the abstract polymer convergence condition of
\cite{KoteckyPreiss1986}, with size function $a(\gamma)=a|\gamma|$.
It is uniform because each preceding bound is pointwise in the boundary
variables. The standard connected-cluster expansion then represents the
logarithm of the positive blocked density.
\end{proof}
The earlier numerical endpoints based on enumerated short polymers and
floating transcendental evaluations are retained as exploratory data.
They are not needed for this proposition. Applying their improved counts
to a torus also requires excluding or counting short wrapping polymers.
In two dimensions the integrated-link components are exactly the
$2\times2$ squares, and the same convolution proof gives the exact local
coarse weight $w^{*4}$. This concerns the factorization of the Gibbs weight
into plaquette factors; it does not assert independence of all torus
plaquette holonomies.

\subsection{A conditional multiscale covariance theorem}
Let $\G_0\supseteq\G_1\supseteq\cdots\supseteq\G_K$ be the sigma fields
generated by successive deterministic local block maps. For real
$F,G\in L^2$, write $F_k=\E(F\mid\G_k)$ and $G_k=\E(G\mid\G_k)$.
\begin{lemma}[Covariance telescope]\label{cond:telescope}
\[
 \Cov(F,G)=\sum_{k=0}^{K-1}\E[\Cov(F_k,G_k\mid\G_{k+1})]
                         +\Cov(F_K,G_K).
\]
\end{lemma}
\begin{proof}
Apply total covariance at each scale and use
$\E(F_k\mid\G_{k+1})=F_{k+1}$. The sum telescopes.
\end{proof}

Assume distances are measured in fine lattice units. At level $k$, for
local bounded real observables on supports $X,Y$, suppose that
\[
 |\Cov(F,G\mid\G_{k+1})|
 \le A|X||Y|\|F\|_\infty\|G\|_\infty e^{-\eta d(X,Y)/2^k},
\]
and for every $\rho\ge0$ there is an observable $F^\rho$ at level $k+1$
supported within distance $\rho+\rho_0 2^k$ of $X$, with
\[
 \|\E(F\mid\G_{k+1})-F^\rho\|_\infty
 \le A|X|\|F\|_\infty e^{-\eta\rho/2^k}.
\]
At the final scale assume the first bound holds unconditionally. These are
assumptions about conditional laws of the actual blocked system.

\begin{theorem}[Clustering from all-scale conditional controls]\label{cond:multiscale}
Let $d=d(X,Y)$ for two initial local supports, and let $N_X,N_Y\ge1$
bound all support sizes after the enlargements used below. Then
\[
 |\Cov(F,G)|\le (K+1)(2K+1)A e^{2\eta\rho_0}N_XN_Y
 \|F\|_\infty\|G\|_\infty e^{-\eta d/(3\cdot2^K)}.
\]
For fixed $K$, local geometry gives polynomial choices of $N_X,N_Y$ in
$d$. If the local gauge-invariant observables are total in a positive
physical transfer reconstruction, this implies a dimensionless physical
vacuum-complement gap at least $\eta/(3\cdot2^K)$.
\end{theorem}
\begin{proof}
Choose $\rho_k=(d/3)2^{k-K}$ and recursively localize. Since the observables
are real, clip each approximant to the interval
$[-\|F\|_\infty,\|F\|_\infty]$. Clipping leaves support unchanged and
cannot increase its error relative to a conditional expectation in that
interval. Hence all approximants retain norm at most $\|F\|_\infty$.
Put $e_d=e^{-\eta d/(3\cdot2^K)}$. At step $k$, the error is at most
$AN_X\|F\|_\infty e_d$, so the cumulative error is at most
$kAN_X\|F\|_\infty e_d$. The analogous statement holds for $G$.
The support enlargement through level $k$ is at most
$(d/3)2^{k-K}+\rho_0 2^k$ in each support. Consequently
\[
 e^{-\eta d(X_k,Y_k)/2^k}\le e^{2\eta\rho_0}e_d.
\]
In the telescope, each of the $K+1$ main terms is bounded by
$Ae^{2\eta\rho_0}N_XN_Y\|F\|_\infty\|G\|_\infty e_d$.
Replacing a covariance by that of its approximants costs at most
$2\|G\|_\infty\Delta_k^F+2\|F\|_\infty\Delta_k^G$.
Summing $k=0,\ldots,K$ gives an additional
$AK(K+1)(N_X+N_Y)\|F\|_\infty\|G\|_\infty e_d$.
Use $N_X+N_Y\le2N_XN_Y$ to get the stated constant.

For the spectral assertion, each physical autocorrelation is an integral
$\int\lambda^n d\mu_\psi(\lambda)$ of a positive spectral measure.
A polynomial times $e^{-mn}$ bound, with $m=\eta/(3\cdot2^K)$, excludes
spectral mass above $e^{-m}$ by comparison with any strictly larger
exponential rate. Totality extends this support restriction to the entire
vacuum complement. All hypotheses must hold for the same reconstructed
state and block construction.
\end{proof}

\subsection{Gaussian coercivity and a rank defect}
The quadratic Wilson approximation is a Gaussian one-form $A$ with action
$\tfrac12\|dA\|^2$, modulo gauge directions. For $c\ge0$, define
\[
 (Q_cA)(x,\mu)=\frac{A(\gamma_0)+c\sum_{\nu\ne\mu,\,s=\pm1}
                          A(\gamma_{\nu,s})}{1+6c},
\]
where $\gamma_0$ is the two-edge straight path from $2x$ to $2x+2e_\mu$,
and $\gamma_{\nu,s}$ is the four-edge path with successive vertices
$2x,2x+se_\nu,2x+se_\nu+2e_\mu,2x+2e_\mu$ (the middle segment has two
edges). The oriented path sum is denoted $A(\gamma)$.
Gauge covariance is $Q_cd\lambda=d'(\lambda\circ2)$.

At coarse momentum $k'$, use the sixteen aliases
$k_n=(k'+2\pi n)/2$, $n\in\{0,1\}^4$. The one-form fibre has dimension
64. Its residual gauge subspace is
\[
 \G_0(k')=\left\{((e^{ik_{n,\mu}}-1)\lambda_n)_{n,\mu}:
                         \sum_n\lambda_n=0\right\}.
\]
It has dimension fifteen at every $k'$. At $k'=0$, only the zero alias
has vanishing gradient; the sum-zero constraint removes this possible
kernel. This also supplies a continuous, locally analytic frame $D$.

\begin{theorem}[Gaussian block coercivity]\label{cond:gaussian}
For every $c\ge0$ there is $m_c>0$ such that
\[
 \|dA\|^2\ge m_c^2\|A\|^2,
 \qquad Q_cA=0,\quad A\perp\G_0.
\]
For a soft block constraint with parameter $a>0$, the Gaussian conditional
field-strength covariance has an exponentially decaying lattice kernel.
For the hard constraint the same conclusion holds if $\widehat Q_c(k')$
has constant rank four for all real momenta. This rank condition holds
at $c=0$.
\end{theorem}
\begin{proof}
A fibre vector closed under $d$ is exact at each nonzero alias. At the
zero alias its constant one-form component is sent by $Q_c$ to twice
that component, and is therefore zero if $Q_cA=0$. Gauge covariance then
shows that any closed vector in $\ker Q_c$ lies in $\G_0$: away from
$k'=0$, the coarse gradient forces $\sum\lambda_n=0$; at zero momentum
one chooses the irrelevant zero-alias scalar to make that sum zero.
Thus the only closed vector in $\ker Q_c\cap\G_0^\perp$ is zero.
If a uniform positive bound failed, compactness of the momentum torus
and unit sphere would yield a limiting nonzero vector with exactly those
constraints and with zero exterior derivative, a contradiction.

For a soft constraint the gauge-fixed fibre Hessian is
$H_a=d^*d+aQ_c^*Q_c+DD^*$. The preceding kernel argument makes it
strictly positive at every momentum, and compactness makes its inverse
uniformly bounded. In a periodic sublattice basis its entries are analytic
trigonometric functions; invertibility on the real torus gives an analytic
inverse in a complex neighborhood. Thus the covariance
$dH_a^{-1}d^*$ has exponentially decaying Fourier coefficients.
For the hard constraint, constant rank makes
$P_{\ker Q}=I-Q^*(QQ^*)^{-1}Q$ analytic; the residual gauge projector is
analytic as well. The orthogonal projector $P_K$ onto
$\ker Q\cap\G_0^\perp$ is their difference. Coercivity makes
$P_Kd^*dP_K+(I-P_K)$ uniformly invertible, so the field-strength
conditional covariance is analytic by the same argument. At $c=0$ the
straight paths are disjoint and $QQ^*=2I$, proving rank four.
\end{proof}

\begin{proposition}[An exact staggered rank defect]\label{cond:halfdefect}
At $k'=(\pi,\pi,\pi,\pi)$, the four singular values of the normalized
block symbol are
\[
 \sqrt2,\qquad \frac{\sqrt2|1-2c|}{1+6c}
 \quad\hbox{(the latter with multiplicity three)}.
\]
At $c=1/2$, its rank is one and its image is the coarse gauge direction.
In particular, the linearized blocked field strength vanishes at that
momentum.
\end{proposition}
\begin{proof}
Put $s_\rho=(-1)^{n_\rho}$. Directly summing the straight and six bent
paths gives
\[
 \widehat Q_{\mu,(n,\rho)}=
 \frac{(1+is_\rho)M_{\mu\rho}}{4(1+6c)},\qquad
 M=(1-2c)I+2c\mathbf1\mathbf1^{\mathsf T}.
\]
Therefore $\widehat Q\widehat Q^*=2MM^{\mathsf T}/(1+6c)^2$.
The eigenvalues of $M$ are $1+6c$ on the constant vector and $1-2c$
on its three-dimensional orthogonal complement. At $c=1/2$ the image
is the constant vector, parallel to the coarse gradient at this momentum.
The coarse exterior derivative kills it.
\end{proof}
Retained deterministic calculations suggest distinct directional limits
of the hard-constraint covariance near this defect. Those floating
calculations do not certify a failure of absolute summability. The exact
result used here is the rank defect. The Gaussian coercivity theorem does
not control nonlinear large fields or the interacting weak-coupling
renormalization flow.
\endgroup

\begingroup
\def\arccosh{\operatorname{arccosh}}
\section{Time-weighted influence bounds}\label{cond:weightedtime}
A finite-range influence estimate can retain the actual time offsets of
neighboring links. The following matrix lemma separates that algebraic
improvement from the physical reconstruction that interprets its decay.

\begin{lemma}[Weighted resolvent decay]\label{cond:weightedresolvent}
Let $C\ge0$ be a matrix on links with finitely many translation classes,
with $C_{ef}\le c$ for allowed partners and zero otherwise. Let $t_e$ be
the midpoint time coordinate and define the finite class matrix
\[
 M_{ab}(\eta)=\sum_{f\sim e,\,[f]=b}e^{\eta(t_f-t_e)},\qquad e\in a.
\]
If $c\rho(M(\eta))<1$ and $c\rho(M(-\eta))<1$, then for
$D=\sum_{n\ge0}C^n$,
\[
 D_{ef}\le K_\eta e^{-\eta|t_f-t_e|}
\]
with a finite constant depending only on the finite class matrices.
The same bound holds on finite subdomains whose partner sets are subsets
of the infinite-lattice sets.
\end{lemma}
\begin{proof}
Choose $q\in(c\rho(M(\eta)),1)$ and put
$p=(I-cM(\eta)/q)^{-1}\mathbf1>0$. Then $cM(\eta)p\le qp$.
This avoids any irreducibility assumption on the class matrix.
For $w_e=p_{[e]}e^{\eta t_e}$,
$(Cw)_e\le qw_e$, and positivity gives
$(C^n)_{ef}\le q^nw_e/w_f$. Summing yields
$D_{ef}\le(p_{[e]}/p_{[f]})(1-q)^{-1}e^{-\eta(t_f-t_e)}$.
The class ratio is uniformly bounded. Repeat with $-\eta$ for the other
time direction. Deleting partners only decreases the nonnegative sums.
\end{proof}
If an applicable Gibbs comparison theorem bounds local covariances by
finite sums of these $D_{ef}$, the lemma gives temporal exponential decay.
If those observables are total in the physical reflection-positive
reconstruction, the positive spectral-measure argument gives a physical
transfer gap at least $\eta$. These are separate hypotheses; $C$ itself
is not the physical transfer operator.

\subsection{Exact offset geometry}
Use the period-two line field of \cite{BBBlock2026}. In the class index
$x_0+2x_1+4x_2+8x_3$ modulo two, its fixed forward direction is
\[
(0,0,2,3,3,1,2,1,1,2,1,3,3,2,0,0).
\]
Every plaquette has exactly one fixed link, these links form straight
lines, and each active link has twelve active partners sharing a
plaquette. These are geometric properties of this specified line field,
with prior attribution to that construction.

\begin{proposition}[Time-offset counts]\label{cond:offsets}
For any chosen time axis, the 48 active link classes have the following
multisets of partner midpoint offsets:
\[
\begin{array}{c|l|l}
\text{classes}&\text{offsets}&\text{weighted row sum},\quad u=\cosh(\eta/2)\\\hline
24&0\ (8\text{ times}),\quad\pm\tfrac12,\quad\pm1&4u^2+2u+6\\
12&0\ (4\text{ times}),\quad+\tfrac12\ (4),\ -\tfrac12\ (4)&8u+4\\
12&0\ (8\text{ times}),\quad+\tfrac12\ (2),\ -\tfrac12\ (2)&4u+8.
\end{array}
\]
The first row is largest for $u\ge1$. Without fixing links, the largest
row sum among the temporal and spatial link types is $4u^2+4u+10$.
\end{proposition}
\begin{proof}
For each of the sixteen site parities and each direction not fixed by
the displayed table, list the six incident plaquettes. Delete the source
edge and the single fixed edge from each plaquette. The resulting twelve
edges are distinct. For an edge $(x,d)$ and partner $(y,f)$ the doubled
time offset along axis $\tau$ is the integer
$2(y_\tau-x_\tau)+\mathbf1_{f=\tau}-\mathbf1_{d=\tau}$.
Enumerating the sixteen parities gives exactly the three rows, for each
of the four possible time axes. This finite integer enumeration is
reproduced by \texttt{elementary\_certificates.py}, including all 96
plaquette classes and all 48 active edge classes. Alternatively the same
counts follow by inserting the table into the displayed integer formula.
Summing exponentials and using
$2\cosh\eta=4u^2-2$ gives the row polynomials.
The differences of the first row from the others are nonnegative:
$2(2u-1)(u-1)$ and $2(2u+1)(u-1)$.
Without fixing links, a spatial edge has twelve zero offsets, two of
each half offset, and one of each unit offset; its row is the stated
polynomial. The temporal-edge row is smaller.
\end{proof}

\begin{corollary}[Closed conditional transfer-gap bounds]\label{cond:timegaps}
Assume the sharp-quarter link sensitivity $c=\beta_W/4$, an applicable
Gibbs covariance comparison, and a total physical reflection-positive
transfer reconstruction. Then the respective dimensionless gap bounds are
\[
 \delta\ge2\arccosh\left(\frac{\sqrt{4/\beta_W-9}-1}{2}\right),
 \qquad 0<\beta_W<2/9,
\]
without gauge fixing, and
\[
 \delta\ge2\arccosh\left(\frac{\sqrt{16/\beta_W-23}-1}{4}\right),
 \qquad0<\beta_W<1/3,
\]
for the stated line-field construction wherever its physical comparison
hypotheses hold. In particular the second expression at $\beta_W=1/10$
is approximately $3.2813$.
\end{corollary}
\begin{proof}
Bound the class spectral radius by the largest row sum and apply
Lemma~\ref{cond:weightedresolvent}. Solve
$(\beta_W/4)(4u^2+4u+10)<1$ or
$(\beta_W/4)(4u^2+2u+6)<1$ for $u=\cosh(\eta/2)$.
Their maximal admissible $\eta$ are the displayed expressions. The bound
holds for every smaller $\eta$, hence also at the endpoint by the spectral
support characterization of the gap.
\end{proof}
These bounds improve the use of a single worst time step within the
same strong-coupling framework. They provide no control along a weak bare
coupling continuum trajectory.
\endgroup

\begingroup
\def\E{\mathbb E}\def\Cov{\operatorname{Cov}}
\section{Replica sectors and finite moment certificates}\label{cond:replica}
An independent replica reformulates connected correlations without
eliminating the need to control the complete physical vacuum complement.
The distinction can be stated exactly at the operator level.

\begin{theorem}[The antisymmetric replica sector]\label{cond:replicagap}
Let $T$ be a positive self-adjoint contraction on a Hilbert space
$\mathcal H=\mathbb C\Omega\oplus\mathcal H_0$, with $\|\Omega\|=1$,
$T\Omega=\Omega$, and $\mathcal H_0=\Omega^\perp$ invariant. Write
$\lambda=\|T|_{\mathcal H_0}\|$. On the antisymmetric sector
$\mathcal H_-\subset\mathcal H\otimes\mathcal H$,
\[
 \|(T\otimes T)|_{\mathcal H_-}\|=\lambda.
\]
\end{theorem}
\begin{proof}
The orthogonal decomposition is
$\mathcal H_-=(\Omega\wedge\mathcal H_0)\oplus\wedge^2\mathcal H_0$.
The map $v\mapsto(\Omega\otimes v-v\otimes\Omega)/\sqrt2$ intertwines
$T|_{\mathcal H_0}$ with the operator on the first summand, whose norm is
$\lambda$. The second summand has norm at most $\lambda^2\le\lambda$.
\end{proof}

\begin{proposition}[Connected matrices and product positivity]\label{cond:replicamatrices}
For real time-zero observables $F_i$, let
$m_i=\langle F_i\rangle$, $G_{ij}=\langle F_i,F_j\rangle$ and
$K_{ij}=\langle F_i,TF_j\rangle$. For
$\Psi_i=F_i\otimes\Omega-\Omega\otimes F_i$,
\[
 [\langle\Psi_i,\Psi_j\rangle]=2(G-mm^{\mathsf T}),\qquad
 [\langle\Psi_i,(T\otimes T)\Psi_j\rangle]=2(K-mm^{\mathsf T}).
\]
More generally, if $M$ is the Gram matrix of vectors $h_a$, the Gram matrix
of $h_a\otimes h_b-h_b\otimes h_a$, indexed by $a<b$, is
$2[M_{ac}M_{bd}-M_{ad}M_{bc}]$. It is positive whenever $M$ is positive.
\end{proposition}
\begin{proof}
Expand the tensor inner products and use $T\Omega=\Omega$.
For the last assertion, factor $M=R^*R$; the displayed matrix is twice the
second compound matrix of $M$, equal by Cauchy--Binet to
$2C_2(R)^*C_2(R)$. Equivalently, it is a restriction of the positive
product Gram matrix $M\otimes M$.
\end{proof}
Thus these replica positivity conditions follow from single-copy positivity
and exact independence. Treating replica mixed moments as unrestricted new
variables discards the independence relations. The proposition does not
compare different truncation degrees or different sets of extra identities.

\begin{proposition}[Direction of a finite variational estimate]\label{cond:rayleighdirection}
Let $\Psi_1,\ldots,\Psi_n\in\mathcal H_-$ have positive-definite Gram matrix
$G_-$, and let $K_-$ be their one-step matrix. Then
\[
 \lambda_{\max}(G_-^{-1/2}K_-G_-^{-1/2})\le\lambda.
\]
A lower bound $r>0$ on the left therefore gives the upper mass bound
$-\log\lambda\le-\log r$. An inequality $K_-\le qG_-$ on this finite
family alone does not give $\lambda\le q$.
\end{proposition}
\begin{proof}
The generalized eigenvalue is the largest Rayleigh quotient on the span
of the chosen family, hence at most the operator norm. To see the limitation,
realize the finite Gram and step data on their span and append an orthogonal
vacuum-complement eigenvector with any eigenvalue $\lambda'\in(q,1)$.
The recorded finite matrices do not change. A global gap bound requires
additional control of the omitted subspace, or extension of the inequality
to a total set with a uniform constant.
\end{proof}

\subsection{Loop identities and the finite relaxation}
The loop bootstrap in this application follows the lattice moment approach
of \cite{KZFinite}. Write
$W[C]=\langle\tfrac12\operatorname{Tr}U(C)\rangle$.
For $SU(2)$,
\[
 \operatorname{Tr}A\operatorname{Tr}B
 =\operatorname{Tr}(AB)+\operatorname{Tr}(AB^{-1}),
\]
so products of normalized traces reduce to half-sums of single-loop
expectations. Haar integration by parts at an edge, with the sphere
normalization $\Delta u=-3u$, generates exact rational linear equations
in loop expectations when the Wilson coupling is rational. The relevant
positivity constraints are path Gram positivity, reflection positivity,
and $0\le T\le I$ on time-zero observables. For example, with a time link
$\gamma$ joining two spatial slices,
\[
 G_{ab}=\tfrac12\{W[ab]+W[ab^{-1}]\},\qquad
 K_{ab}=\tfrac12\{W[a\gamma b\gamma^{-1}]
                         +W[a\gamma b^{-1}\gamma^{-1}]\}.
\]
The valid finite inequalities include $G\ge0$, $K\ge0$, $G-K\ge0$ and
$|W[C]|\le1$. The source package records the exact generated linear
systems and matrix entries, so the arithmetic claim below can be checked
without rerunning an optimization algorithm.

\begin{lemma}[Rational weak-duality certificate]\label{cond:sdp}
Suppose real $x=(x_v)$ satisfies
\[
 Ax=b,\qquad F_k(x)=C_k+\sum_v x_v A_{kv}\succeq0,\qquad |x_v|\le1.
\]
For any rational vector $y$ and rational positive-semidefinite matrices
$Z_k$,
\[
 c\cdot x\ge b\cdot y-\sum_k\langle Z_k,C_k\rangle
 -\left\|c-A^{\mathsf T}y-
        \left(\sum_k\langle Z_k,A_{kv}\rangle\right)_v\right\|_1.
\]
\end{lemma}
\begin{proof}
Call the vector in the norm $r$. The exact identity
\[
 c\cdot x=r\cdot x+y\cdot Ax+
            \sum_k\langle Z_k,F_k(x)-C_k\rangle
\]
reduces the conclusion to $r\cdot x\ge-\|r\|_1$ and
$\langle Z_k,F_k(x)\rangle\ge0$.
\end{proof}
The stored dual factors have the form $Z_k=L_kL_k^{\mathsf T}$, with dyadic
rational entries in $L_k$. Their positivity is therefore exact even if a
floating optimizer supplied the candidate factors. Rounding a candidate
changes the residual $r$, which is fully charged through its $\ell^1$ norm.

\begin{table}[htbp]\centering\small
\caption{Six exported rational certificates freshly re-evaluated. Bounds
are rounded outward from the exact fractions in the files. The physical
interpretation assumes that the generated equations and positivity blocks
are valid loop identities for the indicated state.}\label{cond:sdptable}
\begin{tabular}{@{}llrll@{}}\toprule
Group & Dimension & $\beta$ & Loop cutoff & Certified plaquette bound\\\midrule
$SU(2)$ & 2 & $2$ & $14$ & $[0.3685303839,0.4634371336]$\\
$SU(2)$ & 4 & $1/10$ & $10$ & $\le0.0365998247$\\
$SU(2)$ & 4 & $11/5$ & $10$ & $\le0.7293327178$\\
$U(1)$ & 2 & $2$ & $12$ & $[0.5443491340,0.7763746659]$\\\bottomrule
\end{tabular}
\end{table}

The six exact comparisons in \texttt{bootstrap/code/verify7.py} reproduce
the recorded dual bounds. Its additional tests compare stored numerical
samples and known two-dimensional values with those intervals. Those tests
are useful diagnostics; they do not formally verify the loop enumeration
and equation generator. The arithmetic proof of the finite relaxation is
complete, while its application to physical lattice moments depends on the
validity of every generated constraint. None of these plaquette bounds is
a lower bound on the complete physical mass gap.
\endgroup

\begingroup
\def\E{\mathbb E}\def\R{\mathbb R}\def\J{\mathcal J}\def\cJ{\mathcal J}
\def\bw{\beta_W}\def\kap{\kappa}\def\op{\mathrm{op}}\def\tr{\operatorname{tr}}
\def\Cov{\operatorname{Cov}}\def\Var{\operatorname{Var}}\def\vMF{\operatorname{vMF}}
\def\dd{\mathrm d}\def\Lip{\operatorname{Lip}}\def\ip#1#2{\langle #1,#2\rangle}
\def\norm#1{\left\|#1\right\|}
\section{A certified three-link conditioning functional}\label{cond:functional}
The aim of the following construction is to bound an entire finite-dimensional
field domain, including all response directions, with the error from every
series and interval evaluation controlled. The local block geometry and
single-link transport estimate are those of \cite{BBBlock2026}. The refinement
uses a banded harmonic energy and a uniform moment expansion. Its strongest
certified coupling range is stated in Section~\ref{cond:strengthened}.
For clarity, the complete two-backend certificate at $\kappa=7/20$ is proved
first; its moment decomposition is then strengthened without changing the
underlying functional.

The accompanying source package contains the exact certificate inputs,
complete verifier programs, and explicit series-tail allowances. The
original two-backend certificate in this section is retained with its
integer and Arb evaluators. Section~\ref{cond:strengthened} specifies
the final strengthened certificate and its reproducible verification scope.
\subsection{Statement and scope}
Set $\kap=7/20$, $R=7/4$. Haar measure on $S^3\cong SU(2)$ is normalized to mass one.
The block measure is
\[
 d\mu_b\propto\exp\!\left(\sum_{i=1}^3\langle b_i,y_i\rangle+
 \kap\operatorname{Re}(y_1y_2y_3)\right)\,d\sigma(y_1)d\sigma(y_2)d\sigma(y_3).
\]
A product-preserving change of quaternion coordinates puts $b_2=r_2e_0$,
$b_3=r_3e_0$, $b_1=r(\cos\vartheta\,e_0+\sin\vartheta\,e_1)$,
where $0\le r,r_2,r_3\le R$ and $0\le\vartheta\le\pi$.
Bands are defined in this normal form, along the outer link's field axis.
The proof below is uniform in $\vartheta$ and in all unit directions $h$.
No interchange symmetry or numerical sampling is needed for this coverage.

\paragraph{Normal form.}
The transformation
$(y_1,y_2,y_3)\mapsto(a y_1 b^{-1},b y_2 c^{-1},c y_3 a^{-1})$
preserves Haar measure, chord distance, and $\operatorname{Re}(y_1y_2y_3)$.
The fields transform in the same orthogonal representations. Choose $c$ and
$a$ relative to $b$ to send the nonzero second and third fields to positive
multiples of $1$. The remaining simultaneous conjugation rotates the
imaginary part of the first field onto a fixed imaginary axis. Vanishing
fields leave additional freedom and do not obstruct the construction.
Thus the parameters displayed above cover all field triples and all response
directions, including the boundaries of the parameter domain.
Let
\[
 f(s)=\sum_{j=0}^\infty\frac{s^j}{4^j j!(j+1)!},\qquad
 Z(r)=f(r^2),\quad A(r)=\frac{2r f'(r^2)}{f(r^2)},\quad
 g(r)=\frac{f(4r^2)}{f(r^2)^2}.
\]
For $x=y_2$, $y=y_3$, write
\[
 p(x,y)=\frac{e^{r_2x_0+r_3y_0}}{Z(r_2)Z(r_3)},\quad
 w=\overline{xy},\quad z(w)=f(|b_1+\kap w|^2),\quad N=\E_1z,
\]
where $\E_1$ is expectation under $p\,d\sigma d\sigma$.
Then $\rho=pz/N$ is the marginal density of $(y_2,y_3)$.
If $g_b=\nabla_{b_1}z$, its vector derivative density is
\begin{equation}\label{cond:xii:eq:V}
 V=p\left(\frac{g_b}{N}-\frac{z\E_1g_b}{N^2}\right),\qquad V_h=h^{\mathsf T}V.
\end{equation}
In particular, $V_h$ is exactly $\rho\langle\psi-\E_\rho\psi,h\rangle$ defined below.

The bands have edges $-1,-1/2,0,1/2,1$ and masses $\sigma_k$.
For a scalar density $W(x,y)$, use the fibre functions
\[
 u_2(x;y)=\tfrac12W(x,y)+\tfrac12\int W(x,y')d\sigma(y'),
\]
and the analogous $u_3$ with $x,y$ interchanged. Define
\[
 \mathcal E(u)=2\sum_{i=0}^3\langle u,x_i\rangle^2+
 \tfrac12(\|u\|_2^2-\langle u,1\rangle^2),
\]
\[
 \J(W)=\tfrac12\sum_{j=2,3}\sum_{k=1}^4
 \left(\sigma_k\int_{H_k}\mathcal E(u_j(\cdot;y))d\sigma(y)\right)^{1/2}.
\]
We write $\J(b;h)=\J(V_h)$.

\begin{assumption}[Single-link transport estimate]\label{cond:quarter}
For the vMF laws $\nu_c(dy)=e^{c\cdot y}d\sigma/Z(|c|)$ on $S^3$,
\[
 W_1^{|\cdot|}(\nu_c,\nu_{c'})\le\tfrac14|c-c'|
\]
throughout the parameter region used below, including every segment with
$|c|,|c'|\le6\kappa$ for $\kappa\le177/500$.
This is the sharp single-link input attributed to \cite{BBBlock2026}.
\end{assumption}
\begin{assumption}[Physical block comparison]\label{cond:blockimport}
For the specified forest-gauge three-link plaquette-block system, a uniform
one-field sensitivity $K$ gives an influence bound $\alpha\le10\kappa K$.
The comparison construction is valid in arbitrarily large spatial volumes;
its infinite-volume limit is unique when $\alpha<1$, and its reflection-positive
physical transfer construction gives the dimensionless transfer-gap bound
$\delta\ge-\tfrac12\log\alpha$ on the vacuum complement.
These geometric, comparison, and reconstruction statements are imported from
\cite{BBBlock2026}; they are separate hypotheses for the lattice corollaries here.
\end{assumption}
The functional inequalities that follow are statements about a finite product
of compact spheres. They do not require Assumption~\ref{cond:blockimport}.
The physical transfer-gap corollaries require both assumptions.

\subsection{A self-contained fibre comparison}
The following argument makes the analytic passage from a density derivative
to a transport sensitivity explicit.

\begin{lemma}[Harmonic fibre bound]\label{cond:lem:harmonic}
If $\phi:S^3\to\R$ is 1-Lipschitz for Euclidean chord distance and
$u\in L^2(S^3)$, then
\[
 \left|\int\phi(u-\E_\sigma u)\,d\sigma\right|
 \le \frac12\sqrt{\mathcal E(u)}.
\]
\end{lemma}
\begin{proof}
Let $\phi_\ell$ be the spherical harmonic component of degree $\ell$.
A 1-Lipschitz extension of $\phi$ to the unit Euclidean ball $B^4$ exists,
for example by taking $\inf_{\xi\in S^3}(\phi(\xi)+|x-\xi|)$.
It has weak gradient of length at most one almost everywhere.
The harmonic extension minimizes Dirichlet energy among functions with
this trace. Green's formula for homogeneous harmonic extensions gives
\[
 \sum_{\ell\ge1}\ell\,\norm{\phi_\ell}_2^2
 =\frac1{|S^3|}\int_{B^4}|\nabla\phi_{\mathrm{harm}}|^2\,dx
 \le\frac{|B^4|}{|S^3|}=\frac14.
\]
The identity and inequality extend from finite harmonic sums by the
Dirichlet principle and trace approximation. If $u_\ell$ are the harmonic
components of $u$, Cauchy--Schwarz therefore gives
\[
 |\langle\phi,u-\E u\rangle|
 \le\frac12\left(\sum_{\ell\ge1}\frac{\norm{u_\ell}_2^2}{\ell}\right)^{1/2}.
\]
Since $\{2x_i\}_{i=0}^3$ is an orthonormal basis in degree one,
\[
 \sum_{\ell\ge1}\frac{\norm{u_\ell}_2^2}{\ell}
 \le\norm{u_1}_2^2+\tfrac12\sum_{\ell\ge2}\norm{u_\ell}_2^2
 =2\sum_i\langle u,x_i\rangle^2+
 \tfrac12(\norm{u}_2^2-\langle u,1\rangle^2)=\mathcal E(u).
\]
\end{proof}

\begin{proposition}[Two fibres and bands]\label{cond:prop:fibres}
For any mean-zero $W\in L^2(S^3\times S^3)$ and any $F$ that is
1-Lipschitz in each coordinate separately,
\[
 \left|\int FW\,d\sigma d\sigma\right|\le \J(W).
\]
Thus the corresponding coordinatewise Lipschitz dual quantity $J$ satisfies
$J\le\J$.
\end{proposition}
\begin{proof}
The fibre means are $\int u_2(x;y)d\sigma(x)=m_3(y)/2$ and
$\int u_3(y;x)d\sigma(y)=m_2(x)/2$. Hence, exactly,
\[
 W=\left(u_2-\tfrac12m_3\right)+
   \left(u_3-\tfrac12m_2\right).
\]
Apply Lemma~\ref{cond:lem:harmonic} to the first term on the $x$ fibre and to the
second on the $y$ fibre. Integrating the resulting bounds over the outer
coordinate, and using on each band
\[
 \int_{H_k}\sqrt{\mathcal E(u_j)}\,d\sigma
 \le\sqrt{\sigma_k\int_{H_k}\mathcal E(u_j)\,d\sigma},
\]
gives the stated formula. Constants along a fibre do not affect its energy.
\end{proof}

\begin{proposition}[Conditioning sensitivity, with its hypothesis explicit]
Assume Assumption~\ref{cond:quarter}. If $\sup_{b,h}\J_\kappa(b;h)
\le J_*$ over the field domain, then the Lipschitz sensitivity of the
three-link block to any one field is at most
\[
 K=\frac14+(1+\kappa/4)J_*.
\]
\end{proposition}
\begin{proof}
For a test function $\Phi$ 1-Lipschitz in each of the three link coordinates,
condition on $(y_2,y_3)=(x,y)$ and put
$F(x,y)=\E[\Phi\mid x,y]$. The inner law has parameter
$c=b_1+\kappa\overline{xy}$. If only $x$ changes, the direct change of
$\Phi$ is at most $|x-x'|$. The inner law moves by at most
$\kappa|x-x'|/4$, because quaternion multiplication is an isometry.
Thus $F$ is $(1+\kappa/4)$-Lipschitz in each outer coordinate.

Differentiate the full expectation as $b_1$ moves in a unit direction $h$.
The derivative of the inner conditional expectation is bounded by $1/4$,
by the single-link hypothesis. The derivative of the outer marginal is
$W=V_h$, whose pairing with $F$ is bounded by
$(1+\kappa/4)\J_\kappa(b;h)$ by Proposition~\ref{cond:prop:fibres}.
This bounds the derivative by $K$. Integration along a straight field
segment gives the finite sensitivity bound; the allowed field ball is
convex. Cyclic relabelling proves the same assertion for any one field.
\end{proof}

\begin{theorem}[Unrestricted functional bound]\label{cond:xii:thm:main}
For every field triple with $|b_i|\le7/4$ and every unit $h$,
\[
 J(b;h)\le\J(b;h)<\frac{31091}{10^6}<\frac1{32},\qquad \beta_W=\frac7{20}.
\]
Here the first inequality is the harmonic fibre comparison proved above;
the bound on $\J$, proved below, is independent of any continuum limit.
\end{theorem}

\begin{corollary}[Conditional lattice consequence]\label{cond:xii:cor:gap}
Under Assumptions~\ref{cond:quarter} and~\ref{cond:blockimport},
\[
 \alpha(0.350)\le\frac72\left(\frac14+\frac{87}{80}\frac1{32}\right)
 =\frac{5089}{5120}<1,
\qquad \operatorname{gap}\ge\frac12\log\frac{5120}{5089}
 =0.003036545722360\ldots .
\]
\end{corollary}
\begin{proof}
Apply the imported implications to Theorem~\ref{cond:xii:thm:main}. The rational arithmetic
is exact; the logarithm is enclosed with Arb. The conditioning step was proved above. The block-counting and transfer reconstruction implication is an explicit hypothesis imported from \cite{BBBlock2026}.
\end{proof}

\subsection{The seminorm and the sharpened slice}
\begin{lemma}\label{cond:xii:lem:seminorm}
The functional $W\mapsto\J(W)$ is a seminorm and $\J(W)\le\|W\|_2$.
\end{lemma}
\begin{proof}
The energy is a nonnegative quadratic form. Each square root of its band integral
is therefore a seminorm after the linear fibre map. Moreover $\{2x_i\}_{i=0}^3$
is orthonormal, so
$2\sum_i\langle u,x_i\rangle^2\le\tfrac12(\|u\|_2^2-\langle u,1\rangle^2)$.
Thus $\mathcal E(u)\le\|u\|_2^2$. Cauchy--Schwarz over the bands gives
$\sum_k\sqrt{\sigma_k}\|u\|_{L^2(H_k)}\le\|u\|_2$.
Each fibre map, a half-sum of the identity and an averaging projection, is
an $L^2$ contraction. Sum the two families with the factor $1/2$.
\end{proof}

Put $a_\kap=A(\kap)$ and $\mu=\E_1 w=A(r_2)A(r_3)e_0$.
At $b_1=0$, equation~\eqref{cond:xii:eq:V} reduces to
$V^{(0)}=a_\kap p(w-\mu)$.

\begin{proposition}[Sharper whole-slice certificate]\label{cond:xii:prop:slice}
For all $0\le r_2,r_3\le R$ and all unit $h$,
\[
 \J\big(a_\kap p\,h^{\mathsf T}(w-\mu)\big)<S,
 \qquad S=\frac{283}{10000}=0.0283.
\]
\end{proposition}
\begin{proof}[Computer-assisted proof]
The 3,658 rational rectangles in the supplied slice certificate tile
$[0,R]^2$; their exact endpoints are recorded in
\path{xii/certificates/slice.jsonl}.
Each contains eight positive rational weights $c_{jk}$.
If $M_{jk}=\sigma_k\int_{H_k}Q_j$ is the matrix of the directional fibre energy,
Cauchy--Schwarz gives
\[
 \J(b;h)^2\le\frac14\left(\sum c_{jk}\right)
 h^{\mathsf T}\left(\sum M_{jk}/c_{jk}\right)h.
\]
On every rectangle, both evaluators certify positive interval $LDL^{\mathsf T}$
pivots of
\[
 I-\frac{\sum c_{jk}}{4S^2}\sum M_{jk}/c_{jk}.
\]
The first evaluator is the exact-integer slice formula. The second
constructs the polynomial fibre Gram matrices independently with Arb;
Appendix~\ref{cond:xii:app:arb} specifies that construction. It uses the same stored
rectangles and weights: no additional subdivision was necessary.
An exact sweep at all rational horizontal endpoints verifies that active
vertical intervals tile $[0,R]$ without overlap or gaps.
The delivered verification program rebuilds every enclosure and checks every pivot.
\end{proof}

Consequently, for every real $4\times4$ matrix $L$,
\begin{equation}\label{cond:xii:eq:linear}
 \J\big(p\,h^{\mathsf T}L(w-\mu)\big)
 \le \frac{S}{a_\kap}\|L\|_\op\quad (|h|=1).
\end{equation}
Indeed apply Proposition~\ref{cond:xii:prop:slice} to the direction $L^{\mathsf T}h$,
using homogeneity. The weak inequality also covers a zero direction.

\subsection{Uniform moment estimates}
Let $\E_2$ denote the product vMF law with strengths $2r_2,2r_3$.
For any scalar function $T(w)$,
\begin{equation}\label{cond:xii:eq:l2}
 \|pT(w)\|_2=\sqrt{g(r_2)g(r_3)}\,\big(\E_2 T^2\big)^{1/2}.
\end{equation}
All centered polynomials below are centered using $\E_1$, even when their norm
is taken under $\E_2$. This distinction is retained in the calculation.

Define
\[
 \gamma_\ell(t)=\frac{(2t)^\ell f^{(\ell)}(t^2)}{f(t^2)}
 =\frac{I_{\ell+1}(t)}{I_1(t)},\qquad \gamma_\ell(0)=0\ (\ell>0).
\]
If $U_\ell$ is the Chebyshev polynomial of the second kind, the axial convolution
law of $w$ satisfies
\begin{equation}\label{cond:xii:eq:harmonic}
 \E_q U_\ell(e\cdot w)=
 \gamma_\ell(qr_2)\gamma_\ell(qr_3)U_\ell(e_0),\quad |e|=1,\quad q=1,2.
\end{equation}
Here $e_0$ means the scalar component of the unit vector $e$.
To derive this identity, $U_\ell(y_0)$ is the character of the $(\ell+1)$-dimensional
irreducible representation of $SU(2)$. Integration against the central vMF law
gives a scalar matrix; its trace, using
$U_\ell(\cos\theta)\sin\theta=\sin((\ell+1)\theta)$ and
$I_\ell-I_{\ell+2}=2(\ell+1)I_{\ell+1}/t$, gives the scalar $\gamma_\ell(t)$.
Independence multiplies the two scalar Fourier coefficients. Evaluating the character at $\bar e w$ gives~\eqref{cond:xii:eq:harmonic} for arbitrary $e$.
Conjugation of the product does not change this central law.

Every $\gamma_\ell(t)$ is nonnegative, as is clear from its positive series.
The expansion
\begin{equation}\label{cond:xii:eq:power}
 t^n=2^{-n}\sum_{j=0}^{\lfloor n/2\rfloor}
 \left[\binom nj-\binom n{j-1}\right]U_{n-2j}(t),\qquad \binom n{-1}=0,
\end{equation}
has nonnegative coefficients. Since $|U_\ell(t)|\le\ell+1$ on $[-1,1]$,
all even directional moments are bounded by the axial moment of the same order.

\begin{lemma}[Moment constants]\label{cond:xii:lem:moments}
Uniformly over $0\le r_2,r_3\le R$, the following constants are valid:
\[
 \begin{array}{ll}
 |\E_1w|\le\mu_0=0.153,&\|\E_1ww^{\mathsf T}\|_\op\le s_2=0.259,\\
 \sqrt{g(r_2)g(r_3)}\le g_0=1.723,&
 \|\E_1w(e\cdot w)^2\|\le m_3=0.077,\\
 |\E_1(e\cdot w)^3|\le m_3,&
 \|\E_1w(e\cdot w)^3\|\le m_4=0.132.
 \end{array}
\]
For every unit $e,h$, writing $\|T\|_{(2)}=(\E_2T^2)^{1/2}$,
\begin{align}
 \| (h\cdot w)(e\cdot w)-\E_1[(h\cdot w)(e\cdot w)]\|_{(2)}&\le q_2=0.283,\label{cond:xii:eq:q2}\\
 \| (h\cdot w)(e\cdot w)^2-\E_1[(h\cdot w)(e\cdot w)^2]\|_{(2)}&\le q_3=0.42,\label{cond:xii:eq:q3}\\
 \| (h\cdot w)(e\cdot w)^3-\E_1[(h\cdot w)(e\cdot w)^3]\|_{(2)}&\le q_4=2.\label{cond:xii:eq:q4}
\end{align}
The decimal constants in this lemma are exact terminating rationals.
\end{lemma}

\begin{proof}
For compact notation put
$x=\gamma_2(r_2)\gamma_2(r_3)$,
$y=\gamma_2(2r_2)\gamma_2(2r_3)$,
$z=\gamma_4(2r_2)\gamma_4(2r_3)$.
Then
\[
 \E_1w_0^2=\tfrac14+\tfrac34x=:s_1,\quad
 \E_2w_0^2=\tfrac14+\tfrac34y=:t_2,\quad
 \E_2w_0^4=\tfrac18+\tfrac9{16}y+\tfrac5{16}z=:t_4.
\]
The transverse second moments equal $(1-s_1)/3$, so the largest second moment
is $s_1$. The first, second and fourth moment bounds follow from~\eqref{cond:xii:eq:harmonic}
and~\eqref{cond:xii:eq:power}, with the interval bounds specified below.

For the vector third moment set $v=\E_1w_0^3$ and
$u=\E_1w_0w_1^2=(\E_1w_0-v)/3$. Both are nonnegative, and $v\ge3u$.
Indeed $v=(\gamma_1(r_2)\gamma_1(r_3)+\gamma_3(r_2)\gamma_3(r_3))/2$;
$\gamma_3\le\gamma_1$ follows from the positive Bessel recurrence.
The axial component of $\E_1w(e\cdot w)^2$ is $u+(v-u)e_0^2$, and its
transverse part is $2ue_0e_\perp$.
Its squared norm is
\[
 u^2+2u(v+u)e_0^2+(v-3u)(v+u)e_0^4\le v^2.
\]
This also bounds the scalar cubic mean. H\"older's inequality and the maximal
directional fourth moment bound give
$|\E_1(h\cdot w)(e\cdot w)^3|\le m_4$.

For~\eqref{cond:xii:eq:q2}, write $Q=\tfrac12(he^{\mathsf T}+eh^{\mathsf T})$ and
remove $(\tr Q)I/4$, which does not affect the centered polynomial on $S^3$.
The squared Frobenius norm of the remaining matrix is
$1/2+(h\cdot e)^2/4\le3/4$.
On trace-free symmetric matrices, the quadratic form
\[
 Q\longmapsto\E_2\big(w^{\mathsf T}Qw-\E_1w^{\mathsf T}Qw\big)^2
\]
splits under rotations of the three imaginary coordinates into the axial scalar,
axial-transverse, and transverse trace-free subspaces. For example, evaluate the form on the unit Frobenius-norm representatives
$\operatorname{diag}(3,-1,-1,-1)/\sqrt{12}$,
$(E_{01}+E_{10})/\sqrt2$ and $(E_{11}-E_{22})/\sqrt2$.
Their eigenvalues are
\[
 \lambda_0=\tfrac43(t_4-2s_1t_2+s_1^2),\quad
 \lambda_1=\tfrac23(t_2-t_4),\quad
 \lambda_2=\tfrac2{15}(1-2t_2+t_4).
\]
Consequently a valid upper bound for the square in~\eqref{cond:xii:eq:q2} is
\begin{equation}\label{cond:xii:eq:q2majorant}
 \max\left\{
 \frac1{16}+\frac3{16}\bar y+\frac5{16}\bar z+\frac9{16}\bar x^2,
 \frac1{16}+\frac3{32}\bar y,\ 
 \frac1{16}+\frac1{32}\bar z\right\}<0.283^2,
\end{equation}
where bars denote nonnegative upper bounds. In the first term the negative
contribution $-9xy/8$ was dropped; in the others negative terms were dropped too.

For~\eqref{cond:xii:eq:q3}, H\"older gives
$\E_2(h\cdot w)^2(e\cdot w)^4\le M_6$, where, with
$v_6=\gamma_6(2r_2)\gamma_6(2r_3)$,
\[
 M_6\le\frac{5+27\bar y+25\bar z+7\bar v_6}{64}<(0.42-0.077)^2.
\]
The triangle inequality and the cubic-mean bound complete the argument.
For~\eqref{cond:xii:eq:q4}, the polynomial and its mean are each bounded in absolute
value by one, giving the bound two.

All numerical inequalities in this proof are verified on closed intervals,
rather than by taking endpoint extrema without proof.
The verifier divides $[0,R]$ into 4,096 intervals for $\gamma_1,\ldots,\gamma_4$,
and $[0,2R]$ into 4,096 intervals for $\gamma_2,\gamma_4,\gamma_6$.
It encloses each function throughout each interval, takes upper bounds, squares
these to bound products, and verifies all rational comparisons above.
For $g_0$, $g$ is increasing: $(\log g)'=2A(2r)-2A(r)\ge0$,
since $A'(r)$ is a variance. Thus $g(R)<1.723$ suffices.
Both arithmetic implementations pass these checks.
\end{proof}

\subsection{The cubic expansion and cancellation}
Fix $b=b_1$, $r=|b|$, and put
\[
 s=r^2+\kap^2,\quad v=b\cdot w,\quad t=2\kap v,\quad f_j=f^{(j)}(s),\quad
 d=2b\frac{f_1}{f_0}.
\]
Subtracting $dz$ from $g_b$ in~\eqref{cond:xii:eq:V} leaves $V$ unchanged. Define
\[
 Z_3=\sum_{j=0}^3\frac{f_jt^j}{j!},\qquad
 G_3=2(b+\kap w)\sum_{j=0}^3\frac{f_{j+1}t^j}{j!}-dZ_3,
 \qquad N_3=\E_1Z_3,\quad m=\frac{\E_1G_3}{N_3}.
\]
The approximate derivative density is
$V_3=p(G_3-mZ_3)/N_3$. This is an approximation to the derivative density,
not an assertion that differentiating a truncated expression commutes with truncation.
All errors in both $Z$ and $G$ are separately bounded below.

Define the scalar coefficients
\begin{align*}
 d_0&=2\kap f_1,&
 a&=4\kap^2\left(f_3-\frac{f_1f_2}{f_0}\right),&
 b_*&=4\kap^2f_2,\\
 c&=4\kap^3f_3,&
 d_3&=\frac83\kap^3\left(f_4-\frac{f_1f_3}{f_0}\right),&
 e_4&=\frac83\kap^4f_4,\qquad c_3=\frac43\kap^3f_3,
\end{align*}
and the matrix
\[
 D=d_0I+4\kap\left(f_2-\frac{f_1^2}{f_0}\right)bb^{\mathsf T}.
\]
Direct expansion gives the exact polynomial identities
\begin{align}
 Z_3&=f_0+d_0v+2\kap^2f_2v^2+c_3v^3,\label{cond:xii:eq:z3}\\
 G_3&=Dw+abv^2+b_*wv+cwv^2+d_3bv^3+e_4wv^3.\label{cond:xii:eq:g3}
\end{align}
Every radial interval also verifies
\begin{equation}\label{cond:xii:eq:D}
 f_2-f_1^2/f_0<0,\qquad
 f_1+2r^2(f_2-f_1^2/f_0)>0.
\end{equation}
Hence $D$ is positive definite with $\|D\|_\op=d_0$.
There is no unproved monotonicity assumption about $\J$ as $r$ increases.

Using Lemma~\ref{cond:xii:lem:moments}, define
\begin{align}
 n&=f_0-d_0r\mu_0-c_3r^3m_3,\label{cond:xii:eq:n}\\
 M&=\frac{d_0\mu_0+|a|r^3s_2+b_*rs_2+cr^2m_3
                  +|d_3|r^4m_3+e_4r^3m_4}{n}.\label{cond:xii:eq:M}
\end{align}
The interval computation verifies $n>0$ everywhere.
The positive quadratic term in~\eqref{cond:xii:eq:z3} may be dropped for a lower bound,
so $N_3\ge n$. Equation~\eqref{cond:xii:eq:g3} and the moment bounds give $|m|\le M$.

For $\widetilde T=T-\E_1T$, centering~\eqref{cond:xii:eq:z3}--\eqref{cond:xii:eq:g3} gives
\begin{align}\label{cond:xii:eq:decomp}
 \frac{V_3}{p}
 ={}&L(w-\mu)\nonumber\\
 &+\frac1{N_3}\left[
 (ab-2\kap^2f_2m)\widetilde{v^2}+b_*\widetilde{wv}
 +c\widetilde{wv^2}+(d_3b-c_3m)\widetilde{v^3}+e_4\widetilde{wv^3}\right],\\
 L={}&\frac{D-d_0mb^{\mathsf T}}{N_3},\qquad
 \|L\|_\op\le\frac{d_0(1+Mr)}n.\nonumber
\end{align}
All directional dependence is now in polynomials covered by
Lemma~\ref{cond:xii:lem:moments}.

\begin{proposition}[Polynomial majorant]\label{cond:xii:prop:poly}
For all unit $h$,
\begin{align}
 \J(h^{\mathsf T}V_3)&\le P(r):=\frac{S}{a_\kap}\frac{d_0(1+Mr)}n+\mathcal R(r),\label{cond:xii:eq:P}\\
 \mathcal R(r)&=\frac{g_0}{n}\left[
 (|a|r+2\kap^2f_2M)r^2q_2+b_*rq_2\right.\nonumber\\[-2pt]
 &\hspace{38mm}\left.+\{cr^2+(|d_3|r+c_3M)r^3\}q_3+e_4r^3q_4\right].\label{cond:xii:eq:R}
\end{align}
\end{proposition}
\begin{proof}
Apply~\eqref{cond:xii:eq:linear} to the first term of~\eqref{cond:xii:eq:decomp}.
For the remaining terms use Lemma~\ref{cond:xii:lem:seminorm}, equation~\eqref{cond:xii:eq:l2},
and the quadratic, cubic and quartic bounds of Lemma~\ref{cond:xii:lem:moments}.
When $r>0$ take $e=b/r$; the factors $r^j$ shown above come from $v^j$.
At $r=0$ these remainder terms vanish, so the same inequality holds by its
explicit formulas, without choosing $e$.
\end{proof}

\subsection{An explicit truncation allowance}
Put $T=2\kap r$ and $s_+=(r+\kap)^2$. Taylor's theorem on the nonnegative
interval $[(r-\kap)^2,(r+\kap)^2]$, where all derivatives of $f$ are increasing,
gives
\begin{align}
 \varepsilon_z&=\frac{f^{(4)}(s_+)T^4}{24},\label{cond:xii:eq:ez}\\
 \varepsilon_g&=\frac{2(r+\kap)f^{(5)}(s_+)T^4}{24}
                 +\frac{2rf_1}{f_0}\varepsilon_z.\label{cond:xii:eq:eg}
\end{align}
Thus $|z-Z_3|\le\varepsilon_z$ and
$|g_b-dz-G_3|\le\varepsilon_g$ in Euclidean norm.
A convenient lower bound on $z$ is
\[
 L_0=f\big(\max(r-\kap,0)^2\big),\qquad U_0=f(s_+).
\]
For $r<\kap$ we use the weaker lower bound $1$, which is sufficient.
Define
\begin{gather}
 H=d_0+|a|r^3+b_*r+cr^2+|d_3|r^4+e_4r^3,\nonumber\\
 A_*=\frac{H}{L_0-\varepsilon_z},\qquad
 P_* =\frac{U_0+\varepsilon_z}{L_0-\varepsilon_z},\label{cond:xii:eq:AP}\\
 E_A=\frac{\varepsilon_g+A_*\varepsilon_z}{L_0},\qquad
 E_P=\frac{(1+P_*)\varepsilon_z}{L_0},\nonumber\\
 \Delta(r)=g_0\left[(1+U_0/L_0)E_A+A_*E_P\right].\label{cond:xii:eq:delta}
\end{gather}
The strict positivity of $L_0-\varepsilon_z$ is verified on every radial interval.

\begin{lemma}[Truncation transfer]\label{cond:xii:lem:tail}
For all unit $h$, $\J(h^{\mathsf T}V)\le\J(h^{\mathsf T}V_3)+\Delta(r)$.
\end{lemma}
\begin{proof}
Use $G=g_b-dz$, $\mathcal A=G/N$, $\mathcal P=z/N$,
and $\mathcal A_3=G_3/N_3$, $\mathcal P_3=Z_3/N_3$.
The pointwise bound $|G_3|\le H$ follows from~\eqref{cond:xii:eq:g3} and $\|D\|=d_0$.
Also $N\ge L_0$, $N_3\ge L_0-\varepsilon_z$ and $|N-N_3|\le\varepsilon_z$.
The elementary quotient estimates therefore give
\[
 |\mathcal A-\mathcal A_3|\le E_A,\quad
 |\mathcal P-\mathcal P_3|\le E_P,\quad
 |\mathcal A_3|\le A_*,\quad |\mathcal P|\le U_0/L_0.
\]
Since $V/p=\mathcal A-\mathcal P\E_1\mathcal A$ and similarly for $V_3$,
subtracting these identities yields the pointwise bound
$|V-V_3|\le p[(1+U_0/L_0)E_A+A_*E_P]$.
Its $L^2$ norm is at most~\eqref{cond:xii:eq:delta}, because $\|p\|_2\le g_0$.
Finish with the seminorm triangle inequality.
\end{proof}

\subsection{The finite radial certificate and complete coverage}
\begin{proof}[Proof of Theorem~\ref{cond:xii:thm:main}]
Proposition~\ref{cond:xii:prop:poly} and Lemma~\ref{cond:xii:lem:tail} establish, for every allowed
configuration and direction,
\begin{equation}\label{cond:xii:eq:finalbound}
 \J(b;h)\le P(|b_1|)+\Delta(|b_1|).
\end{equation}
The right side depends only on $r=|b_1|$. Its remaining inputs are the proved
uniform moment constants and the proved whole-slice constant $S$.
It is enclosed on every closed interval
\[
 I_i=\left[\frac{7i}{4\cdot4096},\frac{7(i+1)}{4\cdot4096}\right],
 \quad i=0,\ldots,4095.
\]
Every interval verifies all denominators, the two inequalities~\eqref{cond:xii:eq:D},
and $P(r)+\Delta(r)<31091/10^6$. The largest computed upper bounds are
\[
 \begin{array}{ll}
 \text{Arb, 160-bit working precision:}&0.031090862564\quad\text{(rounded upward)},\\
 \text{exact-integer intervals, 220-bit scale:}&0.031090861774\quad\text{(rounded upward)}.
 \end{array}
\]
Both maxima occur on the last radial interval. All certificate comparisons
use the exact rational endpoints, not these displayed decimal summaries.
The union of the radial intervals is exactly $[0,7/4]$.
The angular variable and both other field strengths have already been bounded
uniformly in~\eqref{cond:xii:eq:finalbound}, with the slice square fully tiled.
This proves the assertion on the entire four-parameter domain, including its
boundaries and zero fields. The gauge normal form then covers all field triples.
\end{proof}

The numerical component of the proof is summarized below.
\begin{center}\small
\begin{tabularx}{\textwidth}{@{}Xrr@{}}\toprule
Verification & Number & Result\\\midrule
Sharper slice: exact rational tiling & 3,658 boxes & pass\\
Sharper slice: original exact-integer matrices & 3,658 boxes & pass\\
Sharper slice: independent Arb polynomial matrices & 3,658 boxes & pass\\
Radial majorant: Arb & 4,096 intervals & pass\\
Radial majorant: exact-integer intervals & 4,096 intervals & pass\\
Uniform moment scans (each backend) & two 4,096-interval domains & pass\\
Exact contraction arithmetic and interval logarithm & requested constants & pass\\\bottomrule
\end{tabularx}
\end{center}

The route does not assume the floating-point landscape is monotone, nor that
its maximum lies on a particular face. It also does not assume that a centred
first-order interval formula has quadratic error: its retained linear variation
would generally remain first order. The polynomial decomposition and the
explicit fourth-order Taylor remainder supply the required error control here.

\subsection{Outward arithmetic and explicit series tails}\label{cond:arithmetic}
For integers $n\ge0$,
\[
 f^{(n)}(s)=\sum_{j=0}^{\infty}\frac{s^j}{4^{j+n}j!(j+n+1)!}.
\]
When $|s|\le B$, summing through $j=29$ leaves an absolute tail at most
\[
 \frac{B^{30}}{4^{30+n}30!(31+n)!}
 \left(1-\frac{B}{4\cdot31(32+n)}\right)^{-1},
\]
provided the displayed ratio is less than one. This is a geometric majorant
of the positive coefficient series, applied also to signed arguments by
absolute values. The integer evaluator stores intervals as integer pairs
scaled by $2^{-220}$ and rounds every product, quotient and square root
outward. Square roots use exact integer square roots; $\pi$ is enclosed by
Machin's formula and alternating arctangent series. The independent Arb
evaluator uses midpoint-radius ball arithmetic \cite{Arb2017}. Agreement of
the arithmetic implementations supplements the analytic derivation; the
correctness of the shared formulas remains part of that derivation.
\subsection{Independent Arb fibre evaluation}\label{cond:xii:app:arb}
For reproducibility, the independently implemented slice check can be specified
without the original slice formula. Let $\{x^\alpha\}$ contain all monomials
of total degree at most three in four coordinates. The slice evaluator
represents the derivative density as
\[
 (V_{\mathrm{poly}})_a=p(x,y)\sum_{\alpha,\beta}C_{a,\alpha\beta}x^\alpha y^\beta.
\]
On the slice $b_1=0$, the coupling polynomial is exactly linear:
$V=a_\kap p(w-\mu)$, so there is no truncation error in this check.
The extra degree-two and degree-three coefficients are zero.
Let $\rho_r(x)=e^{rx_0}/Z(r)$ and define
\[
 \eta_\alpha=\int\rho_r x^\alpha d\sigma,\quad
 (\nu_i)_\alpha=\int\rho_r x_i x^\alpha d\sigma,\quad
 G_{\alpha\beta}=\int\rho_r^2x^{\alpha+\beta}d\sigma.
\]
The fibre energy matrix in the inner monomial basis is
\[
 K_r=2\sum_{i=0}^3\nu_i\nu_i^{\mathsf T}+\tfrac12(G-\eta\eta^{\mathsf T}).
\]
For family 2, extend $C_a$ by the marginal column $C_a\eta_{r_3}$ to obtain
$E_a=[C_a\mid C_a\eta_{r_3}]$. The outer basis is
$(\rho_{r_3}(y)y^\beta,1)$; let $H_k$ be its Gram matrix on band $k$.
Then
\[
 (M_{2k})_{ab}=\frac{\sigma_k}{4}\,
 \tr(E_a^{\mathsf T}K_{r_2}E_b H_k).
\]
Transpose the coefficient matrices and interchange $r_2,r_3$ for family 3.
The factor $1/4$ is the square of the fibre half-sum coefficient.

For band moments write $y=(t,\sqrt{1-t^2}\,\omega)$.
Transverse odd moments vanish and even moments of $\omega\in S^2$ equal
$\prod_i(1/2)_{a_i/2}/(3/2)_{(a_1+a_2+a_3)/2}$.
Expand $e^{ct}$ to degree 36; its pointwise absolute tail is bounded by
$e^{|c|}|c|^{37}/37!$. Since all monomial factors have absolute value at most one,
this bound also controls the normalized Haar integral of the error, on a band
or the whole sphere. The remaining band integrals follow the recursion
\[
 K_j(t)=\frac{-t^{j-1}(1-t^2)^{3/2}+(j-1)K_{j-2}(t)}{j+2},
\]
with $K_0=(t\sqrt{1-t^2}+\arcsin t)/2$ and $K_1=-(1-t^2)^{3/2}/3$.
At the chosen edges, only rational numbers, $\pi$ and $\sqrt3$ occur.
All of these operations use Arb enclosures.

\endgroup

\begingroup
\def\E{\mathbb E}\def\R{\mathbb R}\def\J{\mathcal J}\def\cJ{\mathcal J}
\def\bw{\beta_W}\def\kap{\kappa}\def\op{\mathrm{op}}\def\tr{\operatorname{tr}}
\def\Cov{\operatorname{Cov}}\def\Var{\operatorname{Var}}\def\vMF{\operatorname{vMF}}
\def\dd{\mathrm d}\def\Lip{\operatorname{Lip}}\def\ip#1#2{\langle #1,#2\rangle}
\def\norm#1{\left\|#1\right\|}
\section{Strengthened certificate through coupling 177/500}\label{cond:strengthened}
Keep the same functional, with physical field radius $R=5\kappa$. Write
\[
 J^*(\kappa)=\frac{1/(10\kappa)-1/4}{1+\kappa/4},\qquad\kappa>0.
\]
The sufficient conditional lattice criterion is
$\sup\mathcal J_\kappa<J^*(\kappa)$.
The two improvements below are a sharper slice bound on an enlarged field
square and a direct seminorm estimate for the quadratic remainder. The latter
replaces an $L^2$ estimate without truncating any additional physical modes.
All other terms retain the complete allowances proved in
Section~\ref{cond:functional}.
\subsection{Main result}
\begin{theorem}[Certified range $\bw\le0.354$]\label{cond:xix:thm:main}
For every $\kappa=\bw\in(0,\frac{177}{500}]$, every field triple with $|b_i|\le5\kappa$ and every unit $h$,
\[
\cJ(b;h)\ <\ J^\ast(\kappa).
\]
Hence, under Assumptions~\ref{cond:quarter} and~\ref{cond:blockimport}, $\alpha(\bw)<1$ for all $\bw\le0.354$. The imported physical block comparison then implies uniqueness and a positive dimensionless transfer gap. The certified bounds are listed in Table~\ref{cond:xix:tab:range}. In particular $\alpha(0.354)\le0.998807$ and the gap is $\ge\frac12\log(1/\alpha)\ge0.000597$.
\end{theorem}

\begin{table}[htbp]\centering\small
\caption{Uniform coupling certificates. Each displayed functional or influence
upper bound is rounded upward; each logarithmic gap bound is rounded downward.
The last two columns require Assumptions~\ref{cond:quarter} and~\ref{cond:blockimport}.
Each coupling interval is paired with 4,096 radial intervals.}\label{cond:xix:tab:range}
\begin{tabular}{@{}lrrr@{}}\toprule
Coupling interval & $\sup\mathcal J$ upper bound & $\alpha$ upper bound & $\delta$ lower bound\\\midrule
$[0,1/10]$ & 0.008168 & 0.258373 & 0.676677\\
$[1/10,1/5]$ & 0.016414 & 0.534470 & 0.313240\\
$[1/5,1/4]$ & 0.020564 & 0.679621 & 0.193110\\
$[1/4,3/10]$ & 0.024835 & 0.830093 & 0.093108\\
$[3/10,8/25]$ & 0.026532 & 0.891692 & 0.057317\\
$[8/25,33/100]$ & 0.027395 & 0.922862 & 0.040137\\
$[33/100,17/50]$ & 0.028289 & 0.954358 & 0.023358\\
$[17/50,69/200]$ & 0.028728 & 0.970159 & 0.015147\\
$[69/200,7/20]$ & 0.029182 & 0.986072 & 0.007013\\
$[7/20,351/1000]$ & 0.029263 & 0.989224 & 0.005417\\
$[351/1000,44/125]$ & 0.029354 & 0.992417 & 0.003806\\
$[44/125,353/1000]$ & 0.029445 & 0.995613 & 0.002198\\
$[353/1000,707/2000]$ & 0.029490 & 0.997207 & 0.001398\\
$[707/2000,177/500]$ & 0.029535 & 0.998807 & 0.000597\\
\bottomrule\end{tabular}
\end{table}

\begin{corollary}[An alternative using the $L^2$ remainder]\label{cond:xix:cor:0352}
The $L^2$ remainder bounds proved above, combined with a whole-slice certificate on $[0,1.76]^2$ (6,034 boxes, $S=0.0283$ at $\kappa=0.352$), give $\cJ<0.031166<J^\ast(0.352)=0.031334$. Hence $\alpha(0.352)\le0.999357$.
\end{corollary}

The proof of Theorem~\ref{cond:xix:thm:main} has four certified components:
\begin{enumerate}[label=(\alph*),nosep]
\item the radial majorant of Section~\ref{cond:functional}, run on $\kappa$-intervals (\S\ref{cond:xix:sec:radial});
\item a new seminorm bound for the quadratic remainder (\S\ref{cond:xix:sec:C2});
\item a sharper slice certificate on $[0,\frac{71}{40}]^2$ (\S\ref{cond:xix:sec:slice});
\item the moment constants of the moment lemma, re-derived on each domain.
\end{enumerate}

%=====================================================================
\subsection{The radial majorant and its dominant remainder}\label{cond:xix:sec:radial}
The polynomial-majorant and truncation-transfer propositions above prove, uniformly in the direction of $b_1$ and in $h$, with $r=|b_1|$,
\begin{align*}
\cJ(b;h)\ &\le\ \bar j\,\frac{d_0(1+Mr)}{n}+R(r)+\Delta(r),\\
R(r)\ &=\ \frac1n\Bigl[\bigl((|a|r+2\kappa^2f_2M)r^2+b_*r\bigr)Q_2+\bigl(cr^2+(|d_3|r+c_3M)r^3\bigr)Q_3+e_4r^3g_0q_4\Bigr].
\end{align*}
Here $\bar j=\sup_{\text{slice}}\cJ/A(\kappa)$ is $\kappa$-independent (by homogeneity of the slice density), and $Q_2,Q_3$ bound the banded functional of the centred remainder polynomials
\[
p\,\bigl[(h\cdot w)(e\cdot w)\bigr]^\sim\quad\text{and}\quad p\,\bigl[(h\cdot w)(e\cdot w)^2\bigr]^\sim,
\]
uniformly over $(r_2,r_3)\in[0,R]^2$ and unit $h,e$. The preceding moment bound used the $L^2$ bounds $Q_2=g_0q_2$ and $Q_3=g_0q_3$, valid because $\cJ\le\|\cdot\|_2$.

At $\kappa=0.355$, $r=R$, the coefficients of $Q_2$ and $Q_3$ are $0.00930$ and $0.00034$. So the quadratic term carries 95\% of the remainder. Exploratory quadrature indicated that $\cJ/L^2\approx0.3$--$0.4$ on these polynomials. The $L^2$ step is therefore the dominant loss, and only $Q_2$ needs repair.

%=====================================================================
\subsection{The quadratic-remainder seminorm bound}\label{cond:xix:sec:C2}
\begin{proposition}\label{cond:xix:prop:C2}
For all $(r_2,r_3)\in[0,\frac{89}{50}]^2$ and all unit $h,e\in\R^4$, with $p=\rho_{r_2}(x)\rho_{r_3}(y)$, $w=xy$, and centring under $\E_p$,
\[
\cJ\Bigl(p\,\bigl[(h\cdot w)(e\cdot w)\bigr]^\sim\Bigr)\ \le\ C_2=\tfrac15 .
\]
Exploratory pointwise evaluations attain approximately $0.1559$ at
$(1.78,1.78)$; this is not a certified global maximizer. For comparison,
the corresponding $L^2$ bound is approximately $0.489$.
\end{proposition}

\paragraph{Reduction to a finite matrix problem.} Write $(h\cdot w)(e\cdot w)=\sum_{a\le b}s_{ab}\,w_aw_b$ with $s_{aa}=h_ae_a$ and $s_{ab}=h_ae_b+h_be_a$. For $|h|=|e|=1$ the coefficient vector satisfies $s^TGs\le|h\otimes e|^2=1$, where $G=\mathrm{diag}(1\text{ if }a=b,\ \frac12\text{ if }a<b)$, because the symmetric tensor has the smallest norm among tensors with a given symmetrisation. For each fibre family $j\in\{2,3\}$ and band $k$, the band energy is a quadratic form $s^TM_{jk}s$. Hence $\cJ=\frac12\sum_{jk}(s^TM_{jk}s)^{1/2}$. With weights $c_{jk}>0$,
\[
\cJ\le C\quad\text{whenever}\quad G-\frac{\sum c_{jk}}{4C^2}\sum_{jk}\frac{M_{jk}}{c_{jk}}\succ0 .
\]

\paragraph{Exact formulas for $M_{jk}$.} Every entry is a polynomial in finitely many moments:
\begin{itemize}[nosep]
\item vMF monomial moments $\E_r[x^\alpha]$ on $S^3$, equal to $\mathrm{ES2}(\alpha_\perp)\cdot\frac2\pi\int_{-1}^1e^{rt}t^{\alpha_0}(1-t^2)^{m+1/2}\dd t/Z(r)$;
\item the same moments for the doubled field $2r$, arising from $\rho_r^2=g(r)\rho_{2r}$;
\item band integrals $\int_{H_k}\rho_r^p\,y^\beta\,\dd\sigma$, for $p=1,2$.
\end{itemize}
The fibre function of family 2 is $u(x;y)=\rho_{r_2}(x)\bigl[\frac12\rho_{r_3}(y)\tilde T(x,y)+\frac12\tau(x)\bigr]$, where $\tau(x)=\E_{\rho_{r_3}}\tilde T(x,\cdot)$. Its energy $2\sum_i\ip u{x_i}^2+\frac12(\|u\|^2-\ip u1^2)$ expands into contractions implemented in the file \texttt{remainder\_cert.py}. Family 3 is the same with $x$ and $y$ interchanged. All base integrals use the exact-integer engine (outward-rounded 220-bit integer intervals, explicit series tails).

\paragraph{Independent check of the formulas.} The retained quadrature comparison at five points $(r_2,r_3)$ records agreement of the exact-moment matrices with a completely independent quadrature evaluation of the definition (a separately implemented $S^3$ quadrature) to relative accuracy $\le4.3\times10^{-6}$. The resulting values agree to six digits (Table~\ref{cond:xix:tab:cross}).

\begin{table}[ht]\centering\small
\caption{Cross-check of the exact-moment evaluator of $C_2$ against quadrature.}\label{cond:xix:tab:cross}
\begin{tabular}{lccc}\toprule
$(r_2,r_3)$ & quadrature & exact moments & max.\ rel.\ matrix deviation\\\midrule
$(1.775,1.775)$ & 0.155672 & 0.155672 & $4.2\times10^{-6}$\\
$(1.0,1.5)$ & 0.129400 & 0.129400 & $5.6\times10^{-7}$\\
$(1.775,0.3)$ & 0.126154 & 0.126154 & $1.4\times10^{-6}$\\
$(0.5,0)$ & 0.104266 & 0.104266 & $3.4\times10^{-8}$\\
$(0,0)$ & 0.102062 & 0.102062 & $2.7\times10^{-8}$\\\bottomrule
\end{tabular}
\end{table}

\paragraph{Boxes: first-order Taylor models.} Plain interval evaluation over a box of $(r_2,r_3)$ fails badly. On $[0,0.22]^2$ the enclosure radius is five times the midpoint, because the centring and the energy subtract nearly equal quantities. We therefore hold every quantity as an affine form
\[
c_0+c_2\delta_2+c_3\delta_3+\text{rem},\qquad |\delta_i|\le h_i ,
\]
with exact-integer interval coefficients:
\begin{itemize}[nosep]
\item Base integrals are Taylor-expanded in $r$ with a second-derivative remainder. This uses $\partial_r\int e^{rt}t^j(\cdots)=\int e^{rt}t^{j+1}(\cdots)$.
\item Products and reciprocals send only second-order terms to the remainder.
\end{itemize}
Over a box, the weighted matrix $S(r)=\sum M_{jk}/c_{jk}$ then lies in $S_0+\delta_2S_2+\delta_3S_3+E$, with $\|E\|_2\le\|\mathrm{rem}\|_F$. Hence $G-\lambda S(r)\succ0$ on the box whenever
\[
G-\lambda S_0-\lambda\varepsilon I\succ0,\qquad \varepsilon=h_2\|S_2\|_F+h_3\|S_3\|_F+\|\mathrm{rem}\|_F .
\]
We check this condition exactly. By Weyl's inequality, it suffices that $\lambda_{\min}(\mathrm{mid})>\|\mathrm{rad}\|_F+\lambda\varepsilon$, which we prove by an exact rational $LDL^T$ factorisation of $\mathrm{mid}-\rho I$.

\paragraph{The certificate.} An adaptive quadtree over $[0,\frac{89}{50}]^2$ (64 roots, subdivision on failure) certifies $C_2=\frac15$ on all 385 boxes. The tiling is verified exactly, and every box was recertified from scratch by a separate driver (\texttt{verify\_C2.py --recheck}, all 385 boxes pass again, no failures). Unlike the first two-backend certificate, this part has a single exact-arithmetic backend. Its formulas are cross-checked only in floating point (Table~\ref{cond:xix:tab:cross}), and by an additional quadrature calculation at another point (agreement $\le10^{-13}$ relative). The final complete recalculation is specified below.

%=====================================================================
\subsection{The sharper slice certificate}\label{cond:xix:sec:slice}
On the slice $b_1=0$ the derivative density is exactly $a_\kappa p\,h^T(w-\mu)$, so $\cJ=A(\kappa)\,j(r_2,r_3;h)$ with $j$ independent of $\kappa$ (by homogeneity of the slice density). The final recalculation uses the exact-integer slice formula unchanged on the stored adaptive quadtree, at $\kappa=\frac{71}{200}$ on $[0,\frac{71}{40}]^2$ with target $0.0288$. All 4,273 boxes pass. A separate floating evaluation of the corner gives
approximately $0.028570$; the certificate uses the larger rigorous
upper bound $0.0288$. Therefore
\[
\bar j\ =\ \sup_{[0,1.775]^2,\ h}j\ \le\ \frac{0.0288}{A(0.355)}\ \le\ 0.326207,
\]
valid for every $\kappa\le0.355$. The tiling and every box are re-verified by an independent driver (\texttt{verify\_slice.py}).

For Corollary~\ref{cond:xix:cor:0352}, a separate certificate at $\kappa=0.352$ on $[0,1.76]^2$ with target $0.0283$ has 6,034 boxes. The corner boxes need side $R/4096$. The supplied historical verification records no failures on these 6,034
boxes. The full V5 driver checks this tiling; its fresh box recalculation
uses the stronger 4,273-box certificate needed for the final range.

%=====================================================================

\subsection{Coverage and the range certificate}
The certified slice square is $[0,71/40]^2$ and the quadratic-remainder
square is $[0,89/50]^2$. Both contain every outer-field pair for
$0\le\kappa\le177/500$. The exact rational endpoints in
Table~\ref{cond:xix:tab:range} form a contiguous cover of $[0,177/500]$.
On a coupling interval $[\kappa_-,\kappa_+]$, the radial certificate uses
$0\le r\le5\kappa_+$, split into 4,096 equal closed intervals, and interval
arithmetic in both $\kappa$ and $r$. It recomputes all moment constants on
that enlarged field domain. Every denominator and sign condition in the
polynomial-majorant proof is checked throughout each interval.
The linear coefficient is bounded using the fixed slice ratio
$\bar j=326207/10^6$, so there is no division by $A(0)$.
At $\kappa=0$ the derivative density is zero.

The files \texttt{radial\_00.json} through \texttt{radial\_13.json} record the
exact rational maximum enclosures, constants, and influence bounds.
Every one of the fourteen radial families was freshly recalculated,
including all 4,096 intervals per family. The exact rational results
agree with the supplied scientific records. In particular, for
$\kappa\in[707/2000,177/500]$ they give $\alpha<0.998807$. The logarithm is bounded with the convergent
positive series
\[
 -\log(1-x)=\sum_{n\ge1}\frac{x^n}{n},\qquad0<x<1,
\]
with a geometric majorant for its omitted tail. This also explains why a
nearest-rounded decimal influence bound must not be substituted into the
logarithm as though it were an upper bound.

\paragraph{Recalculation.}
The portable driver \path{code/conditioning/verify_all.py} checks exact
box tilings and arithmetic summaries by default. Its full option freshly
rebuilt all 385 quadratic-remainder boxes, all 4,273 strengthened slice
boxes, and all fourteen radial families: every check passed. The run used
four workers, took 909 seconds elapsed, and accumulated 3,584 seconds
of task wall time. Exact input hashes, per-task outputs and logs are
provided in \path{code/conditioning/fresh}. The default structural pass
does not replace the full calculation.

This is a reproduction with the same exact-arithmetic backend. The
independent numerical quadrature comparisons test formulas but do not
supply outward error certificates. A portable small-grid quadrature
check was additionally run at $(0,0)$ and $(1,1.5)$; its maximal relative
matrix discrepancies were $4.1\times10^{-5}$ and $5.8\times10^{-4}$,
consistent with its lower quadrature resolution. The original two-backend
Arb/integer certificate of Section~\ref{cond:functional} remains available
separately; the final complete recalculation reported here uses the
exact-integer backend.

\endgroup

\begingroup
\def\E{\mathbb E}\def\R{\mathbb R}\def\J{\mathcal J}\def\cJ{\mathcal J}
\def\bw{\beta_W}\def\kap{\kappa}\def\op{\mathrm{op}}\def\tr{\operatorname{tr}}
\def\Cov{\operatorname{Cov}}\def\Var{\operatorname{Var}}\def\vMF{\operatorname{vMF}}
\def\dd{\mathrm d}\def\Lip{\operatorname{Lip}}\def\ip#1#2{\langle #1,#2\rangle}
\def\norm#1{\left\|#1\right\|}
\def\epsfam{\varepsilon_{\rm fam}}
\section{Exact block sensitivity and rigidity}\label{condsens:section}
This section concerns transport on a finite product of compact spheres.
Its exact optimal plans and rigidity statements are independent of the
physical block comparison hypothesis. They identify both attainable
sensitivities and limitations of simple potentials used in sufficient
lattice criteria.
\paragraph{Notation.} Identify SU(2) with the unit quaternions $S^3\subset\R^4$, with $\ip ab=\mathrm{Re}(\bar ab)$, so that $\mathrm{Re}(a\bar b)=\ip ab$ and left/right multiplication and conjugation are isometries. More generally let $G\in\{\mathbb Z_2,\mathrm U(1),\mathrm{SU}(2)\}$ be realised as $S^{p-1}\subset\R^p$, $p\in\{1,2,4\}$ (unit reals, complex numbers, quaternions), with normalised Haar measure $\sigma$. For $\kappa\ge0$, $a\in S^{p-1}$, let $\vMF_\kappa(a)$ be the law $\propto e^{\kappa\ip wa}\sigma(\dd w)$, and for $m\in\R^p$ write $\vMF(m)=\vMF_{|m|}(m/|m|)$. Its mean is $A_p(\kappa)a$, with
\[
A_p(\kappa)=\frac{I_{p/2}(\kappa)}{I_{p/2-1}(\kappa)},\qquad A_1=\tanh,\quad A_2=\frac{I_1}{I_0},\quad A:=A_4=\frac{I_2}{I_1}.
\]
For $\vMF(m)$ on $S^{p-1}$, $p\ge2$, the second moment along any unit $h\perp m$ is $A_p(|m|)/|m|$, with the value $1/p$ at $m=0$.

\paragraph{The block and its sensitivity.} With $\kappa=\bw$ and $n\ge2$ (BB: $G=\mathrm{SU}(2)$, $n=3$),
\[
\mu_b(\dd y)\ \propto\ \exp\Bigl(\sum_{i=1}^n\ip{y_i}{b_i}+\kappa\,\mathrm{Re}(y_1y_2\cdots y_n)\Bigr)\,\sigma^{\otimes n}(\dd y),\qquad d_B(y,y')=\sum_{i=1}^n|y_i-y_i'| .
\]
For a field $b$ and a unit vector $h$, the \emph{sensitivity} and the block constant are
\begin{gather*}
K(b;h)=\sup_{f\ \text{1-Lip for }d_B}\ \frac{\dd}{\dd t}\mu_{b+t(h,0,\dots,0)}(f)\Big|_{t=0}=\sup_f\Cov_{\mu_b}\bigl(f,\ip{y_1}h\bigr),\\
K=\sup_{|b_i|\le5\bw}\ \sup_{|h|=1}K(b;h).
\end{gather*}
By the cyclic invariance of $\mathrm{Re}(y_1\cdots y_n)$ the same $K$ serves every slot. The sufficient condition in Assumption~\ref{cond:blockimport} is $\alpha=10\bw K<1$, written $40\beta K$ with $\beta=\bw/4$. By Kantorovich--Rubinstein duality,
\begin{equation}\label{condsens:eq:KR}
K(b;h)=W_1(\nu_b^+,\nu_b^-),\qquad \nu_b^\pm=\bigl(t-\E_bt\bigr)_\pm\,\mu_b,\quad t=\ip{y_1}h ,
\end{equation}
two positive measures of equal mass, with $W_1$ the (unnormalised) transport cost for $d_B$. We write $S_1(c;h)$ for the single-link quantity, i.e.\ \eqref{condsens:eq:KR} for the law $\vMF(c)$ on one copy of $S^3$. The imported sharp-quarter hypothesis is the statement $S_1(c;h)\le\frac14$.

\subsection{Two transport lemmas}
\begin{lemma}[Free-link reparametrisation]\label{condsens:lem:red}
Fix $j\in\{2,\dots,n\}$ and the links $y_i$, $i\notin\{1,j\}$. The map $y_j\mapsto w:=\overline{y_2\cdots y_n}$ is an isometry of $S^{p-1}$ that preserves $\sigma$, and $\mathrm{Re}(y_1\cdots y_n)=\ip{y_1}{w}$. Consequently, under $\sigma^{\otimes n}$ the variable $w$ is Haar-distributed and independent of $(y_i)_{i\ne j}$. Moving $w$ to $w'$ while keeping all $y_i$, $i\neq j$, fixed costs exactly $|w-w'|$ in $d_B$. Finally, $|w-w'|\le\sum_{i\ge2}|y_i-y_i'|$ for arbitrary moves.
\end{lemma}
\begin{proof}
$w=\overline{P\,y_j\,Q}$ with $P=y_2\cdots y_{j-1}$, $Q=y_{j+1}\cdots y_n$ fixed unit elements. Left and right multiplication by unit elements and conjugation are isometries preserving $\sigma$. Next, $\mathrm{Re}(y_1\bar w)=\ip{y_1}w$. The last claim is $|ab-a'b'|\le|a-a'|+|b-b'|$ for unit elements, iterated.
\end{proof}

\begin{lemma}[Equal-concentration vMF laws]\label{condsens:lem:vmf}
Let $p\ge1$, $\kappa\ge0$ and $a,a'\in S^{p-1}$. For the chordal metric,
\[
W_1\bigl(\vMF_\kappa(a),\vMF_\kappa(a')\bigr)=A_p(\kappa)\,|a-a'| .
\]
For $a\ne a'$ let $e=(a-a')/|a-a'|$ and $R_e w=w-2\ip we e$, so that $R_ea=a'$. An optimal plan is the following \emph{reflection-acceptance kernel}: from $w$, stay with probability $\min\{1,q'(w)/q(w)\}$, otherwise jump to $R_ew$. Here $q,q'$ are the densities of the two laws. The potential $\phi(w)=\ip we$ is optimal.
\end{lemma}
\begin{proof}
\emph{The kernel is a coupling.} $q(R_ew)\propto e^{\kappa\ip{w}{R_ea}}=e^{\kappa\ip w{a'}}$, so $q\circ R_e=q'$ and $q'\circ R_e=q$, and $R_e$ preserves $\sigma$. The image density at $u$ is
\[
q(u)\min\Bigl\{1,\tfrac{q'(u)}{q(u)}\Bigr\}+q(R_eu)\Bigl[1-\min\Bigl\{1,\tfrac{q'(R_eu)}{q(R_eu)}\Bigr\}\Bigr]=\min\{q,q'\}(u)+q'(u)-\min\{q',q\}(u)=q'(u).
\]
\emph{Cost and optimality.} A jump happens only when $q'(w)<q(w)$, i.e.\ $\ip w{a-a'}>0$, i.e.\ $\ip we>0$. Then $|w-R_ew|=2\ip we=\phi(w)-\phi(R_ew)$. Since $\phi$ is 1-Lipschitz, every transported pair is tight, so the plan and $\phi$ are both optimal. The value is $\int\phi\,\dd q-\int\phi\,\dd q'=A_p(\kappa)\ip{a-a'}e=A_p(\kappa)|a-a'|$. For $p=1$ the same computation holds with $R_ew=-w$.
\end{proof}

\begin{remark}
For general $m,m'$ (different concentrations), Assumption~\ref{cond:quarter} gives
\[W_1\bigl(\vMF(m),\vMF(m')\bigr)\le|m-m'|/4\]
for SU(2), as long as the segment $[m,m']$ stays in the assumed parameter domain. Lemma~\ref{condsens:lem:vmf} replaces the factor $\kappa/4$ by the exact $A(\kappa)<\kappa/4$ when the concentrations are equal. This gives an exact improvement when conditioning preserves the concentration.
\end{remark}

%=====================================================================
\begin{lemma}[Exact Wasserstein-dual norm]\label{condsens:lem:transport}
For every unit $h$, the signed measure $\nu=\langle w,h\rangle\,d\sigma^{\otimes2}$ on $(S^3)^2$
has Wasserstein-dual norm exactly $\tfrac14$ for the metric $|y_2-y_2'|+|y_3-y_3'|$:
\[
 \sup\Big\{\int G\,d\nu:\ G\ \text{$1$-Lipschitz}\Big\}=\frac14 .
\]
Write $R=R_h$ for reflection across $h^\perp$.
The supremum is attained by $G=\langle w,h\rangle$, and an optimal plan is the map
$T(y_2,y_3)=(y_2,\,y_3\bar v)$ with $v=(Rw)\bar w$.
\end{lemma}
\begin{proof}
Lower bound: $G=\langle w,h\rangle$ is $1$-Lipschitz (by the product-distance inequality) and
$\int G\,d\nu=\E\langle w,h\rangle^2=\tfrac14$, since $w$ is Haar.

Upper bound. Let $\nu^\pm=\langle w,h\rangle^\pm d\sigma^{\otimes2}$, two positive measures of
equal mass. (i) $T$ preserves $\sigma^{\otimes2}$: for fixed $y_2$ the map $y_3\mapsto w=\bar y_3\bar y_2$
is a Haar-preserving bijection, and in the variable $w$ the map $T$ reads
$w\mapsto\overline{y_2\,y_3\bar v}=v\,w=Rw$, an isometry. (ii) $\langle w\circ T,h\rangle=\langle Rw,h\rangle=-\langle w,h\rangle$,
so $T$ pushes $\nu^+$ forward to $\nu^-$. (iii) The displacement is
$|y_3-y_3\bar v|=|1-\bar v|=|v-1|=|vw-w|=|Rw-w|=2|\langle w,h\rangle|$, and $y_2$ does not move.
By Kantorovich--Rubinstein duality, for every $1$-Lipschitz $G$,
\[
 \int G\,d\nu=\int\big(G-G\circ T\big)\,d\nu^+\le\int d_B(y,Ty)\,d\nu^+
 =\int2\langle w,h\rangle^+\langle w,h\rangle\,d\sigma^{\otimes2}=2\int_{t>0}t^2\,d\sigma=\frac14 .
\]
Equality holds for $G=\langle w,h\rangle$, because $G(y)-G(Ty)=2\langle w,h\rangle=d_B(y,Ty)$ on
the support of $\nu^+$: the plan and the potential certify each other.
\end{proof}

\begin{corollary}\label{condsens:Jzero}
$J(0;h)=A(\kappa)/4$ for every $\kappa>0$ and unit $h$, with a proof that uses neither the banded
functional nor the harmonic fibre comparison.
\end{corollary}
\begin{proof}
At zero field $V_h=A(\kappa)\langle w,h\rangle$ on $(S^3)^2$, so $J(0;h)$ is $A(\kappa)$ times the
norm in Lemma~\ref{condsens:lem:transport}.
\end{proof}

\subsection{The exact zero-field sensitivity}
\begin{theorem}[Exact zero-field sensitivity]\label{condsens:thm:A}
Let $G=S^{p-1}$ with $p\in\{1,2,4\}$, $n\ge2$, $\kappa\ge0$ and $h$ a unit vector. Then
\[
K(0;h)=\frac{1+A_p(\kappa)}{p}.
\]
In particular, for BB's SU(2) block, $K(0)=\frac14\bigl(1+A(\bw)\bigr)$. The optimum is attained by the potential $F_0(y)=\ip{y_1}h+\ip{\overline{y_2\cdots y_n}}h$ and by the following \emph{reflection--Metropolis plan}. With $R=R_h$, $t=\ip{y_1}h>0$ and $s=\ip wh$: move $y_1\mapsto Ry_1$; conditionally, with probability $\theta=(1-e^{-2\kappa ts})\mathbf 1_{s>0}$ also move $w\mapsto Rw$, through the free link $y_n$, keeping $y_2,\dots,y_{n-1}$ fixed. The same statement holds for the two-vector O($p$) dimer $\propto e^{\kappa\ip xw}$ on $S^{p-1}\times S^{p-1}$, for every $p\ge1$, with metric $|x-x'|+|w-w'|$ and perturbation $\ip xh$.
\end{theorem}

\begin{proof}
By Lemma~\ref{condsens:lem:red} with $j=n$, change variables to $(y_1,y_2,\dots,y_{n-1},w)$. Under $\mu_0$, the links $y_2,\dots,y_{n-1}$ are Haar and independent of $(y_1,w)$, and $(y_1,w)$ has density $e^{\kappa\ip{y_1}w}/Z$. Integrating $w$ gives a constant, so $y_1\sim\sigma$, and $w\mid y_1\sim\vMF_\kappa(y_1)$. In particular $\E_0t=0$, $\E_0t^2=1/p$ and $\E_0[s\mid y_1]=A_p(\kappa)t$.

\emph{Lower bound.} $F_0$ is 1-Lipschitz for $d_B$ by Lemma~\ref{condsens:lem:red}, and
\[
\Cov_0(F_0,t)=\E t^2+\E[t\,\E(s\mid y_1)]=(1+A_p)\E t^2=(1+A_p)/p .
\]

\emph{Upper bound.} We glue two couplings. (a) The reflection $R=R_h$ preserves $\sigma$ and $t\circ R=-t$, so it pushes the $y_1$-marginal $t_+\sigma$ of $\nu^+$ onto the $y_1$-marginal $t_-\sigma$ of $\nu^-$, at cost $|y_1-Ry_1|=2t$. (b) Given the pair $(y_1,Ry_1)$, couple $\vMF_\kappa(y_1)$ with $\vMF_\kappa(Ry_1)$ by the kernel of Lemma~\ref{condsens:lem:vmf}. Since $y_1-Ry_1=2th$, here $e=h$ and $R_e=R$. The jump probability is $1-\min\{1,e^{\kappa\ip w{Ry_1-y_1}}\}=(1-e^{-2\kappa ts})\mathbf 1_{s>0}=\theta$. The expected cost is $A_p(\kappa)|y_1-Ry_1|=2A_pt$. (c) Keep $y_2,\dots,y_{n-1}$, and realise $w'$ through $y_n$ (Lemma~\ref{condsens:lem:red}).

Because $\nu^\pm=t_\pm(y_1)\,\sigma(\dd y_1)\,\vMF_\kappa(y_1)(\dd w)\otimes\sigma^{\otimes(n-2)}$, steps (a)--(c) push $\nu^+$ onto $\nu^-$. The total cost is
\[
\int t_+\,(2t+2A_pt)\,\dd\sigma=2(1+A_p)\,\E[t_+^2]=(1+A_p)\,\E t^2=\frac{1+A_p}{p}.
\]
This matches the lower bound. Equivalently, every transported pair satisfies $F_0(x)-F_0(x')=d_B(x,x')$: a jump of $w$ happens only when $s>0$, and then $F_0$ drops by $2t+2s$, which is the displacement. The O($p$) dimer is the case of no spectator links.
\end{proof}

\begin{corollary}[Exact zero-field reach of the criterion]\label{condsens:cor:reach}
At zero field, $\alpha(0)=10\bw K(0)=\frac52\bw(1+A(\bw))$. This is increasing, and it is $<1$ if and only if $\bw<\beta_W^\ast=0.36658958\ldots$, the root of $5\beta(1+A(\beta))=2$. At $\bw=0.35$, $K(0)\approx0.27176419$ and
$1/(10\bw)-K(0)\approx0.01395$.
\end{corollary}

\begin{corollary}[The exact loss of the conditioning route]\label{condsens:cor:loss}
The conditioning estimate proved above bounds $K\le\frac14+(1+\beta)J$ with $\beta=\bw/4$, and $J(0)=A/4$ (by Lemma~\ref{condsens:Jzero}). At zero field it therefore overestimates $K(0)$ by exactly
\[
\Bigl(\tfrac14+(1+\beta)\tfrac A4\Bigr)-\tfrac{1+A}4=\frac{\beta A}4=\frac{\bw A(\bw)}{16},
\]
a relative loss $\beta A/(1+A)=0.70074\%$ at $\bw=0.35$. For $n=3$ and the optimal $f=F_0$, the decomposition $\Cov(f,t)=\E[\Cov(f,t\mid y_{2},y_3)]+\Cov(\E[f\mid y_2,y_3],\E[t\mid y_2,y_3])$ evaluates to $\frac{1-A^2}4+\frac{A(1+A)}4$. The route bounds the first term by $\frac14$, losing $A^2/4$. It bounds the second by $(1+\beta)\frac A4$, losing $A(\beta-A)/4>0$. These sum to $\beta A/4$.
\end{corollary}
\begin{proof}
Given $(y_2,y_3)$, $y_1\sim\vMF_\kappa(w)$, so $\E[t\mid w]=As$ and $\E[\Var(t\mid w)]=\frac14-A^2\E s^2=\frac{1-A^2}4$. Also $\Cov(F_0,t\mid w)=\Var(t\mid w)$, and $\E[F_0\mid w]=(1+A)s$. Finally $A(\kappa)<\kappa/4$ for $\kappa>0$ (termwise Bessel comparison).
\end{proof}

\begin{corollary}[Other groups and lengths]\label{condsens:cor:groups}
For the U(1) block of any length $n\ge2$, $K(0)=\frac12\bigl(1+I_1(\kappa)/I_0(\kappa)\bigr)$. For $\mathbb Z_2$, $K(0)=1+\tanh\kappa$ (chordal metric, $|y-y'|\in\{0,2\}$). For the O(3) dimer, $K(0)=\frac13(1+\coth\kappa-1/\kappa)$.
\end{corollary}

%=====================================================================
\subsection{The faces of the field domain}
\begin{theorem}[Faces]\label{condsens:thm:B}
Let $G=\mathrm{SU}(2)$, $n\ge2$, and perturb slot 1. Suppose $b_j=0$ for some $j\in\{2,\dots,n\}$; the other fields $b_i$, $i\ne1,j$, are arbitrary. Write $r=|b_1|$. Then for every unit $h$:
\begin{enumerate}[label=(\roman*)]
\item $K(b;h)\le\bigl(1+A(\bw)\bigr)\,S_1(b_1;h)$;
\item if $h\perp b_1$, then $K(b;h)=\bigl(1+A(\bw)\bigr)\,\dfrac{A(r)}{r}$, with $A(r)/r:=\frac14$ at $r=0$;
\item if $b_1=0$, then $K(b;h)=K(0)=\frac14(1+A(\bw))$;
\item conditional on Assumption~\ref{cond:quarter} $S_1(c;h)\le\frac14$ for $|c|\le5\bw$: $K(b;h)\le K(0)$.
\end{enumerate}
Consequently, under that single-link assumption on the stated field domain,
\[
\mathcal F_1=\bigcup_{j\ge2}\{b_j=0\},\qquad
\sup_{b\in\mathcal F_1,\,h}K(b;h)=K(0).
\] The value is attained, unconditionally, on all of $\mathcal F_1\cap\{b_1=0\}$, a set of positive dimension. The inequality $\le$ is conditional only on the sharp quarter. For slot $i$, the faces are $\{b_j=0\}$ for some $j\ne i$.
\end{theorem}

\begin{proof}
Use Lemma~\ref{condsens:lem:red} with the free link $j$. In the variables $(y_1,w,(y_i)_{i\ne1,j})$ the density is
\[
\exp\Bigl(\ip{y_1}{b_1}+\kappa\ip{y_1}w+\sum_{i\neq1,j}\ip{y_i}{b_i}\Bigr),
\]
since $b_j=0$. Hence the spectators $(y_i)_{i\ne1,j}$ are independent of $(y_1,w)$. The $y_1$-marginal is $\vMF(b_1)$, because $\int e^{\kappa\ip{y_1}w}\dd w$ is constant, and $w\mid y_1\sim\vMF_\kappa(y_1)$. Since $t$ depends on $y_1$ only, $\nu_b^\pm=(t-\E t)_\pm\vMF(b_1)(\dd y_1)\,\vMF_\kappa(y_1)(\dd w)\otimes(\text{spectators})$.

(i) Let $\pi$ be an optimal coupling of the $y_1$-marginals for $S_1(b_1;h)$. Glue each pair $(y_1,y_1')$ with the kernel of Lemma~\ref{condsens:lem:vmf}, and keep the spectators fixed. The cost is at most $\int\bigl(|y_1-y_1'|+A(\kappa)|y_1-y_1'|\bigr)\dd\pi=(1+A(\kappa))S_1(b_1;h)$.

(ii) If $h\perp b_1$, then $R_hb_1=b_1$, so $\vMF(b_1)$ is $R_h$-invariant, $\E t=0$, and $R_h$ pushes $t_+\vMF(b_1)$ onto $t_-\vMF(b_1)$ at cost $2t$. Thus $S_1\le\E[t^2]=A(r)/r$, and $g=t$ shows equality. For the lower bound on $K$, $F_0$ gives $\Cov(F_0,t)=\E t^2+\E[t\,\E(s\mid y_1)]=(1+A(\kappa))A(r)/r$.

(iii) is (ii) at $r=0$, and (iv) follows from (i).
\end{proof}

%=====================================================================
\subsection{The interior slice and the bond sensitivity}\label{condsens:sec:interior}
The conditioning corner $b=(0,\frac74,\frac74)$ (fields along the identity, $h=e_1$) lies on the slice $b_1=0$ but off the faces.

\begin{proposition}[Slice reduction]\label{condsens:prop:slice}
Let $n=3$ and $b_1=0$. Then $(y_2,y_3)\sim P:=\vMF(b_2)\otimes\vMF(b_3)$ and $y_1\mid(y_2,y_3)\sim\vMF_\kappa(w)$. Define the \emph{bond sensitivity}
\[
N(b_2,b_3;h)=\sup\Bigl\{\Cov_P(G,s):\ G\ \text{1-Lip for }|y_2-y_2'|+|y_3-y_3'|\Bigr\},\qquad s=\ip{\overline{y_2y_3}}h .
\]
Then
\[
\Var(t)+A(\kappa)\,N\ \le\ K(0,b_2,b_3;h)\ \le\ \E\bigl[S_1(\kappa w;h)\bigr]+A(\kappa)\bigl(1+A(\kappa)\bigr)N,
\]
with $\Var(t)=\frac A\kappa+\bigl(1-\frac{4A}\kappa\bigr)\E s^2-A^2(\E s)^2$. Moreover:
\begin{enumerate}[label=(\alph*),nosep]
\item With $x=y_2$ and $z=\bar h\,\overline{y_3}$, $N$ is the sensitivity of two \emph{independent} spins $x\sim\vMF(b_2)$, $z\sim\vMF(\bar h\bar b_3)$ to a ferromagnetic bond $\ip xz$.
\item $N=\frac14$ on the faces $b_2b_3=0$, and $N\ge\Var(s)$ always.
\item If $N>N^\ast:=\frac14+\bigl(1-\frac{4A}\kappa\bigr)\bigl(\frac14-\E s^2\bigr)/A$ (when $\E s=0$), then $K(b)>K(0)$.
\end{enumerate}
\end{proposition}
\begin{proof}
Integrating $y_1$ gives the constant $Z(\kappa)$, which yields the product law $P$.

\emph{Lower bound.} $f=t+G$ is 1-Lipschitz for $d_B$, and $\Cov(t+G,t)=\Var t+\Cov(G,\E[t\mid y_2,y_3])=\Var t+A\Cov(G,s)$.

\emph{Upper bound.} Write $\Cov(f,t)=\E\Cov(f,t\mid y_2,y_3)+A\Cov(\tilde F,s)$ with $\tilde F=\E[f\mid y_2,y_3]$. By Lemma~\ref{condsens:lem:vmf} and Lemma~\ref{condsens:lem:red}, $|\tilde F(y)-\tilde F(y')|\le d+A|w-w'|\le(1+A)d$.

The formula for $\Var t$ uses $\E[t^2\mid w]=\frac A\kappa+(1-\frac{4A}\kappa)s^2$, the vMF second moments. For (a): $s=\mathrm{Re}(y_2y_3h)=\ip{y_2}{\bar h\bar y_3}$, and $y_3\mapsto\bar h\bar y_3$ is an isometry. For (b): if $b_3=0$, then $z$ is Haar and independent of $x$, and the map $z\mapsto R_xz$ (reflection across $x^\perp$) pushes $s_+P$ onto $s_-P$ at cost $2|s|$, which gives $N\le\E s^2=\frac14$; $G=s$ gives equality. The case $b_2=0$ follows by symmetry, and $N\ge\Var(s)$ comes from $G=s$. Part (c) is algebra.
\end{proof}

\begin{remark}
On this slice the conditioning route's Lipschitz factor $1+\beta=1+\bw/4$ can be replaced by $1+A(\bw)$, a small but free improvement. At the faces the upper bound exceeds $K(0)$ by the slack $\E[S_1(\kappa w;h)-\Var(t\mid w)]$. This is at most $A^2/4=1.9\times10^{-3}$ once $S_1\le\frac14$ is used, and numerically $0$ to $5\times10^{-4}$ depending on the direction. So the slice problem is not settled by $N\le\frac14$ alone.
\end{remark}

\paragraph{The J-corner in closed form.} At $b_2=b_3=r\cdot1$, $h=i$, we have $\E s=0$ by conjugation symmetry, and
\[
\E s^2=2m_0m_1+2m_1^2,\qquad m_1=\frac{A(r)}r,\quad m_0=1-3m_1 .
\]
At $r=\frac74$, $\kappa=0.35$ this gives $\E s^2=0.2471211$, $\Var t=0.249985$, and the $F_0$ value $\Var t+A\,\E s^2=0.2714990$. The counterexample threshold of (c) is $N^\ast=0.2501675$. In the clean form of (a) the two fields are orthogonal, $a=r$, $c=-ri$.

\subsection{Exact banded and unbanded conditioning values at zero field}
\begin{proposition}\label{condsens:bandzero}
For every $\kappa\ge0$ and unit $h$,
$J(0;h)=\mathcal J(0;h)=A(\kappa)/4$.
\end{proposition}
\begin{proof}
The exact dual value is Lemma~\ref{condsens:Jzero}. At zero field the
response density is $A\langle\overline{xy},h\rangle$ and its two marginals
vanish. In each fibre $u=(A/2)\langle x,q\rangle$ for a unit vector $q$.
Thus $\langle u,x_i\rangle=Aq_i/8$, and, using $\sum_iq_i^2=1$,
\[
 2\sum_i\langle u,x_i\rangle^2=A^2/32,\quad
 \|u\|_2^2=A^2/16,\quad \langle u,1\rangle=0.
\]
It follows that $\mathcal E(u)=A^2/16$, constant in the outer variable.
The bands have total mass one, and the two fibre families carry the outer
factor $1/2$. Therefore $\mathcal J=A/4$.
\end{proof}
\subsection{A rigidity theorem for the reflection potential}\label{condrig:sec:rigid}
The preceding transport construction shows that the explicit potential
\[
F_0=\ip{y_1}h+\ip{\overline{y_2y_3}}h
\]
is optimal at zero field, and on the faces $b_2b_3=0$ for $h\perp b_1$. There it gives $K=(1+A(\kappa))A(r)/r$. We now show that these are the \emph{only} configurations on the slice and faces where $F_0$ is optimal.

\paragraph{Two-spin form.} For independent $x\sim\vMF(a)$, $z\sim\vMF(c)$ on $S^3$, let
\[
N(a,c)=\sup_G\Cov(G,s),\qquad s=\ip xz ,
\]
with $G$ 1-Lipschitz for $|x-x'|+|z-z'|$. On the slice $b_1=0$ the bond quantity defined above is $N(b_2,\bar h\bar b_3)$.

\begin{lemma}[Tight pairs]\label{condrig:lem:tight}
If $s(u)-s(u')=|x-x'|+|z-z'|$ with $u=(x,z)\ne u'$, then $s(u)>0$ and $u'\in\{T_xu,T_zu\}$. Here $T_x(x,z)=(R_zx,z)$ and $T_z(x,z)=(x,R_xz)$, where $R_v$ is the reflection across $v^\perp$; both flips send $s\mapsto-s$.
\end{lemma}
\begin{proof}
Write $s(u)-s(u')=\ip{x-x'}z+\ip{x'}{z-z'}$. Each term is bounded by the corresponding distance, so both must be tight:
\[
x-x'=\lambda z,\quad z-z'=\mu x',\qquad \lambda,\mu\ge0 .
\]
\emph{Case $\lambda>0$.} Unit length forces $\lambda=2s(u)$, so $x'=R_zx$. If $\mu>0$, unit length would force
$\mu=2\ip z{x'}=-2s(u)<0$, a contradiction. Thus $\mu=0$, and $u'=T_xu$.

\emph{Case $\lambda=0$.} Then $\mu=2s(u)>0$, and $u'=T_zu$.
\end{proof}

\begin{theorem}[Rigidity]\label{condrig:thm:rigid}
Let $\kappa>0$.
\begin{enumerate}[label=(\roman*),nosep]
\item $s$ attains $N(a,c)$ if and only if $a=0$ or $c=0$.
\item On the slice $b_1=0$, $F_0$ attains $K(0,b_2,b_3;h)$ if and only if $b_2=0$ or $b_3=0$. Off these faces (on the slice), $K(0,b_2,b_3;h)>\Cov(F_0,\ip{y_1}h)$ strictly.
\item On the faces $b_2b_3=0$, $F_0$ attains $K(b;h)$ if and only if $h\perp b_1$ (including $b_1=0$).
\end{enumerate}
(At $\kappa=0$ the block decouples and (ii) fails: then $K=\frac14=\Cov(F_0,t)$ for every $b_2,b_3$.)
\end{theorem}

\begin{proof}
\emph{The ``if'' directions.} For (i) use the face case of the slice reduction: with $c=0$, reflect $z\mapsto R_xz$. For (ii) and (iii) use the exact zero-field and face theorems above.

\emph{Complementary slackness.} If a 1-Lipschitz potential $\phi$ attains $W_1(\nu^+,\nu^-)$, then every optimal plan (one exists by compactness) is concentrated on $\phi$-tight pairs, because $d-(\phi(u)-\phi(u'))\ge0$ and its integral vanishes. The measures $\nu^\pm$ are mutually singular, so diagonal pairs carry no mass.

(i) Let $\nu^\pm=(s-\E s)_\pm P$, where $s$ has a positive density on $(-1,1)$.
\begin{itemize}[nosep]
\item \emph{Mean condition.} By Lemma~\ref{condrig:lem:tight}, tight moves go from $\{s>0\}$ to the opposite value $-s$. Suppose $\E s>0$. Then $\nu^-$ has positive mass on $\{-\E s<s<\E s\}$, but a source of such a target would have $s\in(-\E s,\E s)$, outside $\operatorname{supp}\nu^+=\{s\ge\E s\}$. If $\E s<0$, the $\nu^+$-mass on $(\E s,-\E s)$ is stuck in the same way. So $\E s=0$.
\item \emph{Invariant sets.} Identify each 2-plane $\Pi$ with $\mathbb C$ through an oriented orthonormal basis. Write $a_\Pi,c_\Pi\in\mathbb C$ for the projections, $\zeta=\bar a_\Pi c_\Pi$ and $Q=|a_\Pi|^2+|c_\Pi|^2$. Both flips preserve $\Pi=\mathrm{span}(x,z)$ and $|s|$, and map the signed angle $\theta$ from $x$ to $z$ to $\pi-\theta$. Hence for every measurable set $B$ of pairs $(\Pi,|s|)$, the set $A_B=\{(x,z):(\mathrm{span}(x,z),|s|)\in B\}$ is invariant, and a tight plan forces $\nu^+(A_B)=\nu^-(A_B)$.
\item \emph{The imbalance.} Disintegrate the Haar measure over $(\Pi,\theta)$. The Jacobian is $\propto\sin^2\theta$, so the four circles with signed angles $\pm\theta$, $\pi\mp\theta$ carry equal weight. Rigid rotation of a pair inside $\Pi$ gives $\int e^{\ip xa+\ip zc}\dd\phi=2\pi I_0(|W|)$, where:
  \begin{itemize}[nosep]
  \item $|W|^2=Q+u_1$ on the circle $\theta$ and $Q+u_2$ on the circle $-\theta$, the two with $s=\cos\theta>0$;
  \item $|W|^2=Q-u_2$ on $\pi-\theta$ and $Q-u_1$ on $\pi+\theta$, the two with $s<0$;
  \item here $u_{1,2}=2\mathrm{Re}(\zeta e^{\mp i\theta})$.
  \end{itemize}
  Therefore $\nu^+(A_B)-\nu^-(A_B)=\int_B g$, with density
\[
g(\Pi,\theta)=c(\theta)\,\cos\theta\,\bigl[f(Q+u_1)-f(Q-u_1)+f(Q+u_2)-f(Q-u_2)\bigr],
\]
  where $c(\theta)=2\pi\sin^2\theta/(Z(|a|)Z(|c|))\cdot\mathrm{const}>0$ and $f(t)=I_0(\sqrt t)$ is strictly increasing.
\item \emph{Conclusion.} If $\mathrm{Re}\,\zeta=\ip{P_\Pi a}{P_\Pi c}\neq0$ and $\theta$ is small, then $u_1$ and $u_2$ both have the sign of $\mathrm{Re}\,\zeta$, and $g$ has that sign. For $a,c\ne0$ with $\ip ac=0$ (the only case left after the mean condition), the plane $\Pi=\mathrm{span}(\hat a+\hat c,e)$, $e\perp a,c$, gives $\ip{P_\Pi a}{P_\Pi c}=|a||c|/2>0$. By continuity this holds on an open set of planes, which $P$ charges. Taking $B$ there with $\theta$ small gives $\int_Bg\ne0$.
\end{itemize}

(ii) Here $t=\ip{y_1}h$. A pair is $F_0$-tight only if both summands are tight.
\begin{itemize}[nosep]
\item Pairs with $y_1$ fixed keep $t$, so they cannot join $\{t>\E t\}$ to $\{t<\E t\}$. Hence $y_1\mapsto R_hy_1$ with $t>0$.
\item $(y_2,y_3)$ either stays or makes one bond flip (Lemma~\ref{condrig:lem:tight} in the clean coordinates $x=y_2$, $z=\bar h\bar y_3$, where $\ip{\overline{y_2y_3}}h=\ip xz$ and $\ip{y_3}{b_3}=\ip z{\bar h\bar b_3}$).
\item The mean argument forces $\E t=0$.
\item The sets $S^3\times A_B$ are invariant under all tight moves, including the stays. Since $\E[t\mid y_2,y_3]=A(\kappa)s$ at $b_1=0$, their imbalance is $A(\kappa)\int_{A_B}(s-\E s)\,\dd P$. This is $A(\kappa)>0$ times the bond imbalance of (i), with $a=b_2$, $c=\bar h\bar b_3$.
\end{itemize}

(iii) On a face the $y_1$-marginal is $\vMF(b_1)$, so $\E t=A(|b_1|)\ip{\hat b_1}h\neq0$ unless $h\perp b_1$. The mean argument then applies, with the first point of (ii).
\end{proof}

\begin{remark}
\begin{itemize}[nosep]
\item By (i), the continuous-sphere excess $N-\Var(s)$ is strictly positive whenever $a,c\ne0$. Thus an exact bond optimum cannot generally be identified with the variance. This strict inequality alone does not decide whether the block sensitivity exceeds its zero-field value.
\item By (ii), on the slice off the faces, $F_0$ is not optimal, and no coupling of $\nu^+$ and $\nu^-$ built only from reflections of $y_1$, stays and bond reflections exists.
\item The case $b_1\neq0$, $b_2b_3\neq0$ is not covered.
\end{itemize}
\end{remark}

%=====================================================================
\subsection{A quantitative strict improvement over the reflection potential}
\subsubsection{Definitions and the precise claim}
For unit vectors $x,z\in S^3\subset\R^4$, put
\[
s(x,z)=x\cdot z,\qquad
d((x,z),(x',z'))=|x-x'|+|z-z'|.
\]
For a probability law $P$ of $(x,z)$, let
\[
N_P=\sup_{\Lip_d(G)\leq1}\Cov_P(G,s),\qquad V=\Var_P(s).
\]
The potential $s$ is admissible, so $N_P\geq V$. We seek a quantitative lower bound for $N_P-V$. For independent $x\sim\vMF(a)$ and $z\sim\vMF(c)$, this $N_P$ is the quantity $N(a,c)$ of the reflection-potential problem.

The conclusions below use only Euclidean geometry, elementary covariance identities and von Mises--Fisher moments. They do not use the imported lattice uniqueness or transfer-matrix implications for the bond potential. They do not give a continuum Yang--Mills mass gap.

\subsubsection{A general quantitative bound from an explicit potential}
Let $W=x\wedge z\in\Lambda^2\R^4$, with coordinates
\[
W_{ij}=x_i z_j-x_j z_i\quad (i<j),\qquad
\|W\|^2=\sum_{i<j}W_{ij}^2=1-s^2.
\]
The bivector $W$ is unchanged by either of the tight reflections for the bond potential:
\[
(x,z)\mapsto(x-2sz,z),\qquad
(x,z)\mapsto(x,z-2sx).
\]
This invariance motivates the construction, but admissibility follows directly from the next lemma.

\begin{lemma}[A Euclidean Lipschitz embedding]\label{condquant:lem:embedding}
The map
\[
\Phi(x,z)=(s(x,z),W(x,z))\in\R\oplus\Lambda^2\R^4
\]
is $1$-Lipschitz for the metric $d$.
\end{lemma}
\begin{proof}
For fixed $z$ and $p=x-x'$, the difference is $(p\cdot z,p\wedge z)$. Since $|z|=1$,
\[
(p\cdot z)^2+\|p\wedge z\|^2=|p|^2.
\]
Thus $\Phi$ is an isometry when only $x$ changes. The same identity applies when only $z$ changes. Inserting the intermediate point $(x',z)$ and applying the Euclidean triangle inequality proves
\[
\|\Phi(x,z)-\Phi(x',z')\|\leq |x-x'|+|z-z'|.
\]
\end{proof}

\begin{theorem}[Quantitative test-family bound]\label{condquant:thm:general}
Let $P$ be any probability law on $S^3\times S^3$, and let $f$ be a real $1$-Lipschitz function on the closed Euclidean unit ball of $\Lambda^2\R^4$. Set
\[
V=\Var_P(s),\qquad C=\Cov_P(f(W),s).
\]
Then
\begin{equation}\label{condquant:eq:bound}
\boxed{\quad N_P-V\ \geq\ \epsfam(V,C)
:=\sqrt{V^2+C^2}-V
=\frac{C^2}{\sqrt{V^2+C^2}+V}.\quad}
\end{equation}
The rationalized expression applies when its denominator is nonzero. If $V=C=0$, the bound is zero. If $V^2+C^2>0$, the explicit admissible potential
\begin{equation}\label{condquant:eq:potential}
G_*(x,z)=\frac{V\,s(x,z)+C\,f(W(x,z))}{\sqrt{V^2+C^2}}
\end{equation}
has covariance exactly $\sqrt{V^2+C^2}$ with $s$.
\end{theorem}
\begin{proof}
For each $\theta\in\R$, define
\[
G_\theta=\cos\theta\,s+\sin\theta\,f(W).
\]
For any two configurations, Cauchy--Schwarz, the Lipschitz property of $f$, and Lemma~\ref{condquant:lem:embedding} give
\begin{align*}
|G_\theta(x,z)-G_\theta(x',z')|
&\leq\sqrt{(s-s')^2+(f(W)-f(W'))^2}\\
&\leq\sqrt{(s-s')^2+\|W-W'\|^2}\\
&\leq d((x,z),(x',z')).
\end{align*}
Consequently every $G_\theta$ is admissible. Moreover,
\[
\Cov_P(G_\theta,s)=V\cos\theta+C\sin\theta.
\]
The maximum over $\theta$ is $\sqrt{V^2+C^2}$, attained by
\[
\cos\theta_*\!=\frac{V}{\sqrt{V^2+C^2}},\qquad
\sin\theta_*\!=\frac{C}{\sqrt{V^2+C^2}}.
\]
Taking the supremum over all admissible potentials proves the result.
\end{proof}

\begin{remark}[What is exact]
The maximum covariance and increment within this one-parameter family are exact:
\[
\max_\theta\Cov_P(G_\theta,s)=\sqrt{V^2+C^2},\qquad
\max_\theta\{\Cov_P(G_\theta,s)-V\}=\epsfam(V,C).
\]
There is no claim that this family contains a global maximizer for $N_P$. In particular, $N_P-V=\epsfam(V,C)$ is not established.
\end{remark}

\subsubsection{An explicit choice that detects orthogonal nonzero fields}
Use the fixed bivector coordinate order $(12,13,14,23,24,34)$ and set
\begin{equation}\label{condquant:eq:f}
f(W)=W_{13}W_{23}+W_{14}W_{24}.
\end{equation}
Its gradient is
\[
\nabla f=(0,W_{23},W_{24},W_{13},W_{14},0),
\]
so $\|\nabla f\|^2\leq\|W\|^2\leq1$ on the unit ball. The ball is convex, hence $f$ is $1$-Lipschitz there.

If $\mathsf W$ is the skew matrix with entries $x_i z_j-x_j z_i$, then
\[
\mathsf W\mathsf W^T=xx^T+zz^T-s(xz^T+zx^T).
\]
Its $(1,2)$ entry is exactly $f(W)$. Thus an equivalent formula is
\begin{equation}\label{condquant:eq:fexpanded}
f(W)=x_1x_2+z_1z_2-s(x_1z_2+x_2z_1).
\end{equation}

\subsubsection{The needed von Mises--Fisher moments}
Let normalized Haar measure on $S^3$ be $\sigma$. For $x\sim\vMF(r e_1)$,
\[
\frac{dP_r}{d\sigma}(x)=\frac{e^{r x_1}}{Z(r)},\qquad
Z(r)=\frac{2I_1(r)}r,
\]
with $Z(0)=1$. Define, for $r>0$,
\begin{equation}\label{condquant:eq:qtau}
A(r)=\frac{I_2(r)}{I_1(r)},\qquad
q(r)=\frac{A(r)}r,\qquad
\tau(r)=\frac{1-4q(r)}r=\frac{I_3(r)}{rI_1(r)}.
\end{equation}
The removable values are $A(0)=0$, $q(0)=1/4$, and $\tau(0)=0$.

Rotational symmetry about $e_1$ and differentiation of $Z$ give
\begin{align}
\E[x]&=A(r)e_1,\label{condquant:eq:firstmoment}\\
\E[xx^T]&=q(r)I+(1-4q(r))e_1e_1^T.\label{condquant:eq:secondmoment}
\end{align}
For clarity, the third moment needed below can also be derived without introducing a new integral. Since $\E[x_2^2]=q(r)$, exponential-family differentiation gives
\[
q'(r)=\E[x_1x_2^2]-A(r)q(r).
\]
Using $A'(r)=1-3A(r)/r-A(r)^2$ yields
\[
\E[x_1x_2^2]=q'(r)+A(r)q(r)
=\frac{1-4A(r)/r}{r}=\tau(r).
\]
The same value holds with $x_2$ replaced by $x_3$ or $x_4$. Multiplication of $\sum_i x_i^2=1$ by $x_1$ gives
\begin{equation}\label{condquant:eq:thirdmoment}
m_3(r):=\E[x_1^3]=A(r)-3\tau(r).
\end{equation}
The equality to the Bessel ratio in \eqref{condquant:eq:qtau} follows from $I_1(r)-I_3(r)=4I_2(r)/r$.

\begin{corollary}[Closed formula for orthogonal fields]\label{condquant:cor:orthogonal}
Let $x\sim\vMF(r e_1)$ and $z\sim\vMF(u e_2)$ be independent, with $r,u>0$. For the function \eqref{condquant:eq:f},
\begin{equation}\label{condquant:eq:VC}
V=q(r)+q(u)-4q(r)q(u),\qquad C=2\tau(r)\tau(u)>0.
\end{equation}
Consequently
\begin{equation}\label{condquant:eq:orthbound}
\boxed{\begin{aligned}
N(r e_1,u e_2)-\Var(x\cdot z)
&\geq \sqrt{V^2+4\tau(r)^2\tau(u)^2}-V>0,\\
V&=q(r)+q(u)-4q(r)q(u).
\end{aligned}}
\end{equation}
\end{corollary}
\begin{proof}
By independence and orthogonality of the mean directions, $\E[s]=0$. Put $q_r=q(r)$, $q_u=q(u)$. Taking the trace of the product of the second-moment matrices \eqref{condquant:eq:secondmoment} gives
\begin{align*}
V=\E[s^2]
&=4q_rq_u+q_r(1-4q_u)+(1-4q_r)q_u\\
&=q_r+q_u-4q_rq_u.
\end{align*}
Write $\tau_r=\tau(r)$, $\tau_u=\tau(u)$, and similarly for $A$ and $m_3$. Formula \eqref{condquant:eq:fexpanded} gives
\[
C=\E[sf(W)]
=\E[sx_1x_2]+\E[sz_1z_2]
-\E[s^2x_1z_2]-\E[s^2x_2z_1].
\]
The four terms are
\begin{align*}
\E[sx_1x_2]&=A_u\tau_r,\qquad
\E[sz_1z_2]=A_r\tau_u,\\
\E[s^2x_1z_2]&=m_{3,r}\tau_u+\tau_r m_{3,u}+2\tau_r\tau_u,\\
\E[s^2x_2z_1]&=2\tau_r\tau_u.
\end{align*}
For the third line, expand $s^2=\sum_{i,j}x_ix_jz_iz_j$. Only diagonal terms survive the coordinate-reflection symmetries. The two distinguished axes contribute $m_{3,r}\tau_u$ and $\tau_r m_{3,u}$; axes $3,4$ each contribute $\tau_r\tau_u$. In the final line, only $(i,j)=(1,2)$ and $(2,1)$ survive. Therefore
\begin{align*}
C&=(A_u-m_{3,u})\tau_r+(A_r-m_{3,r})\tau_u-4\tau_r\tau_u\\
&=3\tau_u\tau_r+3\tau_r\tau_u-4\tau_r\tau_u
=2\tau_r\tau_u.
\end{align*}
The positive-term power series for $I_3$ and $I_1$ gives $\tau(r),\tau(u)>0$. Apply Theorem~\ref{condquant:thm:general}.
\end{proof}

\subsubsection{Lifting the bound to the block sensitivity}
Consider the slice $b_1=0$, coupling $\kappa>0$, and unit $h$. Write
\[
t=\langle y_1,h\rangle,\qquad
x=y_2,\qquad z=\bar h\,\bar y_3,
\]
so that the bond variable is $s=x\cdot z$ and its reflection potential is $F_0=t+s$. The map from $y_3$ to $z$ is an orthogonal isometry. Integrating out $y_1$ produces a field-independent factor $Z(\kappa)$, so $x,z$ are independent von Mises--Fisher spins with the clean-coordinate fields $a=b_2$ and $c=\bar h\bar b_3$.

The conditional moments are
\begin{equation}\label{condquant:eq:conditional}
\E[t\mid x,z]=A(\kappa)s,\qquad
\E[t^2\mid x,z]=q(\kappa)+(1-4q(\kappa))s^2.
\end{equation}
Let $K$ denote the supremum of $\Cov(F,t)$ over $1$-Lipschitz $F$ for the product chordal sum metric on $(y_1,y_2,y_3)$.

\begin{corollary}[Quantitative block excess]\label{condquant:cor:block}
In the slice above, the potential $F_\theta=t+G_\theta(x,z)$ is admissible, and
\begin{equation}\label{condquant:eq:blockgap}
\boxed{\quad K-\Cov(F_0,t)\ \geq\ A(\kappa)
\bigl(\sqrt{V^2+C^2}-V\bigr).\quad}
\end{equation}
For orthogonal fields as in Corollary~\ref{condquant:cor:orthogonal}, the baseline is
\begin{equation}\label{condquant:eq:baseline}
\Cov(F_0,t)
=q(\kappa)+\bigl(1-4q(\kappa)+A(\kappa)\bigr)V.
\end{equation}
\end{corollary}
\begin{proof}
Admissibility follows from
\[
|F_\theta(y)-F_\theta(y')|
\leq |y_1-y_1'|+|y_2-y_2'|+|y_3-y_3'|.
\]
Conditioning in the covariance and using \eqref{condquant:eq:conditional} gives
\[
\Cov(F_\theta,t)=\Var(t)+A(\kappa)\Cov(G_\theta,s).
\]
At $\theta=0$ this is $\Cov(F_0,t)=\Var(t)+A(\kappa)V$; at $\theta_*$ it is $\Var(t)+A(\kappa)\sqrt{V^2+C^2}$. Taking the supremum defining $K$ proves \eqref{condquant:eq:blockgap}. For orthogonal fields, $\E s=\E t=0$, so \eqref{condquant:eq:conditional} gives $\Var(t)=q(\kappa)+(1-4q(\kappa))V$, which proves \eqref{condquant:eq:baseline}.
\end{proof}

\subsubsection{Worked calculation at an allowed point}
Choose
\[
\kappa=\frac{177}{500}=0.354,\qquad
r=u=\frac{177}{100}=1.77=5\kappa,
\]
with the two clean-coordinate field directions orthogonal. This is a specific allowed point for the bond potential's field domain. 

For equal fields, \eqref{condquant:eq:VC} reduces to
\[
V=2q(r)-4q(r)^2,\qquad C=2\tau(r)^2.
\]
The numerical evaluations of the exact formulas are
\begin{center}
\begin{tabular}{ll}
\toprule
Quantity & Value (rounded)\\
\midrule
$q(1.77)$ & $0.222654418686874111337$\\
$\tau(1.77)$ & $0.061797923871470934832$\\
$V$ & $0.247008876730588880795$\\
$C$ & $0.007637966789648234429$\\
$\sqrt{V^2+C^2}$ & $0.247126938475729558447$\\
$\epsfam=\sqrt{V^2+C^2}-V$ & $0.000118061745140677653$\\
$A(0.354)$ & $0.088041486571540577850$\\
$A(0.354)\epsfam$ & $0.000010394331549415618$\\
\bottomrule
\end{tabular}
\end{center}
Thus the explicit potential is, to the shown precision,
\[
G_*=0.999522262745336957481\,s
+0.030907058683116257640\,f(W).
\]
These rounded coefficients describe the witness; for a proof, the coefficients are the exact expressions in \eqref{condquant:eq:potential}, rather than independent rounded decimals.

The exact baseline formula \eqref{condquant:eq:baseline} evaluates to
\begin{align*}
\Var(t)&\approx0.249984503161749124880,\\
\Cov(F_0,t)&\approx0.271731531865476587741,\\
\Cov(t+G_*,t)&\approx0.271741926197026003359.
\end{align*}
The last number is the achieved covariance of an explicit admissible potential, and hence a lower bound for $K$. It is not an upper bound for $K$ and cannot by itself establish a Dobrushin uniqueness criterion.

The exact rational series evaluator in \texttt{code/conditioning/elementary\_certificates.py} encloses these expressions with outward endpoints. The displayed decimals are descriptive evaluations; the proof uses the exact coefficients and the interval comparisons recorded by that program.

\subsection{Finite-subgroup calculations and their interpretation}
The retained numerical experiments replace $SU(2)$ by the binary
polyhedral groups $2T,2O,2I$, with 24, 48 and 120 elements. For the product
chordal metric their finite transport problem is a linear program, but the
stored solutions use floating-point linear programming. At $\kappa=0.35$,
the reported $2T$ corner value is approximately $0.2729280$, compared with
$0.2717644$ at zero field. The larger subgroup calculations also report
positive corner excess. These values are numerical evidence for the finite
models, not rationally certified inequalities or a limiting theorem for
$SU(2)$.

Only three groups were sampled, and they do not form a controlled
approximation sequence with an available transport-error theorem. Fits of
the excess to different mesh variables give different intercepts.
The analytic rigidity theorem settles the strict failure of the simple
potential on the continuous sphere independently of those fits. It does
not settle the stronger inequality $\sup_bK(b)=K(0)$, because the improved
interior lower bound can remain below $K(0)$.

\subsection{A transverse total-variation estimate}
\begin{proposition}\label{condsens:mad}
If $y\sim\vMF(\kappa e)$ on $S^3$ and $u\perp e$ is unit, then
\[
 \mathbb E|u\cdot y|=\tfrac12\mathbb E\sqrt{1-s^2},\quad s=e\cdot y,
\]
is strictly decreasing for $\kappa>0$, and is at most $4/(3\pi)$.
At $\kappa=0$ the same value holds for every unit direction $u$.
\end{proposition}
\begin{proof}
Conditionally on $s$, the transverse direction is uniform on $S^2$.
Its projection onto $u$ is uniform on $[-1,1]$, giving the factor $1/2$.
The density of $s$ is proportional to $e^{\kappa s}\sqrt{1-s^2}$.
For the even decreasing function $g(s)=\sqrt{1-s^2}$, differentiation gives
$\partial_\kappa\mathbb E g=\Cov(g,s)$.
After folding to $[0,1]$, let $\pi$ have density proportional to
$\cosh(\kappa s)\sqrt{1-s^2}$ and $\rho$ density proportional to
$s\sinh(\kappa s)\sqrt{1-s^2}$.
Then $\Cov(g,s)=\mathbb E[s](\mathbb E_\rho g-\mathbb E_\pi g)$.
The likelihood ratio $d\rho/d\pi$ is proportional to
$s\tanh(\kappa s)$, strictly increasing. The elementary association
inequality, obtained by integrating
$(g(s)-g(t))(s\tanh(\kappa s)-t\tanh(\kappa t))\le0$
against $\pi(ds)\pi(dt)$, makes the covariance strictly negative.
At zero concentration, direct Haar integration yields $4/(3\pi)$.
\end{proof}
The corresponding bound for arbitrary nontransverse directions is not
proved by this argument; numerical scans of that stronger statement remain
numerical evidence.

\endgroup

\begingroup
\def\E{\mathbb E}\def\R{\mathbb R}\def\J{\mathcal J}\def\cJ{\mathcal J}
\def\bw{\beta_W}\def\kap{\kappa}\def\op{\mathrm{op}}\def\tr{\operatorname{tr}}
\def\Cov{\operatorname{Cov}}\def\Var{\operatorname{Var}}\def\vMF{\operatorname{vMF}}
\def\dd{\mathrm d}\def\Lip{\operatorname{Lip}}\def\ip#1#2{\langle #1,#2\rangle}
\def\norm#1{\left\|#1\right\|}
\section{Certified ceilings of the conditioning criterion}\label{cond:ceilings}
Let $K$ be the supremum of the exact block sensitivity, and define the two
sufficient outputs
\[
 \alpha_{\rm band}(\kappa)=10\kappa\left(\tfrac14+(1+\kappa/4)
          \sup_{|b_i|\le5\kappa,|h|=1}\mathcal J(b;h)\right),
\]
\[
 \alpha_{\rm dual}(\kappa)=10\kappa\left(\tfrac14+(1+\kappa/4)
          \sup_{|b_i|\le5\kappa,|h|=1}J(b;h)\right).
\]
The counting criterion itself is $10\kappa K<1$. These three quantities
have different ceilings. Failure of one sufficient criterion says nothing
by itself about closure of the physical transfer gap.
For consistency with the following formulas set $\beta=\kappa/4$ and
$\alpha_{\rm route}=\alpha_{\rm band}$. Equation~\eqref{condceil:eq:route}
below is this same definition.
\begin{equation}\label{condceil:eq:route}
 \alpha_{\mathrm{route}}(\beta_W)=40\beta\left(\tfrac14+(1+\beta)\sup\mathcal J\right).
\end{equation}
\begin{equation}\label{condceil:eq:chain}
 J\le\mathcal J,\qquad K\le\tfrac14+(1+\beta)\sup J,
 \qquad \alpha\le40\beta K.
\end{equation}
\subsection{Three lemmas}
\begin{lemma}[Exact scaling on the slice]\label{condceil:lem:scale}
Let $b_1=0$. Then $\rho$ is the product vMF density $\rho_\times(y_2,y_3)=e^{\langle y_2,b_2\rangle+\langle y_3,b_3\rangle}/(Z(|b_2|)Z(|b_3|))$, independent of $\kappa$, and
\[
 V_h=A(\kappa)\,\rho_\times\big\langle w-\E_\times w,\,h\big\rangle,\qquad
 \J(\kappa;0,b_2,b_3;h)=A(\kappa)\,\Phi(b_2,b_3;h),
\]
where $\Phi$ does not depend on $\kappa$.
\end{lemma}
\begin{proof}
For $b_1=0$ one has $|c|=|\kappa w|=\kappa$ for every $w\in S^3$, so the factor $Z(|c|)$ in
the marginal density is constant and $\rho=\rho_\times$ exactly. Also
$\psi=B(\kappa)\kappa w=A(\kappa)w$ and $m=A(\kappa)\E_\times w$. The bands, the fibre maps
and $\mathcal E$ do not involve $\kappa$, and $\J$ is positively homogeneous of degree one in
$V_h$ (by the seminorm lemma above).
\end{proof}

\begin{lemma}[Monotonicity]\label{condceil:lem:mono}
$A$ is strictly increasing on $[0,\infty)$.
\end{lemma}
\begin{proof}
$A(r)=\frac{d}{dr}\log Z(r)$ and $A'(r)=\operatorname{Var}_{\mathrm{vMF}(r)}(x_0)>0$.
\end{proof}

\begin{lemma}[The corner value is pinned]\label{condceil:lem:corner}
Let $\J_c:=\J(\tfrac7{20};\,0,\tfrac74e_0,\tfrac74e_0;\,e_1)$. Then
\[
 0.0279961\;\le\;\J_c\;\le\;\sup_{|h|=1}\J(\tfrac7{20};0,\tfrac74e_0,\tfrac74e_0;h)\;<\;0.027997 .
\]
\end{lemma}
\begin{proof}[Computer-assisted proof]
The upper bound is supplied by the exact-integer corner certificate. For the lower bound, recall that
$\J(b;h)=\tfrac12\sum_{j=2,3}\sum_{k=1}^4\big(h^{\mathsf T}M_{jk}h\big)^{1/2}$ exactly, where
$M_{jk}=\sigma_k\int_{H_k}Q_j\,d\sigma$ are the eight band matrices.
The supplied \path{slicecert.py} routine \path{slice_matrices}
returns outward-rounded
enclosures of these matrices; called with the zero-width intervals $r_2=r_3=\tfrac74$ it
encloses the matrices at the corner point. Taking the lower end of each enclosed quadratic
form $h^{\mathsf T}M_{jk}h$ for $h=e_1$, the lower end of each square root, and half the sum,
gives the rational number
$\J_c\ge0.02799611967\ldots$ (\texttt{route\_ceiling.py}; exact value recorded in
\texttt{route\_ceiling\_results.json}). The engine uses only exact integers on a $2^{-220}$
grid with floor/ceiling rounding, so every step is an enclosure of the true value.
\end{proof}
This pins the specified corner value; it does not assert that the corner is the global maximizer.

\subsection{The termination theorem}
\begin{theorem}[The route terminates below $\beta_W=0.3559$]\label{condceil:thm:term}
For every $\beta_W\ge\tfrac7{20}$,
\begin{equation}\label{condceil:eq:lower}
 \alpha_{\mathrm{route}}(\beta_W)\;\ge\;10\beta+40\beta(1+\beta)\,\frac{A(\beta_W)}{A(7/20)}\,\J_c,
 \qquad \beta=\beta_W/4,
\end{equation}
and the right side is increasing in $\beta_W$. Consequently
\[
 \alpha_{\mathrm{route}}(\beta_W)\ \ge\ 1.00006>1\quad\text{for all }\beta_W\ge0.3559,
\]
so this banded conditioning route cannot certify $\alpha<1$, hence cannot certify
a transfer-matrix gap, at any coupling $\beta_W\ge0.3559$, however finely the slice,
tube or radial certificates are refined. Moreover $\alpha_{\mathrm{route}}(\beta_W)\ge10\beta=\tfrac52\beta_W$
for every $\beta_W$, so no evaluation of the functional, however sharp, lets this route
certify $\beta_W\ge0.4$.
\end{theorem}
\begin{proof}
For $\beta_W\ge\tfrac7{20}$ the field domain has $R=5\beta_W\ge\tfrac74$, so the corner
configuration $(0,\tfrac74e_0,\tfrac74e_0)$ is admissible and
$\sup\J\ge\J(\beta_W;0,\tfrac74e_0,\tfrac74e_0;e_1)$. By Lemma~\ref{condceil:lem:scale} this equals
$\frac{A(\beta_W)}{A(7/20)}\J_c$. Insert into~\eqref{condceil:eq:route} to get~\eqref{condceil:eq:lower}.
The right side is increasing because $\beta$, $\beta(1+\beta)$ and (Lemma~\ref{condceil:lem:mono})
$A(\beta_W)$ are increasing. At $\beta_W=0.3559$ the exact-integer engine encloses
$A(0.3559)$ and $A(7/20)$; using the lower end of the former, the upper end of the latter and
the lower bound of Lemma~\ref{condceil:lem:corner}, the right side of~\eqref{condceil:eq:lower} is at least
$1.0000636$ (\texttt{route\_ceiling\_results.json}). The last claim is~\eqref{condceil:eq:route}
with $\sup\J\ge0$.
\end{proof}

\begin{remark}
The theorem is about the inequality chain~\eqref{condceil:eq:chain}, which is the stated banded route. The true Dobrushin coefficient $K$ may be smaller than
$\tfrac14+(1+\beta)J$, so a different method could in principle certify somewhat larger
couplings. The next theorem bounds the reach of the exact-dual estimate and of the counting criterion itself.
\end{remark}

\begin{theorem}[Ceilings beyond the banded relaxation]\label{condceil:exact}
The exact conditioning route cannot prove its sufficient output is below
one for $\kappa\ge3641/10000$. The counting criterion using the exact
block sensitivity cannot hold for $\kappa\ge36659/100000$.
More precisely its zero-field threshold is the unique positive root
$\kappa_*$ of
\[
 \frac52\kappa_*(1+A(\kappa_*))=1.
\]
\end{theorem}
\begin{proof}
Zero field is admissible. By Lemma~\ref{condsens:Jzero},
\[
 \alpha_{\rm dual}(\kappa)\ge\frac52\kappa
              (1+(1+\kappa/4)A(\kappa)).
\]
The right side is increasing, since $A'(\kappa)$ is a positive variance.
At $3641/10000$ the positive Bessel series with rational geometric tails
places it strictly above one. The exact zero-field theorem gives
\[
 10\kappa K\ge\frac52\kappa(1+A(\kappa)).
\]
This expression is continuous, strictly increasing from zero, and exceeds
one at $36659/100000$. The same rational series bounds give an isolating
interval for its unique root. Both checks are reproduced in
\texttt{elementary\_certificates.py}; no numerical optimizer is used.
\end{proof}
The upper cutoff $0.3559$ is a proved ceiling for the banded method, not
its exact optimal endpoint. The value $\kappa_*$ is an exact implicit
threshold at zero field, not an exact physical mass and not a proof that
all nonzero fields obey the same sensitivity bound.

\endgroup

\clearpage
\part{Hamiltonian dynamics and spatial refinement}
\section{Kinetic temporal limit with complete representation control}
\label{ham:sec:weakkinetic}

This section constructs the time-continuum limit of the normalized SU(2)
Wilson kinetic operator, including an explicit operator-norm error bound.
There is no representation truncation in the result. The number of spatial
links remains finite. The final proposition shows precisely why this kinetic
limit does not directly approximate the full isotropic Wilson transfer
operator at large coupling.

\subsection{Normalization and a coefficient estimate valid at every coupling}

Write an SU(2) element as $U=x\mathbf1+i\sqrt{1-x^2}\,\widehat n\cdot\sigma$,
where $x=\tfrac12\operatorname{Tr}U\in[-1,1]$.
Normalized Haar measure has $x$-marginal
$\frac2\pi\sqrt{1-x^2}\,dx$.
For $\beta>0$, define the convolution operator on $L^2(\mathrm{SU}(2))$
by the normalized, central density
\[
 q_\beta(U)=\frac{e^{\beta x(U)}}{Z_\beta},\qquad
 Z_\beta=\int e^{\beta x(U)}\,dU=\frac{2I_1(\beta)}\beta.
\]
For the representation $j=(n-1)/2$, $n=1,2,\ldots$, Peter--Weyl
diagonalization gives
\begin{equation}\label{ham:eq:wk:rho}
 \rho_n(\beta)=\frac{I_n(\beta)}{I_1(\beta)}
 =\mathbb E_{\nu_\beta}P_n(x),\qquad
 P_n(x)=\frac{U_{n-1}(x)}n,
\end{equation}
where $U_m$ is the Chebyshev polynomial of the second kind and
$d\nu_\beta\propto e^{\beta x}\sqrt{1-x^2}\,dx$.
Indeed, $\chi_j(U)=U_{n-1}(x)$, and integration of
$\sin\theta\sin(n\theta)$ gives
$\int e^{\beta x}\chi_j(U)\,dU=2nI_n(\beta)/\beta$.
The Bessel and Chebyshev identities used here are the classical integral,
recurrence, and generating-function formulas recorded in the NIST DLMF,
\S\S10.25, 10.29, 10.32, 18.5, and 18.12. The estimates below are proved
explicitly rather than inferred from an asymptotic expansion.

\begin{lemma}[Exact coefficient certificate]\label{ham:lem:wk:coefficient}
For every integer $n\ge1$ and every $\beta>0$, put
\[
 c_n=\frac{n^2-1}{2},\qquad
 d_n=\frac{(n^2-1)(n^2+2)}8,\qquad
 e_n=c_n^2+\frac34c_n.
\]
Then
\begin{align}
 \left|\rho_n(\beta)-1+\frac{c_n}{\beta}\right|
 &\le\frac{d_n}{\beta^2},\label{ham:eq:wk:coefficient}\\
 \left|\rho_n(\beta)-e^{-c_n/\beta}\right|
 &\le\frac{e_n}{\beta^2}.
 \label{ham:eq:wk:coefficient-heat}
\end{align}
Moreover, $1=\rho_1(\beta)>\rho_2(\beta)>\cdots>0$.
\end{lemma}

\begin{proof}
Set $z=1-x$. Under $\nu_\beta$, its density is proportional to
\[
 e^{-\beta z}z^{1/2}\sqrt{1-z/2}\,\mathbf1_{[0,2]}(z).
\]
This is a Gamma distribution of shape $3/2$ and rate $\beta$, reweighted
by the nonnegative decreasing function
$w(z)=\sqrt{1-z/2}\mathbf1_{[0,2]}(z)$.
If $f$ is increasing and $Z,Z'$ are independent Gamma variables, then
\[
 2\operatorname{Cov}(f(Z),w(Z))
 =\mathbb E[(f(Z)-f(Z'))(w(Z)-w(Z'))]\le0.
\]
Consequently the reweighted law is stochastically dominated by the
Gamma law. In particular, with $m_r=\mathbb E_{\nu_\beta}z^r$,
\begin{equation}\label{ham:eq:wk:moments}
 m_1\le\frac3{2\beta},\qquad m_2\le\frac{15}{4\beta^2}.
\end{equation}
Integration by parts of $e^{\beta x}(1-x^2)^{3/2}$ on $[-1,1]$
has zero boundary term and gives
$\beta\mathbb E(1-x^2)=3\mathbb E x$, hence
\[
 (2\beta+3)m_1=3+\beta m_2.
\]
Combining this identity with \eqref{ham:eq:wk:moments} yields
\begin{equation}\label{ham:eq:wk:first-moment}
 0\le\frac3{2\beta}-m_1
 \le\frac9{2\beta(2\beta+3)}\le\frac9{4\beta^2}.
\end{equation}

For completeness, the polynomial estimate needed for the Taylor remainder
holds on the entire interval, including $x=-1$:
\begin{equation}\label{ham:eq:wk:polynomial}
 P_n'(1)=\frac{n^2-1}{3},\qquad
 \sup_{[-1,1]}|P_n''|=\frac{(n^2-1)(n^2-4)}{15}
 \quad(n\ge2).
\end{equation}
To verify the supremum, use
\[
 U_m'(x)=2\sum_{r=0}^{\lfloor(m-1)/2\rfloor}
 (m-2r)U_{m-1-2r}(x).
\]
This follows by differentiating
$\sum_{m\ge0}U_m(x)s^m=(1-2xs+s^2)^{-1}$ and matching coefficients.
All the coefficients in this derivative expansion are nonnegative, and
$|U_\ell(x)|\le\ell+1=U_\ell(1)$, as follows from the SU(2) character
sum or the elementary trigonometric formula. Iterating the derivative
expansion proves $|U_m''(x)|\le U_m''(1)$.
The endpoint values in \eqref{ham:eq:wk:polynomial} follow from the same
generating function, or from
$(1-x^2)U_m''-3xU_m'+m(m+2)U_m=0$ and its derivative at $x=1$.
For $n=1$ both derivatives vanish; for $n=2$ the second derivative vanishes.

Taylor's theorem with the global second-derivative bound gives
\[
 \left|\rho_n-1+\frac{n^2-1}{3}m_1\right|
 \le\frac{(n^2-1)(n^2-4)}{30}m_2\quad(n\ge2).
\]
Equations \eqref{ham:eq:wk:moments} and \eqref{ham:eq:wk:first-moment} now imply
\[
 \left|\rho_n-1+\frac{n^2-1}{2\beta}\right|
 \le\frac{3(n^2-1)}{4\beta^2}
 +\frac{(n^2-1)(n^2-4)}{8\beta^2}
 =\frac{d_n}{\beta^2}.
\]
The case $n=1$ is exact. The inequality
$|e^{-y}-1+y|\le y^2/2$ for $y\ge0$ proves the second assertion because
\[
 d_n+\frac{c_n^2}{2}
 =\frac{(n^2-1)(2n^2+1)}8
 =c_n^2+\frac34c_n=e_n.
\]

Finally, the positive integral representation of $I_\nu$ and
$I_\nu'=I_{\nu+1}+\nu I_\nu/\beta$ give
\[
 \frac{I_{\nu+1}(\beta)}{I_\nu(\beta)}
 =\frac{\int_{-1}^1 u\,e^{\beta u}(1-u^2)^{\nu-1/2}\,du}
 {\int_{-1}^1e^{\beta u}(1-u^2)^{\nu-1/2}\,du}
 \quad(\nu\ge1).
\]
The quotient is strictly less than one since $u<1$ almost everywhere,
and strictly positive by pairing $u$ and $-u$ in the numerator.
Thus $0<I_{\nu+1}/I_\nu<1$, proving the claimed strict ordering.
\end{proof}

\subsection{Uniform operator control over the complete representation space}

Define $G$ spectrally on $L^2(\mathrm{SU}(2))$ by the eigenvalue
$c_n$ on every matrix coefficient of the spin $(n-1)/2$ representation.
Thus $G=2J^2$ in the convention $J^2=j(j+1)$.

\begin{theorem}[All-representation time-continuum limit]\label{ham:thm:wk:norm}
If $\beta\ge9$ and $\beta t\ge4$, then
\begin{equation}\label{ham:eq:wk:norm}
 \left\|C_\beta^{\lfloor\beta t\rfloor}-e^{-tG}\right\|
 \le \min\left\{1,\frac{19/(2t)+9/8}{\beta}\right\}.
\end{equation}
In particular, convergence is in operator norm uniformly for $t\ge t_0>0$
as $\beta\to\infty$. It is strong and uniform for $t$ in compact
subsets of $[0,\infty)$ when applied to each fixed vector.
\end{theorem}

\begin{proof}
Let $m=\lfloor\beta t\rfloor$ and $N=\lceil\sqrt\beta\rceil$.
For $\beta\ge9$,
\[
 N^2+2\le\beta+2\sqrt\beta+3\le2\beta.
\]
For $2\le n\le N$, Lemma~\ref{ham:lem:wk:coefficient} therefore gives
\[
 \rho_n\le1-\frac{c_n}{\beta}+\frac{d_n}{\beta^2}
 \le1-\frac{c_n}{2\beta}\le e^{-c_n/(2\beta)},
\]
where $2d_n/c_n=(n^2+2)/2\le\beta$.
The same exponential upper bound holds for $e^{-c_n/\beta}$.
The telescoping formula for the difference of two powers gives
\begin{align*}
 |\rho_n^m-e^{-mc_n/\beta}|
 &\le \frac{m e_n}{\beta^2}
     e^{-(m-1)c_n/(2\beta)}\\
 &\le\frac{t}{\beta}\left(c_n^2+\frac34c_n\right)e^{-tc_n/4}.
\end{align*}
Here $m\le\beta t$ and $m-1\ge\beta t-2\ge\beta t/2$.
The difference due to rounding the time satisfies
\[
 |e^{-mc_n/\beta}-e^{-tc_n}|
 \le\frac{c_n}{\beta}e^{-mc_n/\beta}
 \le\frac{c_n}{\beta}e^{-3tc_n/4},
\]
since $m/\beta\ge t-1/\beta\ge3t/4$.
Taking a supremum over all real $c\ge0$ bounds the sum by
\[
 \frac1\beta\left(\frac{64}{\mathrm e^2t}
             +\frac3{\mathrm e}+\frac4{3\mathrm e t}\right)
 <\frac1\beta\left(\frac{19}{2t}+\frac98\right).
\]
Indeed, $\mathrm e>1+1+1/2+1/6=8/3$, so the three terms are
bounded respectively by $9/t$, $9/8$, and $1/(2t)$.
The trivial representation has zero error.

For the entire omitted set $n>N$, strict monotonicity of the coefficients
and $c_N\ge(\beta-1)/2\ge4\beta/9$ imply
\[
 0<\rho_n^m\le\rho_N^m\le e^{-mc_N/(2\beta)}
 \le e^{-\beta t/6}.
\]
The heat eigenvalues also satisfy $0<e^{-tc_n}\le e^{-\beta t/6}$.
Because both numbers are nonnegative, their absolute difference is at most
the larger, so the complete tail is bounded by
\[
 e^{-\beta t/6}\le\frac6{\mathrm e\beta t}
 <\frac9{4\beta t}.
\]
This is smaller than the stated bound for the low representations.
Peter--Weyl diagonalization now proves \eqref{ham:eq:wk:norm} for the full
operator. Both operators are positive selfadjoint contractions, so their
norm difference is also at most one.

For strong convergence at the time origin, first take a vector supported
on finitely many representations. Ordinary telescoping without damping,
\eqref{ham:eq:wk:coefficient-heat}, and the rounding bound show uniform
convergence on $[0,T]$ on this finite-dimensional subspace. For an arbitrary
vector, approximate it by such a finite sum; the error on the remaining
vector is at most twice its norm because both operators are contractions.
This proves the final assertion. Uniform operator-norm convergence down to
$t=0$ is neither asserted nor true: if $0<t<1/\beta$, the discrete operator
is $I$ and $\|I-e^{-tG}\|=1$.
\end{proof}

\begin{corollary}[A fixed spatial lattice and physical time]\label{ham:cor:wk:product}
On a finite graph with $E$ links, let $\mathcal C_\beta=C_\beta^{\otimes E}$
and $G_E=\sum_eG_e$. Then
\[
 \left\|\mathcal C_\beta^{\lfloor\beta t\rfloor}-e^{-tG_E}\right\|
 \le\min\left\{1,\frac E\beta\left(\frac{19}{2t}+\frac98\right)\right\}
\]
under the same hypotheses. The bound remains valid on the physical,
locally gauge-invariant Hilbert space.
In the anisotropic Wilson normalization, set
$b=4a/g^2>0$, $\beta_t=b/\epsilon$, and let $s>0$ denote physical time.
If $\epsilon\le\min\{b/9,s/4\}$, then
\begin{equation}\label{ham:eq:wk:physical-time}
 \left\|\mathcal C_{b/\epsilon}^{\lfloor s/\epsilon\rfloor}
          -e^{-sH_0}\right\|_{\rm phys}
 \le\min\left\{1,E\epsilon\left(\frac{19}{2s}+\frac9{8b}\right)\right\},
 \qquad H_0=\frac{g^2}{2a}\sum_eJ_e^2.
\end{equation}
\end{corollary}

\begin{proof}
Tensor telescoping bounds the norm difference of $E$ products of
contractions by the sum of their $E$ norm differences.
Central convolution commutes with both left and right translation on each
link. Thus it commutes with the local gauge action and preserves the
physical Hilbert space; restriction cannot increase the operator norm.
Apply Theorem~\ref{ham:thm:wk:norm} with its dimensionless time $t=s/b$.
Then $G_E/b=(g^2/(2a))\sum_eJ_e^2$.
\end{proof}

\subsection{A complete heat-kernel tail certificate}

\begin{lemma}[Globally positive kernel]\label{ham:lem:wk:heatkernel}
The kernel of $e^{-\tau G}$ is
\[
 k_\tau(U)=\sum_{n\ge1}n\chi_{(n-1)/2}(U)
                    e^{-\tau(n^2-1)/2}.
\]
For every $\tau\ge2$ and every $U\in\mathrm{SU}(2)$,
\begin{equation}\label{ham:eq:wk:kernelbounds}
 \frac34<k_\tau(U)<\frac54.
\end{equation}
All representations are included in this estimate.
\end{lemma}

\begin{proof}
The character bound $|\chi_{(n-1)/2}|\le n$ gives absolute uniform
convergence and
\[
 |k_\tau(U)-1|\le\sum_{n\ge2}n^2e^{-(n^2-1)}.
\]
The ratio of successive summands is at most
$(9/4)e^{-5}<9/512$, and the first summand is $4e^{-3}<1/5$.
The strict inequalities $e^3>20$ and $e^5>128$ follow already from the
positive Taylor sums through degree eight. Therefore the entire tail is
\[
 |k_\tau(U)-1|<\frac{1/5}{1-9/512}=\frac{512}{2515}<\frac14.
\]
The accompanying exact-rational certificate verifies these finite
inequalities without numerical evaluation of any Bessel function.
\end{proof}

\subsection{Why the isotropic interacting operator needs separate control}

For a finite spatial lattice, write
\[
 W(U)=\sum_{p\ \mathrm{spatial}}\left(1-\tfrac12\operatorname{Tr}U_p\right).
\]
Assume the graph has a plaquette configuration with $W>0$; equivalently,
this continuous, gauge-invariant $W$ is nonconstant.
This holds for the usual periodic cubic spatial lattices with side at
least two. In temporal gauge the isotropic transfer operator, up to a
scalar that disappears upon vacuum normalization, is
\[
 T_\beta=M_\beta\mathcal C_\beta M_\beta,
 \qquad M_\beta=e^{-\beta W/2}.
\]
All operators below are restricted to the locally gauge-invariant Hilbert
space. Let $\lambda_{0,\beta}=\|T_\beta\|>0$ and
$A_\beta=T_\beta/\lambda_{0,\beta}$.

\begin{proposition}[Failure of direct isotropic kinetic replacement]
\label{ham:prop:wk:isotropic}
On this fixed physical Hilbert space,
\begin{equation}\label{ham:eq:wk:isotropic}
 \lim_{\beta\to\infty}\|A_\beta-\mathcal C_\beta\|_{\rm phys}=1.
\end{equation}
This concerns the complete physical transfer operator, with its actual
vacuum eigenvalue divided out. It is not a statement about a Monte Carlo
transition operator.
\end{proposition}

\begin{proof}
The coefficient estimate and finite Peter--Weyl approximation give
$\mathcal C_\beta\to I$ strongly as $\beta\to\infty$.
All its nonconstant eigenvalues are positive and the normalized constant
function $\mathbf1$ has eigenvalue one. Therefore
$\mathcal C_\beta\ge|\mathbf1\rangle\langle\mathbf1|$, and
\[
 T_\beta\ge|M_\beta\mathbf1\rangle\langle M_\beta\mathbf1|,
 \qquad
 \lambda_{0,\beta}\ge\int e^{-\beta W(U)}\,dU.
\]
Choose $w_0>0$ such that the gauge-invariant set $\{W\ge w_0\}$ has
positive Haar measure, and let $f$ be its normalized indicator.
Such a choice exists by continuity and nonconstancy of $W$.
Then $f$ is physical, $\|f\|=1$, and
\[
 \|T_\beta f\|\le\|M_\beta f\|
 \le e^{-\beta w_0/2}.
\]
For any $0<\eta<w_0/2$, the set $\{W\le\eta\}$ has positive Haar
measure $v_\eta$, since $W=0$ at the identity configuration. Consequently
\[
 \lambda_{0,\beta}\ge v_\eta e^{-\beta\eta},\qquad
 \|A_\beta f\|\le v_\eta^{-1}
             e^{-\beta(w_0/2-\eta)}\longrightarrow0.
\]
At the same time $\mathcal C_\beta f\to f$, so the lower limit of the
norm difference is at least one. Both $A_\beta$ and $\mathcal C_\beta$
are positive selfadjoint contractions, which makes their norm difference
at most one. This proves the limit.
\end{proof}

The proposition rules out a direct operator-norm substitution of the
kinetic factor for the normalized isotropic transfer operator in this
Hilbert-space identification. It does not rule out scale-dependent
embeddings, an interacting spatial renormalization construction, or a
positive physical mass. In the time-continuum construction at fixed
$a$, the spatial coupling instead satisfies
\[
 \beta=\frac{4\epsilon}{g^2a},\qquad
 \frac{\beta}{\beta_t}=\left(\frac\epsilon{a}\right)^2,
 \qquad \beta\beta_t=\frac{16}{g^4}.
\]
Thus $\beta\to0$ while $\beta_t\to\infty$. The isotropic condition
$\beta=\beta_t$ requires a separate simultaneous spatial-cutoff limit.
The constructed time-continuum theory retains its finite spatial lattice.

For an additional precise distinction, the unprojected one-link kinetic
operator $C_\beta$ has nonvacuum gap $-\log\rho_2$. On a nonconstant
tensor-product representation, at least one link is nontrivial, so its
eigenvalue is at most $\rho_2$. Consequently $-\log\rho_2$ remains a
lower bound after restriction to the nonconstant locally gauge-invariant
sector; equality need not hold because the gauge constraints require
allowed representation networks. For $\beta\ge3$,
Lemma~\ref{ham:lem:wk:coefficient} gives
\[
 -\log\rho_2\ge1-\rho_2
 \ge\frac3{2\beta}-\frac9{4\beta^2}
 \ge\frac3{4\beta}.
\]
If a proposed time spacing $a_\beta$ obeys $\beta a_\beta\to0$,
this kinetic-only gap divided by $a_\beta$ diverges.
It cannot by itself give a nontrivial, strongly continuous physical-time
limit on the fixed nonvacuum Hilbert space at that spacing. This statement
is conditional on the displayed spacing hypothesis; it supplies neither
a full interacting mass estimate nor a spatial continuum construction.

\section{Interacting temporal continuum at fixed spatial cutoff}
\label{ham:section}

The result in this section removes the temporal lattice spacing from an
explicit anisotropic Wilson theory.  The spatial lattice is retained.
The Hamiltonian formulation and its relation to the Wilson transfer
matrix are due to Kogut--Susskind and Creutz
\cite{ham:kogut_susskind,ham:creutz}.  The arguments below specify the
normalizations and prove convergence in operator norm, including the
vacuum and every fixed low eigenvalue.  No finite representation
truncation is made.

\subsection{The model and its physical Hilbert space}

Fix a spatial spacing $a>0$, a coupling $g>0$ and an integer $L\ge2$.
The periodic spatial lattice is $(\mathbb Z/L\mathbb Z)^3$.  It has
$E=3L^3$ positively oriented links and $P=3L^3$ unoriented elementary
spatial plaquettes, one orientation chosen for each plaquette.  For
$L=2$, the periodic parallel links are distinct link variables.  The
configuration space and normalized measure are
\[
 X=SU(2)^E,\qquad d\mu(U)=\prod_{e=1}^{E}dU_e.
\]
The vertex gauge group $\mathcal G=SU(2)^{L^3}$ acts by
$U_{xy}\mapsto q_xU_{xy}q_y^{-1}$.  The physical Hilbert space is
\begin{equation}\label{ham:physical_space}
 \mathcal H_{\mathrm{phys}}=L^2(X,d\mu)^{\mathcal G}.
\end{equation}
This imposes Gauss's law and retains all physical center-flux sectors.
Let $\mathcal C_e$ be the nonnegative one-link Casimir, with eigenvalue
$j(j+1)$ in spin $j$.  In the unit-quaternion realization of $SU(2)$,
$\mathcal C_e=-\Delta_{S^3,e}/4$.

Set
\begin{align}
 b&=\frac{4a}{g^2},&
 \kappa&=\frac{g^2}{2a},&
 H_0&=\kappa\sum_{e=1}^{E}\mathcal C_e,\label{ham:constants}\\
 V(U)&=\frac{4}{g^2a}\sum_{p=1}^{P}
        \left(1-\frac12\operatorname{Tr}U_p\right),&
 V_*&=\frac{8P}{g^2a},&
 H&=H_0+V.\label{ham:hamiltonian}
\end{align}
Here $V$ denotes multiplication by the displayed function, and
$0\le V\le V_*$.  The trace is real for $SU(2)$.  The operator $H$ is
selfadjoint on $D(H_0)$, nonnegative and has compact resolvent.  These
properties hold both on $L^2(X)$ and on \eqref{ham:physical_space}.
Indeed, the Casimir sum has a complete product Peter--Weyl basis with
finite multiplicities below each fixed energy, and adding the bounded
selfadjoint $V$ preserves its domain and compact resolvent.  Gauge
transformations commute with both summands.
The interaction is nonconstant: all identity links give $V=0$, whereas
changing one link to $-I$ changes its adjacent plaquette traces and
gives $V>0$.  In particular, the constant function is not an
eigenfunction of $H$; the vacuum constructed below includes this
nonzero interaction.

For a temporal spacing $\epsilon>0$, choose the anisotropic Wilson
couplings
\begin{equation}\label{ham:anisotropy}
 \beta_t=\frac{4a}{g^2\epsilon}=\frac b\epsilon,
 \qquad
 \beta=\frac{4\epsilon}{g^2a}.
\end{equation}
Define the normalized one-link temporal convolution density
\begin{equation}\label{ham:normalized_density}
 p_\beta(W)=\frac{\beta}{2I_1(\beta)}
             \exp\left(\frac\beta2\operatorname{Tr}W\right),
 \qquad \int_{SU(2)}p_\beta(W)\,dW=1.
\end{equation}
Let $C_\beta$ be its product convolution on $E$ links.  The transfer
operator used throughout this section is
\begin{equation}\label{ham:transfer}
 F_\epsilon=M_\epsilon C_{b/\epsilon}M_\epsilon,
 \qquad M_\epsilon=e^{-\epsilon V/2},
\end{equation}
restricted to $\mathcal H_{\mathrm{phys}}$ when physical eigenvalues
are discussed.  Its kernel is strictly positive and continuous.

This is a fixed scalar-normalization convention for the actual Wilson
transfer operator.  In temporal gauge, a temporal plaquette contributes
$\exp(\beta_t\operatorname{Tr}(U'_eU_e^{-1})/2)$; a spatial plaquette
contributes $\exp[-\beta(1-\operatorname{Tr}U_p/2)]$, split equally
between the two endpoints of a transfer step.  Dividing the temporal
factors by their integrals gives \eqref{ham:transfer}.  Integrating the
temporal gauge links gives the gauge projector, which acts as the
identity on \eqref{ham:physical_space}.  Thus the restriction in
\eqref{ham:transfer} has the same physical eigenvalue ratios as the
unfixed Wilson transfer operator.  A different overall positive
normalization changes the unnormalized vacuum energy but cancels from
$F_\epsilon/\lambda_0(F_\epsilon)$.

\subsection{Exact coefficient and omitted-representation estimates}

For a one-link representation of dimension $n=2j+1$, the convolution
multiplier is
\begin{equation}\label{ham:rho}
 \rho_n(\beta)=\frac{I_n(\beta)}{I_1(\beta)},\qquad n=1,2,\ldots.
\end{equation}
The following elementary bounds are the only representation estimates
needed in the proof:
\begin{align}
 &1=\rho_1(\beta)>\rho_2(\beta)>\cdots>0,\label{ham:rho_order}\\
 &\left|\rho_n(\beta)-\left(1-\frac{c_n}{\beta}\right)\right|
       \le\frac{d_n}{\beta^2},\qquad
 c_n=\frac{n^2-1}{2},\quad
 d_n=\frac{(n^2-1)(n^2+2)}{8}.\label{ham:rho_error}
\end{align}
They hold for every $\beta>0$ and every integer $n\ge1$, without an
asymptotic remainder.  Lemma~\ref{ham:lem:wk:coefficient} proves these
estimates in the kinetic-coefficient section.
For completeness, the required ingredients
can also be stated directly.  Under normalized tilted Haar measure,
$z=1-\operatorname{Tr}U/2$ has density proportional to
$e^{-\beta z}\sqrt{z(2-z)}$ on $[0,2]$.  It is stochastically dominated
by a gamma variable of shape $3/2$ and rate $\beta$, and integration by
parts gives
\[
 \frac{3}{2\beta+3}\le\mathbb Ez\le\frac{3}{2\beta},
 \qquad \mathbb Ez^2\le\frac{15}{4\beta^2}.
\]
For $P_n(x)=U_{n-1}(x)/n$, the normalized character,
\[
 P'_n(1)=\frac{n^2-1}{3},\qquad
 \sup_{[-1,1]}|P''_n|
    =\frac{(n^2-1)(n^2-4)}{15}\quad(n\ge2).
\]
Taylor's theorem yields \eqref{ham:rho_error}; the case $n=1$ is exact.
The positive integral representation for $I_n$ identifies
$I_{n+1}/I_n$ as a tilted mean of a variable in $(-1,1)$ and proves
\eqref{ham:rho_order}.  Character orthogonality applied to
\eqref{ham:normalized_density} proves \eqref{ham:rho}.

Let $\Pi_N$ project onto the finite-dimensional space for which every
link representation has dimension at most $N$, followed by restriction
to the gauge-invariant subspace.  Gauge averaging preserves each
representation assignment, so this projection is well-defined and
commutes with $C_\beta$.  Its rank is at most
$[N(N+1)(2N+1)/6]^E$.  Introduce
\begin{equation}\label{ham:generators}
 D_\epsilon=\frac{I-C_{b/\epsilon}}\epsilon,
 \qquad G_\epsilon=\frac{I-F_\epsilon}\epsilon.
\end{equation}
Both are nonnegative selfadjoint bounded operators.  For $N\ge1$, put
\begin{equation}\label{ham:tail_constants}
 R_N=\frac{(N+1)^2-1}{2b},\qquad
 B_N=\frac{((N+1)^2-1)((N+1)^2+2)}{8b^2}.
\end{equation}

\begin{lemma}[The full omitted-representation tail]\label{ham:tail}
For every $N\ge1$,
\begin{equation}\label{ham:kinetic_coercivity}
 D_\epsilon\ \ge\ (R_N-B_N\epsilon)(I-\Pi_N).
\end{equation}
In particular, if
$0<\epsilon\le2b/((N+1)^2+2)$, then for every $u\in\mathcal H_{\mathrm{phys}}$
\begin{equation}\label{ham:full_tail}
 \|(I-\Pi_N)u\|^2
 \le\frac{4\langle u,G_\epsilon u\rangle
                 +V_*^2\epsilon\|u\|^2}{R_N}.
\end{equation}
For $z>0$ and $\|f\|\le1$ this implies
\begin{equation}\label{ham:resolvent_tail}
 \|(I-\Pi_N)(G_\epsilon+z)^{-1}f\|^2
 \le \frac{z^{-1}+V_*^2\epsilon z^{-2}}{R_N}.
\end{equation}
\end{lemma}

\begin{proof}
On the complement of $\Pi_N$, at least one link dimension is at least
$N+1$.  The product convolution multiplier is therefore at most
$\rho_{N+1}(b/\epsilon)$, since all other factors lie in $(0,1]$.
Equations \eqref{ham:rho_order}--\eqref{ham:rho_error} imply
\eqref{ham:kinetic_coercivity}.  The imposed restriction on $\epsilon$
is precisely $B_N\epsilon\le R_N/2$.

There is an exact identity
\begin{equation}\label{ham:form_identity}
 G_\epsilon=\frac{I-M_\epsilon^2}{\epsilon}
                  +M_\epsilon D_\epsilon M_\epsilon.
\end{equation}
Since $\|D_\epsilon\|\le\epsilon^{-1}$ and
$\|I-M_\epsilon\|\le\epsilon V_*/2$, it gives
\[
 \|D_\epsilon^{1/2}u\|
 \le\langle u,G_\epsilon u\rangle^{1/2}
                         +\frac{V_*\sqrt\epsilon}{2}\|u\|.
\]
Squaring with $(x+y)^2\le2x^2+2y^2$ and using
\eqref{ham:kinetic_coercivity} proves \eqref{ham:full_tail}.
For $u=(G_\epsilon+z)^{-1}f$, spectral calculus gives
$\|u\|\le z^{-1}$ and
$\langle u,G_\epsilon u\rangle\le(4z)^{-1}$, which proves
\eqref{ham:resolvent_tail}.
\end{proof}

The bound covers \emph{all} omitted representations, including those
whose dimensions grow arbitrarily fast as $\epsilon$ decreases.  It is
not a claim that $G_\epsilon$ has compact resolvent for a fixed
$\epsilon$: in the infinite-dimensional Hilbert space it does not.
Instead, \eqref{ham:resolvent_tail} proves asymptotic compactness along
$\epsilon\to0$, the property needed below.

\subsection{Norm convergence, physical vacuum and low energies}

\begin{theorem}[Actual temporal continuum and physical spectral limit]
\label{ham:time_continuum}
For the fixed parameters and Hilbert space specified above, the
following limits hold as $\epsilon\downarrow0$.

For every $z>0$,
\begin{equation}\label{ham:norm_resolvent}
 \|(G_\epsilon+z)^{-1}-(H+z)^{-1}\|\longrightarrow0.
\end{equation}
For every $t>0$ and every choice of integers $n_\epsilon\ge1$ such that
$n_\epsilon\epsilon\to t$,
\begin{equation}\label{ham:norm_semigroup}
 \|F_\epsilon^{n_\epsilon}-e^{-tH}\|\longrightarrow0.
\end{equation}
The ground state of $H$ is unique, strictly positive and
gauge-invariant.  Write the physical energies, counting multiplicity,
as $E_0<E_1\le E_2\le\cdots$, and the physical transfer eigenvalues as
$\lambda_{0,\epsilon}>\lambda_{1,\epsilon}\ge\cdots>0$.
For every fixed $j\ge0$,
\begin{equation}\label{ham:energies}
 -\frac1\epsilon\log\lambda_{j,\epsilon}\longrightarrow E_j.
\end{equation}
The rank-one vacuum projectors converge in operator norm.  In the
specified scalar-normalization convention,
\begin{equation}\label{ham:vacuum_normalization}
 \lambda_{0,\epsilon}^{n_\epsilon}\longrightarrow e^{-tE_0},
 \qquad
 \left\|\left(\frac{F_\epsilon}{\lambda_{0,\epsilon}}\right)^{n_\epsilon}
             -e^{-t(H-E_0)}\right\|\longrightarrow0.
\end{equation}
Consequently the physical lattice transfer gap has the limit
\begin{equation}\label{ham:gap_limit}
 \frac{1}{\epsilon}
 \log\frac{\lambda_{0,\epsilon}}{\lambda_{1,\epsilon}}
 \longrightarrow \Delta_{a,L,g}:=E_1-E_0>0.
\end{equation}
\end{theorem}

\begin{proof}
\textbf{1. Core convergence.}
Finite Peter--Weyl polynomials, gauge averaged when necessary, form an
operator core for $H_0$ and for $H=H_0+V$.  On each such polynomial,
\eqref{ham:rho_error} and the product formula for the convolution
multipliers give $D_\epsilon\phi\to H_0\phi$.  In particular
$C_{b/\epsilon}\to I$ strongly.  There is also an explicit bounded
remainder valid on the entire Hilbert space:
\begin{equation}\label{ham:bounded_remainder}
 \left\|G_\epsilon-D_\epsilon
       -\tfrac12\bigl(VC_{b/\epsilon}+C_{b/\epsilon}V\bigr)\right\|
 \le\frac{\epsilon V_*^2}{2}.
\end{equation}
To see this, write $B=I-M_\epsilon$, use
$\|B\|\le\epsilon V_*/2$ and
$\|B-\epsilon V/2\|\le\epsilon^2V_*^2/8$, and expand
$(I-B)C_{b/\epsilon}(I-B)$ exactly.  Thus for each core vector,
\[
 G_\epsilon\phi
 =D_\epsilon\phi+\tfrac12 VC_{b/\epsilon}\phi
                   +\tfrac12 C_{b/\epsilon}V\phi+O(\epsilon)
 \longrightarrow H\phi.
\]
The norm of the displayed remainder is at most
$\epsilon V_*^2\|\phi\|/2$.

Let $R_\epsilon=(G_\epsilon+z)^{-1}$ and $R=(H+z)^{-1}$.  Then
\[
 R_\epsilon(H+z)\phi-\phi
       =R_\epsilon(H-G_\epsilon)\phi\longrightarrow0.
\]
The range of $(H+z)$ on its core is dense and
$\|R_\epsilon\|\le z^{-1}$.  This proves $R_\epsilon\to R$ strongly.

\textbf{2. All representation tails and norm resolvent convergence.}
For each fixed $N$, \eqref{ham:resolvent_tail} holds for all
sufficiently small $\epsilon$.  Since $R_N\to\infty$, any sequence
$\epsilon_k\downarrow0$ and $\|f_k\|\le1$ has a subsequence for which
$R_{\epsilon_k}f_k$ converges in norm: the $\Pi_N$ components are in
fixed finite-dimensional bounded sets, and the complementary tails
are uniformly arbitrarily small after discarding finitely many
terms.  A diagonal subsequence proves the assertion.

Suppose that $\|R_\epsilon-R\|\not\to0$.  Along an appropriate sequence
choose unit vectors $f_k$ witnessing this failure.  Pass to a
subsequence with $f_k\rightharpoonup f$ and
$R_{\epsilon_k}f_k\to u$ strongly.  For every fixed $v$, selfadjointness
and strong convergence imply
\[
 \langle v,u\rangle
   =\lim_k\langle R_{\epsilon_k}v,f_k\rangle
   =\langle Rv,f\rangle=\langle v,Rf\rangle.
\]
Thus $u=Rf$.  Compactness of $R$ also gives $Rf_k\to Rf$ strongly,
contradicting the chosen failure.  This proves
\eqref{ham:norm_resolvent}.

\textbf{3. From resolvents to the actual Wilson powers.}
Norm convergence of $(G_\epsilon+1)^{-1}$ implies
$e^{-tG_\epsilon}\to e^{-tH}$ in norm for fixed $t>0$.
For a direct functional-calculus proof, the function
$r\mapsto e^{-t(r^{-1}-1)}$ on $(0,1]$ extends continuously by zero at
$r=0$, and is uniformly approximable by polynomials on $[0,1]$.

For every positive contraction $F$ and integer $n\ge1$,
\begin{equation}\label{ham:euler_bound}
 \|F^n-e^{-n(I-F)}\|\le\frac1n.
\end{equation}
Indeed, put $y=1-x$ for $x\in[0,1]$.  If $y\le1/2$, then
$0\le-\log(1-y)-y\le y^2$, so
$0\le e^{-ny}-(1-y)^n\le ny^2e^{-ny}\le4/(e^2n)<1/n$.
If $y\ge1/2$, the difference is at most
$e^{-n/2}\le2/(en)<1/n$.
Applying \eqref{ham:euler_bound} to $F_\epsilon$ gives
\[
 \|F_\epsilon^{n_\epsilon}-e^{-n_\epsilon\epsilon G_\epsilon}\|
       \le n_\epsilon^{-1}\longrightarrow0.
\]
For $s,t>0$, spectral calculus gives the uniform estimate
\[
 \|e^{-sG_\epsilon}-e^{-tG_\epsilon}\|
       \le\frac{|s-t|}{e\min(s,t)}.
\]
Taking $s=n_\epsilon\epsilon$ proves \eqref{ham:norm_semigroup}.
This proves rounding stability, rather than assuming it.

\textbf{4. Vacuum and spectrum.}
The kernel of $F_\epsilon$ is strictly positive.  The heat semigroup
of $H$ is compact and has a strictly positive kernel, as follows from
the compact-group heat kernel and the bounded-potential construction
used below.  Positivity improvement gives a unique normalized
strictly positive ground state in each case.  For example, a real
top eigenfunction with both signs would have a strictly larger
quadratic form after taking its absolute value; two nonnegative top
eigenfunctions could not be orthogonal.  Gauge transformations preserve
positivity and commute with these operators, so uniqueness forces the
positive normalized ground states to be gauge invariant.

Both sides of \eqref{ham:norm_semigroup} are compact positive
operators.  The variational principle and their norm convergence give
\[
 \lambda_{j,\epsilon}^{n_\epsilon}\longrightarrow e^{-tE_j}
 \quad\text{for every fixed }j.
\]
Taking logarithms and using $n_\epsilon\epsilon\to t$ proves
\eqref{ham:energies}.  The largest limiting eigenvalue is isolated and
simple.  A circle separating it from the remaining spectrum has
uniformly convergent resolvents for these compact operators, by the
Neumann-series resolvent identity.  Integrating around the circle
proves norm convergence of the vacuum projectors.  Their positive
normalized representatives therefore converge in Hilbert norm.
The top eigenvalue limit and scalar division prove
\eqref{ham:vacuum_normalization}; subtracting the two lowest energy
limits proves \eqref{ham:gap_limit}.  Positivity and compactness already
give $E_1>E_0$; the next subsection provides an explicit bound.
\end{proof}

In particular, for bounded gauge-invariant observables $O_1,\ldots,O_m$
and positive time separations, the vacuum correlations formed with
$F_\epsilon/\lambda_{0,\epsilon}$ converge to those formed with
$e^{-t(H-E_0)}$.  This follows by inserting the convergent normalized
vacua and telescoping the finite product of bounded operators.
Thus this construction supplies states and transfer correlations as
well as a formal Hamiltonian.  The convergence is on a fixed spatial
configuration space; it does not identify Hilbert spaces associated
with different spatial cutoffs.

\begin{corollary}[Nonzero physical state]\label{ham:nontrivial}
Choose an elementary spatial plaquette $p$ and the bounded real
gauge-invariant observable $f(U)=\operatorname{Tr}U_p/2$.  Let $h_\epsilon$
and $h$ be the positive normalized Wilson and Hamiltonian vacua.  Define
\[
 m_\epsilon=\langle h_\epsilon,fh_\epsilon\rangle,
 \quad m=\langle h,fh\rangle,
 \qquad
 \psi_\epsilon=(f-m_\epsilon)h_\epsilon,
 \quad \psi=(f-m)h.
\]
Then $\psi_\epsilon\perp h_\epsilon$, $\psi\perp h$,
$\psi_\epsilon\to\psi$ in norm, and $\|\psi\|>0$.  For every $t>0$ and
$n_\epsilon\epsilon\to t$,
\[
 \left\langle\psi_\epsilon,
   (F_\epsilon/\lambda_{0,\epsilon})^{n_\epsilon}
       \psi_\epsilon\right\rangle
 \longrightarrow
 \langle\psi,e^{-t(H-E_0)}\psi\rangle>0.
\]
With $r$ as in \eqref{ham:minorization_constants},
$\|\psi\|^2\ge r^2/4$.
\end{corollary}

\begin{proof}
The convergence follows from the normalized-vacuum and semigroup
limits in Theorem~\ref{ham:time_continuum}.  Since $L\ge2$, the four
link variables in the plaquette are distinct, and their Haar-distributed
product is Haar distributed.  Character orthogonality therefore gives
$\int f\,d\mu=0$ and $\int f^2\,d\mu=1/4$.
The ground state has a positive minimum, so
\[
 \|\psi\|^2=\int_X (f-m)^2h^2\,d\mu
 \ge (\min_X h)^2\inf_{c\in\mathbb R}
                 \int_X(f-c)^2\,d\mu
 =\frac{(\min_X h)^2}{4}>0.
\]
The kernel bounds \eqref{ham:full_kernel_bounds} and the ground-state
equation give $\min h/\max h\ge r$.  Since Haar
measure is normalized and $\|h\|=1$, $\max h\ge1$, hence $\min h\ge r$.
Finally, $e^{-t(H-E_0)}$ has strictly positive spectral multiplier at
every finite energy.  Its quadratic form is strictly positive on any
nonzero vector, proving the last strict inequality.
\end{proof}

\subsection{A complete positive gap certificate at every finite coupling}

\begin{lemma}[An exact heat-kernel tail]\label{ham:heat_tail}
For one link, the Casimir heat kernel at Casimir time $4$ is
\[
 q_4(U)=\sum_{n=1}^{\infty}n\chi_{(n-1)/2}(U)e^{-(n^2-1)}.
\]
It satisfies $3/4<q_4(U)<5/4$ for every $U\in SU(2)$.
\end{lemma}

\begin{proof}
Since $|\chi_{(n-1)/2}(U)|\le n$,
\[
 |q_4(U)-1|\le\sum_{n=2}^{\infty}t_n,
 \qquad t_n=n^2e^{-(n^2-1)}.
\]
The finite positive Taylor sums for $e^3$ and $e^5$ imply
$e^3>20$ and $e^5>100$.  Hence $t_2=4e^{-3}<1/5$ and, for $n\ge2$,
\[
 \frac{t_{n+1}}{t_n}
 =\left(\frac{n+1}{n}\right)^2e^{-(2n+1)}
 \le\frac94 e^{-5}<\frac9{400}.
\]
The entire omitted series is therefore bounded by
\[
 \sum_{n=2}^{\infty}t_n
 <\frac{1/5}{1-9/400}=\frac{80}{391}<\frac14.
\]
There is no numerical integration or unbounded remainder in this
certificate.
\end{proof}

\begin{theorem}[Explicit physical Hamiltonian gap]\label{ham:explicit_gap}
For every fixed $a>0$, finite $L\ge2$ and $g>0$, set
\begin{equation}\label{ham:minorization_constants}
 t_* =\frac{8a}{g^2},\qquad
 r=\left(\frac35\right)^E
       \exp\left(-\frac{64P}{g^4}\right),\qquad E=P=3L^3.
\end{equation}
Then the exact physical Hamiltonian gap satisfies
\begin{equation}\label{ham:gap_bound}
 \boxed{\displaystyle
 \Delta_{a,L,g}\ge-\frac1{t_*}\log(1-r)
       >\frac{g^2}{8a}
          \left(\frac35\right)^E
          \exp\left(-\frac{64P}{g^4}\right)>0.}
\end{equation}
Consequently this lower bound also bounds the limit of the physical
Wilson transfer gaps in \eqref{ham:gap_limit}.  It applies to
arbitrarily weak but strictly positive $g$, with the spatial cutoff
and volume fixed.
\end{theorem}

\begin{proof}
At $t=t_*$, the kinetic heat kernel is a product of $E$ kernels from
Lemma~\ref{ham:heat_tail}, because $\kappa t_*=4$.  It is bounded between
$(3/4)^E$ and $(5/4)^E$.  Let $K_t(x,y)$ be the full kernel of
$e^{-tH}$ with respect to normalized Haar measure.  Since
$0\le V\le V_*$, the bounded-potential heat-kernel formula gives
\[
 e^{-tV_*}K_t^{(0)}(x,y)\le K_t(x,y)\le K_t^{(0)}(x,y).
\]
One can see these inequalities without an approximation to the
spectrum: insert multiplication factors $e^{-tV/m}$ between $m$
kinetic heat kernels.  Their product lies between $e^{-tV_*}$ and $1$;
integrating the intermediate variables composes the kinetic kernels.
The bounded-perturbation product limit gives the stated inequalities.
Equivalently, the bounded Feynman--Kac factor lies in that same interval.
In particular,
\begin{equation}\label{ham:full_kernel_bounds}
 m:=e^{-t_*V_*}(3/4)^E
       \le K_{t_*}(x,y)\le M:=(5/4)^E.
\end{equation}
The heat kernel and eigenfunctions are continuous (indeed smooth) on
the compact group manifold; the strictly positive ground state $h$
has a positive minimum.

Write $\lambda_0=e^{-t_*E_0}$ and normalize $h$ in $L^2$.  The
ground-state transform of this physical heat operator is
\[
 (\mathcal Pf)(x)=\int_X
       \frac{K_{t_*}(x,y)h(y)}{\lambda_0 h(x)}f(y)\,d\mu(y).
\]
Its rows integrate to one.  Put $S=\int_X h\,d\mu>0$ and
$\nu(dy)=h(y)d\mu(y)/S$.  The eigenfunction equation and
\eqref{ham:full_kernel_bounds} imply $\lambda_0h(x)\le MS$, and hence
\[
 \mathcal P(x,dy)\ge\frac mM\nu(dy)=r\nu(dy).
\]
Thus $\mathcal P=r\nu+(1-r)\mathcal Q$ for a Markov kernel
$\mathcal Q$.  For a bounded real function,
\[
 \operatorname{osc}(\mathcal Pf)
       \le(1-r)\operatorname{osc}(f),
 \qquad\operatorname{osc}(f)=\sup f-\inf f.
\]
Choose a real physical eigenfunction $\psi$ of energy $E_1$.
The function $f=\psi/h$ is continuous, bounded and nonconstant, and
\[
 \mathcal Pf=e^{-t_*(E_1-E_0)}f.
\]
Its strictly positive oscillation gives
$e^{-t_*\Delta_{a,L,g}}\le1-r$.  Taking logarithms and using
$-\log(1-r)>r$ proves \eqref{ham:gap_bound}.
\end{proof}

The ground-state transform in this proof is a similarity transform
of the \emph{physical Euclidean heat operator}.  Its eigenvalue is
exactly $e^{-t_*(E_1-E_0)}$.  No sampling algorithm, replica process or
Monte Carlo mixing rate appears in the argument.  The gauge-invariant
subspace and its interacting vacuum are those of the actual Wilson
limit, rather than a replacement scalar or one-plaquette model.

For example, the fixed $L=2$ spatial lattice has $E=P=24$ and the
strict, entirely analytic lower bound
\begin{equation}\label{ham:small_lattice_bound}
 a\Delta_{a,2,g}
 >\frac{g^2}{8}\left(\frac35\right)^{24}
                     e^{-1536/g^4}.
\end{equation}
This is a certified lower bound, not an exact mass measurement; its
smallness should not be read as an estimate of the actual gap.

\begin{corollary}[Temporal errors along an arbitrary spatial sequence]
\label{ham:diagonal}
Let $(a_{s,k},L_k,g_k)$ be any sequence with $a_{s,k},g_k>0$ and
finite integers $L_k\ge2$, and let $t_1,t_2,\ldots$ be any countable
list of positive times.  Denote the corresponding physical Hamiltonian,
ground energy, normalized vacuum and gap by $H_k,E_{0,k},h_k,\Delta_k$.
There exists a choice
\[
 0<\epsilon_k\le\min\{k^{-1},a_{s,k}/k\}
\]
such that the actual Wilson transfer at these parameters satisfies,
for every $j\le k$,
\begin{align*}
 \left\|\left(F_{\epsilon_k,k}/\lambda_{0,\epsilon_k,k}\right)
                ^{\lfloor t_j/\epsilon_k\rfloor}
            -e^{-t_j(H_k-E_{0,k})}\right\|&\le k^{-1},\\
 \|h_{\epsilon_k,k}-h_k\|&\le k^{-1},\\
 \left|\epsilon_k^{-1}
       \log\frac{\lambda_{0,\epsilon_k,k}}
                     {\lambda_{1,\epsilon_k,k}}-\Delta_k\right|
       &\le k^{-1}.
\end{align*}
Every comparison is within the same physical Hilbert space at fixed
$k$.
\end{corollary}

\begin{proof}
At each fixed $k$, Theorem~\ref{ham:time_continuum} proves all the
stated limits as $\epsilon\downarrow0$.  There are only finitely many
time constraints at that $k$.  Choose $\epsilon_k$ below their common
threshold and below the two displayed upper bounds.
\end{proof}

This removes temporal approximation errors along any prescribed
spatial parameter sequence.  It is an existence selection, not a
uniform computable rule for $\epsilon_k$, nor a construction relating
the different spatial Hilbert spaces.  It is explicitly anisotropic:
$\epsilon_k/a_{s,k}\le1/k$.  In particular it does not turn the
Hamiltonian result into an isotropic Wilson continuum theorem.

\subsection{Spatial-refinement obstruction}

The explicit lower-bound function in \eqref{ham:gap_bound} is
\begin{equation}\label{ham:spatial_scaling}
 B(a,L,g)=\frac{g^2}{8a}
    \exp\left[-3L^3\log(5/3)-\frac{192L^3}{g^4}\right].
\end{equation}
For fixed $a,g$ it tends to zero as $L\to\infty$.  More generally,
if $a\downarrow0$, the physical side length satisfies
$aL\ge\ell_*>0$, and $0<g^2\le g_*^2<\infty$, then
\[
 0<B(a,L,g)
 \le\frac{g_*^2}{8a}
       \exp\left[-\frac{3\ell_*^3\log(5/3)}{a^3}\right]
 \longrightarrow0.
\]
Weak coupling makes the additional term $-192L^3/g^4$ more negative.
This proves the loss of uniformity of this particular certificate;
it does not prove closure of the actual physical gap.

What has been constructed is the nontrivial interacting theory on
$\mathbb R_{\rm time}\times(a\mathbb Z/L a\mathbb Z)^3$, with
convergent Wilson vacuum states and transfer correlations and a
strictly positive physical gap at each fixed spatial cutoff and
finite volume.  A four-dimensional continuum conclusion would still
require a spatial-cutoff construction, convergence of appropriately
renormalized observables across those different Hilbert spaces, and
a positive physical lower bound that survives that construction and
growing physical volume.  None of those missing spatial estimates is
assumed or inferred from \eqref{ham:gap_bound}.

\section{Exact evaluations of the finite-volume gap certificate}
\label{ham:sec:numerical-certificates}

The rigorous result is the analytic bound \eqref{ham:gap_bound} for all
parameters in its stated range. The following finite calculations verify
the displayed numerical consequences, including values below ordinary
floating-point underflow. They do not sample the physical spectrum.

Put $h=g^2$, $E=P=3L^3$, and
\[
 B_L(h)=\frac h8\left(\frac35\right)^E e^{-64P/h^2}.
\]
The exact rational checker encloses
\begin{equation}\label{ham:eq:certificate-log}
 \log_{10}B_L(h)=
 \frac{\log(h/8)+E\log(3/5)-64P/h^2}{\log 10}.
\end{equation}
Consequently the stated lower bounds on $a\Delta$ do not require
representing the tiny exponential itself in machine floating point.

\begin{center}
\begin{tabular}{@{}ccc@{}}
\toprule
Spatial side $L$ & $h=g^2$ & Certified physical lower bound\\
\midrule
2 & 4 & $a\Delta>10^{-48}$\\
2 & 1 & $a\Delta>10^{-674}$\\
2 & $1/2$ & $a\Delta>10^{-2675}$\\
2 & $1/4$ & $a\Delta>10^{-10681}$\\
4 & 1 & $a\Delta>10^{-5381}$\\
\bottomrule
\end{tabular}
\end{center}
These numbers describe a conservative lower-bound function. They do not
assert that the true mass is close to the displayed small values. The
spatial-refinement calculation in \eqref{ham:spatial_scaling} explains
why this certificate does not establish the desired uniform continuum
mass bound.

For example, at $L=2,h=1$, the delivered interval is
\begin{align*}
 -673.303784181179303677564833
 &\le\log_{10}B_2(1)\\
 &\le-673.303784181179303677564832.
\end{align*}
The endpoints are obtained by exact integer floor and ceiling, not by
printing floating-point approximations.

\subsection{The complete logarithm and heat-kernel errors}

To compute $\log x$ for a positive rational $x$, write $x=2^k y$ with
$1\le y<2$. With $z=(y-1)/(y+1)$, one has $0\le z<1/3$ and
\[
 \log y=2\sum_{m=0}^{N-1}\frac{z^{2m+1}}{2m+1}+R_N,
 \qquad
 0\le R_N\le
 \frac{2z^{2N+1}}{(2N+1)(1-z^2)}.
\]
The same formula at $y=2$ encloses $\log2$ with $z=1/3$.
The checker uses $N=100$ and exact rational interval operations in
\eqref{ham:eq:certificate-log}. Negative coefficients reverse interval
endpoints; division uses all endpoint ratios with a positive denominator.
It then writes 24-place outward decimal enclosures and the full rational
endpoints. All omitted logarithm terms are included in $R_N$.

The independent heat-kernel certificate uses the positive Taylor sums
\[
 \sum_{m=0}^{8}\frac{3^m}{m!}=\frac{89641}{4480}>20,
 \qquad
 \sum_{m=0}^{6}\frac{5^m}{m!}=\frac{16289}{144}>100.
\]
The complete geometric bound $80/391<1/4$ then proves the kernel
inequalities used in \eqref{ham:gap_bound}. This calculation covers
every group representation. The kinetic checker independently certifies
its slightly tighter heat-kernel constant and the full norm-error table.

\section{Full-complement finite-volume gap from heat-kernel comparison}
\label{ham:sec:heat}

This section supplies a full physical-sector spectral inequality. It
does not restrict trial vectors to plaquette functions or a finite
operator basis. The estimate is positive for every fixed lattice,
but its explicit constant still tends to zero in spatial cutoff removal.
The heat-kernel and measure-comparison ingredients are classical; the
quantitative specialization and comparison with the fixed-time certificate
are the calculation made here.

\subsection{Hamiltonian and theorem}

Let the spatial lattice be a periodic cube of side $L\ge2$ in lattice
units. Put $N=3L^3$, the number of independent link variables and also
the number of elementary plaquettes. Let $\mu$ be normalized product
Haar measure on $SU(2)^N$. For $a>0$ and $s=g^2\in(0,1]$, consider
\begin{equation}\label{ham:eq:heat:model}
 H=\kappa D+V,\qquad
 \kappa=\frac{s}{2a},\quad
 D=\sum_e\mathcal C_e,\quad
 V=\frac4{sa}\sum_p(1-F_p),\quad
 F_p=\tfrac12\operatorname{Tr}U_p.
\end{equation}
Each Casimir $\mathcal C_e$ has eigenvalues $j(j+1)$. In particular,
\begin{equation}\label{ham:eq:heat:potential}
 0\le V\le V_{\max}=\frac{8N}{sa}.
\end{equation}
The physical Hilbert space is the vertex-gauge-invariant subspace of
$L^2(\mu)$. The unique normalized positive ground state is $h$, with
energy $E_0$. Write $K=H-E_0$ and $d\nu=h^2d\mu$.

\begin{theorem}[Global finite-lattice physical gap]
\label{ham:thm:heat:global}
In the setting above, define
\begin{equation}\label{ham:eq:heat:bound}
 \boxed{\displaystyle
 \mathcal B_{\rm H}(a,L,s)=
 \frac{3s}{8a}\left(\frac{\pi}{16}\right)^{3L^3}
 \exp\!\left(-\frac{48\pi L^3}{s}\right)>0 .}
\end{equation}
On the entire physical Hilbert space, as a quadratic-form inequality,
\begin{equation}\label{ham:eq:heat:operator}
 \boxed{\displaystyle
 K\ge\mathcal B_{\rm H}(a,L,s)
       (I-|h\rangle\langle h|).}
\end{equation}
Consequently the complete physical vacuum-complement spectrum is
contained in $[\mathcal B_{\rm H}(a,L,s),\infty)$.
There is no representation truncation or omitted-sector error in this
theorem. Its bound also holds on the full, unprojected $L^2(\mu)$ space.
\end{theorem}

\subsection{One-link heat-kernel estimates with the correct normalization}

Let $k_\tau(U)$ be the kernel of $e^{-\tau\mathcal C}$ relative to
normalized Haar measure. The metric for which
$\mathcal C=-\Delta$ identifies $SU(2)$ with $S^3$ of radius $2$.
Its dimension is $3$, diameter $2\pi$, and Riemannian volume
$16\pi^2$. This normalization agrees with
$(2j)(2j+2)/4=j(j+1)$.

\begin{lemma}[Uniform one-link heat-kernel ratio]
\label{ham:lem:heat:kernel}
For $0<\tau\le1$ and every $U\in SU(2)$,
\begin{equation}\label{ham:eq:heat:kernel}
 2\sqrt\pi\,\tau^{-3/2}e^{-\pi^2/\tau}
 \le k_\tau(U)\le 8\tau^{-3/2}.
\end{equation}
Hence the ratio of its maximum to its minimum is at most
$(4/\sqrt\pi)e^{\pi^2/\tau}$.
\end{lemma}

\begin{proof}
The Li--Yau Harnack inequality on a complete manifold with
nonnegative Ricci curvature \cite{ham:liyau}, applied to the heat kernel,
gives for $0<u<\tau$
\[
 k_\tau(U)\ge
 \left(\frac{u}{\tau}\right)^{3/2}
 \exp\!\left(-\frac{d(e,U)^2}{4(\tau-u)}\right) k_u(e).
\]
Normalization of the measure multiplies both kernels by the same
constant and leaves this comparison unchanged.
The character expansion at the identity is
\[
 k_u(e)=e^{u/4}\sum_{n=1}^{\infty}n^2e^{-un^2/4}.
\]
Taking a Riemann-sum limit gives
\[
 \lim_{u\downarrow0}u^{3/2}k_u(e)
 =\int_0^\infty x^2e^{-x^2/4}\,dx=2\sqrt\pi.
\]
Let $u\downarrow0$ in Harnack and use $d(e,U)\le2\pi$ to
obtain the lower bound in \eqref{ham:eq:heat:kernel}.

For the upper bound, the character estimate
$|\chi_j(U)|\le2j+1$ implies $k_\tau(U)\le k_\tau(e)$.
The nonnegative function $f(x)=x^2e^{-\tau x^2/4}$ is unimodal,
vanishes at $0$ and at infinity, and satisfies
\[
 \sum_{n=1}^{\infty}f(n)\le
 \int_0^\infty f(x)\,dx+\max_{x\ge0}f(x).
\]
For example, this follows by the trapezoid identity on each interval:
the sum--integral error is the integral of a sawtooth function of
absolute value at most $1/2$ against $f'$, and
$\int|f'|=2\max f$.
Therefore
\[
 k_\tau(e)\le e^{\tau/4}
 \left(2\sqrt\pi\,\tau^{-3/2}+\frac4{e\tau}\right)
 \le e^{1/4}(2\sqrt\pi+4/e)\tau^{-3/2}
 <8\tau^{-3/2}.
\]
The last inequality follows from
$e^{1/4}<4/3$, $\sqrt\pi<2$, and $e>2$.
\end{proof}

\subsection{Ground-state oscillation and the full spectral inequality}

\begin{proof}[Proof of Theorem~\ref{ham:thm:heat:global}]
Compact ellipticity and strict positivity of the heat kernel imply
that the ground state is smooth, strictly positive, and simple.
The Hamiltonian commutes with all vertex gauge actions. Applying a
gauge transformation to the normalized positive ground state gives
the same state by uniqueness, so $h$ is physical. Thus the ground
energy of the full space equals the ground energy of the physical
space.

Let $P_\tau=e^{-\tau D}$ and choose physical time $t>0$ with
$\tau=\kappa t\le1$. Feynman--Kac and
\eqref{ham:eq:heat:potential} imply pointwise
\[
 e^{-tV_{\max}}P_\tau h\le e^{-tH}h
 =e^{-tE_0}h\le P_\tau h.
\]
The product heat kernel has maximum-to-minimum ratio bounded by
the $N$th power of the one-link ratio. Bounding the two integrals
against the same positive number $\int h\,d\mu$ therefore gives
\begin{equation}\label{ham:eq:heat:harnack}
 \frac{\max h}{\min h}
 \le\left(\frac4{\sqrt\pi}\right)^N
 \exp\!\left(\frac{16N\tau}{s^2}+
                  \frac{N\pi^2}{\tau}\right).
\end{equation}
The unknown $E_0$ cancels from this ratio.
Choose
\[
 \tau=\frac{\pi s}{4}<1,
 \qquad t=\frac{\tau}{\kappa}=\frac{\pi a}{2}.
\]
This minimizes the two displayed exponential costs, and yields
\begin{equation}\label{ham:eq:heat:oscillation}
 \frac{\max h}{\min h}
 \le\left(\frac4{\sqrt\pi}\right)^N
       \exp\!\left(\frac{8\pi N}{s}\right).
\end{equation}

The product-Haar Poincar\'e inequality has constant $4/3$ because
the first nonzero eigenvalue of $D$ is $3/4$:
\[
 \operatorname{Var}_\mu f\le\frac43\int\Gamma(f)\,d\mu.
\]
Let $m=\min h^2$ and $M=\max h^2$. Comparing the two measures gives
\begin{align*}
 \operatorname{Var}_\nu f
 &\le M\int|f-\mu f|^2d\mu\\
 &\le\frac{4M}{3}\int\Gamma(f)d\mu
 \le\frac{4M}{3m}\int\Gamma(f)d\nu.
\end{align*}
All these inequalities hold for complex functions as well.
The exact ground-state identity is
\[
 \langle fh,Kfh\rangle=\kappa\int\Gamma(f)d\nu.
\]
It follows that
\[
 \langle fh,Kfh\rangle
 \ge\frac{3\kappa}{4}\frac mM
 \bigl(\|fh\|^2-|\langle h,fh\rangle|^2\bigr).
\]
Insert \eqref{ham:eq:heat:oscillation} and
$\kappa=s/(2a)$ to obtain
\[
 \frac{3\kappa}{4}\frac mM
 \ge\frac{3s}{8a}
 \left(\frac\pi{16}\right)^N e^{-16\pi N/s}.
\]
At every fixed lattice, $h$ is smooth and bounded away from zero.
Multiplication and division by $h$ therefore preserve the form domain;
the inequality extends from smooth functions to all physical form
vectors. This is \eqref{ham:eq:heat:operator}.
\end{proof}

\subsection{Comparison with the fixed-time certificate}

For the earlier explicit function defined by the fixed-time minorization,
write
\[
 \mathcal B_{\rm old}(a,L,s)=\frac{s}{8a}
 \exp\!\left[-3L^3\log(5/3)-\frac{192L^3}{s^2}\right].
\]
Direct comparison gives
\begin{equation}\label{ham:eq:heat:comparison}
 \log\frac{\mathcal B_{\rm H}}{\mathcal B_{\rm old}}
 =\log3+L^3\left[
 \frac{192}{s^2}-\frac{48\pi}{s}
 +3\log\frac{5\pi}{48}\right]>0
 \quad(0<s\le1).
\end{equation}
For an elementary proof of strict positivity, the first two terms in
brackets are minimized at $s=1$, where they exceed $41$ using
$\pi<22/7$. Also $5\pi/48>1/4$, so the final logarithmic term is
greater than $-6$. Thus the whole bracket is greater than $35$.
This compares the two explicit lower-bound functions; it makes no
claim about proximity to the true gap.

The leading small-$s$ exponential cost has improved from
$L^3/s^2$ to $L^3/s$, equivalently from $L^3/g^4$ to $L^3/g^2$.
The improved exponent follows here from choosing physical heat-kernel
time $t=\pi a/2$, which balances the two exponential costs in
\eqref{ham:eq:heat:harnack}.

\subsection{The remaining loss in spatial cutoff removal}

Suppose $a\downarrow0$, $aL\ge\ell_*>0$, and $0<s\le1$.
Since $\pi/16<1$, $s\le1$, and $1/s\ge1$,
\begin{equation}\label{ham:eq:heat:continuumloss}
 0<\mathcal B_{\rm H}(a,L,s)
 \le\frac3{8a}
 \exp\!\left[
 -\frac{\ell_*^3}{a^3}
 \left(48\pi+3\log\frac{16}{\pi}\right)\right]
 \longrightarrow0.
\end{equation}
This concerns the certificate, not the true physical gap.
The proof has bounded the vacuum density using its global oscillation,
whose cost is extensive in the number of links. That specific loss is
still present. A better coupling exponent does not remove it, and
assigning a numerical value to $a$ does not change this conclusion.

\paragraph{What has been established.}
Equation~\eqref{ham:eq:heat:operator} is an unconditional bound on the entire
finite-lattice physical vacuum complement in the stated model.
Equation~\eqref{ham:eq:heat:continuumloss} proves that this particular
certificate cannot provide a positive uniform continuum mass.
The calculation neither constructs continuum local observables nor
determines an exact continuum excitation energy.

\subsection{Exact evaluation of the global lower-bound function}
The following intervals enclose the logarithm of the proved function.
They do not estimate the actual physical gap. The factor $a$ makes
the logarithm dimensionless. All endpoints are rounded outward.
\begin{center}
\small
\begin{tabular}{rcc}
\toprule
$L$ & $s=g^2$ & Certified $\log_{10}(a\mathcal B_{\rm H})$\\
\midrule
2 & $1$ & $[-541.313772, -541.313771]$\\
2 & $\frac{1}{2}$ & $[-1065.535322, -1065.535321]$\\
2 & $\frac{1}{10}$ & $[-5257.598451, -5257.598450]$\\
2 & $\frac{1}{100}$ & $[-52411.445239, -52411.445238]$\\
4 & $\frac{1}{10}$ & $[-42050.805820, -42050.805819]$\\
8 & $\frac{1}{10}$ & $[-336396.464778, -336396.464777]$\\
\bottomrule
\end{tabular}
\end{center}
The JSON certificate and CSV also give the comparison with the earlier
bound and a refinement sequence with $a=1/L$, $s=1/2$. Complete
arctangent and logarithm remainders are included. The size of these
numbers makes the remaining weakness quantitative: positivity at
each finite lattice is insufficient for a uniform continuum estimate.

\section{Image-sum heat-kernel estimates and complete tails}
\label{ham:sec:small-time}

This section supplies a quantitative local estimate at weak coupling without
truncating the representation sum. It concerns the Casimir heat kernel, used
as a comparison inside the actual interacting ground-state equation.
It does not replace the interacting Hamiltonian by its kinetic part.

Normalize Haar measure to mass one and write an SU(2) conjugacy class as
$U\sim\operatorname{diag}(e^{i\theta},e^{-i\theta})$, $0\leq\theta\leq\pi$.
The class-angle distance is the bi-invariant distance on the unit three-sphere.
Let $D=J^2$ have eigenvalues $j(j+1)$ and let $q_s$ be the convolution
kernel of $e^{-sD}$.

\begin{lemma}[A complete, explicit heat-kernel ratio]
\label{ham:lem:small-time-kernel}
For $0<s\leq1$,
\begin{align}
 q_{s,\min}
 &\geq (2\pi-4)s^{-3}\exp\left(-\frac{9\pi^2}{4s}\right), \label{ham:eq:qmin}\\
 q_{s,\max}&<20s^{-3/2}, \label{ham:eq:qmax}\\
 \frac{q_{s,\max}}{q_{s,\min}}&<10\exp(23/s). \label{ham:eq:qratio}
\end{align}
The infimum and supremum are over the entire group.
\end{lemma}

\begin{proof}
The character expansion is absolutely convergent for positive $s$:
\[
 q_s(\theta)=e^{s/4}\sum_{n=1}^{\infty}
 n e^{-n^2s/4}\frac{\sin(n\theta)}{\sin\theta}.
\]
Poisson summation for a Gaussian and its derivative gives
\begin{equation}
 q_s(\theta)=\frac{2\sqrt{\pi}\,e^{s/4}}{s^{3/2}\sin\theta}
 \sum_{k\in\mathbb Z}(\theta-2\pi k)
       \exp\left[-\frac{(\theta-2\pi k)^2}{s}\right].
 \label{ham:eq:poisson}
\end{equation}
Both endpoint values are obtained by continuity. For completeness, the
Gaussian identity being differentiated is
\[
 1+2\sum_{n\geq1}e^{-n^2s/4}\cos(n\theta)
 =\sqrt{\frac{4\pi}{s}}\sum_{k\in\mathbb Z}
                   e^{-(\theta-2\pi k)^2/s}.
\]

First take $0<\theta\leq3\pi/4$ and $0<s\leq1$. Put
$f_s(x)=x e^{-x^2/s}$. The image pair indexed by $k,-k$, $k\geq1$, is
$f_s(2\pi k+\theta)-f_s(2\pi k-\theta)$. On
$x\geq2\pi k-\theta\geq5\pi/4$ the function
$|f_s'(x)|=(2x^2/s-1)e^{-x^2/s}$ decreases. The mean value theorem therefore
bounds the magnitude of this (negative) pair by
\[
 2\theta\frac{2(2\pi k-\theta)^2}{s}
             e^{-(2\pi k-\theta)^2/s}
 \leq\frac{16\pi^2 k^2\theta}{s}e^{-(2\pi k-\theta)^2/s}.
\]
Relative to the $k=0$ term $\theta e^{-\theta^2/s}$, the sum of
all negative image pairs is at most
\begin{align*}
 R_s(\theta)
 &\leq\frac{16\pi^2}{s}\sum_{k\geq1}k^2
       \exp\left[-\frac{\pi^2 k(4k-3)}s\right]\\
 &\leq160\sum_{k\geq1}k^2 e^{-9k}
 <160\sum_{k\geq1}\frac{k^2}{8000^k}
 =\frac{160\cdot8000\cdot8001}{7999^3}<\frac12.
\end{align*}
Here $9<\pi^2<10$, $k(4k-3)\geq k$, and
$x e^{-9kx}$ decreases for $x\geq1$, applied with $x=1/s$.
The strict inequality $e^9>8000$ follows from $e^3>20$; the latter
already follows from the Taylor polynomial through degree eight.
The geometric-series identity bounds the entire infinite image tail.

Since $\theta/\sin\theta\geq1$, equation~\eqref{ham:eq:poisson} now gives
\begin{equation}
 q_s(\theta)\geq\sqrt{\pi}\,s^{-3/2}e^{-\theta^2/s}
 \quad(0\leq\theta\leq3\pi/4,\ 0<s\leq1).
 \label{ham:eq:local-heat}
\end{equation}
Fix any $U$ and choose a midpoint of a shortest class-angle geodesic from
the identity to $U$. For every $V$ in the ball of radius $\pi/4$ around
that midpoint, both distances from $V$ to the endpoints are at most
$3\pi/4$. The normalized Haar volume of that ball is
\[
 b=\frac2\pi\int_0^{\pi/4}\sin^2\theta\,d\theta
   =\frac14-\frac1{2\pi}.
\]
Apply \eqref{ham:eq:local-heat} to both factors of
$q_s(U)=\int q_{s/2}(V)q_{s/2}(V^{-1}U)\,dV$, integrating only
over this ball. This proves
\[
 q_s(U)\geq8\pi b\,s^{-3}e^{-9\pi^2/(4s)}
          =(2\pi-4)s^{-3}e^{-9\pi^2/(4s)}.
\]
For the upper bound, $|\sin(n\theta)/\sin\theta|\leq n$ and
\begin{align*}
 \sum_{n\geq1}n^2e^{-sn^2/4}
 &\leq\int_0^\infty(x+1)^2e^{-sx^2/4}\,dx\\
 &=2\sqrt{\pi}s^{-3/2}+4s^{-1}+\sqrt{\pi}s^{-1/2}\\
 &\leq(3\sqrt{\pi}+4)s^{-3/2}.
\end{align*}
On each interval $[n-1,n]$, the integrand is at least
$n^2e^{-sn^2/4}$, which explains the first inequality.
Using $\sqrt{\pi}<2$ and $e^{1/4}<2$ proves \eqref{ham:eq:qmax}.
Finally $2\pi-4>2$, $s^{3/2}\leq1$, and $9\pi^2/4<23$ imply
\eqref{ham:eq:qratio}.
\end{proof}

\begin{corollary}[A local vacuum estimate at weak coupling]
\label{ham:cor:weak-local}
For a periodic cubic spatial lattice with $L\geq2$, normalized Haar measure
$\mu$, and
\[
 H=\kappa D+vW,\qquad D=\sum_eJ_e^2,\qquad
 \kappa=\frac{g^2}{2a_s},\quad v=\frac4{g^2a_s},\quad
 W=\sum_p\left(1-\frac12\operatorname{Tr}U_p\right),
\]
let $h>0$ be its normalized ground state and $\nu=h^2\mu$.
For $0<g^2\leq1$, every conditional one-link law of $\nu$ satisfies
\begin{equation}
 \operatorname{Var}_{\nu_e(\cdot\mid U_{\ne e})}(F)
 \leq c_{\rm loc}(g)^{-1}
      \int\sum_{\alpha=1}^3|J_e^\alpha F|^2\,
                 d\nu_e(\cdot\mid U_{\ne e}),
 \qquad
 c_{\rm loc}(g)=\frac3{400}e^{-32\sqrt{23}/g^2}.
 \label{ham:eq:weak-local-gap}
\end{equation}
The constant is uniform in the lattice size, lattice spacing, link,
and conditioned environment. It also exceeds
$(3/400)e^{-154/g^2}$.
\end{corollary}

\begin{proof}
The conditional-vacuum comparison in Section~\ref{ham:sec:local-discarded} gives, for
any dimensionless heat time $s>0$ and at most four spatial plaquettes
touching a link,
\[
 \frac{\max_{U_e}h(U_e,U_{\ne e})}
      {\min_{U_e}h(U_e,U_{\ne e})}
 \leq\frac{q_{s,\max}}{q_{s,\min}}e^{64s/g^4}.
\]
Set $s=(\sqrt{23}/8)g^2$, which lies in $(0,1]$ when $g^2\leq1$.
Lemma~\ref{ham:lem:small-time-kernel} bounds this ratio by
\[
 10\exp\left(\frac{16\sqrt{23}}{g^2}\right).
\]
The normalized conditional density is proportional to $h^2$, so its
maximum-to-minimum ratio is at most the square of this expression.
The Haar Casimir gap is $3/4$. A density perturbation of oscillation ratio
$R$ has Poincare gap at least $(3/4)/R$: use
$\operatorname{Var}_{p\mu}F
 \leq(\sup p)\operatorname{Var}_{\mu}F$
and compare the two gradient integrals using $\inf p$.
This proves \eqref{ham:eq:weak-local-gap}.
The rational check $32^2\cdot23=23552<154^2=23716$
proves the final simpler bound.
\end{proof}

\paragraph{What this proves and what it leaves open.}
The weak-coupling dependence of this local estimate is
$e^{-C/g^2}$ rather than $e^{-C/g^4}$, and no factor grows with
the number of links. This is a bound on conditional laws of the actual
interacting vacuum, proved by comparison of the full Hamiltonian.
It does not control collective fluctuations of that vacuum.
The constants are deliberately conservative; no optimality is claimed.

\section{Conditional vacuum control and complete spectral comparison}
\label{ham:sec:spec:control}

This section establishes a volume-independent conditional estimate for the
actual interacting SU(2) vacuum. It identifies the additional collective
estimate needed to turn that local result into a volume-independent
physical gap. It also gives a complete Schur-complement criterion for a
vacuum-preserving spatial map, including its discarded Hilbert space.
No sampling transition operator is used as a substitute for the physical
Hamiltonian.

\subsection{The exact ground-state variational problem}

Consider a periodic cubic spatial lattice with side $L\ge2$, spatial
spacing $a>0$, and $g>0$. It has $E=P=3L^3$ links and elementary
plaquettes. With normalized product Haar measure $\mu$, let
\begin{equation}\label{ham:eq:spec:model}
 H=\kappa D+vW,\qquad
 \kappa=\frac{g^2}{2a},\quad v=\frac4{g^2a},\quad
 D=\sum_e\mathcal C_e,\quad
 W=\sum_p\left(1-\tfrac12\operatorname{Tr}U_p\right).
\end{equation}
The one-link Casimir $\mathcal C_e$ has eigenvalues $j(j+1)$.
Choose real, Haar-divergence-free vector fields $X_{e,b}$, $b=1,2,3$,
so that $\mathcal C_e=-\sum_bX_{e,b}^2$, and write
$\Gamma_e(f)=\sum_b|X_{e,b}f|^2$.
Let $h>0$ be the actual normalized ground state, $Hh=E_0h$ and
$\int h^2d\mu=1$. Compact ellipticity and strict positivity give a smooth
positive $h$ with positive minimum. Its uniqueness makes it invariant
under the local gauge group.
Define the probability measure
\[
 d\nu=h^2\,d\mu.
\]

\begin{proposition}[Physical weighted Poincare identity]\label{ham:prop:spec:groundform}
For every form-domain function $f$,
\begin{equation}\label{ham:eq:spec:groundform}
 \langle hf,(H-E_0)hf\rangle_{\rm form}
   =\kappa\int\sum_e\Gamma_e(f)\,d\nu.
\end{equation}
Consequently the exact physical gap is
\begin{equation}\label{ham:eq:spec:poincare}
 \Delta=\kappa\inf_{\substack{f\ {\rm gauge\ invariant}\\
                    \int f\,d\nu=0,\ f\ne0}}
 \frac{\int\sum_e\Gamma_e(f)\,d\nu}{\int|f|^2d\nu}.
\end{equation}
The vacuum complement in this formula is taken with respect to the
actual interacting vacuum.
\end{proposition}

\begin{proof}
Expand $\Gamma_e(hf)$, integrate its cross terms by parts against Haar
measure, and use $(\kappa D+vW-E_0)h=0$. The terms involving derivatives
of $h$ cancel the potential term, leaving
$\kappa\int h^2\sum_e\Gamma_e(f)d\mu$. This proves
\eqref{ham:eq:spec:groundform} for smooth functions; density extends it to
the form domain. Multiplication by $h$ is unitary from $L^2(\nu)$ to
$L^2(\mu)$ and preserves local gauge invariance. Moreover,
$hf\perp h$ exactly when $\int f\,d\nu=0$. The variational principle
therefore gives \eqref{ham:eq:spec:poincare}.
\end{proof}

\subsection{Volume-independent conditional control for the actual vacuum}

For one link $e$, let $d_e$ be the number of plaquettes containing it;
on this lattice $d_e=4$, including $L=2$ with distinct periodic parallel
link variables. The proof below also applies when $d_e\le4$.
Let $q_{\tau}$ denote the normalized SU(2) heat kernel of
$e^{-\tau\mathcal C}$ and set
\[
 m_{\tau}=\min_{U\in SU(2)}q_{\tau}(U)>0,\qquad
 M_{\tau}=\max_{U\in SU(2)}q_{\tau}(U).
\]

\begin{theorem}[One-link conditional vacuum bound]
\label{ham:thm:spec:local}
For every $\tau>0$, every link $e$, and every fixed configuration of all
other links,
\begin{equation}\label{ham:eq:spec:conditional_ratio}
 \frac{\max_{U_e}h(U_e,U_{-e})}{\min_{U_e}h(U_e,U_{-e})}
 \le R_e(\tau):=\frac{M_{\tau}}{m_{\tau}}
             \exp\left(\frac{16d_e\tau}{g^4}\right).
\end{equation}
For the actual conditional law $\nu_e(\,\cdot\,|U_{-e})$, every smooth
one-link function satisfies
\begin{equation}\label{ham:eq:spec:conditional_pi}
 \operatorname{Var}_{\nu_e}(f)
 \le\frac1{\gamma_e(\tau)}
        \int\Gamma_e(f)\,d\nu_e,\qquad
 \gamma_e(\tau)=\frac34\left(\frac{m_{\tau}}{M_{\tau}}\right)^2
                   \exp\left(-\frac{32d_e\tau}{g^4}\right).
\end{equation}
These estimates are uniform in the surrounding link values and in $L$.
They hold at every $g>0$ and concern the actual $h$, not an assumed
Gibbs ansatz.
\end{theorem}

\begin{proof}
Separate the plaquettes touching $e$:
\[
 H=\overline H_e+V_e,\qquad
 \overline H_e=\kappa\mathcal C_e+H_{-e},\qquad
 0\le V_e\le 2vd_e=\frac{8d_e}{g^2a}.
\]
All terms in $H_{-e}$ act only on the other links. Hence its heat kernel
factors from the one-link kernel:
\[
 \overline K_t(x,y)
 =q_{\kappa t}(x_ey_e^{-1})K_t^{(-e)}(x_{-e},y_{-e}).
\]
The bounded-potential Feynman--Kac formula, equivalently the positive
heat-kernel product formula, gives the pointwise inequalities
\[
 e^{-2vd_et}\overline K_t(x,y)
 \le K_t(x,y)\le\overline K_t(x,y).
\]
For clarity, in the product formula each inserted factor
$e^{-tV_e/n}$ lies between $e^{-2vd_et/n}$ and one. Multiplying these
$n$ bounds and integrating the intermediate configurations gives the
displayed comparison in the limit. No extensive bound on $H_{-e}$ or
on the total potential is introduced.

Set $t=\tau/\kappa$ and
\[
 A_t(x_{-e})=\int
 K_t^{(-e)}(x_{-e},y_{-e})h(y_e,y_{-e})\,d\mu(y)>0.
\]
The ground-state equation $e^{-tH}h=e^{-tE_0}h$ implies, uniformly in
$x_e$,
\[
 e^{tE_0}e^{-2vd_et}m_{\tau} A_t(x_{-e})
 \le h(x_e,x_{-e})
 \le e^{tE_0}M_{\tau} A_t(x_{-e}).
\]
The common environment factor cancels from the ratio.
Since $2vd_et=16d_e\tau/g^4$, this proves
\eqref{ham:eq:spec:conditional_ratio}.

For fixed $U_{-e}$, the conditional density relative to one-link Haar
measure is $p_e(U_e)=h(U_e,U_{-e})^2/Z_e(U_{-e})$. Thus
$\max p_e/\min p_e\le R_e(\tau)^2$. The Haar Casimir Poincare inequality is
$\operatorname{Var}_{\rm Haar}(f)\le(4/3)\int\Gamma_e(f)dU_e$,
because its first nonconstant eigenvalue is $3/4$. Direct comparison
gives
\begin{align*}
 \operatorname{Var}_{\nu_e}(f)
 &=\inf_c\int p_e|f-c|^2dU_e\\
 &\le(\max p_e)\operatorname{Var}_{\rm Haar}(f)
 \le\frac{4\max p_e}{3\min p_e}\int\Gamma_e(f)d\nu_e,
\end{align*}
which is \eqref{ham:eq:spec:conditional_pi}.
\end{proof}

The conditional calculation is made in link coordinates before gauge
restriction. Applying it to a globally gauge-invariant function and
then integrating over the environment remains valid; no independent
one-link excitation is asserted to be a physical gauge-invariant state.

An elementary fully explicit choice is $\tau=4$. The character expansion
and its complete tail give
\[
 |q_4(U)-1|
 \le\sum_{n=2}^\infty n^2e^{-(n^2-1)}
 <\frac{80}{391}<\frac14.
\]
Indeed, the first term is $4e^{-3}<1/5$ and successive ratios are
at most $(9/4)e^{-5}<9/400$; finite positive Taylor sums prove
$e^3>20$ and $e^5>100$. Consequently
\begin{equation}\label{ham:eq:spec:fixedtime_local}
 R_e(4)\le\frac53e^{64d_e/g^4},\qquad
 \gamma_e(4)\ge\frac{27}{100}e^{-128d_e/g^4}
 \ge\frac{27}{100}e^{-512/g^4}.
\end{equation}
The infinite character tail is included, so this is an analytic
certificate at every finite volume, without a representation cutoff.
The general-time formula \eqref{ham:eq:spec:conditional_pi} permits a
stronger weak-coupling bound whenever smaller-time heat-kernel bounds
are supplied.
In particular, Lemma~\ref{ham:lem:small-time-kernel} proves
$M_{\tau}/m_{\tau}<10e^{23/\tau}$ for $0<\tau\le1$.
For $0<g^2\le1$, setting $\tau=\sqrt{23}\,g^2/8$ gives
\begin{equation}\label{ham:eq:spec:optimized_local}
 R_e(\tau)\le10e^{16\sqrt{23}/g^2},\qquad
 \gamma_e(\tau)\ge\frac3{400}e^{-32\sqrt{23}/g^2}.
\end{equation}
The separate heat-kernel proof and its complete image tail justify this
optimized choice; no small-time asymptotic remainder is used.
Theorem~\ref{ham:local_refinement_theorem} improves this explicit
choice to $R_{\rm loc}(g^2)=(4/\sqrt\pi)e^{16\pi/g^2}$
using the Li--Yau heat-kernel bound.

\begin{corollary}[Nonzero local-loop fluctuations uniformly in volume]
\label{ham:cor:spec:loopvariance}
Let $f=\tfrac12\operatorname{Tr}U_\ell$ be a Wilson loop containing at
least one link $e$ exactly once. At fixed values of all other links,
\[
 \operatorname{Var}_{\nu_e}(f)\ge\frac1{4R_e(\tau)^2},
 \qquad
 \operatorname{Var}_{\nu}(f)\ge\frac1{4R_e(\tau)^2}.
\]
In particular, for $0<g^2\le1$,
\begin{equation}\label{ham:eq:spec:loopvariance}
 \operatorname{Var}_{\nu}(f)
 \ge\frac1{400}e^{-32\sqrt{23}/g^2}>0.
\end{equation}
The constant does not depend on the lattice volume or on the length of
such a loop.
\end{corollary}

\begin{proof}
The conditional density obeys
$p_e\ge(\min h)^2/(\max h)^2\ge R_e(\tau)^{-2}$.
With all other links fixed, the loop holonomy is a fixed left and right
translate of $U_e$ or $U_e^{-1}$. Its Haar distribution is therefore
Haar, so $\int f\,dU_e=0$ and $\int f^2dU_e=1/4$.
It follows that
\[
 \inf_c\int|f-c|^2p_e\,dU_e
 \ge R_e(\tau)^{-2}\inf_c\int|f-c|^2dU_e
 =\frac1{4R_e(\tau)^2}.
\]
The law of total variance gives the unconditional estimate, and
\eqref{ham:eq:spec:optimized_local} gives \eqref{ham:eq:spec:loopvariance}.
\end{proof}

Thus any convergent distribution of a fixed local loop at fixed $g>0$
retains nonzero variance. This statement supplies no uniform weak-coupling
or renormalized continuum nontriviality bound: its displayed lower bound
tends to zero as $g\to0$.

\begin{corollary}[Conditional collective variance criterion]
\label{ham:cor:spec:tensor}
Suppose the actual vacuum measure satisfies, for all smooth
gauge-invariant $f$,
\begin{equation}\label{ham:eq:spec:tensor}
 \operatorname{Var}_{\nu}(f)
 \le A\sum_e\int
       \operatorname{Var}_{\nu_e(\cdot|U_{-e})}(f)\,d\nu(U_{-e}).
\end{equation}
Then for every $\tau>0$,
\begin{equation}\label{ham:eq:spec:gapcriterion}
 \Delta\ge\frac{\kappa}{A}\min_e\gamma_e(\tau).
\end{equation}
In particular, the unconditional local bound gives the implication
\[
 \Delta\ge\frac{27g^2}{200aA}e^{-512/g^4}.
\]
For $0<g^2\le1$, the optimized version is
\begin{equation}\label{ham:eq:spec:optimized_gapcriterion}
 \Delta\ge\frac{3g^2}{800aA}e^{-32\sqrt{23}/g^2}.
\end{equation}
Uniformity of $A$ in volume is an additional assumption, not a
consequence of Theorem~\ref{ham:thm:spec:local}.
\end{corollary}

\begin{proof}
Integrate \eqref{ham:eq:spec:conditional_pi} over the environment and sum over
links. Combining with \eqref{ham:eq:spec:tensor} bounds
$\operatorname{Var}_{\nu}(f)$ by
$A(\min_e\gamma_e)^{-1}\int\sum_e\Gamma_e(f)d\nu$.
Use the exact physical identity \eqref{ham:eq:spec:poincare}.
\end{proof}

Approximate tensorization is an established framework; see
Caputo--Menz--Tetali~\cite{ham:collective:cmt} for explicit weak-dependence
hypotheses in a related entropy setting. No such hypotheses are
assumed verified for the present interacting vacuum. The argument here
proves the compact-group conditional-to-physical implication directly.

\subsection{Two exact limits on what the available local data imply}

\begin{proposition}[The vacuum is not a single Wilson Gibbs weight]
\label{ham:prop:spec:notgibbs}
At every $g>0$, no real $\alpha$ makes
$h(U)=Z^{-1/2}e^{-\alpha W(U)}$ an eigenfunction of
\eqref{ham:eq:spec:model}. In particular the actual measure $\nu$ cannot
be identified with a nearest-plaquette Gibbs weight of that form.
\end{proposition}

\begin{proof}
Each elementary plaquette has four distinct links, and its fundamental
character has Casimir $3/4$ on each. Therefore
\[
 DW=3W-3P.
\]
Writing $\Gamma(W)=\sum_e\Gamma_e(W)$, direct differentiation gives
\[
 \frac{H e^{-\alpha W}}{e^{-\alpha W}}
 =3\kappa\alpha P+(v-3\kappa\alpha)W
                  -\kappa\alpha^2\Gamma(W).
\]
At the identity configuration, $W=\Gamma(W)=0$.
At a configuration with one link equal to $-I$ and all other links
equal to $I$, $W>0$ and $\Gamma(W)=0$: the first trace derivative of
every center-valued plaquette vanishes.
If the displayed expression were constant, these two configurations
would force $v=3\kappa\alpha$, so $\alpha\ne0$.
It would then also force $\Gamma(W)\equiv0$.
This is impossible because $W$ is nonconstant on the connected
configuration manifold. Thus no such eigenfunction exists.
\end{proof}

This proposition excludes only the displayed single-action ansatz.
It does not exclude effective actions containing other interactions,
nor does it prove that a particular collective estimate fails for
the actual vacuum.

\begin{proposition}[Uniform local bounds need not tensorize]
\label{ham:prop:spec:countertensor}
For each odd integer $N\ge3$, define a strictly positive probability law
on $\{-1,+1\}^N$ by
\[
 \pi_N(\sigma)=Z_N^{-1}
       \exp\left[\frac2N\left(\sum_{i=1}^N\sigma_i\right)^2\right].
\]
Every one-site conditional probability ratio is less than $e^8$,
uniformly in $N$ and the environment. Nevertheless, the smallest
constant $A_N$ in the conditional-variance inequality
\[
 \operatorname{Var}_{\pi_N}(F)
 \le A_N\sum_i\mathbb E_{\pi_N}
          \operatorname{Var}_{\pi_N(\cdot|\sigma_{-i})}(F)
\]
satisfies
\begin{equation}\label{ham:eq:spec:countertensor}
 A_N\ge
 \frac{e^{2N-2/N}}
      {N\binom{N-1}{(N-1)/2}}
 \ge\frac{e^{2N-2/N}}{N2^{N-1}}
 >\frac{2^{N+1}}N.
\end{equation}
\end{proposition}

\begin{proof}
If $s=\sum_{j\ne i}\sigma_j$, the conditional odds are
$e^{8s/N}$, whose value and reciprocal are bounded by
$e^{8(N-1)/N}<e^8$. In particular all one-site densities are uniformly
comparable with the uniform two-point law, and have uniform conditional
Poincare estimates by the same direct density comparison used above.

Take $F(\sigma)=\mathbf1_{\{\sum_i\sigma_i>0\}}$.
Oddness of $N$ and spin-flip symmetry give
$\operatorname{Var}_{\pi_N}(F)=1/4$.
The conditional variance at site $i$ vanishes unless the other
$N-1$ spins sum to zero. In that case it is $1/4$, and the two full
configurations have weight $e^{2/N}$ each. Therefore
\[
 \sum_i\mathbb E_{\pi_N}
       \operatorname{Var}(F|\sigma_{-i})
 =\frac{N\binom{N-1}{(N-1)/2}e^{2/N}}{2Z_N}.
\]
The two completely aligned configurations give $Z_N\ge2e^{2N}$.
Dividing the variance by the displayed sum proves the first bound.
The binomial coefficient is at most $2^{N-1}$. Finally, for $N\ge3$,
$2N-2/N\ge3N/2$, and
$e^{3/2}>1+3/2+(3/2)^2/2+(3/2)^3/6=67/16>4$.
This proves the strict final inequality.
\end{proof}

This is an explicit counterexample to an inference from uniform local
conditional bounds alone. It is a different model, and is not a claim
that the SU(2) vacuum has these correlations. Thus
Theorem~\ref{ham:thm:spec:local} is an unconditional advance, while
\eqref{ham:eq:spec:tensor} remains a separate collective task.

\subsection{What a vacuum-preserving spatial compression must certify}

Let $K=H-E_0\ge0$ act on the physical Hilbert space, with unique normalized
vacuum $\Omega$ and closed quadratic form $\mathfrak q$.
Let $U$ be a spatial isometry preserving the vacuum, and put
$P=UU^*$, $Q=I-P$. Assume that $P$ preserves the form domain.
Then $P\Omega=\Omega$. On the vacuum complement, use the orthogonal
decomposition
\[
 \Omega^\perp=(P\mathcal H\cap\Omega^\perp)\oplus Q\mathcal H
              =:\mathcal H_P\oplus\mathcal H_Q.
\]
Write $\mathcal V_P=D(\mathfrak q)\cap\mathcal H_P$ and
$\mathcal V_Q=D(\mathfrak q)\cap\mathcal H_Q$.
Restrictions of $\mathfrak q$ to these spaces are closed forms.
Let $A$ and $C$ be the associated operators, and let
$\mathfrak b(u,v)=\mathfrak q(u,v)$ denote the cross form.
The entire discarded space is $\mathcal H_Q$; it is not replaced by a
finite list of representations.

The variational principle already fixes the direction of inference:
\begin{equation}\label{ham:eq:spec:ritz}
 \inf\sigma(K|_{\Omega^\perp})
 \le\inf_{\substack{u\in\mathcal V_P\\u\ne0}}
                \frac{\mathfrak q(u)}{\|u\|^2}
 =\inf\sigma(A).
\end{equation}
A positive compressed gap is therefore not, by itself, a lower bound
on the full physical gap.

\begin{theorem}[Complete energy-dependent Schur criterion]
\label{ham:thm:spec:schur}
Suppose $C\ge cI$ for a verified $c>0$, and fix $0\le m<c$.
For every $u\in\mathcal V_P$, there exists a unique
$z_m(u)\in\mathcal V_Q$ satisfying
\[
 \mathfrak q(z_m(u),v)-m\langle z_m(u),v\rangle
       =\mathfrak b(u,v)\qquad(v\in\mathcal V_Q).
\]
Define
\[
 \mathfrak s_m(u)=\mathfrak q(u)-m\|u\|^2
          -\bigl[\mathfrak q(z_m(u))-m\|z_m(u)\|^2\bigr].
\]
Then the following are equivalent:
\begin{enumerate}
 \item $K|_{\Omega^\perp}\ge mI$;
 \item $\mathfrak s_m(u)\ge0$ for every $u\in\mathcal V_P$.
\end{enumerate}
If the cross form is represented by $Bu\in\mathcal H_Q$,
the correction is exactly
\begin{equation}\label{ham:eq:spec:schur}
 \mathfrak s_m(u)
 =\langle u,(A-m)u\rangle_{\rm form}
     -\langle Bu,(C-m)^{-1}Bu\rangle.
\end{equation}
\end{theorem}

\begin{proof}
Since $C\ge cI$ and $m<c$, the form
$\mathfrak c_m(v,w)=\mathfrak q(v,w)-m\langle v,w\rangle$
is a positive complete inner product on $\mathcal V_Q$, equivalent to
its original form norm. The Cauchy--Schwarz inequality for
$\mathfrak q\ge0$ and
$\mathfrak q(v)\le(1-m/c)^{-1}\mathfrak c_m(v)$ make
$v\mapsto\mathfrak b(u,v)$ continuous in that norm.
The Riesz representation theorem gives the unique $z_m(u)$.
Completing the square gives the exact identity
\[
 \mathfrak q(u+v)-m\|u+v\|^2
 =\mathfrak s_m(u)+
      \mathfrak c_m(v+z_m(u),v+z_m(u)).
\]
Its right side is nonnegative for all $u,v$ precisely when
$\mathfrak s_m\ge0$: necessity follows by choosing $v=-z_m(u)$.
The bounded inverse of $C-m$ gives \eqref{ham:eq:spec:schur} whenever
$Bu$ is an ordinary Hilbert-space vector.
\end{proof}

This is the quadratic-form version of the Feshbach--Schur correction;
see Dusson--Sigal--Stamm~\cite{ham:global:dusson2021} for its standard
isospectral formulation. The proof here explicitly includes the
vacuum complement, form domains, and all discarded modes.

\begin{corollary}[One sufficient quantitative certificate]
\label{ham:cor:spec:schur}
Suppose $A\ge a_0I$ and $C\ge c_0I$, where $a_0,c_0>0$.
Assume the actual residual is represented by $Bu$ and satisfies on the
whole retained form domain
\[
 \|Bu\|^2\le\alpha\,\mathfrak q(u)+\beta\|u\|^2,
 \qquad\alpha,\beta\ge0.
\]
If $\alpha a_0+\beta<a_0c_0$, then
\begin{equation}\label{ham:eq:spec:schurbound}
 \Delta\ge
 \frac{a_0+c_0-
       \sqrt{(a_0-c_0)^2+4(\alpha a_0+\beta)}}2>0.
\end{equation}
For a bounded cross operator $\|B\|\le\eta$, this applies with
$\alpha=0$, $\beta=\eta^2$.
\end{corollary}

\begin{proof}
For $m<c_0$,
\[
 \langle Bu,(C-m)^{-1}Bu\rangle
 \le\frac{\alpha\mathfrak q(u)+\beta\|u\|^2}{c_0-m}.
\]
If $c_0-m>\alpha$, the Schur form is bounded below by
\[
 \left[a_0-m-\frac{\alpha a_0+\beta}{c_0-m}\right]\|u\|^2.
\]
The smaller root of
$(a_0-m)(c_0-m)=\alpha a_0+\beta$ is the positive number
in \eqref{ham:eq:spec:schurbound}. For $m$ below that root the displayed
coefficient is nonnegative and $c_0-m>\alpha$.
Theorem~\ref{ham:thm:spec:schur}, followed by $m$ increasing to the root,
proves the result, including the case of zero coupling correction.
\end{proof}

If a finite spectral calculation of the correction is used, its omitted
part also needs a certificate. For example, if $\Pi$ is a reducing
projection of $C$ and $C\ge\Lambda I$ on $(I-\Pi)\mathcal H_Q$, where
$\Lambda>m$, then
\[
 \bigl\langle (I-\Pi)Bu,(C-m)^{-1}(I-\Pi)Bu\bigr\rangle
 \le\frac{\|(I-\Pi)Bu\|^2}{\Lambda-m}.
\]
An unverified cutoff cannot replace this estimate. In particular, neither
coarse vacuum preservation nor an attractive compressed gap proves
$C\ge c_0$, controls $B$, or supplies the displayed tail denominator.

\subsection{Explicit hidden-mode examples with the vacuum handled exactly}

First, take
\[
 K_\varepsilon=\operatorname{diag}(0,1,\varepsilon)
 \quad\hbox{on }\mathbb C^3,\qquad0<\varepsilon<1,
\]
and let $U:\mathbb C^2\to\mathbb C^3$ include the first two coordinates.
The vacuum is preserved exactly, $U^*K_\varepsilon U=\operatorname{diag}(0,1)$,
and the retained space is exactly invariant:
$QK_\varepsilon U=0$. Even
$U^*e^{-tK_\varepsilon}U=e^{-t\,\operatorname{diag}(0,1)}$
holds for every $t\ge0$. Nevertheless the full gap is
$\varepsilon$, while the compressed gap is one.
The entire missing information is the discarded eigenvalue.

There is an infinite-dimensional compact-group version. On two
SU(2) class-function rotors, let
\[
 K_\varepsilon=\frac43\bigl(\mathcal C_1+\varepsilon\mathcal C_2\bigr).
\]
Its unique vacuum is the constant function. The map
$Uf(U_1,U_2)=f(U_1)$ preserves that vacuum, has exactly vanishing
residual, and compresses to $(4/3)\mathcal C_1$, with gap one.
The physical class-function state
$\chi_{1/2}(U_2)$ has energy $\varepsilon$, so the full gap is
$\varepsilon$. These two rotors are a counterexample to a proposed
spectral inference, not a replacement for the specified Wilson model.

Second, knowledge that both diagonal blocks are uniformly gapped is
also insufficient without the cross correction. With the same vacuum
and retained space, set
\[
 \widetilde K_\varepsilon=
 \begin{pmatrix}
 0&0&0\\
 0&1&1-\varepsilon\\
 0&1-\varepsilon&1
 \end{pmatrix}.
\]
Both $A=C=1$, but the nonvacuum eigenvalues are
$\varepsilon$ and $2-\varepsilon$.
The Schur form is
\[
 S_m=1-m-\frac{(1-\varepsilon)^2}{1-m},
\]
and the quantitative bound \eqref{ham:eq:spec:schurbound} is exact:
$\Delta=1-|1-\varepsilon|=\varepsilon$.
Thus complete discarded-mode and residual control have distinct roles.
These examples do not rule out a successful renormalization construction;
they specify what such a construction must actually verify.

\section{Complete local discarded-sector bound}
\label{ham:sec:local-discarded}

The discarded operator in a spatial compression requires a separate lower bound. Such a bound can be obtained when any fixed finite set of link variables is eliminated.
The bound includes every representation of those links and is
independent of the total spatial volume.

Let $X=SU(2)^E$ be a finite periodic cubic configuration space, $L\ge2$,
with normalized Haar measure $\mu$. Use
\[
 H=\kappa D+vW,\qquad
 \kappa=\frac{g^2}{2a},\quad v=\frac4{g^2a},\quad
 D=\sum_e J_e^2,\qquad K=H-E_0.
\]
Let $h>0$ be the normalized actual ground state and
$d\nu=h^2d\mu$. Suppose $R(g)$ is a verified uniform one-link
amplitude ratio:
\begin{equation}
 \frac{\max_{U_e}h(U_e,U_{-e})}
      {\min_{U_e}h(U_e,U_{-e})}\le R(g).
 \label{ham:eq:delete-local-ratio}
\end{equation}
The actual-vacuum comparison recalled below supplies
\[
 R(g)=\frac53e^{256/g^4}\quad(g>0).
\]
For $0<g^2\le1$, the complete small-time estimate reproduced in
Section~\ref{ham:sec:small-time} improves this to
$R(g)=10e^{16\sqrt{23}/g^2}$.

Choose a nonempty set $S$ of $q$ links, and write $U=(y,z)$ for the
retained and deleted coordinates. Define
\[
 w(y)=\int h(y,z)^2d\mu_S(z),\qquad
 h_{\mathrm{eff}}(y)=\sqrt{w(y)},\qquad
 (J_S\psi)(y,z)=h(y,z)\frac{\psi(y)}{h_{\mathrm{eff}}(y)}.
\]
Then $J_S$ is an isometry preserving the actual vacuum. In the
ground-state representation its range projection is
\[
 P_S(hF)=h\,\mathbb E_\nu[F\mid y],\qquad Q_S=I-P_S.
\]
Thus the full discarded space consists of vectors $hF$ with
$\mathbb E_\nu[F\mid y]=0$.

\begin{theorem}[Full discarded-sector gap for a fixed local deletion]
\label{ham:thm:local-discarded}
Let $C_S$ be the selfadjoint operator associated with the restriction
of the closed form of $K$ to $Q_S\mathcal H_{\mathrm{phys}}$.
Then
\begin{equation}
 \boxed{C_S\ge c_S(a,g)I,\qquad
 c_S(a,g)=\frac{3\kappa}{4}R(g)^{-2q}>0.}
 \label{ham:eq:delete-gap}
\end{equation}
In particular,
\begin{equation}
 C_S\ge
 \frac{3g^2}{8a}\left(\frac35\right)^{2q}
                 e^{-512q/g^4}I
 \qquad(g>0).
 \label{ham:eq:delete-gap-allg}
\end{equation}
For $0<g^2\le1$, the sharper estimate is
\begin{equation}
 C_S\ge\frac{3g^2}{8a}\,100^{-q}
               e^{-32q\sqrt{23}/g^2}I.
 \label{ham:eq:delete-gap-weak}
\end{equation}
These are actual bounds for the entire discarded operator, not
hypotheses on a finite matrix approximation. They are uniform in
the surrounding box and retained configuration when $q$ is fixed.
\end{theorem}

\begin{proof}
Changing the $q$ deleted coordinates one at a time in
\eqref{ham:eq:delete-local-ratio} proves that the amplitude ratio on
each full fiber is at most $R^q$. Therefore the conditional
density $p_y(z)=h(y,z)^2/w(y)$ has
\[
 \frac{\max_zp_y(z)}{\min_zp_y(z)}\le R^{2q}.
\]
The product Haar Casimir $\sum_{e\in S}J_e^2$ has first
nonconstant eigenvalue $3/4$, independent of $q$: its spectrum
is the sum of the one-link spectra, and exactly one spin-$1/2$
factor realizes the smallest positive value.
Direct density comparison consequently gives
\[
 \int\sum_{e\in S}|\nabla_eF|^2p_y\,d\mu_S
 \ge\frac34R^{-2q}\operatorname{Var}_{p_y\mu_S}(F).
\]
For $hF\in Q_S\mathcal H$, the conditional mean vanishes, so the
variance on the right is its conditional squared norm.
The exact ground-state identity now yields
\begin{align*}
 \langle hF,KhF\rangle_{\mathrm{form}}
 &=\kappa\int\sum_e|\nabla_eF|^2d\nu\\
 &\ge\kappa\int\sum_{e\in S}|\nabla_eF|^2d\nu
 \ge\frac{3\kappa}{4}R^{-2q}\|hF\|^2.
\end{align*}
This proves the form lower bound and the asserted operator inequality.

The domain and gauge assertions do not require a gauge fixing.
For a fixed finite box, $h$, $h_{\mathrm{eff}}$ and their inverses
are smooth bounded multipliers. Integration over the compact
deleted variables maps first Sobolev functions to first Sobolev
functions; differentiating the numerator of $J_S^*$ proves this
directly. Hence $P_S$ preserves the form domain. Its intersection
with $Q_S\mathcal H$ is dense there and is closed in the form norm,
so the restricted form defines $C_S$.
Every vertex gauge transformation preserves the retained-coordinate
sigma-algebra, and $\nu$ is invariant. Conditional expectation
therefore commutes with the gauge action. Thus $P_S,Q_S$ preserve
the full physical sector, and $Q_Sh=0$. Restriction to that sector
does not weaken the proved lower bound.
\end{proof}

\paragraph{Scope of the improvement.}
For any fixed $q$, this supplies an unconditional $c_0$ for the
full complementary block in a local Schur step, independently of
total volume. A bound on the retained-to-discarded coupling is
still necessary to obtain a lower bound on the full gap.
If $q$ grows with the number of eliminated fine links, the
explicit bound again decays exponentially in $q$.
For fixed $q$, positivity along a spatial-cutoff trajectory still requires comparison of the exponential coupling dependence with $a^{-1}$. The theorem supplies a local elimination estimate; it does not complete spatial renormalization.

\section{Local vacuum and discarded-sector refinement}
\label{ham:local_refinement}

The local Feynman--Kac comparison can be combined with the sharper
one-link heat-kernel estimate to improve every weak-coupling bound
that uses the conditional oscillation of the actual vacuum.
This refinement retains a constant independent of the surrounding
spatial volume. It remains a local estimate: the collective
variance inequality needed for a full volume-uniform gap is a
separate condition.

\begin{theorem}[Improved actual-vacuum local bounds]
\label{ham:local_refinement_theorem}
Let $L\ge2$, $a>0$, $0<s=g^2\le1$, and let $h>0$ be the
normalized ground state of
\[
 H=\frac{s}{2a}\sum_e\mathcal C_e+
       \frac4{sa}\sum_p(1-F_p),\qquad d\nu=h^2d\mu.
\]
On the periodic cubic spatial lattice, for every link $e$ and
every fixed exterior configuration,
\begin{equation}\label{ham:local_refinement_ratio}
 \frac{\max_{U_e}h}{\min_{U_e}h}
 \le R_{\rm loc}(s):=\frac4{\sqrt\pi}
                         \exp\left(\frac{16\pi}{s}\right).
\end{equation}
The conditional Casimir Poincar\'e inequality holds with
\begin{equation}\label{ham:local_refinement_gamma}
 \gamma_{\rm loc}(s)=\frac{3\pi}{64}
                         \exp\left(-\frac{32\pi}{s}\right),
 \qquad
 \operatorname{Var}_{\nu_e}(f)
 \le\gamma_{\rm loc}(s)^{-1}\int\Gamma_e f\,d\nu_e.
\end{equation}
If $F$ is the normalized trace of a loop containing a link exactly
once, then
\begin{equation}\label{ham:local_refinement_variance}
 \operatorname{Var}_\nu F\ge\frac\pi{64}
                        \exp\left(-\frac{32\pi}{s}\right).
\end{equation}
All three bounds include the full representation space.
\end{theorem}

\begin{proof}
For a link incident on $d_e\le4$ plaquettes, write
$H=\kappa\mathcal C_e+H_{-e}+V_e$, where $\kappa=s/(2a)$ and
$0\le V_e\le8d_e/(sa)$. At physical time $t=\tau/\kappa$,
the positive heat-kernel comparison and the ground-state equation
give, exactly as in Theorem~\ref{ham:thm:spec:local},
\[
 \frac{\max_{U_e}h}{\min_{U_e}h}
 \le\frac{k_{\tau,\max}}{k_{\tau,\min}}
                     \exp\left(\frac{16d_e\tau}{s^2}\right).
\]
The environmental integral and the unknown ground energy cancel in
this ratio. Lemma~\ref{ham:lem:heat:kernel} bounds the heat-kernel
ratio by $(4/\sqrt\pi)e^{\pi^2/\tau}$ for $0<\tau\le1$.
Choose $\tau=\pi s/8\le\pi/8<1$. Then, using $d_e\le4$,
\[
 \frac{16d_e\tau}{s^2}+\frac{\pi^2}{\tau}
 \le\frac{64\tau}{s^2}+\frac{\pi^2}{\tau}
 =\frac{16\pi}{s}.
\]
This proves \eqref{ham:local_refinement_ratio}; no small-time
asymptotic expansion or representation truncation was used.

The conditional density relative to one-link Haar measure has
maximum-to-minimum ratio at most $R_{\rm loc}^2$. Comparison with
the exact Haar Casimir gap $3/4$ gives
$\gamma_{\rm loc}=(3/4)R_{\rm loc}^{-2}$, which is
\eqref{ham:local_refinement_gamma}. Its normalized density is also
at least $R_{\rm loc}^{-2}$. At fixed exterior links, the loop
holonomy is a translated or inverse-translated Haar variable; its
normalized trace has Haar mean zero and variance $1/4$.
Conditional variance comparison, followed by total variance, gives
$\operatorname{Var}_\nu F\ge(4R_{\rm loc}^2)^{-1}$.
\end{proof}

The same ratio passes to the exact blocked vacuum marginal by
varying the last fine link in each ordered coarse path. Thus
\eqref{ham:local_refinement_ratio} also sharpens the local input
for the vacuum-adapted spatial maps.
It is strictly stronger than $10e^{16\sqrt{23}/s}$: the exact
inequalities $\pi>3$ and $\pi^2<10<23$ give
$4/\sqrt\pi<10$ and $\pi<\sqrt{23}$.
The polynomial loop-variance estimate of
Theorem~\ref{ham:thm:fisher:variance} is a different, generally
stronger marginal norm estimate; it does not replace the conditional
density control proved here.

\begin{corollary}[Every representation in a fixed deleted link set]
\label{ham:local_refinement_discarded}
Let $S$ be any nonempty set of $q$ deleted link variables in the
same finite spatial lattice. In the actual ground-state
representation define
\[
 P_S(hf)=h\,\mathbb E_\nu[f\mid U_{S^c}],\qquad Q_S=I-P_S.
\]
Let $C_S$ be the selfadjoint operator associated with the
restriction of the closed form of $H-E_0$ to
$Q_S\mathcal H_{\rm phys}$. Then
\begin{equation}\label{ham:local_refinement_discarded_bound}
 \boxed{\displaystyle
 C_S\ge c_{S,\rm loc}(a,s)I,\qquad
 c_{S,\rm loc}(a,s)=\frac{3s}{8a}
       \left(\frac\pi{16}\right)^q
       \exp\left(-\frac{32\pi q}{s}\right)>0.}
\end{equation}
The bound is uniform in the exterior configuration and total
volume for fixed $q$. The discarded space includes all
representations and all gauge-invariant vectors satisfying the
indicated conditional mean-zero condition.
\end{corollary}

\begin{proof}
Changing the $q$ coordinates successively gives a conditional
amplitude ratio at most $R_{\rm loc}^q$, hence a density ratio
at most $R_{\rm loc}^{2q}$. The product-Haar Casimir gap is
$3/4$ independently of $q$. For a vector $hf$ with
$\mathbb E_\nu[f\mid U_{S^c}]=0$, conditional comparison and the
exact ground-state form give
\[
 \langle hf,(H-E_0)hf\rangle_{\rm form}
 =\frac{s}{2a}\int\sum_e\Gamma_e f\,d\nu
 \ge\frac{3s}{8a}R_{\rm loc}^{-2q}\|hf\|^2.
\]
The conditional projection preserves the form domain at each
finite lattice: the vacuum and its positive marginal square root
are smooth bounded multipliers, and integration over the compact
deleted coordinates preserves first Sobolev regularity. Gauge
invariance of $\nu$ and preservation of the retained-coordinate
sigma algebra imply that this projection commutes with gauge
transformations. Its discarded range is orthogonal to the actual
vacuum. The restricted closed form therefore defines $C_S$, and
the proved inequality gives its asserted spectral floor.
Substitution of $R_{\rm loc}$ yields
\eqref{ham:local_refinement_discarded_bound}.
\end{proof}

For the conditional-block collective criterion, this refinement
also permits the seed
\[
 \gamma_B=\frac34\left(\frac\pi{16}\right)^{|B|}
                       e^{-32\pi|B|/s}.
\]
The conditional projection angles, retained-to-discarded coupling,
and control as $a\to0$ remain distinct estimates. A lower bound
on $C_S$ is not by itself a lower bound on the entire physical
vacuum complement.

\subsection{Exact evaluations and complete arithmetic errors}

The accompanying calculation evaluates logarithms of the explicit
bound functions to avoid numerical underflow. It uses Machin's
identity for $\pi$, with the signed first omitted term of each
alternating arctangent series. After binary scaling to $1\le x<2$,
it evaluates
\[
 \log x=2\sum_{j=0}^{N-1}\frac{z^{2j+1}}{2j+1}+r_N,
 \quad z=\frac{x-1}{x+1},\quad
 0\le r_N\le\frac{2z^{2N+1}}{(2N+1)(1-z^2)}.
\]
Every endpoint is a rational number. Intermediate rounding is
outward to $40$ decimal places, and the printed endpoints are
outward to $12$ decimal places. The certificate records all
formulas, rational endpoint pairs and complete series-tail rules.
For example, the enclosures for $\log_{10}(a c_{S,\rm loc})$ are
\begin{center}
\begin{tabular}{cccc}
\toprule
$s$ & $q$ & Lower endpoint & Upper endpoint\\
\midrule
$1$ & $1$ & $-44.792982165173$ & $-44.792982165172$\\
$1/2$ & $1$ & $-88.754055483776$ & $-88.754055483775$\\
$1/4$ & $1$ & $-176.375172125318$ & $-176.375172125317$\\
$1$ & $4$ & $-177.894022463876$ & $-177.894022463875$\\
$1/4$ & $4$ & $-702.416602330471$ & $-702.416602330470$\\
\bottomrule
\end{tabular}
\end{center}
These intervals enclose certified lower-bound functions. They are
not two-sided enclosures of the true discarded-sector gap or
measurements of a physical mass.

\begin{figure}[htbp]
\centering
\includegraphics[width=\textwidth]{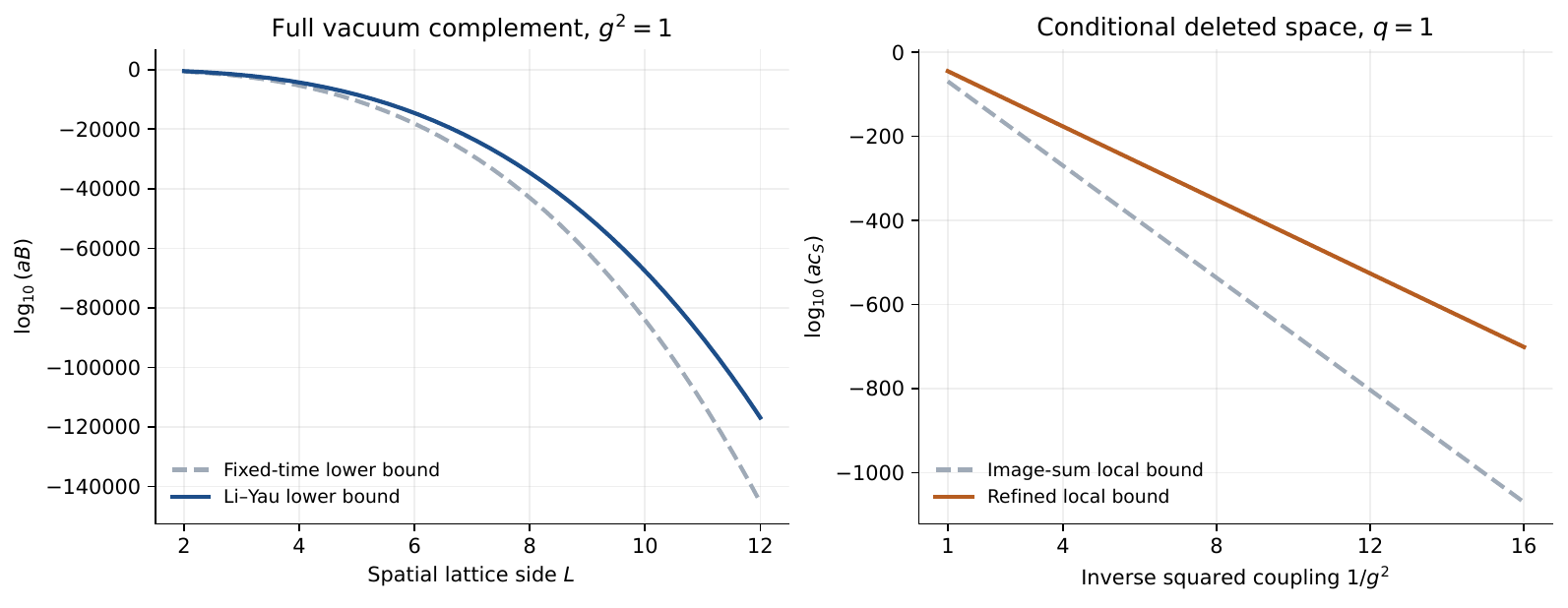}
\caption{Comparison of explicit analytic lower-bound functions. Left:
full-complement bounds at $g^2=1$ versus spatial lattice side $L$.
Right: conditional deleted-space bounds for one deleted link versus
$1/g^2$. The latter bounds are independent of the surrounding volume;
they apply to the indicated conditional mean-zero sector. The curves
are numerical illustrations of the proved formulas. Exact outward
rational enclosures for selected values are given in the tables and
calculation data. Neither panel estimates the true physical gap.}
\label{ham:fig:explicit_bound_comparison}
\end{figure}

\section{Exact spatial refinement maps and their eliminated dynamics}
\label{ham:srmap:section}

This section constructs maps between different spatial cutoffs and
computes their compression and discarded terms exactly.  The bare Haar
map and the map adapted to the true fine vacuum have different coarse
vacua.  Neither compression is identified with a continuum Hamiltonian.
All calculations retain the full representation space.

\subsection{Spatial geometry, gauge symmetry and Haar isometry}

Fix an integer $L\ge2$, an integer refinement factor $r\ge2$, and a
coarse spacing $a>0$.  The coarse graph is the periodic cubic lattice
$\Lambda_c=(\mathbb Z/L\mathbb Z)^3$ and the fine graph is
$\Lambda_f=(\mathbb Z/(rL)\mathbb Z)^3$, of spacing $a_f=a/r$.
They describe the same physical three-torus.  Each graph has three
positively oriented links and three elementary plaquettes per vertex.
Write
\[
 E_c=P_c=3L^3,\qquad E_f=P_f=3(rL)^3,\qquad
 X_c=SU(2)^{E_c},\quad X_f=SU(2)^{E_f},
\]
and let $\mu_c,\mu_f$ be normalized product Haar measures.  Opposite
orientations refer to inverse group elements, not additional variables.

For a coarse positive link $e=(x,i)$, its fine skeleton path is
\[
 {\cal P}_e=\{(rx+s\widehat i,i):s=0,\ldots,r-1\}.
\]
These paths are pairwise disjoint as sets of fine link variables.
They may meet at vertices.  In fine integer coordinates, a direction
$i$ link belongs to the skeleton precisely when its two transverse
coordinates are multiples of $r$.  Define the blocking map
\begin{equation}\label{ham:srmap:block}
 (BU)_e=U_{e,1}U_{e,2}\cdots U_{e,r}.
\end{equation}
The factors are in their path order.  The Haar pullback is
\begin{equation}\label{ham:srmap:haar_embedding}
 (I\psi)(U)=\psi(BU).
\end{equation}

\begin{proposition}[Exact Haar and gauge structure]\label{ham:srmap:isometry}
The map $I:L^2(X_c,\mu_c)\to L^2(X_f,\mu_f)$ is an isometry.  Its
adjoint is Haar conditional expectation on $B$:
\[
 I^*F(v)=\mathbb E_{\mu_f}[F\mid BU=v].
\]
It intertwines the fine gauge action with its restriction to the
coarse vertices.  In particular it maps the full coarse gauge-invariant
Hilbert space into the full fine gauge-invariant Hilbert space.
No center-flux sector is removed.
\end{proposition}

\begin{proof}
For each disjoint skeleton path, replace its last variable by its
ordered product $v_e$.  Given the other $r-1$ variables, the last is
$(U_{e,1}\cdots U_{e,r-1})^{-1}v_e$.  Haar invariance makes this a
smooth global product-coordinate change with Jacobian one.  Together
with all unused fine links, these remaining coordinates form
$Z=SU(2)^{E_f-E_c}$, and
\begin{equation}\label{ham:srmap:haar_disintegration}
 X_f\simeq X_c\times Z,\qquad d\mu_f=d\mu_c\,d\mu_Z.
\end{equation}
Consequently $B_*\mu_f=\mu_c$, proving the isometry and adjoint.

Under a fine vertex gauge transformation $q$, all intermediate
endpoint factors on a path cancel, leaving
\[
 B(q\cdot U)_{(x,i)}
 =q_{rx}(BU)_{(x,i)}q_{r(x+\widehat i)}^{-1}.
\]
Every coarse gauge transformation can be extended to the fine
vertices.  This proves equivariance and the assertion about the
physical subspaces.
\end{proof}

\subsection{Exact electric compression and erased magnetic expectation}

Choose an orthonormal anti-Hermitian Lie-algebra basis $T_1,T_2,T_3$
in the convention for which the Casimir has spin-$j$ eigenvalue
$j(j+1)$.  Let $X_{\ell,\alpha}$ denote differentiation under left
multiplication $U_\ell\mapsto e^{tT_\alpha}U_\ell$.  Set
\[
 D_c=-\sum_{e,\alpha}X_{e,\alpha}^2,\qquad
 D_f=-\sum_{\ell,\alpha}X_{\ell,\alpha}^2.
\]
These are nonnegative selfadjoint sums of link Casimirs.  With one
orientation of each elementary plaquette, define
\[
 f_p(U)=\tfrac12\operatorname{Tr}U_p,\qquad
 W_f(U)=\sum_{p\in\Lambda_f}(1-f_p(U)).
\]
For a finite coupling $g_f>0$, the fine Hamiltonian is
\begin{equation}\label{ham:srmap:fine_H}
 H_f=\kappa_fD_f+v_fW_f,\qquad
 \kappa_f=\frac{g_f^2}{2a_f},\quad
 v_f=\frac{4}{g_f^2a_f}.
\end{equation}
It acts on the gauge-invariant subspace when physical spectra are
discussed.  On this fixed finite graph it is selfadjoint, elliptic with
bounded smooth potential, and has compact resolvent and a unique
normalized strictly positive gauge-invariant ground state.

\begin{theorem}[Exact bare compression and discarded magnetic norm]
\label{ham:srmap:bare}
On smooth cylinder states, and subsequently on the corresponding
operator domains,
\begin{align}
 D_fI&=rID_c,\label{ham:srmap:electric_intertwining}\\
 I^*W_fI&=P_f\,\mathbf1,\label{ham:srmap:magnetic_mean}\\
 I^*(W_f-P_f)^2I&=\frac{P_f}{4}\,\mathbf1.\label{ham:srmap:magnetic_variance}
\end{align}
The cylinder subspace $\operatorname{Ran}I$ reduces $D_f$.
If $P_0=II^*$ and $Q_0=\mathbf1-P_0$, then
\begin{align}
 I^*H_fI&=r\kappa_fD_c+v_fP_f\,\mathbf1,\label{ham:srmap:bare_H}\\
 Q_0H_fI&=v_f(W_f-P_f)I,\label{ham:srmap:bare_residual}\\
 \|Q_0H_fI\psi\|^2&=\frac{v_f^2P_f}{4}\|\psi\|^2.
 \label{ham:srmap:bare_residual_norm}
\end{align}
The last two expressions extend to bounded maps from the entire coarse
Hilbert space.  All these statements remain true on the physical
gauge-invariant subspaces.
\end{theorem}

\begin{proof}
For a path variable numbered $s$, put
$A_{e,s}=U_{e,1}\cdots U_{e,s-1}$, with $A_{e,1}=\mathbf1$.
Let $R_{\beta\alpha}(A)$ be the orthogonal matrix of
$\operatorname{Ad}_A$ in the chosen Lie-algebra basis.  Differentiation
of the ordered product gives
\begin{equation}\label{ham:srmap:derivative}
 X_{e,s,\alpha}(F\circ B)
 =\sum_{\beta=1}^3R_{\beta\alpha}(A_{e,s})
                        (X_{e,\beta}F)\circ B.
\end{equation}
The coefficient $R(A_{e,s})$ is independent of the current variable
$U_{e,s}$, so its derivative under $X_{e,s,\alpha}$ is zero.
Summing a second derivative over $\alpha$ and using orthogonality
therefore gives exactly one coarse Casimir for each of the $r$ path
variables.  Derivatives at unused fine links vanish.  This proves
\eqref{ham:srmap:electric_intertwining}.  Equivalently, for every smooth $F$,
\begin{equation}\label{ham:srmap:pointwise_energy}
 \sum_{\ell,\alpha}|X_{\ell,\alpha}(F\circ B)|^2
 =r\left(\sum_{e,\beta}|X_{e,\beta}F|^2\right)\circ B
\end{equation}
pointwise, before any integration or approximation.

For the domain and reducing statement, expand a coarse vector in
product Peter--Weyl eigenvectors of $D_c$.  Their isometric images
are $D_f$ eigenvectors with $r$ times the eigenvalue by the identity
just proved.  Completeness in $\operatorname{Ran}I$ shows that this
range and its orthogonal complement are spectral invariant for
$D_f$.  Equivalently the identity
$e^{-tD_f}I=Ie^{-rtD_c}$ holds on the entire Hilbert space.
Closedness gives the operator-domain intertwining.

Every elementary fine plaquette has at least two links outside the
skeleton.  In an $i,j$ plaquette, its two $i$ links have transverse
$j$ coordinates differing by one; they cannot both be multiples of
$r\ge2$.  The two $j$ links give a second unused link by the same
argument.  Such variables are absent from $BU$ and are independent
Haar variables under \eqref{ham:srmap:haar_disintegration}.
Integration over any one of them yields
\[
 \mathbb E_{\mu_f}[f_p\mid B]=0,\qquad
 \mathbb E_{\mu_f}[f_p^2\mid B]=\tfrac14.
\]
The first identity is the vanishing Haar integral of a fundamental
matrix coefficient.  The second is
$\int(\operatorname{Tr}U/2)^2\,dU=1/4$, since the real fundamental
character has squared Haar norm one.

Two distinct elementary fine plaquettes share at most one unoriented
link when $rL\ge4$.  In different coordinate planes their common
edges can lie on only one unit line segment; in the same plane two
distinct unit squares share at most one edge.  The periodic
identification at side length at least four creates no additional
coincidences between these unit squares.  Since each plaquette has
at least two unused links, for $p\ne q$ there is an unused link
belonging to $p$ but not $q$.  Integrating it proves
\[
 \mathbb E_{\mu_f}[f_pf_q\mid B]=0.
\]
The conditional mean and second moment of
$W_f-P_f=-\sum_pf_p$ are consequently $0$ and $P_f/4$.  This proves
\eqref{ham:srmap:magnetic_mean}--\eqref{ham:srmap:magnetic_variance}.
The reducing electric subspace and these identities give
\eqref{ham:srmap:bare_H}--\eqref{ham:srmap:bare_residual_norm}.
Finally $W_f$, the gauge projectors and $D_f$ preserve the physical
spaces, so restriction does not change any of the computations.
\end{proof}

The magnetic interaction is thus erased by the \emph{compressed bare
expectation}, while the coupling to the discarded space has the
nonzero exact norm
\begin{equation}\label{ham:srmap:bare_residual_size}
 \|Q_0H_fI\|
 =\frac{v_f\sqrt{P_f}}2
 =\frac{2\sqrt{3}\,r^{5/2}L^{3/2}}{g_f^2a}.
\end{equation}
No exact interacting eigenvector of $H_f$ can lie entirely in
$\operatorname{Ran}I$: if $H_fI\psi=E I\psi$, its $Q_0$ component
would contradict \eqref{ham:srmap:bare_residual_norm}.  In particular
the true fine vacuum is absent from this bare cylinder subspace.
This is an obstruction to this specified map and compression, not
to spatial renormalization in general.

The compressed electric coefficient is
$r\kappa_f=r^2g_f^2/(2a)$.  Matching it to an unmodified coarse
coefficient $g_c^2/(2a)$ by compression alone would require
$g_c^2=r^2g_f^2$.  This is not a renormalization-group law: the full
eliminated dynamics can also alter the effective kinetic operator,
as the nonzero discarded term already shows.

\subsection{The exact isometry adapted to the actual fine vacuum}

Let $h_f>0$ be the actual normalized ground state of
\eqref{ham:srmap:fine_H}, of energy $E_f^0$, and set
$K_f=H_f-E_f^0$.  Define the true fine-vacuum probability measure and
its coarse marginal by
\begin{equation}\label{ham:srmap:vacuum_marginal}
 d\nu_f=h_f^2\,d\mu_f,\qquad
 w(v)=\mathbb E_{\mu_f}[h_f^2\mid BU=v],\qquad
 h_{\mathrm{eff}}(v)=\sqrt{w(v)},\qquad d\nu_c=w\,d\mu_c.
\end{equation}
The fiber coordinates \eqref{ham:srmap:haar_disintegration} make these
ordinary compact integrals.  Thus $w$ and $h_{\mathrm{eff}}$ are
smooth, strictly positive, normalized and gauge invariant.

\begin{theorem}[Vacuum-adapted spatial map]\label{ham:srmap:vacuum_map}
The map
\begin{equation}\label{ham:srmap:vacuum_embedding}
 (U_h\psi)(U)=h_f(U)\,
       \frac{\psi(BU)}{h_{\mathrm{eff}}(BU)}
\end{equation}
is an isometry $L^2(\mu_c)\to L^2(\mu_f)$, is gauge equivariant,
intertwines multiplication by every bounded coarse observable with
its pullback, and satisfies
$U_hh_{\mathrm{eff}}=h_f$.  Its adjoint and range projection are
\begin{align}
 U_h^*F(v)&=\frac{\mathbb E_{\mu_f}[h_fF\mid B=v]}
                              {h_{\mathrm{eff}}(v)},\label{ham:srmap:adjoint}\\
 P_hF(U)&=h_f(U)\left(
       \frac{\mathbb E_{\mu_f}[h_fF\mid B]}{w}\right)(BU),
 \qquad P_h=U_hU_h^*.\label{ham:srmap:projection}
\end{align}
In the weighted representation $F=h_fG$, this projection is exactly
conditional expectation:
\[
 h_f^{-1}P_h(h_fG)
       =\bigl(\mathbb E_{\nu_f}[G\mid B]\bigr)\circ B.
\]
Both $U_h:H^1(X_c)\to H^1(X_f)$ and
$U_h^*:H^1(X_f)\to H^1(X_c)$ are bounded at these fixed finite
parameters.  In particular, $P_h$ preserves the full form domain
of $K_f$, including its gauge-invariant restriction.
\end{theorem}

\begin{proof}
The squared norm of \eqref{ham:srmap:vacuum_embedding} is
$\int h_f^2|\psi\circ B|^2/(w\circ B)\,d\mu_f
=\int|\psi|^2\,d\mu_c$.  The same disintegration with two test
vectors proves the adjoint and projection formulas.  Gauge
invariance of the fine vacuum and equivariance of $B$ imply gauge
invariance of its coarse marginal and equivariance of the map.
The multiplication intertwining and the vacuum identity follow
directly from the formula.

For the domain assertion, pullback and the smooth positive
multipliers in $U_h$ preserve first Sobolev regularity.  For
$i=(e,\beta)$ define
$Y_i=\sum_{s,\alpha}R_{\beta\alpha}(A_{e,s})X_{e,s,\alpha}$.
Each coefficient is independent of its own differentiated variable,
so this vector field is Haar divergence free, and
$Y_i(F\circ B)=r(X_iF)\circ B$.  If
$n(v)=\mathbb E_{\mu_f}[h_fF\mid B=v]$, integration by parts gives
the weak derivative identity
\[
 X_i n=\frac1r\,\mathbb E_{\mu_f}[Y_i(h_fF)\mid B].
\]
Haar conditional expectation contracts $L^2$.  The vector-field
coefficients, $h_f$, $h_{\mathrm{eff}}^{-1}$ and their needed first
derivatives are bounded on their compact spaces.  The identity
therefore bounds the $H^1$ norm of $U_h^*F=n/h_{\mathrm{eff}}$ by a
finite constant times the $H^1$ norm of $F$.  Density extends this
argument from smooth functions and proves the projection assertion.
\end{proof}

Here $h_{\mathrm{eff}}$ is the \emph{induced coarse vacuum}.  It has
not been identified with the vacuum of a coarse Wilson Hamiltonian
at some scalar coupling.  The following exact criterion prevents
silently making that identification.

\begin{proposition}[When a specified coarse vacuum can be preserved]
\label{ham:srmap:vacuum_criterion}
Let $h_c>0$ be a specified normalized coarse wavefunction.  There
exists an isometry $J:L^2(\mu_c)\to L^2(\mu_f)$ such that
$Jh_c=h_f$ and
\[
 J M_F=M_{F\circ B}J
 \quad\hbox{for every bounded coarse multiplication observable }F
\]
if and only if $h_c^2=w$ almost everywhere.  When it exists, this
map is uniquely \eqref{ham:srmap:vacuum_embedding}.
\end{proposition}

\begin{proof}
The two required identities force
$\int Fh_c^2\,d\mu_c=\int(F\circ B)h_f^2\,d\mu_f=\int Fw\,d\mu_c$
for every bounded $F$, which is the necessary marginal equality.
When it holds the displayed map provides sufficiency.  The vectors
$Fh_c$ with bounded $F$ are dense in $L^2(\mu_c)$, and the
intertwining identity forces their images to be
$(F\circ B)h_f$.  This proves uniqueness.
\end{proof}

\begin{corollary}[Local vacuum control survives exact spatial blocking]
\label{ham:srmap:local_inheritance}
Every coarse marginal in \eqref{ham:srmap:vacuum_marginal} obeys the same
one-link vacuum ratio bound as the fine vacuum in
Theorem~\ref{ham:thm:spec:local}.  In particular, for all fixed values of
the other coarse links,
\begin{equation}\label{ham:srmap:local_ratio}
 \frac{\max_{v_e}h_{\mathrm{eff}}(v_e,v_{-e})}
      {\min_{v_e}h_{\mathrm{eff}}(v_e,v_{-e})}
 \le R_*(g_f):=\frac53\exp\left(\frac{256}{g_f^4}\right).
\end{equation}
This bound is independent of $L$, $a_f$ and $r$.
The coarse conditional one-link Poincare constant is consequently
at least $(3/4)R_*(g_f)^{-2}$.  If
$F(v)=\operatorname{Tr}v_{\ell}/2$ is a coarse Wilson loop
containing some coarse link exactly once, then
\begin{equation}\label{ham:srmap:coarse_loop_variance}
 \operatorname{Var}_{\nu_c}(F)
 \ge\frac1{4R_*(g_f)^2}
 =\frac9{100}\exp\left(-\frac{512}{g_f^4}\right)>0.
\end{equation}
Any sharper one-fine-link ratio bound in Theorem~\ref{ham:thm:spec:local}
passes to the coarse marginal in the same manner.
\end{corollary}

\begin{proof}
In the fiber coordinates \eqref{ham:srmap:haar_disintegration}, varying
one coarse holonomy $v_e$ while keeping the fiber coordinates and
all other coarse holonomies fixed changes only the last fine link
on that path.  The fine one-link ratio bound holds pointwise in
every such environment.  Squaring it and integrating the fiber
coordinates gives the same squared ratio for $w$, hence the same
ratio for $h_{\mathrm{eff}}$.  Equation
\eqref{ham:eq:spec:fixedtime_local}, with at most four incident
plaquettes, gives \eqref{ham:srmap:local_ratio}.
The normalized conditional coarse density thus has maximum-to-minimum
ratio at most $R_*^2$ and minimum at least $R_*^{-2}$.  Comparison
with the Haar Casimir gap $3/4$ gives the stated conditional
Poincare constant.  Holding all other coarse links fixed, the
chosen loop is a translated or inverse-translated single Haar
variable, with Haar mean zero and second moment $1/4$.  The
conditional minimum-density bound and the law of total variance
give \eqref{ham:srmap:coarse_loop_variance}.
\end{proof}

\subsection{Exact compressed form, induced potential and residual}

The positive fine-vacuum ground-state identity is
\begin{equation}\label{ham:srmap:ground_form}
 \langle h_fF,K_fh_fG\rangle
 =\kappa_f\sum_{\ell,\alpha}
       \int h_f^2\,\overline{X_{\ell,\alpha}F}
                         X_{\ell,\alpha}G\,d\mu_f
\end{equation}
for smooth $F,G$.  It follows by product differentiation,
integration by parts and $(H_f-E_f^0)h_f=0$; the terms without
derivatives of $F,G$ cancel by that eigenfunction equation.
It extends by closure to the corresponding first Sobolev form
domains.

\begin{theorem}[Complete effective form and its discarded coupling]
\label{ham:srmap:effective}
Put $\gamma=r\kappa_f$.  The closed compressed form is
\begin{equation}\label{ham:srmap:effective_form}
 q_{\mathrm{eff}}[\psi]
 :=\|K_f^{1/2}U_h\psi\|^2
 =\gamma\sum_{e,\alpha}
     \int_{X_c}w\,
       \left|X_{e,\alpha}
                \left(\frac{\psi}{h_{\mathrm{eff}}}\right)\right|^2
          d\mu_c .
\end{equation}
Its associated selfadjoint operator on the coarse space is
\begin{equation}\label{ham:srmap:effective_H}
 S_{\mathrm{eff}}
 =\gamma D_c+V_{\mathrm{eff}},\qquad
 V_{\mathrm{eff}}
 =-\gamma\,\frac{D_ch_{\mathrm{eff}}}{h_{\mathrm{eff}}}.
\end{equation}
It is nonnegative, has compact resolvent, and has the unique
normalized vacuum $h_{\mathrm{eff}}$ of energy zero.  On smooth
vectors, $U_h^*K_fU_h=S_{\mathrm{eff}}$.

For an exact expression for all discarded terms, use the prefix
matrices in \eqref{ham:srmap:derivative} and define, with $i=(e,\beta)$,
\begin{align}
 A_{e,\beta}(U)
 &=\sum_{s=1}^r\sum_{\alpha=1}^3
      R_{\beta\alpha}(A_{e,s})
                      X_{e,s,\alpha}\log h_f(U),\label{ham:srmap:A}\\
 Z_{e,\beta}(U)
 &=A_{e,\beta}(U)
       -r(X_{e,\beta}\log h_{\mathrm{eff}})(BU),\label{ham:srmap:Z}\\
 \Sigma_{ij}(v)
 &=\mathbb E_{\nu_f}[Z_iZ_j\mid BU=v].\label{ham:srmap:Sigma}
\end{align}
Then
\begin{equation}\label{ham:srmap:centered_drift}
 \mathbb E_{\nu_f}[Z_i\mid B]=0,
\end{equation}
and the full residual $R_h=(\mathbf1-P_h)K_fU_h$ is
\begin{equation}\label{ham:srmap:vacuum_residual}
 R_h\psi
 =-2\kappa_fh_f\sum_i Z_i
             \left[X_i\left(\frac{\psi}{h_{\mathrm{eff}}}\right)\right]
                    \circ B .
\end{equation}
In particular its entire squared norm, including all fiber and
representation degrees of freedom, is
\begin{equation}\label{ham:srmap:residual_covariance}
 \|R_h\psi\|^2
 =4\kappa_f^2\int_{X_c}w(v)
       \sum_{i,j}\overline{X_iF(v)}\,\Sigma_{ij}(v)\,X_jF(v)\,d\mu_c(v),
 \qquad F=\psi/h_{\mathrm{eff}}.
\end{equation}
These residual identities hold first on smooth vectors; the bound
below extends $R_h$ to the entire retained form domain as an
$L^2$ representative of the cross form with the discarded space.
The matrix $\Sigma$ is smooth, positive semidefinite and bounded at
these fixed finite parameters.  With
$\sigma_*=\sup_v\|\Sigma(v)\|_{\mathrm{op}}$,
\begin{equation}\label{ham:srmap:residual_form_bound}
 \|R_h\psi\|^2
 \le\frac{4\kappa_f\sigma_*}{r}\,
                    q_{\mathrm{eff}}[\psi].
\end{equation}
\end{theorem}

\begin{proof}
Apply \eqref{ham:srmap:ground_form} with $F\circ B$ and use the
pointwise identity \eqref{ham:srmap:pointwise_energy}.  Disintegration
against $h_f^2$ gives \eqref{ham:srmap:effective_form}.  Integration by
parts on the coarse compact group gives
\[
 q_{\mathrm{eff}}[\psi]
 =\gamma\langle\psi,D_c\psi\rangle
      -\gamma\int\frac{D_ch_{\mathrm{eff}}}{h_{\mathrm{eff}}}
                         |\psi|^2\,d\mu_c,
\]
which proves \eqref{ham:srmap:effective_H}.  All multipliers are smooth
and bounded, since $h_{\mathrm{eff}}$ has a positive minimum.
The first Sobolev form domain is unchanged by multiplication or
division by $h_{\mathrm{eff}}$, so this form is closed.
Compact resolvent follows from the coarse Casimir.  Its zero form
vectors have $\psi/h_{\mathrm{eff}}$ constant on the connected
configuration space, proving the vacuum assertion.
The smooth functions form an operator core; all formulas also
restrict to the gauge-invariant spaces.

To identify the mean of $A_{e,\beta}$ without assuming any closure
of the interaction, introduce the fine vector field
\[
 Y_{e,\beta}
 =\sum_{s,\alpha}R_{\beta\alpha}(A_{e,s})X_{e,s,\alpha}.
\]
Each coefficient is independent of its own differentiated link, so
$Y_{e,\beta}$ is divergence free for product Haar measure.
Equation \eqref{ham:srmap:derivative} gives
$Y_{e,\beta}(F\circ B)=r(X_{e,\beta}F)\circ B$, and
$Y_{e,\beta}\log h_f=A_{e,\beta}$.  Integration by parts therefore
gives, for every smooth coarse test function $F$,
\[
 r\int wX_{e,\beta}F\,d\mu_c
 =-2\int wF\,\mathbb E_{\nu_f}[A_{e,\beta}\mid B]\,d\mu_c
 =-r\int (X_{e,\beta}w)F\,d\mu_c.
\]
Thus
$\mathbb E_{\nu_f}[A_{e,\beta}\mid B]
  =(r/2)X_{e,\beta}\log w
  =rX_{e,\beta}\log h_{\mathrm{eff}}$,
which proves \eqref{ham:srmap:centered_drift}.

The fine ground-transformed operator acts on a pulled-back smooth
function as
\[
 h_f^{-1}K_f\bigl(h_f(F\circ B)\bigr)
 =\kappa_f\left[
       r(D_cF)\circ B-2\sum_i A_i(X_iF)\circ B\right].
\]
Its conditional mean in the fine-vacuum measure is
$h_{\mathrm{eff}}^{-1}S_{\mathrm{eff}}
 (h_{\mathrm{eff}}F)$, pulled back by $B$.  Subtracting this mean
gives \eqref{ham:srmap:vacuum_residual}.  Squaring and conditioning proves
\eqref{ham:srmap:residual_covariance}.  The covariance-matrix bound and
\eqref{ham:srmap:effective_form} give \eqref{ham:srmap:residual_form_bound}.
\end{proof}

The potential \eqref{ham:srmap:effective_H} is an exact function determined
by the true fine vacuum.  An identification with a Wilson plaquette
potential at a single coarse coupling would require the additional
pointwise identity
\[
 -\gamma\frac{D_ch_{\mathrm{eff}}}{h_{\mathrm{eff}}}
       =v_cW_c-E_c^0.
\]
No such identity follows from the marginal construction.
Furthermore $S_{\mathrm{eff}}$ is a compression, not the full
dynamical effective Hamiltonian.  Eliminating its orthogonal
complement feeds the residual \eqref{ham:srmap:vacuum_residual} back into
the coarse resolvent.  That term can change kinetic as well as
potential contributions; ignoring it is not an approximation
certified by the present formulas.

The physical gap of the compression has the elementary variational
relation
\begin{equation}\label{ham:srmap:gap_direction}
 \Delta_f\ \le\
 \inf_{\psi\perp h_{\mathrm{eff}},\ \|\psi\|=1}
                         q_{\mathrm{eff}}[\psi]
 =\Delta_{\mathrm{eff}}.
\end{equation}
Indeed $U_h$ is an isometry preserving the actual vacuum, so the
right-hand minimization is over a subset of the full fine vacuum
complement.  A lower bound on $\Delta_{\mathrm{eff}}$ alone
therefore gives no lower bound on $\Delta_f$.  The discarded-sector
and residual estimates in the spectral-comparison section are
needed to change this direction into a useful full-space bound.
The quantities $\sigma_*$, the omitted-sector spectrum and their
spatial-scaling behavior have not been bounded uniformly here.

\subsection{Exact finite-scale consistency}

Consider three aligned nested spatial lattices $c,m,f$ with blocking
maps $B_{cm}$ and $B_{mf}$.  Ordered products give
$B_{cf}=B_{cm}\circ B_{mf}$ exactly.  Start with one fine-vacuum
measure $\nu_f$, and define $\nu_m=(B_{mf})_*\nu_f$ and
$\nu_c=(B_{cm})_*\nu_m$.  Their positive square-root Haar densities
$h_f,h_m,h_c$ give maps of the form \eqref{ham:srmap:vacuum_embedding},
and direct cancellation proves
\begin{equation}\label{ham:srmap:composition}
 U_{mf}U_{cm}=U_{cf}.
\end{equation}
The corresponding compressed forms also agree under this further
compression.  This consistency is exact when all intermediate
states are marginals of the \emph{same} fine state.  It does not
assert equality between those marginals and the separately chosen
vacua of intermediate Wilson Hamiltonians.

\section{Subsequential equal-time spatial holonomy construction}
\label{ham:srmap:equal_time}

There is an unconditional compactness construction for spatial
holonomies.  It is useful to specify exactly what it produces and
which physical conclusions it does not supply.

Fix a physical three-torus with initial spacing $a_0$ and initial
side length $L_0\ge2$.  Choose integers $r_k\ge2$, set
$R_k=r_1\cdots r_k$, and let
$a_k=a_0/R_k$ and $L_k=L_0R_k$.  Let $B_{jk}:X_k\to X_j$ be the
aligned ordered-path blocking maps for $k\ge j$.
At each level choose any finite $g_k>0$ and take the actual positive
normalized ground state $h_k$ of the corresponding finite spatial
Hamiltonian.  Write $\nu_k=h_k^2\mu_k$.

\begin{theorem}[Subsequential vacuum holonomy measure]\label{ham:srmap:compactness}
There is a subsequence $k_\ell\to\infty$ and probability measures
$\nu_j^\infty$ on every fixed $X_j$ such that
\[
 (B_{j,k_\ell})_*\nu_{k_\ell}
       \ \Longrightarrow\ \nu_j^\infty
 \quad\hbox{for every fixed }j,
 \qquad
 (B_{j,j+1})_*\nu_{j+1}^\infty=\nu_j^\infty.
\]
Each limiting measure is gauge invariant.  These measures determine
a unique gauge-invariant probability measure $\nu_\infty$ on the
compact inverse-limit configuration space
\begin{equation}\label{ham:srmap:inverse_limit}
 X_\infty
 =\{(v_0,v_1,\ldots):B_{j,j+1}v_{j+1}=v_j
                           \text{ for every }j\}.
\end{equation}
For every continuous cylinder observable $F\circ\pi_j$,
\begin{equation}\label{ham:srmap:equal_time_limit}
 \int F\,d\nu_j^\infty
 =\lim_{\ell\to\infty}
       \langle h_{k_\ell},(F\circ B_{j,k_\ell})h_{k_\ell}\rangle.
\end{equation}
The same holds for every finite product of such observables.
If, in addition, $g_k\ge g_{\min}>0$, every fixed coarse Wilson
loop that contains an edge exactly once satisfies
\begin{equation}\label{ham:srmap:limit_nontriviality}
 \operatorname{Var}_{\nu_j^\infty}(F)
 \ge\frac9{100}\exp\left(-\frac{512}{g_{\min}^4}\right)>0.
\end{equation}
In particular, fixed-$g$ refinement has an actual nontrivial
gauge-invariant equal-time limit along this subsequence.
\end{theorem}

\begin{proof}
Each $X_j$ is a compact metric space, so its probability measures
are sequentially compact for weak convergence.  This can also be
proved directly by taking convergent subsequences of the integrals
against a countable dense subset of $C(X_j)$ and extending the
resulting positive normalized functional.  Apply this compactness
successively at $j=0,1,\ldots$ and take a diagonal subsequence.
Continuity of the blocking maps passes
$B_{j,j+1}B_{j+1,k}=B_{jk}$ to the weak limits and proves consistency.
Gauge invariance passes to the limit because each gauge action is
continuous and the blocking maps are equivariant.

The space \eqref{ham:srmap:inverse_limit} is a closed subset of a
countable product of compact metric spaces.  The blocking maps are
surjective, so every coordinate projection $\pi_j$ is surjective.
On its continuous cylinder algebra define
$\mathcal L(F\circ\pi_j)=\int F\,d\nu_j^\infty$.
Consistency makes this definition independent of the cylinder
representation.  It is positive, normalized and bounded in
supremum norm.  The cylinder algebra contains constants and
separates points, so its uniform closure is $C(X_\infty)$.
The positive-functional representation theorem gives a unique
probability measure $\nu_\infty$.  This proves the assertions and
the equal-time convergence.  Gauge equivariance gives invariance
under compatible vertex gauge transformations.
Finally, apply Corollary~\ref{ham:srmap:local_inheritance} to each direct
blocking map $B_{j,k_\ell}$.  Its variance lower bound is uniform
when $g_{k_\ell}\ge g_{\min}$.  Both $F$ and $F^2$ are continuous
and bounded, so weak convergence passes the variance inequality
to the limit and proves \eqref{ham:srmap:limit_nontriviality}.
\end{proof}

The consistent limiting measures give actual weighted Hilbert
isometries by pullback,
$L^2(X_j,\nu_j^\infty)\to L^2(X_{j+1},\nu_{j+1}^\infty)$.
Their cylinder subspaces are dense in $L^2(X_\infty,\nu_\infty)$,
and the constant function is the normalized vacuum vector for this
equal-time probability representation.
The construction is on generalized spatial holonomies.  The
unrestricted compactness argument supplies no positive smooth
limiting Haar density and by itself does not exclude singular
finite-level limits.  If $g_k\ge g_{\min}$, the inherited one-link
ratios instead bound every fixed-level limiting density between
$R_*(g_{\min})^{-2E_j}$ and $R_*(g_{\min})^{2E_j}$.  To see this,
change the $E_j$ coarse coordinates one at a time, normalize the
resulting global density-ratio bound, and pass the two measure
inequalities to the weak limit.  Even then no derivative bound
has been proved, so the smooth density formulas cannot simply be
differentiated after taking the limit.

This theorem supplies actual subsequential equal-time vacuum
moments, and nonzero coarse-loop variance when the couplings stay
bounded away from zero.  For unrestricted weak-coupling sequences
$g_k\to0$, the displayed lower bound degenerates and does not
establish continuum nontriviality.  The theorem does not prove
that different subsequences give the same state or that
renormalized local fields exist.  It
does not construct a strongly continuous physical time semigroup
on the limiting Hilbert space, a continuum Hamiltonian, Euclidean
rotation symmetry, an infinite-volume limit, or a positive
continuum spectral gap.
Along the intended weak-coupling sequence, compactness and the
finite-volume positive densities do not supply a uniform positive
variance or a uniform coercive spectral estimate after spatial
refinement.

The precise remaining spectral task is to control the full
vacuum-complement energy, including the covariance residual and
discarded sector, under compatible spatial limits and
renormalized observables.  The equal-time compactness construction
does not supply those estimates.  It can be combined with local
conditional vacuum estimates, but it cannot replace their missing
global and dynamical control.

\section{Local physical states, exact loop energies, and spectral tightness}
\label{ham:sec:states}

The results here concern the actual interacting vacuum of the spatial
SU(2) Hamiltonian. They provide volume-independent bounds on energies
and time correlations of specified physical states. They also prove a
restricted failure of spectral tightness under spatial refinement at
fixed coupling. Neither positive spectral weight at some positive
energies nor a finite-state variational energy is a lower bound on the
entire physical spectrum.

\subsection{A volume-independent weak-coupling energy estimate}

Use a periodic cubic lattice with $L\ge2$, $E=P=3L^3$ links and
plaquettes, normalized product Haar measure $\mu_L$, and
\begin{equation}\label{ham:eq:states:H}
 H_{a,L,g}=\kappa D+vW,\qquad
 \kappa=\frac{g^2}{2a},\quad v=\frac4{g^2a},\quad
 D=\sum_e\mathcal C_e,\quad
 W=\sum_p(1-f_p),\quad f_p=\tfrac12\operatorname{Tr}U_p.
\end{equation}
The one-link Casimir has spin-$j$ eigenvalues $j(j+1)$.
Let $h_L>0$ be the unique normalized ground state, $E_{0,L}$ its
energy, $K_L=H_{a,L,g}-E_{0,L}$, and
$d\nu_L=h_L^2d\mu_L$.
The full ground state is gauge invariant: the strictly positive heat
kernel makes the ground state unique and positive, and each local
gauge transformation preserves positivity and commutes with $H$.
Thus the full and physical ground energies coincide.

\begin{lemma}[An actual weak-coupling vacuum estimate]\label{ham:lem:states:trial}
For every $a,g>0$ and every $L\ge2$,
\begin{equation}\label{ham:eq:states:groundupper}
 E_{0,L}\le P\min\left\{\frac3a,\frac4{g^2a}\right\}.
\end{equation}
For every individual link and elementary plaquette,
\begin{align}
 \langle h_L,\mathcal C_e h_L\rangle
 &\le M_g:=\min\{6/g^2,\,8/g^4\},\label{ham:eq:states:electric}\\
 \nu_L(1-f_p)&\le\frac34g^2,\label{ham:eq:states:plaquette_deficit}\\
 \operatorname{Var}_{\nu_L}(f_p)&\le\frac32g^2.
 \label{ham:eq:states:plaquette_variance_upper}
\end{align}
All bounds are uniform in the lattice volume.
\end{lemma}

\begin{proof}
On one link put $x=\operatorname{Tr}U/2$, let
$Z_\beta=\int e^{\beta x}dU$, and define the normalized trial
wavefunction $\phi_\beta=Z_\beta^{-1/2}e^{\beta x/2}$.
Write $r_\beta=\int x\phi_\beta^2dU\in(0,1)$.
The positive tilt of the even Haar marginal proves $r_\beta>0$.
The identities
\[
 \Gamma(x):=\sum_b|X_bx|^2=\frac{1-x^2}{4},\qquad
 \beta\,\mathbb E_\beta(1-x^2)=3\mathbb E_\beta x
\]
give the exact trial electric energy
\begin{equation}\label{ham:eq:states:trial_electric}
 \langle\phi_\beta,\mathcal C\phi_\beta\rangle
 =\frac{\beta^2}{16}\mathbb E_\beta(1-x^2)
 =\frac{3\beta r_\beta}{16}.
\end{equation}
The second identity follows by integrating
$[e^{\beta x}(1-x^2)^{3/2}]'$ on $[-1,1]$; its boundary term is zero.

Here is a complete bound on the tilted first moment. For $z=1-x$,
its density is proportional to
$e^{-\beta z}z^{1/2}\sqrt{1-z/2}\,\mathbf1_{[0,2]}(z)$.
This is the Gamma law of shape $3/2$ and rate $\beta$, reweighted
by a decreasing nonnegative function. If $Z,Z'$ are independent
Gamma variables and $w$ is that decreasing weight, then
\[
 2\operatorname{Cov}(Z,w(Z))
 =\mathbb E[(Z-Z')(w(Z)-w(Z'))]\le0.
\]
The reweighted mean is consequently at most the original mean:
\begin{equation}\label{ham:eq:states:tilted_moment}
 1-r_\beta=\mathbb E_\beta z\le\frac3{2\beta}.
\end{equation}
Equivalently $r_\beta=I_2(\beta)/I_1(\beta)$, but no numerical Bessel
evaluation or asymptotic expansion is needed.

Use the product trial $\Phi_\beta=\prod_e\phi_\beta(U_e)$.
Conjugation invariance gives $\mathbb E_\beta U_e=r_\beta I$,
so the four distinct links of an elementary plaquette have
$\langle\Phi_\beta,f_p\Phi_\beta\rangle=r_\beta^4$.
Since $1-r_\beta^4\le4(1-r_\beta)\le6/\beta$ and $E=P$,
\[
 \frac{\langle\Phi_\beta,H\Phi_\beta\rangle}{P}
 \le\frac{3g^2\beta}{32a}+\frac{24}{g^2a\beta}.
\]
Choose $\beta=16/g^2$. The right side is exactly $3/a$.
The trial need not satisfy Gauss's law: its expectation bounds the
full minimum, already proved equal to the physical minimum.
Alternatively the constant trial gives $E_{0,L}\le vP$, yielding
\eqref{ham:eq:states:groundupper}.

Uniqueness makes $h_L$ invariant under translations and cubic
coordinate permutations, so every link has the same Casimir
expectation and every elementary plaquette the same deficit
expectation. Positivity of both terms of $H$ therefore gives
\[
 \kappa E\,\langle\mathcal C_e\rangle_{\nu_L}\le E_{0,L},
 \qquad
 vP\,\nu_L(1-f_p)\le E_{0,L}.
\]
These are \eqref{ham:eq:states:electric} and
\eqref{ham:eq:states:plaquette_deficit}. Finally
$\operatorname{Var}(f_p)\le\nu_L[(1-f_p)^2]
\le2\nu_L(1-f_p)$ because $-1\le f_p\le1$.
\end{proof}

In particular the centered bare plaquette state has norm tending
to zero as $g\to0$, uniformly in volume. This is a statement about
an unrenormalized fixed lattice observable. It does not identify
a renormalized local field or a fixed-physical-size continuum loop.

\subsection{Uniform first and second energy moments of local states}

Let $X_{e,b}$ be real divergence-free derivatives with
$\mathcal C_e=-\sum_bX_{e,b}^2$. For a smooth local function $F$ set
$\Gamma(F)=\sum_{e,b}|X_{e,b}F|^2$. If $F$ is gauge invariant, the
centered state
\[
 \psi_{F,L}=(F-\nu_LF)h_L
\]
is physical and orthogonal to the actual vacuum.

\begin{lemma}[Uniform local-state second moment]\label{ham:lem:states:secondmoment}
Suppose $F$ depends on $s$ links, with its finite support embedded
in the periodic boxes under consideration. Then
\begin{align}
 \|\psi_{F,L}\|^2&=\operatorname{Var}_{\nu_L}(F),\label{ham:eq:states:norm}\\
 \langle\psi_{F,L},K_L\psi_{F,L}\rangle
 &=\kappa\int\Gamma(F)d\nu_L,\label{ham:eq:states:form}\\
 \|K_L\psi_{F,L}\|
 &\le B_g(F):=
 \kappa\left[\|DF\|_\infty+
              2\sqrt{sM_g\,\|\Gamma(F)\|_\infty}\right].
 \label{ham:eq:states:secondmoment}
\end{align}
In particular the weaker bound
$\kappa[\|DF\|_\infty+(4\sqrt2/g^2)
\sqrt{s\|\Gamma(F)\|_\infty}]$ also holds.
\end{lemma}

\begin{proof}
The first identity is exact centering. The ground-state quadratic-form
identity gives \eqref{ham:eq:states:form}. Direct differentiation gives
\begin{equation}\label{ham:eq:states:commutator}
 K_L(Fh_L)=\kappa\left[(DF)h_L
             -2\sum_{e,b}(X_{e,b}F)(X_{e,b}h_L)\right].
\end{equation}
The vacuum multiple removed in $\psi_{F,L}$ contributes zero.
Pointwise Cauchy--Schwarz bounds the squared modulus of the sum
by $\Gamma(F)\sum_{e\in\operatorname{supp}F,b}|X_{e,b}h_L|^2$.
Its integral is at most
$\|\Gamma(F)\|_\infty\,sM_g$ by
Lemma~\ref{ham:lem:states:trial}. The triangle inequality proves
\eqref{ham:eq:states:secondmoment}. Smoothness and compactness justify
these operator-domain identities on every finite box.
\end{proof}

This is stronger than a form bound alone. If the physical spectral
measure of $\psi_{F,L}$ is
$\eta_{F,L}(B)=\langle\psi_{F,L},\mathbf1_B(K_L)\psi_{F,L}\rangle$,
then
\begin{equation}\label{ham:eq:states:moment_tails}
 \int E^2d\eta_{F,L}(E)\le B_g(F)^2,\qquad
 \int_{E>R}E\,d\eta_{F,L}(E)\le\frac{B_g(F)^2}{R}.
\end{equation}
Thus first energy moments are uniformly integrable in volume at
fixed $a,g$.
When an actual limiting time-correlation construction has been
established and the equal-time marginals converge, its limiting
state obeys
\[
 \psi_F\in D(H_\infty),\qquad
 \|H_\infty\psi_F\|\le B_g(F),\qquad
 \langle\psi_F,H_\infty\psi_F\rangle
       =\kappa\int\Gamma(F)d\nu_\infty.
\]
Indeed weak convergence of the spectral measures and
\eqref{ham:eq:states:moment_tails} pass the first moment exactly;
lower semicontinuity bounds the second moment. This implication
uses the actual limiting correlations, rather than postulating
a limiting Hamiltonian from the formal expression for $H$.

\subsection{Exact Wilson-loop identities and nonconstant time correlations}

Let $\ell$ be a closed path using $n$ distinct unoriented link
variables, each once, and put $F=\operatorname{Tr}U_\ell/2$.
Choose a volume-independent bound $R_g\ge1$ on the ratio of the
maximum and minimum of $h_L$ as one link varies and all others
are fixed. The established actual-vacuum comparison allows
\begin{equation}\label{ham:eq:states:R}
 R_g=\frac53e^{256/g^4}\quad(g>0).
\end{equation}
For $0<g^2\le1$ the sharper certified choice
$R_g=10e^{16\sqrt{23}/g^2}$ is available.
Either may be used, or their minimum where both apply.
In the formulas below define
\begin{equation}\label{ham:eq:states:constants}
 v_*=\frac1{4R_g^2},\qquad
 A_g=\frac34+\sqrt{M_g},\qquad
 Q=\frac{\kappa n}{4},\qquad
 S=(\kappa n A_g)^2.
\end{equation}

\begin{theorem}[Actual loop-state spectral bounds]\label{ham:thm:states:loop}
Writing $m_L=\nu_LF$, $v_L=\nu_L(F^2)-m_L^2$, and
$\psi_L=(F-m_L)h_L$, one has
\begin{align}
 DF&=\frac{3n}{4}F,&
 \Gamma(F)&=\frac n4(1-F^2),\label{ham:eq:states:loop_derivatives}\\
 v_L&\ge v_*,&
 \nu_L(1-F^2)&\ge\frac3{4R_g^2},\label{ham:eq:states:loop_variance}\\
 q_L:=\langle\psi_L,K_L\psi_L\rangle
 &=Q[1-\nu_L(F^2)],&
 \frac{3\kappa n}{16R_g^2}\le q_L&\le Q,\label{ham:eq:states:loop_energy}\\
 \langle\psi_L,K_L^2\psi_L\rangle&\le S.&&
 \label{ham:eq:states:loop_second}
\end{align}
The normalized spectral probability measure
$\rho_L=\eta_{F,L}/v_L$ therefore has
\begin{equation}\label{ham:eq:states:normalized_tail}
 \rho_L((R,\infty))
 \le\min\left\{\frac{\kappa nR_g^2}{R},
               \frac{4R_g^2S}{R^2}\right\}.
\end{equation}
At every $t\ge0$, the actual connected physical correlation
$C_L(t)=\langle\psi_L,e^{-tK_L}\psi_L\rangle$ satisfies
\begin{equation}\label{ham:eq:states:correlation_lower}
 C_L(t)\ge v_L e^{-tq_L/v_L}
       \ge v_*e^{-t\kappa nR_g^2}>0,\qquad
 0\le v_L-C_L(t)\le tQ.
\end{equation}
\end{theorem}

\begin{proof}
Fix all loop links except one. Its holonomy is a left and right
translate of that link or its inverse. The normalized fundamental
trace has Casimir $3/4$ and squared gradient $(1-F^2)/4$.
Bi-invariance preserves these formulas for each of the $n$ distinct
links, proving \eqref{ham:eq:states:loop_derivatives}.

For the chosen link, the conditional density $p$ relative to Haar
measure obeys $p\ge R_g^{-2}$: its unnormalized density is $h_L^2$
and the normalizing Haar integral is at most its maximum.
The Haar mean of $F$ is zero and its second moment is $1/4$.
Consequently
\[
 \operatorname{Var}(F\mid\hbox{other links})
 \ge R_g^{-2}\inf_c\int(F-c)^2dU=\frac1{4R_g^2}.
\]
Total variance gives the first inequality in
\eqref{ham:eq:states:loop_variance}.
Likewise the nonnegative function $1-F^2$ has Haar integral
$3/4$, proving the second.
The ground-state form identity now gives
\eqref{ham:eq:states:loop_energy}. Apply
Lemma~\ref{ham:lem:states:secondmoment} with $s=n$,
$\|DF\|_\infty=3n/4$ and $\|\Gamma(F)\|_\infty=n/4$
to obtain \eqref{ham:eq:states:loop_second}.

The first normalized moment is at most $Q/v_*=\kappa nR_g^2$
and the second at most $S/v_*=4R_g^2S$. Markov's inequality gives
\eqref{ham:eq:states:normalized_tail}.
Jensen's inequality for the convex function $E\mapsto e^{-tE}$
gives $C_L(t)/v_L\ge e^{-tq_L/v_L}$.
The function $v\mapsto ve^{-tQ/v}$ is increasing for $v>0$,
so $v_L\ge v_*$ and $q_L\le Q$ prove the uniform lower bound.
Finally $0\le1-e^{-tE}\le tE$ proves the last estimate.
\end{proof}

\begin{proposition}[A positive spectral weight away from zero]
\label{ham:prop:states:positive_weight}
For every box and loop in Theorem~\ref{ham:thm:states:loop}, set
\begin{equation}\label{ham:eq:states:positive_weight}
 E_*=\frac{3\kappa n}{32R_g^2},\qquad
 p_g=\frac9{1024R_g^4A_g^2}>0.
\end{equation}
Then
\begin{equation}\label{ham:eq:states:weight}
 \rho_L([E_*,\infty))\ge p_g,\qquad
 \frac{C_L(t)}{v_L}
 \le1-p_g(1-e^{-tE_*})\quad(t>0).
\end{equation}
Together with \eqref{ham:eq:states:normalized_tail}, this proves
uniform spectral tightness in volume and nonconstant dynamics
for a fixed local loop, or a family with uniformly bounded $n$,
in any actual fixed-$a,g$ thermodynamic correlation limit.
\end{proposition}

\begin{proof}
For any nonnegative random variable $X$ with finite second moment,
write $m=\mathbb EX$. Splitting its mean below and above $m/2$,
then applying Cauchy--Schwarz, gives
\[
 \frac m2
 \le\mathbb E[X\mathbf1_{\{X\ge m/2\}}]
 \le(\mathbb EX^2)^{1/2}\mathbb P(X\ge m/2)^{1/2}.
\]
Apply this elementary Paley--Zygmund estimate to $E$ under $\rho_L$:
\[
 \rho_L([q_L/(2v_L),\infty))
 \ge\frac{q_L^2}{4v_L
                    \langle\psi_L,K_L^2\psi_L\rangle}.
\]
Since $v_L\le1$, the energy threshold is at least $E_*$.
The probability is at least
\[
 \frac{(3\kappa n/(16R_g^2))^2}{4S}=p_g.
\]
Integrating $e^{-tE}\le1$ below the threshold and
$e^{-tE}\le e^{-tE_*}$ above it proves \eqref{ham:eq:states:weight}.
\end{proof}

For explicit localization in a finite energy interval, the same
estimates give mass at least $p_g/2$ in
\[
 \left[E_*,\,\frac{64\sqrt2}{3}\kappa nR_g^3A_g^2\right].
\]
The upper endpoint makes the second-moment tail at most $p_g/2$.
These statements do not exclude spectral weight arbitrarily close
to zero. In an infinite-volume limit additional zero-energy
components can occur even though every finite-volume vacuum is
unique. Positive weight away from zero proves nonconstant
dynamics, not vacuum uniqueness or a positive lower spectral edge.

\subsection{Obstruction to bare fixed-coupling spatial refinement}

\begin{proposition}[Fixed-coupling bare-loop spectral non-tightness]
\label{ham:prop:states:fixedg_escape}
Fix $g>0$. Consider any sequence of the specified periodic
Hamiltonians with $a_k\downarrow0$, arbitrary finite $L_k\ge2$,
and nontrivial loops with $n_k\ge1$ distinct link variables.
For their centered, normalized loop states, the physical energy
spectral measures are not tight. More precisely, for every finite
$R>0$,
\[
 \liminf_{k\to\infty}\rho_k((R,\infty))\ge p_g>0.
\]
For every fixed physical time $t>0$,
\begin{equation}\label{ham:eq:states:discontinuity}
 \limsup_{k\to\infty}\frac{C_k(t)}{v_k}\le1-p_g,
 \qquad \frac{C_k(0)}{v_k}=1.
\end{equation}
Thus these normalized bare-loop states cannot have a
norm-preserving, strongly continuous physical-time correlation
limit under this fixed-coupling refinement.
\end{proposition}

\begin{proof}
The constants $R_g,A_g,p_g$ do not depend on $a,L$ or $n$.
But
$E_{*,k}=3g^2n_k/(64a_kR_g^2)\to\infty$.
Proposition~\ref{ham:prop:states:positive_weight} places at least
$p_g$ of the probability above this diverging threshold.
This proves non-tightness and \eqref{ham:eq:states:discontinuity}.
A unit vector acted on by a strongly continuous semigroup has a
correlation tending to one as $t\downarrow0$, in contradiction
with that inequality at every $t>0$.
\end{proof}

This conclusion uses a positive high-energy weight, not merely a
divergent first moment. It applies, in particular, to the bare
ordered-product refinement of a fixed coarse loop. It does not
apply to a weak-coupling sequence $g_k\to0$ by this proof, because
the constants $p_{g_k}$ can tend to zero. It also does not exclude
operator mixing, smoothing, or other renormalization constructions.
The physical times in \eqref{ham:eq:states:discontinuity} are those
of \eqref{ham:eq:states:H}; a new scale-dependent redefinition of time
would be a different construction.

\begin{proposition}[What scalar renormalization can and cannot change]
\label{ham:prop:states:scalar}
Replace a loop observable by
$G_k=Z_kF_k+c_kI$, where $Z_k\ne0$ and $c_k$ are arbitrary scalars.
After actual-vacuum centering, its state is
$Z_k(F_k-\nu_kF_k)h_k$.
Its normalized spectral measure and normalized connected time
correlation are exactly unchanged. Therefore such multiplicative
normalization and identity subtraction cannot turn the non-tight
family in Proposition~\ref{ham:prop:states:fixedg_escape} into a tight,
nontrivial normalized family.
\end{proposition}

\begin{proof}
Centering removes $c_k$ and multiplies the original centered
vector by $Z_k$. Every spectral measure and every correlation is
multiplied by $|Z_k|^2$, which cancels on dividing by the squared
norm.
\end{proof}

The converse inference from a divergent first moment is false.
For example,
\[
 \rho_k=(1-k^{-1})\delta_1+k^{-1}\delta_{k^2},\qquad k\ge2,
\]
is a tight probability family converging to $\delta_1$, while
$\int E\,d\rho_k=1-k^{-1}+k\to\infty$.
Its correlations differ from $e^{-t}$ by at most $1/k$, uniformly
for $t\ge0$. These measures are realized by normalized
vacuum-orthogonal vectors for
$\operatorname{diag}(0,1,k^2)$.
Thus failure of a first-moment bound is not itself a spectral
escape theorem. The second-moment argument above supplies the
missing quantitative weight in the fixed-$g$ case.

\subsection{Weak-coupling estimate for spatially smeared plaquettes}

For finitely many distinct elementary plaquettes and real weights
$w_p$, set $O_w=\sum_pw_pf_p$. Each link belongs to at most four
of these plaquettes.

\begin{proposition}[Uniform smeared-state energy estimate]
\label{ham:prop:states:smearing}
For the actual centered state $\psi_w=(O_w-\nu_LO_w)h_L$,
\begin{equation}\label{ham:eq:states:smearing}
 \langle\psi_w,K_L\psi_w\rangle
 \le6\kappa g^2\sum_pw_p^2
 =\frac{3g^4}{a}\sum_pw_p^2,
 \qquad
 \operatorname{Var}_{\nu_L}(O_w)
 \le\frac32g^2\left(\sum_p|w_p|\right)^2.
\end{equation}
In particular, an average over $N$ plaquettes has energy at most
$3g^4/(aN)$. A normalized-energy conclusion additionally requires
the connected variance of that average.
\end{proposition}

\begin{proof}
For a link touching $d_e\le4$ selected plaquettes, Cauchy--Schwarz
and the one-link trace identity give
\[
 \Gamma_e(O_w)
 \le\frac{d_e}{4}\sum_{p\ni e}w_p^2(1-f_p^2).
\]
Sum over links. Each plaquette has four links, so
$\Gamma(O_w)\le4\sum_pw_p^2(1-f_p^2)$.
Since $1-f_p^2\le2(1-f_p)$,
Lemma~\ref{ham:lem:states:trial} gives
$\nu_L\Gamma(O_w)\le6g^2\sum_pw_p^2$.
The physical form identity proves the energy bound.
The triangle inequality in the Hilbert space of centered
random variables, followed by
\eqref{ham:eq:states:plaquette_variance_upper}, proves the variance
upper bound.
\end{proof}

Uniform one-site conditional variances do not alone give the
missing lower connected variance for an average. A fully rational
counterexample makes the inference precise.
For even $N$, put
\[
 \pi_N(\sigma)=Z_N^{-1}\,2^{-|\sum_i\sigma_i|},
 \qquad \sigma_i\in\{-1,1\}.
\]
Given the other spins, the conditional odds are exactly $4$ or
$1/4$, so every conditional spin variance is $16/25$.
Nevertheless symmetry and comparison with the central binomial
coefficient give
\begin{align*}
 \operatorname{Var}_{\pi_N}\left(\sum_i\sigma_i\right)
 &\le8\sum_{m\ge1}m^2\,4^{-m}\\
 &=8\frac{(1/4)(1+1/4)}{(1-1/4)^3}
 =\frac{160}{27}.
\end{align*}
Indeed $Z_N\ge\binom{N}{N/2}$, and the multiplicity at total
magnetization $2m$ is at most that central coefficient.
Consequently the average has variance at most $160/(27N^2)$.
This different model shows why the available local estimates do
not by themselves establish a lower variance of order $1/N$.
It makes no assertion about the sign or size of actual SU(2)
plaquette correlations.

\subsection{Exact scale covariance and a dimensionful mass value}

\begin{proposition}[Scale covariance of the full physical spectrum]
\label{ham:prop:states:scale}
Under the canonical identification of the same finite graph's
Haar Hilbert spaces,
\begin{equation}\label{ham:eq:states:scale}
 H_{a,L,g}=\frac1aH_{1,L,g},\qquad
 h_{a,L,g}=h_{1,L,g},\qquad
 \Delta_{a,L,g}=\frac1a\Delta_{1,L,g}.
\end{equation}
Changing every spatial and temporal length by a common constant
$s>0$ leaves all dimensionless lattice data unchanged and divides
every energy by $s$.
\end{proposition}

\begin{proof}
Both coefficients in \eqref{ham:eq:states:H} contain the same factor
$a^{-1}$, while $D,W$, Haar measure and the gauge constraints
depend on the graph, not on the assigned physical length.
The spectral theorem and positive normalized ground-state
uniqueness prove all identities. In particular
$e^{-tK_{sa,L,g}}=e^{-(t/s)K_{a,L,g}}$.
\end{proof}

Consequently dimensionless bare data alone cannot select one
numerical mass in a specified physical unit. If a continuum
construction with a mass $M$ and an inverse-length reference
scale $\Lambda$ were established, a common dilation would replace
them by $M/s$ and $\Lambda/s$ while preserving $M/\Lambda$.
Specifying a reference scale removes this dimensionful ambiguity;
it does not determine the dimensionless ratio.
Reference-length scale setting in SU(2) gauge theory is discussed,
for example, by Sommer, arXiv:hep-lat/9310022.
No value of the unknown continuum ratio is inferred here, and
this dimensional observation is separate from proving a
positive continuum gap.

\section{Polynomial lower bounds for actual physical loop states}
\label{ham:sec:fisher}

This section replaces the exponentially small lower bound on the norm of
a centered Wilson-loop state by a polynomial bound in the bare coupling.
The argument uses the electric energy of the actual vacuum, without a
pointwise comparison of its density to Haar measure. All representations
and all spectator links are retained.

\subsection{Model and the electric input}

On a periodic cubic spatial lattice with $L\ge2$, let $E=P=3L^3$ be
the numbers of links and elementary plaquettes. With normalized product
Haar measure $\mu$, use
\begin{equation}\label{ham:eq:fisher:model}
 H=\kappa D+vW,\qquad
 \kappa=\frac{g^2}{2a},\quad v=\frac4{g^2a},\quad
 D=\sum_e\mathcal C_e,\quad W=\sum_p(1-f_p),\quad
 f_p=\tfrac12\operatorname{Tr}U_p.
\end{equation}
The one-link Casimir $\mathcal C_e=-\sum_{b=1}^3X_{e,b}^2$ has
spin-$j$ eigenvalues $j(j+1)$. Let $h>0$ be the normalized ground
state, $E_0$ its energy, $K=H-E_0$, and $d\nu=h^2d\mu$.
Ellipticity on the compact configuration space, the strictly positive
heat kernel, and the bounded real potential give a smooth, strictly
positive, unique ground state. Gauge transformations preserve positivity
and commute with $H$, so uniqueness makes $h$ gauge invariant.

The actual, volume-independent input is
\begin{equation}\label{ham:eq:fisher:input}
 \int\sum_b|X_{e,b}h|^2d\mu\le
 M_g:=\min\{6/g^2,8/g^4\},\qquad
 \nu(1-f_p)\le\frac34g^2.
\end{equation}
The complete variational proof is Lemma~\ref{ham:lem:states:trial}.

\subsection{A radial uncertainty estimate retaining every spectator}

\begin{lemma}[Radial reduction and interval probabilities]
\label{ham:lem:fisher:radial}
Let $h$ be any smooth normalized nonnegative wavefunction on the
finite configuration space. Let $F=\operatorname{Tr}U_\ell/2$ be
the trace of a loop that contains a chosen link $e$ exactly once,
with either orientation. Suppose
\[
 \int\sum_b|X_{e,b}h|^2d\mu\le M.
\]
There is a nonnegative function $u\in H_0^1(0,\pi)$ such that
$\|u\|_2=1$, the law of $F$ under $h^2\mu$ is the law of
$\cos\Theta$ for density $u(\theta)^2d\theta$, and
\begin{equation}\label{ham:eq:fisher:radial-energy}
 \|u'\|_2^2\le Q:=1+4M,\qquad \|u\|_\infty^2\le\sqrt Q.
\end{equation}
Every interval $J\subset\mathbb R$ of length $d$ therefore obeys
\begin{equation}\label{ham:eq:fisher:interval}
 \nu(F\in J)\le\min\left\{1,\pi\sqrt{Qd/2}\right\}.
\end{equation}
These are unconditional marginal estimates; no conditional electric
bound is being assumed.
\end{lemma}

\begin{proof}
Write $\xi$ for all other links. Conditional on $\xi$, the full loop
holonomy is $A(\xi)U_e^{\pm1}B(\xi)$. Changing from $U_e$ to this
holonomy preserves Haar measure and the one-link Casimir: left and
right multiplication and inversion are isometries. Write the resulting
wavefunction as $f_\xi(U)$.

Use the SU(2) coordinates
\[
 U=\cos\theta I+i\sin\theta\,\omega\cdot\sigma,
 \qquad 0\le\theta\le\pi,\quad\omega\in S^2.
\]
With normalized sphere measure $d\omega$,
$dU=(2/\pi)\sin^2\theta\,d\theta\,d\omega$. The Casimir
normalization in \eqref{ham:eq:fisher:model} gives the energy density
\[
 \sum_b|X_bf|^2=\tfrac14\left(
 |\partial_\theta f|^2+\sin^{-2}\theta|\nabla_{S^2}f|^2\right).
\]
Define
\[
 b(\theta)^2=\int |f_\xi(\theta,\omega)|^2d\omega\,d\mu(\xi),
 \qquad u(\theta)=\sqrt{2/\pi}\,\sin\theta\,b(\theta).
\]
Then $\int u^2=1$. Cauchy--Schwarz when differentiating $b^2$
gives
\[
 |b'(\theta)|^2\le
 \int|\partial_\theta f_\xi(\theta,\omega)|^2
           d\omega\,d\mu(\xi).
\]
The same conclusion at zeros of $b$ follows by approximation.
The smooth finite-dimensional input makes $u$ vanish at both
endpoints. Expanding its derivative and integrating the cross term
exactly gives
\[
 \int_0^\pi |u'|^2d\theta
 =1+\frac2\pi\int_0^\pi\sin^2\theta\,|b'|^2d\theta
 \le1+4M.
\]
This proves the first assertion, including the contributions of all
spectator variables. The endpoint values imply
\[
 |u(\theta)|^2
 =\tfrac12\left(\int_0^\theta (u^2)'
                         -\int_\theta^\pi (u^2)'\right)
 \le\int_0^\pi|uu'|\le\|u\|_2\|u'\|_2\le\sqrt Q.
\]

For $0\le\theta\le\phi\le\pi$, put $d_\theta=\phi-\theta$.
Since $(\theta+\phi)/2\in[d_\theta/2,\pi-d_\theta/2]$,
\begin{equation}\label{ham:eq:fisher:cosine}
 \cos\theta-\cos\phi
 =2\sin\frac{\theta+\phi}{2}\sin\frac{d_\theta}{2}
 \ge2\sin^2\frac{d_\theta}{2}
 \ge\frac{2d_\theta^2}{\pi^2}.
\end{equation}
Thus the inverse image of an interval of width $d$ under cosine has
length at most $\pi\sqrt{d/2}$. Integrating the density bound proves
\eqref{ham:eq:fisher:interval}.
\end{proof}

\begin{theorem}[Polynomial norm lower bound]
\label{ham:thm:fisher:variance}
Under Lemma~\ref{ham:lem:fisher:radial},
\begin{equation}\label{ham:eq:fisher:variance}
 \boxed{\displaystyle
 \operatorname{Var}_{h^2\mu}(F)
 \ge\frac1{\pi^4(1+4M)^2}.}
\end{equation}
In particular, the actual physical vacuum of
\eqref{ham:eq:fisher:model} satisfies this bound with $M=M_g$,
uniformly in the spatial volume and in the loop length.
For $s=g^2\in(0,1]$,
\begin{equation}\label{ham:eq:fisher:weak-variance}
 \operatorname{Var}_{\nu}(F)
 \ge\frac{s^2}{\pi^4(24+s)^2}
 >\frac{2401s^2}{234256(24+s)^2}
 >\frac{s^2}{62500}.
\end{equation}
\end{theorem}

\begin{proof}
Let $m_\theta=\int\theta u^2d\theta$. The zero boundary values
give the exact integration-by-parts identity
\[
 1=-2\int_0^\pi(\theta-m_\theta)u(\theta)u'(\theta)d\theta.
\]
Consequently $\operatorname{Var}(\Theta)\ge1/(4Q)$ by
Cauchy--Schwarz. For an independent copy $\Theta'$,
\[
 \mathbb E(\Theta-\Theta')^4
 =2\mathbb E(\Theta-m_\theta)^4
       +6\operatorname{Var}(\Theta)^2
 \ge8\operatorname{Var}(\Theta)^2.
\]
The independent-pair variance identity and
\eqref{ham:eq:fisher:cosine} therefore imply
\begin{align*}
 \operatorname{Var}(\cos\Theta)
 &=\tfrac12\mathbb E(\cos\Theta-\cos\Theta')^2\\
 &\ge\frac2{\pi^4}\mathbb E(\Theta-\Theta')^4
 \ge\frac{16}{\pi^4}\operatorname{Var}(\Theta)^2
 \ge\frac1{\pi^4Q^2}.
\end{align*}
For $s\le1$, $M_g=6/s$, so $Q=(24+s)/s\le25/s$.
The elementary strict bound $\pi<22/7$ gives the middle rational
bound in \eqref{ham:eq:fisher:weak-variance}; $\pi^4<100$ gives its
last bound. No angular, representation, or volume cutoff was used.
\end{proof}

\subsection{A polynomial lower bound on the exact loop energy}

\begin{lemma}[Transverse fluctuation lower bound]
\label{ham:lem:fisher:transverse}
Under the same one-link electric assumption, put
$A=\nu(1-F^2)$. Then
\begin{align}
 A&\ge
 \frac{2M+3-\sqrt{M(4M+3)}}{4(M+1)}
 =\frac9{4(2M+3+\sqrt{M(4M+3)})}\label{ham:eq:fisher:Aexact}\\
 &\ge\frac9{16M+15}.
 \label{ham:eq:fisher:Arational}
\end{align}
For the actual vacuum and $s=g^2\le1$ this gives
\begin{equation}\label{ham:eq:fisher:Aweak}
 \nu(1-F^2)\ge\frac{9s}{96+15s}\ge\frac{3s}{37}.
\end{equation}
\end{lemma}

\begin{proof}
For the chosen link the exact fundamental-representation identities
are
\[
 \mathcal C_eF=\tfrac34F,\qquad
 \Gamma_eF=\tfrac14(1-F^2),\qquad
 \mathcal C_e(F^2)=2F^2-\tfrac12,\qquad
 \Gamma_e(F^2)=F^2(1-F^2).
\]
Integration by parts against $h^2$ and Cauchy--Schwarz give
\[
 |\tfrac32-2A|
 =\left|2\int h\sum_bX_{e,b}(F^2)X_{e,b}h\,d\mu\right|
 \le2\sqrt{M\nu(F^2(1-F^2))}
 \le2\sqrt{MA(1-A)}.
\]
In the last step,
$\nu(F^2(1-F^2))=A-\nu((1-F^2)^2)\le A-A^2$.
Squaring yields
\[
 4(M+1)A^2-(4M+6)A+\tfrac94\le0.
\]
The smaller root is \eqref{ham:eq:fisher:Aexact}. Rationalization gives
its second expression. Finally
$\sqrt{M(4M+3)}\le2M+3/4$ because the difference of the squares
is $9/16$, so the denominator there is at most $16M+15$.
\end{proof}

If the loop has $n$ distinct links, the centered physical state
$\psi_F=(F-\nu F)h$ has the exact norm and energy
\begin{equation}\label{ham:eq:fisher:state-energy}
 \|\psi_F\|^2=\operatorname{Var}_{\nu}F,\qquad
 \langle\psi_F,K\psi_F\rangle
 =\kappa\int\Gamma(F)d\nu
 =\frac{\kappa n}{4}\nu(1-F^2).
\end{equation}
Indeed, expand $\langle Fh,(H-E_0)Fh\rangle$ and use
$(H-E_0)h=0$; the cross terms cancel to leave
$\kappa\int h^2\Gamma(F)d\mu$. Centering does not change this
form. Each distinct link contributes $(1-F^2)/4$.
Gauge invariance of $F$ and $h$ proves that these are physical
vacuum-complement states. The lower bound on this specified state's
energy is not a lower bound on the full spectrum.

\begin{corollary}[Actual plaquette states, uniformly in volume]
\label{ham:cor:fisher:plaquette}
Let $F=f_p$ be an elementary plaquette and $s=g^2\in(0,1]$.
Write $V=\|\psi_F\|^2$ and $q=\langle\psi_F,K\psi_F\rangle$.
Then
\begin{equation}\label{ham:eq:fisher:plaquette-bounds}
 \frac{s^2}{62500}<V\le\frac32s,\qquad
 \frac{3s^2}{74a}\le q\le\frac{3s^2}{4a},\qquad
 \frac{s}{37a}\le\frac qV<\frac{46875}{a}.
\end{equation}
The sharper input bounds
\begin{equation}\label{ham:eq:fisher:sharper-upper}
 A+V=1-(\nu F)^2\le U(s):=\frac32s-\frac9{16}s^2,
 \qquad A\ge\frac{9s}{96+15s}
\end{equation}
are also valid. In particular $V\le U(s)-9s/(96+15s)$.
\end{corollary}

\begin{proof}
The mean bound in \eqref{ham:eq:fisher:input} gives
$\nu F\ge1-3s/4\ge1/4$. Hence
$A+V=1-(\nu F)^2\le U(s)\le3s/2$.
Equations \eqref{ham:eq:fisher:weak-variance} and
\eqref{ham:eq:fisher:Aweak}, with $n=4$ and $\kappa=s/(2a)$,
give every bound displayed. Division by the positive variance is
legitimate by Theorem~\ref{ham:thm:fisher:variance}.
\end{proof}

The same statements hold for fixed-$a,g$ thermodynamic vacuum limits
constructed from these finite boxes: all equal-time quantities above
are expectations of bounded local functions, and the already proved
local second-moment control passes the exact energy in
\eqref{ham:eq:fisher:state-energy}. The constants do not deteriorate
with the spatial volume.

\subsection{What scalar normalization now gives}

For the centered plaquette state the explicit scalar choice
$\Psi_s=s^{-1}\psi_F=g^{-2}\psi_F$ obeys
\begin{equation}\label{ham:eq:fisher:scalar-norm}
 \frac1{62500}<\|\Psi_s\|^2\le\frac{3}{2s}.
\end{equation}
This proves a uniform positive lower norm. The displayed upper bound
does not prove boundedness of this scalar-renormalized family as
$s\downarrow0$.

Exact variance normalization,
$\widehat\psi_F=\psi_F/\sqrt V$, always gives a physical unit
vector. If $\widehat\rho$ is its spectral probability for $aK$, then
\begin{equation}\label{ham:eq:fisher:lattice-tail}
 \int x\,d\widehat\rho(x)<46875,\qquad
 \widehat\rho((R,\infty))\le\frac{46875}{R},\qquad
 \langle\widehat\psi_F,e^{-\tau aK}\widehat\psi_F\rangle
 \ge e^{-46875\tau}.
\end{equation}
The tail inequality is Markov's inequality for the spectral measure;
the last inequality is Jensen's inequality. Thus the normalized
plaquette spectral measures are tight in lattice energy units,
uniformly in volume and weak coupling. These statements do not
give uniform tightness in physical energy units: for $K$ itself the
proved first-moment upper bound is $46875/a$. A divergent upper
bound is not evidence that the actual measures escape. No assertion
of physical spectral escape follows from this estimate.

\subsection{Sharpness and a precise limit of the local inputs}

The powers in the Fisher lower bounds cannot be improved using only
an upper bound on the one-link electric energy. To see this within
the physical Hilbert space, fix a simple loop and consider the
smooth gauge-invariant wavefunction
\[
 h_\beta=Z_\beta^{-1/2}e^{\beta F/2}.
\]
The Haar marginal of $F$ is the usual one-link trace marginal, since
a link occurring once can be changed to the full loop holonomy.
The variational identities proved above therefore give, for each
link in the loop,
\[
 \langle h_\beta,\mathcal C_eh_\beta\rangle
 =\frac{3\beta r_\beta}{16}\le\frac{3\beta}{16},\qquad
 \mathbb E_\beta(1-F^2)=\frac{3r_\beta}{\beta}\le\frac3\beta.
\]
The same decreasing Gamma tilt, now with the increasing function
$z^2$, gives
\[
 \operatorname{Var}_\beta F\le\mathbb E_\beta(1-F)^2
 \le\frac{15}{4\beta^2}.
\]
Choosing $\beta=16M/3$ gives electric energy at most $M$ and
variance at most $135/(1024M^2)$. Thus no universal lower variance
bound with a slower decay than $M^{-2}$ can hold for this class.
Likewise the order $M^{-1}$ for $\nu(1-F^2)$ is optimal for this
class. These comparison states are not asserted to be the
interacting Wilson-Hamiltonian vacuum.

There is also a fully explicit reason why the electric bound and
the chosen plaquette's mean bound alone do not provide the missing
$O(s^2)$ upper variance. For $s\in(0,1]$, set $\beta=8/s$,
$w=s/4$, and let
\[
 h_s^2=(1-w)\rho_\beta^+ +w\rho_\beta^-,\qquad
 \rho_\beta^\pm=Z_\beta^{-1}e^{\pm\beta F}.
\]
For a chosen plaquette $F$, this is a positive smooth
gauge-invariant wavefunction on the same configuration space.
Fisher information is convex: pointwise Cauchy--Schwarz gives
\[
 \left|\nabla\sqrt{(1-w)h_+^2+wh_-^2}\right|^2
 \le(1-w)|\nabla h_+|^2+w|\nabla h_-|^2.
\]
Consequently its electric expectation for each link in that
plaquette is at most $3/(2s)$; the other links have zero electric
expectation. Its chosen plaquette mean is
$(1-s/2)r_\beta$, so
\[
 \mathbb E_{h_s^2\mu}(1-F)
 \le\frac{3s}{16}+\frac s2=\frac{11s}{16}<\frac{3s}{4}.
\]
The variance decomposition of a mixture gives
\begin{equation}\label{ham:eq:fisher:mixture}
 \operatorname{Var}_{h_s^2\mu}F
 \ge4w(1-w)r_\beta^2
 \ge s\left(1-\frac s4\right)
          \left(1-\frac{3s}{16}\right)^2
 \ge\frac{507}{1024}s.
\end{equation}
It satisfies the two local inequalities used in the attempted
fourth-moment inference but violates an $O(s^2)$ upper variance.
It is not a counterexample for the actual Yang--Mills vacuum and
does not assert the mean bound for every other plaquette. It
isolates exactly why additional ground-state information is needed
for that inference. The bounded nonlinear renormalization developed
in the following analysis uses a different consequence of
Lemma~\ref{ham:lem:fisher:radial} and does not require such a fourth
moment estimate.

\section{Bounded renormalized plaquette states and their time correlations}
\label{ham:sec:bounded}

The polynomial uncertainty estimate permits an explicit bounded
renormalization with a uniformly nonzero centered norm. A complete
physical-time test is essential: for the construction below it proves
that part of the normalized spectral weight escapes to ultraviolet
energies when the spatial cutoff is removed. Thus this section supplies
an actual positive norm result and an actual obstruction for that same
family, rather than identifying a nonzero equal-time norm with a
continuum state.

Put $s=g^2\in(0,1]$, let $F=f_p$ be one elementary plaquette, and define
\begin{equation}\label{ham:eq:bounded:observable}
 Y_s=\frac{1-F}{s},\qquad
 G_s=\frac1{1+Y_s}=\frac{s}{s+1-F},\qquad
 \psi_s=(G_s-\nu G_s)h.
\end{equation}
All quantities refer to the actual interacting Hamiltonian and vacuum
of \eqref{ham:eq:fisher:model}. The function $G_s$ is smooth, gauge invariant,
and bounded between zero and one. The vector $\psi_s$ is physical and
orthogonal to the actual finite-volume vacuum.

\begin{theorem}[Uniformly nonzero bounded renormalized plaquette states]
\label{ham:thm:bounded:norm}
For every $a>0$, $L\ge2$, and $0<s\le1$,
\begin{align}
 v_s:=\|\psi_s\|^2
 &=\operatorname{Var}_\nu(G_s),&
 \frac1{8192000000}&\le v_s\le\frac14,\label{ham:eq:bounded:norm}\\
 q_s:=\langle\psi_s,K\psi_s\rangle,&&
 \frac1{896000a}&\le q_s\le\frac{27}{256a},\label{ham:eq:bounded:energy}\\
 \|K\psi_s\|&\le\frac{15}{4a}.\label{ham:eq:bounded:second}
\end{align}
The constants are independent of coupling, spacing, and spatial volume
in the stated range, apart from the explicitly displayed physical
energy powers of $a^{-1}$. No vacuum or representation truncation is
used. In particular, the centered state norm has both a strictly
positive uniform lower bound and a finite uniform upper bound along
any weak-coupling spatial-refinement sequence.
\end{theorem}

\begin{proof}
The input \eqref{ham:eq:fisher:input} gives
\[
 Y_s\ge0,\qquad \nu Y_s\le\frac34.
\]
Lemma~\ref{ham:lem:fisher:radial}, with
$Q=1+4M_g=(24+s)/s\le25/s$, gives the complete interval estimate
\begin{equation}\label{ham:eq:bounded:anti}
 \nu(Y_s\in I)\le \frac{5\pi}{\sqrt2}\sqrt{|I|}
 \quad\text{for every interval }I.
\end{equation}
The interval inverse image is clipped at the support if necessary,
which can only decrease its probability.

Let $Y_s'$ be an independent copy, set $R=3$ and $\delta=1/1000$.
Markov's inequality gives
$\mathbb P(Y_s\le3)\ge3/4$, so
\[
 \mathbb P(Y_s\le3,\ Y_s'\le3)\ge\frac9{16}.
\]
For every fixed value of $Y_s'$, the event
$|Y_s-Y_s'|\le\delta$ is an interval of length at most $2\delta$.
Thus \eqref{ham:eq:bounded:anti} and $\pi^2<10$ give
\[
 \mathbb P(|Y_s-Y_s'|\le\delta)
 \le5\pi\sqrt\delta<\frac12.
\]
The event that both variables are at most three and are separated by
more than $\delta$ has probability greater than $1/16$. On this event,
\[
 \left|\frac1{1+Y_s}-\frac1{1+Y_s'}\right|
 =\frac{|Y_s-Y_s'|}{(1+Y_s)(1+Y_s')}
 >\frac\delta{16}.
\]
The independent-pair variance identity therefore yields
\[
 \operatorname{Var}(G_s)
 >\frac12\frac1{16}\left(\frac\delta{16}\right)^2
 =\frac1{8192000000}.
\]
The non-strict version in \eqref{ham:eq:bounded:norm} follows; the upper
bound $1/4$ holds for any random variable in $[0,1]$.

For the energy, use the exact plaquette identities
$DF=3F$, $\Gamma(F)=1-F^2$. Chain differentiation gives
\begin{align}
 \Gamma(G_s)
 &=\frac{Y_s(2-sY_s)}{s(1+Y_s)^4},\label{ham:eq:bounded:gamma}\\
 DG_s
 &=\frac{3-(1+3s)Y_s-sY_s^2}{s(1+Y_s)^3}.
 \label{ham:eq:bounded:laplace}
\end{align}
The maximum of $y/(1+y)^4$ for $y\ge0$ is $27/256$, attained at
$y=1/3$. Since $0\le2-sY_s\le2$ on the support,
\[
 \|\Gamma(G_s)\|_\infty\le\frac{27}{128s}.
\]
The physical form identity gives the upper bound in
\eqref{ham:eq:bounded:energy}.

For the lower bound choose
\[
 d=\frac{49}{72000},\qquad R_0=\frac43.
\]
Markov's inequality and \eqref{ham:eq:bounded:anti} give
\[
 \mathbb P(Y_s\le R_0)\ge\frac7{16},\qquad
 \mathbb P(Y_s<d)<\sqrt{125d}=\frac7{24}.
\]
Consequently $\mathbb P(d\le Y_s\le4/3)>7/48$.
On this interval $2-sY_s\ge2/3$, and
$(1+Y_s)^4\le(7/3)^4=2401/81$. Equation~\eqref{ham:eq:bounded:gamma}
therefore gives
\[
 \nu\Gamma(G_s)
 \ge\frac7{48}\frac{54d}{2401s},\qquad
 q_s=\frac{s}{2a}\nu\Gamma(G_s)
 \ge\frac1{896000a}.
\]

To obtain the full second moment, the numerator of
\eqref{ham:eq:bounded:laplace} is at most three. Since $0<s\le1$,
its sum with $3(1+Y_s)^3$ is at least
$6+5Y_s+8Y_s^2+3Y_s^3\ge0$. Hence
$\|DG_s\|_\infty\le3/s$.
For any smooth local $G$ supported on $r$ links, direct use of
$Kh=0$ gives
\[
 K(Gh)=\kappa\left[(DG)h-2\sum_{e,b}(X_{e,b}G)(X_{e,b}h)\right],
\]
and Cauchy--Schwarz with the actual electric bound implies
\[
 \|K(Gh)\|\le\kappa\left[
       \|DG\|_\infty+2\sqrt{rM_g\|\Gamma(G)\|_\infty}\right].
\]
Here $r=4$, $M_g=6/s$. Centering removes a vacuum multiple and does
not change $K(G_sh)$. Therefore
\[
 \|K\psi_s\|
 \le\frac{s}{2a}
 \left[\frac3s+2\sqrt{4\frac6s\frac{27}{128s}}\right]
 =\frac{15}{4a}.
\]
All formulas hold on the operator domain at finite cutoff, because
the group manifold is compact and the functions are smooth.
\end{proof}

\begin{theorem}[A physical-time obstruction for this same family]
\label{ham:thm:bounded:escape}
Let $a_k\downarrow0$, $0<g_k^2\le1$, and let $L_k\ge2$ be arbitrary.
The normalized physical energy spectral measures $\rho_k$ of
$\widehat\psi_k=\psi_{g_k^2}/\sqrt{v_{g_k^2}}$ satisfy
\begin{equation}\label{ham:eq:bounded:escape}
 \boxed{\displaystyle
 \rho_k\!\left(\left[\frac1{448000a_k},\infty\right)\right)
 \ge p_*:=\frac{16}{225\cdot896000^2}
 =\frac1{11289600000000}>0.}
\end{equation}
Thus this particular family is not tight in physical energy, even
when $g_k\to0$ and $a_kL_k\to\infty$. It cannot have a
norm-preserving strongly continuous physical-time correlation limit.
\end{theorem}

\begin{proof}
Under $\rho_k$, the first two moments are
$q_k/v_k$ and $\|K_k\psi_k\|^2/v_k$. The elementary
Paley--Zygmund estimate gives
\[
 \rho_k\left([q_k/(2v_k),\infty)\right)
 \ge\frac{q_k^2}{4v_k\|K_k\psi_k\|^2}.
\]
Since $v_k\le1/4$, the threshold is at least $2q_k$ and the
probability is at least $q_k^2/\|K_k\psi_k\|^2$.
Theorem~\ref{ham:thm:bounded:norm} proves \eqref{ham:eq:bounded:escape}.
The threshold tends to infinity while the probability stays fixed,
which proves non-tightness. For every fixed $t>0$,
\[
 \langle\widehat\psi_k,e^{-tK_k}\widehat\psi_k\rangle
 \le1-p_*\left(1-e^{-t/(448000a_k)}\right).
\]
Its limsup is at most $1-p_*$ at every positive time, whereas
the equal-time norm is one. A strongly continuous semigroup acting
on a unit vector has correlation tending to one as $t\downarrow0$,
so such a norm-preserving limiting state is impossible for this family.
The proof uses an actual positive escaping spectral weight; it does
not infer non-tightness merely from a divergent mean or upper bound.
\end{proof}

\paragraph{Exact scope.}
The norm construction is unconditional for the stated SU(2) vacuum.
It is a bounded nonlinear renormalization of a shrinking elementary
plaquette. The spectral theorem shows why it cannot be promoted to
the requested four-dimensional local state merely because its norm
stays nonzero. This does not exclude renormalized observables with
appropriate space-time smearing, other operator mixing, or a
continuum theory with a mass gap. It does not prove that the physical
mass gap closes. The escaping energies concern this ultraviolet
state family, not the lowest energy in the full vacuum complement.

Multiplicative rescaling and additive identity subtraction leave
the normalized spectral measures of this family unchanged. Thus
they cannot repair \eqref{ham:eq:bounded:escape}. A construction that
does repair it must supply additional physical-time and locality
control for a different family of states.

\paragraph{Reproduction.}
The standard-library script \texttt{certify\_bounded\_states.py}
checks every rational constant above, the polynomial inequalities,
the complete interval-probability accounting, and the first/second
moment spectral-weight calculation. It records exact fractions and
outward decimal intervals. These are certified bounds on the actual
Hamiltonian states, not measured masses or numerical samples
of the unknown vacuum.

\section{Infinite-volume dynamics at fixed spatial cutoff}
\label{ham:thermo:section}

Fix the spatial spacing $a>0$ and the coupling $g>0$ throughout this
section. The physical side length $aL$ tends to infinity, while the
spatial ultraviolet cutoff remains $a$. We construct the limit from
the actual finite periodic Hamiltonian vacua. The Hilbert space and
time semigroup are reconstructed from all positive-time histories,
so the argument does not assume that an equal-time limit is already
cyclic or that a Markov property survives the limit.
The reflection-positive construction follows the
Osterwalder--Schrader framework \cite{OS1,OS2};
the time-continuity and domain estimates needed here are proved
directly for this lattice Hamiltonian.

\subsection{Finite models and a uniform bound for every local history}

Let $L\ge3$ run through odd integers, and let $\Lambda_L$ be the
periodic cubic spatial lattice. Its link configuration space is
$X_L=SU(2)^{3L^3}$ with normalized product Haar measure $\mu_L$.
Use
\begin{equation}\label{ham:thermo:model}
 H_L=\kappa D_L+vW_L,\qquad
 \kappa=\frac{g^2}{2a},\quad v=\frac4{g^2a},\quad
 D_L=-\sum_{e,b}X_{e,b}^2,\quad
 W_L=\sum_p\left(1-\tfrac12\operatorname{Tr}U_p\right).
\end{equation}
The $X_{e,b}$ are the Haar-divergence-free link vector fields in the
Casimir convention $j(j+1)$. Compact ellipticity, bounded smooth
potential and positivity of the heat kernel give a normalized
smooth strictly positive ground state $h_L$ and energy $E_{0,L}$.
Its uniqueness implies gauge, translation and cubic symmetry.
Write
\[
 K_L=H_L-E_{0,L}\ge0,\qquad d\nu_L=h_L^2\,d\mu_L,\qquad
 \mathcal H_{L,\mathrm{phys}}=L^2(\mu_L)^{\mathcal G_L}.
\]
All operator insertions used below are gauge invariant, so their
vectors and evolution lie in this physical subspace. The full
positive kernel is only an exact representation of those same
physical correlations.

A smooth local observable $F$ is a smooth bounded function of
finitely many links of the infinite lattice $\mathbb Z^3$.
We take it to be gauge invariant unless stated otherwise.
For sufficiently large $L$ its support embeds without periodic
identifications. Define the finite, volume-independent constants
\begin{equation}\label{ham:thermo:local_energy_constant}
 \Gamma(F)=\sum_{e,b}|X_{e,b}F|^2,\qquad
 \mathcal E(F)=\kappa\|\Gamma(F)\|_\infty,\qquad M(F)=\|F\|_\infty.
\end{equation}
Here and below an estimate for a fixed finite family of local
observables is required only for sufficiently large $L$.

The exact ground-state form is
\begin{equation}\label{ham:thermo:ground_form}
 q_L[h_Lf]
 :=\|K_L^{1/2}(h_Lf)\|^2
 =\kappa\int\sum_{e,b}|X_{e,b}f|^2\,d\nu_L.
\end{equation}
This identity follows by product differentiation, Haar integration
by parts and $H_Lh_L=E_{0,L}h_L$. It holds first on smooth functions
and then on the closed form domain.

\begin{lemma}[Uniform multiplication and history energy]\label{ham:thermo:history_energy}
For every form-domain vector $\xi$ and smooth local $F$,
\begin{equation}\label{ham:thermo:multiplication_energy}
 q_L[M_F\xi]^{1/2}
 \le M(F)q_L[\xi]^{1/2}
                       +\mathcal E(F)^{1/2}\|\xi\|.
\end{equation}
Both the norm and form norm decrease under $e^{-tK_L}$ for $t\ge0$.
For a positive-time history
\[
 \Phi=\prod_{j=1}^mF_j(X(t_j)),\qquad
 0\le t_1\le\cdots\le t_m,
\]
define its finite physical vector
\begin{equation}\label{ham:thermo:history_vector}
 V_L\Phi
 =e^{-t_1K_L}M_{F_1}e^{-(t_2-t_1)K_L}M_{F_2}
                 \cdots e^{-(t_m-t_{m-1})K_L}M_{F_m}h_L.
\end{equation}
Then
\begin{align}
 \|V_L\Phi\|&\le N_\Phi:=\prod_jM(F_j),\label{ham:thermo:history_norm}\\
 q_L[V_L\Phi]^{1/2}
 &\le C_\Phi:=
       \sum_j\left(\prod_{i\ne j}M(F_i)\right)\mathcal E(F_j)^{1/2}.
 \label{ham:thermo:history_energy_bound}
\end{align}
For a finite linear combination of histories, use the corresponding
sums of $N_\Phi,C_\Phi$ weighted by the absolute coefficients.
All these constants are independent of $L$ and of the time
separations.
\end{lemma}

\begin{proof}
Write $\xi=h_Lf$ and use
$\nabla(Ff)=F\nabla f+f\nabla F$ in \eqref{ham:thermo:ground_form}.
The triangle inequality in the weighted gradient space gives
\eqref{ham:thermo:multiplication_energy}. Spectral calculus proves the
two contraction assertions. Start at the vacuum, whose norm is
one and whose form energy is zero, and apply the factors of
\eqref{ham:thermo:history_vector} from right to left. At each
multiplication step the previous energy is multiplied by $M(F_j)$
and a new term bounded by the norm of the remaining vector times
$\mathcal E(F_j)^{1/2}$ is added. This proves the displayed sum.
The form seminorm and Hilbert norm satisfy the triangle inequality,
which proves the linear-combination statement.
\end{proof}

\subsection{One subsequence for all Euclidean time correlations}

The ground-state transform
\[
 \mathcal P_{L,t}f=h_L^{-1}e^{-tK_L}(h_Lf)
\]
is a positive Markov operator, preserves $1$, and is selfadjoint in
$L^2(\nu_L)$. Its transition kernel is obtained from the actual
physical heat kernel by this displayed identity. Thus it defines
consistent stationary reversible finite-dimensional path
distributions, with
\begin{equation}\label{ham:thermo:finite_moments}
 \mathbb E_L\prod_{j=1}^mF_j(X(t_j))
 =\left\langle h_L,
  M_{F_1}e^{-(t_2-t_1)K_L}M_{F_2}\cdots
     e^{-(t_m-t_{m-1})K_L}M_{F_m}h_L\right\rangle
\end{equation}
when $t_1\le\cdots\le t_m$. Stationarity removes the absolute first
time. These probabilities are an auxiliary exact representation
of the quantum correlations, not the transition law of a sampling
algorithm.

Let $\mathcal X=SU(2)^{E(\mathbb Z^3)}$, a compact metrizable space.
Extend each finite torus configuration periodically to
$\mathbb Z^3$. At rational times, the finite path law is then a
probability measure $\mathbb P_L$ on the single compact metrizable
space $\mathcal X^{\mathbb Q}$. The usual extension of consistent
finite-dimensional distributions gives this measure; no temporal
regularity assumption is needed to define it.

\begin{theorem}[Actual thermodynamic Euclidean kernels]\label{ham:thermo:kernels}
There exists a subsequence $L_n\to\infty$ such that every
finite product of smooth local observables, at every tuple of real
times, has a limit in \eqref{ham:thermo:finite_moments}.
For each fixed finite list of observables, convergence is uniform
when the time tuple ranges over a compact subset of $\mathbb R^m$.
One and the same subsequence works for all these observables and
times. The limiting kernels are stationary, time-reflection
invariant, positive as probability moments, and reflection positive.
They inherit spatial lattice translation and cubic symmetry and
are gauge invariant for physical insertions.
\end{theorem}

\begin{proof}
Compactness of probability measures on $\mathcal X^{\mathbb Q}$
gives a subsequence $\mathbb P_{L_n}\Rightarrow\mathbb P$.
It therefore gives convergence of all continuous finite
space-time cylinder functions at rational times, without a
separate subsequence for each observable.

For any local smooth $F$, stationarity and reversibility give
\begin{align}
 \mathbb E_L|F(X(t))-F(X(s))|^2
 &=2\langle Fh_L,(I-e^{-|t-s|K_L})Fh_L\rangle\nonumber\\
 &\le2|t-s|q_L[Fh_L]
 \le2|t-s|\mathcal E(F).
 \label{ham:thermo:increment}
\end{align}
Consequently, telescoping a product and applying Cauchy--Schwarz
proves, with no ordering requirement on the two time tuples,
\begin{equation}\label{ham:thermo:multitime_modulus}
 \left|\mathbb E_L\prod_jF_j(X(t_j))
       -\mathbb E_L\prod_jF_j(X(s_j))\right|
 \le\sum_j\left(\prod_{i\ne j}M(F_i)\right)
                    \sqrt{2\mathcal E(F_j)|t_j-s_j|}.
\end{equation}
The bound is uniform in volume. Approximation by rational time
tuples extends each limiting kernel to all real times. A finite
rational mesh on a compact time set and
\eqref{ham:thermo:multitime_modulus} prove uniform convergence there.
Continuous local observables also follow by uniform smooth
approximation; gauge averaging preserves that approximation for
the invariant algebra.

Stationarity, reflection symmetry and spatial symmetries pass to
the rational-time measure and then to real times. A fixed local
gauge transformation and a fixed cylinder test function can be
realized in all sufficiently large tori with no extra periodic
copies in their supports; their invariance consequently also
passes to the limit. Positivity of arbitrary finite moment
matrices passes by entrywise convergence.
Reflection positivity follows more specifically from
\eqref{ham:thermo:OS_finite} below, which represents each finite
positive-time reflection matrix as a physical Hilbert-space
Gram matrix.
\end{proof}

The limiting equal-time law on $\mathcal X$ will be denoted by
$\nu_\infty$. The proof constructs its time correlations jointly;
equal-time compactness by itself would not have done so.

\subsection{Physical Hilbert space, semigroup and complete domain control}

Let $\mathcal D_+$ be the algebra of finite linear combinations of
positive-time histories with rational $t_j\ge0$ and smooth local
gauge-invariant factors. Time reflection is
$(\theta X)(t)=X(-t)$, and time translation on a history adds $s$
to each of its insertion times; denote this map by $\tau_s$.
For $\Phi,\Psi\in\mathcal D_+$ set
\[
 (\Phi,\Psi)_{\mathrm{OS}}
   =\mathbb E_{\mathbb P}
             [\,\overline{\Phi(\theta X)}\,\Psi(X)\,].
\]
The finite-volume Markov kernels and their reversibility give the
exact identities
\begin{equation}\label{ham:thermo:OS_finite}
 (\Phi,\Psi)_{\mathrm{OS},L}
      =\langle V_L\Phi,V_L\Psi\rangle,\qquad
 V_L(\tau_s\Phi)=e^{-sK_L}V_L\Phi,\qquad
 V_L(F(X(0))\Phi)=M_FV_L\Phi.
\end{equation}
For example, reflection reverses the negative-time factor order;
the two propagations from their first insertion times to zero
combine to give the inner product in \eqref{ham:thermo:OS_finite}.
This also proves the formulas when insertions occur at time zero.

\begin{theorem}[The fixed-cutoff infinite-volume physical theory]
\label{ham:thermo:construction}
Quotient $\mathcal D_+$ by its OS-null vectors and complete it to
a Hilbert space $\mathcal H_\infty$. This construction produces:
\begin{enumerate}
 \item a normalized vector $\Omega$ represented by the constant
 history $1$;
 \item a bounded $*$-representation $\pi$ of the uniform closure
 of the continuous local gauge-invariant multiplication algebra,
 with $\|\pi(F)\|\le\|F\|_\infty$;
 \item a strongly continuous semigroup
 \[
 T(t)=e^{-tH_\infty},\qquad t\ge0,
 \]
 of positive selfadjoint contractions, with $H_\infty\ge0$
 selfadjoint and $H_\infty\Omega=0$.
\end{enumerate}
The vectors
\begin{equation}\label{ham:thermo:total_history_vectors}
 T(t_1)\pi(F_1)T(t_2-t_1)\pi(F_2)\cdots
                   T(t_m-t_{m-1})\pi(F_m)\Omega
\end{equation}
span a dense subspace. Their inner products and all their
time-ordered vacuum correlations are the actual limits in
Theorem~\ref{ham:thermo:kernels}. Spatial lattice translations act
unitarily, leave $\Omega$ invariant and commute with $T(t)$.

Every finite history vector lies in $D(H_\infty^{1/2})$ and obeys
\begin{equation}\label{ham:thermo:limiting_history_energy}
 \|H_\infty^{1/2}[\Phi]\|\le C_\Phi.
\end{equation}
The span of histories whose earliest insertion time is strictly
positive is an operator core for $H_\infty$ and a form core.
Multiplication by smooth local $F$ preserves the entire form
domain and satisfies
\begin{equation}\label{ham:thermo:limiting_multiplication}
 \|H_\infty^{1/2}\pi(F)\xi\|
 \le M(F)\|H_\infty^{1/2}\xi\|
                   +\mathcal E(F)^{1/2}\|\xi\|.
\end{equation}
\end{theorem}

\begin{proof}
Reflection positivity follows by taking the limit of the Gram
identity \eqref{ham:thermo:OS_finite}. It gives a positive
semidefinite sesquilinear form and its Cauchy--Schwarz inequality,
so the quotient is well-defined. The norm is bounded by the
ordinary path $L^2$ norm because reflection preserves
$\mathbb P$:
$\|\Phi\|_{\mathrm{OS}}^2\le\|\Phi\|_{L^2(\mathbb P)}^2$.
The resulting Hilbert space is separable, for example by uniform
approximation of local smooth observables by a countable set of
gauge-averaged Peter--Weyl polynomials at rational times.

For rational $s\ge0$, the identities
\eqref{ham:thermo:OS_finite} imply, after taking limits,
\begin{align*}
 \|\tau_s\Phi\|_{\mathrm{OS}}&\le\|\Phi\|_{\mathrm{OS}},\\
 (\tau_s\Phi,\Psi)_{\mathrm{OS}}
       &=(\Phi,\tau_s\Psi)_{\mathrm{OS}},\\
 (\Phi,\tau_s\Phi)_{\mathrm{OS}}&\ge0.
\end{align*}
The first inequality preserves the null space. Thus time
translations descend to positive selfadjoint contractions $T(s)$
on the quotient and its completion; the rational semigroup law
holds algebraically.

This semigroup is genuinely continuous at zero. Spectral
calculus and $(1-e^{-x})^2\le x$ for $x\ge0$ give
\[
 \|(e^{-sK_L}-I)V_L\Phi\|^2
       \le s\,q_L[V_L\Phi]\le s C_\Phi^2.
\]
Taking the limit proves
\begin{equation}\label{ham:thermo:history_continuity}
 \|(T(s)-I)[\Phi]\|\le\sqrt{s}\,C_\Phi
 \qquad(s\in\mathbb Q,\ s\ge0).
\end{equation}
Density and contractivity extend continuity to every vector.
Completing rational times therefore gives a unique strongly
continuous semigroup on all $t\ge0$, retaining selfadjointness,
positivity and contractivity. The spectral theorem for such a
semigroup gives its nonnegative selfadjoint generator $H_\infty$.
The constant history has norm one and is fixed by every shift,
so $H_\infty\Omega=0$.

The third identity in \eqref{ham:thermo:OS_finite} gives
$\|F(X(0))\Phi\|_{\mathrm{OS}}\le M(F)\|\Phi\|_{\mathrm{OS}}$.
It thus defines a bounded operator $\pi(F)$ preserving the null
space. Products and adjoints are inherited from multiplication,
proving the $*$-representation and its uniform extension.
Equations \eqref{ham:thermo:OS_finite} imply
\eqref{ham:thermo:total_history_vectors} and the required dense span.
They also identify every rational-time matrix element with its
finite-volume limit. Strong continuity and
Theorem~\ref{ham:thermo:kernels} identify all real-time separations.
Spatial symmetry acts on histories, preserves their Gram forms
and commutes with their time shifts, giving the asserted unitaries.

For domain control, the finite spectral inequality
\[
 \frac{\langle V_L\Phi,(I-e^{-sK_L})V_L\Phi\rangle}{s}
       \le C_\Phi^2
\]
passes to the limit. Monotone convergence in the spectral measure
as $s\downarrow0$ proves \eqref{ham:thermo:limiting_history_energy}.
For every $u>0$ and integer $p\ge1$,
\[
 \|H_\infty^p T(u)\|
       \le \sup_{\lambda\ge0}\lambda^pe^{-u\lambda}
       =\left(\frac{p}{eu}\right)^p.
\]
Thus a history shifted by positive rational $u$ belongs to every
generator power domain. To prove it is a core, take
$\xi\in D(H_\infty)$ and first approximate $\xi$ in graph norm by
$T(u)\xi$ as rational $u\downarrow0$. At fixed $u>0$, approximate
$\xi$ in Hilbert norm by history vectors $\xi_n$; boundedness of
$(I+H_\infty)T(u)$ makes $T(u)\xi_n\to T(u)\xi$ in graph norm.
The same argument with $H_\infty^{1/2}$ proves the form-core
assertion.

To pass the multiplication inequality to the whole form domain,
first use a positive-initial-time history $\xi=T(u)[\Phi]$.
Its finite representatives are $e^{-uK_L}V_L\Phi$, whose second
spectral moments are bounded uniformly by $N_\Phi^2/(eu)^2$.
Their Laplace transforms and masses converge by
\eqref{ham:thermo:OS_finite}. Spectral tightness and uniqueness of the
Laplace transform therefore give weak convergence of their
spectral measures. The uniform second moments make their first
moments uniformly integrable, so
\[
 q_L[e^{-uK_L}V_L\Phi]\longrightarrow
                       \|H_\infty^{1/2}\xi\|^2.
\]
For $M_F e^{-uK_L}V_L\Phi$, convergence of the masses and
Laplace transforms, together with the uniform first-moment bound
from \eqref{ham:thermo:multiplication_energy}, gives tightness and
lower semicontinuity of its first moment.
Applying \eqref{ham:thermo:multiplication_energy} and the
preceding exact energy limit proves
\eqref{ham:thermo:limiting_multiplication} on this form core.
Closure of the form and boundedness of $\pi(F)$ extend it to all
$D(H_\infty^{1/2})$.
\end{proof}

The construction proves that $\Omega$ is a vacuum vector. It does
not prove that $\ker H_\infty=\mathbb C\Omega$, that the limiting
state is pure, or that it clusters. The finite-volume uniqueness
of $h_L$ does not justify any of those conclusions.
The equal-time vectors $\pi(F)\Omega$ give a subspace of
$\mathcal H_\infty$; cyclicity is asserted for the complete histories
in \eqref{ham:thermo:total_history_vectors}, not assumed for that
smaller subspace.

\subsection{Exact local energy and nontrivial time dependence}

The uniform second-moment estimate in
Lemma~\ref{ham:lem:states:secondmoment} is particularly useful here.
For each smooth local $F$, with $s(F)$ support links, it gives
\begin{equation}\label{ham:thermo:second_moment}
 \|K_L(Fh_L)\|
 \le \mathcal B(F):=
 \kappa\left[\|DF\|_\infty+
            2\sqrt{s(F)M_g\|\Gamma(F)\|_\infty}\right],
\end{equation}
where $M_g$ is a fixed, volume-independent bound on the one-link
vacuum electric expectation.
Already the constant trial state and cubic symmetry give
$M_g=8/g^4$: $E_{0,L}\le vP_L$, $P_L=E_L$, and
$\kappa\langle h_L,\mathcal C_eh_L\rangle\le v$.
The sharper bound $M_g=\min(8/g^4,6/g^2)$ proved in
Lemma~\ref{ham:lem:states:trial} can be used without changing the
construction.

\begin{proposition}[Local operator domains and the exact energy form]
\label{ham:thermo:local_domain}
For every smooth local physical observable $F$,
$\pi(F)\Omega\in D(H_\infty)$ and
$\|H_\infty\pi(F)\Omega\|\le\mathcal B(F)$.
For two such observables,
\begin{equation}\label{ham:thermo:exact_local_energy}
 \left\langle H_\infty^{1/2}\pi(F)\Omega,
                H_\infty^{1/2}\pi(G)\Omega\right\rangle
 =\kappa\int\sum_{e,b}
            \overline{X_{e,b}F}\,X_{e,b}G\,d\nu_\infty.
\end{equation}
These are the actual limiting local physical energies.
\end{proposition}

\begin{proof}
Let $\rho_{L,F}$ be the spectral measure of $Fh_L$ for $K_L$.
Its mass and Laplace transform converge to those of
$\pi(F)\Omega$ for $H_\infty$. The uniform second moment
\eqref{ham:thermo:second_moment} makes the measures tight and gives a
second-moment bound for their weak limit. Uniqueness of the
Laplace transform identifies that limit with the reconstructed
spectral measure. The latter therefore has finite second
moment at most $\mathcal B(F)^2$, proving the operator-domain
statement. Its first moment converges, because
\[
 \int_{[R,\infty)}\lambda\,d\rho_{L,F}(\lambda)
       \le\frac{\mathcal B(F)^2}{R}.
\]
The finite first moment is exactly
$\kappa\int\Gamma(F)d\nu_L$ by
\eqref{ham:thermo:ground_form}, whose limit is
$\kappa\int\Gamma(F)d\nu_\infty$ since $\Gamma(F)$ is a continuous
local observable. Polarization proves
\eqref{ham:thermo:exact_local_energy}.
\end{proof}

For explicit nontriviality, let
$F=\operatorname{Tr}U_\ell/2$ be a simple finite Wilson loop
containing $n$ distinct links. The actual-vacuum local ratio
bound, uniform in $L$, can be taken as
\[
 R(g)=\frac53e^{256/g^4}.
\]
The loop estimates in Theorem~\ref{ham:thm:states:loop} therefore pass
to the thermodynamic limit. If
\[
 \psi=(\pi(F)-\nu_\infty(F))\Omega,\qquad
 V=\|\psi\|^2,\qquad Q=\|H_\infty^{1/2}\psi\|^2,
\]
then
\begin{equation}\label{ham:thermo:nontrivial_bounds}
 V\ge\frac1{4R(g)^2}>0,\qquad
 Q=\frac{\kappa n}{4}\bigl(1-\nu_\infty(F^2)\bigr)
       \ge\frac{3\kappa n}{16R(g)^2}>0,\qquad
 Q\le\frac{\kappa n}{4}.
\end{equation}
Consequently, for every finite $t>0$,
\begin{equation}\label{ham:thermo:nonconstant_correlation}
 0<
 V\exp\left(-\frac{tQ}{V}\right)
 \le \langle\psi,e^{-tH_\infty}\psi\rangle
 <V.
\end{equation}
The non-strict lower inequality is Jensen's inequality for the
normalized positive spectral measure. Strictness of the upper
inequality follows from $Q>0$, which forces nonzero positive-energy
spectral weight. A useful weaker lower bound is
$[4R(g)^2]^{-1}\exp[-\kappa nR(g)^2t]$.
Thus the infinite-volume fixed-cutoff theory has nonzero physical
states and genuinely time-dependent correlations. A possible
zero-energy component of $\psi$ is allowed; these inequalities
do not assert a positive lower edge of the entire spectrum on
$\Omega^\perp$.

\subsection{Physical real-time continuation on the same subsequence}

The reconstructed generator defines the strongly continuous unitary
group $e^{-itH_\infty}$. Its local vacuum correlations also agree
with limits of the finite-volume real-time correlations along the
same subsequence. The argument can be given without a
point-norm continuity claim for every local bounded operator.

\begin{proposition}[Real-time local correlation limit]\label{ham:thermo:real_time}
For smooth local gauge-invariant $F_0,\ldots,F_m$ and real
$t_1,\ldots,t_m$,
\begin{align}
 &\left\langle h_{L_n},
 M_{F_0}e^{-it_1K_{L_n}}M_{F_1}\cdots
               e^{-it_mK_{L_n}}M_{F_m}h_{L_n}\right\rangle
 \nonumber\\
 &\hspace{1cm}\longrightarrow
 \left\langle\Omega,
 \pi(F_0)e^{-it_1H_\infty}\pi(F_1)\cdots
               e^{-it_mH_\infty}\pi(F_m)\Omega\right\rangle.
 \label{ham:thermo:real_time_limit}
\end{align}
For each fixed finite list of observables, convergence is uniform
on compact sets of the real time parameters.
\end{proposition}

\begin{proof}
The Gram matrices of the maps
$[\Phi]\mapsto V_{L_n}\Phi$ converge by construction; the map is
used on formal histories, so a limiting null vector is represented
by vectors whose norms tend to zero. Positive rational heat
functions of $H_\infty$ correspond exactly to adding a rational
time shift to a history.

For a history $\Phi$, its finite spectral measures satisfy
\[
 \rho_{L,\Phi}([R,\infty))\le C_\Phi^2/R,
\]
and the same bound holds for its limiting vector by
\eqref{ham:thermo:limiting_history_energy}. Fix a real $t$.
Choose a continuous cutoff in energy, equal to one on $[0,R]$
and zero beyond $2R$, and multiply $e^{-it\lambda}$ by it.
The resulting function extends continuously to the one-point
compactification $[0,\infty]$, and is uniformly approximable
by polynomials in $e^{-\lambda}$, by the change of variable
$x=e^{-\lambda}$ and polynomial approximation on $[0,1]$.
The error from removing the cutoff is at most
$2C_\Phi/\sqrt R$ in Hilbert norm, uniformly in $L$ and in the
limiting space. Polynomial heat functions of a history are
again finite linear combinations of positive-time histories.
It follows that $e^{-itK_L}V_L\Phi$ can be approximated, uniformly
for large $L$, by representatives of histories that approximate
$e^{-itH_\infty}[\Phi]$ in the limiting Hilbert space.

Multiplication by $F$ has norm at most $M(F)$ in every space and
maps a history to the history with $F$ inserted at time zero.
Induction through a finite word of multiplication operators and
unitaries therefore gives simultaneous history approximations
of its finite and limiting vectors. Convergence of history
Gram matrices proves \eqref{ham:thermo:real_time_limit}.
Finally
$|e^{-ix}-1|^2\le2|x|$ gives
$\|(e^{-itK_L}-e^{-isK_L})\xi\|^2
\le2|t-s|q_L[\xi]$.
The history-energy induction also applies to words of unitaries,
since they preserve the form norm. It therefore gives a uniform
square-root continuity bound for the real-time words. A finite
time mesh proves compact-uniform convergence.
\end{proof}

\subsection{Extent of the fixed-cutoff construction}

The physical spatial size tends to infinity in this construction:
$aL_n\to\infty$. At each fixed $a,g$, it supplies a stationary
infinite-volume state, all-time Euclidean and real-time local
correlations, a nonnegative selfadjoint physical Hamiltonian, a
dense operator core, exact local energies and nontrivial
time-dependent gauge-invariant states.
It is a vacuum representation on
$\mathbb R_{\mathrm{time}}\times a\mathbb Z^3$.

The subsequence can depend on $a$ and $g$. No uniqueness of phase,
clustering, or full vacuum-complement spectral gap has been
proved. A continuum claim would require compatible control as
$a\downarrow0$ and the coupling and local observables are
renormalized. The local ratio and energy constants displayed
here are finite at fixed $a,g$ but do not provide those uniform
weak-coupling estimates. In particular, neither a positive first
spectral moment nor a nonzero amount of positive-energy weight
excludes other spectral weight arbitrarily close to zero.
No exact continuum mass is determined by the construction.

\clearpage
\part{Collective spectrum and vacuum approximation}
\section{Conditional projection angles and collective physical-gap criteria}
\label{ham:sec:collective:bridge}

Nonzero local states and a collective spectral lower bound are different
requirements. This section supplies a sufficient collective estimate in
terms of the \emph{actual vacuum's} conditional projection angles. It also
gives an overlapping-block hierarchy for which summable losses suffice.
The projection argument is part of the established subspace-angle
framework \cite{ham:collective:kassabov}; conditional variance factorization
is likewise an established framework \cite{ham:collective:cmt}. All constants
and implications used here are proved below. Neither reference supplies
the unverified SU(2) weak-coupling angles.

\subsection{The actual measure, physical sector, and conditional blocks}

On a finite periodic spatial lattice, write
\[
 H=\kappa D+vW,\qquad \kappa=\frac{g^2}{2a},\quad
 v=\frac4{g^2a},\qquad
 D=-\sum_{e,b}X_{e,b}^2 .
\]
Let $h>0$ be its normalized ground state, $E_0$ its energy, and
$d\nu=h^2d\mu$ its actual vacuum probability measure, where $\mu$ is
normalized product Haar measure. The exact physical form is
\begin{equation}\label{ham:eq:collective:physical}
 q[hf]=\kappa\int\sum_e\Gamma_e(f)\,d\nu,\qquad
 \Gamma_e(f)=\sum_b|X_{e,b}f|^2.
\end{equation}
Multiplication by $h$ is unitary from $L^2(\nu)$ onto the Haar Hilbert
space. The physical subspace corresponds to the gauge-invariant
functions; its vacuum complement corresponds to $\nu f=0$.

For any link set $B$, let
\[
 P_B f=\mathbb E_\nu[f\mid U_{B^c}],\qquad
 Q_B=I-P_B,\qquad
 V_B(f)=\|Q_Bf\|_{L^2(\nu)}^2 .
\]
Thus $V_B(f)$ is the mean conditional variance when links in $B$ are
integrated out. In particular, if $\mathcal E$ is the full link set,
$P_{\mathcal E}=\Pi$, the projection onto constants, and
$V_{\mathcal E}(f)=\operatorname{Var}_\nu(f)$.
All these are conditional expectations of the actual $h^2d\mu$.
No local Gibbs formula for that density is assumed.

The density $h^2$ is smooth and strictly positive on the compact
configuration space. Consequently
\begin{equation}\label{ham:eq:collective:intersection}
 \operatorname{ran}P_A\cap\operatorname{ran}P_B
 =\operatorname{ran}P_{A\cup B}.
\end{equation}
Indeed, equivalence with product Haar measure and Fubini's theorem show
that a function depending only on $U_{A^c}$ and also only on $U_{B^c}$
depends only on $U_{(A\cup B)^c}$.
Furthermore each $P_B$ commutes with every gauge transformation.
The gauge action preserves $\nu$ and the sigma algebra generated by
$U_{B^c}$, so this follows from the defining integral identity for
conditional expectation. The projection estimates below therefore
restrict to the physical sector. We prove the stronger estimates on
the full $L^2(\nu)$ when possible.

Suppose block constants $\gamma_B>0$ have been certified so that
\begin{equation}\label{ham:eq:collective:blockpi}
 V_B(f)\le\gamma_B^{-1}
     \int\sum_{e\in B}\Gamma_e(f)\,d\nu .
\end{equation}
It suffices to prove the corresponding Poincare inequality for the
conditional law of $U_B$, uniformly for almost every exterior
configuration. The full Sobolev form domain is covered by closure;
no finite representation basis is involved.

\paragraph{A fully analytic seed available for the actual SU(2) vacuum.}
Let $R(g)$ bound the ratio of the maximum to the minimum of $h$ when one
link is varied and every other link is fixed. A proved,
completely explicit choice at every $g>0$ is
\begin{equation}\label{ham:eq:collective:R}
 R(g)=\frac53\exp\left(\frac{256}{g^4}\right).
\end{equation}
For completeness, removing the at most four incident plaquettes writes
$H=\kappa\mathcal C_e+H_{-e}+V_e$ with
$0\le V_e\le32/(g^2a)$. At $t=4/\kappa$ the positive-kernel comparison
and the ground-state equation give a one-link ratio at most
$(M_4/m_4)e^{256/g^4}$. The SU(2) character expansion has complete tail
\[
 |q_4(U)-1|
 \le\sum_{n=2}^{\infty}n^2e^{-(n^2-1)}
 <\frac{80}{391}<\frac14 .
\]
Its first term is below $1/5$, and successive ratios are below
$9/400$, by $e^3>20$ and $e^5>100$; these two inequalities follow from
finite positive Taylor sums. Hence $M_4/m_4<5/3$.
This recalls an analytic all-representation certificate.

Changing links in $B$ successively gives a conditional density ratio
at most $R(g)^{2|B|}$. Product Haar measure on $SU(2)^{|B|}$ has
Casimir Poincare constant $3/4$. Comparing its maximum and minimum
density therefore proves the unconditional block seed
\begin{equation}\label{ham:eq:collective:seed}
 \gamma_B=\frac34R(g)^{-2|B|}
\end{equation}
in \eqref{ham:eq:collective:blockpi}, uniformly in the total volume and in
the exterior configuration. A stronger certified one-link ratio can
be substituted without changing any argument. This seed by itself
is not a collective inequality.

\subsection{The exact two-projection estimate}

\begin{lemma}[Pair angle and variance factorization]
\label{ham:lem:collective:two}
Let $P,Q$ be orthogonal projections on a Hilbert space, and let $R$
project onto their range intersection. Put
$c=\|PQ-R\|\le1$. Then
\begin{align}
 (I-P)+(I-Q)&\ge(1-c)(I-R),\label{ham:eq:collective:pairvar}\\
 (I-P)(I-Q)+(I-Q)(I-P)
 &\ge-c\bigl((I-P)+(I-Q)\bigr).\label{ham:eq:collective:anti}
\end{align}
Consequently, for conditional projections and $c_{AB}<1$,
\begin{equation}\label{ham:eq:collective:two_variance}
 V_{A\cup B}(f)\le
 \frac{V_A(f)+V_B(f)}{1-c_{AB}},\qquad
 c_{AB}:=\|P_AP_B-P_{A\cup B}\|.
\end{equation}
\end{lemma}

\begin{proof}
Set $P'=P-R$, $Q'=Q-R$. They are projections and
$\|P'Q'\|=c$. For vectors in their respective ranges,
$|\langle u,v\rangle|\le c\|u\|\|v\|$. It follows that, with
$S=P'+Q'$,
\[
 \langle x,S^2x\rangle
 =\|P'x+Q'x\|^2
 \le(1+c)(\|P'x\|^2+\|Q'x\|^2)
 =(1+c)\langle x,Sx\rangle .
\]
The spectral theorem gives $S\le(1+c)I$. On $(\operatorname{ran}R)^\perp$,
$D:=(I-P)+(I-Q)=2I-S\ge(1-c)I$, while $D$ vanishes on
$\operatorname{ran}R$. This proves \eqref{ham:eq:collective:pairvar}
and also $D^2\ge(1-c)D$. Expanding $D^2$ proves
\eqref{ham:eq:collective:anti}. Apply \eqref{ham:eq:collective:intersection}
to obtain \eqref{ham:eq:collective:two_variance}.
\end{proof}

\subsection{Finite collective threshold}

\begin{theorem}[Weighted conditional-angle criterion for the full physical gap]
\label{ham:thm:collective:angle}
Let $B_1,\ldots,B_m$ cover all links. For $i\ne j$, suppose certified
symmetric bounds $c_{ij}\ge0$ obey
\[
 \|P_{B_i}P_{B_j}-P_{B_i\cup B_j}\|\le c_{ij}.
\]
Let $\gamma_i>0$ satisfy \eqref{ham:eq:collective:blockpi} for $B_i$.
If positive weights $w_i$ give
\begin{equation}\label{ham:eq:collective:rho}
 \rho:=\min_i\left(w_i-\sum_{j\ne i}c_{ij}w_j\right)>0,
 \qquad
 C:=\max_e\sum_{i:e\in B_i}\frac{w_i}{\gamma_i},
\end{equation}
then every form-domain function satisfies
\begin{equation}\label{ham:eq:collective:globalpi}
 \operatorname{Var}_\nu(f)\le
 \frac{C}{\rho}\int\sum_e\Gamma_e(f)\,d\nu .
\end{equation}
In particular the actual physical Hamiltonian obeys
\begin{equation}\label{ham:eq:collective:gap}
 H-E_0\ge m_{\rm coll}(I-|h\rangle\langle h|)
 \quad\hbox{on the physical sector},\qquad
 m_{\rm coll}:=\frac{g^2}{2a}\frac{\rho}{C}>0 .
\end{equation}
The estimate controls all physical states and all representation
sectors.
\end{theorem}

\begin{proof}
Set $Q_i=I-P_{B_i}$ and $G=\sum_iw_iQ_i$, a bounded positive operator.
By Lemma~\ref{ham:lem:collective:two},
\begin{align*}
 G^2
 &=\sum_iw_i^2Q_i+
       \sum_{i<j}w_iw_j(Q_iQ_j+Q_jQ_i)\\
 &\ge\sum_iw_i\left(w_i-\sum_{j\ne i}c_{ij}w_j\right)Q_i
 \ge\rho G .
\end{align*}
Thus $\operatorname{spec}G\subset\{0\}\cup[\rho,\infty)$.
Also $\ker G=\bigcap_i\operatorname{ran}P_{B_i}$, which is the constants
by the cover and \eqref{ham:eq:collective:intersection}. Consequently
$G\ge\rho(I-\Pi)$.
On the other hand, the block Poincare inequalities imply
\[
 \langle f,Gf\rangle
 =\sum_iw_iV_{B_i}(f)
 \le\int\sum_e
       \left(\sum_{i:e\in B_i}\frac{w_i}{\gamma_i}\right)
                  \Gamma_e(f)\,d\nu
 \le C\int\sum_e\Gamma_e(f)\,d\nu .
\]
Combining the two bounds proves \eqref{ham:eq:collective:globalpi}.
The unitary ground-state identity \eqref{ham:eq:collective:physical} and
restriction to gauge-invariant functions give
\eqref{ham:eq:collective:gap}. No conditional projection is identified with
the physical time-transfer operator; $G$ is only an auxiliary positive
form used in a proved comparison.
\end{proof}

This condition has a finite matrix certificate. Form the nonnegative
symmetric matrix $\mathsf C=(c_{ij})$, with $c_{ii}=0$. If its spectral
radius is below one, then
\[
 w=(I-\mathsf C)^{-1}{\boldsymbol 1}
   =\sum_{n=0}^{\infty}\mathsf C^n{\boldsymbol 1}>0
\]
gives $\rho=1$. Conversely, any positive $w$ with
$\mathsf Cw<w$ implies spectral radius below one: the weighted
maximum norm bounds it by
$\max_i(\mathsf Cw)_i/w_i<1$. An exact positive vector $w$ and exact
positive residual $(I-\mathsf C)w$ are therefore a sufficient
finite-arithmetic certificate; no numerical eigenvalue rounding is
needed.

For single-link blocks with common $\gamma$ and
$\max_i\sum_{j\ne i}c_{ij}\le1-\varepsilon$, the simple choice $w_i=1$
gives the explicit implication
\begin{equation}\label{ham:eq:collective:single}
 \Delta_{\rm phys}\ge\frac{g^2}{2a}\gamma\varepsilon .
\end{equation}
Uniform local constants without a positive collective residual
$\varepsilon$ do not prove this conclusion.

\subsection{Complete two-block dependence estimates}

Fix exterior links $z$ outside $A\cup B$, write
$x=U_{A\setminus B}$ and $y=U_{B\setminus A}$, and integrate the overlap
$U_{A\cap B}$ out of the actual conditional measure. Let
$p_z(x,y)$ be the resulting density and $p_z^x,p_z^y$ its marginals.
If either exclusive set is empty, the pair angle is zero. Otherwise
\begin{equation}\label{ham:eq:collective:maxcorr}
 c_{AB}=\mathop{\rm ess\,sup}_z
 \sup_{\substack{\mathbb E_z u(x)=\mathbb E_z v(y)=0\\
             \|u\|_{L^2(p_z^x)}=\|v\|_{L^2(p_z^y)}=1}}
       |\mathbb E_z[\overline{u(x)}v(y)]| .
\end{equation}
Indeed, after removing the common exterior projection,
$P_AP_B$ is precisely the conditional-expectation operator from centered
$x$-functions to centered $y$-functions. It acts fiber by fiber over
$z$; its norm is the essential supremum of these fiber norms. Taking
input functions already in $\operatorname{ran}P_B$ attains the same
operator norm, so this is an equality, not just a correlation bound
for a selected list of observables.

Define the normalized joint-density ratio
$r_z=p_z/(p_z^xp_z^y)$. Three useful sufficient certificates are
\begin{align}
 c_{AB}^2
 &\le\mathop{\rm ess\,sup}_z
       \int |r_z-1|^2p_z^xp_z^y\,dx\,dy,\label{ham:eq:collective:chi}\\
 |r_z-1|\le\eta\ \hbox{for all relevant }z,x,y
 &\ \Longrightarrow\ c_{AB}\le\eta,\label{ham:eq:collective:relative}\\
 r_z\ge\delta>0\ \hbox{uniformly},\quad \delta\le1
 &\ \Longrightarrow\ c_{AB}\le1-\delta.\label{ham:eq:collective:minor}
\end{align}
To prove \eqref{ham:eq:collective:chi}, subtract one from the kernel when
acting on a centered function, then apply Cauchy--Schwarz and integrate.
Equation \eqref{ham:eq:collective:relative} follows immediately. For
\eqref{ham:eq:collective:minor}, write the joint measure as
$\delta p_z^xp_z^y+(1-\delta)\widetilde p_z$. The remainder has the same
marginals; its conditional-expectation operator is an $L^2$ contraction
by Jensen's inequality. On centered functions the independent part
vanishes, proving the bound.

These tests concern every centered function of the exclusive blocks.
A finite table of Wilson-loop covariances gives no upper bound on
\eqref{ham:eq:collective:maxcorr} unless the omitted subspace is also bounded.
Similarly, weak correlation of selected observables, a fitted
correlation length, or a sampling synchronization rate does not certify
the required operator norm. For the quantum Hamiltonian, $h^2d\mu$
must not be replaced by the four-dimensional classical Wilson density
or by a product of local plaquette factors.

\subsection{A multiscale criterion with a summable overlap cost}

The all-pairs condition can be stringent. An alternative uses only two
blocks at a time, but requires their conditional angles throughout an
explicit hierarchy.

\begin{theorem}[Averaged overlapping-block hierarchy]
\label{ham:thm:collective:hierarchy}
Let $\mathfrak B_0,\ldots,\mathfrak B_J$ be families of link sets,
with $\mathcal E\in\mathfrak B_J$.
For every leaf $R\in\mathfrak B_0$, assume
\[
 V_R(f)\le\gamma_0^{-1}
       \int\sum_{e\in R}\Gamma_e(f)\,d\nu .
 \]
For each $R\in\mathfrak B_j$, choose finitely many covers
$A_\alpha\cup B_\alpha=R$ with
$A_\alpha,B_\alpha\in\mathfrak B_{j-1}$ and probabilities
$p_\alpha\ge0$, $\sum_\alpha p_\alpha=1$. Suppose
\begin{align}
 \|P_{A_\alpha}P_{B_\alpha}-P_R\|
 &\le\eta_j<1
 \quad\hbox{for every cover},\label{ham:eq:collective:levelangle}\\
 \sum_\alpha p_\alpha
   \bigl(\mathbf1_{e\in A_\alpha}+\mathbf1_{e\in B_\alpha}\bigr)
 &\le1+\omega_j
 \quad(e\in R).\label{ham:eq:collective:overlap}
\end{align}
Then the full physical gap satisfies
\begin{equation}\label{ham:eq:collective:hierarchy_gap}
 \Delta_{\rm phys}\ge
 \frac{g^2}{2a}\gamma_0
       \prod_{j=1}^J\frac{1-\eta_j}{1+\omega_j}.
\end{equation}
If an arbitrarily deep family has the same seed $\gamma_0>0$ and
$\sum_j(\eta_j+\omega_j)<\infty$, the product has a strictly positive
lower bound independent of $J$, hence independent of the total
spatial volume covered by the hierarchy.
\end{theorem}

\begin{proof}
Let $C_0=\gamma_0^{-1}$ and
$C_j=C_{j-1}(1+\omega_j)/(1-\eta_j)$. The two-projection lemma,
averaged over covers, gives
\[
 V_R(f)\le\frac1{1-\eta_j}
        \sum_\alpha p_\alpha
               (V_{A_\alpha}(f)+V_{B_\alpha}(f)).
\]
Using the inductive estimates for the children and then
\eqref{ham:eq:collective:overlap} bounds this by
$C_j\int\sum_{e\in R}\Gamma_e(f)d\nu$. At the root it is a global
Poincare inequality; \eqref{ham:eq:collective:physical} gives
\eqref{ham:eq:collective:hierarchy_gap}.
For the last assertion use
\[
 \log\prod_j\frac{1-\eta_j}{1+\omega_j}
 \ge-\sum_j\left(\frac{\eta_j}{1-\eta_j}+\omega_j\right)>-\infty.
\]
The sum is finite because $\eta_j\to0$ and each of the finitely many
remaining $\eta_j$ is strictly below one.
\end{proof}

This hierarchy separates an exact combinatorial cost from a measure
estimate. For example, in the interval of $6m$ integer sites, use cuts
$q\in\{2m,\ldots,4m\}$ with equal weights and
\[
 A_q=\{0,\ldots,q+s-1\},\qquad
 B_q=\{q-s,\ldots,6m-1\},\qquad 1\le s\le m .
\]
Each child contains at most $5m$ sites, and any site belongs to the
overlap of at most $2s$ cuts. Hence the averaged multiplicity is at
most $1+2s/(2m+1)$. Tensoring with transverse coordinates preserves
this count. With $s$ of order $\sqrt m$, the overlap loss decreases
as $m^{-1/2}$. This is only a geometry check: decay of the actual
conditional angle across the integrated buffer still has to be proved,
and periodic covers must account for all periodic connections.

A complete infinite-hierarchy certificate must also bound its tail.
For example, under the illustrative assumptions
$\eta_j,\omega_j\le2^{-j-2}$ for all $j\ge1$,
\[
 \prod_{j=1}^{\infty}\frac{1-\eta_j}{1+\omega_j}
 \ge 1-\sum_{j=1}^{\infty}(\eta_j+\omega_j)
 \ge\frac12 .
\]
This follows from $\prod(1-u_j)\ge1-\sum u_j$ for $0\le u_j\le1$.
After verifying the first $J$ factors, the corresponding tail factor
is at least $1-2^{-J-1}$. These rational inequalities illustrate a
complete tail certificate; the displayed values of $\eta_j$ are not
claimed for SU(2).

\subsection{Gaussian example separating local and collective control}

The following exactly soluble quantum system tests the constants and
the inference, not the dynamics of SU(2). Let $N\ge2$, $0<\epsilon\le1$,
\[
 A_N=I-\frac{1-\epsilon}{N}{\boldsymbol1}{\boldsymbol1}^{T},\qquad
 h_N(x)=Z^{-1/2}\exp\left(-\frac14x^TA_Nx\right).
\]
The positive Hamiltonian on $L^2(\mathbb R^N,dx)$ is
\[
 K_N=-\Delta+\frac14x^TA_N^2x-\frac12\operatorname{Tr}A_N.
\]
It has actual normalized vacuum $h_N$, ground-state form
$q[h_Nf]=\int|\nabla f|^2h_N^2dx$, and exact gap $\epsilon$.
To see this, orthogonally diagonalize $A_N$: it has one eigenvalue
$\epsilon$ and $N-1$ eigenvalues one. The resulting independent
one-dimensional oscillators have excitation energies $n\epsilon$
and $n$, respectively. Equivalently, the Gaussian Poincare inequality
has constant $\min\operatorname{spec}A_N=\epsilon$, attained by
$f=N^{-1/2}\sum_i x_i$.

For the actual measure $\nu_N=h_N^2dx$, put
\[
 d=\frac{N-1+\epsilon}{N},\qquad
 r=\frac{1-\epsilon}{N-1+\epsilon}.
\]
Each one-coordinate conditional Gaussian has Poincare constant $d$.
For distinct coordinates, the conditional two-coordinate precision
matrix is
$\left(\begin{smallmatrix}d&-(1-\epsilon)/N\\
-(1-\epsilon)/N&d\end{smallmatrix}\right)$, so its correlation is $r$.
Its maximal $L^2$ correlation is exactly $r$. Indeed, after
standardizing the two Gaussians, conditional expectation sends the
$n$th normalized Hermite polynomial to $r^n$ times that polynomial.
This follows by conditioning the generating function
$e^{tx-t^2/2}$; completeness of the Hermite basis then gives the
centered operator norm $r$.

Thus Theorem~\ref{ham:thm:collective:angle}, with single-coordinate blocks
and $w_i=1$, has
\[
 \rho=1-(N-1)r=\frac{\epsilon}{d},\qquad C=d^{-1},
 \qquad \frac{\rho}{C}=\epsilon .
\]
The collective estimate is exact in this family.
Yet when $\epsilon=1/N$, the conditional constants stay above $1/2$,
while
\[
 \operatorname{Var}_{\nu_N}(x_i)=2-\frac1N,\qquad
 \int|\partial_i h_N|^2dx=\frac d4\le\frac14,\qquad
 \operatorname{gap}K_N=\frac1N\longrightarrow0 .
\]
Consequently nonzero bounded local state norms, uniformly bounded
local vacuum Fisher energy, and uniformly good conditional one-site
Poincare inequalities can all coexist with a vanishing collective
physical gap. The disappearing quantity is the collective projection
residual. This example uses the exact ground state of the displayed
Hamiltonian, not a Monte Carlo transition rate.

\subsection{Renormalized states, all histories, and the remaining SU(2) input}

If either collective theorem gives $m_L>0$ on the finite physical
space, then for \emph{every} finite family of physical vectors
$\xi_i$ let
$\xi_i^\circ=\xi_i-\langle h_L,\xi_i\rangle h_L$ and define
\[
 (G_0)_{ij}=\langle\xi_i^\circ,\xi_j^\circ\rangle,\qquad
 (G_t)_{ij}=\langle\xi_i^\circ,e^{-t(H_L-E_{0,L})}
                                     \xi_j^\circ\rangle .
\]
The physical spectral theorem gives the matrix inequality
\begin{equation}\label{ham:eq:collective:gram}
 0\le G_t\le e^{-m_Lt}G_0\qquad(t\ge0)
\end{equation}
in the positive-semidefinite order. It remains true after any
finite linear renormalization or mixing of these states, since
congruence $G\mapsto Z^*GZ$ preserves that order. A nonzero limiting
Gram norm supplies a nonzero state; it does not have to be inferred
from the gap estimate.

In the fixed-$a,g$ infinite-volume construction of Section~\ref{ham:thermo:section},
take the $\xi_i$ to be the full positive-time history family. Those
history Gram matrices and their time shifts converge along the same
subsequence and the history span is dense. Therefore a bound
$\inf_Lm_L\ge m>0$, obtained from the local-to-collective inputs above,
would pass \eqref{ham:eq:collective:gram} to the entire limiting physical
vacuum complement. It would also rule out additional vacuum vectors.
There is no assumption here that equal-time local observables alone
are cyclic or that a thermodynamic mixture is pure.

For a spatial-refinement sequence, the additional quantitative
condition would be, respectively,
\begin{align}
 \liminf_k\frac{g_k^2}{2a_k}\frac{\rho_k}{C_k}
 &\ge M>0,\label{ham:eq:collective:scaling}\\
 \text{or}\qquad
 \liminf_k\frac{g_k^2}{2a_k}\gamma_{0,k}
       \prod_{j=1}^{J_k}\frac{1-\eta_{j,k}}{1+\omega_{j,k}}
 &\ge M>0.\label{ham:eq:collective:scaling_hierarchy}
\end{align}
These inequalities do not construct a spatial continuum state or
its time correlations. They specify sufficient \emph{collective}
inputs in physical units, separately from the nonzero-state and
convergence requirements.

For the actual SU(2) vacuum the analytic block seeds
\eqref{ham:eq:collective:seed} are available, and the improved local
Fisher-based norm estimates provide a different, useful type of
nontriviality control. No uniform weak-coupling bound on
\eqref{ham:eq:collective:maxcorr}, no verified summable hierarchy of
angles, and no positive limit in \eqref{ham:eq:collective:scaling} or
\eqref{ham:eq:collective:scaling_hierarchy} has been supplied here.
Finite projection matrices or a few observed covariances would not
fill that omission without their full-space remainder bounds.
The bridge is a proved, explicitly checkable local-to-collective
criterion and a sharp diagnosis of the missing estimate; it is not
an application of that criterion to an unconstructed SU(2) continuum.

\paragraph{Exact reproduction.}
The Python script
\texttt{certify\_collective\_bridge.py} checks rational projection
identities and positive-semidefinite remainders, the Gaussian
conditional-angle and sharp-gap algebra, finite overlap geometry,
and complete geometric tail inequalities for the illustrative
hierarchy. It does not compute the unknown interacting SU(2)
conditional angles, approximate its vacuum, or certify a continuum
mass.

\section{Uniform Gram control and a physical-time obstruction for finite operator mixing}
\label{ham:sec:mixing}

This section gives a quantitative result for a specified family of
renormalized point-plaquette observables in the actual SU(2) Hamiltonian
vacuum. It treats every complex, cutoff-dependent linear combination
of a fixed polynomial basis at once, and then extends the certificate
to linear combinations on edge-disjoint elementary plaquettes.
It does not concern arbitrary local composite operators,
cross-products between different plaquettes, or overlapping link
supports. It gives necessary resource bounds for the specified
averages, not a sufficient construction of physical-scale smeared fields.

\subsection{Model and self-contained vacuum inputs}

Let the spatial lattice be periodic and cubic, with $L\ge2$. Its link
and plaquette counts are both $3L^3$. Use normalized product Haar
measure $\mu$ and the Hamiltonian
\begin{equation}\label{ham:eq:mixing:model}
 \begin{gathered}
 H=\kappa D+vW,\qquad
 \kappa=\frac{s}{2a},\quad v=\frac4{sa},\quad
 s=g^2\in(0,1],\\
 D=\sum_e\mathcal C_e,\quad
 W=\sum_p(1-F_p),\quad
 F_p=\tfrac12\operatorname{Tr}U_p.
 \end{gathered}
\end{equation}
Here $a>0$, $\mathcal C_e=-\sum_{b=1}^3X_{e,b}^2$ has eigenvalues
$j(j+1)$, and the physical Hilbert space is the vertex-gauge-invariant
subspace of $L^2(\mu)$. The smooth strictly positive normalized vacuum
is $h$, its energy is $E_0$, and
$K=H-E_0$, $d\nu=h^2d\mu$.
The full Hamiltonian has a unique positive ground state by its
strictly positive heat kernel and compact elliptic realization.
Gauge transformations commute with $H$ and preserve positivity,
so uniqueness makes $h$ gauge invariant and identifies the full
ground energy with the physical ground energy.

The actual-vacuum input is Lemma~\ref{ham:lem:states:trial}.

Fix one elementary plaquette and abbreviate
\begin{equation}\label{ham:eq:mixing:G}
 F=F_p,\qquad Y=\frac{1-F}{s},\qquad
 G=\frac1{1+Y}=\frac{s}{s+1-F}.
\end{equation}
The symbol $G$ here denotes an observable, not a transfer or sampling
operator.

\begin{lemma}[Radial anticoncentration with all spectators retained]
\label{ham:lem:mixing:anti}
For every interval $I\subset\mathbb R$,
\begin{equation}\label{ham:eq:mixing:Yanti}
 \nu Y\le\frac34,\qquad
 \nu(Y\in I)\le\frac{5\pi}{\sqrt2}\sqrt{|I|}.
\end{equation}
\end{lemma}

\begin{proof}Apply Lemma~\ref{ham:lem:fisher:radial} with $M=6/s$ and $F=1-sY$, and Lemma~\ref{ham:lem:states:trial} for the mean. All spectator variables remain included.\end{proof}

The positive Laplacian and gradient identities are
\begin{align}
 \Gamma(G)
 &=\frac{Y(2-sY)}{s(1+Y)^4},&
 \|\Gamma(G)\|_\infty&\le\frac{27}{128s},
 \label{ham:eq:mixing:gamma}\\
 DG&=\frac{3-(1+3s)Y-sY^2}{s(1+Y)^3},&
 \|DG\|_\infty&\le\frac3s.
 \label{ham:eq:mixing:DG}
\end{align}
They follow by differentiating $G$ and using
$DF=3F$, $\Gamma(F)=1-F^2$.
For the first upper bound use
$\max_{y\ge0}y/(1+y)^4=27/256$.
For the second, the numerator is at most three, and its sum with
$3(1+Y)^3$ is at least $6+5Y+8Y^2+3Y^3\ge0$.

\subsection{A polynomial estimate from a fixed-probability interval}

Put
\begin{equation}\label{ham:eq:mixing:middle}
 \mathcal A=\left\{\frac1{2000}\le Y\le\frac43\right\},
 \qquad I_*=\left[\frac37,\frac{2000}{2001}\right],
 \qquad \gamma_*=\frac1{1024}.
\end{equation}
Markov's inequality gives $\nu(Y\le4/3)\ge7/16$.
Since $\pi^2<10$, \eqref{ham:eq:mixing:Yanti} gives
$\nu(Y<1/2000)<1/4$. Therefore
\begin{equation}\label{ham:eq:mixing:middlebounds}
 \nu(\mathcal A)\ge\frac3{16},\qquad
 G\in I_*\text{ on }\mathcal A,\qquad
 \Gamma(G)\ge\frac{\gamma_*}{s}\text{ on }\mathcal A.
\end{equation}
For the last bound put $\phi(y)=y(2-y)/(1+y)^4$.
Since $s\le1$, $\Gamma(G)\ge\phi(Y)/s$ on $\mathcal A$.
The derivative
\[
 \phi'(y)=\frac{2(y^2-4y+1)}{(1+y)^5}
\]
has just one zero in $(0,4/3)$, a maximum at $2-\sqrt3$.
Thus $\phi$ takes its minimum on $[1/2000,4/3]$ at an endpoint.
Both endpoint values exceed $1/1024$:
\[
 \phi(1/2000)=\frac{15996000000}{16032024008001}>\frac1{1024},
 \qquad
 \phi(4/3)=\frac{72}{2401}>\frac1{1024}.
\]
For every interval $J$,
\begin{equation}\label{ham:eq:mixing:Ganti}
 \nu(\mathcal A\cap\{G\in J\})\le27\sqrt{|J|}.
\end{equation}
Indeed, on $I_*$ the derivative of $x\mapsto x^{-1}-1$
has absolute value at most $49/9$. Combining this with
\eqref{ham:eq:mixing:Yanti} gives constant $35\pi/(3\sqrt2)<27$.

For $d\ge1$ define the exact rational constants
\begin{equation}\label{ham:eq:mixing:epsilon}
 \delta_d=\frac1{165888d^2},\qquad
 \varepsilon_d=\frac{d!\,\delta_d^d}{2\,4^d}
 =\frac{d!}{2(663552d^2)^d}.
\end{equation}

\begin{lemma}[Uniform polynomial sublevel estimate]
\label{ham:lem:mixing:polynomial}
Let $q$ be a complex polynomial of degree at most $d$,
with at least one monomial coefficient of modulus at least $\eta>0$.
Then
\begin{equation}\label{ham:eq:mixing:sublevel}
 \nu\bigl(\mathcal A\cap
       \{|q(G)|>\varepsilon_d\eta\}\bigr)\ge\frac3{32}.
\end{equation}
\end{lemma}

\begin{proof}
Suppose the sublevel set
$S=\{x\in I_*:|q(x)|\le\varepsilon_d\eta\}$
contained $d+1$ points
$x_0<\cdots<x_d$ with consecutive separations at least $\delta_d$.
Lagrange interpolation expresses $q$ as
\[
 q(x)=\sum_{j=0}^d q(x_j)
    \frac{\prod_{\ell\ne j}(x-x_\ell)}
         {\prod_{\ell\ne j}(x_j-x_\ell)}.
\]
The numerator's coefficient $\ell^1$ norm is at most $2^d$,
since $x_\ell\in[0,1]$. Its denominator has modulus at least
$\delta_d^d j!(d-j)!$. Thus every coefficient of $q$ has
modulus at most
\[
 \varepsilon_d\eta\,2^d\delta_d^{-d}
 \sum_{j=0}^d\frac1{j!(d-j)!}
 =\varepsilon_d\eta\frac{4^d}{d!\delta_d^d}
 =\frac{\eta}{2},
\]
a contradiction.

A greedy maximal separated subset therefore contains at most
$d$ points, and intervals of radius $\delta_d$ about those points
cover $S$. Equation~\eqref{ham:eq:mixing:Ganti} gives
\[
 \nu(\mathcal A\cap\{G\in S\})
 \le27d\sqrt{2\delta_d}=\frac3{32}.
\]
Subtract this from $\nu(\mathcal A)\ge3/16$ to obtain
\eqref{ham:eq:mixing:sublevel}. The proof uses moduli throughout
and therefore covers complex coefficients.
\end{proof}

\subsection{Uniform matrix norms and spectral-tail theorem}

Define
\begin{align}
 V_d&=\frac3{32}\varepsilon_d^2,&
 Q_d&=\frac{3}{65536}\varepsilon_d^2
      =\frac{\gamma_*}{2}V_d,\label{ham:eq:mixing:VQ}\\
 A_d&=\frac{d(d+1)}2,&
 B_d&=\frac{d(d+1)(d-1)}3,\label{ham:eq:mixing:AB}\\
 R_d&=\frac{15}{4}A_d+\frac{27}{256}B_d,&
 \ell_d&=\frac{Q_d}{d^2},\quad
 E_d=\frac{\ell_d}{2},\quad
 p_d=\frac{Q_d^2}{4d^2R_d^2}.
 \label{ham:eq:mixing:tailconstants}
\end{align}
The constants are conservative, explicit, strictly positive, and
independent of $a,s,L$.

\begin{theorem}[All mixtures of a fixed point-plaquette basis]
\label{ham:thm:mixing:main}
Let $d\ge1$ and
$p_c(x)=\sum_{j=1}^d c_jx^j$ for arbitrary
$c\in\mathbb C^d$. Put
\[
 \psi_c=(p_c(G)-\nu p_c(G))h,\qquad
 \eta=\max_{1\le j\le d}|c_j|.
\]
For every $a>0$, $L\ge2$, and $s\in(0,1]$,
\begin{align}
 V_d\eta^2&\le\|\psi_c\|^2\le d^2\eta^2,
 \label{ham:eq:mixing:v}\\
 \langle\psi_c,K\psi_c\rangle
 &\ge\frac{Q_d}{a}\eta^2,
 \label{ham:eq:mixing:q}\\
 \|K\psi_c\|&\le\frac{R_d}{a}\eta .
 \label{ham:eq:mixing:second}
\end{align}
Let $\psi_j=(G^j-\nu G^j)h$ and let $\mathsf V_d$ be their
Gram matrix. Then
\begin{equation}\label{ham:eq:mixing:Gram}
 \frac{V_d}{d}I_d\le\mathsf V_d\le dI_d.
\end{equation}
For the matrix-valued physical spectral measure
\[
 \mathsf M_d(S)_{ij}
 =\langle\psi_i,\mathbf1_S(K)\psi_j\rangle,
\]
the simultaneous spectral-tail inequality is
\begin{equation}\label{ham:eq:mixing:matrixtail}
 \boxed{\displaystyle
 \mathsf M_d([E_d/a,\infty))\ge p_d\,\mathsf V_d
 }
\end{equation}
in the positive-semidefinite order. It controls every coefficient
vector, including coefficient vectors that depend on the cutoffs.
\end{theorem}

\begin{proof}
The zero vector is immediate, so assume $\eta>0$.
The polynomial $p_c-\nu p_c(G)$ has degree at most $d$ and
has at least one nonconstant coefficient of modulus $\eta$.
Apply Lemma~\ref{ham:lem:mixing:polynomial} and square its lower
threshold to get the first inequality in \eqref{ham:eq:mixing:v}.
The upper follows from
$\operatorname{Var}_\nu p_c(G)\le\nu|p_c(G)|^2\le d^2\eta^2$.
In Euclidean coefficient norm, the sharper upper is
$\nu|p_c(G)|^2\le d\|c\|_2^2$; combining it with
$\eta^2\ge\|c\|_2^2/d$ proves \eqref{ham:eq:mixing:Gram}.

The derivative $p_c'$ has a coefficient of modulus at least
$\eta$: if $|c_j|=\eta$, its corresponding derivative coefficient
has modulus $j\eta$. It also has degree at most $d$.
The exact ground-state identity, the chain rule, and
Lemma~\ref{ham:lem:mixing:polynomial} give
\[
 \langle\psi_c,K\psi_c\rangle
 =\frac{s}{2a}\nu\!\left(|p_c'(G)|^2\Gamma(G)\right)
 \ge\frac{s}{2a}\frac{\gamma_*}{s}
       \frac3{32}\varepsilon_d^2\eta^2
 =\frac{Q_d}{a}\eta^2.
\]
All vectors are physical, and centering subtracts the actual vacuum.

For the second moment, on $[0,1]$,
$|p_c'|\le A_d\eta$ and $|p_c''|\le B_d\eta$.
The positive-Laplacian chain rule is
$D(p_c(G))=p_c'(G)DG-p_c''(G)\Gamma(G)$, also for complex
$p_c$. Hence
\[
 \|D(p_c(G))\|_\infty
 \le\frac{\eta}{s}\left(3A_d+\frac{27}{128}B_d\right),
 \qquad
 \|\Gamma(p_c(G))\|_\infty
 \le\frac{27A_d^2\eta^2}{128s}.
\]
The exact product rule retains every vacuum derivative:
\[
 K(ph)=\kappa\left[(Dp)h-
       2\sum_{e,b}(X_{e,b}p)(X_{e,b}h)\right].
\]
Here $p=p_c(G)$ depends on four links. Cauchy--Schwarz and
Lemma~\ref{ham:lem:states:trial} therefore imply
\begin{align*}
 \|K\psi_c\|
 &\le\frac{s}{2a}\left[
 \frac{\eta}{s}\left(3A_d+\frac{27}{128}B_d\right)
 +2\sqrt{4\,\frac6s\,
              \frac{27A_d^2\eta^2}{128s}}\right]\\
 &=\frac{\eta}{a}\left(\frac{15}{4}A_d+
                         \frac{27}{256}B_d\right).
\end{align*}
The operator-domain calculation is legitimate at each cutoff
because $p$, $G$, and $h$ are smooth on a compact manifold.

Write $v_c=\|\psi_c\|^2$ and $q_c=\langle\psi_c,K\psi_c\rangle$.
For $c\ne0$, the normalized physical spectral measure $\rho_c$
has first moment $q_c/v_c$ and second moment
$\|K\psi_c\|^2/v_c$. Paley--Zygmund gives
\[
 \rho_c([q_c/(2v_c),\infty))
 \ge\frac{q_c^2}{4v_c\|K\psi_c\|^2}
 \ge\frac{Q_d^2}{4d^2R_d^2}=p_d.
\]
Also $q_c/(2v_c)\ge Q_d/(2d^2a)=E_d/a$.
Thus $\rho_c([E_d/a,\infty))\ge p_d$ for every $c\ne0$,
which is exactly \eqref{ham:eq:mixing:matrixtail}.
\end{proof}

\begin{corollary}[Arbitrary mixing within this fixed basis cannot repair physical time]
\label{ham:cor:mixing:continuum}
Fix $d$. Along any sequence $a_k\to0$, $s_k\in(0,1]$,
and arbitrary $L_k\ge2$, no sequence of centered states chosen from
\[
 \operatorname{span}\{G_{s_k}-\nu_kG_{s_k},\ldots,
                 G_{s_k}^{d}-\nu_kG_{s_k}^{d}\}h_k
\]
whose norms converge to a positive finite value can have a
norm-preserving strongly continuous physical-time correlation limit.
This allows arbitrary complex cutoff-dependent
coefficients, arbitrary scalar normalization, and arbitrary identity
subtractions.
More generally, the same obstruction survives every cutoff-dependent
matrix mixing of this finite family.
\end{corollary}

\begin{proof}
Define the physical correlation matrix
$\mathsf C_d(t)_{ij}=\langle\psi_i,e^{-tK}\psi_j\rangle$.
The spectral theorem and \eqref{ham:eq:mixing:matrixtail} give
\begin{equation}\label{ham:eq:mixing:time}
 \mathsf C_d(t)\le
 \left[1-p_d(1-e^{-tE_d/a})\right]\mathsf V_d
 \qquad(t\ge0).
\end{equation}
For every nonzero state in the family its normalized correlation
has limsup at most $1-p_d$ at every fixed $t>0$, whereas its
equal-time norm is one. This is incompatible with a strongly
continuous semigroup on a limiting unit vector.
If the state norms converge to a positive finite value, the same
contradiction follows after normalization.
For a mixing matrix $Z_k$, congruence by $Z_k$ preserves
\eqref{ham:eq:mixing:matrixtail} and \eqref{ham:eq:mixing:time}.
Identity subtraction changes no centered vector.
\end{proof}

\paragraph{Interpretation and restrictions.}
This is a quantitative ultraviolet obstruction for one explicitly
specified class of point operators in the actual interacting vacuum.
The constants are uniform in weak coupling and spatial volume.
The result does not show that the physical gap closes, does not
exclude finite-energy states outside this family, and does not
exclude a continuum theory. It says that any successful repair must
leave the fixed-degree polynomial family, for example through
physical-scale smearing, different local variables, or increasing
operator complexity. None of those alternatives is proved sufficient
by this theorem.

\subsection{A necessary degree growth for uniformly bounded physical energy}

The theorem also produces a quantitative resource statement.
For every nonzero $\psi_c$ of degree at most $d$,
\begin{equation}\label{ham:eq:mixing:rayleigh}
 \frac{\langle\psi_c,K\psi_c\rangle}{\|\psi_c\|^2}
 \ge\frac{\ell_d}{a},\qquad
 \ell_d=
 \frac{3(d!)^2}
 {262144\,d^2(663552d^2)^{2d}}.
\end{equation}

\begin{corollary}[Necessary operator-degree growth]
\label{ham:cor:mixing:degree}
Let $\psi_k$ be nonzero centered states of the preceding polynomial
form with degree at most $d_k\ge1$, and let $a_k\to0$.
Suppose their normalized physical energy is uniformly bounded
by some $E_*>0$:
\[
 \frac{\langle\psi_k,K_k\psi_k\rangle}{\|\psi_k\|^2}\le E_*.
\]
Writing $N_k=\log(1/(a_kE_*))$, one necessarily has, for large $k$,
\begin{equation}\label{ham:eq:mixing:degreefinite}
 N_k\le
 2d_k\log d_k+
 2d_k(1+\log663552)+2\log d_k+
 \log(262144/3).
\end{equation}
In particular,
\begin{equation}\label{ham:eq:mixing:degreeasymptotic}
 \boxed{\displaystyle
 \liminf_{k\to\infty}\frac{d_k\log N_k}{N_k}\ge\frac12 .
 }
\end{equation}
Equivalently, the degree must be at least
$(1/2-o(1))\log(1/(a_kE_*))/\log\log(1/(a_kE_*))$.
All logarithms have dimensionless arguments.
\end{corollary}

\begin{proof}
Equation~\eqref{ham:eq:mixing:rayleigh} implies
$\ell_{d_k}\le a_kE_*$. Taking logarithms and using
\[
 \log(d!)=\sum_{j=1}^d\log j
 \ge\int_1^d\log x\,dx=d\log d-d+1
\]
gives \eqref{ham:eq:mixing:degreefinite}, with the harmless negative
constant $-2$ discarded.
If \eqref{ham:eq:mixing:degreeasymptotic} failed, there would be a
subsequence and $c<1/2$ with
$d_k\le cN_k/\log N_k$. On it
$2d_k\log d_k\le(2c+o(1))N_k$, and the other terms on the
right of \eqref{ham:eq:mixing:degreefinite} are $o(N_k)$.
This contradicts that inequality.
\end{proof}

The coefficient $1/2$ is a consequence of these explicit conservative
constants; it is not claimed to be the optimal physical complexity
threshold. A bounded first energy moment is a sufficient spectral
tightness condition and is stronger than an arbitrary strongly
continuous correlation limit. Without the first-moment hypothesis,
Corollary~\ref{ham:cor:mixing:continuum} forces $d_k\to\infty$
for a nonzero norm-preserving limit, but gives no universal rate.
The probabilities $p_d$ decrease with $d$, so the fixed-degree
spectral-escape conclusion cannot simply be applied with $d=d_k\to\infty$.

\subsection{Disjoint plaquettes: a support--degree tradeoff without independence}
\label{ham:sec:mixing:disjoint}

The same estimates yield a collective statement about a larger
specified family. Let $p_1,\ldots,p_n$ be elementary plaquettes with
pairwise disjoint sets of link variables. They may share vertices,
and no independence or sign condition on their vacuum covariances
is assumed. Put
\[
 G_i=\frac{s}{s+1-F_{p_i}},\qquad
 f_i=\sum_{j=1}^d c_{ij}G_i^j,\qquad
 \Psi=\sum_{i=1}^n(f_i-\nu f_i)h,\qquad
 \eta_i=\max_j|c_{ij}|,\quad S=\sum_i\eta_i^2 .
\]
Products between different plaquettes and plaquettes sharing a link
are outside the following assertion.

\begin{theorem}[Disjoint-support matrix certificate and resource inequality]
\label{ham:thm:mixing:disjoint}
For the preceding family,
\begin{align}
 \frac{Q_d^2}{R_d^2n}S
 &\le\|\Psi\|^2\le nd^2S,\label{ham:eq:mixing:disjointnorm}\\
 \langle\Psi,K\Psi\rangle
 &\ge\frac{Q_d}{a}S,\qquad
 \|K\Psi\|^2\le\frac{R_d^2n}{a^2}S.
 \label{ham:eq:mixing:disjointmoments}
\end{align}
The Gram matrix of the $nd$ individual centered vectors
$(G_i^j-\nu G_i^j)h$ obeys
\begin{equation}\label{ham:eq:mixing:disjointGram}
 \frac{Q_d^2}{dR_d^2n}I_{nd}
 \le\mathsf V_{n,d}\le nd\,I_{nd}.
\end{equation}
Its matrix spectral measure satisfies
\begin{equation}\label{ham:eq:mixing:disjointtail}
 \mathsf M_{n,d}([E_d/(na),\infty))
 \ge\frac{p_d}{n^2}\mathsf V_{n,d}.
\end{equation}
In particular, every nonzero state in this family has
\begin{equation}\label{ham:eq:mixing:disjointRayleigh}
 \frac{\langle\Psi,K\Psi\rangle}{\|\Psi\|^2}
 \ge\frac{\ell_d}{na}.
\end{equation}
If the normalized physical mean energy is at most $E_*>0$, then
$naE_*\ge\ell_d$. Thus for fixed $d$ one necessarily has
$n\ge\ell_d/(aE_*)$.
More generally the explicit necessary tradeoff is
\begin{equation}\label{ham:eq:mixing:tradeoff}
 \log\frac1{aE_*}
 \le\log n+
 2d\log d+2d(1+\log663552)+2\log d+
 \log(262144/3).
\end{equation}
\end{theorem}

\begin{proof}
Since each link derivative hits at most one summand, the exact
ground-state form gives
\[
 \langle\Psi,K\Psi\rangle
 =\frac{s}{2a}\nu\Gamma\left(\sum_i f_i\right)
 =\sum_i\frac{s}{2a}\nu\Gamma(f_i).
\]
This additivity uses disjoint link supports, not independence of
the $f_i$ under $\nu$. The one-plaquette bounds give the energy
lower bound in \eqref{ham:eq:mixing:disjointmoments}.
The triangle inequality and Cauchy--Schwarz give
\[
 \|\Psi\|\le d\sum_i\eta_i\le d\sqrt{nS},
 \qquad
 \|K\Psi\|\le\frac{R_d}{a}\sum_i\eta_i
 \le\frac{R_d}{a}\sqrt{nS}.
\]
Finally
$\langle\Psi,K\Psi\rangle^2\le\|\Psi\|^2\|K\Psi\|^2$
proves the norm lower bound. Since
$S\ge d^{-1}\sum_{i,j}|c_{ij}|^2$, this proves the
Gram lower bound. For its upper bound, use
$\|(f_i-\nu f_i)h\|\le\sqrt d\,\|c_i\|_2$ before summing
over $i$; the result is
$\|\Psi\|^2\le nd\sum_{i,j}|c_{ij}|^2$.

For the normalized spectral measure of any nonzero $\Psi$,
Paley--Zygmund and the preceding bounds give tail probability at least
$Q_d^2/(4R_d^2n^2d^2)=p_d/n^2$ above the threshold
$Q_d/(2nd^2a)=E_d/(na)$. This proves the matrix inequality.
Dividing the energy bound by the norm upper bound gives
\eqref{ham:eq:mixing:disjointRayleigh}. Its logarithm and the elementary
factorial estimate used in Corollary~\ref{ham:cor:mixing:degree}
give \eqref{ham:eq:mixing:tradeoff}.
\end{proof}

For linear mixing ($d=1$), the constants can be made substantially
stronger without using the polynomial sublevel lemma. The exact
single-plaquette bounds are
\[
 \operatorname{Var}_\nu G_i\le\frac14,\qquad
 \langle(G_i-\nu G_i)h,K(G_i-\nu G_i)h\rangle
 \ge\frac1{5120a},\qquad
 \|K(G_i h)\|\le\frac{15}{4a}.
\]
To verify the middle constant directly, use the stronger layer-cake
calculation below. Set
\[
 A=\frac7{16},\qquad R=\frac43,\qquad
 T=\frac{A^2}{125}=\frac{49}{32000}.
\]
Markov's inequality gives $\nu(Y\le R)\ge A$, and
\eqref{ham:eq:mixing:Yanti} gives $\nu(Y\le t)\le\sqrt{125t}$.
On $[0,R]$ one has
\[
 \phi(y)\ge\frac74\min(y,T).
\]
Indeed, for $0\le y\le T<1/100$,
$\phi(y)/y\ge(199/100)/(101/100)^4>7/4$ (interpreted by continuity
at zero). On $[T,R]$ the minimum of $\phi$ is attained at an
endpoint by its single-maximum property, and
$\phi(R)=72/2401>7T/4$, while $\phi(T)\ge7T/4$.
The nonnegative physical gradient therefore satisfies
$\Gamma(G)\ge(7/(4s))\min(Y,T)\mathbf1_{\{Y\le R\}}$.
Layer cake gives
\begin{align*}
 \nu\bigl(\min(Y,T)\mathbf1_{\{Y\le R\}}\bigr)
 &=\int_0^T\nu(t<Y\le R)\,dt\\
 &\ge\int_0^T\bigl(A-\sqrt{125t}\bigr)\,dt
 =\frac{AT}{3}=\frac{343}{1536000}.
\end{align*}
Consequently
\[
 \frac{s}{2a}\nu\Gamma(G)
 \ge\frac{2401}{12288000a}
 >\frac1{5120a},
\]
where the last comparison is $2401>2400$ in a common denominator.
The other two bounds follow from $0<G_i\le1$ and
\eqref{ham:eq:mixing:second} with $A_1=1$, $B_1=0$.

Let
\[
 p_*=\frac1{368640000}.
\]
Repeating the preceding proof gives the explicit improved linear
certificate
\begin{equation}\label{ham:eq:mixing:linear}
 \boxed{\begin{gathered}
 \frac{p_*}{n}I_n\le\mathsf V_{n,1}\le\frac n4 I_n,
 \qquad
 \mathsf M_{n,1}\left(
       \left[\frac1{2560na},\infty\right)\right)
       \ge\frac{p_*}{n^2}\mathsf V_{n,1},\\
 \frac{\langle\Psi,K\Psi\rangle}{\|\Psi\|^2}
 \ge\frac1{1280na},\qquad
 \text{mean energy}\le E_*\ \Longrightarrow\
 n\ge\frac1{1280aE_*}.
 \end{gathered}}
\end{equation}
The full vacuum correlations remain present in these matrices.

For bounded $n,d$ and state norms converging to a positive finite value,
\eqref{ham:eq:mixing:disjointtail} rules out a
norm-preserving continuous-physical-time limit even if the chosen
plaquette positions vary with the cutoff. For growing $n$, the
tail probability can tend to zero and no general obstruction of that
kind follows. Equations~\eqref{ham:eq:mixing:tradeoff} and
\eqref{ham:eq:mixing:linear} are necessary resource bounds under the
explicit bounded-mean-energy hypothesis; they do not prove that the
indicated support growth is sufficient.

\paragraph{Exact calculation and attribution boundary.}
The companion Python script
checks all rational constants, the interpolation factorial identity,
the probability budget, the derivative sums, and the matrix-tail
arithmetic. It produces exact fractions, outward scientific decimal
intervals, and a conservative degree-threshold table.
It does not approximate the vacuum, truncate its representations,
or certify historical novelty. The sublevel interpolation and
spectral-moment tools are standard; the stated application and
constants are the calculation established here.

\section{Anchored-graph collective overlap estimate}
\label{ham:sec:overlap:graph}

The collective estimate below permits selected plaquettes with no private link. It concerns the specified plaquette states, rather than the whole physical vacuum complement.

\subsection{The per-link identity is pointwise}

Use the finite periodic cubic spatial lattice with $L\ge2$, independent
link variables in $\mathrm{SU}(2)$, and normalized product Haar measure.
Links reused in reversed orientation count as the same variable.
Write
\[
 H=\frac{s}{2a}D+\frac4{sa}W,\qquad
 K=H-E_0,\qquad d\nu=h^2d\mu,
 \qquad 0<s\le1,
\]
where $h$ is the actual normalized positive physical vacuum. The
Casimir normalization is $\mathcal C_e=-\sum_{b=1}^3X_{e,b}^2$ with
eigenvalues $j(j+1)$, and
\[
 \Gamma_e f=\sum_b|X_{e,b}f|^2,
 \qquad \Gamma f=\sum_e\Gamma_e f.
\]
For a real elementary plaquette trace $F_p=\tfrac12\operatorname{Tr}U_p$,
every one of its four distinct link variables satisfies
\begin{equation}\label{ham:eq:overlap:perlink}
 \Gamma_e F_p=\frac{1-F_p^2}{4},\qquad
 \mathcal C_eF_p=\frac34F_p.
\end{equation}
Indeed, after left/right translation or inversion of the chosen link,
the loop holonomy is $U=x_0I+i\boldsymbol{x}\cdot\boldsymbol{\sigma}$.
The derivatives generated by $i\sigma_b/2$ have squared sum
$|\boldsymbol{x}|^2/4=(1-x_0^2)/4$. Bi-invariance preserves this sum
under the spectator-dependent changes of link coordinate.

Consequently, for any complex differentiable scalar function
$f_i=p_i(G_i)$, where
\[
 G_i=\frac{s}{s+1-F_{p_i}},
\]
the value $\Gamma_e f_i$ is the same for each of its four links.
In particular,
\begin{equation}\label{ham:eq:overlap:quarter}
 \Gamma_e f_i=\frac14\Gamma f_i\quad(e\in p_i).
\end{equation}
This is a pointwise identity. The private-link proof does not need
spatial symmetry of $h$ to assign one quarter of the single-plaquette
energy to a chosen link. The spatial symmetry used earlier to bound
the actual vacuum's electric energy is a separate input.

\begin{proposition}[Private links at arbitrary overlap multiplicity]
\label{ham:overlap:private_general}
If every selected plaquette has at least $r\ge1$ links belonging to no other selected plaquette, then
\[\Gamma\!\left(\sum_i p_i(G_i)\right)\ge\frac r4\sum_i\Gamma(p_i(G_i)).\]
No bound on the number of plaquettes using the remaining links is required.
\end{proposition}
\begin{proof}
Keep only the nonnegative gradient energies on private links. Each of these contains exactly one summand and contributes one quarter of its full plaquette gradient energy by \eqref{ham:eq:overlap:quarter}. Summing proves the inequality.
\end{proof}

\subsection{The overlap graph and its exact quadratic form}

Select $N$ distinct elementary plaquettes, and assume each underlying
link variable belongs to at most two selected plaquettes. Form the
undirected multigraph with one vertex per plaquette and one edge per
link shared by two selected plaquettes. Let $A_{ij}$ count these shared
links, and set
\[
 r_i=4-\sum_j A_{ij},\qquad
 \mathcal L=4I_N-A.
\]
Thus $r_i$ is the number of private links of plaquette $i$; multiple
shared links between a pair, which can occur at small periodic $L$,
are counted with their multiplicity. There are no graph self-loops.
For every $z\in\mathbb C^N$,
\begin{equation}\label{ham:eq:overlap:graphform}
 z^*\mathcal Lz
 =\sum_{\{i,j\}\text{ shared link}}|z_i-z_j|^2
    +\sum_i r_i|z_i|^2.
\end{equation}

\begin{lemma}[Anchored overlap graph]
\label{ham:lem:overlap:graph}
The matrix $\mathcal L$ is positive definite if and only if every
connected graph component contains a vertex with a private link.
When this condition holds, its smallest eigenvalue $\lambda$ satisfies
\begin{equation}\label{ham:eq:overlap:pathbound}
 \lambda\ge\frac{2}{N(N+1)}>0.
\end{equation}
For arbitrary complex scalar functions $p_i$ as above, pointwise in
the full link configuration,
\begin{equation}\label{ham:eq:overlap:pointwise}
 \boxed{\displaystyle
 \Gamma\!\left(\sum_i p_i(G_i)\right)
 \ge\frac{\lambda}{4}\sum_i\Gamma(p_i(G_i)).}
\end{equation}
The inequality holds with $\lambda=\lambda_{\min}(\mathcal L)\ge0$
even when the anchoring condition fails, but then it gives no positive
constant.
\end{lemma}

\begin{proof}
Equation~\eqref{ham:eq:overlap:graphform} proves nonnegativity. Its kernel
consists of vectors constant on each connected component and zero on
every component containing an $r_i>0$. This proves the definiteness
criterion.

For the quantitative bound, root a component of size $n$ at a vertex
with $r_i\ge1$, and join every vertex $v$ to the root by a shortest
path of length $d_v$. For $Q=z^*\mathcal Lz$, Cauchy--Schwarz on that
path gives
\[
 |z_v|^2\le(d_v+1)
 \left(|z_{\rm root}|^2+
 \sum_{\text{path}}|z_i-z_j|^2\right)
 \le(d_v+1)Q.
\]
The ordered distances obey $d_{(k)}\le k-1$: a shortest path to
distance $d$ already contains vertices at all smaller distances.
Hence $\sum_v(d_v+1)\le n(n+1)/2$. Applying this separately to
components yields $\|z\|^2\le N(N+1)Q/2$.

For the gradient assertion, set
$x_i=\sqrt{\Gamma_e(p_i(G_i))}\ge0$, with any link $e\in p_i$.
This is independent of the choice of $e$ by
\eqref{ham:eq:overlap:quarter}. A private link contributes $x_i^2$ to
the gradient of the sum. On a shared link, the two vectors of complex
Lie derivatives have lengths $x_i,x_j$, and the reverse triangle
inequality gives
\[
 \sum_b|X_{e,b}p_i(G_i)+X_{e,b}p_j(G_j)|^2
 \ge(x_i-x_j)^2.
\]
Summing over links proves
$\Gamma(\sum_i p_i(G_i))\ge x^T\mathcal Lx
\ge\lambda\sum_i x_i^2$.
Since $\Gamma(p_i(G_i))=4x_i^2$, this is the desired inequality.
No independence, covariance sign, or gradient-alignment assumption
has been used.
\end{proof}

The preceding proof is the genuine extension beyond the private-link
argument. It uses shared-link gradients instead of discarding all of
them. If every plaquette has at least $r$ private links, then
$\lambda\ge r$ by \eqref{ham:eq:overlap:graphform}; the private-link
constant is recovered or improved.

\subsection{Physical matrix and spectral consequences}

For clarity, the single-plaquette inputs being used are stated
explicitly. They are the proved actual-vacuum estimates in the
single-plaquette calculation. For a degree at most $d$ polynomial
$p_i(x)=\sum_{j=1}^d c_{ij}x^j$, put
$\eta_i=\max_j|c_{ij}|$. Then
\begin{equation}\label{ham:eq:overlap:singleinputs}
 \begin{gathered}
 \|(p_i(G_i)-\nu p_i(G_i))h\|\le d\eta_i,\qquad
 \|K(p_i(G_i)h)\|\le\frac{R_d}{a}\eta_i,\\
 q(p_i):=\langle(p_i-\nu p_i)h,K(p_i-\nu p_i)h\rangle
 =\frac{s}{2a}\nu\Gamma p_i
 \ge\frac{Q_d}{a}\eta_i^2,
 \end{gathered}
\end{equation}
where the argument $G_i$ is implicit in the definition of $q(p_i)$ and
\[
 \begin{gathered}
 \varepsilon_d=\frac{d!}{2(663552d^2)^d},\qquad
 Q_d=\frac3{65536}\varepsilon_d^2,\\
 R_d=\frac{15}{4}\frac{d(d+1)}2
       +\frac{27}{256}\frac{d(d+1)(d-1)}3,\\
 \ell_d=\frac{Q_d}{d^2},\qquad
 E_d=\frac{\ell_d}{2},\qquad
 p_d=\frac{Q_d^2}{4d^2R_d^2}.
 \end{gathered}
\]
Let $S=\sum_i\eta_i^2$ and
$\Psi=\sum_i(p_i(G_i)-\nu p_i(G_i))h$.

\begin{theorem}[Matrix certificate with no private link at some vertices]
\label{ham:thm:overlap:matrix}
Under the anchored-graph hypotheses, for every coefficient array,
\begin{equation}\label{ham:eq:overlap:moments}
 \begin{gathered}
 q:=\langle\Psi,K\Psi\rangle\ge\frac{\lambda Q_d}{4a}S,
 \qquad
 v:=\|\Psi\|^2\le Nd^2S,\\
 w:=\|K\Psi\|^2\le\frac{R_d^2N}{a^2}S,
 \qquad
 v\ge\frac{\lambda^2Q_d^2}{16R_d^2N}S.
 \end{gathered}
\end{equation}
For the $Nd$ centered basis vectors, their Gram and spectral matrices
therefore obey
\begin{equation}\label{ham:eq:overlap:matrixbounds}
 \boxed{\begin{gathered}
 \frac{\lambda^2Q_d^2}{16dR_d^2N}I_{Nd}
 \le\mathsf V\le NdI_{Nd},\\
 \mathsf M\!\left(\left[\frac{\lambda E_d}{4Na},\infty\right)\right)
 \ge\frac{\lambda^2p_d}{16N^2}\mathsf V.
 \end{gathered}}
\end{equation}
Every nonzero $\Psi$ in the family satisfies
\begin{equation}\label{ham:eq:overlap:rayleigh}
 \frac{q}{v}\ge\frac{\lambda\ell_d}{4Na}.
\end{equation}
Coefficients may be complex and may depend arbitrarily on the cutoff,
volume, coupling, or selected geometry.
\end{theorem}

\begin{proof}
Integrate \eqref{ham:eq:overlap:pointwise} against the actual $\nu$ and
use \eqref{ham:eq:overlap:singleinputs} to obtain the lower energy bound.
The triangle inequality followed by Cauchy--Schwarz gives the upper
bounds on $v,w$. The operator inequality $q^2\le vw$ then gives the
lower bound on $v$. Since
$S\ge d^{-1}\sum_{i,j}|c_{ij}|^2$, this proves the Gram lower bound.
For the upper Gram bound, Cauchy--Schwarz over all $Nd$ bounded
monomials gives $v\le Nd\sum_{i,j}|c_{ij}|^2$.

The normalized spectral measure of any nonzero $\Psi$ has first
moment $q/v$ and second moment $w/v$. Paley--Zygmund yields
\[
 \rho_\Psi([q/(2v),\infty))\ge\frac{q^2}{4vw}
 \ge\frac{\lambda^2p_d}{16N^2},\qquad
 \frac{q}{2v}\ge\frac{\lambda E_d}{4Na}.
\]
Since this holds for every coefficient array, it is precisely the
spectral matrix inequality. Division of the first lower bound in
\eqref{ham:eq:overlap:moments} by the upper bound on $v$ gives
\eqref{ham:eq:overlap:rayleigh}.
\end{proof}

For the linear family $d=1$, retain the sharper earlier inputs
$\operatorname{Var}_\nu G_i\le1/4$,
$q(G_i)\ge1/(5120a)$, and $\|KG_ih\|\le15/(4a)$.
With $p_*=1/368640000$, the sharper consequences are
\begin{equation}\label{ham:eq:overlap:linear}
 \boxed{\begin{gathered}
 \frac{\lambda^2p_*}{16N}I_N\le\mathsf V\le\frac N4I_N,\\
 \mathsf M\!\left(\left[\frac{\lambda}{10240Na},\infty\right)\right)
 \ge\frac{\lambda^2p_*}{16N^2}\mathsf V,\qquad
 \frac qv\ge\frac{\lambda}{5120Na}.
 \end{gathered}}
\end{equation}
At fixed bounded $N,d$, the path bound keeps the tail floor positive.
Thus state sequences whose norms converge to a positive finite value
cannot have a norm-preserving strongly continuous physical-time
correlation limit inside these bounded families. If $N$ or $d$ grows,
the displayed tail floor may vanish. The estimates do not then prove
non-tightness or retained norm after a physical-energy filter.

\subsection{An open patch with an interior plaquette having no private link}

Take an $m\times n$ rectangular array of coplanar plaquettes, with
$1\le m,n<L$ so the selected patch does not wrap around the torus.
Its graph is the Cartesian product of two path graphs. Sine vectors
diagonalize each path adjacency, giving exactly
\begin{equation}\label{ham:eq:overlap:rectangle}
 \lambda_{m,n}
 =4-2\cos\frac{\pi}{m+1}-2\cos\frac{\pi}{n+1}>0.
\end{equation}
For $m,n\ge3$, interior plaquettes have zero private links.
The $3\times3$ example has
$\lambda_{3,3}=4-2\sqrt2>0$, so it is not covered by the assumption
that every plaquette has a private link.

For a square patch with $N=m^2$, the elementary inequality
$1-\cos x\ge2x^2/\pi^2$ for $0\le x\le\pi$ gives
\[
 \lambda_{m,m}\ge\frac8{(m+1)^2}.
\]
The linear family consequently satisfies
\begin{equation}\label{ham:eq:overlap:squareresource}
 \frac qv\ge\frac1{640a\,m^2(m+1)^2}
 \ge\frac1{2560aN^2}.
\end{equation}
If its normalized physical mean energy is at most $E_*>0$, a
necessary condition is therefore
\[
 N\ge\frac1{\sqrt{2560aE_*}}.
\]
This condition concerns the indicated square-patch ansatz and is not
sufficient. The more exact condition is
$m^2aE_*\ge\lambda_{m,m}/5120$. It is inappropriate to transfer the
disjoint-family $N=\Omega(a^{-1})$ estimate to overlapping patches:
the geometry changes the available collective constant.

\subsection{Why the same pointwise argument stops at closed sheets}

Select all $L^2$ $xy$-plaquettes in one periodic $xy$-plane. Every
selected link is shared twice and the graph has no anchor, so
$\lambda=0$. This is not just a defect of the graph estimate: a
positive pointwise comparison to the sum of individual gradient
energies is false for this family.

Set $\theta=2\pi/L^2$ and choose diagonal link matrices in the plane:
\[
 \begin{gathered}
 U_x(x,y)=e^{-i\theta y\sigma_3},\\
 U_y(x,y)=I\quad(y<L-1),\qquad
 U_y(x,L-1)=e^{i\theta Lx\sigma_3}.
 \end{gathered}
\]
All other links may be set to $I$. Periodicity and
$e^{i\theta L^2\sigma_3}=I$ give
$U_p=e^{i\theta\sigma_3}$ for every selected plaquette. Put
$s=1-\cos\theta\in(0,1]$. Then every selected $Y_p=(1-F_p)/s$ equals
one, so it lies strictly inside the favorable interval used in the
single-plaquette proof, and $G_p=1/2$.

For $b=1,2$, all link derivatives of these traces vanish at this
diagonal configuration. For $b=3$, the two incident plaquettes have
opposite orientation at a shared link and the same trace derivative
magnitude. Their $G$ derivatives cancel because their $F_p$ values
are equal. Therefore
\begin{equation}\label{ham:eq:overlap:cancellation}
 \boxed{\displaystyle
 \Gamma\!\left(\sum_pG_p\right)=0,\qquad
 \sum_p\Gamma G_p
 =\frac{L^2(2-s)}{16s}>0.}
\end{equation}
The ratio of these two continuous functions is arbitrarily small in
an open neighborhood of this configuration. That neighborhood can be
chosen inside the same favorable $Y$ interval. Since the actual
finite-volume vacuum density $h^2$ is strictly positive, every such
nonempty open neighborhood has positive $\nu$ measure. Thus no
strictly positive pointwise comparison constant can hold almost
everywhere, even on this favorable event.

This does not disprove a lower bound after integration against the
actual vacuum. Such a bound would need additional information about
the joint vacuum law or a different collective argument. At $L=2$,
$s=1$, every matrix entry in this example lies in
$\mathbb Q+i\mathbb Q$; the companion checker verifies all link
derivatives and cancellations exactly, with
$\sum_p\Gamma G_p=1/4$.

\paragraph{Scope.}
The positive estimate uses an explicit incidence condition and
the actual vacuum single-plaquette inputs. It does not cover arbitrary
three- or four-plaquette link multiplicities, cross-products between
different plaquettes, all physical operators, or a continuum mass
gap. The closed-sheet example excludes a proposed pointwise extension,
not a continuum construction or an integrated spectral inequality.

\section{Collective estimates with three and four plaquettes on a link}
\label{ham:sec:overlap3}

The private-link theorem requires a private link on every selected
plaquette. The subsequent anchored-graph theorem permits zero-private
interior plaquettes but assumes that each link belongs to at most two
selected plaquettes. This section proves an extension beyond both
conditions. The result concerns the specified plaquette-state family;
it is not a lower bound on the entire physical vacuum complement.

\subsection{A pointwise estimate valid before taking a vacuum expectation}

Select $N$ distinct elementary plaquettes on the periodic cubic spatial
lattice. Each plaquette has four distinct link variables; reversed use
of a link counts as the same variable. For
\[
 F_i=\tfrac12\operatorname{Tr}U_{p_i},\qquad
 G_i=\frac{s}{s+1-F_i},\qquad f_i=p_i(G_i),
\]
the scalar functions satisfy $p_i\in C^1([s/(s+2),1];\mathbb C)$.
They may in particular be polynomials with cutoff-dependent coefficients;
the spectral consequences below use the stated smooth functions.
Write
$\Gamma_e f=\sum_b|X_{e,b}f|^2$ and $\Gamma f=\sum_e\Gamma_e f$.
The exact per-link identity is
\begin{equation}\label{ham:eq:overlap3:quarter}
 \Gamma_e f_i=\frac14\Gamma f_i\quad(e\in p_i).
\end{equation}
Indeed, bi-invariance and the generators $i\sigma_b/2$ give
$\Gamma_e F_i=(1-F_i^2)/4$ on each plaquette link, and the chain rule
proves \eqref{ham:eq:overlap3:quarter}. No symmetry or independence of
the actual vacuum is used here. At each configuration, put
\[
 u_i=\sqrt{\Gamma f_i/4},\qquad F=\sum_i f_i.
\]
Thus the derivative vector of $f_i$ on every one of its links has
norm $u_i$, independently of its direction in the complex derivative
space.

One immediate extension retains the links of selected multiplicity
one or two and discards the nonnegative energy of all other links.
Let $r_i$ count private links, let $A^{(2)}_{ij}$ count links shared
by precisely $i,j$, and set
\[
 L_{\rm ret}=\operatorname{diag}
       \left(r_i+\sum_jA^{(2)}_{ij}\right)-A^{(2)}.
\]
The reverse triangle inequality gives
\begin{equation}\label{ham:eq:overlap3:retained}
 \Gamma F\ge u^*L_{\rm ret}u.
\end{equation}
This matrix is positive definite exactly when every connected
component of the retained pairwise graph contains a private-link
anchor. The statement follows from
\[
 z^*L_{\rm ret}z=\sum_{\text{retained shared }\{i,j\}}
 |z_i-z_j|^2+\sum_i r_i|z_i|^2.
\]
The two examples below have a zero row in $L_{\rm ret}$, so even
this extension cannot certify them. The next argument can.

\subsection{Free-link elimination and an explicit integer certificate}

Call a plaquette family \emph{eliminable} if it can be ordered
$p_1,\ldots,p_N$ so that each $p_i$ has a link $e_i$ belonging to no
other plaquette in the remaining family $p_i,\ldots,p_N$. The link
may belong to plaquettes removed earlier. Such an order is found by
repeatedly removing a plaquette incident on a currently private
link. A successful order is a finite, directly verifiable certificate.

For that order, define the strictly lower triangular integer matrix
\[
 A_{ij}=\mathbf1_{\{j<i,\ e_i\in p_j\}},\qquad
 R=(I-A)^{-1}=I+A+\cdots+A^{N-1}.
\]
All entries of $R$ are nonnegative integers. Define
\begin{equation}\label{ham:eq:overlap3:lambdaelim}
 \lambda_{\rm elim}=\frac1{\|R\|_2^2}>0.
\end{equation}

\begin{theorem}[Collective estimate from link elimination]
\label{ham:thm:overlap3:elimination}
For every eliminable plaquette family, pointwise in the full link
configuration and for all the scalar functions above,
\begin{equation}\label{ham:eq:overlap3:elimination}
 \boxed{\displaystyle
 \Gamma\!\left(\sum_i f_i\right)
 \ge\frac{\lambda_{\rm elim}}4\sum_i\Gamma f_i.}
\end{equation}
One may replace $\lambda_{\rm elim}$ by the fully rational lower bound
$1/\sum_{i,j}R_{ij}^2$, or by any rigorous lower enclosure of
$1/\lambda_{\max}(R^*R)$.
\end{theorem}

\begin{proof}
Put $v_i=\sqrt{\Gamma_{e_i}F}$. On $e_i$, only $f_i$ and previously
removed terms can contribute. The triangle inequality gives
\[
 u_i\le v_i+\sum_{j<i}A_{ij}u_j.
\]
Forward substitution, using the nonnegative entries, yields the
componentwise inequality $u\le Rv$. The selected pivot links $e_i$
are distinct: if $e_i=e_j$ with $i<j$, then $e_i$ would have belonged
to the still-present plaquette $p_j$ when $p_i$ was removed.
Consequently
\[
 \sum_i\Gamma f_i=4\|u\|_2^2
 \le4\|R\|_2^2\|v\|_2^2
 \le4\|R\|_2^2\Gamma F.
\]
This proves \eqref{ham:eq:overlap3:elimination}. Finally
$\|R\|_2^2\le\|R\|_{\rm F}^2=\sum_{i,j}R_{ij}^2$.
\end{proof}

Every family satisfying the earlier anchored pairwise criterion is
eliminable: removing an anchor makes an adjacent pairwise link
private, and the removal propagates through the component. The
converse fails in the examples below. Elimination is a sufficient
geometric condition, not a necessary condition for every possible
physical collective inequality. Its generic constant also need not
dominate an earlier constant when both methods apply; independent
certificates can be combined by taking the largest valid one.

\subsection{Two three-dimensional overlap configurations}

Take $L\ge4$ and let $c=(1,1,1)$. Write $p(x;xy)$, $p(x;xz)$,
and $p(x;yz)$ for the positively oriented elementary faces based
at $x$. Start with the central face $p(c;xy)$ and the four planar
neighbors
\[
 p(c-e_x;xy),\quad p(c+e_x;xy),\quad
 p(c-e_y;xy),\quad p(c+e_y;xy).
\]
Add the four upper vertical faces
\[
 p(c;xz),\quad p(c+e_y;xz),\quad
 p(c;yz),\quad p(c+e_x;yz).
\]
This is the nine-face family. Adding the translations of those four
vertical faces by $-e_z$ gives the thirteen-face family. Let $b=1$
for the upper-only family and $b=2$ for the upper-and-lower family,
so $N=5+4b$.

Every central edge belongs to $b+2$ selected faces: the central
face, one planar neighbor, and $b$ vertical faces. Thus the central
face has no private link and no retained multiplicity-two link.
Its row in $L_{\rm ret}$ is zero. The four planar neighbors each
have three private links, and each vertical face has one private
horizontal edge. Remove these outer faces first and the central
face last. This gives an elimination certificate.

For that order, $R$ is the identity except that its final row has
$b+1$ additional unit entries. Direct diagonalization of the
two-dimensional nontrivial singular block gives
\[
 \lambda_{\rm elim}
 =\frac{b+3-\sqrt{(b+1)(b+5)}}2.
\]
It equals $2-\sqrt3$ for nine faces and
$(5-\sqrt{21})/2$ for thirteen faces. These are already positive
collective constants for geometries not covered by the prior
private-link or anchored pairwise theorem.

\begin{figure}[htbp]
\centering
\includegraphics[width=\textwidth]{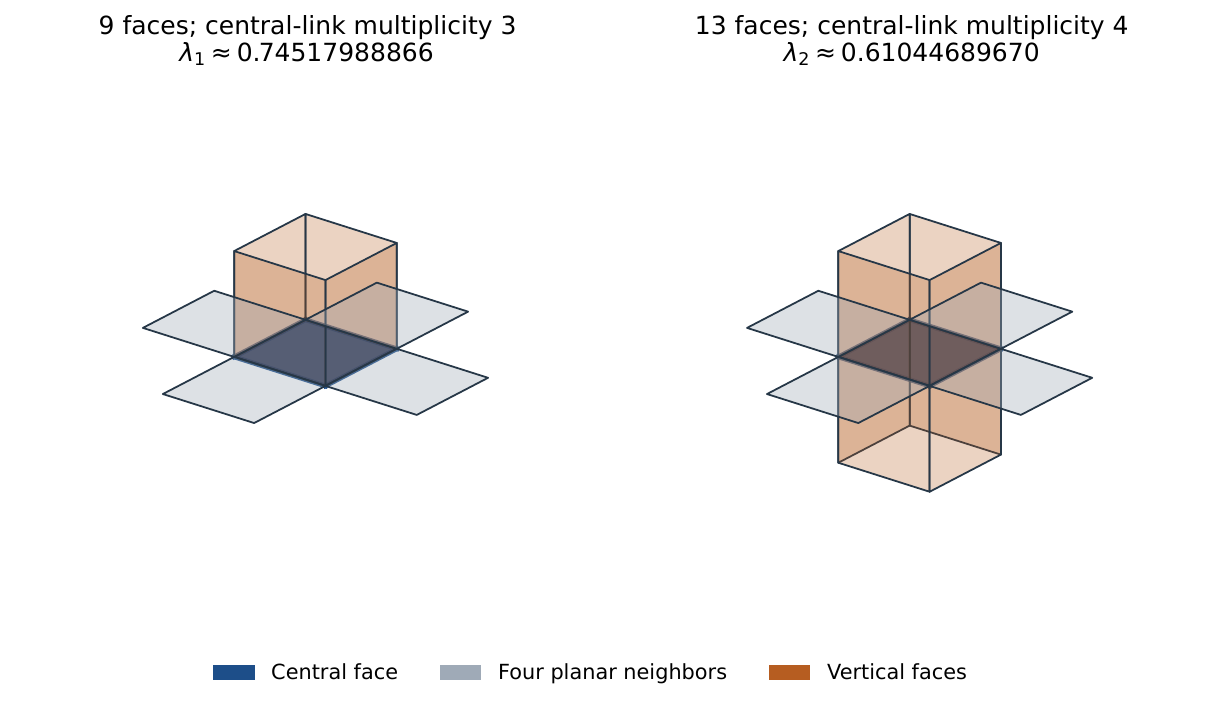}
\caption{The exact nine- and thirteen-face families. The central face
has four shared edges and no private edge. Every central edge belongs
to three selected faces on the left and four on the right. The weighted
constant $\lambda_b$ controls the complete pointwise derivative form
through \eqref{ham:eq:overlap3:weighted}; it is not a mass or a
full-Hilbert-space gap.}
\label{ham:fig:overlap_geometries}
\end{figure}

\subsection{A stronger constant using all four central edges}

The generic elimination argument uses only one pivot per face.
For these two families, all four central edges and all private
edges give a stronger bound.

\begin{theorem}[Weighted estimate for the nine- and thirteen-face families]
\label{ham:thm:overlap3:weighted}
For $b\in\{1,2\}$, let $\lambda_b$ be the unique solution in $(0,1)$ of
\begin{equation}\label{ham:eq:overlap3:root}
 \lambda\left(1+\frac1{3-\lambda}
                 +\frac b{1-\lambda}\right)=4.
\end{equation}
Equivalently, it is the unique root in $(0,1)$ of
\begin{equation}\label{ham:eq:overlap3:cubic}
 \lambda^3-(9+b)\lambda^2+(20+3b)\lambda-12=0.
\end{equation}
On the corresponding $5+4b$ face family,
\begin{equation}\label{ham:eq:overlap3:weighted}
 \boxed{\displaystyle
 \Gamma\!\left(\sum_i f_i\right)
 \ge\frac{\lambda_b}{4}\sum_i\Gamma f_i.}
\end{equation}
In particular, $\lambda_1>37/50$ and $\lambda_2>3/5$.
\end{theorem}

\begin{proof}
Let $u_0$ be the central derivative norm, $p_j$ the norm of the
planar neighbor on central edge $j$, and $v_{j\ell}$ the norms of
its $b$ vertical neighbors. Let $y_j=\sqrt{\Gamma_{e_j}F}$ be the
full derivative norm on central edge $j$. The triangle inequality,
with no directional or orientation assumption, gives
\[
 u_0\le y_j+p_j+\sum_{\ell=1}^b v_{j\ell}.
\]
The private links and the four distinct central links supply
\begin{equation}\label{ham:eq:overlap3:starenergy}
 \Gamma F\ge
 3\sum_{j=1}^4p_j^2+
 \sum_{j=1}^4\sum_{\ell=1}^b v_{j\ell}^2+
 \sum_{j=1}^4 y_j^2.
\end{equation}
For $0<\lambda<1$, weighted Cauchy--Schwarz gives
\[
 u_0^2\le D_b(\lambda)
 \left(y_j^2+(3-\lambda)p_j^2
            +(1-\lambda)\sum_\ell v_{j\ell}^2\right),
 \qquad D_b(\lambda)=1+\frac1{3-\lambda}+\frac b{1-\lambda}.
\]
After summing over $j$, \eqref{ham:eq:overlap3:starenergy} implies
\[
 \Gamma F\ge
 \lambda\left(\sum_jp_j^2+\sum_{j,\ell}v_{j\ell}^2\right)
 +\frac4{D_b(\lambda)}u_0^2.
\]
Set $\lambda=\lambda_b$, so that $4/D_b(\lambda_b)=\lambda_b$,
and use \eqref{ham:eq:overlap3:quarter}. The function
$\lambda D_b(\lambda)$ is strictly increasing from zero to infinity
on $(0,1)$, which proves existence and uniqueness. Multiplication
by the positive denominators gives \eqref{ham:eq:overlap3:cubic}.
Substitution of $37/50$ and $3/5$ into
$\lambda D_b(\lambda)<4$ proves the stated rational floors.
\end{proof}

The constant $\lambda_b$ is optimal for this particular scalar
norm relaxation. To check this, set every $y_j=1$, every
$p_j=(3-\lambda_b)^{-1}$, every
$v_{j\ell}=(1-\lambda_b)^{-1}$, and $u_0=D_b(\lambda_b)$.
All the scalar constraints are saturated, and the right side of
\eqref{ham:eq:overlap3:starenergy} equals $\lambda_b$ times the sum
of squared face norms. This does not assert that the scalar
configuration is realized by an SU(2) link configuration, nor that
$\lambda_b$ is the optimal physical constant.

\subsection{Actual-vacuum spectral consequences}

Return to the stated Hamiltonian with $0<s\le1$ and actual vacuum
$h$. For the single-plaquette vectors
$\psi_i=(G_i-\nu G_i)h$, the single-plaquette theorem supplies
\[
 \|\psi_i\|^2\le\frac14,\qquad
 \langle\psi_i,K\psi_i\rangle\ge\frac1{5120a},\qquad
 \|K\psi_i\|\le\frac{15}{4a}.
\]
For any family admitting a pointwise constant $\lambda$ above,
let $\mathsf V$ be the Gram matrix of these $N$ vectors and let
$\mathsf M(S)_{ij}=\langle\psi_i,\mathbf1_S(K)\psi_j\rangle$.
Put $p_*=1/368640000$. For arbitrary complex coefficients,
\begin{align}
 \frac{\lambda^2p_*}{16N}I_N
 &\le\mathsf V\le\frac N4I_N,
 \label{ham:eq:overlap3:gram}\\
 \mathsf M\!\left(\left[\frac{\lambda}{10240Na},\infty\right)\right)
 &\ge\frac{\lambda^2p_*}{16N^2}\mathsf V,
 \label{ham:eq:overlap3:tail}\\
 \frac{\langle\Psi,K\Psi\rangle}{\|\Psi\|^2}
 &\ge\frac{\lambda}{5120Na}\quad(\Psi\ne0).
 \label{ham:eq:overlap3:rayleigh}
\end{align}
For clarity, the proof is unchanged but its input is now genuinely
collective. If $S_c=\sum_i|c_i|^2$, then
\[
 q\ge\frac{\lambda}{20480a}S_c,
 \quad v\le\frac N4S_c,
 \quad w\le\frac{225N}{16a^2}S_c.
\]
Use $q^2\le vw$ for the lower Gram estimate, and
$\rho_\Psi([q/(2v),\infty))\ge q^2/(4vw)$ for the matrix-tail
estimate. These arguments apply simultaneously to all complex
linear combinations. They require no independence of the vacuum.

For degree-$d$ polynomials, the proved $Q_d,R_d,\ell_d,E_d,p_d$
constants remain unchanged: replace the earlier private-link
parameter $r$ by any proved $\lambda$ above. In particular the
Gram lower bound is $\lambda^2Q_d^2/(16dR_d^2N)$, the spectral
threshold is $\lambda E_d/(4Na)$, the tail fraction is
$\lambda^2p_d/(16N^2)$, and the Rayleigh lower bound is
$\lambda\ell_d/(4Na)$.

\subsection{Certified calculations and scope}

The checker constructs every actual torus link incidence, verifies
the private counts and higher multiplicities, and verifies the
elimination order and integer inverse $R$. Positive definiteness
is checked by exact rational $LDL^*$ pivots. Bisection encloses
the smallest retained-matrix eigenvalue, the largest eigenvalue
of $R^*R$, and the weighted roots, without floating-point arithmetic.
Every reported decimal is rounded outward. The companion JSON
records matrices, pivots, exact rational endpoints, and source
hashes; the CSV contains the derived constants.

For the two central examples, the resulting intervals are
\[
 \begin{aligned}
 0.7451798886612789&<\lambda_1<0.7451798886612790,\\
 0.6104468967029451&<\lambda_2<0.6104468967029453.
 \end{aligned}
\]
Thus the nine-face ansatz has a dimensionless Rayleigh floor
$a\,q/v>0.0000161714385560173$, and the thirteen-face ansatz has
$a\,q/v>0.0000091713776547918$. These are selected-family bounds,
not numerical approximations to the global mass gap.

The method still does not certify a closed unanchored periodic
sheet: it has no free link, the retained matrix has a constant
kernel vector, and elimination cannot begin. The previously
established closed-sheet cancellation remains relevant. This
section broadens the admissible overlapping families; it neither
constructs nonzero continuum states nor supplies a uniform gap
on the full physical vacuum complement. The ingredients are
elementary norm inequalities and finite incidence calculations.
Historical priority for this application has not been established.

\section{Sharp electric-cutoff obstruction for Wilson-loop expectations}
\label{ham:sec:cutoff}

This section gives a necessary test for an approximation made by retaining
only SU(2) electric representations $j\le J$. It does not assume that the
truncated state approximates the vacuum, and it allows arbitrary
entanglement with the rest of the lattice. The test is a sharp bound on
the original Wilson-loop observable, followed by an explicit consequence
of the actual-vacuum plaquette estimate.

The underlying ingredients are standard SU(2) harmonic analysis and
orthogonal-polynomial recurrences. The known difficulty of fixed electric
cutoffs at weak coupling is discussed in
\cite{ham:cutoff:dad,ham:cutoff:tong,ham:cutoff:ciavarella}. The calculation below
provides a self-contained sharp observable bound and an explicit
volume-independent application. Historical priority for that combined
formulation has not been established.

\subsection{The full one-link spectral edge}

Let $P_J$ be the Peter--Weyl projection on $L^2(SU(2),dU)$ that retains
every matrix element of each spin
$j=0,\tfrac12,\ldots,J$, where $J$ is a nonnegative half-integer.
Put
\[
 N=2J+1,\qquad
 x(U)=\tfrac12\operatorname{Tr}U,\qquad
 c_J=\cos\frac{\pi}{2J+2}=\cos\frac{\pi}{N+1}.
\]
Thus $N$ is the number of retained spin labels, not the dimension of the
full one-link space. The latter is
\begin{equation}\label{ham:eq:cutoff:dimension}
 D_J=\sum_{r=1}^N r^2=\frac{N(N+1)(2N+1)}6 .
\end{equation}

\begin{theorem}[Exact compressed Wilson-coordinate edge]
\label{ham:thm:cutoff:coordinate}
On $\operatorname{ran}P_J$,
\begin{equation}\label{ham:eq:cutoff:edge}
 -c_J I\le P_JxP_J\le c_J I,
 \qquad
 \lambda_{\max}(P_JxP_J)=c_J.
\end{equation}
For $J=0$, the compressed operator is zero. For every $J$, its upper
edge is attained by the normalized class function
\begin{equation}\label{ham:eq:cutoff:saturator}
 \phi_J(U)=
 \sqrt{\frac2{N+1}}
 \sum_{r=1}^N
 \sin\frac{r\pi}{N+1}\,
 \chi_{(r-1)/2}(U).
\end{equation}
\end{theorem}

\begin{proof}
Identify $SU(2)$ with the unit three-sphere by
\[
 U=xI+i\sqrt{1-x^2}\,\omega\cdot\sigma,
 \qquad -1\le x\le1,\quad\omega\in S^2.
\]
The spin-$j$ matrix elements are the degree-$2j$ spherical harmonics
on $S^3$: their Casimir eigenvalue is
$j(j+1)=(2j)(2j+2)/4$. Hence $\operatorname{ran}P_J$ consists of
all three-sphere harmonics of degrees at most $N-1$.

Separation into ordinary $S^2$ harmonics gives the orthogonal blocks
\[
 (1-x^2)^{\ell/2} C_n^{\ell+1}(x)Y_{\ell m}(\omega),
 \quad
 0\le\ell\le N-1,\quad -\ell\le m\le\ell,\quad
 0\le n\le N-1-\ell .
\]
This decomposition follows either by separating the $S^3$ Laplacian or
by inserting the Gegenbauer differential equation. Each displayed
function has total three-sphere degree $n+\ell$, and
\[
 \sum_{\ell=0}^{N-1}(2\ell+1)(N-\ell)
 =\sum_{r=1}^N r^2,
\]
so the blocks exhaust the full cutoff space, including all angular
sectors.

Within a block, put $\lambda=\ell+1$. The radial measure is proportional
to $(1-x^2)^{\lambda-1/2}dx$. In the corresponding orthonormal
Gegenbauer basis, multiplication by $x$ has zero diagonal and
nearest-neighbor coefficients
\begin{equation}\label{ham:eq:cutoff:gegenbauer}
 b_n^{(\lambda)}
 =
 \frac12
 \sqrt{\frac{(n+1)(n+2\lambda)}
 {(n+\lambda)(n+\lambda+1)}},
 \qquad n\ge0.
\end{equation}
For completeness, the unnormalized recurrence is
\[
 2(n+\lambda)xC_n^\lambda
 =(n+1)C_{n+1}^\lambda+(n+2\lambda-1)C_{n-1}^\lambda,
\]
and its squared norms are proportional to
$\Gamma(n+2\lambda)/[n!(n+\lambda)]$.
Their ratio gives \eqref{ham:eq:cutoff:gegenbauer}.
These standard recurrence and normalization formulas are also recorded
in the orthogonal-polynomial references \cite{DLMF}.

Since
\[
 (n+\lambda)(n+\lambda+1)-(n+1)(n+2\lambda)
 =\lambda(\lambda-1)\ge0,
\]
all off-diagonal entries satisfy $0<b_n^{(\lambda)}\le1/2$.
The block has size $m_\ell=N-\ell$.
For a complex vector $v$ in that block,
\[
 \langle v,(P_JxP_J)v\rangle
 \le
 \sum_{n=0}^{m_\ell-2}|v_n||v_{n+1}|.
\]
The expression on the right is the quadratic form of the
$m_\ell$-vertex path matrix whose off-diagonal entries are $1/2$.
Its eigenvectors are
$\sin((n+1)r\pi/(m_\ell+1))$, and its largest eigenvalue is
$\cos(\pi/(m_\ell+1))$. Therefore
\[
 \lambda_{\max}\bigl((P_JxP_J)_{\ell,m}\bigr)
 \le
 \cos\frac{\pi}{N-\ell+1}
 \le
 \cos\frac{\pi}{N+1}=c_J.
\]
The same absolute-value argument bounds the negative edge by $-c_J$.
For the block $\ell=0$, all coefficients in
\eqref{ham:eq:cutoff:gegenbauer} equal $1/2$ and its size is $N$.
Its orthonormal radial basis is the character sequence
$\chi_{(r-1)/2}$, so the first path eigenvector is exactly
\eqref{ham:eq:cutoff:saturator}. The identity
$\sum_{r=1}^N\sin^2(r\pi/(N+1))=(N+1)/2$ proves its normalization.
This also proves sharpness and the $J=0$ boundary case.
\end{proof}

\subsection{Spectators, entanglement, and gauge-invariant attainment}

Let $F=\tfrac12\operatorname{Tr}U_\ell$ be the trace of a loop
containing a chosen link $e$ exactly once, with either orientation.
Let $P_{e,J}$ retain the same electric cutoff on that link and act as
the identity on every other link.

\begin{theorem}[Sharp loop bound with arbitrary spectators]
\label{ham:thm:cutoff:loop}
Every normalized pure or mixed state $\sigma$ supported in
$\operatorname{ran}P_{e,J}$ obeys
\begin{equation}\label{ham:eq:cutoff:loop}
 \left|\operatorname{Tr}(\sigma F)\right|\le c_J,
 \qquad
 \operatorname{Tr}(\sigma(1-F))\ge1-c_J.
\end{equation}
The conclusion holds in particular after restriction to the physical
gauge-invariant sector. For an elementary plaquette the upper bound on
$\operatorname{Tr}(\sigma F)$ is attained by a normalized
gauge-invariant state that also has spin at most $J$ on every lattice
link.
\end{theorem}

\begin{proof}
Write $\xi$ for all spectator links. At fixed $\xi$, the loop holonomy
is $A(\xi)U_e^{\pm1}B(\xi)$.
Left multiplication, right multiplication, and inversion preserve
Haar measure and preserve each spin subspace. Thus the corresponding
unitary change of variable preserves $P_J$.

If $\psi=P_{e,J}\psi$, almost every fiber $\psi_\xi(U_e)$ belongs
to $\operatorname{ran}P_J$. Apply
Theorem~\ref{ham:thm:cutoff:coordinate} in every fiber and integrate
over $\xi$. This proves
$|\langle\psi,F\psi\rangle|\le c_J\|\psi\|^2$ without any
factorization assumption on $\psi$. A spectral decomposition of a
positive trace-class density operator proves the mixed-state version.

The cutoff $P_{e,J}$ commutes with both endpoint gauge actions because
it retains complete SU(2) representations. Therefore the bound remains
valid on the physical sector.

For an elementary plaquette, replace $U$ in
\eqref{ham:eq:cutoff:saturator} by $U_p$.
The resulting wavefunction $\Phi_J=\phi_J(U_p)$ is gauge invariant
because each character is a class function. Its norm is one: integration
over any one of the four distinct plaquette links reduces the
character inner product to normalized Haar orthogonality.
Each summand carries spin $(r-1)/2\le J$ on each plaquette link;
every other link carries spin zero. Character multiplication
$F\chi_j=(\chi_{j+1/2}+\chi_{j-1/2})/2$, with
$\chi_{-1/2}=0$, then gives $\langle\Phi_J,F\Phi_J\rangle=c_J$.
No claim that this saturator is a ground state is made.
\end{proof}

\subsection{A robust necessary cutoff for the actual vacuum}

For the actual Hamiltonian vacuum in this model, put $s=g^2$ and
$\delta_{\rm vac}=\nu(1-F_p)$. The established variational input is
$\delta_{\rm vac}\le3s/4$. Let $\sigma_J$ be any normalized physical
state with cutoff $j\le J$ on at least one plaquette link.
If it reproduces this plaquette deficit with absolute tolerance
$\varepsilon\ge0$,
\[
 \left|\operatorname{Tr}(\sigma_J(1-F_p))-\delta_{\rm vac}\right|
 \le\varepsilon,
\]
then, with $D=3s/4+\varepsilon$,
\begin{equation}\label{ham:eq:cutoff:necessary}
 \boxed{\displaystyle
 1-\cos\frac{\pi}{2J+2}\le D.}
\end{equation}
For $0<D<1$ this is equivalent to the explicit necessary bound
\begin{equation}\label{ham:eq:cutoff:J}
 \boxed{\displaystyle
 J\ge\frac{\pi}{2\arccos(1-D)}-1,}
\end{equation}
with $J$ rounded upward to the next nonnegative half-integer.
For $D\ge1$ this test gives no positive cutoff requirement.

More directly, every such truncated state has the certified observable
bias
\begin{equation}\label{ham:eq:cutoff:bias}
 \left|
 \operatorname{Tr}(\sigma_J(1-F_p))-\delta_{\rm vac}
 \right|
 \ge
 \left(1-\cos\frac{\pi}{2J+2}-\frac34s\right)_+.
\end{equation}
This follows immediately from Theorem~\ref{ham:thm:cutoff:loop};
no numerical approximation to the actual vacuum is used.

If $\varepsilon=o(g^2)$ along weak coupling, then
\begin{equation}\label{ham:eq:cutoff:asymptotic}
 \liminf_{g\to0} gJ(g)\ge\frac{\pi}{\sqrt6}.
\end{equation}
Indeed $\arccos(1-D)\sim\sqrt{2D}$ and
$D=(3/4)g^2(1+o(1))$.
For a fixed tolerance $\eta\ge0$ on the rescaled observable
$Y=(1-F_p)/g^2$, the same argument gives
\[
 \liminf_{g\to0} gJ(g)\ge\frac{\pi}{\sqrt{6+8\eta}}.
\]
This distinction matters: fixed absolute error for a plaquette
approaching one is weaker than fixed error for its rescaled deficit.

In the conventional full one-link Peter--Weyl basis,
\eqref{ham:eq:cutoff:dimension} additionally implies
\begin{equation}\label{ham:eq:cutoff:dimension_asymptotic}
 \liminf_{g\to0}g^3D_{J(g)}
 \ge\frac{4\pi^3}{9\sqrt6}
 \quad\hbox{when }\varepsilon=o(g^2).
\end{equation}
This is a dimension requirement for that specific unreduced link basis.
It is not a universal qubit lower bound: gauge reduction, other bases,
different encodings, and changed effective observables require separate
analysis.

\subsection{Exact numerical table and certificate}

The table uses $\varepsilon=0$ and lists the smallest half-integer $J$
not ruled out by the necessary condition
$1-c_J\le3s/4$.
It does not assert that this cutoff is sufficient to approximate the
vacuum or its spectrum.

\begin{center}
\begin{tabular}{rcc}
\hline
$s=g^2$ & Necessary $J$ at least & Full one-link dimension $D_J$\\
\hline
$1$ & $1/2$ & $5$\\
$1/2$ & $1$ & $14$\\
$1/4$ & $2$ & $55$\\
$1/8$ & $3$ & $140$\\
$1/16$ & $9/2$ & $385$\\
$1/32$ & $13/2$ & $1015$\\
$1/100$ & $12$ & $5525$\\
\hline
\end{tabular}
\end{center}

The standard-library checker uses only integers and exact fractions.
It encloses $\pi$ through Machin's identity
$\pi=16\arctan(1/5)-4\arctan(1/239)$ with complete alternating-series
remainder bounds. It then encloses every required cosine by a finite
Taylor sum and its full alternating remainder.
For each row it proves that the listed $J$ passes the necessary scalar
test while $J-\tfrac12$ fails. Monotonicity of $c_J$ certifies
minimality. Exact fractions and outward decimal intervals are included.
No floating-point eigenvalue calculation or unbounded representation
tail enters this certificate.

\paragraph{Interpretation.}
This result supplies a falsifiable accuracy test for electric-basis
SU(2) approximations, independently of their optimization, sampling,
or time-evolution algorithm. A calculation violating
\eqref{ham:eq:cutoff:necessary} cannot reproduce even the specified
actual-vacuum plaquette expectation to the stated tolerance.
The result does not estimate a physical mass gap, construct a continuum
state, or exclude an alternative finite-dimensional formulation with
different observables and a separately justified continuum limit.

\section{Extensive relative entropy of the interacting vacuum}
\label{ham:sec:entropy}

This section quantifies how the actual interacting vacuum probability
measure differs from product Haar measure. The entropy below is a classical
relative entropy of configuration probabilities. It is not a von Neumann
entropy or an entanglement entropy of the pure vacuum state.

\subsection{Model and a self-contained vacuum input}

On the periodic cubic spatial lattice with $L\ge2$, let $a>0$ and
$0<s=g^2\le1$. With normalized product Haar measure $\mu$, use
\[
 H=\kappa D+V,\qquad \kappa=\frac{s}{2a},\qquad
 D=\sum_e\mathcal C_e,\qquad V=\frac4{sa}\sum_p(1-F_p),
 \qquad F_p=\tfrac12\operatorname{Tr}U_p.
\]
Each Casimir has spin-$j$ eigenvalue $j(j+1)$. There are $3L^3$ links
and plaquettes. The compact elliptic operator has a unique smooth positive
normalized ground state $h$. Positivity and gauge invariance imply that
$h$ is physical. Set $K=H-E_0$ and $d\nu=h^2d\mu$.

The actual-vacuum input is Lemma~\ref{ham:lem:states:trial}.

\subsection{A complete Haar Laplace bound}

Let $p_1,\ldots,p_n$ be edge-disjoint elementary plaquettes and write
$X_i=1-F_{p_i}$. Under $\mu$, their holonomies are independent Haar
SU(2) variables. A single $X\in[0,2]$ has density
\[
 \frac2\pi\sqrt{2x-x^2}\,dx.
\]
Consequently, for every $t>0$,
\begin{equation}\label{ham:eq:entropy:laplace}
 \mathbb E_\mu e^{-tX}
 \le\frac{2\sqrt2}{\pi}\int_0^\infty x^{1/2}e^{-tx}\,dx
 =\sqrt{\frac2\pi}\,t^{-3/2}.
\end{equation}
No approximation of the vacuum and no independence assumption under
$\nu$ is made. Independence is used only for the reference Haar measure.

\begin{theorem}[Vacuum relative entropy and density oscillation]
\label{ham:thm:entropy}
Put
\[
 C_0=\frac{e^{3/2}}{2\sqrt\pi},\qquad
 b(s)=\log\frac1{C_0s^{3/2}}
 =\frac32\log\frac1s+\log(2\sqrt\pi)-\frac32.
\]
For every edge-disjoint selected family of $n$ plaquettes,
\begin{equation}\label{ham:eq:entropy:main}
 \boxed{\displaystyle
 D_{\rm KL}(\nu\Vert\mu)
 :=\int h^2\log h^2\,d\mu
 \ge n\,[b(s)]_+.}
\end{equation}
Writing $\rho_{\min}=\min h^2$ and $\rho_{\max}=\max h^2$,
\begin{equation}\label{ham:eq:entropy:ratio}
 \boxed{\displaystyle
 \frac{\rho_{\min}}{\rho_{\max}}
 \le\min\{1,(C_0s^{3/2})^n\}.}
\end{equation}
One may take $n=L\lfloor L/2\rfloor^2\ge L^3/9$ on every periodic
cubic lattice with $L\ge2$.
\end{theorem}

\begin{proof}
For every bounded real $f$, nonnegativity of the relative entropy
with respect to the tilted probability law
$e^f d\mu/(\mu e^f)$ gives the variational inequality
\[
 D_{\rm KL}(\nu\Vert\mu)\ge\nu f-\log\mu e^f.
\]
Set $f=-t\sum_iX_i$. The preceding lemma and
\eqref{ham:eq:entropy:laplace} give
\[
 D_{\rm KL}(\nu\Vert\mu)
 \ge n\left[-\frac34ts-\frac12\log\frac2\pi
                      +\frac32\log t\right].
\]
The displayed lower bound is maximized at $t=2/s$. Substitution
gives $n b(s)$; nonnegativity of relative entropy gives its positive part.
Furthermore,
\[
 D_{\rm KL}(\nu\Vert\mu)\le\log\rho_{\max},\qquad
 \rho_{\min}\le1
\]
by normalization, proving \eqref{ham:eq:entropy:ratio}.

For the packing, select the $xy$-plaquettes based at
$(2i,2j,z)$ with $0\le i,j<\lfloor L/2\rfloor$ and $0\le z<L$.
Their underlying link variables are disjoint. Thus the number is
$L\lfloor L/2\rfloor^2$. Since $\lfloor L/2\rfloor\ge L/3$
for $L\ge2$, the claimed lower bound follows.
\end{proof}

\subsection{A rigorous limitation of the global Haar comparison}

Let $d_{\rm Haar,phys}$ be the exact first positive eigenvalue of $D$
on the gauge-invariant Haar space. The standard global density-ratio
Poincar\'e comparison gives the certificate
\[
 B_{\rm cmp}
 =\frac{s}{2a}\,d_{\rm Haar,phys}
       \frac{\rho_{\min}}{\rho_{\max}}.
\]
An elementary plaquette trace has Haar mean zero and satisfies
$DF_p=3F_p$, so $d_{\rm Haar,phys}\le3$. Therefore
\begin{equation}\label{ham:eq:entropy:comparison}
 0<B_{\rm cmp}\le\frac{3s}{2a}(C_0s^{3/2})^n.
\end{equation}
The elementary bounds $e<11/4$ and $\pi>3$ imply
$C_0<4/3$: after squaring it suffices to check
$1331/768<16/9$. Hence for $0<s\le1/4$,
\begin{equation}\label{ham:eq:entropy:six}
 \boxed{\displaystyle
 B_{\rm cmp}\le\frac3{8a}\,6^{-L^3/9}.}
\end{equation}
If $a\downarrow0$ and $aL\ge\ell_*>0$, this gives
\[
 B_{\rm cmp}\le\frac3{8a}
 \exp\!\left(-\frac{\ell_*^3\log6}{9a^3}\right)\longrightarrow0.
\]
This proves that even the exact vacuum extrema and exact Haar
Poincar\'e constant cannot rescue that specific global comparison.
It does not upper-bound the true physical gap, and it does not exclude
local, conditional, or adapted-reference estimates using additional
vacuum structure. In particular, extensive relative entropy is
compatible with a positive physical spectral gap.

\subsection{Certified evaluations and their interpretation}

The companion checker evaluates $n b(s)$ and the logarithms of the
right sides of \eqref{ham:eq:entropy:ratio} and
\eqref{ham:eq:entropy:comparison}. It uses exact fractions, a complete
alternating remainder for the Machin formula for $\pi$, and a complete
positive-series remainder for every logarithm. Its intervals enclose
the explicit bound functions. They are not measurements or two-sided
enclosures of the unknown vacuum entropy or true spectral gap.

The entropy variational principle and the elementary Laplace method
are established tools. The explicit SU(2) constants, their application
to the actual-vacuum input, and the strengthened obstruction are
derived here. Historical priority for this combination is not claimed.

\section{Vacuum fidelity under a hard electric cutoff}
\label{ham:sec:fidelity}

The sharp cutoff test bounded an error in a Wilson-loop expectation.
This section gives a different conclusion: an upper bound on the overlap
of the actual interacting vacuum with \emph{every} state in a specified
hard electric cutoff. Arbitrary spectator entanglement and mixed states
are included. The result uses only the actual-vacuum plaquette mean
estimate; it assumes neither independent vacuum plaquettes nor an
approximate vacuum wavefunction.

\subsection{Model, fidelity convention, and vacuum input}

On the periodic cubic spatial lattice, $L\ge2$, use normalized product
Haar measure $\mu$ and
\[
 H=\frac{s}{2a}D+\frac4{sa}\sum_p(1-F_p),\qquad
 F_p=\frac12\operatorname{Tr}U_p,\qquad 0<s=g^2\le1.
\]
The positive normalized actual vacuum is $h$, and $d\nu=h^2d\mu$.
The analytic input is
\begin{equation}\label{ham:eq:fidelity:vacuuminput}
 \nu(1-F_p)\le\frac34s.
\end{equation}
It follows from the product-link variational bound $E_0\le9L^3/a$,
positivity of the two Hamiltonian terms, and cubic symmetry of the
unique vacuum. Lemma~\ref{ham:lem:states:trial} supplies the complete proof.

Select $n\ge1$ edge-disjoint elementary plaquettes and choose one link
from each. Impose the standard Peter--Weyl cutoff $j\le J$ on these
chosen links, where $J\in\tfrac12\mathbb N_0$. Let $P$ be the product
of these selected-link projections, with the identity on all other links.
The complete one-link retained dimension is
\begin{equation}\label{ham:eq:fidelity:dimension}
 d_J=\sum_{r=1}^{2J+1}r^2
 =\frac{(2J+1)(2J+2)(4J+3)}6.
\end{equation}
Extra cutoffs on other links only restrict the state space further.

For a density operator $\sigma=P\sigma P$, define the squared fidelity
with the pure vacuum and the trace distance by
\[
 F(h,\sigma)=\langle h,\sigma h\rangle,\qquad
 D_{\rm tr}(h,\sigma)=\frac12\bigl\||h\rangle\langle h|-\sigma\bigr\|_1.
\]
The maximal squared fidelity is exactly
\begin{equation}\label{ham:eq:fidelity:maximum}
 F_{\max}(P)=\sup_{\sigma=P\sigma P}F(h,\sigma)=\|Ph\|^2.
\end{equation}
Indeed, $P|h\rangle\langle h|P$ has its only nonzero eigenvalue
$\|Ph\|^2$, achieved by the normalized vector $Ph$. The projection
commutes with vertex gauge transformations, so $Ph$ is physical and the
same supremum is obtained within the physical state space.

There is also an exact subspace-constrained distance identity:
\begin{equation}\label{ham:eq:fidelity:subspacedistance}
 \inf_{\sigma=P\sigma P}D_{\rm tr}(h,\sigma)
 =\sqrt{1-\|Ph\|^2}.
\end{equation}
For a proof, put $p=\|Ph\|^2$, $u=Ph/\sqrt p$, and
$v=(I-P)h/\sqrt{1-p}$ when $0<p<1$. Pinching onto
$\operatorname{span}\{u,v\}$ and its complement leaves the vacuum
state fixed and can only decrease trace distance. A supported $\sigma$
then has the form $q|u\rangle\langle u|\oplus\tau$, with
$\operatorname{Tr}\tau=1-q$. Its distance from the vacuum is
\[
 \frac12\left[1-q+
 \sqrt{(1-q)^2+4q(1-p)}\right]\ge\sqrt{1-p}.
\]
Writing $r=\sqrt{1-p}$, the last inequality is immediate if
$1-q\ge2r$; otherwise squaring leaves
$4(1-q)r(1-r)\ge0$. The normalized $Ph$ attains equality, and the
cases $p=0,1$ follow directly. This identity uses the support constraint;
it is not an inequality for arbitrary pairs of mixed states. Thus every
certified $F_{\max}(P)\le u\le1$ also gives
\begin{equation}\label{ham:eq:fidelity:geometricdistance}
 D_{\rm tr}(h,\sigma)\ge\sqrt{1-u}.
\end{equation}

\subsection{A concentration bound including arbitrary spectators}

Put $X_i=1-F_{p_i}$ and, for $0<\epsilon<1$, define
\[
 A_\epsilon=\left\{\sum_{i=1}^n X_i\le\frac{3ns}{4\epsilon}\right\}.
\]
Markov's inequality and \eqref{ham:eq:fidelity:vacuuminput} give
\begin{equation}\label{ham:eq:fidelity:vacuumprob}
 \nu(A_\epsilon)\ge1-\epsilon.
\end{equation}
Set
\begin{equation}\label{ham:eq:fidelity:constant}
 C=\frac{e^{3/2}}{2\sqrt\pi},\qquad
 B=C d_Js^{3/2},\qquad
 b_\epsilon=\min\{1,B^n\epsilon^{-3n/2}\}.
\end{equation}

\begin{theorem}[Cutoff concentration, fidelity, and trace distance]
\label{ham:thm:fidelity:main}
For every density operator $\sigma=P\sigma P$, whether physical or
not, and every $0<\epsilon<1$,
\begin{align}
 \operatorname{Tr}(\sigma\mathbf1_{A_\epsilon})
 &\le b_\epsilon,\label{ham:eq:fidelity:cutoffprob}\\
 F(h,\sigma)&\le
 \min\{1,(\sqrt{b_\epsilon}+\sqrt\epsilon)^2\},
 \label{ham:eq:fidelity:simple}\\
 D_{\rm tr}(h,\sigma)&\ge(1-\epsilon-b_\epsilon)_+.
 \label{ham:eq:fidelity:distance}
\end{align}
If $b_\epsilon<1-\epsilon$, the sharper fidelity bound is
\begin{equation}\label{ham:eq:fidelity:binary}
 F(h,\sigma)\le
 \left(\sqrt{(1-\epsilon)b_\epsilon}
       +\sqrt{\epsilon(1-b_\epsilon)}\right)^2.
\end{equation}
Otherwise the corresponding binary bound is $1$.
The inequalities include every representation in the selected cutoff,
all untruncated spectator representations, and arbitrary correlations
between the links.
\end{theorem}

\begin{proof}
Keep the nonselected links as spectator variables $\xi$. On the chosen
link of plaquette $i$, change coordinates from $U_i$ to its complete
plaquette holonomy
\[
 V_i=A_i(\xi)U_i^{\pm1}B_i(\xi).
\]
Edge-disjointness ensures that $A_i,B_i$ depend on none of the other
chosen links. Each change preserves Haar measure. Left and right
translations and inversion preserve every Casimir eigenspace, so the
fiberwise unitary coordinate change also preserves the selected-link
Peter--Weyl projection. In the new variables, $X_i=1-\operatorname{Tr}V_i/2$.
The spectators have not been fixed, approximated, or projected away.

For any $t>0$, multiplication-operator order gives
\[
 \mathbf1_{A_\epsilon}
 \le \exp\!\left(\frac{3nst}{4\epsilon}\right)
       \exp\!\left(-t\sum_iX_i\right).
\]
Compression by $P$ preserves this order. In the holonomy coordinates,
the compressed exponential is the tensor product of $n$ copies of
$P_Je^{-tX}P_J$, tensored with the spectator identity. Each one-link
compression is positive, and therefore its norm is at most its finite
trace. The projection kernel has constant diagonal $d_J$: for the
orthonormal matrix coefficients $\sqrt{2j+1}\,D^j_{mn}$, unitarity gives
\[
 \sum_{j\le J}(2j+1)\sum_{m,n}|D^j_{mn}(U)|^2
 =\sum_{j\le J}(2j+1)^2=d_J.
\]
Consequently
\[
 \|P_Je^{-tX}P_J\|
 \le d_J\int e^{-tX}\,dU.
\]
For a Haar SU(2) holonomy, $X\in[0,2]$ has density
$(2/\pi)\sqrt{x(2-x)}$. Hence the complete integral, without a cutoff,
satisfies
\[
 \int e^{-tX}\,dU
 \le\frac{2\sqrt2}{\pi}\int_0^\infty e^{-tx}x^{1/2}\,dx
 =\sqrt{\frac2\pi}\,t^{-3/2}.
\]
Taking $t=2\epsilon/s$ yields
\[
 \|P\mathbf1_{A_\epsilon}P\|
 \le\left[C d_J(s/\epsilon)^{3/2}\right]^n.
\]
The trivial probability upper bound $1$ proves
\eqref{ham:eq:fidelity:cutoffprob}, for all density operators supported in
$P$, even when their spectators are entangled.

For a unit vector $f\in\operatorname{Ran}P$, split its inner product
with $h$ over $A_\epsilon$ and its complement. Cauchy--Schwarz gives
\[
 |\langle f,h\rangle|
 \le\sqrt{\nu(A_\epsilon)\|\mathbf1_{A_\epsilon}f\|^2}
  +\sqrt{\nu(A_\epsilon^c)\|\mathbf1_{A_\epsilon^c}f\|^2}
 \le\sqrt{b_\epsilon}+\sqrt\epsilon.
\]
Convex decomposition into pure states proves \eqref{ham:eq:fidelity:simple}
for mixed $\sigma$. If $b_\epsilon<1-\epsilon$, the preceding binary
expression is maximized, subject to the known probabilities, at
$\nu(A_\epsilon)=1-\epsilon$ and
$\|\mathbf1_{A_\epsilon}f\|^2=b_\epsilon$. This proves
\eqref{ham:eq:fidelity:binary} and again extends by convexity. Finally,
the two-outcome measurement $\{A_\epsilon,A_\epsilon^c\}$ gives
\[
 D_{\rm tr}(h,\sigma)
 \ge\nu(A_\epsilon)-\operatorname{Tr}(\sigma\mathbf1_{A_\epsilon}),
\]
which proves \eqref{ham:eq:fidelity:distance}.
\end{proof}

\subsection{Closed formulas and exact rational evaluation}

For $\eta=3n/2$, define positive numbers
\begin{equation}\label{ham:eq:fidelity:optima}
 \epsilon_F^{3n+2}=\eta^4B^{2n},\qquad
 \epsilon_D^{3n+2}=\eta^2B^{2n}.
\end{equation}
Optimizing the elementary bounds before their trivial caps gives
\begin{align}
 F_{\max}(P)&\le
 \min\{1,(1+1/\eta)^2\epsilon_F\},
 \label{ham:eq:fidelity:optimizedF}\\
 D_{\rm tr}(h,\sigma)&\ge
 \left[1-(1+1/\eta)\epsilon_D\right]_+.
 \label{ham:eq:fidelity:optimizedD}
\end{align}
To see this, differentiate
$\sqrt\epsilon+B^{n/2}\epsilon^{-3n/4}$ for fidelity, and
$\epsilon+B^n\epsilon^{-3n/2}$ for trace distance. At their respective
optima the second term is $\sqrt\epsilon/\eta$ and $\epsilon/\eta$.
Whenever a displayed result is nontrivial, its optimizing $\epsilon$
lies in $(0,1)$. In the remaining cases the caps give valid trivial
bounds. These formulas optimize the elementary estimates, not the
sharper optional binary fidelity estimate.

For entirely rational numerical certification, use
\begin{equation}\label{ham:eq:fidelity:rationalconstant}
 C<\frac43.
\end{equation}
Indeed, $e<11/4$ and $\pi>3$ imply
$9e^3<9(11/4)^3<192<64\pi$, which proves
$C^2=e^3/(4\pi)<16/9$.
For explicit remainder control,
$e\le1+1+1/2+1/6+(1/24)/(1-1/5)=87/32<11/4$;
the geometric term bounds the whole exponential-series tail.
The eight-term lower alternating sum for $4\arctan1$ exceeds $3$,
proving the other constant inequality by exact rational arithmetic.
Writing
\[
 Q=\frac{16}{9}d_J^2s^3,
\]
the certified optimizing roots are those of the rational numbers
$\eta^4Q^n$ and $\eta^2Q^n$, of degree $3n+2$.
The companion checker brackets these roots by exact integer inequalities.
It uses no floating-point eigenvalue solver, vacuum surrogate, or
special-function evaluation. Decimal endpoints are rounded outward.
It additionally encloses the square root in
\eqref{ham:eq:fidelity:geometricdistance} and takes the larger of that
distance lower bound and \eqref{ham:eq:fidelity:optimizedD}.
All representation effects outside the selected cutoff remain included
through the analytic trace and spectator-identity argument above.

\subsection{A uniform obstruction and a necessary cutoff scale}

For the single chosen-link case $n=1$, choose $\epsilon=B^{2/5}$ when
$B<1$. Then $b_\epsilon=B^{2/5}$, giving
\begin{equation}\label{ham:eq:fidelity:single}
 F_{\max}(P)\le\min\{1,4B^{2/5}\},\qquad
 D_{\rm tr}(h,\sigma)\ge(1-2B^{2/5})_+.
\end{equation}
Every larger selected cutoff is a subprojection of a one-link cutoff,
so these bounds already suffice uniformly in the spectator space,
number of links, and spatial lattice size. Since
\[
 d_J\le\frac{(2J+2)^3}{3},
\]
one has $B\to0$ whenever $g\to0$ and $gJ\to0$. Therefore
\begin{equation}\label{ham:eq:fidelity:orthogonality}
 \boxed{\displaystyle
  g\to0,\quad gJ\to0
  \quad\Longrightarrow\quad
  F_{\max}(P)\to0,\qquad
  \inf_{\sigma=P\sigma P}D_{\rm tr}(h,\sigma)\to1.}
\end{equation}
This is a statement about approximation of the finite-lattice vacuum
along a weak-coupling sequence. It is not a construction of continuum
states. The condition $g\to0$ is essential: $gJ\to0$ alone would also
hold for $J=0$ and fixed $g$, which is not the asserted limit.

More generally define
\[
 c_n=(1+2/(3n))^2(3n/2)^{4/(3n+2)}.
\]
Equation~\eqref{ham:eq:fidelity:optimizedF} reads
$F_{\max}\le\min\{1,c_nB^{2n/(3n+2)}\}$.
If a cutoff state has squared fidelity at least $f>0$, then necessarily
\begin{equation}\label{ham:eq:fidelity:necessarydimension}
 d_J\ge C^{-1}g^{-3}
 \left(\frac f{c_n}\right)^{(3n+2)/(2n)}.
\end{equation}
Replacing $C^{-1}$ by $3/4$ gives a weaker rationally certified version.
Along any sequence with $n\to\infty$, $g\to0$, and
$\liminf F(h,\sigma)\ge f>0$, it follows that
\begin{equation}\label{ham:eq:fidelity:necessaryJ}
 \boxed{\displaystyle
 \liminf gJ\ge
 \left(\frac{3}{8C}\right)^{1/3}\sqrt f
 =\left(\frac{3\sqrt\pi}{4e^{3/2}}\right)^{1/3}\sqrt f
 \ge\left(\frac9{32}\right)^{1/3}\sqrt f.}
\end{equation}
For the proof, $c_n\to1$ and $2n/(3n+2)\to2/3$, so the fidelity
assumption forces $\liminf C d_Jg^3\ge f^{3/2}$. It also forces
$J\to\infty$, after which $d_J\sim(8/3)J^3$ gives the stated constant.
Use any $f'<f$ first if the lower fidelity is only a limit inferior.

On a periodic cube an explicit family has
$n=L\lfloor L/2\rfloor^2$: take the $xy$ plaquettes based at
$(2i,2j,z)$ for $0\le i,j<\lfloor L/2\rfloor$ and all $z$.
They are edge disjoint, including at $L=2$, and $n\ge L^3/9$.
An all-link hard cutoff therefore satisfies every bound above for
this family.

At fixed $s,J$, increasing $n$ alone does not make the proved fidelity
bound tend to zero. Its asymptotic upper function is
$\min\{1,B^{2/3}\}$. The exponential cutoff-probability bound for
$A_\epsilon$ must not be relabeled as exponential fidelity decay at
fixed coupling. The vacuum input used here controls a mean, not a
vacuum large-deviation rate.

\paragraph{Scope and attribution.}
The argument uses standard Peter--Weyl projection kernels, exponential
Markov bounds, and fidelity estimates for a binary measurement. The
quantitative specialization combines those tools with the actual
SU(2) vacuum input. Historical priority is not established. These
necessary conditions apply to the stated standard hard electric
cutoff and the original vacuum. They do not bound coupling-adapted
mixed bases, arbitrary effective encodings, or universal qubit counts.
The fixed-fidelity condition is a different accuracy requirement from
the earlier $o(g^2)$ Wilson-observable error criterion; the numerical
constants from those two criteria should not be ranked as improvements
of the same statement. No continuum mass gap is inferred.

\section{Certified numerical results}
\label{ham:sec:data}

Every displayed number in this section is an outward-rounded evaluation
of a proved bound function. Actual overlaps, entropies, or physical
excitation energies have not been numerically measured.

\subsection{Maximum squared fidelity and minimum trace distance}

The cutoff is imposed on one link of each of
$n=L\lfloor L/2\rfloor^2$ edge-disjoint plaquettes. An all-link
cutoff is a smaller state space, so the same bounds apply to it.
The numbers use the rational upper constant $4/3>C_0$, not an
uncertified floating-point value of $C_0$.

\begin{center}\small
\begin{tabular}{rrrrrr}
\toprule
$L$ & $g^2$ & $J$ & $d_J$ & $F_{\max}\le$ & $D_{\rm tr}\ge$\\
\midrule
2 & $\tfrac{1}{100}$ & $1$ & 14 & 0.420699006799 & 0.761118251786 \\
4 & $\tfrac{1}{100}$ & $1$ & 14 & 0.109485641547 & 0.943670683265 \\
8 & $\tfrac{1}{100}$ & $1$ & 14 & 0.076124297340 & 0.961184531014 \\
8 & $\tfrac{1}{100}$ & $4$ & 285 & 0.561669147888 & 0.662065595022 \\
8 & $\tfrac{1}{100}$ & $10$ & 3311 & 1.000000000000 & 0.000000000000 \\
2 & $\tfrac{1}{10000}$ & $4$ & 285 & 0.060024686280 & 0.969523240423 \\
4 & $\tfrac{1}{10000}$ & $4$ & 285 & 0.009055990017 & 0.995461706939 \\
8 & $\tfrac{1}{10000}$ & $4$ & 285 & 0.005752322989 & 0.997119690414 \\
8 & $\tfrac{1}{10000}$ & $10$ & 3311 & 0.029257709656 & 0.985262548940 \\
8 & $\tfrac{1}{1000000}$ & $4$ & 285 & 0.000058912298 & 0.999970543417 \\
8 & $\tfrac{1}{1000000}$ & $10$ & 3311 & 0.000299642230 & 0.999850167660 \\
8 & $\tfrac{1}{1000000}$ & $20$ & 23821 & 0.001109115088 & 0.999445288603 \\
\bottomrule
\end{tabular}
\end{center}

The $F_{\max}=1$ row is a trivial upper bound. It does not say that
the vacuum is exactly representable at that cutoff. Likewise the
distance lower bounds are guaranteed floors, not actual distances.

\subsection{Necessary standard electric cutoff sizes}

For $L=8$, the following is the smallest nonnegative half-integer
$J$ not excluded by the stated rational fidelity bound at the target
$f$. All smaller cutoffs are rigorously excluded. The listed cutoff
itself is not proved sufficient. These are full one-link Peter--Weyl
dimensions, not lower bounds for arbitrary encodings or quantum-memory
requirements.

\begin{center}
\begin{tabular}{rrrr}
\toprule
$g^2$ & Target squared fidelity $f$ & Necessary $J\ge$ & $d_J$ there\\
\midrule
$\tfrac{1}{100}$ & $\tfrac{9}{10}$ & $\tfrac{11}{2}$ & 650 \\
$\tfrac{1}{100}$ & $\tfrac{99}{100}$ & $6$ & 819 \\
$\tfrac{1}{10000}$ & $\tfrac{9}{10}$ & $\tfrac{119}{2}$ & 583220 \\
$\tfrac{1}{10000}$ & $\tfrac{99}{100}$ & $\tfrac{125}{2}$ & 674751 \\
$\tfrac{1}{1000000}$ & $\tfrac{9}{10}$ & $601$ & 581054614 \\
$\tfrac{1}{1000000}$ & $\tfrac{99}{100}$ & $\tfrac{1261}{2}$ & 670768775 \\
\bottomrule
\end{tabular}
\end{center}

\subsection{Higher-multiplicity collective constants}

The interval endpoints below enclose the scalar certificate constants.
The nine- and thirteen-face examples use all four central links. The
closed cube and periodic sheet are negative controls for these methods;
zero is not an assertion that their actual physical gap is zero.

\begin{center}\small
\begin{tabular}{lrrl}
\toprule
Family & Faces & Max.\ multiplicity & Certified $\lambda$ interval\\
\midrule
Planar cross & 5 & 2 & $[1.999999999999, 2.000000000000]$ \\
One triple link & 6 & 3 & $[1.697224362268, 1.697224362269]$ \\
One quadruple link & 7 & 4 & $[1.697224362268, 1.697224362269]$ \\
Four triple links & 9 & 3 & $[0.745179888661, 0.745179888662]$ \\
Four quadruple links & 13 & 4 & $[0.610446896702, 0.610446896703]$ \\
Closed cube & 6 & 2 & $[0.000000000000, 0.000000000000]$ \\
Periodic sheet & 16 & 2 & $[0.000000000000, 0.000000000000]$ \\
\bottomrule
\end{tabular}
\end{center}

The full CSV carries twenty decimal places, the exact matrix witnesses,
and the resulting Gram, tail-fraction, and selected Rayleigh constants.
For the main two examples, the certified dimensionless Rayleigh floors
are $0.00001617143855601733$ and $0.00000917137765479184$.

\subsection{Vacuum concentration and the comparison-method ceiling}

For $a=1/L$, the table gives a lower bound on
$D_{\rm KL}(\nu\Vert\mu)$, in nats, and an upper bound on the logarithm
of the explicit comparison ceiling
$U=(3s/(2a))(C_0s^{3/2})^n$. Since $B_{\rm cmp}\le U$, the final
column bounds $\log_{10}B_{\rm cmp}$ from above. It does not bound
the actual gap from above. All rows have $b(s)>0$.

\begin{center}\small
\begin{tabular}{rrrrr}
\toprule
$L$ & $g^2$ & $n$ & $D_{\rm KL}\ge$ & $\log_{10}U\le$\\
\midrule
2 & $\tfrac{1}{4}$ & 2 & 3.689907330328 & -1.727445128904 \\
4 & $\tfrac{1}{4}$ & 16 & 29.519258642631 & -12.643959879314 \\
8 & $\tfrac{1}{4}$ & 128 & 236.154069141053 & -102.083287852238 \\
16 & $\tfrac{1}{4}$ & 1024 & 1889.232553128428 & -819.705121605285 \\
4 & $\tfrac{1}{16}$ & 16 & 62.790323309509 & -27.695459662513 \\
8 & $\tfrac{1}{16}$ & 128 & 502.322586476072 & -218.280866178535 \\
4 & $\tfrac{1}{100}$ & 16 & 106.772278439468 & -47.592460096115 \\
8 & $\tfrac{1}{100}$ & 128 & 854.178227515748 & -371.885709525942 \\
4 & $\tfrac{1}{10000}$ & 16 & 217.296362903182 & -97.592460096115 \\
8 & $\tfrac{1}{10000}$ & 128 & 1738.370903225461 & -757.885709525942 \\
16 & $\tfrac{1}{10000}$ & 1024 & 13906.967225803693 & -6042.338914934206 \\
\bottomrule
\end{tabular}
\end{center}

The inequalities distinguish three objects: the true physical gap,
the valid but possibly weak certificate $B_{\rm cmp}$, and the proved
upper ceiling $U$ on that certificate. Only the latter two are bounded
by the density-ratio argument.

\section{The finite-torus sectors invisible to contractible loop states}
\label{ham:sec:center:scope}

There is an exact structural reason why proving ever larger
contractible-plaquette certificates does not, by itself, control the
entire physical finite-torus Hilbert space. This section states that
reason as a theorem, supplies explicitly nonzero missing vectors,
and distinguishes it from the infinite-volume local-vacuum question.

\subsection{Three commuting center transformations}

For each spatial direction $\alpha=1,2,3$, define $z_\alpha$ on link
configurations by multiplying by $-I$ every positively oriented
$\alpha$-link whose starting coordinate obeys $x_\alpha=L-1$.
All other links remain unchanged. On $L^2(\mu)$ let
\[
 (Z_\alpha\psi)(U)=\psi(z_\alpha U).
\]
The $Z_\alpha$ are commuting self-adjoint unitaries with
$Z_\alpha^2=I$. They commute with every gauge transformation, because
the inserted sign is central, and therefore preserve the physical
Hilbert space. Each elementary plaquette contains either zero or two
links crossing the relevant cut. Its holonomy trace is unchanged.
Haar measure, the link Casimirs, the magnetic potential, and the
Hamiltonian are consequently invariant:
\begin{equation}\label{ham:eq:center:commute}
 [Z_\alpha,H]=[Z_\alpha,K]=0.
\end{equation}
Uniqueness and positivity of the full vacuum imply
$Z_\alpha h=h$: applying $Z_\alpha$ gives another normalized positive
ground state. This uses the finite lattice only and does not assume
spontaneous symmetry breaking or its absence in an infinite system.

For $\chi\in\{0,1\}^3$, define
\begin{equation}\label{ham:eq:center:projection}
 P_\chi=\frac18\sum_{b\in\{0,1\}^3}
 (-1)^{\chi\cdot b}Z_1^{b_1}Z_2^{b_2}Z_3^{b_3}.
\end{equation}
Finite character orthogonality proves that these are pairwise
orthogonal projectors summing to the identity. On the physical
Hilbert space,
\begin{equation}\label{ham:eq:center:decomposition}
 \mathcal H_{\rm phys}
 =\bigoplus_{\chi\in\{0,1\}^3}\mathcal H_\chi,
 \qquad \mathcal H_\chi=P_\chi\mathcal H_{\rm phys},
 \qquad h\in\mathcal H_0.
\end{equation}
All eight subspaces reduce $K$.

For an oriented closed lattice loop $C$, a cut transformation changes
its fundamental trace by
\begin{equation}\label{ham:eq:center:loopcharge}
 Z_\alpha F_C=(-1)^{w_\alpha(C)}F_C,
\end{equation}
where $w_\alpha(C)$ is its winding number modulo two in direction
$\alpha$. Reversed crossings have the same sign modulo two. In
particular, every contractible loop is neutral.

\begin{theorem}[Contractible observables miss seven physical sectors]
\label{ham:thm:center:blindness}
Let $\mathcal A_{\rm ctr}$ be the algebra generated by multiplication
by contractible Wilson-loop traces, bounded functions of those
traces, and the identity. Then
\begin{equation}\label{ham:eq:center:neutralclosure}
 \overline{\mathcal A_{\rm ctr}h}\subset\mathcal H_0.
\end{equation}
The same inclusion holds if one also applies bounded spectral
functions of $K$ or forms finite products of such functions with the
specified observables. Every nontrivial $\mathcal H_\chi$ contains an
explicit nonzero smooth gauge-invariant vector orthogonal to this
entire closure. Consequently, arbitrary polynomial degree, spatial
support growth, operator mixing, and physical-energy filtering within
this neutral family cannot make it dense in the full physical
vacuum complement on a finite torus.
\end{theorem}

\begin{proof}
Every generating observable commutes with all $Z_\alpha$, and the
same is true of bounded functions of $K$ by
\eqref{ham:eq:center:commute}. Since $h$ is neutral, every generated
vector lies in the closed subspace $\mathcal H_0$.

At the origin choose the three straight fundamental winding loops
$C_1,C_2,C_3$, each of length $L$, and write their based holonomies
as $P_1,P_2,P_3$. For each nonzero $\chi$, concatenate the loops with
$\chi_\alpha=1$ in the fixed order $1,2,3$, and set
\begin{equation}\label{ham:eq:center:missingvectors}
 F_\chi=\tfrac12\operatorname{Tr}
 (P_1^{\chi_1}P_2^{\chi_2}P_3^{\chi_3}),\qquad
 \psi_\chi=F_\chi h.
\end{equation}
All factors are based at the origin, so the product trace is gauge
invariant. The concatenated loop contains $L|\chi|$ distinct link
variables, each exactly once, even though its base vertex is revisited.
Equation~\eqref{ham:eq:center:loopcharge} gives
$Z_\alpha\psi_\chi=(-1)^{\chi_\alpha}\psi_\chi$.
At the identity configuration $F_\chi=1$; continuity and $h>0$
imply $\psi_\chi\ne0$. Distinct characters give mutually orthogonal
vectors, and all seven are orthogonal to $\mathcal H_0$.
\end{proof}

\subsection{A quantitative norm for the missing winding states}

The nonzero conclusion has a volume-independent quantitative version.
It uses only the actual one-link electric energy, not an assumed law
for the global holonomies.

\begin{lemma}[One-link second-moment lower bound]
\label{ham:lem:center:norm}
Let $F=\tfrac12\operatorname{Tr}U_C$ for a loop containing a link
$e$ exactly once, let $h$ be real, smooth, normalized, and set
$M_e=\int\Gamma_e h\,d\mu$. Then
\begin{equation}\label{ham:eq:center:normgeneric}
 \nu F^2\ge
 \frac1{4(\sqrt{M_e+1}+\sqrt{M_e})^2}
 \ge\frac1{16M_e+8}.
\end{equation}
If $F$ is center charged and $h$ is center invariant, then
$\nu F=0$, so this is also the centered-state squared norm.
\end{lemma}

\begin{proof}
The one-link identities \eqref{ham:eq:overlap:perlink} hold for any loop
containing that chosen link exactly once. Integrating the divergence
of $Fh^2\nabla_eF$ against Haar measure gives, with $v=\nu F^2$,
\[
 1-4v=-8\int Fh\sum_b(X_{e,b}F)(X_{e,b}h)\,d\mu.
\]
Cauchy--Schwarz and $\Gamma_eF=(1-F^2)/4$ imply
\[
 |1-4v|
 \le4\sqrt{M_e\,\nu[F^2(1-F^2)]}
 \le4\sqrt{M_ev}.
\]
Thus $1\le4v+4\sqrt{M_ev}$. Solving the quadratic inequality in
$\sqrt v$ gives the first bound. The second follows from
$\sqrt{M_e(M_e+1)}\le M_e+1/2$.
For a charged loop, invariance of $\nu$ under a transformation that
changes $F$ to $-F$ forces $\nu F=0$.
\end{proof}

For the actual Hamiltonian vacuum with $0<s\le1$, the established
variational estimate is $M_e\le6/s$ for every link. It follows that
the seven vectors in \eqref{ham:eq:center:missingvectors} obey
\begin{equation}\label{ham:eq:center:gram}
 \boxed{\displaystyle
 \frac{s}{96+8s}I_7
 \le\big(\langle\psi_\chi,\psi_{\chi'}\rangle\big)_{
       \chi,\chi'\ne0}
 \le I_7.}
\end{equation}
The matrix is exactly diagonal by the sector characters. In
particular, the omitted directions cannot be dismissed as identically
null finite-volume states. This bound alone does not supply nonzero
continuum states: it may vanish as $s\to0$, and normalization does
not supply tightness of physical spectral measures.

For reference, the chosen winding state's exact Dirichlet energy is
\[
 \langle\psi_\chi,K\psi_\chi\rangle
 =\frac{sL|\chi|}{8a}(1-\nu F_\chi^2).
\]
This identity and the norm lower bound give only the variational
upper estimate
$\inf\operatorname{spec}(K|_{\mathcal H_\chi})
\le L|\chi|(12+s)/a$. They do not provide a positive lower bound
on any charged sector.

\subsection{The exact target for a global finite-torus gap}

The finite-lattice operator has compact resolvent, and its vacuum is
unique. Define
\[
 m_0=\inf\operatorname{spec}
       (K|_{\mathcal H_0\cap h^\perp}),\qquad
 m_\chi=\inf\operatorname{spec}(K|_{\mathcal H_\chi})
       \quad(\chi\ne0).
\]
Each is positive for fixed finite parameters, and the full physical
gap is exactly
\begin{equation}\label{ham:eq:center:globalmin}
 \boxed{\displaystyle
 m_{\rm phys}=\min\left(m_0,\min_{\chi\ne0}m_\chi\right).}
\end{equation}
Consequently a global inequality on the whole physical vacuum
complement is equivalent to the corresponding inequality in all
eight blocks, with the vacuum removed from the neutral block.
The contractible-loop estimates supply no charged-block lower
bound. Inclusion in the neutral block also does not, by itself,
prove completeness or a lower gap throughout that block.

There is an exact information-theoretic illustration of the
distinction. At fixed parameters and any $\epsilon>0$, define
\[
 \widetilde K_\epsilon
 =K+\sum_{\chi\ne0}(\epsilon-m_\chi)P_\chi.
\]
This is a self-adjoint nonnegative operator with the same unique
vacuum and the same neutral restriction as $K$. Its full gap is
$\min(m_0,\epsilon)$. All vacuum correlations made from neutral
bounded observables and physical-time evolution are exactly the
same as for $K$, since they never leave $\mathcal H_0$.
Thus even exact knowledge of those correlations alone cannot
determine the gap in sectors invisible to them. This modified
operator contains nonlocal sector projectors and is not asserted
to be the Wilson Hamiltonian or a new local gauge theory. It is a
logical counterexample to an inference from the restricted data,
not a counterexample to a Yang--Mills gap.

\subsection{Passage from finite-torus sectors to infinite volume}

The preceding statements concern the full gauge-invariant Hilbert
space on a periodic finite spatial lattice. A winding loop extends
around the entire torus. When the physical torus grows without
bound, these vectors need not survive as states in the vacuum
representation generated by bounded-region local observables.
The local infinite-volume vacuum theory may therefore ask for a gap
in a different, smaller representation. The finite-torus sector
obstruction does not rule out that construction or its mass gap.

It does identify a precise missing condition for the stronger route
posed here: a certificate for the full finite-torus physical
complement must control the charged blocks as well as the neutral
one. If the intended target is instead the local infinite-volume
vacuum representation, the construction must specify that
representation, establish its nontrivial local states and time
correlations, and prove the required inequality on a dense form
core there. Neither an overlap certificate nor a sector
decomposition supplies these continuum steps.

\paragraph{Attribution boundary.}
Electric center sectors and winding-loop charges are established
lattice-gauge structures. No historical novelty is claimed for the
decomposition. The application here makes the missing blocks in
the proposed certificate explicit and adds the quantitative
one-link norm estimate \eqref{ham:eq:center:normgeneric}. The companion
checker certifies the finite character and incidence arithmetic;
the operator and integration-by-parts assertions are proved above.

\section{Finite-matrix certification of the full physical spectrum}
\label{ham:sec:global:schur}

The following construction controls every omitted electric representation.
Unlike a Gram inequality on a selected operator family, it is an eigenvalue
enclosure for the entire gauge-invariant Hilbert space of a fixed spatial
lattice. Its ingredients are standard operator comparison, min--max, and
Schur-complement methods; the specialization below makes their required
inputs and representation tails explicit for the Wilson Hamiltonian.
Related rigorous Feshbach--Schur discretization methods are described in
\cite{ham:global:dusson2020,ham:global:dusson2021}. No historical priority for the
general spectral method is claimed.

\subsection{The physical space and a complete electric cutoff}

Let the spatial lattice be the periodic cubic lattice with $L\ge2$,
$N_e=N_p=3L^3$, and normalized product Haar measure. On its full
gauge-invariant Hilbert space set
\begin{equation}\label{ham:eq:global:model}
 H=\kappa D+V,\qquad
 \kappa=\frac{s}{2a},\quad
 D=\sum_e\mathcal C_e,\quad
 V=\frac4{sa}\sum_p(1-F_p),\quad
 a,s>0.
\end{equation}
Here $\mathcal C_e$ has spin-$j$ eigenvalue $j(j+1)$, and
$F_p=\operatorname{Tr}U_p/2$. The bounded multiplication operator obeys
\begin{equation}\label{ham:eq:global:potential}
 0\le V\le\frac{8N_p}{sa}I.
\end{equation}
The operator has compact resolvent. Its full product-space ground state
is unique and strictly positive by positivity improvement of the heat
semigroup; invariance makes that ground state gauge invariant. Thus the
physical ground state $h$ is unique as well. Write the physical eigenvalues,
counting multiplicities, as $E_0<E_1\le E_2\le\cdots$.

For $m\ge0$ an integer, put
\[
 R_m=\frac m4,\qquad P_m=\mathbf1_{[0,R_m]}(D),\qquad Q_m=I-P_m.
\]
These projections act on the \emph{entire} physical space. In particular,
they include every allowed intertwiner and every electric center sector,
not just states generated by contractible plaquettes. Spin networks give
an orthonormal electric eigenbasis. Since
$j(j+1)=k(k+2)/4$ for $k=2j\in\mathbb N_0$, all electric eigenvalues are
integer multiples of $1/4$. Consequently
\begin{equation}\label{ham:eq:global:tailfloor}
 Q_m DQ_m\ge\frac{m+1}{4}Q_m,
 \qquad \tau_m=\frac{s(m+1)}{8a}
\end{equation}
is a certified omitted-sector energy floor. A better exactly enumerated
electric floor may replace it. Each $P_m$ is finite rank and $P_m\uparrow I$.

\subsection{Wilson multiplication has a finite representation halo}

Multiplication by a fundamental plaquette trace changes the spin on each
of its four distinct links by $+1/2$ or $-1/2$, subject to the usual
admissibility conditions. This follows from the exact tensor-product rule
$j\otimes\tfrac12=(j+\tfrac12)\oplus(j-\tfrac12)$, with the negative
spin omitted. Thus $Q_mVP_m$ has finite representation support.

There is an explicit energy bound for that support. Raising all four
spins gives the largest possible increase, bounded by
\[
 \sum_{e\in p}(j_e+3/4)
 \le 2\sqrt{R_m+1}+1.
\]
Indeed, $\sum_{e\in p}(j_e+1/2)^2\le R_m+1$, and Cauchy--Schwarz
controls their sum. Therefore
\begin{equation}\label{ham:eq:global:halo}
 VP_m\mathcal H_{\rm phys}
 \subset\mathbf1_{[0,R_m+2\sqrt{R_m+1}+1]}(D)\mathcal H_{\rm phys}.
\end{equation}
Choose an integer $\widehat m$ satisfying
$\widehat m/4\ge R_m+2\sqrt{R_m+1}+1$, and put
\[
 P_h=P_{\widehat m}-P_m,\qquad P_o=I-P_{\widehat m}.
\]
All three spaces are physical and electric-invariant. Define finite
matrices in their complete orthonormal electric bases by
\begin{equation}\label{ham:eq:global:matrices}
 A=P_mHP_m,\qquad B=P_mVP_h,\qquad D_h=P_hDP_h,
 \qquad
 \mathcal M_m=
 \begin{pmatrix}A&B\\B^*&\kappa D_h\end{pmatrix}.
\end{equation}
Every displayed entry is a finite SU(2) recoupling calculation. For rational
$a,s$, standard normalized Clebsch--Gordan coefficients give algebraic
entries that can be treated exactly or by certified intervals. No
approximation of the vacuum is required to define these matrices.

\begin{theorem}[Complete tail control and a full-complement gap certificate]
\label{ham:thm:global:finite}
Let $\mu_j$ be the ordered eigenvalues of $\mathcal M_m$, counting
multiplicities, and put
$\tau_o=s(\widehat m+1)/(8a)$. Then, for every available index $j$,
\begin{equation}\label{ham:eq:global:lower}
 E_j\ge\min\{\mu_j,\tau_o\}.
\end{equation}
For $j<\operatorname{rank}P_m$, if $\alpha_j$ is the $j$th eigenvalue
of $A$, then also $E_j\le\alpha_j$.

In particular, let $u_0$ be any rigorous upper bound on $E_0$, obtained
from a normalized physical trial state or from the lowest Ritz eigenvalue.
If a certified lower bound $\underline\mu_1$ satisfies
\begin{equation}\label{ham:eq:global:positivecriterion}
 b:=\min\{\underline\mu_1,\tau_o\}>u_0,
\end{equation}
then every physical vector in the form domain with $\psi\perp h$ obeys
\begin{equation}\label{ham:eq:global:fullgap}
 \boxed{\displaystyle
 \langle\psi,(H-E_0)\psi\rangle
 \ge(b-u_0)\|\psi\|^2.}
\end{equation}
It follows that
$\langle\psi,e^{-t(H-E_0)}\psi\rangle
\le e^{-t(b-u_0)}\|\psi\|^2$ for every $t\ge0$.
The conclusion is about the full physical vacuum complement.
\end{theorem}

\begin{proof}
Consider the self-adjoint operator
\[
 H^-_m=H-Q_mVQ_m.
\]
Because $V\ge0$, one has $H^-_m\le H$ in quadratic-form order.
The kinetic term is block diagonal under electric projections. The halo
identity implies $P_mVP_o=0$. All magnetic terms entirely inside $Q_m$
have been removed. Hence the exact decomposition is
\[
 H^-_m=\mathcal M_m\oplus\kappa D|_{P_o\mathcal H_{\rm phys}}.
\]
The second summand is bounded below by $\tau_o$. Its spectrum may have
arbitrary multiplicities, but no eigenvalue lies below that floor.
Ordering the spectrum of this direct sum gives
$E_j(H^-_m)\ge\min\{\mu_j,\tau_o\}$. Min--max and $H\ge H^-_m$
prove the lower enclosure. Restriction to the finite trial space $P_m$
gives the upper Ritz enclosure.

Uniqueness identifies the orthogonal complement of the actual vacuum
with the physical spectral subspace starting at $E_1$. Thus
$H-E_0\ge E_1-E_0\ge b-u_0$ on that complement. Functional calculus
gives the semigroup statement. The proof does not replace $h$ by a
truncated ground state and does not discard an unbounded representation
remainder.
\end{proof}

A smaller halo may be used by retaining the electric-invariant span of
$Q_mVP_m$ alone. In that case all other electric modes must still be
counted through a certified floor, for which $\tau_m$ is always valid.
The smaller matrix does not justify forgetting low electric modes
orthogonal to the selected halo.

\subsection{An equivalent Schur test and finite-arithmetic error bounds}

For any real $z<\tau_m$, every omitted electric denominator is positive.
The finite Schur matrix is
\begin{equation}\label{ham:eq:global:schur}
 \mathcal S_m(z)=A-zI-B(\kappa D_h-zI)^{-1}B^*.
\end{equation}
All terms here are finite. Block completion of squares, together with
positivity of the omitted electric block, shows that if
$\mathcal S_m(z)$ has at most one strictly negative eigenvalue, then
$H$ has at most one eigenvalue below $z$. If also $u_0<z$, it follows
that $E_1\ge z$ and the full gap is at least $z-u_0$.
For rational matrices, exact $LDL^*$ inertia is one way to check this;
algebraic or interval enclosures give the same rigorous test. Zero
pivots must be treated by an exact pivoting or Sturm rule, not silently
rounded away.

Suppose a Hermitian numerical matrix $\widehat{\mathcal M}$ has a
rigorously bounded residual $\|\mathcal M_m-\widehat{\mathcal M}\|
\le\epsilon_M$. Then
$\mu_1\ge\lambda_1(\widehat{\mathcal M})-\epsilon_M$.
Entrywise interval bounds imply such a residual, for example through
the largest absolute row sum of a symmetric entrywise error bound.
Any error in the finite eigenvalue calculation must be included as
well. The analytic representation remainder is already controlled by
the theorem; it is separate from this finite-arithmetic error.

\subsection{A simpler norm certificate and finite-volume completeness}

Let $\alpha_0\le\alpha_1$ be the first two Ritz eigenvalues and choose
$\beta_m\ge\|P_mVQ_m\|$. Since the midpoint of the potential range
is a scalar, one always has the explicit bound
\begin{equation}\label{ham:eq:global:couplingnorm}
 \|P_mVQ_m\|\le\frac{4N_p}{sa}.
\end{equation}
The right side can be chosen for $\beta_m$; the sharper finite matrix
norm can instead be enclosed from $B$.
For every $z<\tau_m$,
\[
 \mathcal S_m(z)\ge
 A-zI-\frac{\beta_m^2}{\tau_m-z}I.
\]
Solving the resulting scalar quadratic yields
\begin{equation}\label{ham:eq:global:scalar}
 E_1\ge\frac{\alpha_1+\tau_m-
 \sqrt{(\tau_m-\alpha_1)^2+4\beta_m^2}}2.
\end{equation}
The lower bound can be replaced by zero if negative. A positive value
of the right side minus $\alpha_0$ is therefore another complete
physical gap certificate. If $\tau_m>\alpha_1$, the weaker convenient
form is
\begin{equation}\label{ham:eq:global:scalarerror}
 E_1-E_0\ge
 \alpha_1-\alpha_0-
 \frac{\beta_m^2}{\tau_m-\alpha_1}.
\end{equation}
Certified upper/lower endpoints are substituted in the conservative
directions when the Ritz eigenvalues themselves are enclosed.

At a fixed finite lattice and fixed $a,s>0$, this is an eventually
successful finite procedure when the parameters are rational or supplied
with arbitrarily refinable certified enclosures. The operator estimates
themselves hold for all real $a,s>0$. Completeness does not assume an
unproved omitted-sector bound. Electric spectral projections approximate the form domain of
$D$, and $V$ is bounded, so the Ritz eigenvalues satisfy
$\alpha_j\to E_j$ for each fixed $j$. Meanwhile $\tau_m\to\infty$
and $\beta_m$ is bounded uniformly in $m$. The right side of
\eqref{ham:eq:global:scalarerror} therefore converges to $E_1-E_0>0$.
With increasing finite precision, a strict positive certificate is
eventually obtained. This proves completeness of the procedure at
fixed parameters; it does not give a small computational budget.

\paragraph{Why this does not finish the continuum step.}
The projection must contain the complete physical electric space, not
only a small chosen operator basis. Its dimension grows quickly with
volume and electric threshold. A scalar test using the unconditional
bound \eqref{ham:eq:global:couplingnorm} has an extensive coupling scale
$4N_p/(sa)$. Even the requirement $z<\tau_m$ puts the electric cutoff
above the absolute energy threshold being tested. The vacuum energy
cannot simply be subtracted from $D$ using an unknown or truncated
vacuum. No estimate here proves that the resulting positive lower
gap remains bounded away from zero as $a\to0$, $s\to0$, and physical
volume grows. Proving that property requires actual bounds along such
a sequence, as well as a compatible nontrivial continuum construction.

\paragraph{A full three-dimensional scalar check.}
There is a simple nontruncated check on the periodic $L=2$ lattice.
It has $N_p=24$. Every nonconstant physical electric eigenstate has
$D\ge3/2$: an invariant spin network with a nonzero edge cannot have
a vertex of support degree one, so its support contains at least a
two-edge cycle; each nonzero spin contributes at least $3/4$.
A straight two-link winding loop attains $D=3/2$.
Thus $E_1\ge3s/(4a)$ by $V\ge0$, while the constant Haar trial gives
$E_0\le96/(sa)$. At $s=16$ this certifies
\begin{equation}\label{ham:eq:global:strongexample}
 E_1-E_0\ge\frac6a
\end{equation}
on the full physical complement, with every representation and sector
included. This is a strong-coupling example, not a weak-coupling or
continuum result. The following separate one-plaquette calculation
illustrates much tighter spectral enclosures in a smaller model.

\section{Certified infinite-representation one-plaquette benchmark}
\label{ham:sec:global:oneplaquette}

This section computes rigorous spectral intervals for a \emph{single
plaquette graph}: four distinct links forming one cycle, gauge invariance
at all four vertices, and no external charges. It is a reduced model,
not a three-dimensional spatial box or a four-dimensional continuum
theory. Its purpose is to demonstrate a complete physical-spectrum
calculation in which the infinite electric tail is certified. The
single-plaquette character and Mathieu descriptions have established
precedents, including \cite{ham:global:zache}; the numerical bounds below
are independently calculated in the normalization stated here.

\subsection{The exact infinite physical Jacobi matrix}

Gauge invariance reduces the physical Hilbert space to class functions
of the plaquette holonomy. Its complete orthonormal basis is
$e_n(U)=\chi_{n/2}(U)$, $n=0,1,\ldots$. All four links carry the same
spin, so $De_n=n(n+2)e_n$. The fundamental character recurrence gives
$Fe_n=(e_{n+1}+e_{n-1})/2$, with $e_{-1}=0$. Hence the dimensionless
operator $J_s=aH$ has entries
\begin{equation}\label{ham:eq:global:jacobi}
 (J_s)_{nn}=\frac{s}{2}n(n+2)+\frac4s,
 \qquad (J_s)_{n,n+1}=(J_s)_{n+1,n}=-\frac2s.
\end{equation}
It is self-adjoint on the electric domain and has compact resolvent,
because its diagonal grows quadratically and its remaining terms are
bounded. These are all physical states of this graph, not just a
chosen variational operator family.

Let $A_N$ retain indices $0,\ldots,N-1$, set $b=2/s$,
$t_n=sn(n+2)/2$, and define the $(N+1)$-dimensional lower matrix
\begin{equation}\label{ham:eq:global:jacobilower}
 J_N^-=
 \begin{pmatrix}
 A_N&-b\,e_{N-1}\\
 -b\,e_{N-1}^*&t_N
 \end{pmatrix}.
\end{equation}
The last diagonal is $t_N$, without its magnetic constant $4/s$.

\begin{theorem}[All-tail spectral enclosure for the one-plaquette graph]
\label{ham:thm:global:jacobienclosure}
Let $\mu_j^{(N)}$ and $\alpha_j^{(N)}$ be the ordered eigenvalues of
$J_N^-$ and $A_N$, respectively. For $j<N$, the exact infinite physical
eigenvalues $\epsilon_j=aE_j$ obey
\begin{equation}\label{ham:eq:global:jacobienclosure}
 \min\{\mu_j^{(N)},t_{N+1}\}
 \le\epsilon_j\le\alpha_j^{(N)}.
\end{equation}
Thus any rigorous endpoints
$\underline\epsilon_j,\overline\epsilon_j$ obtained from this formula
give the physical gap enclosure
\begin{equation}\label{ham:eq:global:gapinterval}
 \frac{\underline\epsilon_1-\overline\epsilon_0}{a}
 \le E_1-E_0\le
 \frac{\overline\epsilon_1-\underline\epsilon_0}{a}.
\end{equation}
\end{theorem}

\begin{proof}
The plaquette potential $4(1-F)/s$ is nonnegative. Remove only its
compression to the complement of indices $0,\ldots,N-1$, preserving
the coupling across the boundary. The resulting lower operator is
exactly
\[
 J_N^-\oplus\operatorname{diag}(t_{N+1},t_{N+2},\ldots).
\]
All of the removed quadratic form is nonnegative. Min--max proves the
lower bound, and the complete finite physical trial space gives the
upper bound. This accounts for every omitted representation, including
its coupling through the boundary. There is no assumed decay of the
true eigenvector and no unestimated tail.
\end{proof}

\subsection{Exact integer inertia and all rounding errors}

The companion checker uses only Python's standard library. At a rational
spectral test point $z=p/q$ and rational $s=r/t$, multiplication of either
matrix minus $zI$ by the positive integer $2trq$ gives integer off-diagonal
entries of magnitude $4t^2q$. Its diagonal entries are
\[
 r^2q\,n(n+2)+8t^2q-2trp
\]
for indices in $A_N$. At the added lower-matrix index $n=N$, the term
$8t^2q$ is absent. The leading principal determinants satisfy the exact
Sturm recurrence
\[
 P_0=1,\qquad P_1=d_0,\qquad
 P_{k+1}=d_kP_k-b_{\rm int}^2P_{k-1}.
\]
The number of negative eigenvalues is the number of sign changes in
$P_0,P_1,\ldots$, with zero entries omitted. Consecutive zero entries
are impossible because the off-diagonal entries are nonzero; the rule
also counts an endpoint zero eigenvalue correctly as nonnegative.
This is the exact tridiagonal inertia rule, not a floating-point
eigenvalue calculation.

Rational bisection encloses the first two eigenvalues of each finite
matrix. At each reported lower endpoint, the lower matrix has at most
$j$ eigenvalues below that endpoint. At each reported upper endpoint,
the Ritz matrix has at least $j+1$ eigenvalues below that endpoint.
The JSON record includes the exact endpoints, determinant sign strings,
counts, electric tail floor, and outward decimal intervals.

The calculation starts with $N=\max\{16,\lceil2/s\rceil\}$ and doubles
$N$ only if necessary to reduce the final gap interval width below
$10^{-12}$. It uses 64 exact bisections per finite eigenvalue in the
listed parameter range. The interval width includes both the finite
bisection error and the entire analytic representation remainder.

\subsection{Reproduced dimensionless gap intervals}

The following endpoints are outward rounded to twelve decimal places.
The machine-readable certificate retains narrower exact rational
intervals. The column $N$ is the size of the Ritz matrix; the lower
matrix has size $N+1$, and the largest retained Ritz spin is $(N-1)/2$.

\begin{center}
\begin{tabular}{rcrr}
\hline
$s=g^2$ & $N$ & Lower bound on $a(E_1-E_0)$ & Upper bound\\
\hline
$1$ & $16$ & $3.269099185520$ & $3.269099185521$\\
$1/2$ & $16$ & $3.663550052480$ & $3.663550052481$\\
$1/4$ & $32$ & $3.838313950841$ & $3.838313950842$\\
$1/8$ & $32$ & $3.920572875168$ & $3.920572875169$\\
$1/16$ & $64$ & $3.960618500959$ & $3.960618500960$\\
$1/32$ & $64$ & $3.980389784555$ & $3.980389784556$\\
$1/100$ & $200$ & $3.993741967366$ & $3.993741967367$\\
\hline
\end{tabular}
\end{center}

The apparent weak-coupling approach to $4$ in this table is not used
as a proof of any limiting value. It is a feature of this reduced
model and must not be identified with the continuum SU(2) mass gap.
The table entries are deterministic mathematical enclosures, not
Monte Carlo measurements.

\subsection{Exact special-function identification}

There is also an exact special-function description of this reduced
model. Write $U=\cos\theta\,I+i\sin\theta\,\omega\cdot\sigma$ and
apply the unitary transformation $u(\theta)=\sqrt{2/\pi}\sin\theta f(\theta)$.
Then $J_s$ becomes the Dirichlet operator on $(0,\pi)$
\[
 -\frac s2\frac{d^2}{d\theta^2}-\frac s2+
 \frac4s(1-\cos\theta).
\]
With $x=\theta/2$, its eigenvalue equation is Mathieu's equation
$v''+(A-2q\cos2x)v=0$ on $(0,\pi/2)$ with zero boundary values,
where $q=-16/s^2$ and $A=8\epsilon/s-32/s^2+4$.
Denoting its ordered Dirichlet characteristic values by the conventional
$b_{2r}(q)$ gives
\[
 \epsilon_{r-1}=\frac s8\bigl(b_{2r}(-16/s^2)-4\bigr)+\frac4s,
 \qquad
 a(E_1-E_0)=\frac s8\bigl(b_4(-16/s^2)-b_2(-16/s^2)\bigr).
\]
No numerical special-function implementation is used in the certificate;
the integer Sturm calculation and positive-tail comparison suffice.

\clearpage
\part{Continuum construction and spectral passage}
\section{A physical-scale block criterion and an exact obstruction to naive blocking}
\label{sec:physical-block}

The preceding comparison theorems concern microscopic links or links remaining after a specified gauge fixing. A continuum application needs decay measured in physical distance. This section proves an explicit sufficient criterion at a fixed physical block scale, including all support and normalization constants. It then gives a compact-group example in which replacing individual variables by large blocks makes a total-variation comparison fail, even though the underlying correlations have finite range. These results clarify what a successful block construction would have to improve.

The comparison framework is classical; generalized block formulations are developed by Rebeschini and van Handel \cite{RVH}. The finite proof below is included to specify the constants and hypotheses used here. The criterion is not asserted to hold for the four-dimensional Wilson theory along a continuum trajectory.

\subsection{Exact block variables and conditional influence}

Let \(I\) be a finite index set. At index \(i\) the variable takes values in a compact metric space \(K_i\), with a fixed reference probability measure. Assume the joint law \(\mu\) has a strictly positive continuous density relative to the product reference measure. Write \(\gamma_i(\,\cdot\,|x)\) for its conditional distribution at \(i\) given all other coordinates. For a bounded function \(F\), define its coordinate oscillation by
\[
 \delta_iF=\sup\{|F(x)-F(y)|:\ x_k=y_k\ \hbox{for all }k\ne i\}.
 \tag{PB.1}
\]
The index \(i\) may represent an entire disjoint block of microscopic links. No independence within that block is assumed.

Suppose a nonnegative matrix \(C\), with \(C_{ii}=0\), satisfies
\[
 \|\gamma_i(\,\cdot\,|x)-\gamma_i(\,\cdot\,|y)\|_{\rm TV}
 \le C_{ij}
 \quad\hbox{whenever }x_k=y_k\ (k\ne j),
 \qquad
 \sup_i\sum_j C_{ij}\le q<1.
 \tag{PB.2}
\]
Total variation is one half of the \(L^1\) distance between densities. The matrix need not be symmetric. In the statements using a logarithmic rate we choose \(0<q<1\); a smaller actual row sum can always be bounded by such a positive \(q\).

For later use, set
\[
 D=(I-C)^{-1}=\sum_{n=0}^{\infty}C^n.
 \tag{PB.3}
\]
The series converges in the maximum-row-sum norm and has nonnegative entries. The same definitions apply to a finite part of an infinite block lattice, with fixed exterior variables entering the conditional densities.

\subsection{A complete finite comparison proof}

\noindent\textbf{Lemma PB.1 (comparison for disjoint block variables).}
Let \(\nu\) be another probability law with positive jointly measurable conditional densities \(\widetilde\gamma_i\) on the same product space. Define
\[
 b_i=\sup_x\|\gamma_i(\,\cdot\,|x)
                   -\widetilde\gamma_i(\,\cdot\,|x)\|_{\rm TV}.
\]
Under (PB.2), every bounded measurable \(F\) satisfies
\[
 |\mu F-\nu F|\le\sum_{i,j}\delta_iF\,D_{ij}b_j.
 \tag{PB.4}
\]

\textit{Proof.} Apply Lemma SC.3 with each disjoint block as one coordinate. Its measurable maximal-coupling proof uses only conditional densities relative to fixed reference probability measures, and therefore applies verbatim to the present compact block spaces. The row bound (PB.2) gives the same resolvent (PB.3), proving (PB.4). \hfill\(\square\)

\noindent\textbf{Lemma PB.2 (covariance and path length).}
Under the same assumptions,
\[
 |\operatorname{Cov}_{\mu}(F,G)|
 \le\frac14\sum_{i,j}\delta_iF\,D_{ij}\delta_jG
 \tag{PB.5}
\]
for bounded real \(F,G\). If the directed support of \(C\) is contained in the edge set of a graph \(\mathcal G\), then
\[
 D_{ij}\le\frac{q^{d_{\mathcal G}(i,j)}}{1-q},
 \tag{PB.6}
\]
where an absent path gives a zero right side. Therefore, for supports \(A,B\),
\[
 |\operatorname{Cov}_{\mu}(F,G)|
 \le
 \frac{q^{d_{\mathcal G}(A,B)}}{4(1-q)}
 \left(\sum_{i\in A}\delta_iF\right)
 \left(\sum_{j\in B}\delta_jG\right).
 \tag{PB.7}
\]

\textit{Proof.} The exponential-tilt argument in Theorem LC.3 applies to the present block coordinates and proves (PB.5), with the same factor $1/4$.

A nonzero matrix product contributing to \((C^n)_{ij}\) is a path of length \(n\). Thus \((C^n)_{ij}=0\) when \(n<d_{\mathcal G}(i,j)\), while every entry is at most the row sum \(q^n\). Sum the geometric series to obtain (PB.6), and then sum over the supports to obtain (PB.7).
\hfill\(\square\)

\subsection{The physical-distance conversion for the Wilson lattice}

For the ordinary four-dimensional lattice, represent a positively oriented microscopic link by \((x,\mu)\), where \(x\in\mathbb Z^4\) is its initial vertex and \(\mu\in\{1,2,3,4\}\). For an integer \(L\ge1\), put all four links with initial vertex \(x\) in block
\[
 i(x)=\bigl(\lfloor x_1/L\rfloor,\ldots,\lfloor x_4/L\rfloor\bigr).
 \tag{PB.8}
\]
Every microscopic link belongs to exactly one block. A full block has \(4L^4\) link variables, so its state space is the compact product \(SU(2)^{4L^4}\). The Wilson measure remains an exact measure on this block product space; grouping variables has not integrated out any field.

\noindent\textbf{Lemma PB.3 (block interaction geometry).}
For this partition, the conditional law of a block in the nearest-neighbor Wilson plaquette action depends on at most twenty other blocks. Their displacement vectors belong to
\[
 \{\pm e_\mu:\ 1\le\mu\le4\}
 \ \cup\
 \{e_\mu-e_\nu:\ \mu\ne\nu\}.
 \tag{PB.9}
\]
In particular, if block centers are measured in physical units at microscopic spacing \(a\), each nonzero influence step moves their \(\ell^\infty\) distance by at most \(aL\).

\textit{Proof.}
An elementary plaquette in directions \(\mu,\nu\) uses positive-link variables with initial vertices \(x,x+e_\mu,x+e_\nu\). The block index of \(x+e_\mu\) is either \(i(x)\) or \(i(x)+e_\mu\), and similarly for \(\nu\). Thus every difference of two block indices belonging to that plaquette is in (PB.9) or is zero. Only plaquettes containing variables in the conditioned block can produce a dependence on another block. There are eight axial displacements and twelve ordered mixed displacements. The physical centers of adjacent blocks consequently differ by at most \(aL\) in the maximum norm. Near an ordinary fixed exterior boundary the number of free neighbors can only decrease.
\hfill\(\square\)

This twenty-neighbor count concerns block variables grouped by initial vertices. It is different from the eighteen-neighbor count for an individual microscopic link. Small periodic quotients can identify indices or plaquettes and need a separate incidence check. The following statement is made for ordinary boxes and their infinite-lattice limits.

\noindent\textbf{Theorem PB.4 (a cutoff-uniform sufficient block criterion).}
Fix a physical length \(\ell>0\) and a number \(0<q<1\). For every microscopic spacing \(0<a\le\ell\), choose
\[
 L_a=\lceil\ell/a\rceil,\qquad \ell_a=aL_a\in[\ell,2\ell].
\]
Suppose that, for the actual Wilson block conditionals at a specified coupling \(\beta(a)\), there are interdependence matrices \(C^{(a)}\) satisfying (PB.2), with the same \(q\), uniformly in the finite volume and all exterior configurations. Let
\[
 m_{\rm block}=\frac{-\log q}{2\ell}>0.
 \tag{PB.10}
\]
For bounded real local functions \(F_a,G_a\), let \(r\) be the minimum physical \(\ell^\infty\) distance between their microscopic link-midpoint supports. Then
\[
 \boxed{
 |\operatorname{Cov}(F_a,G_a)|
 \le
 \frac{e^{-m_{\rm block}r}}{4q(1-q)}
 \left(\sum_i\delta_iF_a\right)
 \left(\sum_j\delta_jG_a\right).
 }
 \tag{PB.11}
\]
Here \(\delta_i\) is the oscillation when the entire \(i\)-th block changes. The bound is independent of microscopic spacing and volume whenever the two displayed oscillation sums are bounded independently of those regulators.

If \(F_a,G_a\) depend on at most \(N_F,N_G\) blocks respectively, then
\[
 |\operatorname{Cov}(F_a,G_a)|
 \le\frac{N_FN_G}{q(1-q)}
       \|F_a\|_\infty\|G_a\|_\infty e^{-m_{\rm block}r}.
 \tag{PB.12}
\]
A support contained in a fixed physical cube of side \(R\) meets no more than
\[
 \bigl(\lceil R/\ell\rceil+2\bigr)^4
 \tag{PB.13}
\]
blocks. Thus the support counts in (PB.12) can remain uniform when \(a\downarrow0\).

\textit{Proof.}
The center of the block containing a link midpoint is at maximum-norm distance at most \(\ell_a/2\) from that midpoint. If \(s\) denotes the minimum center distance between supporting blocks, the triangle inequality gives \(s\ge r-\ell_a\).
A path of \(n\) influence edges moves the center distance by at most \(n\ell_a\), so its graph length is at least \(s/\ell_a\). Since \(0<q<1\),
\[
 q^{d_{\mathcal G}(A,B)}
 \le q^{-1}\exp\!\left(-\frac{-\log q}{\ell_a}r\right)
 \le q^{-1}e^{-m_{\rm block}r}.
\]
The first inequality remains valid when \(r<\ell_a\), because its right side is then at least one. Apply (PB.7). The inequality
\(\delta_iF_a\le2\|F_a\|_\infty\) gives (PB.12).
In each coordinate, an interval of length \(R\) intersects at most
\(\lceil R/\ell_a\rceil+2\) block intervals; multiplying four such bounds proves (PB.13).
\hfill\(\square\)

For example, the stronger condition \(C^{(a)}_{ij}\le c<1/20\) for every allowed neighboring pair implies the hypothesis with \(q=20c\). This is a sufficient way to check the row condition, not a claim that the actual coefficients obey this estimate.

At each fixed \(a\), the same argument gives a unique infinite-volume Gibbs state if the conditional kernels have their usual continuous finite-range specification. For completeness, compactness gives subsequential limits of finite-volume laws. The finite-volume conditional identities pass to such limits for continuous local test functions. For two exterior conditions, the forcing in Lemma PB.1 is
\(b_i\le\sum_{j\notin I}C_{ij}\). Paths contributing to \(Db\) must reach the boundary, which gives the bound
\[
 |\mu_I^\eta F-\mu_I^\xi F|
 \le
 \sum_{i\in\operatorname{supp}F}
 \delta_iF\,
 \frac{q^{d_{\mathcal G}(i,I^c)}}{1-q}.
 \tag{PB.14}
\]
It tends to zero along an exhaustion for every fixed local \(F\).
Integrating this bound against the two exterior distributions proves uniqueness of the infinite-volume state. For a fixed finite support, apply (PB.14) uniformly over all functions with supremum norm at most one; each oscillation is then at most two. This proves that the corresponding finite-volume local marginals are Cauchy in total variation. Their total-variation limits agree with their weak limits. Consequently expectations, products and covariances of bounded measurable local observables converge, and the displayed covariance inequalities pass to the infinite-volume state.

\subsection{Uniform moments in place of uniform supremum norms}

Renormalized fields usually do not have a useful uniform supremum norm. A moment assumption can replace it, at a cost in the certified decay exponent.

\noindent\textbf{Corollary PB.5 (an explicit \(L^p\) version).}
Assume the hypotheses of Theorem PB.4, with support counts bounded by \(N_F,N_G\). Let \(p>2\), and suppose real local observables satisfy
\[
 \sup_a\|F_a\|_{L^p}\le B_F,\qquad
 \sup_a\|G_a\|_{L^p}\le B_G.
\]
Then
\[
 \boxed{
 |\operatorname{Cov}(F_a,G_a)|
 \le
 \left(\frac{N_FN_G}{q(1-q)}+2\right)
 B_FB_G
 \exp\!\left[-\frac{p-2}{p+2}m_{\rm block}r\right].
 }
 \tag{PB.15}
\]
The same assertion holds in finite volume if the moment bounds hold there uniformly.

\textit{Proof.}
The case of a zero moment bound is immediate. Otherwise let
\[
 u=\exp\!\left(\frac{2m_{\rm block}r}{p+2}\right),\quad
 R_F=B_Fu,\quad R_G=B_Gu,
\]
and truncate \(F_a,G_a\) to the intervals \([-R_F,R_F]\) and \([-R_G,R_G]\), denoting the results by \(F_a^R,G_a^R\).
For any real \(X\),
\[
 \|X-\max(-R,\min(X,R))\|_2^2
 \le \mathbb E\bigl(|X|^2\mathbf1_{\{|X|>R\}}\bigr)
 \le R^{2-p}\mathbb E|X|^p.
\]
Therefore
\[
 \|F_a-F_a^R\|_2\le B_Fu^{1-p/2},\qquad
 \|G_a-G_a^R\|_2\le B_Gu^{1-p/2}.
\]
Cauchy--Schwarz for centered variables and the fact that truncation cannot increase the \(L^2\) norm imply
\[
 |\operatorname{Cov}(F_a,G_a)-\operatorname{Cov}(F_a^R,G_a^R)|
 \le2B_FB_Gu^{1-p/2}.
\]
Apply (PB.12) to the truncated variables. Its contribution is at most
\[
 \frac{N_FN_G}{q(1-q)}B_FB_Gu^2e^{-m_{\rm block}r}.
\]
The chosen value of \(u\) makes both powers equal to
\(\exp[-(p-2)m_{\rm block}r/(p+2)]\), proving (PB.15).
\hfill\(\square\)

This corollary supplies a correlation estimate for the actual observables to which its moment bounds apply. It does not prove those moment bounds. A fixed physical support controls the number of blocks but not a normalization factor such as \(a^{-4}\), a subtraction, an operator-mixing coefficient, or a reflected norm. Those factors still have to be estimated for the same renormalized family.

\subsection{An exact compact-group obstruction to naive block total variation}

The following model shows why the sufficient block hypothesis cannot be inferred just by making blocks large. It is deliberately solvable, and is not claimed to be the four-dimensional Wilson plaquette theory.

Let \(G=SU(2)\), \(w(U)=\tfrac12\operatorname{Tr}U\), and \(\beta>0\). At each integer site \(i\), place a block containing \(2N\) group variables,
\[
 X_i=(U_{i,1}^{-},\ldots,U_{i,N}^{-},
      U_{i,1}^{+},\ldots,U_{i,N}^{+}).
\]
Give a finite interval the nearest-neighbor density
\[
 d\mu\ \propto\
 \prod_{i,k}\exp\!\left\{
   \beta\,w\bigl(U_{i,k}^{+}(U_{i+1,k}^{-})^{-1}\bigr)
 \right\}
 \prod_{i,k}dh(U_{i,k}^{-})\,dh(U_{i,k}^{+}),
 \tag{PB.16}
\]
with free endpoint variables. Each variable occurs in at most one interacting pair.

\noindent\textbf{Theorem PB.6 (block comparison can fail in a finite-range independent-pair model).}
The model (PB.16) has a unique infinite-volume law, and functions supported on blocks at graph distance at least two are independent. Nevertheless its exact whole-block total-variation interdependence matrix satisfies, for each interior block,
\[
 C_{i,i-1}+C_{i,i+1}
 \ge 2\bigl(1-Z(\beta)^{-N}\bigr),
 \qquad
 Z(\beta)=\frac{2I_1(\beta)}{\beta}.
 \tag{PB.17}
\]
For every fixed \(\beta>0\), the lower bound tends to two as \(N\to\infty\). In particular the whole-block row-sum criterion fails whenever
\[
 N>\frac{\log2}{\log Z(\beta)}.
 \tag{PB.18}
\]
The elementary condition \(N\beta^2>8\) is sufficient for this failure.

\textit{Proof.}
The factors in (PB.16) involve disjoint pairs of variables. Hence the law is a product of independent pair laws and independent endpoint Haar variables. This also defines the unique infinite-volume Gibbs law: every finite component is either one pair or one isolated variable, and its conditional distribution is uniquely specified. A block \(i\) uses pair components crossing only the interfaces \(i-1,i\) and \(i,i+1\). Blocks whose indices differ by at least two use disjoint pair components, proving independence.

Now condition on every block except \(i\). Its \(N\) right variables have independent conditional densities
\[
 U\longmapsto Z(\beta)^{-1}
       \exp\{\beta w(UV^{-1})\},
 \qquad V=U_{i+1,k}^{-}.
\]
Compare two exterior configurations that agree except in block \(i+1\), with all its relevant \(V\)'s equal to \(I\) in the first and to \(-I\) in the second. Since \(-I\) is central in \(SU(2)\), the two conditional densities for one right variable are
\[
 p_+(U)=e^{\beta w(U)}/Z(\beta),\qquad
 p_-(U)=e^{-\beta w(U)}/Z(\beta).
\]
Their affinity is exactly
\[
 \int\sqrt{p_+p_-}\,dh=Z(\beta)^{-1}.
\]
The affinity for \(N\) independent right variables is \(Z(\beta)^{-N}\); all left-variable conditionals are identical in the two configurations and make no change to this value.

For any densities \(p,q\),
\[
 \|p-q\|_{\rm TV}
 =1-\int\min(p,q)
 \ge1-\int\sqrt{pq}.
\]
It follows that \(C_{i,i+1}\ge1-Z(\beta)^{-N}\). The same construction at the left interface proves the identical bound for \(C_{i,i-1}\), establishing (PB.17). These are separate suprema, so no assertion about simultaneously maximizing both influences is needed.

Since \(Z(\beta)>1\), (PB.18) makes the right side of (PB.17) exceed one. Finally the positive series gives
\[
 Z(\beta)\ge1+\beta^2/8,\qquad
 Z(\beta)^N\ge(1+\beta^2/8)^N\ge1+N\beta^2/8.
\]
If \(N\beta^2>8\), then \(Z(\beta)^N>2\), which suffices.
\hfill\(\square\)

For a completely explicit instance, \(\beta=1/3\) and \(N=73\) already force the block row sum to exceed one by integer arithmetic, even though correlations vanish between non-neighboring blocks. This does not contradict any single-link result: changing an entire neighboring block changes many independent channels at once. It demonstrates a loss of information caused by using a worst-case whole-block total variation.

The example does not disprove a multiscale Yang--Mills argument. It rules out the general inference that large blocks must have a small whole-block total-variation influence merely because the original local model mixes well. Overlapping update schemes, weighted transport costs, observables adapted to interior regions, and exact renormalized effective interactions can behave differently. Each requires its own quantitative estimate rather than the replacement of a microscopic variable by a large tuple with no further analysis.

\subsection{Robustness of a finite block certificate under action errors}

A certificate for an approximate block kernel is useful only if its error can be transferred to the actual conditional law. The following bound keeps both the allowed error and the interaction degree explicit.

\noindent\textbf{Lemma PB.7 (sharp bounded-tilt discrepancy).}
Let \(Q\) be a probability measure and let
\[
 dP(u)=\frac{e^{r(u)}}{Q(e^r)}\,dQ(u),\qquad
 \sup r-\inf r\le\varepsilon.
\]
Then
\[
 \|P-Q\|_{\rm TV}\le\tanh(\varepsilon/4).
 \tag{PB.19}
\]
The constant is sharp over all such probability measures and tilts.

\textit{Proof.}
There is nothing to prove when \(\varepsilon=0\). Otherwise, subtract a constant from \(r\), set \(R=e^\varepsilon>1\), and put \(L=e^r\), so that \(1\le L\le R\). Write \(m=Q(L)\). Then
\[
 \|P-Q\|_{\rm TV}
 =\frac{Q((L-m)_+)}{m}.
\]
The convex function \(x\mapsto(x-m)_+\) is bounded above on \([1,R]\) by the chord joining its endpoint values. Therefore
\[
 Q((L-m)_+)
 \le\frac{R-m}{R-1}\,Q(L-1)
 =\frac{(R-m)(m-1)}{R-1}.
\]
It follows that
\[
 \|P-Q\|_{\rm TV}
 \le\frac{R+1-m-R/m}{R-1}
 \le\frac{R+1-2\sqrt R}{R-1}
 =\frac{\sqrt R-1}{\sqrt R+1}
 =\tanh(\varepsilon/4).
\]
The middle inequality is \(m+R/m\ge2\sqrt R\). Equality is obtained on a two-point space by choosing \(L\) to take only the values \(1,R\), with probabilities making \(Q(L)=\sqrt R\). This also proves sharpness.
\hfill\(\square\)

\noindent\textbf{Proposition PB.8 (an explicit stability margin for block comparison).}
For every block \(i\) and exterior configuration \(x\), suppose its actual conditional law and a reference probability kernel satisfy
\[
 d\gamma_i(u|x)
 =\frac{e^{r_i(u,x)}}{\widehat\gamma_i(e^{r_i(\,\cdot\,,x)}|x)}
       \,d\widehat\gamma_i(u|x),
 \qquad
 \sup_u r_i(u,x)-\inf_u r_i(u,x)\le\varepsilon.
 \tag{PB.20}
\]
Assume both kernels depend only on exterior indices in a specified set \(N_i\), with \(|N_i|\le z\), and that verified reference influence coefficients obey
\[
 \|\widehat\gamma_i(\,\cdot\,|x)
      -\widehat\gamma_i(\,\cdot\,|y)\|_{\rm TV}
 \le \widehat C_{ij},
 \qquad
 \sup_i\sum_{j\in N_i}\widehat C_{ij}\le q_0<1
\]
whenever \(x,y\) differ only at index \(j\). Then the actual kernels have a comparison matrix with row sums bounded by
\[
 \boxed{q_{\rm true}\le q_0+2z\tanh(\varepsilon/4).}
 \tag{PB.21}
\]
In particular, for \(z\ge1\), the strict condition
\[
 \varepsilon<
 4\operatorname{arctanh}\!\left(\frac{1-q_0}{2z}\right)
 \tag{PB.22}
\]
makes the actual row bound smaller than one. No compatibility of the reference kernels with a separate joint Gibbs law is required for this implication.

If the true conditional laws have additional dependence outside \(N_i\), the same conclusion holds with an extra \(t\) on the right side of (PB.21), provided that independently verified coefficients \(T_{ij}\) bound their true influences there and satisfy
\[
 \sup_i\sum_{j\notin N_i}T_{ij}\le t.
 \tag{PB.23}
\]
In this variant the condition is \(q_0+2z\tanh(\varepsilon/4)+t<1\).

\textit{Proof.}
For \(j\in N_i\), insert the two reference kernels between the actual kernels at exterior configurations \(x,y\). The triangle inequality and Lemma PB.7 give
\[
 \|\gamma_i(\,\cdot\,|x)-\gamma_i(\,\cdot\,|y)\|_{\rm TV}
 \le\tanh(\varepsilon/4)+\widehat C_{ij}
                         +\tanh(\varepsilon/4).
\]
Outside the common dependency set the influence is zero. Summing proves (PB.21), and solving its strict inequality proves (PB.22). In the additional-tail variant, use \(T_{ij}\) outside \(N_i\) and sum it separately. The proof involves only individual probability kernels and therefore never needs a joint reference measure.
\hfill\(\square\)

For the twenty-neighbor partition of Lemma PB.3, a reference certificate with row bound \(q_0\) remains a contracting certificate when
\[
 q_0+40\tanh(\varepsilon/4)<1.
\]
This is a concrete analytic margin that an exact or interval-assisted calculation could check. It is not enough to bound a few action coefficients numerically. The oscillation in (PB.20) is over the full block variable, uniformly over all exterior configurations. All omitted interactions have to be included either in that oscillation estimate and the common dependency graph, or in the separately controlled tail (PB.23). A renormalization step can generate long-range or multi-block interactions; their absence must not be assumed just because the original Wilson action had a local plaquette form.

A scalar bound on the global partition function is also insufficient for (PB.20). The proposition needs control of conditional densities, up to constants independent of the variable \(u\). Conversely, exterior-dependent additive constants in \(r_i\) cancel from the conditional normalization and need not be bounded. This is why the oscillation, rather than an arbitrary absolute action offset, is the appropriate error quantity.

\subsection{What would be needed for a continuum application}

Theorem PB.4 replaces a desired infinite-distance correlation estimate by a local conditional-kernel inequality at a specified physical scale. For each fixed cutoff it is a statement about finitely many block variables and boundary fields. It is not one finite computation whose validity automatically extends to all cutoffs.

The missing model-specific task is to construct exact block or effective variables for the relevant continuum trajectory and prove, with constants independent of \(a\),
\[
 \sup_i\sum_j C^{(a)}_{ij}\le q<1,
 \tag{PB.24}\label{pb:eq:required-row-bound}
\]
together with the required observable bounds. Even in the finite-range Wilson action, the boundary configurations in these suprema range over compact sets of increasing dimension. A proof at a list of couplings, on a finite boundary grid, or at one block size does not establish \eqref{pb:eq:required-row-bound}. An interval computation would require a valid finite reduction with controlled covering error, and a separate argument giving uniformity over the unbounded cutoff sequence.

The microscopic transport estimate in the companion section controls fields of radius bounded by a fixed constant. Its four-dimensional single-link range \(\beta<2/9\) therefore supplies no direct estimate on a trajectory with \(\beta(a)\to\infty\). A forest-gauge comparison may improve a fixed-spacing constant, but any such constant remains separate from \eqref{pb:eq:required-row-bound}. A gauge transformation is not an ultraviolet renormalization step; it does not by itself give a controlled effective action at fixed physical resolution.

For an exact renormalization map, three additional points matter. First, the effective conditional law must actually equal the pushforward or marginal law being studied, including all generated interactions. Second, the mapped observables and their supports must be followed through the map. Third, any approximation error must be summable or otherwise controlled over all scales. If a coarse observable is nonlocal in the fine variables, its nominal coarse label is not proof of physical locality. If the effective state space has constraints, a product-space comparison theorem needs an appropriate representation or a different update construction.

Once the actual block condition, renormalized moment bounds, a nonzero limiting reflected norm, and the required reconstruction properties have all been proved, (PB.11) or (PB.15) supplies physical-distance decay with an explicit positive exponent. The spectral criteria elsewhere in the manuscript can then be applied to a total family of physical states. Neither a Gibbs-sampler update count nor the existence of the block comparison theorem is substituted for physical Euclidean time.

The completed result of this section is therefore precise: a sufficient physical-scale criterion, its full proof with constants, and a solvable obstruction to one naive way of trying to satisfy it. The required estimate \eqref{pb:eq:required-row-bound} for four-dimensional Yang--Mills at weak coupling is not established here.

\hypertarget{continuum_attempts-continuum-construction-weak-coupling-control-compactness-and-exact-elimination}{%
\section{Continuum construction: weak-coupling control, compactness and exact elimination}\label{continuum_attempts-continuum-construction-weak-coupling-control-compactness-and-exact-elimination}}

This section develops three parts of a possible continuum argument that can be proved without assuming the desired continuum theory. The first is an estimate for the actual Wilson measure at weak bare coupling: on ordinary periodic four-dimensional boxes, the mean plaquette defect is bounded uniformly in volume by a constant times \((1+\log\beta)/\beta\). The second is a collection of compactness criteria that state precisely which uniform bounds would produce limits of renormalized observables. The third is an exact calculation of coarse-graining and its errors in models where every integration can be justified.

These results have different logical roles. The Wilson estimate is unconditional in its displayed finite-lattice setting. The compactness results are theorems about probability laws; their hypotheses must be verified for the chosen Yang--Mills observables. Gaussian elimination is exact for Gaussian measures, and the nonlinear extensions retain explicit integrability assumptions. Keeping these roles visible makes it possible to identify a failed estimate instead of transferring a conclusion from one model to another.

The methods are standard probability, compact Lie group geometry and finite-dimensional integration, written out here with the constants and limiting operations required by this investigation. Bałaban's four-dimensional small-field effective-action work supplies an important constructive context \cite{Balaban4}; the present elementary estimates do not reproduce that construction. Chatterjee's work obtains substantially more precise leading free-energy information in specified lattice regimes \cite{Chatterjee2016}. Neither reference is used as an unstated premise of the proofs below.

\hypertarget{continuum_attempts-a-volume-uniform-weak-coupling-estimate-for-the-wilson-action}{%
\subsection{A volume-uniform weak-coupling estimate for the Wilson action}\label{continuum_attempts-a-volume-uniform-weak-coupling-estimate-for-the-wilson-action}}

Let \(G\) be a connected compact Lie group of positive dimension \(d_G\), with normalized Haar measure \(h\). Fix a faithful smooth unitary representation \(\rho:G\to U(N)\). Its normalized real character is

\[w(g)=\frac1N\operatorname{Re}\operatorname{Tr}\rho(g).\]

For a complex \(N\times N\) matrix define the normalized Hilbert--Schmidt norm

\[\|A\|_*^2=\frac1N\operatorname{Tr}(A^*A).\]

For every unitary matrix \(V\),

\[\frac12\|V-I\|_*^2=1-\frac1N\operatorname{Re}\operatorname{Tr}V.\]

Indeed, \((V-I)^*(V-I)=2I-V-V^*\), and taking normalized traces gives the identity. In particular, \(0\le1-w(g)\le2\). This nonnegative defect will be denoted by \(s(g)\).

Choose a bi-invariant Riemannian metric on \(G\). Such a metric is obtained by averaging any inner product on the Lie algebra under the adjoint action and extending it by translations. Write \(\operatorname{dist}_G\) for its distance, retaining \(d_G\) for the group dimension. Smoothness of \(\rho\) and compactness give a finite constant \(L_\rho\) satisfying

\[\|\rho(g)-I\|_*\le L_\rho\operatorname{dist}_G(g,e).\]

To justify the global inequality, let \(M\) be the maximum operator norm of the differential of \(\rho\) on the compact manifold. Integrating the derivative along a length-minimizing geodesic bounds the norm of its matrix increment by \(M\) times the geodesic length. The representation takes values in a finite-dimensional Euclidean matrix space, so the ordinary fundamental theorem of calculus applies to that path. One can take \(L_\rho=M\).

There are constants \(\varepsilon_0\in(0,1]\) and \(c_G\in(0,1]\) such that

\[h\{g:\operatorname{dist}_G(g,e)\le\varepsilon\}\ge c_G\varepsilon^{d_G}
\qquad(0<\varepsilon\le\varepsilon_0).\]

Here is the local reason. The exponential map is a diffeomorphism on a sufficiently small Lie algebra ball. In these coordinates Haar measure has a continuous strictly positive density, whose value at the origin is positive. On a smaller closed ball that density has a positive minimum. The integral over the Euclidean ball is therefore at least a positive constant times its radius to the power \(d_G\). The Riemannian and coordinate distances agree to first order and are uniformly comparable on this smaller neighborhood. Decreasing the constant, and replacing it by its minimum with one, gives the displayed estimate. Neither \(c_G\) nor \(\varepsilon_0\) depends on a lattice volume.

For a finite set of oriented links \(E\) and a finite set of elementary plaquettes \(P\), define

\[S(U)=\sum_{p\in P}s(U_p),\qquad
Z_\beta=\int_{G^E}e^{-\beta S(U)}\prod_{e\in E}h(dU_e),\qquad
d\mu_\beta=Z_\beta^{-1}e^{-\beta S(U)}\,dh^E.\]

Each elementary plaquette is a product of four link variables or their inverses. We consider free boundary boxes or ordinary periodic boxes, so the configuration with every link equal to the identity has zero action. Arbitrarily frustrated fixed exterior links are not covered by the small-ball argument. All action terms and link variables counted in \(P,E\) belong to the displayed model.

\textbf{Theorem CT.1 (weak-coupling plaquette-defect bound).} Put \(K_\rho=8L_\rho^2\) and suppose \(\beta\ge\varepsilon_0^{-2}\). Then

\[Z_\beta\ge
e^{-K_\rho|P|}\,c_G^{|E|}\,\beta^{-d_G|E|/2}.\]

With

\[A_\beta=K_\rho|P|-|E|\log c_G+
\frac{d_G|E|}{2}\log\beta,\]

the action obeys

\[\mu_\beta\{S\ge u\}\le\min\{1,e^{A_\beta-\beta u}\}
\qquad(u\ge0),\]

and

\[\mathbb E_\beta S\le\frac{A_\beta+1}{\beta}.\]

For the four-dimensional periodic box \((\mathbb Z/L\mathbb Z)^4\), with \(L\ge3\) and each elementary unoriented plaquette counted once,

\[\mathbb E_\beta[1-w(U_p)]
\le\frac{K_\rho-\frac23\log c_G+
\frac{d_G}{3}\log\beta+\frac1{6L^4}}{\beta}
\qquad(p\in P).\]

\textbf{Proof.} Let \(B_\varepsilon\) be the group ball in the Haar small-ball estimate. If all four links of a plaquette lie in \(B_\varepsilon\), then their inverse variables also lie there, because the metric is bi-invariant. For unitary matrices \(V_1,\ldots,V_4\), telescoping gives

\[V_1V_2V_3V_4-I
=(V_1-I)V_2V_3V_4+(V_2-I)V_3V_4+(V_3-I)V_4+(V_4-I).\]

Left and right multiplication by a unitary matrix preserve \(\|\cdot\|_*\). The triangle inequality therefore yields

\[\|\rho(U_p)-I\|_*\le4L_\rho\varepsilon,
\qquad s(U_p)\le8L_\rho^2\varepsilon^2=K_\rho\varepsilon^2.\]

On the event \(B_\varepsilon^E\), the total action is at most \(K_\rho|P|\varepsilon^2\). The product Haar measure of that event is at least \((c_G\varepsilon^{d_G})^{|E|}\). Restricting the positive partition integral to this event and taking \(\varepsilon=\beta^{-1/2}\le\varepsilon_0\) proves the lower bound for \(Z_\beta\).

On the event \(S\ge u\), the integrand \(e^{-\beta S}\) is at most \(e^{-\beta u}\), and the underlying Haar probability of any event is at most one. Thus

\[\mu_\beta\{S\ge u\}
\le Z_\beta^{-1}e^{-\beta u}
\le e^{A_\beta-\beta u}.\]

The probability is also at most one. The conditions \(c_G\le1\) and \(\beta\ge\varepsilon_0^{-2}\ge1\) give \(A_\beta\ge0\). Tonelli's theorem for the nonnegative random variable \(S\) gives

\[\mathbb E_\beta S
=\int_0^\infty\mu_\beta\{S\ge u\}\,du
\le\int_0^{A_\beta/\beta}1\,du+
\int_{A_\beta/\beta}^{\infty}e^{A_\beta-\beta u}\,du
=\frac{A_\beta+1}{\beta}.\]

The periodic four-dimensional box has \(4L^4\) positively oriented links and \(6L^4\) elementary plaquettes. Translations and coordinate permutations preserve product Haar measure and the action and act transitively on the geometric plaquettes; inversion causes no change in \(w\). Every plaquette therefore has the same expected defect. Divide the total-action estimate by \(6L^4\) and use \(|E|/|P|=2/3\). This proves the last formula. \(\square\)

The result is uniform in periodic volume. For every \(\eta>0\), Markov's inequality also gives

\[\sup_{L\ge3}\mu_{\beta,L}\{1-w(U_p)\ge\eta\}
\le\frac{K_\rho-\frac23\log c_G+
\frac{d_G}{3}\log\beta+\frac1{486}}{\eta\beta}.\]

The harmless last constant uses \(6L^4\ge486\). One may retain the sharper \(1/(6L^4)\) instead. The identity connecting defect and normalized matrix distance gives

\[\sup_{L\ge3}\mathbb E_{\beta,L}\|\rho(U_p)-I\|_*^2
=O\!\left(\frac{1+\log\beta}{\beta}\right).\]

This is a real weak-coupling statement about the interacting Wilson model. Its meaning is also limited: unrescaled plaquette holonomies become close to the identity. A continuum curvature field, if defined through shrinking loops and suitable subtraction or normalization, is a different observable. The estimate does not bound that renormalized field uniformly after multiplication by the necessary inverse powers of the lattice spacing.

\textbf{Corollary CT.1a (fixed-spacing Gibbs limits at weak coupling).} For each fixed \(\beta>0\), periodic Wilson measures have subsequences converging locally weakly to infinite-volume Gibbs measures on \(G^{E(\mathbb Z^4)}\). These limits are invariant under translations and the coordinate symmetries of the lattice. If \(\beta\ge\varepsilon_0^{-2}\), every such limit \(\mu_\beta\) satisfies

\[\mathbb E_{\mu_\beta}[1-w(U_p)]
\le\frac{K_\rho-\frac23\log c_G+\frac{d_G}{3}\log\beta}{\beta}.\]

Uniqueness at these large values of \(\beta\) is not asserted.

\textbf{Proof.} Extend a periodic configuration periodically to all links of \(\mathbb Z^4\). Its law is a probability measure on the countable product of the compact metrizable space \(G\). That product is compact and metrizable, for example by a weighted sum of bounded coordinate metrics. Its probability laws are therefore tight, and a sequence of periods \(L\to\infty\) has a weakly convergent subsequence.

To verify the Gibbs property, fix a finite link set \(\Lambda\), a continuous function \(F\) depending on its links, and a continuous cylinder function \(H\) depending on a finite set of links outside \(\Lambda\). Take \(L\) sufficiently large that reduction modulo \(L\) is injective on the union of these supports and all links in plaquettes meeting \(\Lambda\). The torus conditional-density formula then gives

\[\mu_{\beta,L}(HF)=\mu_{\beta,L}(H\gamma_\Lambda F),\]

where \(\gamma_\Lambda\) is the ordinary finite-volume Wilson specification on \(\mathbb Z^4\). The function \(\gamma_\Lambda F\) depends only on a finite neighborhood of \(\Lambda\) and is continuous there. Indeed, its numerator and strictly positive denominator are integrals over a compact finite product of a jointly continuous bounded density; uniform continuity justifies continuity of the integrals, and positivity permits division. Both sides therefore pass to the weak limit.

For each fixed finite \(\Lambda\), continuous functions on \(G^\Lambda\) determine finite measures, and continuous cylinder functions generate the exterior Borel sigma-algebra. Applying the monotone-class theorem first to the exterior tests and then to the interior tests extends the limiting identity to all bounded measurable tests. This is the Dobrushin--Lanford--Ruelle conditional expectation identity. Notice that the periodically extended finite-volume law itself was not declared a Gibbs measure on the whole infinite lattice: its repeated coordinates create global constraints. Only the finite local identities, valid once the quotient is injective on their supports, were passed to the limit.

Translations and coordinate symmetries act continuously on the configuration product, with inversion on links when an orientation is reversed. Each periodic law is invariant under these transformations, because the action and product Haar measure are invariant. Their weak limits inherit the same invariances. Finally, \(1-w(U_p)\) is a continuous bounded local function. Pass Theorem CT.1 to the convergent subsequence and let \(L\to\infty\) in its coefficient. This proves the estimate. \(\square\)

The corollary removes the volume cutoff at fixed lattice spacing. It supplies actual weak-coupling infinite-volume states and a local bound on those states. It neither removes the lattice spacing nor proves that distinct possible Gibbs states coincide. Those are different assertions from the compactness argument just established.

\hypertarget{continuum_attempts-what-the-local-action-estimate-does-and-does-not-control}{%
\subsection{What the local-action estimate does and does not control}\label{continuum_attempts-what-the-local-action-estimate-does-and-does-not-control}}

The preceding expectation estimate is stronger than a finite-dimensional Laplace statement with an uncontrolled volume-dependent constant. Its coefficient after division by the number of plaquettes remains bounded in \(L\). However, it controls a local positive action density, not a connected correlation at a physical separation. This distinction can be made quantitative.

\textbf{Proposition CT.2 (fixed-loop consequence).} Let \(C\) be the boundary of a planar rectangular union of \(A\) elementary plaquettes inside one lattice box. Suppose

\[\max_{p\text{ in the rectangle}}\mathbb E s(U_p)\le\epsilon.\]

Then

\[\mathbb E\|\rho(U_C)-I\|_*^2\le2A^2\epsilon,
\qquad \mathbb E[1-w(U_C)]\le A^2\epsilon.\]

\textbf{Proof.} A rectangular disk can be built by successively adjoining plaquettes along a connected boundary path, row by row. When one plaquette is attached, compare the old boundary holonomy with the new one, based at the same vertex. The new boundary is obtained by replacing one boundary segment with the other part of the plaquette boundary. Its holonomy is consequently the product of the old boundary holonomy and a conjugate of the new plaquette holonomy or its inverse, with the location of multiplication determined by the common base path. Induction expresses \(U_C\) as a product of \(A\) conjugated plaquette holonomies, each possibly inverted. This is a group identity; commutativity is not used.

If these factors are \(Q_1,\ldots,Q_A\), unitary invariance and repeated telescoping give

\[\|\rho(U_C)-I\|_*\le\sum_{i=1}^A\|\rho(Q_i)-I\|_*
=\sum_{p}\|\rho(U_p)-I\|_*.\]

Conjugation and inversion preserve the last norm. By the scalar Cauchy--Schwarz inequality,

\[\|\rho(U_C)-I\|_*^2
\le A\sum_p\|\rho(U_p)-I\|_*^2
=2A\sum_p s(U_p).\]

Expectation and the assumed plaquette estimate give the first conclusion. The normalized trace identity gives the second. \(\square\)

For fixed \(A\), Theorem CT.1 forces the right side to zero at weak coupling. At a fixed physical area, however, \(A\) grows proportionally to \(a^{-2}\) as lattice spacing \(a\) decreases. The same right side can then be of order \(a^{-4}(1+\log\beta(a))/\beta(a)\). No decay follows from the displayed bound along a logarithmically growing inverse-coupling trajectory. This calculation identifies the quantifier that changes: a bound for every fixed lattice loop does not control a family of loops whose lattice area tends to infinity.

A second limitation comes from topology. On a periodic lattice choose one direction and a group element \(g\). Assign the same link value \(g\) to every link in that direction and identity to all other links. Every plaquette holonomy is identity, because a plaquette containing that direction has product \(gIg^{-1}I\). Yet a loop winding once around that periodic direction has holonomy \(g^L\), which need not be identity. Thus a zero plaquette action does not eliminate global holonomy sectors. The example is exact for nonabelian as well as abelian groups; only one nontrivial group element is being used. It is not a statement that this single configuration dominates the Gibbs law. It shows why local action control is not a global gauge-coordinate theorem.

There is a useful way to formulate the missing improvement from \(O((\log\beta)/\beta)\) to \(O(1/\beta)\) without differentiating an asymptotic expression illegally. Let \(F(\beta)=\log Z_\beta\). Since \(S\) is bounded on a finite lattice, differentiation under the integral gives

\[F'(\beta)=-\mathbb E_\beta S,
\qquad F''(\beta)=\operatorname{Var}_\beta(S)\ge0.\]

Therefore \(F\) is convex, and for \(0<\theta<1\),

\[\mathbb E_\beta S
\le\frac{F(\theta\beta)-F(\beta)}{(1-\theta)\beta}.\]

To check the sign, convexity gives \(F(\theta\beta)\ge F(\beta)+F'(\beta)(\theta\beta-\beta)\); rearranging is exactly the displayed inequality.

\textbf{Proposition CT.3 (free-energy increment criterion).} Suppose that for a family of finite models and all sufficiently large \(\beta\),

\[\log\frac{Z_{\theta\beta}}{Z_\beta}\le C|P|\]

with fixed \(\theta\in(0,1)\) and \(C<\infty\), independent of model size. Then

\[\frac{\mathbb E_\beta S}{|P|}\le\frac{C}{(1-\theta)\beta}.\]

In particular, it suffices to prove two-sided estimates

\[\left|\log Z_\beta+\alpha_\Lambda\log\beta\right|\le B|P|,
\qquad 0\le\alpha_\Lambda\le A|P|,\]

uniformly both at \(\beta\) and \(\theta\beta\). In that case one may take \(C=A\log(1/\theta)+2B\).

\textbf{Proof.} The first assertion is the convexity inequality divided by \(|P|\). For the second, subtract the two displayed free-energy expressions. The leading logarithmic terms give \(\alpha_\Lambda\log(1/\theta)\), and the two remainders contribute at most \(2B|P|\). The coefficient bound completes the estimate. \(\square\)

The proposition is a conditional improvement criterion, not an assertion that these two-sided estimates have been proved here for the required four-dimensional trajectory. It makes the intended next calculation explicit: matching the leading exponent in upper and lower partition estimates can remove a logarithmic loss, whereas a one-sided lower bound cannot do so automatically. Even the improved local estimate would still require renormalized-observable and physical-distance analysis before it could address a mass gap.

\hypertarget{continuum_attempts-countably-many-observables-a-genuine-probability-law-limit}{%
\subsection{Countably many observables: a genuine probability-law limit}\label{continuum_attempts-countably-many-observables-a-genuine-probability-law-limit}}

A lattice observable can be a smeared plaquette expression, a Wilson loop, a finite product of such quantities, or another prescribed gauge-invariant function. Before speaking about a field-valued limit, fix a countable collection \(\mathcal O_1,\mathcal O_2,\ldots\) and specify a real representative \(X_{j,k}\) at each cutoff \(j\). Complex observables can be replaced by their real and imaginary parts. The laws at different cutoffs may live on completely different original configuration spaces; only their observable vectors are compared.

Put \(X_j=(X_{j,1},X_{j,2},\ldots)\), considered as a random element of \(\mathbb R^{\mathbb N}\) with its product topology and Borel sigma-algebra. A convenient complete separable metric is

\[d(x,y)=\sum_{k=1}^{\infty}2^{-k}\min\{1,|x_k-y_k|\}.\]

Convergence in this metric is exactly coordinatewise convergence. The following theorem uses the standard probability compactness theorem that tight probability laws on a complete separable metric space have weakly convergent subsequences \cite{Prokhorov1956}. All its hypotheses are established explicitly below.

\textbf{Theorem CT.4 (observable-law compactness and moment passage).} Suppose, for every coordinate \(k\) and positive integer \(p\),

\[M_{k,p}:=\sup_j\mathbb E|X_{j,k}|^p<\infty.\]

Then a subsequence of the joint laws of \(X_j\) converges weakly on \(\mathbb R^{\mathbb N}\) to a probability law \(\nu\). Along that subsequence, every real-coefficient polynomial \(P\) in any finite set of coordinates satisfies

\[\mathbb E P(X_j)\longrightarrow\int P(x)\,\nu(dx).\]

The limit has finite moments of every order in every coordinate. Its finite-dimensional laws are consistent, and the limiting expectation is positive on nonnegative cylinder functions and on polynomial squares. The convergence statement also holds for complex-coefficient polynomials by applying its real version separately to the real and imaginary parts; the square positivity statement then uses \(|P|^2\).

\textbf{Proof.} Fix \(\eta>0\). Choose \(R_k>0\) so large that \(M_{k,2}/R_k^2\le\eta2^{-k}\). Markov's inequality and countable subadditivity give

\[\sup_j\mathbb P\{X_j\notin K_\eta\}
\le\sum_{k=1}^{\infty}\sup_j\mathbb P\{|X_{j,k}|>R_k\}
\le\eta,\]

where \(K_\eta=\prod_k[-R_k,R_k]\). This set is compact in the product metric. To see this without an uncountable-product theorem, take any sequence in \(K_\eta\). Successively extract convergent subsequences in the first coordinate, the second, and so on, and take the diagonal subsequence. Its coordinates converge to a point of \(K_\eta\). The metric tails are bounded by \(\sum_{k>m}2^{-k}\), so convergence of the first \(m\) coordinates proves convergence in the metric. Sequential compactness and compactness coincide in a metric space.

Thus the laws are tight. The probability compactness theorem supplies a weakly convergent subsequence and a probability law \(\nu\) on the same space. Coordinate projections are continuous, so their finite-dimensional laws converge to the corresponding marginals of \(\nu\). Those marginals are automatically consistent because they come from one law, rather than from unrelated choices of subsequences.

For fixed \(k,p\), the function \(\min\{|x_k|^p,R\}\) is bounded and continuous. Weak convergence gives convergence of its expectations. Taking the supremum over \(R\) by monotone convergence yields

\[\int|x_k|^p\,d\nu\le M_{k,p}.\]

Let \(P\) be a polynomial in coordinates \(k_1,\ldots,k_m\), of degree \(D\). There is a finite constant \(C_P\) such that

\[|P(x)|^2\le C_P\left(1+\sum_{i=1}^m|x_{k_i}|^{2D}\right).\]

For example, bound every monomial by a constant times \(1+\max_i|x_{k_i}|^D\), square the finite sum, and use \(\max_i a_i\le\sum_i a_i\). The assumed moments therefore give a uniform second-moment bound for \(P(X_j)\) and the same bound under \(\nu\).

Let \(\chi_R:\mathbb R\to[-R,R]\) be clipping to \([-R,R]\). The cylinder function \(\chi_R(P(x))\) is bounded and continuous. Its expectations converge. Moreover,

\[\mathbb E|P(X_j)-\chi_R(P(X_j))|
\le\mathbb E[|P(X_j)|\mathbf1_{\{|P(X_j)|>R\}}]
\le R^{-1}\mathbb E|P(X_j)|^2.\]

The same estimate holds for the limit law. First let \(j\) tend to infinity with \(R\) fixed, then let \(R\to\infty\). This proves polynomial expectation convergence. Positivity and normalization hold because \(\nu\) is a probability law; equivalently, expectation of every polynomial square is the limit of nonnegative expectations. \(\square\)

The hypothesis can be weakened: coordinate tightness suffices for the probability-law conclusion, and appropriate uniform integrability suffices for each chosen polynomial. Requiring all coordinate moments is convenient when the intended object is a full correlation hierarchy. It is not a hidden theorem that Wilson renormalizations satisfy these bounds.

This theorem makes no assertion of uniqueness of the limit. A sequence of deterministic one-coordinate laws alternating between zero and one has all moments bounded, but has two different subsequential limits. Establishing a universal continuum theory requires an additional uniqueness or comparison argument. It also makes no assertion that the coordinates form a distribution on spacetime. A countable list may satisfy no linearity relation in the test functions, no continuity estimate, and no transformation law. Those issues are addressed by additional structure rather than by calling the product law a quantum field.

There is a simple stability principle for algebraic relations. If a polynomial \(Q\) in finitely many coordinates vanishes almost surely at every cutoff, then Theorem CT.4 gives \(\int Q^2d\nu=0\), so \(Q=0\) almost surely in the limit. This preserves any exact finite polynomial relation included in the chosen coordinate representation. For approximate relations, the same conclusion follows if \(\mathbb E|Q(X_j)|^2\to0\). Thus linearity on a countable rational test-function space can be encoded and passed to the limit. Extending it continuously to all smooth test functions still needs a bound in a test-function norm.

\hypertarget{continuum_attempts-a-distribution-valued-compactness-criterion-with-explicit-tails}{%
\subsection{A distribution-valued compactness criterion with explicit tails}\label{continuum_attempts-a-distribution-valued-compactness-criterion-with-explicit-tails}}

On a compact spacetime torus, or on a fixed compact box with a chosen basis, a countable Fourier or eigenfunction expansion gives a concrete way to obtain that missing continuity. The argument is elementary Hilbert-scale compactness and does not use any Yang--Mills equation.

Let \(\lambda_k\ge1\) increase to infinity. For \(s\in\mathbb R\), define the real Hilbert sequence space

\[\mathcal H^{-s}=\left\{x:\|x\|_{-s}^2=
\sum_{k=1}^{\infty}\lambda_k^{-s}|x_k|^2<\infty\right\}.\]

If the coordinates correspond to Laplacian eigenfunctions and \(\lambda_k\) to one plus the eigenvalue, this is the usual negative-Sobolev convention. The exact power is a convention; the key fact is that increasing \(s\) weakens the norm. Write \(P_nx=(x_1,\ldots,x_n,0,\ldots)\).

\textbf{Theorem CT.5 (Hilbert-scale tightness).} Let \(\delta>0\) and let \(X_j\) be random elements of \(\mathcal H^{-s}\) satisfying

\[\sup_j\mathbb E\|X_j\|_{-s}^2\le C.\]

Then their laws are tight on \(\mathcal H^{-s-\delta}\). Every subsequential weak limit \(X\) on that space satisfies \(\mathbb E\|X\|_{-s}^2\le C\), interpreted as the increasing limit of finite partial sums. Furthermore,

\[\sup_j\mathbb E\|(I-P_n)X_j\|_{-s-\delta}^2
\le C\lambda_{n+1}^{-\delta}.\]

If all finite coordinate laws converge to those of a specified \(\mathcal H^{-s-\delta}\)-valued random element \(X\), and \(X\) has the same \(\mathcal H^{-s}\) second-moment bound, then \(X_j\) converges weakly to \(X\) in \(\mathcal H^{-s-\delta}\) without further subsequence extraction.

\textbf{Proof.} For every sequence \(x\),

\[\|(I-P_n)x\|_{-s-\delta}^2
=\sum_{k>n}\lambda_k^{-s-\delta}|x_k|^2
\le\lambda_{n+1}^{-\delta}\|x\|_{-s}^2.\]

Expectation gives the displayed uniform tail bound. The closed ball \(B_R=\{x:\|x\|_{-s}\le R\}\) is compact when considered as a subset of \(\mathcal H^{-s-\delta}\). Indeed, a sequence in \(B_R\) has a subsequence with all coordinates converging by diagonal extraction. The uniform tail estimate for deterministic elements of \(B_R\) shows that this subsequence is Cauchy in the weaker norm. Completeness gives a limit, and convergence of each finite partial sum implies by monotone convergence that its stronger norm is at most \(R\). Thus the weaker-space limit remains in \(B_R\), proving compactness.

Markov's inequality gives \(\mathbb P\{X_j\notin B_R\}\le C/R^2\), uniformly in \(j\). Compactness of \(B_R\) proves tightness. For a weak limit, finite partial stronger norms are continuous functions of finitely many coordinates; after clipping them to be bounded, pass to the limit and then increase the clipping level and the number of coordinates. This proves the stronger-norm moment bound for \(X\).

For the last assertion, let \(F\) be a bounded function that is Lipschitz with constant one for the weaker Hilbert norm. Cauchy--Schwarz and the tail estimate give

\[|\mathbb EF(X_j)-\mathbb EF(P_nX_j)|
\le C^{1/2}\lambda_{n+1}^{-\delta/2}.\]

The same estimate holds for \(X\). For fixed \(n\), convergence of the finite coordinate laws gives \(\mathbb EF(P_nX_j)\to\mathbb EF(P_nX)\). Consequently

\[\limsup_j|\mathbb EF(X_j)-\mathbb EF(X)|
\le2C^{1/2}\lambda_{n+1}^{-\delta/2}.\]

Let \(n\to\infty\). Bounded Lipschitz functions determine weak convergence on a separable metric space, which proves the claim. Alternatively, tightness and uniqueness of all finite coordinate limits show that every subsequential limit is the specified law. \(\square\)

The estimate is directly usable once a field representative is specified. Suppose \(\Phi_j(f)=\sum_kX_{j,k}f_k\) and \(f\) is smooth. The weighted Cauchy--Schwarz inequality gives

\[|\Phi_j(f)|\le\|X_j\|_{-s}\left(\sum_k\lambda_k^s|f_k|^2\right)^{1/2}.\]

Thus almost every \(X_j\) defines a continuous linear functional on the corresponding positive-Sobolev test space. For a Fourier basis, smooth test functions belong to every such space, so the law is genuinely distribution-valued. This is stronger than merely assigning arbitrary random variables to a list of test functions.

Uniform control on each compact box can be combined through a countable exhaustion, but compatibility must be retained. One can choose restrictions of a single lattice field to nested boxes, prove tightness in each local weaker space, and use a diagonal subsequence. Continuous restriction maps ensure that the limiting restrictions agree. An alternative is to define a local negative-Sobolev topology by a countable set of compactly supported cutoffs. Either construction requires uniform estimates for the same observable normalization across boxes; independently renormalizing each box would not automatically produce compatible restrictions.

For Yang--Mills, raw connection coordinates are gauge dependent. A use of this theorem must say whether \(X_j\) represents a gauge-fixed field with a proved gauge treatment or a specified gauge-invariant composite observable. The compactness theorem does not choose a gauge, remove singular composite products, or identify the resulting distribution as curvature. These are mathematical definitions and estimates to be supplied before invoking it.

\hypertarget{continuum_attempts-why-bounded-action-is-insufficient-for-renormalized-tightness}{%
\subsection{Why bounded action is insufficient for renormalized tightness}\label{continuum_attempts-why-bounded-action-is-insufficient-for-renormalized-tightness}}

The action estimate gives \(\mathbb E s(U_p)=O((\log\beta)/\beta)\), while a smooth classical expansion suggests \(s(U_p)\) is of order \(a^4|F|^2\) under a suitable field convention. Multiplication by \(a^{-4}\) therefore transforms the proved upper bound into \(O(a^{-4}(\log\beta)/\beta)\), which diverges along many natural trajectories. This calculation does not prove that a correctly renormalized energy density diverges; it proves that the present upper bound does not control it.

The distinction survives even in one dimension of probability. Let \(Y_n\) equal \(n\) with probability \(n^{-2}\) and zero otherwise. Then \(\mathbb E|Y_n|=n^{-1}\to0\), but \(\mathbb E|Y_n|^2=1\), and \(\mathbb E|Y_n|^3=n\to\infty\). An \(L^1\) concentration statement alone does not control higher correlations. If instead \(Z_n\) is an independent centered Gaussian with variance \(n^{-1}\), then \(Z_n\to0\) in every fixed finite moment, yet \(\sqrt n Z_n\) has a nondegenerate limit. Choosing a normalization changes the question and its required estimates.

For a spatial family there is an additional accumulation issue. Suppose \(N_a\) bounded local observables are averaged with weights \(c_{a,x}\). The exact variance is

\[\operatorname{Var}\left(\sum_xc_{a,x}Y_{a,x}\right)
=\sum_{x,y}c_{a,x}c_{a,y}\operatorname{Cov}(Y_{a,x},Y_{a,y}).\]

Bounding only diagonal variances does not estimate the double sum sharply. Perfectly correlated variables give variance of order \(N_a^2\) for their unnormalized sum; independent variables of equal variance give order \(N_a\). The two models have identical one-point laws. Consequently, a one-plaquette concentration theorem cannot determine the normalization or tightness of a smeared continuum observable. The missing covariance estimates are not a technical rephrasing of the proved action bound.

An exact example makes this point without asymptotics. Let \(B\) be a centered variance-one variable and put \(Y_x=B\) for every site in a finite set of size \(N\). Compare this with independent copies \(\widetilde Y_x\). Both have \(\mathbb EY_x^2=\mathbb E\widetilde Y_x^2=1\). But

\[\operatorname{Var}\left(N^{-1}\sum_xY_x\right)=1,
\qquad
\operatorname{Var}\left(N^{-1}\sum_x\widetilde Y_x\right)=N^{-1}.\]

Neither conclusion can be selected from the local second moment alone. In gauge theory, the action constrains neighboring links through plaquettes, but the long-distance covariance structure is precisely additional information to be proved.

\hypertarget{continuum_attempts-exact-gaussian-block-integration-and-what-an-effective-action-means}{%
\subsection{Exact Gaussian block integration and what an effective action means}\label{continuum_attempts-exact-gaussian-block-integration-and-what-an-effective-action-means}}

The simplest exact renormalization step is integration of some coordinates of a Gaussian measure. It illustrates both the mechanism and several mistakes that an interacting calculation must avoid.

Let \(x\in\mathbb R^m\), \(y\in\mathbb R^n\), and let

\[Q=\begin{pmatrix}A&B\\B^{\mathsf T}&C\end{pmatrix}\]

be a real symmetric positive-definite matrix. Define the quadratic action

\[S_0(x,y)=\frac12x^{\mathsf T}Ax+x^{\mathsf T}By+\frac12y^{\mathsf T}Cy.\]

The fine variables are \(y\) and the retained variables are \(x\). Write

\[S=A-BC^{-1}B^{\mathsf T}.\]

\textbf{Theorem CT.6 (exact elimination with normalization).} The matrix \(C\) and the Schur complement \(S\) are positive definite, and

\[\int_{\mathbb R^n}e^{-S_0(x,y)}dy
=(2\pi)^{n/2}(\det C)^{-1/2}e^{-\frac12x^{\mathsf T}Sx}.\]

The marginal law of \(x\) is centered Gaussian with covariance \(S^{-1}\). Conditional on \(x\), the variable \(y\) has mean \(-C^{-1}B^{\mathsf T}x\) and covariance \(C^{-1}\). The full partition function is

\[\int e^{-S_0(x,y)}dxdy=(2\pi)^{(m+n)/2}(\det Q)^{-1/2},\]

with \(\det Q=\det C\det S\).

\textbf{Proof.} Positivity of \(C\) follows by testing \(Q\) on vectors \((0,y)\). Completing the square gives

\[S_0(x,y)=\frac12x^{\mathsf T}Sx+
\frac12(y+C^{-1}B^{\mathsf T}x)^{\mathsf T}C(y+C^{-1}B^{\mathsf T}x).\]

For \(x\ne0\), substitute \(y=-C^{-1}B^{\mathsf T}x\) in the positive quadratic form of \(Q\) to obtain \(x^{\mathsf T}Sx>0\). Diagonalize \(C\) by an orthogonal matrix, translate \(y\), and use the one-dimensional Gaussian integral in each coordinate. This proves the integral and conditional law. Integrating once more in \(x\) gives the partition function with determinant \(\det C\det S\). Finally,

\[\begin{pmatrix}I&-BC^{-1}\\0&I\end{pmatrix}
Q
\begin{pmatrix}I&0\\-C^{-1}B^{\mathsf T}&I\end{pmatrix}
=\begin{pmatrix}S&0\\0&C\end{pmatrix}.\]

The triangular matrices have determinant one, proving the determinant identity independently. \(\square\)

If \(Q\ge m^2I\) as quadratic forms, then

\[x^{\mathsf T}Sx=\inf_y(x,y)^{\mathsf T}Q(x,y)
\ge m^2\|x\|^2.\]

So this particular coordinate marginal retains the same Euclidean quadratic lower bound. This is not automatically a physical mass-gap statement: the quadratic form concerns retained coordinate amplitudes, not a reconstructed time-translation Hamiltonian. A further rescaling of \(x\) also changes the numerical lower bound. If \(x=Rz\) for an invertible matrix \(R\), the new precision matrix is \(R^{\mathsf T}SR\) and its least eigenvalue depends on the singular values of \(R\).

Observables have to be transformed together with the action. For example, a linear fine observable \(O(x,y)=u^{\mathsf T}x+v^{\mathsf T}y\) has conditional expectation

\[\mathbb E[O\mid x]=(u-BC^{-1}v)^{\mathsf T}x.\]

For its square,

\[\mathbb E[O^2\mid x]=
\big((u-BC^{-1}v)^{\mathsf T}x\big)^2+v^{\mathsf T}C^{-1}v.\]

The second term is absent if one simply substitutes the minimizing fine field into the observable. Thus minimizing an action and integrating a probability measure are different operations even in the exactly solvable Gaussian case. Composite observables acquire additional terms under elimination.

For a numerical symbolic example, take \(m=n=1\) and

\[Q=\begin{pmatrix}2&-1\\-1&2\end{pmatrix}.\]

Then \(S=3/2\), the retained variance is \(2/3\), and \(y\) conditioned on \(x\) has mean \(x/2\) and variance \(1/2\). Hence \(\mathbb E[y^2\mid x]=x^2/4+1/2\). Dropping the final \(1/2\) would give an incorrect two-point function, although the minimizing fine field \(y=x/2\) is correct. This small calculation is an exact test that any proposed coarse-graining rule must pass.

\hypertarget{continuum_attempts-nonlinear-elimination-an-exact-identity-and-a-controlled-expansion}{%
\subsection{Nonlinear elimination: an exact identity and a controlled expansion}\label{continuum_attempts-nonlinear-elimination-an-exact-identity-and-a-controlled-expansion}}

Add a measurable interaction \(V(x,y)\) and suppose the relevant integrals are finite. Let \(\gamma_x\) be the conditional Gaussian law of \(y\) from Theorem CT.6. The exact effective interaction is

\[V_{\mathrm{eff}}(x)=-\log\mathbb E_{\gamma_x}e^{-V(x,Y)}.\]

Up to the explicit Gaussian determinant factor, the retained density is

\[\exp\left[-\frac12x^{\mathsf T}Sx-V_{\mathrm{eff}}(x)\right].\]

This identity follows by factoring the completed-square Gaussian integral; it does not replace \(V_{\mathrm{eff}}\) by a finite polynomial. Even when \(V\) is a polynomial, integrating \(e^{-V}\) generally generates infinitely many terms in an expansion of its logarithm.

An observable \(O\) transforms to

\[O_{\mathrm{eff}}(x)=
\frac{\mathbb E_{\gamma_x}[O(x,Y)e^{-V(x,Y)}]}
{\mathbb E_{\gamma_x}e^{-V(x,Y)}}.\]

The denominator must not be discarded. Fubini's theorem proves that the expectation of \(O_{\mathrm{eff}}\) in the exact retained measure equals the original expectation, whenever \(O\) is integrable. This equality is the mathematical content of preserving observables through one exact elimination step.

The following bounded-interaction lemma makes one perturbative estimate fully rigorous. Its boundedness hypothesis is deliberately explicit and is not asserted for an arbitrary polynomial field interaction.

\textbf{Proposition CT.7 (third-order control of an effective interaction).} Let \(W\) be a real random variable on a probability space with \(|W|\le M\). For real \(t\), define

\[F(t)=-\log\mathbb Ee^{-tW}.\]

Then

\[F(t)=t\mathbb EW-\frac{t^2}{2}\operatorname{Var}(W)+R_3(t),
\qquad |R_3(t)|\le\frac43M^3|t|^3.\]

The bound is uniform over the underlying probability law, including a conditional Gaussian law depending on retained variables.

\textbf{Proof.} Boundedness permits repeated differentiation under the expectation for every real \(t\). Let \(\mathbb E_t\) denote expectation under the tilted probability proportional to \(e^{-tW}\). Direct quotient differentiation gives

\[F'(t)=\mathbb E_tW,\qquad
F''(t)=-\operatorname{Var}_t(W),\qquad
F'''(t)=\mathbb E_t[(W-\mathbb E_tW)^3].\]

Since \(|W-\mathbb E_tW|\le2M\), one has \(|F'''(t)|\le8M^3\). Taylor's theorem with integral remainder gives \(|R_3(t)|\le8M^3|t|^3/6=\frac43M^3|t|^3\). The first two derivatives at zero give the stated coefficients. \(\square\)

For an unbounded \(W\), the derivative identities require exponential integrability and uniform control of the tilted centered third moment. Merely knowing all ordinary moments is insufficient to justify a nonzero radius of convergence of its moment-generating function. The lognormal distribution is a familiar source of this distinction, but a simpler series argument also suffices: tails can decay faster than every inverse power while slower than every exponential. In that situation all polynomial moments exist, whereas \(\mathbb Ee^{t|W|}=\infty\) for every \(t>0\).

Even the bounded lemma needs a size estimate under repeated elimination. If \(W\) is a sum of \(V\) local terms each bounded by one, the crude choice \(M=V\) gives a remainder of order \(V^3|t|^3\). It does not yield an extensive bound proportional to \(V\). A cluster or connected-cumulant argument would have to exploit locality to replace that cubic volume growth. This is precisely where a formal cumulant series can conceal the missing uniform estimate. The exact identity remains valid, but its approximation is not automatically useful.

\hypertarget{continuum_attempts-accumulation-over-scales-and-why-errors-need-a-summable-budget}{%
\subsection{Accumulation over scales and why errors need a summable budget}\label{continuum_attempts-accumulation-over-scales-and-why-errors-need-a-summable-budget}}

A continuum construction typically requires a number of scale transformations tending to infinity. An estimate that is small at each individual scale can still fail after their errors accumulate. The following deterministic lemma describes the bookkeeping without assuming any field theory.

\textbf{Theorem CT.8 (stability of approximate iterations).} Let \((X,d)\) be a complete metric space, and let \(R_k:X\to X\) be maps with Lipschitz constants \(L_k\ge0\). Suppose exact and approximate sequences satisfy

\[x_{k+1}=R_kx_k,\qquad
d(y_{k+1},R_ky_k)\le\varepsilon_k.\]

Then for \(n>0\),

\[d(x_n,y_n)
\le\left(\prod_{j=0}^{n-1}L_j\right)d(x_0,y_0)
+\sum_{k=0}^{n-1}\varepsilon_k\prod_{j=k+1}^{n-1}L_j.\]

An empty product is one. If every \(L_j\le1\), the accumulated error is at most \(d(x_0,y_0)+\sum_{k<n}\varepsilon_k\). If \(L_j\le L<1\), it is at most \(L^nd(x_0,y_0)+\sum_{k<n}L^{n-1-k}\varepsilon_k\).

\textbf{Proof.} The triangle inequality and Lipschitz bound give

\[d(x_{k+1},y_{k+1})\le L_kd(x_k,y_k)+\varepsilon_k.\]

Substitute this inequality successively into itself. Induction on \(n\) gives the stated product-and-sum expression; the two special cases follow by bounding each product. \(\square\)

For cutoff constructions, an additional Cauchy comparison is often needed. Suppose a common observable functional \(F_N\) is obtained after \(N\) scales and

\[|F_{N+1}-F_N|\le\varepsilon_N,\qquad
\sum_{N=0}^{\infty}\varepsilon_N<\infty.\]

Then \(F_N\) converges and

\[|F_\infty-F_N|\le\sum_{k=N}^{\infty}\varepsilon_k.\]

This follows by telescoping finite differences, then using completeness of the scalar field. The same proof works in a complete Banach space. It supplies an actual convergence mechanism once the per-scale estimates are proved in a norm controlling the observables of interest.

The summability condition cannot be replaced by \(\varepsilon_N\to0\). Take \(F_N=\sum_{k=1}^Nk^{-1}\). Its increments tend to zero, but the sequence diverges. Nor does a bounded total error automatically preserve a positive lower bound: if a target squared norm starts at \(c>0\), a perturbation estimate of size \(2c\) gives no nontriviality information. To preserve positivity one needs an error budget smaller than the known positive margin.

There is a parallel normalization issue. If each coarse-graining step rescales an observable by \(z_k>0\), then after \(N\) steps the factor is \(\prod_{k<N}z_k\). Uniform closeness of each \(z_k\) to one does not control that product. Its logarithm is \(\sum_{k<N}\log z_k\). For example, \(z_k=1+1/(k+1)\) has product \(N+1\), while \(z_k=1-1/(k+2)\) has product \(1/(N+1)\). Both factors approach one. A construction must track these products alongside action errors, because they determine whether observable norms survive, diverge, or vanish.

These elementary statements do not claim that a Yang--Mills renormalization map is a contraction in a suitable norm. Such a norm, the map, and the constants must be defined and estimated. They identify the exact place where a phrase such as ``the small errors disappear in the limit'' needs a proof.

\hypertarget{continuum_attempts-exact-classical-scaling-and-the-nonabelian-term}{%
\subsection{Exact classical scaling and the nonabelian term}\label{continuum_attempts-exact-classical-scaling-and-the-nonabelian-term}}

Classical power counting can be made into an exact statement for smooth fields. Its exactness in that setting does not turn it into an estimate for random distribution-valued fields. The following calculation identifies the scaling distinction without using a perturbative quantum series.

Represent the Lie algebra of \(G\) by anti-Hermitian matrices, and let \(A=(A_1,\ldots,A_d)\) be a smooth compactly supported connection on \(\mathbb R^d\). Define

\[F_{\mu\nu}(A)=\partial_\mu A_\nu-\partial_\nu A_\mu+[A_\mu,A_\nu],
\qquad
\mathcal E(A)=\frac12\sum_{\mu<\nu}\int_{\mathbb R^d}\|F_{\mu\nu}(A)(x)\|_*^2\,dx.\]

Every integral exists because the coefficients are smooth and compactly supported. For \(\lambda>0\), set \(A^{(\lambda)}_\mu(x)=\lambda A_\mu(\lambda x)\).

\textbf{Proposition CT.9 (classical critical scaling and a commutator plaquette).} The curvature and energy satisfy

\[F_{\mu\nu}(A^{(\lambda)})(x)=\lambda^2F_{\mu\nu}(A)(\lambda x),
\qquad
\mathcal E(A^{(\lambda)})=\lambda^{4-d}\mathcal E(A).\]

In particular, the classical energy is unchanged by this scaling in dimension four. For fixed anti-Hermitian matrices \(X,Y\), the unitary commutator plaquette

\[U(a)=e^{aX}e^{aY}e^{-aX}e^{-aY}\]

satisfies, as \(a\to0\),

\[U(a)=I+a^2[X,Y]+O(a^3),
\qquad
1-\frac1N\operatorname{Re}\operatorname{Tr}U(a)
=\frac{a^4}{2}\|[X,Y]\|_*^2+O(a^5).\]

The matrix remainders are uniform for \((X,Y)\) in any fixed bounded set. If \(X,Y\) commute, \(U(a)=I\) exactly for every \(a\).

\textbf{Proof.} The ordinary chain rule gives \(\partial_\mu A^{(\lambda)}_\nu(x)=\lambda^2(\partial_\mu A_\nu)(\lambda x)\). Bilinearity of the matrix commutator gives \([A^{(\lambda)}_\mu,A^{(\lambda)}_\nu](x)=\lambda^2[A_\mu,A_\nu](\lambda x)\). This proves the curvature identity. Substitution \(y=\lambda x\) in the energy integral contributes the Jacobian \(\lambda^{-d}\), while the squared curvature contributes \(\lambda^4\). The scaling formula follows.

For the second assertion, expand each matrix exponential to second order:

\[e^{\pm aX}=I\pm aX+\frac{a^2X^2}{2}+O(a^3),
\qquad
e^{\pm aY}=I\pm aY+\frac{a^2Y^2}{2}+O(a^3).\]

Multiplication cancels the first-order terms and the pure second-order squares, leaving \(a^2(XY-YX)\). In operator norm, the series remainder estimate \(\sum_{k\ge3}|a|^k\|X\|_{\mathrm{op}}^k/k!\le |a|^3\|X\|_{\mathrm{op}}^3e^{|a|\|X\|_{\mathrm{op}}}/6\) proves the asserted uniformity on bounded sets; finite products preserve it. The normalized Hilbert--Schmidt norm is at most the operator norm, so the same remainder order holds in \(\|\cdot\|_*\). Write \(U(a)-I=a^2[X,Y]+a^3R(a)\), with bounded \(R(a)\). Squaring its normalized Hilbert--Schmidt norm gives

\[\|U(a)-I\|_*^2=a^4\|[X,Y]\|_*^2+O(a^5).\]

Since \(U(a)\) is unitary, the exact trace-defect identity at the beginning of this section proves the remaining formula. If \(X,Y\) commute, their exponentials commute, and the four factors cancel directly. \(\square\)

Writing \(D_{\mu\nu}A=\partial_\mu A_\nu-\partial_\nu A_\mu\), the integrand expands exactly as

\[\|F_{\mu\nu}(A)\|_*^2
=\|D_{\mu\nu}A\|_*^2
+2\operatorname{Re}\langle D_{\mu\nu}A,[A_\mu,A_\nu]\rangle_*
+\|[A_\mu,A_\nu]\|_*^2.\]

Under the displayed scaling, all three integrated terms acquire the same factor \(\lambda^{4-d}\). In four dimensions this elementary calculation gives no strictly positive scaling power that would make the nonlinear terms disappear. It explains the word ``critical'' in this particular classical sense. It does not prove a beta-function formula, quantum existence, confinement or a spectral gap.

The commutator calculation also shows concretely what is lost by replacing every nonabelian plaquette with an abelian one: constant noncommuting link generators already produce a leading curvature term. Conversely, it shows why a smooth-field expansion is not yet a uniform probabilistic estimate. Its remainder constant depends on bounds for the generators. A random lattice field whose coordinate size or higher moments grow as the cutoff is removed need not satisfy the same uniform bound. Controlling the exceptional configurations and the normalization of composite observables is necessary before integrating the expansion and passing to a continuum limit. The exact finite-dimensional algebra remains valid, but the limiting interchange requires additional estimates.

\hypertarget{continuum_attempts-removing-the-two-regulators-in-a-justified-order}{%
\subsection{Removing the two regulators in a justified order}\label{continuum_attempts-removing-the-two-regulators-in-a-justified-order}}

A lattice calculation has both an ultraviolet cutoff and a physical-volume cutoff. Let \(n\) index decreasing lattice spacings \(a_n\downarrow0\), and let \(R\) index increasing physical boxes. The number of lattice sites in a physical box depends on both \(n\) and \(R\); treating that number as a single parameter can obscure the order of limits. For a fixed observable supported in a fixed compact physical region, denote its normalized expectation by \(f_{n,R}\). The same notation can represent a finite correlation, provided its normalization and support have already been fixed.

\textbf{Theorem CT.10 (uniform removal of the volume regulator).} Suppose that for each \(n\) the limit \(f_{n,\infty}=\lim_{R\to\infty}f_{n,R}\) exists. Suppose there is a deterministic function \(\epsilon(R)\to0\) such that

\[\sup_n|f_{n,R}-f_{n,\infty}|\le\epsilon(R).\]

Suppose also that for every fixed \(R\), the limit \(g_R=\lim_{n\to\infty}f_{n,R}\) exists. Then \(g_R\) has a limit \(g_\infty\) as \(R\to\infty\), \(f_{n,\infty}\to g_\infty\), and for every joint sequence \(n_j\to\infty\), \(R_j\to\infty\),

\[f_{n_j,R_j}\longrightarrow g_\infty.\]

The conclusion and proof hold for values in any complete normed space, with absolute values replaced by norms.

\textbf{Proof.} For fixed \(R,S\) and every \(n\),

\[|f_{n,R}-f_{n,S}|
\le|f_{n,R}-f_{n,\infty}|+|f_{n,S}-f_{n,\infty}|
\le\epsilon(R)+\epsilon(S).\]

Pass to the \(n\to\infty\) limit to obtain \(|g_R-g_S|\le\epsilon(R)+\epsilon(S)\). Thus \(g_R\) is Cauchy and has a limit \(g_\infty\). Sending \(S\to\infty\) gives \(|g_R-g_\infty|\le\epsilon(R)\).

For each fixed \(R\),

\[|f_{n,\infty}-g_\infty|
\le|f_{n,\infty}-f_{n,R}|+|f_{n,R}-g_R|+|g_R-g_\infty|
\le2\epsilon(R)+|f_{n,R}-g_R|.\]

Take the upper limit in \(n\) and then let \(R\to\infty\). This proves \(f_{n,\infty}\to g_\infty\). Finally,

\[|f_{n_j,R_j}-g_\infty|
\le\epsilon(R_j)+|f_{n_j,\infty}-g_\infty|\longrightarrow0.\]

Every step used only triangle inequalities and completeness, so the normed-space statement follows as well. \(\square\)

Without the uniform estimate, the two iterated limits need not agree. Define \(f_{n,R}=\mathbf1_{\{R\ge n\}}\) for positive integer indices. At fixed \(n\), the large-volume limit is one. At fixed \(R\), the ultraviolet-index limit is zero. Consequently

\[\lim_{n\to\infty}\lim_{R\to\infty}f_{n,R}=1,
\qquad
\lim_{R\to\infty}\lim_{n\to\infty}f_{n,R}=0.\]

The diagonal \(R=n\) has value one, whereas \(R=\lfloor n/2\rfloor\) has value zero. Every scalar is bounded by one; the failure comes entirely from lack of uniformity. Boundedness, therefore, is not enough to exchange the regulators.

The theorem can be used without an explicit infinite-volume limit at the outset if one proves the equivalent Cauchy estimate

\[\sup_n\sup_{S\ge R}|f_{n,S}-f_{n,R}|\le\epsilon(R).\]

Completeness gives \(f_{n,\infty}\) for every \(n\), and passing \(S\to\infty\) supplies the displayed hypothesis of CT.10. This formulation often matches a boundary-influence estimate more directly. The observable must remain supported a controlled physical distance from the boundary; a support moving toward the boundary is a different family.

In a lattice strong-coupling argument, influence might decay exponentially in a graph distance measured in lattice steps. A continuum use of CT.10 requires translating that estimate to the actual observable normalization and physical geometry uniformly in \(n\). The strong-coupling constants can cease to apply along the intended bare-coupling trajectory. Even when they do apply, they can force the triviality mechanism proved elsewhere in the manuscript. The regulator theorem itself neither repairs that loss of applicability nor identifies a nontrivial limit; it ensures that a valid uniform estimate is used with the correct order of quantifiers.

For a correlation hierarchy, CT.10 should be applied to each member of a countable family whose moment and continuity bounds are sufficient to recover the desired distributions. One can allow the function \(\epsilon\) to depend on that fixed member. A single rate for all unbounded composite observables is not required by this elementary interchange theorem. A common physical decay rate across a total family of centered states is a different requirement, arising in the spectral-gap theorem. Distinguishing these constants avoids demanding an unnecessarily strong compactness estimate while inadvertently weakening the essential mass-gap estimate.

\hypertarget{continuum_attempts-a-constructive-route-stated-in-verifiable-stages}{%
\subsection{A constructive route stated in verifiable stages}\label{continuum_attempts-a-constructive-route-stated-in-verifiable-stages}}

The actual Wilson weak-coupling estimate and the abstract compactness criteria can be connected only through further analytic work. A precise route is to choose a sequence of physical lattices, specify renormalized gauge-invariant representatives for a countable dense family of test observables, and prove uniform moment estimates in a local distribution topology. Theorems CT.4 and CT.5 then supply subsequential probability laws and convergence of suitable correlation functions. Their proofs show exactly which constants must be independent of cutoff and which may depend on a fixed test observable or compact region.

The next requirements concern the limiting law, rather than its mere existence as a probability measure. Algebraic consistency, reflection positivity, Euclidean covariance and the regularity assumptions of a reconstruction theorem must be checked for the same limiting hierarchy. Exact symmetries can pass to a limit when the corresponding transformed observables converge, but a lattice symmetry group is smaller than the full Euclidean group. Restoring the missing continuous transformations is a separate assertion. The spectral and reflection sections give the necessary positive-form and semigroup arguments once their hypotheses are available.

Nontriviality requires at least one centered positive-time observable class with strictly positive finite reflected norm. A nonzero equal-point variance does not provide that norm. A practical strategy would be to prove a lower bound for a specific smeared reflected two-point function and an upper bound that ensures finiteness, with both bounds stable under the cutoff removal. The stability lemmas above explain how a known positive margin can survive controlled approximation errors. They do not manufacture the initial lower bound.

Finally, a physical mass gap requires decay on a total class of reconstructed centered states, with a positive rate in physical time and constants compatible with the continuum limit. The weak-coupling action estimate does not establish that decay. It also does not exclude it. The conclusion of the present calculation is narrower and useful: the actual Wilson family has volume-uniform local concentration at large \(\beta\), while the remaining continuum task is explicitly a problem about renormalized multipoint control, nonzero reflected norms and physical-distance decay.

There are therefore proved additions to the investigation: an unconditional weak-coupling local-action estimate, exact fixed-loop consequences, probability-law and distribution-valued compactness criteria, exact Gaussian coarse-graining with transformed observables, a bounded nonlinear remainder estimate, and a scale-error convergence criterion. The hypotheses needed to apply the latter tools to four-dimensional continuum Yang--Mills are left visible. None of them is renamed as an established Yang--Mills estimate merely because it would complete the intended argument.

\hypertarget{gap_transfer-gap-transfer-continuum-scaling-and-precise-obstructions}{%
\section{Gap transfer, continuum scaling, and precise obstructions}\label{gap_transfer-gap-transfer-continuum-scaling-and-precise-obstructions}}

The distinction is between a regulator-dependent gap and a positive, finite mass in a nontrivial continuum quantum field theory. A lattice estimate does not supply the continuum theory whose spectrum one wishes to study. Once that theory and sufficiently strong uniform correlation estimates exist, the final spectral deduction is short.

\hypertarget{gap_transfer-exponential-decay-on-a-total-set-implies-a-hamiltonian-gap}{%
\subsection{Exponential decay on a total set implies a Hamiltonian gap}\label{gap_transfer-exponential-decay-on-a-total-set-implies-a-hamiltonian-gap}}

Let \(\mathcal H\) be a complex Hilbert space, \(H\ge0\) self-adjoint, and \(\Omega\in\ker H\) a unit vector. A set \(\mathcal D\subset\Omega^\perp\) is \textbf{total} if its complex linear span is dense in \(\Omega^\perp\).

\textbf{Theorem GT.1 (positive spectral-measure criterion).} Suppose \(m>0\) is fixed and \(\mathcal D\subset\Omega^\perp\) is total. For every \(v\in\mathcal D\), suppose there are finite constants \(C_v\) and \(t_v\ge0\) such that

\[0\le\langle v,e^{-tH}v\rangle\le C_v e^{-mt}\qquad(t\ge t_v).\]

Then

\[\ker H=\mathbb C\Omega,\qquad \sigma(H|_{\Omega^\perp})\subset[m,\infty).\]

If \(\Omega^\perp\ne\{0\}\), the bottom of the excitation spectrum is finite as well as positive. Conversely, these kernel and spectral inclusions imply

\[0\le\langle v,e^{-tH}v\rangle\le e^{-mt}\|v\|^2\]

for every \(v\in\Omega^\perp\) and every \(t\ge0\).

\textbf{Proof.} Let \(E_H\) denote the spectral resolution of \(H\). For \(v\in\mathcal D\), define the finite positive measure \(\mu_v(S)=\langle v,E_H(S)v\rangle\). The spectral theorem gives

\[\langle v,e^{-tH}v\rangle=\int_{[0,\infty)}e^{-t\lambda}\,d\mu_v(\lambda).\]

For any \(b\in[0,m)\), positivity gives

\[e^{-tb}\mu_v([0,b])\le\int e^{-t\lambda}\,d\mu_v(\lambda)\le C_v e^{-mt}.\]

Letting \(t\to\infty\) yields \(\mu_v([0,b])=0\), hence \(E_H([0,b])v=0\). Taking a countable sequence \(b\uparrow m\) gives \(E_H([0,m))v=0\). This bounded projection therefore vanishes on the closure of the span of \(\mathcal D\), namely \(\Omega^\perp\). Since it acts as the identity on \(\Omega\), its range is exactly \(\mathbb C\Omega\). This proves the gap and vacuum uniqueness.

For finiteness, take \(0\ne w\in\Omega^\perp\). Its spectral measure has mass \(\|w\|^2>0\). Some finite interval \([0,n]\) has positive measure, because their increasing union is \([0,\infty)\). Thus the excitation spectrum contains a finite value. Conversely, when the spectral measure is supported in \([m,\infty)\), integration of \(e^{-t\lambda}\le e^{-tm}\) proves the asserted bound. \(\square\)

The constants \(C_v\) need not be uniform over the total set. The rate \(m\) is common. The same proof works if the bound holds along an unbounded sequence of times for each \(v\).

Density is essential. On

\[\mathcal H=\mathbb C\Omega\oplus\mathbb C e\oplus L^2((0,1),d\lambda),\]

let \(H\) be zero on \(\Omega\), one on \(e\), and multiplication by \(\lambda\) on the last summand. The probe \(e\) has correlation \(e^{-t}\), yet the full Hamiltonian is gapless. Thus one field, one Wilson loop, or finitely many probes cannot establish a full gap without a totality argument.

\hypertarget{gap_transfer-a-precise-lattice-to-continuum-gap-transfer-theorem}{%
\subsection{A precise lattice-to-continuum gap-transfer theorem}\label{gap_transfer-a-precise-lattice-to-continuum-gap-transfer-theorem}}

Let \(a_j\downarrow0\). Suppose cutoff \(j\) has a Hilbert space \(\mathcal H_j\), a unit vacuum \(\Omega_j\), and a positive self-adjoint contraction \(T_j\) with \(T_j\Omega_j=\Omega_j\). The transfer matrix advances \textbf{physical Euclidean time by \(a_j\)}. Put \(Q_j=I-|\Omega_j\rangle\langle\Omega_j|\).

If a transfer matrix is only self-adjoint, its negative spectrum must be addressed first; using its square changes the physical time step to \(2a_j\). The positivity assumption below is explicit.

\textbf{Theorem GT.2 (gap transfer under an established semigroup limit).} Assume:

\begin{enumerate}
\def\labelenumi{\arabic{enumi}.}
\tightlist
\item
  A continuum triple \((\mathcal H,H,\Omega)\) has already been constructed, with \(H\ge0\) self-adjoint, \(\|\Omega\|=1\) and \(H\Omega=0\).
\item
  There is a total set \(\mathcal D\subset\Omega^\perp\) and, for each \(v\in\mathcal D\), centered states \(v_j\in\Omega_j^\perp\) satisfying \(\|v_j\|^2\to\|v\|^2\).
\item
  For every \(v\in\mathcal D\) and fixed \(t>0\),
\end{enumerate}

\[\langle v_j,T_j^{\lfloor t/a_j\rfloor}v_j\rangle\longrightarrow\langle v,e^{-tH}v\rangle.\]

\begin{enumerate}
\def\labelenumi{\arabic{enumi}.}
\setcounter{enumi}{3}
\tightlist
\item
  There are \(m_j\ge0\) such that
\end{enumerate}

\[\|T_jQ_j\|\le e^{-a_jm_j},\qquad\liminf_jm_j\ge m>0.\]

Then \(\ker H=\mathbb C\Omega\) and \(\sigma(H|_{\Omega^\perp})\subset[m,\infty)\). If some \(v\in\mathcal D\) is nonzero, the excitation threshold is finite.

\textbf{Proof.} Since \(T_j\) is self-adjoint and fixes \(\Omega_j\), it commutes with \(Q_j\). Its positivity and the spectral theorem imply

\[0\le\langle v_j,T_j^nv_j\rangle\le e^{-a_jm_jn}\|v_j\|^2\qquad(n\ge0).\]

For \(0<m'<m\), eventually \(m_j\ge m'\). Since \(a_j\lfloor t/a_j\rfloor\to t\), passing to the assumed limit gives

\[0\le\langle v,e^{-tH}v\rangle\le e^{-m't}\|v\|^2.\]

Let \(m'\uparrow m\) and apply Theorem GT.1. \(\square\)

The proof does not require embeddings of all cutoff spaces into one space. Nor does it define \(-\log T_j\) on a possible kernel: it uses positive contractions directly.

If \(q_j=\|T_jQ_j\|\in(0,1)\), the dimensionless and physical gaps are

\[m_{\mathrm{lat},j}=-\log q_j,\qquad m_{\mathrm{phys},j}=\frac{m_{\mathrm{lat},j}}{a_j}.\]

The relevant premise is therefore \(\liminf m_{\mathrm{lat},j}/a_j>0\), not positivity at each regulator. Physical spatial volume must also tend to infinity, or the estimate must be uniform enough to take that limit separately.

\textbf{Correlation-only version.} Assumption 4 may be replaced by: for each \(v\), there are \(C_v,t_v<\infty\) independent of \(j\) such that

\[0\le\langle v_j,T_j^nv_j\rangle\le C_v e^{-ma_jn}\qquad(a_jn\ge t_v).\]

For each fixed \(t>t_v\), assumption 3 then supplies the hypothesis of Theorem GT.1. Cutoff-dependent prefactors or physical starting times must be controlled before passing to the limit.

\hypertarget{gap_transfer-what-reconstruction-still-requires}{%
\subsection{What reconstruction still requires}\label{gap_transfer-what-reconstruction-still-requires}}

Theorem GT.2 is not a construction of the continuum triple it assumes. For a Euclidean Yang--Mills programme, one must separately:

\begin{itemize}
\tightlist
\item
  Specify the renormalized smeared gauge-invariant observables corresponding to local curvature polynomials and their covariant derivatives.
\item
  Obtain the full hierarchy of limiting Schwinger distributions with the regularity and growth conditions of an applicable reconstruction theorem.
\item
  Establish reflection positivity, restored Euclidean symmetry and the remaining axioms. Closed positivity inequalities pass to limits if all the necessary entries converge; that fact does not establish convergence or the other axioms.
\item
  Establish a total family of limiting observable states and a nonzero centered state.
\item
  Identify the resulting theory as four-dimensional Yang--Mills, including the required short-distance properties.
\end{itemize}

The applicable Osterwalder--Schrader reconstruction hypotheses must hold for the limiting correlations \cite{OS1,OS2}. The required Yang--Mills construction must identify its local observable theory, nontrivial physical states and time-translation dynamics. A finite lattice integral or a Gaussian comparison model supplies only its own stated conclusions.

\hypertarget{gap_transfer-fixed-positive-lattice-decay-can-force-a-trivial-limit}{%
\subsection{Fixed positive lattice decay can force a trivial limit}\label{gap_transfer-fixed-positive-lattice-decay-can-force-a-trivial-limit}}

If \(m_{\mathrm{lat},j}\ge\kappa>0\), then \(m_{\mathrm{phys},j}\ge\kappa/a_j\to\infty\). A fixed lattice gap does not shrink to zero under this scaling.

\textbf{Theorem GT.3 (fixed lattice-rate obstruction).} Retain the continuum triple, total set and correlation convergence from Theorem GT.2. Suppose that for each \(v\in\mathcal D\) there are \(K_j(v)>0\) such that

\[0\le\langle v_j,T_j^nv_j\rangle\le K_j(v)e^{-\kappa n}\qquad(n\ge0),\]

where \(\kappa>0\) is fixed, and

\[\limsup_j a_j\log K_j(v)<\infty.\]

Then every \(v\in\mathcal D\) is zero and \(\mathcal H=\mathbb C\Omega\).

\textbf{Proof.} Choose \(A<\infty\) such that \(a_j\log K_j(v)\le A\) eventually. Choose \(t>\max\{0,A/\kappa\}\). Since \(\lfloor t/a_j\rfloor\ge t/a_j-1\),

\[K_j(v)e^{-\kappa\lfloor t/a_j\rfloor}
\le e^\kappa\exp\left(\frac{A-\kappa t}{a_j}\right)\longrightarrow0.\]

Thus \(\langle v,e^{-tH}v\rangle=0\). Since \(e^{-t\lambda}>0\) at every finite \(\lambda\ge0\), its integral against the positive spectral measure of a nonzero vector cannot vanish. Hence \(v=0\). Totality proves the claim. \(\square\)

Polynomially growing prefactors satisfy the hypothesis. This theorem does not assert existence of a fixed-coupling continuum limit. It proves that \textbf{if} the specified correlation and semigroup limit exists, a fixed positive lattice decay rate with the stated amplitude control precludes a nontrivial reconstructed theory.

\hypertarget{gap_transfer-an-explicit-consequence-for-the-actual-wilson-strong-coupling-regime}{%
\subsection{An explicit consequence for the actual Wilson strong-coupling regime}\label{gap_transfer-an-explicit-consequence-for-the-actual-wilson-strong-coupling-regime}}

Use the four-dimensional hypercubic Wilson model with normalized plaquette density

\[d\mu_{\Lambda,\beta}(U)=Z_{\Lambda,\beta}^{-1}
\exp\left(\beta\sum_p\frac{\operatorname{ReTr}\rho(U_p)}N\right)\prod_{e\in\Lambda}dU_e,\]

with fixed exterior links when applicable. Theorem SC.1 establishes, for real bounded local observables supported on link sets \(A,B\),

\[|\operatorname{Cov}(F,G)|
\le\frac{\min\{|A|,|B|\}\|F\|_\infty\|G\|_\infty}{1-q}\,q^{d(A,B)},
\qquad q=18\beta<1.\]

The distance is in the graph where links sharing a plaquette are adjacent. The bound holds uniformly in finite volume and for the unique infinite-volume Gibbs measure. The normalization is the same as in Theorem SC.1.

\textbf{Proposition GT.4 (separated covariance vanishes faster than every power).} Fix \(0<q_0<1\). Suppose \(0<q_a=18\beta_a\le q_0\). Let the link-center supports of \(F_a,G_a\) have Euclidean separation at least \(r>0\), independent of lattice spacing \(a\). Suppose fixed \(C<\infty,p\ge0\) satisfy

\[\min\{|A_a|,|B_a|\}\|F_a\|_\infty\|G_a\|_\infty\le Ca^{-p}.\]

With \(\kappa=-\log q_0>0\),

\[|\operatorname{Cov}(F_a,G_a)|
\le\frac C{1-q_0}a^{-p}e^{-\kappa r/a}=o(a^N)\]

for every fixed \(N\ge0\).

\textbf{Proof.} Centers of two edges on one plaquette are at Euclidean distance at most \(a\): opposite edges have distance \(a\) and adjacent perpendicular edges have distance \(a/\sqrt2\). Thus any path of \(k\) interaction-graph steps moves a center at most \(ka\), so \(d(A_a,B_a)\ge r/a\). Apply the covariance bound with \(q_a\le q_0\) and \(1-q_a\ge1-q_0\). Finally

\[\log(a^{-p-N}e^{-\kappa r/a})=-(p+N)\log a-\kappa r/a\longrightarrow-\infty.\]

This proves the claimed rate. \(\square\)

One can instead define support using full edges, with an \(O(a)\) adjustment, or replace a fixed separation by a slightly smaller fixed positive separation for small \(a\).

\textbf{Corollary GT.5 (conditional triviality of an OS limit).} Use the reflection-invariant infinite-volume Gibbs measures supplied by Theorem SC.1, or reflection-symmetric finite-volume approximants. Suppose a reflection-invariant, reflection-positive continuum functional \(S\) is obtained as a limit of these Wilson expectations. Let \(\mathcal A_+\) be a real family of Euclidean observables, each supported in a compact set with time coordinate at least \(\varepsilon_F>0\). Suppose each \(F\in\mathcal A_+\) has representatives \(F_a\) such that:

\begin{enumerate}
\def\labelenumi{\arabic{enumi}.}
\tightlist
\item
  Its lattice support lies at times at least \(\varepsilon_F/2\) for small \(a\).
\item
  The polynomial support/norm bound of Proposition GT.4 holds for \(F_a\) and its time reflection.
\item
  Its one-point and reflected two-point functions converge to those of \(S\).
\item
  The complex span of the centered OS classes \([F-S(F)]\) is dense in the vacuum orthocomplement of the reconstructed Hilbert space.
\end{enumerate}

Then that Hilbert space is one-dimensional.

\textbf{Proof.} Let \(\theta\) reflect Euclidean time. The supports of \(F_a\) and \(\theta F_a\) are separated by at least \(\varepsilon_F\). Proposition GT.4 gives \(\operatorname{Cov}(\theta F_a,F_a)\to0\). Reflection symmetry and the assumed limits imply

\[\|[F-S(F)]\|_{\mathrm{OS}}^2
=S((\theta F)F)-S(F)^2=0.\]

All centered classes in the specified total family are therefore null. Their span and closure vanish. For complex observables apply the argument to real and imaginary parts, or use the corresponding finite linear combination of real covariance bounds in the sesquilinear reflected form. \(\square\)

The total family must include the necessary polynomial products, with correct smearing and renormalization. A calculation for one plaquette field alone is insufficient. This statement also allows a Euclidean random distribution with only \textbf{contact} correlations: white noise, for example, can have nonzero coincident covariance while all its centered strictly positive-time classes are null in OS reconstruction. The conclusion is about the physical reconstructed Hilbert space, not merely the existence of a random distribution.

This exact consequence does not extend strong-coupling estimates to the weak-bare-coupling regime or establish the required renormalization trajectory. It identifies why fixed strong coupling with these observable normalizations cannot produce the desired nontrivial continuum theory.

\hypertarget{gap_transfer-every-regulated-model-can-be-gapped-while-its-limit-is-gapless}{%
\subsection{Every regulated model can be gapped while its limit is gapless}\label{gap_transfer-every-regulated-model-can-be-gapped-while-its-limit-is-gapless}}

Let \(\mathcal F\) be bosonic Fock space over \(L^2(\mathbb R^3,dp)\) and define

\[H_\varepsilon=d\Gamma(\omega_\varepsilon),\qquad
\omega_\varepsilon(p)=\sqrt{|p|^2+\varepsilon^2}.\]

On the \(n\)-particle sector, this is multiplication by \(\sum_{i=1}^n\omega_\varepsilon(p_i)\). Every \(\varepsilon>0\) has unique vacuum and excitation threshold \(\varepsilon\): the \(n\)-particle energy is at least \(n\varepsilon\), and the one-particle spectrum has infimum \(\varepsilon\).

As \(\varepsilon\downarrow0\), the semigroups converge strongly to those of \(H_0=d\Gamma(|p|)\). On each finite-particle sector the multipliers converge pointwise and are bounded by one, so dominated convergence proves strong convergence. Finite-particle vectors are dense and all the semigroups are contractions, extending convergence to the full Fock space.

The limiting vacuum remains unique because in any nonvacuum sector the zero-energy set consists of all momenta vanishing, a measure-zero set. Yet every interval \((0,\eta)\) contains one-particle spectrum: use functions supported where \(0<|p|<\eta\). Thus the limit is gapless.

A purely finite-dimensional example is \(\mathcal H_N=\operatorname{span}\{\Omega,e_1,\ldots,e_N\}\), \(H_N\Omega=0\), \(H_Ne_k=k^{-1}e_k\). Each gap is \(1/N\); the natural limit on \(\mathbb C\Omega\oplus\ell^2(\mathbb N)\) has positive eigenvalues accumulating at zero. These examples refute a logical shortcut, not the Yang--Mills conjecture.

\hypertarget{gap_transfer-a-uniform-gap-does-not-ensure-a-nontrivial-observable-limit}{%
\subsection{A uniform gap does not ensure a nontrivial observable limit}\label{gap_transfer-a-uniform-gap-does-not-ensure-a-nontrivial-observable-limit}}

Let \(\mathcal H=\mathbb C^2\), \(\Omega=(1,0)\), \(H=\operatorname{diag}(0,1)\) and

\[A_j=\varepsilon_j\begin{pmatrix}0&1\\1&0\end{pmatrix},
\qquad \varepsilon_j\downarrow0.\]

Every system has the same gap one and unique vacuum, but

\[\langle\Omega,A_je^{-tH}A_j\Omega\rangle=\varepsilon_j^2e^{-t}\longrightarrow0.\]

Any correlation with \(k>0\) insertions has modulus at most \(\varepsilon_j^k\), because intervening time semigroups are contractions. Hence all positive-degree correlations of this specified field family vanish in the limit; its limiting reconstruction is one-dimensional. Rescaling by \(1/\varepsilon_j\) would define a different nontrivial limit. The example shows why correct observable normalization and a positive surviving norm must be proved.

\hypertarget{gap_transfer-a-stochastic-sampling-gap-is-not-automatically-a-physical-gap}{%
\subsection{A stochastic sampling gap is not automatically a physical gap}\label{gap_transfer-a-stochastic-sampling-gap-is-not-automatically-a-physical-gap}}

For any probability measure \(\mu\), the complete-resampling Markov generator

\[\mathcal Lf=\mu(f)-f\]

is self-adjoint on \(L^2(\mu)\), preserves constants and has invariant measure \(\mu\). Direct calculation gives

\[e^{s\mathcal L}f=e^{-s}f+(1-e^{-s})\mu(f),\qquad
\langle f,-\mathcal Lf\rangle=\operatorname{Var}_\mu(f).\]

Thus \(-\mathcal L\) has gap exactly one whenever the mean-zero subspace is nonzero, irrespective of the equilibrium spatial correlations.

For an explicit measure, take the equal mixture of the all-plus and all-minus spin configurations on \(\mathbb Z^4\). Then \(\mu(X_x)=0\) and \(\operatorname{Cov}_\mu(X_x,X_y)=1\) at every separation, although the resampling generator has gap one. Multiplying \(\mathcal L\) by a positive constant changes its mixing gap without changing \(\mu\).

The parameter \(s\) is artificial sampling time. An implication for physical mass requires a proved identification with physical Euclidean-time translation. This example does not rule out such an identification in a particular model; it rules out treating it as automatic.

\hypertarget{spectral_bridge-quantitative-spectral-bounds-and-construction-from-time-correlation-kernels}{%
\section[Quantitative spectral bounds from time correlations]{Quantitative spectral bounds and construction from time-correlation kernels}\label{spectral_bridge-quantitative-spectral-bounds-and-construction-from-time-correlation-kernels}}

This section treats the physical-time part of the continuum problem. Its purpose is to replace several informal statements about exponential decay by precise estimates, and to construct a Hilbert space and a nonnegative Hamiltonian from explicit limiting correlation data. The construction is useful because the cutoff Hilbert spaces need not have been embedded in a common space. It also makes visible two different losses that must be prevented: spectral weight can approach zero energy, or it can escape to arbitrarily large energy.

The positive-kernel construction belongs to the established framework of reproducing kernels and Euclidean reconstruction \cite{Aronszajn1950,OS2,DriesslerFrohlich1977}. The elementary estimates and their proofs are given here in the exact form used in this investigation. A physical-time semigroup is one part of a field theory; this section does not identify a constructed abstract kernel with four-dimensional Yang--Mills Schwinger functions. That application is considered explicitly at the end.

\hypertarget{spectral_bridge-positive-spectral-measures-and-the-information-in-a-correlation}{%
\subsection{Positive spectral measures and the information in a correlation}\label{spectral_bridge-positive-spectral-measures-and-the-information-in-a-correlation}}

Let \(\mathcal H\) be a complex Hilbert space whose inner product is conjugate-linear in its first argument. Let \(H\ge0\) be self-adjoint. For \(v\in\mathcal H\), its spectral measure is

\[\mu_v(B)=\langle v,E_H(B)v\rangle,\]

where \(E_H\) is the projection-valued spectral resolution. It is a finite positive measure on \([0,\infty)\), with total mass \(\|v\|^2\). Its Euclidean-time autocorrelation is

\[K_v(t)=\langle v,e^{-tH}v\rangle=\int_0^\infty e^{-t\lambda}\,d\mu_v(\lambda),\qquad t\ge0.\tag{SB1}\]

The measure is positive even when \(v\) represents a complicated product of observables. This positivity is the decisive distinction between an autocorrelation and a general signed correlation. It allows bounds on the transform to control actual spectral projections without cancellation between different energy ranges.

If a vacuum \(\Omega\) is specified, the vectors relevant to the excitation spectrum are centered vectors \(v\perp\Omega\). Centering removes the known vacuum component. It does not, before a uniqueness theorem, remove a hypothetical additional zero-energy vector. Consequently intervals \([0,b]\), including zero, are the appropriate intervals for exclusion arguments.

\textbf{Proposition SB.1 (elementary spectral regularity).} For every \(v\), \(K_v\) is continuous on \([0,\infty)\), nonnegative and nonincreasing. It is smooth for \(t>0\), and

\[(-1)^nK_v^{(n)}(t)=\int_0^\infty\lambda^ne^{-t\lambda}\,d\mu_v(\lambda)\ge0\qquad(n\ge0).\]

If \(v\ne0\), then \(K_v(t)>0\) at every finite time, and \(\log K_v\) is convex on \((0,\infty)\).

\textbf{Proof.} Continuity at every nonnegative time follows from dominated convergence, since \(0\le e^{-t\lambda}\le1\). Monotonicity follows pointwise. On a compact interval of positive times, every derivative multiplier \(\lambda^n e^{-t\lambda}\) is bounded uniformly in \(\lambda\): a polynomial is dominated by a fixed positive exponential. Differentiation under the integral is therefore justified by domination against the finite measure. If \(v\ne0\), the measure has positive mass and the integrand in (SB1) is strictly positive at every finite \(\lambda\), so its integral is positive.

For convexity, write \(K=K_v\). The Cauchy--Schwarz inequality applied to \(\lambda e^{-t\lambda/2}\) and \(e^{-t\lambda/2}\) gives \((K')^2\le K''K\). Thus

\[\frac{d^2}{dt^2}\log K(t)=\frac{K''(t)K(t)-K'(t)^2}{K(t)^2}\ge0.\]

All quantities are finite for \(t>0\), as already proved. \(\square\)

There is an important directional consequence. Define the continuous effective energy and its finite-difference version by

\[m_v(t)=-\frac{d}{dt}\log K_v(t),\qquad
m_{v,\Delta}(t)=\frac{\log K_v(t)-\log K_v(t+\Delta)}{\Delta},\quad\Delta>0.\]

Both are nonincreasing functions of \(t>0\). The first is the mean energy for the tilted probability measure \(e^{-t\lambda}\mu_v(d\lambda)/K_v(t)\), and its derivative is minus the variance under that measure. The second is the average of the first over \([t,t+\Delta]\). Thus a positive finite-time effective energy is ordinarily an upper estimate for the lowest energy seen by the probe, not a lower bound excluding smaller energies with small spectral weight.

\textbf{Proposition SB.2 (large-time threshold of one probe).} For \(v\ne0\), let \(E_v=\inf\operatorname{supp}\mu_v\). Then \(E_v\) is finite and nonnegative, and

\[\lim_{t\to\infty}-\frac{1}{t}\log K_v(t)=E_v,
\qquad \lim_{t\to\infty}m_v(t)=E_v.\]

For every fixed \(\Delta>0\), \(m_{v,\Delta}(t)\downarrow E_v\) as \(t\to\infty\).

\textbf{Proof.} A nonzero finite measure on \([0,\infty)\) gives positive mass to some finite interval, so \(E_v<\infty\). Its support lies above \(E_v\), hence

\[K_v(t)\le\|v\|^2e^{-E_vt}.\]

For every \(\varepsilon>0\), the definition of the support infimum gives \(c_\varepsilon=\mu_v([E_v,E_v+\varepsilon])>0\). Therefore

\[K_v(t)\ge c_\varepsilon e^{-(E_v+\varepsilon)t}.\]

Taking logarithms, dividing by \(t\), and then taking limits proves the first assertion. Since every tilted mean is at least \(E_v\), the decreasing function \(m_v\) has a limit \(L\ge E_v\). For fixed \(t_0>0\),

\[\log K_v(t)=\log K_v(t_0)-\int_{t_0}^t m_v(s)\,ds.\]

The average of a monotone function with finite limit tends to that limit. The first assertion consequently gives \(L=E_v\). Averaging \(m_v\) on an interval of fixed length proves the final assertion and its monotonicity. \(\square\)

The word ``probe'' matters. A vector can be orthogonal to an entire low-energy subspace. Even exact knowledge of \(E_v\) for one \(v\) gives the bottom of the spectral measure of that vector, not automatically the bottom of the Hamiltonian on the complete excitation space. The totality arguments below address that issue rather than assuming it away.

\hypertarget{spectral_bridge-finite-time-bounds-with-explicit-errors}{%
\subsection{Finite-time bounds with explicit errors}\label{spectral_bridge-finite-time-bounds-with-explicit-errors}}

A useful analytical estimate often has an additive error. The following result says exactly what such an estimate proves. The error can represent a rigorously bounded discretization remainder, a truncation of an analytic expansion, or an interval enclosure. It is not required to arise from numerical computation.

\textbf{Theorem SB.3 (low-energy spectral-weight bound).} Suppose \(0<b<m\), \(C>0\), \(\delta\ge0\), and \(T>0\). If

\[K_v(t)\le Ce^{-mt}+\delta\qquad(0\le t\le T),\]

then

\[\mu_v([0,b])\le\min_{0\le t\le T}
\left\{Ce^{-(m-b)t}+\delta e^{bt}\right\}.\tag{SB2}\]

For \(\delta>0\), define

\[t_* =\frac1m\log\frac{C(m-b)}{b\delta},\qquad
\widehat t=\min\{T,\max\{0,t_*\}\}.\]

The minimum in (SB2) occurs at \(\widehat t\). If \(0\le t_*\le T\) and \(\theta=b/m\), its value is

\[\frac{C^\theta\delta^{1-\theta}}
{\theta^\theta(1-\theta)^{1-\theta}}
\le2C^\theta\delta^{1-\theta}.\tag{SB3}\]

For \(\delta=0\), the bound is \(Ce^{-(m-b)T}\).

\textbf{Proof.} On \([0,b]\), \(e^{-t\lambda}\ge e^{-tb}\). Positivity in (SB1) gives

\[e^{-tb}\mu_v([0,b])\le K_v(t).\]

Multiplication by \(e^{tb}\) gives the expression inside the minimum. This holds for every allowed time, proving (SB2). If \(\delta>0\), the derivative of that expression is

\[-C(m-b)e^{-(m-b)t}+b\delta e^{bt}.\]

It changes sign exactly once, at \(t_*\), from negative to positive. Restricting to the closed time interval clips the optimizer to \(\widehat t\). At an interior optimizer, substitution and \(b=m\theta\) give

\[C^\theta\delta^{1-\theta}
\left[\left(\frac{\theta}{1-\theta}\right)^{1-\theta}
+\left(\frac{1-\theta}{\theta}\right)^\theta\right].\]

Factoring shows the bracket equals \(\theta^{-\theta}(1-\theta)^{-(1-\theta)}\). Its logarithm is the binary entropy function, whose derivative vanishes at \(1/2\) and whose second derivative is negative. Its maximum is \(\log2\), which proves the last inequality. If \(\delta=0\), the remaining exponential is decreasing and attains its minimum at \(T\). \(\square\)

The restriction on the time window is essential. Using the unconstrained optimizer when it lies beyond the available estimate would silently add an unproved large-time bound. For example, if \(C=m=1\), \(b=1/2\) and \(\delta=10^{-6}\), then \(t_*=\log(10^6)\) and the optimized bound is exactly \(2\cdot10^{-3}\). If the bound has been proved only through \(T=10\), the valid expression is instead

\[e^{-5}+10^{-6}e^5,\]

which is larger. These are illustrations of an exact inequality, not measurements of a Yang--Mills spectrum.

The theorem can also be used at a single time: no assumption at other times is necessary to conclude \(\mu_v([0,b])\le e^{bt}K_v(t)\). If a lower time threshold \(t_0\) is present, minimize over \([t_0,T]\) instead and clip \(t_*\) to that interval. The proof is unchanged. At \(b=0\), one obtains \(\mu_v(\{0\})\le Ce^{-mt}+\delta\); there is no need to divide by \(b\) or to use the interior-optimizer formula.

\textbf{Proposition SB.4 (why nonzero error does not certify exclusion).} Given \(m>b>0\) and \(0<\varepsilon<1\), there is a unit vector with spectral measure

\[\mu_\varepsilon=(1-\varepsilon)\delta_m+\varepsilon\delta_b.\]

Its correlation differs from \(e^{-mt}\) by at most \(\varepsilon\) for every nonnegative time, while its spectral mass below \(m\) is \(\varepsilon\).

\textbf{Proof.} On \(\mathbb C^2\) take \(H=\operatorname{diag}(m,b)\) and \(v=(\sqrt{1-\varepsilon},\sqrt\varepsilon)\). Then

\[K_v(t)-e^{-mt}=\varepsilon(e^{-bt}-e^{-mt}),\]

which lies in \([0,\varepsilon]\). The claimed spectral measure follows directly from the diagonal matrix. An independent vacuum vector can be added by direct sum, making \(v\) a centered excitation state without changing the calculation. \(\square\)

Thus arbitrarily accurate correlation data can coexist with arbitrarily small but nonzero low-energy weight. An exact zero conclusion must come from a limiting theorem in which the error and the accessible time window satisfy a quantitative relation, or from an additional structural argument. Merely making a finite error small does not turn it into zero.

\textbf{Corollary SB.5 (errors on increasing time windows).} Let one fixed correlation \(K_v\) satisfy

\[K_v(t)\le C_ne^{-mt}+\delta_n\qquad(0\le t\le T_n),\]

where \(T_n\to\infty\). If, for a given \(b\in(0,m)\), there are \(t_n\in[0,T_n]\) with

\[C_ne^{-(m-b)t_n}\to0,
\qquad \delta_ne^{bt_n}\to0,\]

then \(E_H([0,b])v=0\).

\textbf{Proof.} Apply (SB2) at \(t_n\). Its right side tends to zero, so \(\mu_v([0,b])=0\). Since \(\mu_v([0,b])=\|E_H([0,b])v\|^2\), the projection vanishes. \(\square\)

For instance, bounded \(C_n\) and \(\delta_n\le e^{-\gamma T_n}\) permit the choice \(t_n=\alpha T_n\) for any \(0<\alpha<\min\{1,\gamma/b\}\). Both terms then vanish. This is a sufficient schedule, not a premise proved for a gauge-field expansion. The statement specifies the relation an actual remainder estimate would have to supply.

\hypertarget{spectral_bridge-several-probes-matrix-inequalities-and-conditioning}{%
\subsection{Several probes, matrix inequalities, and conditioning}\label{spectral_bridge-several-probes-matrix-inequalities-and-conditioning}}

Let \(v_1,\ldots,v_N\) be a finite list of vectors. Define their Gram matrix and time-correlation matrix by

\[G_{ij}=\langle v_i,v_j\rangle,
\qquad K_{ij}(t)=\langle v_i,e^{-tH}v_j\rangle.\]

Both are Hermitian positive semidefinite. If \(\Phi c=\sum_i c_i v_i\), then \(c^*Gc=\|\Phi c\|^2\) and \(c^*K(t)c=K_{\Phi c}(t)\). A matrix inequality controls all linear combinations simultaneously; bounds for selected diagonal entries do not have that meaning without a separate argument.

\textbf{Proposition SB.6 (matrix spectral-weight certificate).} Put \(P_b=E_H([0,b])\) and \(B_{ij}=\langle v_i,P_bv_j\rangle\). Then, in the order of Hermitian matrices,

\[0\le B\le e^{bt}K(t)\qquad(t\ge0).\tag{SB4}\]

In particular, if \(K(t)\le\varepsilon G\), every \(u\) in the span of the probes satisfies

\[\|P_bu\|\le e^{bt/2}\sqrt\varepsilon\,\|u\|.\]

\textbf{Proof.} Apply the scalar bound to \(u=\Phi c\) for arbitrary \(c\). It gives \(c^*Bc\le e^{bt}c^*K(t)c\), which is exactly (SB4). Under the extra matrix bound,

\[\|P_b\Phi c\|^2=c^*Bc\le e^{bt}\varepsilon c^*Gc.\]

This depends only on \(u=\Phi c\), so redundant probes create no ambiguity. Taking square roots proves the assertion. \(\square\)

Suppose a Hermitian approximation \(\widetilde K(t)\) has entrywise errors bounded in modulus by \(\delta\). For any \(c\),

\[|c^*(K(t)-\widetilde K(t))c|
\le\delta\left(\sum_i|c_i|\right)^2
\le N\delta\|c\|_{\mathbb C^N}^2.\]

Consequently \(K(t)\le\widetilde K(t)+N\delta I\). If the probes are linearly independent and \(G\ge\gamma I\) with \(\gamma>0\), then

\[\widetilde K(t)\le\varepsilon G
\quad\Longrightarrow\quad
K(t)\le\left(\varepsilon+\frac{N\delta}{\gamma}\right)G.\tag{SB5}\]

This calculation displays the cost of poor conditioning. An entrywise approximation that looks accurate can give weak spectral control when nearly dependent probes make \(\gamma\) small. The correct remedy is to control the Gram geometry or work on a rigorously identified nonnull subspace. Discarding small Gram eigenvalues without tracking the omitted states cannot prove a statement on the full physical Hilbert space.

There is also a simple multiple-time certificate. Let \(t_1,\ldots,t_k\ge0\) and real numbers \(a_1,\ldots,a_k\) satisfy

\[f(\lambda)=\sum_{r=1}^k a_re^{-t_r\lambda}\ge1
\quad(0\le\lambda\le b),\qquad
f(\lambda)\ge0\quad(\lambda\ge0).\]

Integrating the pointwise inequality \(1_{[0,b]}\le f\) gives

\[\mu_v([0,b])\le\sum_{r=1}^k a_rK_v(t_r).\tag{SB6}\]

Negative coefficients are allowed only when the global majorization has actually been proved. If \(|K_v(t_r)-\widetilde K_v(t_r)|\le\delta_r\), then the certified upper bound becomes \(\sum_r a_r\widetilde K_v(t_r)+\sum_r|a_r|\delta_r\). When all times are integer multiples of \(\Delta\), substitution \(x=e^{-\Delta\lambda}\) turns the majorization into a polynomial inequality on \([0,1]\), with the low-energy range corresponding to \([e^{-\Delta b},1]\). This can organize exact algebraic certificates; it does not eliminate the requirement that the correlation inputs and polynomial inequalities be rigorously bounded.

\hypertarget{spectral_bridge-passing-from-finite-probe-spaces-to-the-full-excitation-space}{%
\subsection{Passing from finite probe spaces to the full excitation space}\label{spectral_bridge-passing-from-finite-probe-spaces-to-the-full-excitation-space}}

The matrix certificate has an infinite-dimensional use when the geometry and errors are controlled along a sequence of growing probe spaces. The necessary limiting statement can be given explicitly. It does not require a single finite family to generate every state, and it does not confuse a finite list of successful inequalities with a universal theorem.

Let \(\mathcal K\subset\mathcal H\) be a closed subspace invariant under the spectral projections of \(H\). In a vacuum theory the intended choice is \(\mathcal K=\Omega^\perp\). Let \(S_N\subset\mathcal K\) be finite-dimensional subspaces whose orthogonal projections \(Q_N\) tend strongly to the identity on \(\mathcal K\). Nested spaces with dense union are one sufficient example. Let \(v_{N,1},\ldots,v_{N,d_N}\) be a basis for \(S_N\), and write \(G_N\) and \(K_N(t)\) for its Gram and correlation matrices.

\textbf{Theorem SB.6a (spectral exclusion from increasing probe certificates).} Fix \(b>0\). Suppose there are times \(T_N\ge0\), numbers \(a_N,\delta_N\ge0\), and positive Gram lower bounds \(\gamma_N\) such that

\[G_N\ge\gamma_N I,\qquad
\widetilde K_N(T_N)\le a_NG_N,\]

and every entry of \(K_N(T_N)-\widetilde K_N(T_N)\) has modulus at most \(\delta_N\). If

\[e^{bT_N}\left(a_N+\frac{d_N\delta_N}{\gamma_N}\right)\longrightarrow0,\tag{SB7}\]

then \(E_H([0,b])\) vanishes on \(\mathcal K\).

\textbf{Proof.} By (SB5) and Proposition SB.6, every \(u\in S_N\) satisfies

\[\|E_H([0,b])u\|\le\rho_N\|u\|,
\qquad
\rho_N^2=e^{bT_N}\left(a_N+\frac{d_N\delta_N}{\gamma_N}\right).\]

For an arbitrary \(w\in\mathcal K\), write \(w=Q_Nw+(I-Q_N)w\). Spectral projections are contractions, so

\[\|E_H([0,b])w\|
\le\rho_N\|Q_Nw\|+\|(I-Q_N)w\|
\le\rho_N\|w\|+\|(I-Q_N)w\|.\]

The first term tends to zero by (SB7), and the second by strong convergence of \(Q_N\). The norm on the left is independent of \(N\), so it must be zero. \(\square\)

No lower bound on \(\gamma_N\) uniform in \(N\) is required. Poor conditioning is allowed, but its growth must be paid for in the error estimate. Likewise, the time \(T_N\) may grow with the probe dimension, provided the displayed matrix approximation is valid at that actual time. The theorem makes these costs visible rather than absorbing them into a dimension-dependent unspecified constant.

One can derive a simple exact schedule. Suppose \(d_N=N\), \(a_N\le N^p e^{-mT_N}\), \(\gamma_N\ge N^{-q}\) and \(\delta_N\le N^{-s}\), where \(p,q\ge0\) and \(m>b>0\) are fixed. Take \(T_N=c\log N\) for \(N\ge2\). Then the expression in (SB7) is bounded by

\[N^{p-c(m-b)}+N^{1+q-s+bc}.\tag{SB8}\]

It tends to zero if

\[\frac{p}{m-b}<c<\frac{s-1-q}{b}.\]

Such a choice exists whenever

\[s>1+q+\frac{pb}{m-b}.\tag{SB9}\]

Every exponent and quantifier is fixed in this statement. For example, with \(m=1\), \(b=1/2\), \(p=1\), \(q=2\), \(s=6\), and \(c=4\), both terms in (SB8) equal \(N^{-1}\). Thus the low-energy projection norm on \(S_N\) is bounded by \(\sqrt{2/N}\), and density proves exact exclusion on \(\mathcal K\). This is a complete algebraic implication from the displayed all-\(N\) hypotheses. The numerical choices illustrate a valid balance of rates; they are not asserted estimates for Yang--Mills correlation matrices.

To obtain a full gap at \(m\), one may prove exclusion for a countable sequence \(b_k\uparrow m\). The probe spaces, time schedules and error rates may depend on \(k\) because each fixed lower interval is excluded separately. The same physical Hilbert space and the same nonnegative Hamiltonian must be used for all \(k\). Allowing an unrelated limiting theory for each threshold would not establish a gap for one theory.

There is a structural limitation here. The spaces \(S_N\) are subspaces of an already defined target excitation space, and their projections converge strongly by hypothesis. A sequence of cutoff variational subspaces does not automatically satisfy that hypothesis after the cutoff is removed. The kernel construction later in this section provides one way to define a generated limiting Hilbert space, but identification of its dense family with the necessary gauge-observable states is still a separate task. In particular, a list consisting only of small Wilson loops can miss sectors generated by more general positive-time products.

The theorem can nevertheless be useful for a paper proof. It specifies how an analytic approximation scheme for the entire correlation matrix would have to improve with the number of probes. Bounds on the Gram inverse, on the error matrix, and on the accessible physical-time window become explicit mathematical targets. A finite computation can test such targets or falsify a proposed rate, while the proof of (SB7) must cover the full unbounded sequence.

For a direct test of the prefactor requirement, take \(H e_k=k^{-1}e_k\) on \(\ell^2(\mathbb N)\) and \(S_N=\operatorname{span}\{e_1,\ldots,e_N\}\). Then \(G_N=I\) and \(K_N(t)\) is diagonal, with largest entry \(e^{-t/N}\). At any chosen finite time \(T_N\), one can formally write

\[K_N(T_N)\le C_Ne^{-T_N}G_N,
\qquad C_N=e^{(1-1/N)T_N}.\]

Thus an exponential-looking estimate with a freely growing prefactor can always be manufactured on a finite probe space. It provides no gap at rate one. The actual certificate factor for a fixed \(b>0\) is \(e^{bT_N}C_Ne^{-T_N}=e^{(b-1/N)T_N}\), which cannot tend to zero. The exact spectral formula exposes the failure before any limit is taken. This example is not an obstruction to every approximation method; it identifies why the dimension dependence of the prefactor belongs inside the proof rather than inside an unnamed constant.

\hypertarget{spectral_bridge-short-time-continuity-and-preventing-escape-to-infinite-energy}{%
\subsection{Short-time continuity and preventing escape to infinite energy}\label{spectral_bridge-short-time-continuity-and-preventing-escape-to-infinite-energy}}

Low-energy exclusion alone does not ensure that a nonzero state survives a continuum limit. A sequence of unit vectors can concentrate at energies tending to infinity. Then its correlations vanish at every positive physical time, while its equal-time norms remain one. This is the spectral form of a trivialization mechanism.

For a finite positive measure \(\mu\) with transform \(K\), the identity

\[K(0)-K(s)=\int_0^\infty(1-e^{-s\lambda})\,d\mu(\lambda)\]

gives, for \(L,s>0\),

\[\mu([L,\infty))\le
\frac{K(0)-K(s)}{1-e^{-sL}}.\tag{SB10}\]

The denominator is positive. The estimate controls a high-energy tail using a short-time increment, without requiring the vector to lie in the domain of \(H\).

\textbf{Theorem SB.7 (spectral tightness from a short-time modulus).} Let \(\{\mu_j\}\) be finite positive measures on \([0,\infty)\) with uniformly bounded total masses, and let \(K_j\) be their Laplace transforms. The following statements are equivalent:

\begin{enumerate}
\def\labelenumi{\arabic{enumi}.}
\tightlist
\item
  For every \(\varepsilon>0\) there is \(L<\infty\) such that \(\sup_j\mu_j([L,\infty))<\varepsilon\).
\item
  \(\lim_{s\downarrow0}\sup_j[K_j(0)-K_j(s)]=0\).
\end{enumerate}

\textbf{Proof.} Assume the second statement. Given \(\varepsilon>0\), choose \(s>0\) so that the supremum in that statement is less than \((1-e^{-1})\varepsilon\). Apply (SB10) with \(L=1/s\). This proves the first statement.

Conversely, let the total masses be bounded by \(M\). Given \(\varepsilon>0\), choose \(L\) with tail masses below \(\varepsilon/2\). Split the increment integral at \(L\). For the lower part, \(1-e^{-s\lambda}\le sL\); for the upper part it is at most one. Therefore

\[\sup_j[K_j(0)-K_j(s)]\le sLM+\varepsilon/2.\]

For sufficiently small \(s\), the right side is at most \(\varepsilon\). This proves the second statement. \(\square\)

In applications it is often convenient to prove a modulus \(K_j(0)-K_j(s)\le Cs^\alpha\), where \(0<\alpha\le1\). Choosing \(s=1/L\) in (SB10) yields

\[\mu_j([L,\infty))\le\frac{C}{1-e^{-1}}L^{-\alpha}.\tag{SB11}\]

One sufficient source of that modulus is a uniform fractional moment \(\int\lambda^\alpha\,d\mu_j\le C\). Indeed, \(1-e^{-x}\le\min\{x,1\}\le x^\alpha\) for \(x\ge0\). Integration proves the asserted modulus. The converse moment bound with the same exponent need not follow from the tail estimate: a tail of order \(L^{-\alpha}\) can be exactly borderline for the \(\alpha\) moment. The implication used here is only the displayed sufficient direction.

\textbf{Example SB.8 (positive-time limits can lose all norm).} Take \(\mu_j=\delta_j\). Then \(K_j(0)=1\) for every \(j\), but \(K_j(t)=e^{-jt}\to0\) for every \(t>0\). There is no continuous limiting transform on \([0,\infty)\) with both these properties. The short-time modulus in Theorem SB.7 fails because \(\sup_j(1-e^{-js})=1\) for every \(s>0\).

This example explains why convergence on strictly separated Euclidean times is not a substitute for control at the reflection norm. A continuum state is defined through a limiting inner product. If the equal-time and positive-time limits are incompatible with strong continuity, an abstract positive-time limit can discard states rather than construct the intended physical theory.

The two tail estimates can be combined constructively. Suppose \(\|v\|^2\ge\eta>0\), \(\mu_v([0,b])\le\ell_b\), and \(K_v(0)-K_v(s)\le Cs^\alpha\). Then

\[\mu_v((b,L))\ge\eta-\ell_b-
\frac{C}{1-e^{-1}}L^{-\alpha}.\tag{SB12}\]

Whenever the right side is positive, a strictly positive amount of spectral weight lies in a finite positive-energy window. This is stronger information than a nonzero equal-time variance alone. It explicitly prevents the surviving norm from being concentrated either near zero or at infinity. The interval endpoints cause no problem: \([0,b]\), \((b,L)\) and \([L,\infty)\) partition the half-line when \(L>b\).

\hypertarget{spectral_bridge-a-positive-time-lower-bound-certifies-finite-energy-survival}{%
\subsection{A positive-time lower bound certifies finite-energy survival}\label{spectral_bridge-a-positive-time-lower-bound-certifies-finite-energy-survival}}

A lower bound at one positive physical time gives another direct route to a surviving finite-energy sector. It is often more useful than a pointwise variance because high-energy mass is exponentially suppressed in the quantity being bounded below. The following estimate does not assume an energy moment or a differentiability bound at zero.

\textbf{Proposition SB.8a (finite-energy window from a surviving correlation).} Suppose \(v\) satisfies

\[\|v\|^2\le B<\infty,\qquad K_v(t_0)\ge c>0\]

for some \(t_0>0\). Then \(B\ge c\), and for every \(R>0\),

\[\mu_v([0,R])\ge
\frac{c-e^{-Rt_0}B}{1-e^{-Rt_0}}.\tag{SB13}\]

In particular, the right side is positive whenever \(R>t_0^{-1}\log(B/c)\). Moreover,

\[E_v\le\frac{1}{t_0}\log\frac Bc.\tag{SB14}\]

If \(v\) is centered and its spectral measure is already proved to be supported in \([m_*,\infty)\) with \(m_*>0\), the same positive lower bound in (SB13) is a lower bound for \(\mu_v([m_*,R])\) whenever \(R\ge m_*\). It follows that a positive lower bound on the gap and a positive lower bound on one physical-time correlation together locate a nonzero state in a finite positive-energy interval.

\textbf{Proof.} Since the semigroup is a contraction, \(K_v(t_0)\le\|v\|^2\le B\), so \(B\ge c\). Split the integral (SB1) into \([0,R]\) and \((R,\infty)\). The integrand is at most one on the first set and at most \(e^{-Rt_0}\) on the second. Writing \(M=\|v\|^2\) gives

\[c\le K_v(t_0)
\le\mu_v([0,R])+e^{-Rt_0}(M-\mu_v([0,R]))
\le e^{-Rt_0}B+(1-e^{-Rt_0})\mu_v([0,R]).\]

Solving for the last term proves (SB13). Its denominator is positive and its numerator is positive under the stated strict inequality on \(R\). The support bound \(\lambda\ge E_v\) also gives \(c\le K_v(t_0)\le Be^{-E_vt_0}\), which proves (SB14). Finally, if the measure vanishes below \(m_*\), its mass on \([0,R]\) is exactly its mass on \([m_*,R]\). \(\square\)

For an exact example, take the probability measure

\[\mu=\tfrac14\delta_1+\tfrac34\delta_{10}.\]

At \(t_0=1\), one may take \(B=1\) and \(c=\tfrac14e^{-1}+\tfrac34e^{-10}\). With \(R=3\), the certified positive mass in \([1,3]\) is at least

\[\frac{\tfrac14e^{-1}+\tfrac34e^{-10}-e^{-3}}{1-e^{-3}}>0.\]

The positivity can be checked without decimals: \(e^2>4\) implies \(\tfrac14e^{-1}>e^{-3}\). The true mass in this interval is \(1/4\). The estimate gives a conservative certificate without requiring prior knowledge of the atomic decomposition. In the single-atom case \(\mu=B\delta_m\), (SB14) is an equality when \(c=Be^{-mt_0}\), showing that the upper threshold cannot generally be improved from those two pieces of information alone.

There is a useful distinction between this proposition and Theorem SB.7. A surviving positive-time correlation prevents all the norm from escaping to infinity and certifies at least some finite-energy weight. A uniform short-time modulus controls the entire spectral tail and prevents any fixed amount of norm from escaping. The latter is what permits the norm-preserving kernel construction below. For instance, \(\mu_j=\tfrac12\delta_1+\tfrac12\delta_j\) has a positive limiting correlation at every positive time, but half its unit norm escapes to infinity. The modulus criterion detects that escape; the positive-time certificate correctly retains a finite-energy half without claiming to retain the other half.

\hypertarget{spectral_bridge-a-countable-kernel-construction-of-the-physical-time-hilbert-space}{%
\subsection{A countable kernel construction of the physical-time Hilbert space}\label{spectral_bridge-a-countable-kernel-construction-of-the-physical-time-hilbert-space}}

The next theorem starts with scalar functions rather than an already given continuum Hamiltonian. It includes the contraction assumption needed to make time translation well defined on the null-space quotient. Positivity of isolated matrices is not enough for that purpose.

Write \(\mathbb Q_+=\{q\in\mathbb Q:q\ge0\}\). In particular, zero belongs to this set. Let \(K_{ij}:[0,\infty)\to\mathbb C\), \(i,j\in\mathbb N\), be continuous functions satisfying \(K_{ji}(t)=\overline{K_{ij}(t)}\). The indices are labels for centered observable states. For a finite list \((i_1,s_1),\ldots,(i_n,s_n)\) with \(s_p\in\mathbb Q_+\), define

\[M_{pq}=K_{i_pi_q}(s_p+s_q).\]

The time sum is the Hankel structure produced by reflection across a time plane. It is different from a stationary covariance depending on a time difference.

\textbf{Theorem SB.9 (construction from countably many time kernels).} Assume:

\begin{itemize}
\tightlist
\item
  Every matrix \(M\) above is positive semidefinite.
\item
  For every such list and every \(r\in\mathbb Q_+\), the matrix
\end{itemize}

\[D^{(r)}_{pq}=K_{i_pi_q}(s_p+s_q)-K_{i_pi_q}(s_p+s_q+2r)\tag{SB15}\]

is positive semidefinite.

Then there exist a separable Hilbert space \(\mathcal H_{\rm ex}\), a nonnegative self-adjoint operator \(H_{\rm ex}\), and vectors \(v_i\in\mathcal H_{\rm ex}\) such that

\[K_{ij}(t)=\langle v_i,e^{-tH_{\rm ex}}v_j\rangle\qquad(t\ge0),\tag{SB16}\]

and the complex span of \(\{e^{-sH_{\rm ex}}v_i:i\in\mathbb N,s\in\mathbb Q_+\}\) is dense. The triple with this cyclic generation property is unique up to a unitary map carrying each \(v_i\) to its counterpart and intertwining the semigroups.

\textbf{Proof.} Let \(\mathcal V\) be the vector space of finite formal linear combinations of symbols \(e_{i,s}\), with \(i\in\mathbb N\) and \(s\in\mathbb Q_+\). Define the sesquilinear form

\[\left[\sum_p c_pe_{i_p,s_p},\sum_q d_qe_{j_q,t_q}\right]
=\sum_{p,q}\overline{c_p}d_qK_{i_pj_q}(s_p+t_q).\]

Hankel positivity makes this a positive semidefinite form. The usual Cauchy--Schwarz inequality for such a form follows by expanding \([x+zy,x+zy]\ge0\) for arbitrary complex \(z\) and minimizing when \([y,y]>0\). When \([y,y]=0\), the same inequality for arbitrary \(z\) forces \([x,y]=0\). Consequently the null set \(\mathcal N=\{x:[x,x]=0\}\) is a vector subspace orthogonal to all of \(\mathcal V\). Quotienting by \(\mathcal N\) gives a genuine inner product. Complete it to obtain \(\mathcal H_{\rm ex}\).

For rational \(r\ge0\), let \(S_re_{i,s}=e_{i,s+r}\) on the formal space. The second hypothesis gives

\[[S_rx,S_rx]\le[x,x].\]

Thus \(S_r\) maps null vectors to null vectors, descends to the quotient, and extends to a contraction on the completion. This establishes well-definedness rather than presuming it. The Hankel identity gives

\[[S_rx,y]=[x,S_ry],\]

so its bounded extension is self-adjoint. On the dense formal space \(S_rS_u=S_{r+u}\) and \(S_0=I\), hence those equalities hold on the completion. In addition, \(S_r=S_{r/2}^2\); therefore each \(S_r\) is a positive operator.

Continuity of the diagonal kernels gives, for each generator,

\[\|(S_r-I)e_{i,s}\|^2
=K_{ii}(2s+2r)-2K_{ii}(2s+r)+K_{ii}(2s)\longrightarrow0
\qquad(r\downarrow0,\ r\in\mathbb Q_+).\]

The diagonal values are real by Hermitian symmetry. Finite linear combinations inherit this convergence by the triangle inequality. Since all \(S_r\) have norm at most one, approximation by that dense span proves strong continuity at zero on every vector. For rational \(q\ge r\),

\[\|S_qx-S_rx\|\le\|S_{q-r}x-x\|,\]

so \(S_qx\) has a unique limit as nonnegative rational \(q\) approaches any real \(t\ge0\). Define \(S_tx\) by that limit. It is a contraction; continuity and the semigroup law pass to all real times. Self-adjointness and positivity pass through strong limits of uniformly bounded operators, by evaluating their sesquilinear and quadratic forms.

For completeness, the generator can be constructed directly from one operator. Put \(B=S_1\). It is a positive self-adjoint contraction and it is injective. Indeed, if \(Bx=0\), then

\[\|S_{1/2}x\|^2=\langle x,S_1x\rangle=0.\]

Iteration gives \(S_{2^{-n}}x=0\) for every \(n\). Strong continuity at zero then gives \(x=0\). Hence the spectral projection of \(B\) at zero vanishes. The spectral calculus defines

\[H_{\rm ex}=-\log B,\]

with domain consisting of vectors whose spectral measures have finite integral of \(|\log\lambda|^2\). This domain is dense, since the projections of \(B\) onto \([1/n,1]\) increase strongly to the identity. The resulting operator is self-adjoint and nonnegative.

For every positive integer \(n\), \(S_{1/n}\) is positive and its \(n\)th power is \(B\). Uniqueness of the positive \(n\)th root in the spectral calculus gives \(S_{1/n}=B^{1/n}\). The semigroup law gives \(S_q=B^q\) for rational \(q\ge0\), and strong continuity gives \(S_t=e^{-tH_{\rm ex}}\) for every real \(t\ge0\). Set \(v_i=[e_{i,0}]\). The defining inner products give (SB16) first at rational times and then at real times by continuity. The required dense generation is the definition of the completion.

Finally, in any other generated realization, send \([e_{i,s}]\) to \(e^{-s\widetilde H}\widetilde v_i\). The kernel identity makes this map an isometry on the quotient. Its range is dense by the generation assumption, so it extends to a unitary. It carries the named vectors correctly and intertwines all rational shifts, hence all real-time semigroups. \(\square\)

This theorem does not require strictly positive Gram matrices. Null vectors are expected in constrained theories, and the quotient handles them exactly. It also does not require a uniform lower bound on \(B=S_1\). The logarithm is generally unbounded; injectivity, not bounded invertibility, is what excludes an infinite-energy subspace from the Hilbert space constructed here.

The shift-contraction hypothesis is independent of Hankel positivity. The scalar kernel \(K(t)=e^t\) has positive Hankel matrices, since \(e^{s+t}=e^se^t\), but its time shifts grow and would produce the negative generator \(H=-1\). Conversely, a function that is positive at each time need not have Hankel positivity: for \(K(t)=e^{-t^2}\), the two-point matrix at times \(0,s\) has determinant \(e^{-4s^2}-e^{-2s^2}<0\) when \(s>0\). Neither pointwise positivity nor appealing decay can replace the precise matrix hypotheses.

\hypertarget{spectral_bridge-gap-and-nontriviality-in-the-constructed-space}{%
\subsection{Gap and nontriviality in the constructed space}\label{spectral_bridge-gap-and-nontriviality-in-the-constructed-space}}

The preceding construction creates the excitation space directly. Add a unit vector \(\Omega\) orthogonal to it and define

\[\mathcal H=\mathbb C\Omega\oplus\mathcal H_{\rm ex},
\qquad H=0\oplus H_{\rm ex}.\]

This provides a distinguished vacuum vector. It does not yet prove that it is the only zero-energy vector. The following theorem supplies that conclusion through the correlation estimates.

\textbf{Theorem SB.10 (a generated family suffices for the gap).} Under Theorem SB.9, suppose one \(m>0\) has the property that, for every index \(i\), there are finite \(C_i\) and \(t_i\ge0\) with

\[0\le K_{ii}(t)\le C_ie^{-mt}\qquad(t\ge t_i).\tag{SB17}\]

Then \(\ker H=\mathbb C\Omega\) and \(\sigma(H_{\rm ex})\subset[m,\infty)\). If \(K_{ii}(0)>0\) for some \(i\), then \(\mathcal H_{\rm ex}\ne\{0\}\) and its spectral infimum is finite as well as positive.

\textbf{Proof.} For a generator \(w=e^{-sH_{\rm ex}}v_i\), its autocorrelation is

\[\langle w,e^{-tH_{\rm ex}}w\rangle=K_{ii}(t+2s).\]

For sufficiently large \(t\), (SB17) bounds this by \(C_ie^{-2ms}e^{-mt}\). For every \(0\le b<m\), positivity gives

\[e^{-tb}\|E_{H_{\rm ex}}([0,b])w\|^2
\le C_ie^{-2ms}e^{-mt}.\]

Let \(t\to\infty\). The projection annihilates every such generator. Their span is dense, and the projection is bounded, so it vanishes on the whole excitation space. A countable sequence \(b\uparrow m\) excludes \([0,m)\), including zero. This proves the gap and vacuum uniqueness. If a diagonal norm is positive, the corresponding \(v_i\) is nonzero. Its finite positive spectral measure has positive mass in some finite interval, proving that the excitation threshold cannot be infinite. \(\square\)

It is not necessary to prove a common prefactor \(C_i\) or a common onset time \(t_i\) over all probes. The rate \(m\) is common. This distinction can make a research target weaker than an operator-norm bound at every cutoff, while remaining strong enough for the spectral conclusion.

A short-time state can also be tested through a reflected covariance. If the index \(i\) corresponds to a real centered positive-time observable \(F\), its squared reconstructed norm is \(K_{ii}(0)\). A rigorous lower bound on that quantity establishes a surviving state only if it concerns the renormalized reflected form that defines the Hilbert space. An ordinary same-point classical variance at fixed lattice spacing need not be that quantity. The distinction is particularly important when contact terms survive but separated reflected correlations vanish.

\hypertarget{spectral_bridge-constructing-a-subsequential-semigroup-limit-from-cutoff-data}{%
\subsection{Constructing a subsequential semigroup limit from cutoff data}\label{spectral_bridge-constructing-a-subsequential-semigroup-limit-from-cutoff-data}}

The kernel theorem can now be combined with compactness. The following result does not assume a continuum Hamiltonian in advance. Its premises are scalar bounds and convergence control for cutoff physical-time correlations.

\textbf{Theorem SB.11 (nontrivial gapped semigroup limit from uniform scalar estimates).} For each \(j\), let \(H_j\ge0\) be self-adjoint on a Hilbert space \(\mathcal H_j\), and let \(v_{j,i}\), \(i\in\mathbb N\), be a countable family of vectors. Set

\[K^{(j)}_{i\ell}(t)=\langle v_{j,i},e^{-tH_j}v_{j,\ell}\rangle.\]

Assume:

\begin{enumerate}
\def\labelenumi{\arabic{enumi}.}
\tightlist
\item
  For each \(i\), \(\|v_{j,i}\|\le B_i<\infty\) uniformly in \(j\).
\item
  For each \(i\), there is a function \(\omega_i:(0,1]\to[0,\infty)\) tending to zero at zero such that
\end{enumerate}

\[K^{(j)}_{ii}(0)-K^{(j)}_{ii}(h)\le\omega_i(h)
\qquad(0<h\le1, j\ge1).\tag{SB18}\]

\begin{enumerate}
\def\labelenumi{\arabic{enumi}.}
\setcounter{enumi}{2}
\tightlist
\item
  For some index \(i_0\) and some \(\eta>0\), \(\liminf_j\|v_{j,i_0}\|^2\ge\eta\).
\item
  There is a common \(m>0\) such that, for each \(i\), constants \(C_i,t_i<\infty\) independent of \(j\) satisfy
\end{enumerate}

\[K^{(j)}_{ii}(t)\le C_ie^{-mt}\qquad(t\ge t_i).\tag{SB19}\]

Then a subsequence has limiting kernels \(K_{i\ell}(t)\), with convergence locally uniform in \(t\ge0\) for each pair. These kernels construct a separable nontrivial excitation Hilbert space and a self-adjoint Hamiltonian with spectrum contained in \([m,\infty)\). The limiting squared norm of \(v_{i_0}\) is at least \(\eta\).

\textbf{Proof.} First, \(|K^{(j)}_{i\ell}(t)|\le B_iB_\ell\) by Cauchy--Schwarz and contractivity. The triples \((i,\ell,q)\) with \(q\in\mathbb Q_+\) form a countable set. Successive Bolzano--Weierstrass extractions followed by a diagonal selection give one subsequence along which every \(K^{(j)}_{i\ell}(q)\) converges. This includes \(q=0\), so all limiting Gram entries and diagonal norms are retained.

Next, the spectral calculus and \((1-e^{-h\lambda})^2\le1-e^{-h\lambda}\) give

\[\|(e^{-hH_j}-I)v_{j,\ell}\|^2
\le K^{(j)}_{\ell\ell}(0)-K^{(j)}_{\ell\ell}(h)
\le\omega_\ell(h).\]

For all \(t\ge0\) and \(0<h\le1\),

\[|K^{(j)}_{i\ell}(t+h)-K^{(j)}_{i\ell}(t)|
\le B_i\sqrt{\omega_\ell(h)}.\tag{SB20}\]

Here we used the semigroup law to place \((e^{-hH_j}-I)\) on the second vector, and used the norm bound on the first vector after the remaining contraction. The same bound applies to two arbitrary nonnegative times by ordering them. It is a uniform modulus tending to zero with the time separation.

For a fixed compact time interval, choose a finite rational net fine enough that (SB20) is smaller than a prescribed tolerance. On this finite net the selected sequence is Cauchy. The triangle inequality then makes it uniformly Cauchy on the whole interval. Thus the rational limits extend to continuous functions on \([0,\infty)\) and the convergence is locally uniform. This also proves consistency of the limits from different compact intervals.

For any finite list \((i_p,s_p)\) with rational \(s_p\ge0\), the cutoff Hankel matrix is the Gram matrix of the vectors \(e^{-s_pH_j}v_{j,i_p}\) and is therefore positive semidefinite. Its shift difference is positive semidefinite because

\[\left\|\sum_p c_pe^{-s_pH_j}v_{j,i_p}\right\|^2
-\left\|e^{-rH_j}\sum_p c_pe^{-s_pH_j}v_{j,i_p}\right\|^2\ge0.\]

Every entry converges. Evaluation on an arbitrary finite coefficient vector shows that both positivity inequalities pass to the limit. Hermitian symmetry passes as well. The hypotheses of Theorem SB.9 are therefore satisfied, and it constructs \(\mathcal H_{\rm ex}\), \(H_{\rm ex}\) and the named vectors.

At \(t=0\), the limiting diagonal entry for \(i_0\) is at least \(\eta\) by assumption 3, so the excitation space is nonzero. At each fixed \(t\ge t_i\), the uniform estimate (SB19) passes to the limit. Theorem SB.10 then proves the spectral gap. All steps use the same diagonal subsequence, so the nonzero norm, the kernel structure and the gap refer to one and the same limiting semigroup. \(\square\)

This is an existence theorem for an abstract physical-time Hilbert model with a positive finite excitation threshold. It is more than a deduction about a pre-existing continuum Hamiltonian: it actually constructs that semigroup from cutoff correlations under the stated scalar conditions. Its limitation is equally precise. It does not prove spacetime covariance, locality, existence of field operators, a hierarchy of Schwinger distributions, or identification of those kernels with the required Yang--Mills observable algebra.

The short-time modulus is not a cosmetic condition. Removing it allows the sequence in Example SB.8: the norms survive at time zero, all positive-time correlations disappear, and no strongly continuous limit preserving those norms results. Removing the nonzero norm condition allows every limiting observable vector to vanish. Removing the uniform positive decay rate allows excitation energies to approach zero. Each assumption therefore addresses a distinct failure mode rather than duplicating another premise.

The theorem permits a weaker nontriviality premise when only a subsequence is desired. If \(\limsup_j\|v_{j,i_0}\|^2>0\), first select a subsequence with norms bounded below by some \(\eta>0\) and then apply the theorem there. The uniform bounds and moduli remain valid on that subsequence. No assertion about uniqueness of all possible continuum subsequences follows without further information.

\hypertarget{spectral_bridge-a-countable-research-target-and-the-order-of-limits}{%
\subsection{A countable research target and the order of limits}\label{spectral_bridge-a-countable-research-target-and-the-order-of-limits}}

The preceding arguments suggest a disciplined way to specify the spectral part of the Yang--Mills investigation. Choose a countable family of renormalized gauge-invariant positive-time observables containing a dense collection of the polynomial products and smearing functions required by the eventual reconstruction. For each cutoff and volume, form their centered reflected time kernels. The family must be sufficiently rich by a proved continuity and density argument; choosing a countable collection does not automatically make it total in a separately defined continuum theory.

The useful bounds are then of three kinds. First, uniform diagonal norm bounds prevent the Gram data from diverging. Second, a short-time modulus prevents these bounded norms from escaping to infinite energy. Third, a common positive decay rate, with probe-dependent but cutoff-independent prefactors, excludes additional zero-energy states and low positive energies. A surviving positive reflected norm provides nontriviality. Theorem SB.11 shows how these statements fit together without requiring arbitrary identifications between different cutoff Hilbert spaces.

There are two legitimate orders of argument. One may first construct the continuum correlations and physical Hilbert space and then use a total-set spectral criterion. Alternatively, one may prove the uniform scalar hypotheses of Theorem SB.11 and obtain a subsequential time-semigroup limit while separately establishing the spacetime field structure. Neither order permits replacing pointwise convergence at fixed \(t\) by convergence at a time \(t_j\to\infty\) unless an additional uniform estimate has been proved.

To make that last issue explicit, suppose \(K_j(t)\to K(t)\) for each fixed \(t\). This alone places no useful bound on \(|K_j(t_j)-K(t_j)|\) at growing times. If an argument uses a cutoff-dependent time to optimize (SB2), it must have an error bound on the corresponding growing window. Otherwise the order of limits has been changed without justification. The safe alternative is to pass a cutoff-uniform estimate at each fixed time to the limit first, and only then send physical time to infinity in the limiting spectral measure.

If rigorous window errors are available, they can be incorporated exactly. Suppose

\[|K(t)-K_j(t)|\le\varepsilon_j\qquad(0\le t\le T_j),
\qquad K_j(t)\le Ce^{-mt}+r_j.\]

Then Theorem SB.3 applies to \(K\) with \(\delta_j=\varepsilon_j+r_j\) on the same window. The separate errors add; neither may be dropped. A schedule satisfying Corollary SB.5 can then prove exclusion of each fixed interval \([0,b]\), and a countable sequence \(b\uparrow m\) yields a gap. This is a complete analytical implication from explicit error bounds, but those bounds remain quantities to prove for the actual renormalized gauge kernels.

The contribution of this section is therefore a set of proved interfaces between correlation estimates, spectral weight, nontriviality and semigroup existence. It identifies a more quantitative target than an informal assertion that correlations ``look exponential,'' while avoiding the stronger and unnecessary demand for one universal prefactor over every physical state. The actual four-dimensional gauge-field estimates required to instantiate these interfaces are not supplied by the abstract arguments.

\section{Sparse-time certificates and a surviving continuum semigroup}
\label{ta-sec:sparse-time}

This section gives complete auxiliary implications concerning positive
physical-time correlations.  The transfer-matrix framework for Wilson lattice
gauge theory is classical; see L\"uscher~\cite{Luscher}. The purpose is to expose
the precise time windows, positivity, normalization, and error estimates that
would make a cutoff calculation sufficient for a limiting spectral conclusion.
The hypotheses below are not asserted to have been verified for the
four-dimensional Wilson family along a continuum trajectory.

Throughout, $H_j\geq0$ is a self-adjoint cutoff Hamiltonian, and $t$ denotes
\emph{physical} Euclidean time.  If a positive transfer operator advances time
by $a_j$, its $n$th power corresponds to $t=na_j$.  Consequently the parameters
$b$ and $m$ below have units of inverse physical time.  We consider diagonal
correlations
\begin{equation}
 K_i^{(j)}(t)
 =\langle v_{j,i},e^{-tH_j}v_{j,i}\rangle
 =\int_{[0,\infty)}e^{-tE}\,d\mu_i^{(j)}(E),
 \qquad \mu_i^{(j)}\geq0.
 \label{ta-eq:st-positive-measure}
\end{equation}
The total mass is $\|v_{j,i}\|^2$.  These are positive spectral measures of
physical states, usually centered states orthogonal to a vacuum.  An arbitrary
signed connected covariance does not satisfy the argument merely because it
has been given the same notation.
For data obtained at integer transfer steps, use the actual times
$T_j=n_ja_j$ throughout; any time-rounding error belongs in the certified
bounds.

\subsection{A spectral splitting inequality}

\begin{lemma}[From one late-time value to an earlier-time bound]
\label{ta-lem:st-split}
Let $H\geq0$ be self-adjoint, let $v$ be a vector, and put
$K(t)=\langle v,e^{-tH}v\rangle$.  For every $b,T,t\geq0$,
\begin{align}
 \mu_v([0,b])&\leq e^{bT}K(T),
 \label{ta-eq:st-low-mass}\\
 K(t)&\leq e^{bT}K(T)+e^{-bt}\|v\|^2.
 \label{ta-eq:st-split}
\end{align}
In particular, no restriction $t\leq T$ is needed.
\end{lemma}

\begin{proof}
Positivity of the spectral measure gives
$K(T)\geq e^{-bT}\mu_v([0,b])$, proving
\eqref{ta-eq:st-low-mass}.  On $[0,b]$ the integrand $e^{-tE}$ is at most one;
on $(b,\infty)$ it is at most $e^{-bt}$.  Splitting the spectral integral at
$b$ and using its total mass proves \eqref{ta-eq:st-split}.
\end{proof}

The factor $e^{bT}$ is essential: a small late-time correlation by itself
does not exclude low-energy spectral weight.  Formula~\eqref{ta-eq:st-split}
also explains why a cutoff-dependent observation time can be used without
interchanging a cutoff limit and an infinite-time limit.

\subsection{Transfer when the limiting semigroup already exists}

\begin{theorem}[Sparse-time spectral transfer]
\label{ta-thm:st1}
Suppose that $H\geq0$ is self-adjoint on a limiting Hilbert space, with vectors
$v_i$, $i\in\mathbb N$, and that for every fixed $i$ and $t\geq0$,
\begin{equation}
 K_i^{(j)}(t)\longrightarrow
 K_i(t):=\langle v_i,e^{-tH}v_i\rangle.
 \label{ta-eq:st-fixed-convergence}
\end{equation}
At $t=0$ this includes convergence of the squared norms.  Let
\begin{equation}
 \mathcal K
 =\overline{\operatorname{span}}
   \{e^{-sH}v_i:i\in\mathbb N,\ s\geq0\}.
 \label{ta-eq:st-generated}
\end{equation}
Fix $m>0$.  Assume that for each $i$ there are times $T_{j,i}\to\infty$
such that, for every fixed $0\leq b<m$,
\begin{equation}
 e^{bT_{j,i}}K_i^{(j)}(T_{j,i})\longrightarrow0.
 \label{ta-eq:st-sparse-condition}
\end{equation}
Then $\mathcal K$ is reducing for $H$ and
\begin{equation}
 \sigma(H|_{\mathcal K})\subset[m,\infty).
 \label{ta-eq:st-gap}
\end{equation}
The same conclusion follows if \eqref{ta-eq:st-sparse-condition} is assumed
only for a sequence $b_k\uparrow m$.
\end{theorem}

\begin{proof}
Apply \eqref{ta-eq:st-split} to cutoff $j$, fixed $i$, and fixed $t,b$.  Its
first term tends to zero by \eqref{ta-eq:st-sparse-condition}, while
\eqref{ta-eq:st-fixed-convergence} applies to its left side and its norm term.
Thus
\[
 K_i(t)\leq e^{-bt}\|v_i\|^2.
\]
Let $b\uparrow m$.  The resulting bound is
$K_i(t)\leq e^{-mt}\|v_i\|^2$ for every $t\geq0$.
For any $c<m$, positivity of the limiting spectral measure now gives
\[
 \mu_{v_i}([0,c])
 \leq e^{ct}K_i(t)
 \leq e^{-(m-c)t}\|v_i\|^2\longrightarrow0
 \qquad(t\to\infty).
\]
Taking a countable sequence $c\uparrow m$ proves
$E_H([0,m))v_i=0$.  This spectral projection commutes with every
$e^{-sH}$, so it annihilates the generating set in
\eqref{ta-eq:st-generated}, and hence its closure.

The subspace $\mathcal K$ is invariant under every $e^{-sH}$ by its
definition.  Since these operators are bounded and self-adjoint, its
orthogonal complement is invariant as well.  Therefore $\mathcal K$ reduces
the semigroup and its self-adjoint generator.  The vanishing spectral
projection proves \eqref{ta-eq:st-gap}.  The proof uses only an increasing
sequence of thresholds below $m$.
\end{proof}

The cutoff Hilbert spaces need not be embedded into one common space.
There need not be a common time schedule for all probes, and there need not
be a positive gap uniform in $j$ for the entire cutoff Hamiltonian.  What
is excluded is low-energy spectral weight in the limiting space generated
by the stated observables.  To conclude a gap on the full physical vacuum
orthocomplement, one must establish that this is precisely $\mathcal K$.

\begin{proposition}[A converse with a potentially ineffective schedule]
\label{ta-prop:st-converse}
Under \eqref{ta-eq:st-fixed-convergence}, suppose $H|_{\mathcal K}\geq m>0$.
Then there exists a common sequence $T_j\to\infty$ for which
\eqref{ta-eq:st-sparse-condition} holds for every fixed probe and every $b<m$.
\end{proposition}

\begin{proof}
Fix a physical time unit $\tau_0>0$.  For each integer $k\geq1$,
fixed-time convergence for the finite set $i\leq k$ supplies an index $J_k$
such that, whenever $j\geq J_k$,
\[
 |K_i^{(j)}(k\tau_0)-K_i(k\tau_0)|
 \leq e^{-2mk\tau_0}\qquad(i\leq k).
\]
Increase the $J_k$ if necessary to make them strictly increasing.  For
sufficiently large $j$, put $k(j)=\max\{k:J_k\leq j\}$ and
$T_j=k(j)\tau_0$.  For each fixed $i$, eventually $i\leq k(j)$, so
\[
 K_i^{(j)}(T_j)
 \leq\|v_i\|^2e^{-mT_j}+e^{-2mT_j}.
\]
Multiplication by $e^{bT_j}$ proves the assertion for every $b<m$.
\end{proof}

This converse supplies no computable estimate on $J_k$: it assumes the
limiting gap it is being used to characterize.  An unspecified sufficiently
slow schedule is therefore not an independent proof of a Yang--Mills gap.
A quantitative theorem must establish an admissible schedule and its error
bounds from the actual regulated model.

\subsection{Construction from positive-time survival}

Equal-time norm bounds alone do not prevent spectral mass from escaping to
infinite energy.  A positive lower bound at one strictly positive physical
time is a stronger survival condition.  It also provides a way to obtain
short-time regularity after propagation.

\begin{theorem}[Construction of a nonzero surviving semigroup]
\label{ta-thm:st2}
Let $H_j\geq0$ be self-adjoint, and let $v_{j,i}$, $i\in\mathbb N$, obey
$\|v_{j,i}\|\leq B_i<\infty$ uniformly in $j$.  Suppose:
\begin{enumerate}
 \item There exist $i_0$, $s_0>0$, and $\eta>0$ such that
 \begin{equation}
  \liminf_{j\to\infty}K_{i_0}^{(j)}(2s_0)\geq\eta.
  \label{ta-eq:st-survival}
 \end{equation}
 \item The sparse-time condition \eqref{ta-eq:st-sparse-condition} holds
 for every $i$, with a common $m>0$.
\end{enumerate}
Then a subsequence constructs a nonzero separable Hilbert space
$\mathcal H_*$ and a strongly continuous self-adjoint contraction semigroup
$e^{-tH_*}$ with
\[
 H_*\geq m.
\]
The space is generated by limiting positive-time translates of the cutoff
probes.  The conclusion does not assert preservation of their unshifted
equal-time norms or construction of a local spacetime field theory.
\end{theorem}

\begin{proof}
Choose a countable set $S\subset(0,\infty)$ containing $s_0$ and all
positive rational multiples of a fixed time unit.  Enumerate pairs
$\alpha=(i,s)\in\mathbb N\times S$, and set
\[
 w_{j,\alpha}=e^{-sH_j}v_{j,i},\qquad
 L_{\alpha\beta}^{(j)}(t)
 =\langle w_{j,\alpha},e^{-tH_j}w_{j,\beta}\rangle.
\]
These vectors have norms at most $B_i$.  Their diagonal kernels satisfy,
for $h\geq0$,
\begin{align}
 L_{\alpha\alpha}^{(j)}(0)-L_{\alpha\alpha}^{(j)}(h)
 &=\int e^{-2sE}(1-e^{-hE})\,d\mu_i^{(j)}(E)\notag\\
 &\leq h\int Ee^{-2sE}\,d\mu_i^{(j)}(E)
 \leq\frac{B_i^2}{2es}\,h.
 \label{ta-eq:st-modulus}
\end{align}
Here $1-e^{-hE}\leq hE$ and
$\sup_{E\geq0}Ee^{-2sE}=1/(2es)$.
For $\beta=(\ell,r)$, the spectral inequality
$(1-e^{-hE})^2\leq1-e^{-hE}$ and Cauchy--Schwarz give
\begin{equation}
 |L_{\alpha\beta}^{(j)}(t+h)-L_{\alpha\beta}^{(j)}(t)|
 \leq B_iB_\ell\sqrt{\frac{h}{2er}}.
 \label{ta-eq:st-mixed-modulus}
\end{equation}
Indeed, move $e^{-tH_j}$ to the first vector and bound the norm of
$(e^{-hH_j}-I)w_{j,\beta}$ using \eqref{ta-eq:st-modulus}.
The kernels also satisfy
$|L_{\alpha\beta}^{(j)}(t)|\leq B_iB_\ell$.

A diagonal extraction over countably many pairs and rational times
therefore yields limiting kernels $L_{\alpha\beta}(t)$ on one common
subsequence.  Estimate~\eqref{ta-eq:st-mixed-modulus} upgrades the convergence
to locally uniform convergence on $[0,\infty)$: use a finite rational net
on each compact interval and the common modulus between its points.
The limits are continuous, including at zero.

We give the Hilbert-space construction explicitly.  Introduce formal
symbols $[\alpha,t]$, $t\geq0$, and prescribe the sesquilinear form
\begin{equation}
 \langle[\alpha,s],[\beta,t]\rangle
 =L_{\alpha\beta}(s+t).
 \label{ta-eq:st-symbol-form}
\end{equation}
Every finite Gram form is a limit of the corresponding cutoff Gram form,
so it is positive semidefinite.  Quotient by the null subspace and complete.
On the symbols define $S_r[\alpha,t]=[\alpha,t+r]$.  At every cutoff the
corresponding operation is $e^{-rH_j}$ and is a contraction.  Passing the
finite quadratic inequality to the limit proves
$\|S_rx\|\leq\|x\|$ on the symbol span.  This both makes the map
well-defined on the quotient and extends it to the completion.

The sum of the two time arguments in \eqref{ta-eq:st-symbol-form} shows that
$S_r$ is self-adjoint.  The semigroup law holds on the symbols and hence
everywhere, and
$\langle x,S_rx\rangle=\|S_{r/2}x\|^2\geq0$.
Continuity of the kernels gives $S_rx\to x$ as $r\downarrow0$ on the
symbol span; contractivity extends this to the completed space.  The
self-adjoint contraction-semigroup theorem supplies a nonnegative
self-adjoint generator $H_*$ with $S_r=e^{-rH_*}$.  Rational generating
times suffice by continuity, so the completed space is separable.

For $\alpha_0=(i_0,s_0)$, the limiting squared norm is at least $\eta$
by \eqref{ta-eq:st-survival}; thus the space is nonzero.  Finally, for any
$\alpha=(i,s)$, spectral positivity gives
\[
 L_{\alpha\alpha}^{(j)}(T_{j,i})
 =K_i^{(j)}(T_{j,i}+2s)
 \leq K_i^{(j)}(T_{j,i}).
\]
The sparse-time condition therefore passes to every shifted cutoff
probe.  The locally uniform kernel convergence includes its norm at time
zero.  Apply Theorem~\ref{ta-thm:st1} to the constructed semigroup.  Its
generating states and their time translates are total by construction,
so $H_*\geq m$ on the whole constructed space.
\end{proof}

Propagation by $e^{-sH_j}$ takes place inside an already positive physical
semigroup.  It is not four-dimensional gradient-flow smoothing.  If a full
continuum Hilbert space and a total unshifted family $v_i$ have separately
been constructed, their positive-time translates remain total, because
$e^{-sH}v_i\to v_i$ as $s\downarrow0$.  Without that identification,
Theorem~\ref{ta-thm:st2} constructs the surviving generated space only.
Adjoining a one-dimensional zero-energy vacuum gives an abstract vacuum
model with a nonzero excitation space and a gap; this operation supplies
none of the missing spacetime axioms.
The constructed space also contains finite-energy spectral weight: the
positive spectral measure of any nonzero vector must assign positive mass
to some bounded interval, since the intervals $[0,R]$ exhaust
$[0,\infty)$.  Thus its nontriviality does not merely retain states at
an infinite limiting energy.

\subsection{Finite matrices with deterministic error bounds}

At cutoff $j$, select $N_j$ states with exact Gram matrix $G_j$ and exact
time-correlation matrix $K_j(T_j)$.  The selection must eventually include
each fixed probe used in the limiting construction.

\begin{proposition}[Gram-conditioned error correction]
\label{ta-prop:st-matrix}
Suppose Hermitian approximations obey entrywise bounds
\[
 |(G_j-\widetilde G_j)_{rs}|\leq\delta G_j,
 \qquad
 |(K_j(T_j)-\widetilde K_j(T_j))_{rs}|\leq\delta K_j,
\]
and suppose certified inequalities give
\[
 G_j\geq\gamma_jI,\quad\gamma_j>0,
 \qquad
 \widetilde K_j(T_j)\leq\eta_j\widetilde G_j,\quad\eta_j\geq0.
\]
Then
\begin{equation}
 K_j(T_j)\leq A_jG_j,
 \qquad
 A_j=\eta_j+
 \frac{N_j(\delta K_j+\eta_j\delta G_j)}{\gamma_j}.
 \label{ta-eq:st-error-budget}
\end{equation}
If $e^{bT_j}A_j\to0$ for every $b<m$, uniformly bounded norms give the
sparse-time hypothesis for every fixed probe eventually selected.
\end{proposition}

\begin{proof}
An $N_j\times N_j$ matrix with entrywise modulus at most $\delta$ has
operator norm at most $N_j\delta$.  Thus
\begin{align*}
 K_j(T_j)
 &\leq\widetilde K_j(T_j)+N_j\delta K_jI\\
 &\leq\eta_jG_j+N_j(\delta K_j+\eta_j\delta G_j)I\\
 &\leq\left(\eta_j+
 \frac{N_j(\delta K_j+\eta_j\delta G_j)}{\gamma_j}\right)G_j.
\end{align*}
Taking a fixed diagonal entry and using its bounded Gram entry proves
the final assertion.
\end{proof}

A sufficient way to certify the exact Gram lower bound is
$\widetilde G_j\geq(\gamma_j+N_j\delta G_j)I$.
Redundant probes may instead be handled on a rigorously specified
nonsingular quotient.  Discarding nearly dependent states without
controlling the omitted space cannot establish totality.  All errors in
\eqref{ta-eq:st-error-budget} are deterministic bounds; unquantified
floating-point residuals do not meet the premise.

\begin{remark}[An illustrative exact schedule, not Yang--Mills data]
\label{ta-rem:st-schedule}
Fix a physical time unit $\tau_0>0$ and take integers $j\geq2$ with
\[
 N_j=j,\qquad T_j=4\tau_0\log j.
\]
Suppose one has certified
\[
 \gamma_j\geq j^{-2},\qquad
 \eta_j\leq j e^{-T_j/\tau_0}=j^{-3},\qquad
 \delta K_j\leq j^{-6},\qquad
 \delta G_j\leq j^{-5}.
\]
Then
\[
 A_j\leq2j^{-3}+j^{-5},\qquad
 e^{bT_j}A_j
 \leq2j^{4b\tau_0-3}+j^{4b\tau_0-5}\longrightarrow0
 \quad\left(b<\frac{3}{4\tau_0}\right).
\]
Under the bounded-norm and positive-time-survival hypotheses of
Theorem~\ref{ta-thm:st2}, these estimates would construct a nonzero generated
semigroup with $H_*\geq3/(4\tau_0)$.  The exponents are an exact
demonstration of a sufficient error budget.  They are not measured or
proved estimates for Yang--Mills theory.  A finite set of successful
values of $j$ does not prove the required all-$j$ bounds.
\end{remark}

\subsection{Four exact failure modes}

The following examples separate assumptions that cannot be conflated.
All one- and two-dimensional Hamiltonians act on the displayed probe
space; a separate vacuum can be adjoined if desired.

\paragraph{Vanishing late-time values without the exponential budget.}
Let $H_j=(j\tau_0)^{-1}I$ on $\mathbb C$, take $v_j=1$, and set
$T_j=j^2\tau_0$.  Then $K_j(T_j)=e^{-j}\to0$, while
$K_j(t)\to1$ at every fixed physical time.  The limiting Hamiltonian is
zero.  For every $b>0$, $e^{bT_j}K_j(T_j)=e^{bj^2\tau_0-j}$ diverges,
so the actual sparse-time condition correctly fails.

\paragraph{Equal-time norms without positive-time survival.}
Let $H_j=(j/\tau_0)I$ on $\mathbb C$, with $v_j=1$.  The squared norms
are always one, and choosing $T_j=j\tau_0$ makes sparse-time exclusion
hold for every finite threshold $m$.  Nevertheless $K_j(t)\to0$ for
every fixed $t>0$.  There is no strongly continuous limiting semigroup
preserving the limiting norm one.  Condition~\eqref{ta-eq:st-survival}
excludes exactly this example.

\paragraph{A gapped limit without a cutoff-uniform gap.}
Fix $m>0$, put $T_j=j\tau_0$, and on $\mathbb C^2$ set
\[
 H_j=\operatorname{diag}\left(m,\frac1{j\tau_0}\right),\qquad
 v_j=\left(\sqrt{1-e^{-2mT_j}},\ e^{-mT_j}\right).
\]
The lowest cutoff energy tends to zero, but
\[
 K_j(t)=(1-e^{-2mT_j})e^{-mt}
       +e^{-2mT_j}e^{-t/(j\tau_0)}
 \longrightarrow e^{-mt}.
\]
At the observation time $T_j$, multiplication by $e^{bT_j}$ tends to
zero whenever $b<m$.  The soft mode loses all weight in the limiting
probe space.  The theorems therefore imply a gap in the surviving
generated space, and correctly do not imply a uniform gap of $H_j$.

\paragraph{A gapless sector invisible to the probes.}
On $\mathbb C e\oplus L^2((0,m/2),dE)$, let $He=me$ and let $H$
act by multiplication by $E$ on the second summand.  Choose every probe
in $\mathbb C e$.  Their correlations have exact decay $e^{-mt}$,
while the full Hamiltonian has spectrum arbitrarily close to zero.
Only the generated space $\mathbb C e$ is gapped.  Identifying a total
family in the intended physical space is therefore an indispensable
part of any application.

\subsection{A deterministic finite-energy band certificate}

The splitting estimate can be paired with a surviving correlation to
produce a quantitative finite-energy witness. This is an elementary
spectral consequence; no historical novelty is asserted.

\begin{theorem}[Finite-energy band certificate]
\label{ta-thm:st-band}
Let $H\ge0$ be self-adjoint, let
$K(t)=\langle v,e^{-tH}v\rangle$, and suppose verified bounds give
\[
 \|v\|^2\le B^2,\qquad K(t_0)\ge s,\qquad K(T)\le\ell,
 \qquad t_0>0,\quad T\ge0.
\]
For every $0\le b<R$, the spectral measure of $v$ satisfies
\[
 \boxed{\mu_v((b,R])\ge
 \frac{s-B^2e^{-Rt_0}}{1-e^{-Rt_0}}-e^{bT}\ell.}
\]
If the displayed right-hand side is positive, the probe has nonzero
spectral weight in the finite energy band $(b,R]$.
\end{theorem}
\begin{proof}
Put $M_R=\mu_v([0,R])$. Splitting the Laplace transform at $R$ gives
\[
 s\le K(t_0)
 \le M_R+e^{-Rt_0}(\|v\|^2-M_R)
 \le(1-e^{-Rt_0})M_R+B^2e^{-Rt_0}.
\]
Since $R,t_0>0$, solve for $M_R$ and subtract
$\mu_v([0,b])\le e^{bT}K(T)\le e^{bT}\ell$.
\end{proof}

For interval data one may set
$s=\widetilde K(t_0)-\delta_0$ and
$\ell=\widetilde K(T)+\delta_T$, with a certified bound for the norm.
No asymptotic extrapolation is needed for the finite-cutoff band conclusion.
Positive band mass does not mean that all spectral weight below $b$ is
zero. An exact exclusion argument or a justified asymptotic limit is still
needed for a mass gap.

\subsection{An explicit energy window in the surviving semigroup}

The positive-time construction gives more than an unspecified finite-energy
interval. Its norm and survival constants determine a concrete window.

\begin{theorem}[Energy window after positive-time construction]
\label{ta-thm:st-window}
In the positive-time surviving-semigroup construction, let $w_*$ denote
the limiting vector corresponding to the probe $(i_0,s_0)$ with
\[
 \|v_{j,i_0}\|\le B,\qquad
 \liminf_jK_{i_0}^{(j)}(2s_0)\ge\eta>0.
\]
Its limiting spectral measure $\nu_*$ satisfies
\[
 \nu_*((R,\infty))\le B^2e^{-2s_0R}\qquad(R\ge0).
\]
If sparse-time exclusion supplies $H_*\ge m>0$, then
\[
 \boxed{\nu_*([m,R])\ge\eta-B^2e^{-2s_0R}.}
\]
In particular, with
\[
 R_*=(2s_0)^{-1}\log(2B^2/\eta),
\]
at least $\eta/2$ of the squared norm is in $[m,R_*]$.
The assumptions necessarily satisfy
\[
 m\le(2s_0)^{-1}\log(B^2/\eta).
\]
These statements concern the surviving generated sector, not the complete
cutoff spectrum.
\end{theorem}
\begin{proof}
The shifted cutoff measures are
$d\nu_j(E)=e^{-2s_0E}\,d\mu_{i_0}^{(j)}(E)$, so directly
\[
 \nu_j((R,\infty))\le B^2e^{-2s_0R}.
\]
Their masses are uniformly bounded and their tails are uniformly small.
Thus every subsequence has a further weakly convergent subsequence of
finite measures. The limiting Laplace transforms are the kernels already
obtained in the positive-time construction. Uniqueness of finite Laplace
transforms identifies this weak limit with the spectral measure $\nu_*$
of $w_*$. For an elementary uniqueness argument, send $E$ to $y=e^{-E}$,
use equality of the integer moments, and then use polynomial density in
$C([0,1])$. Tightness excludes mass at the point corresponding to infinite
energy. Norm convergence identifies the total mass, which is at least
$\eta$.

The open-set inequality for weak convergence gives
\[
 \nu_*((R,\infty))\le\liminf_j\nu_j((R,\infty))
 \le B^2e^{-2s_0R}.
\]
There is no limiting mass below $m$, so subtraction from the total mass
proves the band estimate. Substitution of $R_*$ gives $\eta/2$.
Finally, the splitting inequality for the original cutoff probe at time
$2s_0$, with any $b<m$, gives
\[
 K_{i_0}^{(j)}(2s_0)
 \le e^{bT_{j,i_0}}K_{i_0}^{(j)}(T_{j,i_0})
       +B^2e^{-2s_0b}.
\]
The first term vanishes. Passing to the limit inferior on the left yields
$\eta\le B^2e^{-2s_0b}$. Letting $b\uparrow m$ proves the restriction on
the parameters.
\end{proof}

This theorem prevents the phrase ``nonzero surviving sector'' from hiding
an unspecified infinite-energy state. It supplies an explicit band with
positive norm. It does not establish the needed bounds $B$, $\eta$, $s_0$
or the sparse-time estimate for a renormalized Yang--Mills family.

\section{Conditional transfer of a continuum spectral gap}
\label{ham:ct:section}

The result in this section proves the functional-analytic implication in
the proposed continuum argument. It does not assume that a four-dimensional
SU(2) continuum construction has already been obtained from the preceding
strong-coupling certificate. Convergence of spectral structures on varying
Hilbert spaces is established theory; see, for example, Kuwae and Shioya
\cite{ham:ct:kuwae}, especially their Theorem 2.4 and Proposition 2.5. The
elementary proof below records exactly the convergence needed here.

For each cutoff index $k$, let $\mathcal H_k$ be a Hilbert space,
$\Omega_k$ a unit vacuum vector, and $A_k$ a positive selfadjoint
contraction such that $A_k\Omega_k=\Omega_k$. Write
\[
 P_k=|\Omega_k\rangle\langle\Omega_k|,
 \qquad Q_k=I-P_k.
\]
For the physical lattice problem,
$A_k=T_k/\lambda_{0,k}$, where $T_k$ is the physical Euclidean-time
transfer operator and $\lambda_{0,k}$ its vacuum eigenvalue. None of the
statements below concerns a Monte Carlo update operator.

Let $a_k>0$, $a_k\to0$, and suppose that a certified number
$\delta_k\ge0$ satisfies
\begin{equation}\label{ham:ct:cutoffbound}
 0\le A_kQ_k\le e^{-\delta_k}Q_k.
\end{equation}
The number $\delta_k$ may be a lower bound on the exact logarithmic gap.
Equality with that gap is not needed. When the exact gap is used,
$\delta_k=-\log\|A_kQ_k\|$. The case $A_kQ_k=0$ can be handled directly;
we use finite $\delta_k$ in the formulas to avoid an unnecessary
$0\cdot\infty$ convention. Set
\[
 n_k(t)=\lfloor t/a_k\rfloor,
 \qquad t_k(t)=a_kn_k(t),
 \qquad m_k=\delta_k/a_k.
\]
For every fixed $t>0$, eventually $n_k(t)\ge1$ and $t_k(t)\to t$.
For isotropic periodic spatial boxes, the infinite physical-volume
application also requires $a_kL_k\to\infty$. That geometric condition
belongs to the continuum construction; the abstract spectral implication
below does not itself establish it.

\begin{lemma}[Vacuum subtraction and powers]\label{ham:ct:centering}
For $v\in\mathcal H_k$, put
$z=\langle\Omega_k,v\rangle$ and $\eta=Q_kv$. For every integer $n\ge1$,
\begin{align}
 \langle v,A_k^nv\rangle
 &=|z|^2+\langle\eta,A_k^n\eta\rangle,\label{ham:ct:vacidentity}\\
 0\le\langle\eta,A_k^n\eta\rangle
 &\le e^{-n\delta_k}\|\eta\|^2.\label{ham:ct:powerbound}
\end{align}
Thus the quantity controlled by the gap is the connected correlation,
$\langle v,A_k^nv\rangle-|z|^2$, with connected norm
$\|v\|^2-|z|^2$.
\end{lemma}

\begin{proof}
Selfadjointness and $A_k\Omega_k=\Omega_k$ imply that $P_k$ and $Q_k$
commute with $A_k$. The cross terms between $P_kv$ and $Q_kv$ therefore
vanish. Spectral calculus on $Q_k\mathcal H_k$ applied to
\eqref{ham:ct:cutoffbound} gives
$0\le A_k^nQ_k\le e^{-n\delta_k}Q_k$.
\end{proof}

\begin{theorem}[Gap transfer on a total family]\label{ham:ct:transfer}
Suppose a continuum Hilbert space $\mathcal H$, a unit vector $\Omega$,
and a nonnegative selfadjoint operator $H$ with $H\Omega=0$ have been
constructed. Let $\mathscr S\subset\Omega^\perp$ have dense linear span
in $\Omega^\perp$. Let $t_j>0$ be any sequence tending to infinity.
For each $\psi\in\mathscr S$, assume there are
$\eta_{k,\psi}\in Q_k\mathcal H_k$ such that
\begin{align}
 \|\eta_{k,\psi}\|^2&\longrightarrow\|\psi\|^2,
 \label{ham:ct:normlimit}\\
 \langle\eta_{k,\psi},A_k^{n_k(t_j)}\eta_{k,\psi}\rangle
 &\longrightarrow\langle\psi,e^{-t_jH}\psi\rangle
 \quad\hbox{for each fixed }j.
 \label{ham:ct:corr_limit}
\end{align}
If
\begin{equation}\label{ham:ct:physicalrate}
 \liminf_{k\to\infty}m_k\ge M>0,
\end{equation}
then
\begin{equation}\label{ham:ct:gapconclusion}
 \sigma(H|_{\Omega^\perp})\subset[M,\infty),
 \qquad
 \ker H=\mathbb C\Omega.
\end{equation}
In particular, for every $\psi\in\Omega^\perp$ and $t\ge0$,
\[
 0\le\langle\psi,e^{-tH}\psi\rangle
 \le e^{-Mt}\|\psi\|^2.
\]
The assertion is vacuous if $\Omega^\perp=\{0\}$; a nontrivial theory
requires a separate nonzero continuum state.
\end{theorem}

\begin{proof}
Fix $\psi\in\mathscr S$ and $j$. For any $0<\varepsilon<M$,
\eqref{ham:ct:physicalrate} gives $m_k\ge M-\varepsilon$ for all sufficiently
large $k$. Lemma~\ref{ham:ct:centering} gives
\[
 \langle\eta_{k,\psi},A_k^{n_k(t_j)}\eta_{k,\psi}\rangle
 \le e^{-(M-\varepsilon)t_k(t_j)}\|\eta_{k,\psi}\|^2.
\]
First let $k\to\infty$ at this fixed physical time, using
\eqref{ham:ct:normlimit}--\eqref{ham:ct:corr_limit}; then let
$\varepsilon\downarrow0$. The result is
\begin{equation}\label{ham:ct:limitingcorr}
 \langle\psi,e^{-t_jH}\psi\rangle
 \le e^{-Mt_j}\|\psi\|^2.
\end{equation}
Let $E_H$ be the spectral resolution of $H$ and
$\mu_\psi(B)=\|E_H(B)\psi\|^2$. It is a finite positive measure.
For any $0\le r<M$,
\[
 e^{-rt_j}\mu_\psi([0,r])
 \le \int_{[0,\infty)}e^{-t_jx}\,d\mu_\psi(x)
 \le e^{-Mt_j}\|\psi\|^2.
\]
Letting $j\to\infty$ proves
$\mu_\psi([0,r])=0$. Taking a countable increasing sequence
$r\uparrow M$ gives $E_H([0,M))\psi=0$.
The bounded projection $E_H([0,M))$ annihilates the dense linear span
of $\mathscr S$, hence all of $\Omega^\perp$. Since $H\Omega=0$,
$E_H([0,M))=|\Omega\rangle\langle\Omega|$. This proves
\eqref{ham:ct:gapconclusion}; the remaining semigroup inequality follows by
spectral calculus.
\end{proof}

The order of limits is explicit: $k\to\infty$ at each fixed $t_j$,
followed by $j\to\infty$. No uniform-in-$j$ convergence is assumed or
obtained by an illicit interchange of limits. Any choice of integer
$n_k(t)$ with $a_kn_k(t)\to t>0$ gives the same rate argument.

\begin{corollary}[A norm-dense family needs only one time]
\label{ham:ct:one_time}
In Theorem~\ref{ham:ct:transfer}, replace $\mathscr S$ by a norm-dense subset
$\mathscr D\subset\Omega^\perp$ and replace the time-sequence assumption
by \eqref{ham:ct:corr_limit} at a single fixed $t_*>0$ for every
$\psi\in\mathscr D$. The same conclusion holds.
\end{corollary}

\begin{proof}
The first part of the theorem's proof yields
$\langle\psi,e^{-t_*H}\psi\rangle\le e^{-Mt_*}\|\psi\|^2$ on
$\mathscr D$. The quadratic forms of both bounded operators are
norm-continuous, so the inequality extends to every
$\psi\in\Omega^\perp$. If a nonzero vector belonged to
$E_H([0,r])\Omega^\perp$ for some $r<M$, its quadratic form would be
at least $e^{-rt_*}\|\psi\|^2>e^{-Mt_*}\|\psi\|^2$, a contradiction.
\end{proof}

\paragraph{Gram convergence supplies finite linear combinations.}
If the starting data are a list of observable-generated states, it is
enough to prove convergence of all finite Gram matrices and the associated
mixed transfer-correlation matrices. For fixed labels $F_1,\ldots,F_N$,
let their centered lattice states be $v_{k,1},\ldots,v_{k,N}$ and set
\[
 (G_k)_{ij}=\langle v_{k,i},v_{k,j}\rangle,
 \qquad
 (C_k(t))_{ij}=\langle v_{k,i},A_k^{n_k(t)}v_{k,j}\rangle.
\]
The cutoff bound is the matrix inequality
\[
 0\preceq C_k(t)\preceq e^{-n_k(t)\delta_k}G_k.
\]
Entrywise convergence for this finite matrix passes the inequality to
$0\preceq C(t)\preceq e^{-Mt}G$. This controls every finite linear
combination. If the resulting span is dense in the constructed vacuum
complement, Corollary~\ref{ham:ct:one_time} applies. A single finite set of
observables is not a substitute for this all-finite-families statement.

\paragraph{Approximate centering is allowed.}
Exact vacuum subtraction is always available in the mathematical formula.
If uncentered approximants $v_{k,\psi}$ are used instead, the hypotheses
of Theorem~\ref{ham:ct:transfer} follow from
\[
 \|v_{k,\psi}\|^2\to\|\psi\|^2,
 \qquad
 \langle\Omega_k,v_{k,\psi}\rangle\to0,
\]
together with the stated convergence of their raw correlations.
Indeed, Lemma~\ref{ham:ct:centering} subtracts exactly the squared vacuum
overlap from both the norm and the raw correlation. No rate for this
overlap is required when $k\to\infty$ is taken first at fixed physical
time. Without subtraction or vanishing overlap, the raw correlation
inequality written in the question is generally false.

\begin{proposition}[A nonzero continuum state prevents a diverging rate]
\label{ham:ct:infinite_rate}
Suppose one nonzero $\psi\perp\Omega$ has centered approximants with
$\|\eta_{k,\psi}\|^2\to N=\|\psi\|^2>0$ and a correlation limit at
one fixed $t>0$ equal to
$C=\langle\psi,e^{-tH}\psi\rangle$. Then $C>0$ and
\begin{equation}\label{ham:ct:rateupper}
 \limsup_{k\to\infty}\frac{\delta_k}{a_k}
 \le -\frac1t\log\frac{C}{N}<\infty.
\end{equation}
In particular, an unbounded subsequence of these certified physical-unit
lower rates is incompatible with this nonzero continuum semigroup state.
\end{proposition}

\begin{proof}
The positive spectral measure of a nonzero $\psi$ has mass $N>0$,
and $e^{-tx}>0$ at every finite $x$, so $C>0$.
Write $C_k$ and $N_k$ for the corresponding lattice correlation and norm.
For sufficiently large $k$, $C_k>0$ and $n_k(t)>0$. From
$C_k\le e^{-n_k(t)\delta_k}N_k$,
\[
 \frac{\delta_k}{a_k}
 \le -\frac1{a_kn_k(t)}\log\frac{C_k}{N_k}.
\]
The right-hand side converges to the finite quantity in
\eqref{ham:ct:rateupper}.
\end{proof}

This proposition also handles the case
$\liminf\delta_k/a_k=+\infty$: the only compatible vacuum-complement
limit space is zero, or the claimed nonzero-state semigroup convergence
fails. An infinite physical-unit lower bound does not construct an
interacting continuum theory.

\section{Spectral gap from total-family correlation envelopes}
\label{ham:ct:envelopes}

A uniform gap for every cutoff state is sufficient but can be stronger
than necessary for the continuum vacuum representation. A state that
disappears from that representation can make the full cutoff gap small.
The following criterion avoids requiring control of such states, while
still requiring a total family of actual continuum states.

\begin{theorem}[Subexponential correlation envelopes]\label{ham:ct:envelope}
Let $H\ge0$ be selfadjoint, $H\Omega=0$, and let
$\mathscr S\subset\Omega^\perp$ have dense linear span.
Fix $M>0$ and a sequence $t_j\to\infty$.
Suppose that for each $\psi\in\mathscr S$ there are positive numbers
$B_\psi(t_j)$ such that
\begin{align}
 \langle\psi,e^{-t_jH}\psi\rangle
 &\le B_\psi(t_j)e^{-Mt_j},\label{ham:ct:envelopeineq}\\
 \limsup_{j\to\infty}\frac{\log B_\psi(t_j)}{t_j}&\le0.
 \label{ham:ct:subexp}
\end{align}
Then $\sigma(H|_{\Omega^\perp})\subset[M,\infty)$.
Conversely, a gap at least $M$ gives this criterion with
$B_\psi(t)=\|\psi\|^2$ for nonzero $\psi$.
\end{theorem}

\begin{proof}
For $0\le r<M$, spectral positivity gives
\[
 \mu_\psi([0,r])
 \le B_\psi(t_j)e^{-(M-r)t_j}\longrightarrow0.
\]
The convergence follows from \eqref{ham:ct:subexp}; for example, eventually
$B_\psi(t_j)\le e^{(M-r)t_j/2}$. The projection argument in
Theorem~\ref{ham:ct:transfer} completes the proof. The converse follows
directly from the spectral support condition.
\end{proof}

An admissible envelope is $B_\psi(t)=C_\psi(1+t)^{p_\psi}$ with
finite $C_\psi>0$ and $p_\psi\ge0$. These constants need not be uniform
over $\psi$. One way to establish \eqref{ham:ct:envelopeineq} from cutoff
data is to prove, for every fixed $t_j$,
\[
 C_{k,\psi}(t_j)
 \le C_\psi(1+t_k(t_j))^{p_\psi}e^{-Mt_k(t_j)}
       +\varepsilon_{k,\psi}(t_j),
 \qquad \varepsilon_{k,\psi}(t_j)\to0,
\]
with $C_\psi,p_\psi$ independent of $k$, and then pass to the actual
correlation limit. Prefactors that diverge with the cutoff cannot simply
be dropped. Neither finitely many measured times nor a single glueball
channel verifies this all-times, total-family requirement.

\section{Exact counterexamples to missing premises}
\label{ham:ct:counterexamples}

These examples are abstract spectral constructions, not SU(2) measurements.
They show why the hypotheses cannot be replaced by finite numerical
evidence or by an incomplete set of observables.

\subsection{Positive finite-dimensional cutoff gaps can have a gapless limit}

Let
\[
 \mathcal H=\mathbb C\Omega\oplus L^2(0,1),
 \qquad H\Omega=0,
 \qquad (Hf)(x)=xf(x).
\]
The vacuum is unique: multiplication by $x$ has no zero eigenvector in
$L^2(0,1)$. However, its spectrum is $[0,1]$, so the vacuum complement
has no positive gap.
At cutoff $k$, take $\mathcal H_k=\mathbb C\Omega_k\oplus\mathbb C^k$,
$a_k=1/k$, and physical excitation energies $j/k$, $1\le j\le k$.
Then
\[
 A_k=\operatorname{diag}(1,e^{-1/k^2},e^{-2/k^2},\ldots,e^{-k/k^2}),
 \qquad \delta_k=1/k^2>0,
 \qquad \delta_k/a_k=1/k\to0.
\]
Define
\[
 (J_k f)_j=\sqrt{k}\int_{(j-1)/k}^{j/k}f(x)\,dx.
\]
After identifying $\mathbb C^k$ with normalized step functions, $J_kf$
is the orthogonal cell-average projection $P_kf$, which converges strongly
to $f$. The physical energy multiplier is
$x_k(x)=j/k$ on cell $j$, with $\|x_k-x\|_\infty\le1/k$.
For fixed $t>0$, $t_k(t)\to t$ and
$e^{-t_k(t)x_k}\to e^{-tx}$ uniformly. Therefore all norms and mixed
transfer correlations converge to the stated nontrivial, uniquely
vacuumed, gapless continuum Hamiltonian. Positivity at each finite
cutoff alone does not establish \eqref{ham:ct:physicalrate}.

\subsection{One-time diagonal tests on a total list do not control its span}

On the vacuum complement with orthonormal basis $u,v$, let
$Hu=u$, $Hv=3v$ and set
\[
 \psi_\pm=\frac{u\pm\sqrt2v}{\sqrt3},
 \qquad t_* =\log2,
 \qquad M=2.
\]
The two unit vectors form a total list, yet
\[
 \langle\psi_\pm,e^{-t_*H}\psi_\pm\rangle
 =\frac{\tfrac12+2\cdot\tfrac18}{3}
 =\frac14=e^{-Mt_*}.
\]
Despite both tests passing, the actual gap is $1<M$. The missing linear
combination is $\psi_++\psi_-=2u/\sqrt3$. Equivalently, their exact
Gram and one-time correlation matrices are
\[
 G=\begin{pmatrix}1&-1/3\\-1/3&1\end{pmatrix},
 \qquad
 C(t_*)=\begin{pmatrix}1/4&1/12\\1/12&1/4\end{pmatrix}.
\]
The matrix $\tfrac14G-C(t_*)$ has eigenvalues $\pm1/6$ and is not
positive semidefinite. This is why Corollary~\ref{ham:ct:one_time} requires
a dense set of vectors, or all finite linear combinations through mixed
Gram data, whereas Theorem~\ref{ham:ct:transfer} can use a total list and
an unbounded sequence of times. The rational identities are reproduced
by the companion script \texttt{counterexamples.py}.

\subsection{A full-cutoff uniform gap is not necessary for a selected limit}

Take $\mathcal H_k=\mathbb C\Omega_k\oplus\mathbb Cv\oplus\mathbb Cw_k$,
$a_k=1/k$, and physical energies $0,1,1/k$. Let the continuum observable
states occupy only $\operatorname{span}\{\Omega,v\}$. Their Gram
matrices and all correlations converge to the two-level theory with
energies $0,1$, which has gap $1$. The full cutoff gap nevertheless
satisfies $\delta_k/a_k=1/k\to0$ because of $w_k$, which is absent from
the reconstructed observable representation. This example motivates
the total-family envelope theorem. It does not establish that any
particular SU(2) cutoff states decouple.

\subsection{A fixed positive lattice gap can destroy time continuity}

Let $A_k=\operatorname{diag}(1,e^{-c})$ with $c>0$, and $a_k=1/k$.
For the normalized excited vector $v$,
\[
 \langle v,A_k^{\lfloor kt\rfloor}v\rangle
 \longrightarrow0\quad(t>0),
 \qquad \|v\|^2=1.
\]
The candidate limit equals the vacuum projection at every positive time
but the identity at time zero. It is not strongly continuous and cannot
be $e^{-tH}$ for a selfadjoint Hamiltonian on this nonzero excitation
space. Alternatively, discarding the excitation leaves a trivial
one-dimensional vacuum theory. This is the explicit obstruction captured
by Proposition~\ref{ham:ct:infinite_rate}.

\paragraph{Status.}
The transfer, dense-family and envelope results above are fully proved
conditional statements. They identify what an actual SU(2) continuum
construction and a physical-rate estimate would imply. They do not
prove that the required limits or weak-coupling estimates exist for
four-dimensional SU(2). The abstract spectral principles and their
reconstruction antecedents are established mathematics; no historical
priority claim is made here.

\section{Conditional reconstruction of the physical time Hilbert space}
\label{ham:ctr:section}

The gap-transfer theorem starts with a constructed continuum Hamiltonian.
The following lemma spells out sufficient correlation-limit hypotheses
that construct its time Hilbert space and semigroup. It does not construct
the required SU(2) limits. Nor does time-semigroup reconstruction alone
prove the locality, Euclidean covariance, field regularity, gauge-observable
structure or other properties needed for a four-dimensional relativistic
Yang--Mills theory. Full Euclidean reconstruction has additional hypotheses;
the corrected Osterwalder--Schrader framework is the relevant classical
reference \cite{OS2}.

Use the physical $A_k$, vacua $\Omega_k$, and projectors $Q_k$ from
Section~\ref{ham:ct:section}. For an index set $\mathscr F$ of renormalized
observable labels, suppose centered vectors
$u_k(F)\in Q_k\mathcal H_k$ have been specified. For $t\ge0$, define
\[
 C_{k;F,G}(t)
 =\langle u_k(F),A_k^{\lfloor t/a_k\rfloor}u_k(G)\rangle.
\]
No boundedness of the underlying observable is required at this abstract
stage, but the vectors and their limiting norms must be finite.

\begin{lemma}[Rounding at positive physical time]\label{ham:ctr:rounding}
For a positive contraction $A$ and integers $0\le r\le s$,
\begin{equation}\label{ham:ctr:roundbound}
 \|A^s-A^r\|
 \le\frac{s-r}{r+1}.
\end{equation}
Consequently, if $a_k\to0$ and two integer sequences satisfy
$a_kr_k\to t>0$ and $a_ks_k\to t$, then
$\|A_k^{r_k}-A_k^{s_k}\|\to0$.
\end{lemma}

\begin{proof}
Spectral calculus gives
$\|A^j(I-A)\|\le\sup_{0\le x\le1}x^j(1-x)\le1/(j+1)$.
Telescope $A^r-A^s$ and sum these bounds. For the last assertion, put
$r=\min(r_k,s_k)$ and $s=\max(r_k,s_k)$ in
\eqref{ham:ctr:roundbound}. Its right-hand side equals
$|a_ks_k-a_kr_k|/[a_k(\min(r_k,s_k)+1)]\to0$.
\end{proof}

\begin{theorem}[Time reconstruction from complete Gram limits]
\label{ham:ctr:reconstruction}
Assume the following three premises.
\begin{enumerate}
\item For every $F,G\in\mathscr F$ and every fixed $t\ge0$, the finite
limit
\begin{equation}\label{ham:ctr:limits}
 C_{F,G}(t)=\lim_{k\to\infty}C_{k;F,G}(t)
\end{equation}
exists. In particular, all mixed equal-time Gram limits exist and
$\|u_k(F)\|^2\to C_{F,F}(0)<\infty$.
\item For every $F\in\mathscr F$ there is continuity at the time origin:
\begin{equation}\label{ham:ctr:continuity}
 \lim_{t\downarrow0}C_{F,F}(t)=C_{F,F}(0).
\end{equation}
\item For at least one label $F_*$,
\begin{equation}\label{ham:ctr:nonzero}
 C_{F_*,F_*}(0)>0.
\end{equation}
\end{enumerate}
Then there exist a nonzero Hilbert space $\mathcal K$, vectors
$\eta_F\in\mathcal K$, and a nonnegative selfadjoint operator $H_*$
such that
\begin{equation}\label{ham:ctr:realization}
 C_{F,G}(t)=\langle\eta_F,e^{-tH_*}\eta_G\rangle
 \quad(t\ge0),
\end{equation}
and the linear span of
$\{e^{-sH_*}\eta_F:F\in\mathscr F,\ s\ge0\}$ is dense in
$\mathcal K$. Adjoining a unit vector $\Omega$ gives
\[
 \mathcal H=\mathbb C\Omega\oplus\mathcal K,
 \qquad H=0\oplus H_*.
\]
The vacuum need not be unique until a positive gap on $\mathcal K$ is
proved. The realization is unique up to a unitary map preserving every
time-shifted vector and hence preserving the semigroup.

Moreover, \eqref{ham:ctr:limits} has the same value if
$\lfloor t/a_k\rfloor$ is replaced by any integers $n_k\ge0$ with
$a_kn_k\to t$, including $t=0$.
\end{theorem}

\begin{proof}
\emph{Positive Gram kernel.}
Start with the complex vector space of finite formal sums of symbols
$[F,s]$, where $F\in\mathscr F$ and $s\ge0$. Define
\[
 \langle[F,s],[G,t]\rangle_0=C_{F,G}(s+t).
\]
For a finite sum $X=\sum_i c_i[F_i,s_i]$, its cutoff representative is
\[
 W_kX=\sum_i c_i A_k^{\lfloor s_i/a_k\rfloor}u_k(F_i).
\]
The two exponents
$\lfloor s_i/a_k\rfloor+\lfloor s_j/a_k\rfloor$ and
$\lfloor(s_i+s_j)/a_k\rfloor$ differ by at most one. If
$s_i+s_j>0$, Lemma~\ref{ham:ctr:rounding} makes their operator difference
tend to zero. If $s_i+s_j=0$, both exponents equal zero exactly.
The norms of each $u_k(F_i)$ are bounded by \eqref{ham:ctr:limits}, so
\begin{equation}\label{ham:ctr:gramlimit}
 \lim_{k\to\infty}\langle W_kX,W_kY\rangle
 =\langle X,Y\rangle_0
\end{equation}
for all finite sums $X,Y$. Hence $\langle\cdot,\cdot\rangle_0$ is
positive semidefinite. Quotient by its null space and complete to obtain
$\mathcal K$. Write $\eta_F$ for the class of $[F,0]$.
Premise~\eqref{ham:ctr:nonzero} makes this Hilbert space nonzero.

\emph{Contractive time shifts.}
On formal symbols put $R_u[F,s]=[F,s+u]$ for $u\ge0$.
For fixed $u>0$, the powers in
$W_kR_uX$ and $A_k^{\lfloor u/a_k\rfloor}W_kX$ differ termwise by
at most one, and their smaller exponents are at least
$\lfloor u/a_k\rfloor$. The bound
\eqref{ham:ctr:roundbound}, the finite number of terms and the bounded
norms of their vectors imply
\[
 \|W_kR_uX-A_k^{\lfloor u/a_k\rfloor}W_kX\|\longrightarrow0.
\]
Therefore \eqref{ham:ctr:gramlimit} and $\|A_k\|\le1$ give
$\|R_uX\|_0\le\|X\|_0$. At $u=0$ the statement is immediate.
Every $R_u$ preserves the null space and extends to a contraction $S_u$
on $\mathcal K$.

\emph{Semigroup and symmetry.}
The exact identity on labels gives $S_uS_v=S_{u+v}$ and $S_0=I$.
The equality
\[
 C_{F,G}(s+u+t)=C_{F,G}(s+t+u)
\]
shows that $S_u$ is symmetric on the dense formal span, hence selfadjoint
on $\mathcal K$. Since $S_u=S_{u/2}^2$, it is also positive.

\emph{Strong continuity.}
For a base vector,
\[
 \|(S_u-I)\eta_F\|^2
 =C_{F,F}(2u)-2C_{F,F}(u)+C_{F,F}(0)\longrightarrow0
\]
by \eqref{ham:ctr:continuity}. For a shifted vector, commutation and
contractivity give
\[
 \|(S_u-I)S_s\eta_F\|
 \le\|(S_u-I)\eta_F\|\longrightarrow0.
\]
Finite sums form a dense subspace. Uniform contractivity extends
continuity at zero to every vector; the semigroup property then gives
strong continuity at all times.

\emph{Selfadjoint generator.}
An elementary spectral argument completes this step. If $S_1x=0$, then
$\|S_{1/2}x\|^2=\langle x,S_1x\rangle=0$. Repeating gives
$S_{2^{-n}}x=0$ for every $n$, and continuity at zero implies $x=0$.
Thus $S_1$ is injective. Spectral calculus defines the densely defined
nonnegative selfadjoint operator $H_*=-\log S_1$.
For every positive integer $n$, $S_{1/n}$ is the unique positive
$n$th root of $S_1$. Hence
$S_{m/n}=S_1^{m/n}=e^{-(m/n)H_*}$ for all nonnegative rational times.
Strong continuity extends the identity to every real $t\ge0$.
The defining inner product gives \eqref{ham:ctr:realization}; the formal
span gives the stated totality. Any other realization with that totality
has the same Gram kernel on its time-shifted vectors, so the
correspondence of labels extends to a unitary intertwiner.

\emph{Rounding at the origin.}
At $t>0$, Lemma~\ref{ham:ctr:rounding} already proves invariance under
any integer sequence with $a_kn_k\to t$. Suppose now $a_kn_k\to0$.
For any fixed $\varepsilon>0$, eventually
$n_k\le\lfloor\varepsilon/a_k\rfloor$. Spectral calculus gives
\begin{align*}
 \|(I-A_k^{n_k})u_k(F)\|^2
 &\le\langle u_k(F),(I-A_k^{n_k})u_k(F)\rangle\\
 &\le C_{k;F,F}(0)-C_{k;F,F}(\varepsilon).
\end{align*}
Take the limit superior in $k$, then let $\varepsilon\downarrow0$.
Premises~\eqref{ham:ctr:limits}--\eqref{ham:ctr:continuity} force the result to
zero. Cauchy--Schwarz and boundedness of $\|u_k(G)\|$ prove the same
rounding assertion for mixed correlations at $t=0$.
\end{proof}

\begin{corollary}[The gap in the reconstructed time theory]
\label{ham:ctr:gap}
If, in addition to Theorem~\ref{ham:ctr:reconstruction}, the physical cutoff
operators satisfy \eqref{ham:ct:cutoffbound} and
$\liminf\delta_k/a_k\ge M>0$, then
\[
 \sigma(H_*)\subset[M,\infty),
 \qquad \ker(0\oplus H_*)=\mathbb C\Omega.
\]
\end{corollary}

\begin{proof}
For a finite formal sum $X$, the centered cutoff vectors $W_kX$ have
norm limit $\|X\|_0$. Their physical-time correlations converge to
$\langle X,S_tX\rangle$: expand the finite sum and use
Lemma~\ref{ham:ctr:rounding} on the three summed exponents. Thus the dense
formal span satisfies the hypotheses of Corollary~\ref{ham:ct:one_time}
at any fixed $t>0$.
\end{proof}

\paragraph{What has and has not been constructed.}
The proof above constructs a nonzero time Hilbert space and a selfadjoint
Hamiltonian from the stated complete family of limiting correlations.
It also proves that a certified physical-rate bound passes to that
Hamiltonian. It does not produce the correlation limits in
\eqref{ham:ctr:limits}, verify the continuity condition
\eqref{ham:ctr:continuity}, or exhibit a nonzero SU(2) state satisfying
\eqref{ham:ctr:nonzero} along $\beta_k\to\infty$, $a_k\to0$ and
$a_kL_k\to\infty$. Those are mathematical premises to be proved for
the actual renormalized observables. Additional spatial and field-theory
axioms remain necessary to identify this abstract time theory with
four-dimensional continuum Yang--Mills.

\section{What rescaling and exact time blocking can establish}
\label{ham:obs:section}

The following results identify limitations of specific proposed extensions
of the existing lattice certificates. They do not exclude a continuum
Yang--Mills construction, a different strong-coupling estimate, or a
controlled renormalization argument. All lengths below are numerical values
in one fixed physical reference unit; changing that unit does not change
any assertion about a zero, finite, or infinite limiting rate.

\subsection{Exact time blocking preserves the physical spectral rate}

\begin{proposition}[Exact blocking identity]\label{ham:obs:blocking}
Let $A$ be a positive selfadjoint contraction on a Hilbert space, with a
normalized vacuum $\Omega$, $A\Omega=\Omega$, and vacuum projector
$P=|\Omega\rangle\langle\Omega|$. Put $Q=I-P$ and
\[
 q=\|AQ\|,\qquad \delta=-\log q,
\]
where $-\log0=+\infty$. For every integer $b\ge1$, the exact transfer
operator across $b$ time steps is $A^{[b]}=A^b$, and
\[
 \|A^{[b]}Q\|=q^b,
 \qquad \delta^{[b]}=b\delta.
\]
If one original step has physical duration $a$, the blocked step has
duration $a^{[b]}=ba$. Consequently,
\[
 \boxed{\displaystyle
 \frac{\delta^{[b]}}{a^{[b]}}=\frac{\delta}{a}.}
\]
The statement includes volume-dependent operators and block sizes that
vary along a sequence, as an identity at each sequence index.
\end{proposition}

\begin{proof}
Selfadjointness and $A\Omega=\Omega$ give $AQ=QA$, so the restriction
$A_Q=A|_{Q\mathcal H}$ is positive and selfadjoint. The spectral theorem
gives $\|A_Q^b\|=\|A_Q\|^b$. Taking logarithms proves the assertion.
For an unnormalized transfer operator $T$ with vacuum eigenvalue
$\lambda_0$, the normalized blocked operator is
$T^b/\lambda_0^b=(T/\lambda_0)^b$, giving precisely the same identity.
\end{proof}

Thus exact grouping of time steps does not convert a fixed-cutoff spectral
rate into a different rate in physical units. Multiplying $T$ by an overall
normalization likewise leaves eigenvalue ratios unchanged. A claimed
effective coupling after blocking needs an identity or a controlled
approximation for the blocked operator; the blocking identity alone does
not supply one.

\subsection{A divergent physical lower rate is incompatible with nonzero
continuum states}

\begin{proposition}[Collapse under a divergent lower rate]
\label{ham:obs:collapse}
For each $k$, let $A_k$ be a positive selfadjoint contraction with an
orthogonal vacuum projector $P_k$, $A_kP_k=P_k$, and $Q_k=I-P_k$. Suppose
\[
 \|A_kQ_k\|\le e^{-d_k},\qquad a_k\downarrow0,
 \qquad \frac{d_k}{a_k}\longrightarrow+\infty.
\]
Let $u_k$ satisfy $\sup_k\|u_k\|<\infty$ and
$\|P_ku_k\|\to0$. Then, for every $t>0$ and integers $n_k(t)$ with
$a_kn_k(t)\to t$,
\[
 \langle u_k,A_k^{n_k(t)}u_k\rangle\longrightarrow0.
\]
In particular, these correlations cannot converge to
$\langle u,e^{-tH}u\rangle$ for a nonzero vector $u$ and a nonnegative
selfadjoint continuum Hamiltonian $H$, even at one fixed $t>0$.
\end{proposition}

\begin{proof}
The orthogonal vacuum decomposition and positivity give
\[
 0\le\langle u_k,A_k^{n_k(t)}u_k\rangle
 \le\|P_ku_k\|^2
   +e^{-n_k(t)d_k}\|Q_ku_k\|^2.
\]
Since $n_k(t)d_k=(a_kn_k(t))(d_k/a_k)\to+\infty$, the right-hand side
tends to zero. For nonzero $u$, however, its spectral measure for $H$
is a nonzero positive finite measure on $[0,\infty)$, and
\[
 \langle u,e^{-tH}u\rangle
 =\int_{[0,\infty)}e^{-tE}\,d\mu_u(E)>0.
\]
This proves the contradiction. Alternatively, if convergence is assumed
for all $t>0$ to a strongly continuous semigroup, vanishing of every
such autocorrelation contradicts its limit $\|u\|^2>0$ as $t\downarrow0$.
\end{proof}

This proposition allows arbitrary field-renormalization factors inside
$u_k$: the condition is on the norms of the already renormalized states.
It does not forbid every Euclidean random-field limit. It says that a
limit of the specified correlations either has no nonzero physical vacuum
complement or fails the stated Hilbert-space convergence assumptions.

\paragraph{Application to the existing volume-uniform certificate.}
Recall the proved bound
\[
 \delta_L(\beta)\ge K(\beta),\qquad
 K(\beta)=\operatorname{arcosh}
 \frac{1-9\beta+12\beta^2}{\beta(1+3\beta)},
 \qquad L\ge2,\quad0<\beta<1/9.
\]
If $0<\beta_k\le\beta_*<1/9$, then
\[
 K(\beta_k)\ge K(\beta_*)>0.
\]
Indeed, the argument of arcosh minus one is
$(1-9\beta)(1-\beta)/[\beta(1+3\beta)]$; its numerator is no smaller
and its denominator no larger when $\beta\le\beta_*$. Hence
$K(\beta_k)/a_k\to+\infty$ whenever $a_k\to0$, independently of the
growth of $L_k$. Proposition~\ref{ham:obs:collapse} applies.
Assigning successively smaller values to $a$ while remaining uniformly
inside this strong-coupling region therefore does not yield the required
nontrivial continuum quantum dynamics.

\subsection{Approaching the boundary of the sufficient criterion}

\begin{proposition}[Exact boundary asymptotics of the lower-bound function]
\label{ham:obs:boundary}
Set $\epsilon=1-9\beta\in(0,1)$. Then
\[
 \cosh K(\beta)-1
 =\frac{3\epsilon(8+\epsilon)}{(1-\epsilon)(4-\epsilon)}
 =:y(\epsilon).
\]
For all $0<\epsilon<1$,
\[
 \frac{12(8+\epsilon)}{8+14\epsilon+5\epsilon^2}
 \le\frac{K(\beta)^2}{\epsilon}
 \le\frac{6(8+\epsilon)}{(1-\epsilon)(4-\epsilon)}.
\]
Consequently,
\[
 \boxed{\displaystyle
 K(\beta)\sim\sqrt{12(1-9\beta)}
 \quad\text{as }\beta\uparrow1/9.}
\]
If $a_k\to0$ and $K(\beta_k)/a_k$ stays bounded, it is necessary that
$\beta_k\to1/9$ and $1-9\beta_k=O(a_k^2)$.
\end{proposition}

\begin{proof}
The formula for $y$ follows by substituting $\beta=(1-\epsilon)/9$.
For $y>0$,
\[
 \operatorname{arcosh}(1+y)
 =\int_0^y\frac{ds}{\sqrt{s(2+s)}}.
\]
Bounding $2+s$ between $2$ and $2+y$ gives
\[
 \frac{4y}{2+y}
 \le[\operatorname{arcosh}(1+y)]^2\le2y.
\]
Substitution of the exact rational $y(\epsilon)$ proves the two displayed
bounds. Both bounds for $K^2/\epsilon$ tend to $12$. If $K/a$ is bounded,
then $K\to0$, which forces $\epsilon\to0$. The same two bounds then
give $\epsilon=O(K^2)=O(a^2)$.
\end{proof}

This is a necessary compatibility condition for this particular lower
bound. It does not identify $\beta=1/9$ as a physical critical point.
The vanishing of a sufficient bound is not a proof that the actual gap
vanishes. Approaching this finite coupling also does not supply the
weak-bare-coupling continuum direction of the Wilson theory.

\subsection{The generic finite-box bound vanishes in physical units}

\begin{proposition}[Loss of the generic certificate under cutoff removal]
\label{ham:obs:generic}
Let $a_k\to0$, let $L_k$ be positive integers, and suppose that eventually
\[
 a_kL_k\ge\ell_*>0,\qquad\beta_k\ge\beta_*>0.
\]
For the existing finite-box lower-bound function
\[
 B_k=-\log\tanh(3\beta_kL_k^3)
 =2\operatorname{atanh}(e^{-6\beta_kL_k^3}),
\]
one has
\[
 \boxed{\displaystyle \frac{B_k}{a_k}\longrightarrow0.}
\]
In fact, put $c=6\beta_*\ell_*^3$. For each integer $N\ge1$,
\[
 \frac{B_k}{a_k}
 \le\frac{8N!}{3c^N}\,a_k^{3N-1}
\]
once $6\beta_kL_k^3\ge\log2$.
\end{proposition}

\begin{proof}
Write $x_k=6\beta_kL_k^3\ge c/a_k^3$. For $0\le z\le1/2$, the
positive series for atanh gives
\[
 2\operatorname{atanh}z
 \le\frac{2z}{1-z^2}\le\frac83z.
\]
Also $e^x\ge x^N/N!$ for every $x>0$. Therefore
\[
 \frac{B_k}{a_k}
 \le\frac{8e^{-x_k}}{3a_k}
 \le\frac{8N!}{3a_kx_k^N}
 \le\frac{8N!}{3c^N}\,a_k^{3N-1}.
\]
Taking $N=1$ already proves the limit; larger $N$ quantify its rapid
decay.
\end{proof}

The hypotheses include the requested situation $\beta_k\to\infty$ and
$a_kL_k\to\infty$, and even a box whose physical size remains bounded
below by a fixed positive length. The conclusion concerns $B_k$, which
is only a lower bound for the actual logarithmic gap $\delta_k$.
It does not imply $\delta_k/a_k\to0$. It proves that this particular
generic finite-volume estimate cannot provide a positive continuum
lower bound along any such isotropic sequence.

\section{Nonclosure of Wilson convolution under time blocking}
\label{ham:obs:kineticsection}

This section concerns a precisely specified component: normalized
one-link kinetic convolution on $SU(2)$, restricted if desired to class
functions. It is not a statement that the complete interacting blocked
Wilson transfer operator equals this component. The purpose is to show,
with both an exact proof and a numerical certificate, why an effective
Wilson coupling cannot simply be assigned after blocking.

Let
\[
 p_\beta(U)=\frac{\beta}{2I_1(\beta)}
 e^{\beta\operatorname{Tr}U/2},\qquad\beta>0,
\]
and let $C_\beta$ be convolution with $p_\beta$ in normalized Haar measure.
At $\beta=0$, use the continuous extension $p_0=1$.
The spin-$j$ character $\chi_j$ is a normalized class function with
eigenvalue
\[
 C_\beta\chi_j=
 \frac{I_{2j+1}(\beta)}{I_1(\beta)}\chi_j.
\]
For completeness, writing $n=2j+1$ and $\operatorname{Tr}U=2\cos\theta$,
the scalar Fourier coefficient before normalization is
\[
 \frac1n\frac2\pi\int_0^\pi
 e^{\beta\cos\theta}\sin\theta\sin(n\theta)\,d\theta
 =\frac{I_{n-1}(\beta)-I_{n+1}(\beta)}{n}
 =\frac{2I_n(\beta)}\beta.
\]
This gives the stated eigenvalue after dividing by $2I_1(\beta)/\beta$.
The Bessel integral and recurrence used here are the standard identities
in NIST DLMF 10.32.3 and 10.29.1.

\subsection{Exact failure of closure under convolution powers}

\begin{proposition}[Wilson one-link convolution is not closed under blocking]
\label{ham:obs:nonclosure}
For every finite $\beta>0$ and integer $b\ge2$, there is no finite
$\gamma\ge0$ such that
\[
 C_\beta^b=C_\gamma.
\]
The same assertion holds if both operators are restricted to the
conjugation-invariant class-function subspace.
\end{proposition}

\begin{proof}
The case $\gamma=0$ is excluded by any nonconstant character:
$C_0\chi_j=0$ for $j>0$, whereas every $I_{2j+1}(\beta)$ is positive.
For $\gamma>0$, equality would imply
\[
 \left(\frac{I_n(\beta)}{I_1(\beta)}\right)^b
 =\frac{I_n(\gamma)}{I_1(\gamma)}
 \qquad(n=1,2,\ldots).
\]
Taking successive ratios gives
\[
 \left(\frac{I_{n+1}(\beta)}{I_n(\beta)}\right)^b
 =\frac{I_{n+1}(\gamma)}{I_n(\gamma)}.
\]
Here the required large-order limit has a direct proof from the positive
series, without an asymptotic remainder assumption. For fixed $x>0$,
\[
 I_n(x)=\frac{(x/2)^n}{n!}S_n(x),\qquad
 S_n(x)=\sum_{m=0}^\infty
 \frac{(x^2/4)^m}{m!(n+1)\cdots(n+m)}.
\]
The empty product is one, and
\[
 1\le S_n(x)\le
 \exp\!\left(\frac{x^2}{4(n+1)}\right)\longrightarrow1.
\]
It follows that
\[
 (n+1)\frac{I_{n+1}(x)}{I_n(x)}\longrightarrow\frac{x}{2}.
\]
Multiplying the preceding equality of ratios by $n+1$, its left side
therefore tends to zero when $b\ge2$, while the right side tends to
$\gamma/2>0$. This contradiction proves the proposition.
\end{proof}

The proof uses every representation through its large-order limit.
It does not assume a representation cutoff. It refutes exact closure of
this one-link family; a claim about a complete interacting renormalization
step still requires analysis of that full operator and all generated
interactions.

\subsection{A positive operator-norm distance, certified without sampling}

Write
\[
 r(x)=\frac{I_2(x)}{I_1(x)},\qquad
 s(x)=\frac{I_3(x)}{I_1(x)},\qquad r(0)=s(0)=0.
\]

\begin{lemma}[Strict monotonicity of the two ratios]\label{ham:obs:monotonic}
Both $r$ and $s$ are strictly increasing on $(0,\infty)$.
\end{lemma}

\begin{proof}
For $\nu=1,2$, let
\[
 Z_\nu(x)=\int_{-1}^{1}e^{xt}(1-t^2)^{\nu-1/2}\,dt,
 \qquad R_\nu(x)=I_{\nu+1}(x)/I_\nu(x).
\]
The positive integral representation of $I_\nu$ shows that $Z_\nu(x)$
is a positive constant times $x^{-\nu}I_\nu(x)$.
Differentiating the positive Bessel series gives
$I_\nu'(x)=I_{\nu+1}(x)+(\nu/x)I_\nu(x)$, hence
\[
 R_\nu(x)=\frac{d}{dx}\log Z_\nu(x),\qquad
 R_\nu'(x)=\operatorname{Var}_{\nu,x}(t)>0.
\]
The tilted distribution has positive density on $(-1,1)$, so its variance
is strictly positive. Its mean is positive for $x>0$ by symmetry of the
untilted weight. Thus $r=R_1$ is strictly increasing, and
$s=R_1R_2$ is the product of two positive strictly increasing functions.
The integral representation is NIST DLMF 10.32.2; the derivative identity
is DLMF 10.29.2. Their use here is exact.
\end{proof}

\begin{proposition}[Two-character norm-distance certificate]
\label{ham:obs:distance}
Fix $\beta>0$ and $\eta>0$ such that
\[
 d_1=r(\beta)^2-r(\eta)>0,\qquad
 d_2=s(\eta)-s(\beta)^2>0.
\]
Then, on $L^2(SU(2))$ and on its class-function subspace,
\[
 \boxed{\displaystyle
 \inf_{\gamma\ge0}\|C_\beta^2-C_\gamma\|
 \ge\min(d_1,d_2)>0.}
\]
\end{proposition}

\begin{proof}
The normalized characters of spins $1/2$ and $1$ give
\[
 \|C_\beta^2-C_\gamma\|
 \ge\max\!\left\{
 |r(\beta)^2-r(\gamma)|,
 |s(\beta)^2-s(\gamma)|\right\}.
\]
If $0\le\gamma\le\eta$, monotonicity gives a lower bound $d_1$ from
the first character. If $\gamma\ge\eta$, the second gives $d_2$.
Every $\gamma\ge0$ belongs to at least one of these cases.
Other representations cannot invalidate a norm lower bound witnessed
by these two normalized vectors.
\end{proof}

The standard-library rational checker supplies the following strict
lower bounds. The displayed decimal $\eta$ values are exact rational
inputs, not rounded evaluations of an unknown optimizer.
\begin{center}
\begin{tabular}{@{}cc>{\raggedright\arraybackslash}p{0.43\textwidth}@{}}
\toprule
$\beta$ & Exact rational split $\eta$ &
Lower bound for $\inf_{\gamma\ge0}\|C_\beta^2-C_\gamma\|$\\
\midrule
$1/10$ & $0.0024975729830188664$ & $0.00000008651687270520$\\
$177/500$ & $0.0309541343067307$ & $0.00001307871009344142$\\
$27/10$ & $1.1654038422471582$ & $0.00791096303511152187$\\
\bottomrule
\end{tabular}
\end{center}
These are lower bounds, not exact distances or claims of global
optimization. They concern the normalized one-link kinetic component.
They do not quantify the distance of the full interacting blocked
Wilson action from any proposed effective action.

\subsection{Complete arithmetic certificate and reproducibility}

For integer $n\ge0$ and positive rational $x$, the checker uses
\[
 I_n(x)=\sum_{k=0}^{\infty}
 t_k,\qquad
 t_k=\frac{(x/2)^{n+2k}}{k!(n+k)!}.
\]
After summing through $k=m$, set
\[
 \rho=\frac{(x/2)^2}{(m+2)(n+m+2)}.
\]
When $\rho<1$, positivity and decreasing successive term ratios give
\[
 \sum_{k=0}^{m}t_k
 < I_n(x)
 \le\sum_{k=0}^{m}t_k+\frac{t_{m+1}}{1-\rho}.
\]
The complete omitted tail is bounded before any ratio is formed. For
positive intervals $[a_-,a_+]$ and $[b_-,b_+]$, interval division uses
$[a_-/b_+,a_+/b_-]$. Every series, ratio, square, difference, and printed
decimal is evaluated using exact rational arithmetic. The script stops
when each absolute Bessel tail bound is less than $10^{-70}$ and stores
the exact partial sums, first omitted terms, ratio bounds, and complete
tail bounds as rational numerators and denominators.

The certificate for the convolution calculation also checks the rational
identities and provides instances of the exact bounds in
Proposition~\ref{ham:obs:boundary}. Those finite arithmetic instances do not
replace the analytic proof for all $\epsilon$.

\paragraph{Primary identity sources.}
NIST Digital Library of Mathematical Functions, equations 10.25.2,
10.29.1--2, and 10.32.2--3:
\url{https://dlmf.nist.gov/10.25.E2},
\url{https://dlmf.nist.gov/10.29.E2}, and
\url{https://dlmf.nist.gov/10.32.E2}.
The certificates above specialize these exact identities to the stated convolution comparison.

\section{Full-operator error budget for an approximate blocked transfer}
\label{ham:sec:comparison}

The following result gives a concrete sufficient error budget for a proposed
blocking or spectral approximation. It does not assert that this budget has
been established for weak-coupling SU(2). All operators in the comparison
must act on the same full physical Hilbert space. A compression to a finite
basis supplies only part of the required hypothesis.

\begin{theorem}[Normalized full-operator comparison]\label{ham:thm:blockcompare}
Let $A$ be a positive selfadjoint compact contraction on a physical Hilbert
space of dimension at least two, with largest eigenvalue $1$ and a simple vacuum. Let $b\ge1$ be an
integer, let $a>0$, and write
\[
 S=A^b,\qquad \tau=ba,\qquad
 \delta=-\log\lambda_1(A).
\]
Here $\lambda_1$ denotes the second eigenvalue, counting multiplicity.
If it is zero, use $-\log0=+\infty$ and $e^{-\infty}=0$.
Suppose $B$ is a positive selfadjoint compact operator on the same Hilbert
space, its largest eigenvalue is $1$, and
\[
 \lambda_1(B)\le q<1,\qquad \|S-B\|\le\varepsilon,
 \qquad q+\varepsilon<1.
\]
Then
\begin{equation}\label{ham:eq:blockcompare}
 \boxed{\displaystyle
 \frac{\delta}{a}\ge
 -\frac{1}{\tau}\log(q+\varepsilon).}
\end{equation}
The vacuum eigenvectors of $S$ and $B$ need not agree.
\end{theorem}

\begin{proof}
Let $v$ be the normalized leading eigenvector of $B$. For every unit
$x\perp v$,
\[
 \langle x,Sx\rangle
 \le\langle x,Bx\rangle+\varepsilon\le q+\varepsilon.
\]
The min--max principle therefore gives
$\lambda_1(S)\le q+\varepsilon$. Positivity implies
$\lambda_1(S)=\lambda_1(A)^b=e^{-b\delta}$.
Taking logarithms and dividing by $\tau=ba$ proves the claim.
The use of min--max, rather than identifying the approximate and exact
vacuum vectors, is what makes the vacuum treatment valid.
\end{proof}

\begin{corollary}[Accuracy needed in physical time]\label{ham:cor:blockbudget}
For a sequence of comparisons as in Theorem~\ref{ham:thm:blockcompare}, the
condition
\begin{equation}\label{ham:eq:budget}
 q_k+\varepsilon_k\le e^{-M\tau_k},\qquad M>0,
\end{equation}
implies $\delta_k/a_k\ge M$.
If $\tau_k\to0$, $q_k\le e^{-M\tau_k}$, and
$\varepsilon_k=o(\tau_k)$, then
\[
 \liminf_k\frac{\delta_k}{a_k}\ge M.
\]
More generally, if $\varepsilon_k/\tau_k\to c<M$, the guaranteed lower
limit is at least $M-c$. If $\tau_k$ instead stays at a fixed positive
physical scale, the finite margin in \eqref{ham:eq:budget} is sufficient;
one does not need to demand $o(a_k)$ error for that blocked step.
\end{corollary}

\begin{proof}
The first assertion is immediate from \eqref{ham:eq:blockcompare}.
For the second and third,
\[
 q_k+\varepsilon_k
 \le 1-(M-c)\tau_k+o(\tau_k).
\]
The right side is below one for all sufficiently large $k$, and
$-\log(1-u)/u\to1$ as $u\to0$. The same argument with any fixed
$\mu<M$ applies if only a liminf lower bound for the approximate rate is
known. Taking $\mu\uparrow M$ completes that version.
\end{proof}

Thus an error tending to zero is not, by itself, the correct requirement.
At a vanishing blocked time step, the error must be compared with that
physical time step. For example, the exact diagonal operators
\[
 B_k=\operatorname{diag}(1,1-M\tau_k),\qquad
 S_k=\operatorname{diag}(1,1-(M-c)\tau_k)
\]
have $\|S_k-B_k\|=c\tau_k\to0$ but limiting physical gaps $M$ and $M-c$.
This proves that the order-$\tau_k$ loss can actually occur.

\subsection{Including a finite basis and its complete complement}

Suppose $P$ is a finite-rank orthogonal projection, $Q=I-P$, and $B$ is
supported on $P\mathcal H$. To apply the theorem to $S$, it is enough to
prove, on the full Hilbert space,
\[
 \|P(S-B)P\|\le e_P,\qquad
 \|QSQ\|\le d_Q,\qquad
 \|PSQ\|\le\eta.
\]
Then
\begin{equation}\label{ham:eq:fullnormbudget}
 \|S-B\|\le
 \frac{e_P+d_Q+\sqrt{(e_P-d_Q)^2+4\eta^2}}{2}.
\end{equation}
Indeed, for $z=x+y$ with $x=Pz$ and $y=Qz$, the absolute quadratic form
of $S-B$ is bounded by
\[
 e_P\|x\|^2+2\eta\|x\|\|y\|+d_Q\|y\|^2.
\]
Its maximum on $\|x\|^2+\|y\|^2=1$ is the displayed two-by-two matrix
eigenvalue. Selfadjointness then gives the operator-norm bound.
This explicitly accounts for every discarded representation and for its
coupling to the retained subspace. A finite matrix eigenvalue without the
$d_Q$ and $\eta$ estimates is not a full spectral certificate.

\subsection{When vacuum normalization is not known exactly}

For positive compact $T$ and $B$ with $\|T-B\|\le\varepsilon$, denote
the two largest eigenvalues of $B$ by $\nu_0\ge\nu_1$.
Rayleigh--Ritz and min--max give
\[
 \lambda_0(T)\ge\nu_0-\varepsilon,\qquad
 \lambda_1(T)\le\nu_1+\varepsilon.
\]
Therefore, whenever $\nu_0-\varepsilon>\nu_1+\varepsilon>0$,
\begin{equation}\label{ham:eq:unnormalized}
 \log\frac{\lambda_0(T)}{\lambda_1(T)}
 \ge\log\frac{\nu_0-\varepsilon}{\nu_1+\varepsilon}>0.
\end{equation}
An independently certified lower bound on $\lambda_0(T)$ can replace
$\nu_0-\varepsilon$. An approximate vacuum energy cannot be silently
treated as exact when forming $T/\lambda_0$; its error belongs in this
certificate. Equations \eqref{ham:eq:fullnormbudget} and \eqref{ham:eq:unnormalized}
state one complete way to close the finite-basis and vacuum-normalization
errors if the necessary estimates can actually be established.

\subsection{What remains unverified for SU(2)}

No $B_k$, $q_k$, and full operator errors $\varepsilon_k$ satisfying
\eqref{ham:eq:budget} have been constructed here along a weak-coupling,
growing-volume SU(2) sequence. The theorem supplies a precise sufficient
criterion and error budget; it is not evidence that those bounds hold.
Even a successful spectral comparison must be combined with the continuum
construction and nontriviality hypotheses in the preceding sections.

\section{Why a finite-error correlator fit cannot certify the missing gap}
\label{ham:sec:hiddenmode}

This section is an exact finite-dimensional counterexample, not an SU(2)
simulation or a Yang--Mills mass estimate. It identifies a concrete
limitation of using numerical correlator agreement in place of a uniform
spectral theorem.

Choose $0<m<M$ and $0<\epsilon<1/2$. On $\mathbb C^3$, let
\[
 H=\operatorname{diag}(0,m,M),\quad
 \Omega=(1,0,0),\quad
 \psi_\epsilon=(0,\sqrt\epsilon,\sqrt{1-\epsilon}).
\]
The vacuum is unique and the exact physical gap is $m$. The normalized,
centered correlator of $\psi_\epsilon$ is
\[
 C_\epsilon(t)
 =\epsilon e^{-mt}+(1-\epsilon)e^{-Mt}.
\]
Nevertheless,
\begin{equation}\label{ham:eq:hiddenabsolute}
 0\le C_\epsilon(t)-e^{-Mt}
 =\epsilon(e^{-mt}-e^{-Mt})\le\epsilon
 \qquad\text{for every }t\ge0.
\end{equation}
Thus even a uniform absolute error tolerance over all times can hide a
smaller positive gap. The lower-energy term eventually dominates whenever
it has nonzero spectral weight. The crossover of the two contributions
occurs at
\[
 t_* = \frac{1}{M-m}\log\frac{1-\epsilon}{\epsilon}.
\]
The instantaneous effective rate is
\[
 m_{\rm eff}(t)=
 \frac{m\epsilon e^{-mt}+M(1-\epsilon)e^{-Mt}}
 {\epsilon e^{-mt}+(1-\epsilon)e^{-Mt}}
 \longrightarrow m.
\]
This is why the continuum implication uses a true exponential upper bound
on a total family of states, with controlled prefactors and limiting
norms. Ordinary finite-time fits and additive Monte Carlo error bars do
not supply that hypothesis.

For a reproducible illustration only, the companion rational checker uses
$M=1$, $m=1/10$, and $\epsilon=10^{-6}$. It certifies the displayed
crossover and sample correlator values using positive exponential and
logarithm series with explicit tails. These numbers describe the
counterexample above and have no SU(2) mass interpretation.

\section{Two-time comparison without a prefactor assumption}
Let $a>0$ be the physical lattice spacing. Suppose $T_a$ is a positive
self-adjoint contraction on a Hilbert space, $T_a\Omega_a=\Omega_a$,
$\|\Omega_a\|=1$, and $P_a$ projects onto $\mathbb C\Omega_a$.
Assume
\begin{equation}\label{ham:x14:transfer}
 \|T_a(I-P_a)\|\le e^{-g_a},\qquad g_a\ge0.
\end{equation}
The number $g_a$ can be a certified lower bound for the dimensionless
transfer-energy gap; it need not be its exact value.
For an arbitrary centered vector $v_a\perp\Omega_a$ define
\[
 C_a(n)=\langle v_a,T_a^n v_a\rangle,\qquad n=0,1,2,\ldots .
\]
The vector may depend on $a$ and its norm may diverge as $a\to0$.

\begin{lemma}[Two-time inequality]\label{ham:x14:two_time}
For integers $n_2\ge n_1\ge0$,
\begin{equation}\label{ham:x14:ratio}
 0\le C_a(n_2)\le e^{-g_a(n_2-n_1)}C_a(n_1).
\end{equation}
\end{lemma}
\begin{proof}
On $\Omega_a^\perp$, the spectral measure of $T_a$ is supported in
$[0,e^{-g_a}]$. For every spectral value $\lambda$ there,
$\lambda^{n_2}\le e^{-g_a(n_2-n_1)}\lambda^{n_1}$.
Integrate against the positive spectral measure of $v_a$.
When $n_1=0$, $T_a^0$ is the identity, including on the kernel of $T_a$.
\end{proof}

\begin{theorem}[Necessary scaling of any lattice gap certificate]\label{ham:x14:necessary}
Fix physical times $0<s<t$, and integers $n_a(s),n_a(t)$ with
$an_a(s)\to s$ and $an_a(t)\to t$. Suppose
\[
 C_a(n_a(s))\longrightarrow C(s)<\infty,\qquad
 C_a(n_a(t))\longrightarrow C(t)>0.
\]
Then $C(s)\ge C(t)>0$ and
\begin{equation}\label{ham:x14:necessary_rate}
 \limsup_{a\downarrow0}\frac{g_a}{a}
 \le\frac{\log(C(s)/C(t))}{t-s}<\infty .
\end{equation}
In particular a fixed positive lower bound $g_a\ge g_0>0$ is incompatible
with these two nonzero finite continuum correlations.
\end{theorem}
\begin{proof}
For sufficiently small $a$, $n_a(t)>n_a(s)$ and $C_a(n_a(t))>0$.
The contraction property implies $C_a(n_a(s))\ge C_a(n_a(t))$, so the
same ordering holds in the limit. Taking logarithms in
\eqref{ham:x14:ratio} gives
\[
 \frac{g_a}{a}
 \le
 \frac{\log\!\left(C_a(n_a(s))/C_a(n_a(t))\right)}
 {a(n_a(t)-n_a(s))}.
\]
The right side converges to that in \eqref{ham:x14:necessary_rate}.
\end{proof}

This necessary scaling conclusion permits arbitrary cutoff-dependent norms
and scalar renormalizations: only the two positive-time limits are assumed.
For a nonzero vector in a strongly continuous semigroup theory, both such
autocorrelations are finite and strictly positive. It therefore applies to
any proposed continuum identification with those limits.

\subsection{Fixed-prefactor collapse and physical-block contraction}
\begin{proposition}[Fixed lattice decay under polynomial rescaling]\label{ham:prop:scaling}
Let $a\downarrow0$, $d>0$, $\delta>0$, $0<C<\infty$,
$p<\infty$ and $c<\infty$ be fixed. Suppose numbers $C_a$ obey
\[
 |C_a|\le C a^{-p}e^{-\delta n_a},\qquad
 n_a\ge d/a-c.
\]
Then $C_a\to0$.
\end{proposition}
\begin{proof}
The right side is at most $Ce^{\delta c}a^{-p}e^{-\delta d/a}$.
Its logarithm is bounded above by a constant plus
$p\log(1/a)-\delta d/a$, which tends to $-\infty$.
\end{proof}

\begin{theorem}[A sufficient physical-block transfer criterion]\label{ham:x13:physical_block}
For each $a>0$, let $H_a\ge0$ be self-adjoint,
$T_a=e^{-aH_a}$, $H_a\Omega_a=0$, $\norm{\Omega_a}=1$, and let
$P_a$ project onto $\mathbb C\Omega_a$.
Suppose integers $n_a\ge1$, numbers $q\in(0,1)$ and $L_+<\infty$
satisfy, for all sufficiently small $a$,
\[
 an_a\le L_+,\qquad
 \norm{T_a^{n_a}(I-P_a)}\le q.
\]
Then the gap of $H_a$ on $\Omega_a^\perp$ is at least
$m_0=-\log q/L_+>0$.

Suppose in addition a continuum pair $(H,\Omega)$ with
$H\ge0$, $\ker H=\mathbb C\Omega$ has been constructed.
For a dense $D\subset\Omega^\perp$, suppose every $v\in D$ has
$v_a\perp\Omega_a$ with $\sup_a\norm{v_a}^2=C_v<\infty$ and, for all $t\ge0$,
\[
 \langle v,e^{-tH}v\rangle
 =\lim_{a\downarrow0}
 \langle v_a,T_a^{\lfloor t/a\rfloor}v_a\rangle .
\]
Then the continuum gap is at least $m_0$.
\end{theorem}
\begin{proof}
Functional calculus on $\Omega_a^\perp$ gives
$e^{-an_a\lambda}\le q$ for every spectral value $\lambda$ there.
Thus
\[
 \lambda\ge\frac{-\log q}{an_a}\ge\frac{-\log q}{L_+}=m_0.
\]
In particular,
$\langle v_a,T_a^kv_a\rangle\le C_ve^{-m_0ak}$
for every integer $k\ge0$. Let $k=\lfloor t/a\rfloor$ and pass to
the assumed limit; $ak\to t$. Theorem~\ref{ham:ct:envelope} now applies.
\end{proof}

\begin{lemma}[A finite upper mass witness]
If a reconstructed theory has a normalized vector
$v\perp\Omega$ in the quadratic-form domain of $H$, with
$\langle v,Hv\rangle\le E<\infty$, and its gap is at least $m_0>0$,
then its gap $m$ satisfies $m_0\le m\le E$.
\end{lemma}
\begin{proof}
The spectral measure of $v$ is a probability measure supported in
$[m,\infty)$, so its first moment is at least $m$.
\end{proof}

\section{Local-curvature obstruction to Wilson-family closure}
A plausible next idea is to integrate out fine variables until the
effective theory has a coupling in the certified range. This requires a
theorem about the \emph{actual effective interaction}. A change of scale
does not by itself preserve the tested three-link law.

\begin{proposition}[Even one convolution leaves the Wilson family]\label{ham:x14:convolution}
On $SU(2)\simeq S^3$, let
\[
 k_\kappa(g)=\frac{e^{\kappa\operatorname{Re}g}}{f(\kappa^2)},\qquad
 f(z)=\sum_{j\ge0}\frac{z^j}{4^j j!(j+1)!}.
\]
For $\kappa\ne0$, the convolution $k_\kappa*k_\kappa$ is not
$k_{\kappa'}$ for any real $\kappa'$.
\end{proposition}
\begin{proof}
Write $t=\operatorname{Re}g$ and identify quaternions with $\mathbb R^4$.
Since $\operatorname{Re}(x^{-1}g)=x\cdot g$ on the unit sphere,
\[
 (k_\kappa*k_\kappa)(g)
 =\frac{\int_{S^3}e^{\kappa x\cdot(e_0+g)}\,d\sigma(x)}
 {f(\kappa^2)^2}
 =\frac{f(2\kappa^2(1+t))}{f(\kappa^2)^2}.
\]
Here $|e_0+g|^2=2(1+t)$ and Haar measure is normalized.
The series gives
$f(0)=1$, $f'(0)=1/8$, $f''(0)=1/96$.
Consequently, with derivatives taken from the interval $[-1,1]$,
\begin{align*}
 \left.\frac{d^2}{dt^2}\log(k_\kappa*k_\kappa)(t)\right|_{t=-1}
 &=4\kappa^4\left(f''(0)-f'(0)^2\right)\\
 &=4\kappa^4\left(\frac1{96}-\frac1{64}\right)
 =-\frac{\kappa^4}{48}\ne0.
\end{align*}
In contrast, $\log k_{\kappa'}(t)=\kappa't-\log f(\kappa'^2)$
has identically zero second derivative.
\end{proof}

This is an exact elementary convolution model, not a computation of the
full four-dimensional renormalized action. Its role is to disprove the
shortcut that closure under integration can simply be assumed.
The actual four-dimensional effective action and its additional
interactions require their own bounds before a certificate for the
original Wilson law can be applied.

\subsection{A shifted-correlation spectral criterion}
\begin{proposition}[An exact continuum criterion]\label{ham:x14:criterion}
Let $H\ge0$ be self-adjoint with $\ker H=\mathbb C\Omega$, and let
$D\subset\Omega^\perp$ be dense. Fix any $s_0>0$ and write
$C_v(u)=\langle v,e^{-uH}v\rangle$.
For $\Delta>0$, the following are equivalent:
\begin{enumerate}
\item The spectrum on $\Omega^\perp$ is contained in $[\Delta,\infty)$.
\item For every $v\in D$ and every $t\ge0$,
\[
 C_v(s_0+t)\le e^{-\Delta t}C_v(s_0).
\]
\end{enumerate}
\end{proposition}
\begin{proof}
The first statement implies the second by integrating
$e^{-(s_0+t)\lambda}\le e^{-\Delta t}e^{-s_0\lambda}$.
Conversely, fix $0<\eta<\Delta$. Positivity gives
\[
 C_v(s_0+t)\ge
 e^{-(s_0+t)\eta}\|E_H((0,\eta])v\|^2.
\]
The second statement and $t\to\infty$ show that this projection vanishes
on $D$, hence on $\Omega^\perp$ by density.
Let $\eta\uparrow\Delta$. The zero eigenspace was specified in the
hypotheses.
\end{proof}

Proposition~\ref{ham:x14:criterion} is an equivalence, not a proof of its
second condition for Yang--Mills. It must hold for a dense physical family
and every time, not just the observables and times sampled numerically.

\section{Cutoff gaps, surviving spectra, and physical scale calibration}
%=====================================================================

\subsection{Lattice mass gap}

For each coupling under discussion, choose an infinite-volume Wilson Gibbs measure $\mu_\beta$ on $\Z^4$ that is invariant under lattice translations and reflection positive for time reflections. Outside a proved uniqueness range, this choice is part of the data and is not silently assumed to be independent of the limiting sequence. A weak limit of periodic-box measures has these properties \cite{OSLattice}. The Osterwalder--Seiler construction \cite{OSLattice} builds a Hilbert space $\HH_\beta$ from gauge-invariant local functions supported at positive times. It comes with a vacuum vector $\Omega=[1]$ and a transfer matrix $T_\beta$, a positive self-adjoint contraction with $T_\beta\Omega=\Omega$; see also L\"uscher \cite{Luscher}.

\begin{definition}\label{fg-def:gap}
The \emph{lattice mass gap} is $m(\beta):=-\log\|T_\beta|_{\Omega^\perp}\|\in[0,\infty]$. Along a trajectory of spacings $a_j\downarrow0$ and couplings $\beta_j$, the \emph{physical gap} is $\Delta_j:=m(\beta_j)/a_j$.
\end{definition}

\subsection{Effective mass: an upper bound from any two correlations}

\begin{lemma}[Effective mass]\label{fg-lem:effmass}
Let $\psi\in\Omega^\perp$ and $C(n):=\langle\psi,T_\beta^n\psi\rangle$. Then:
\begin{enumerate}[label=(\roman*)]
\item $C(n+1)\le e^{-m(\beta)}C(n)$;
\item for $n\ge1$, $C(n)^2\le C(n-1)\,C(n+1)$;
\item if $C(n)>0$ for all $n$, then $m_{\rm eff}(n):=\log\bigl(C(n)/C(n+1)\bigr)$ is non-increasing in $n$, satisfies $m(\beta)\le m_{\rm eff}(n)$ for every $n$, and converges to $-\log\max\operatorname{supp}\mu_\psi$.
\end{enumerate}
\end{lemma}
\begin{proof}
$\Omega^\perp$ reduces $T_\beta$. Let $\mu_\psi$ be the spectral measure of $\psi$ for $T_\beta|_{\Omega^\perp}$. It is supported in $[0,\lambda_1]$ with $\lambda_1=e^{-m(\beta)}$, and $C(n)=\int\lambda^n\,d\mu_\psi$. Then (i) follows from $\lambda^{n+1}\le\lambda_1\lambda^n$, and (ii) is Cauchy--Schwarz for $\lambda^{(n-1)/2}\cdot\lambda^{(n+1)/2}$. By (ii), the ratio $C(n)/C(n+1)$ is non-increasing; by (i) it is at least $e^{m(\beta)}$. Finally $C(n)^{1/n}\to\max\operatorname{supp}\mu_\psi$, because $L^n$ norms converge to the $L^\infty$ norm. So the Ces\`aro means of $m_{\rm eff}$, and hence the monotone sequence itself, converge to $-\log\max\operatorname{supp}\mu_\psi$.
\end{proof}

\subsection{Positive-time survival gives an upper bound on cutoff gaps}

\begin{proposition}[Survival and cutoff spectral upper bounds]\label{fg-prop:twosided}
Let $a_j\downarrow0$, let $\beta_j$ be couplings, and let $\psi_j\in\Omega_j^\perp\subset\HH_{\beta_j}$ satisfy $\|\psi_j\|\le B$. Write $C_j(n)=\langle\psi_j,T_{\beta_j}^n\psi_j\rangle$.
\begin{enumerate}[label=(\alph*)]
\item \emph{Survival forces $m=O(a)$.} Suppose that for some physical time $\tau>0$ and $n_j:=\lceil\tau/a_j\rceil$ we have $\liminf_j C_j(n_j)\ge\eta>0$. Then $\limsup_j m(\beta_j)/a_j\le\tau^{-1}\log(B^2/\eta)$.
\item \emph{A lattice-unit gap kills survival.} If $m(\beta_j)\ge m_0>0$ for all $j$, then $C_j(\lceil\tau/a_j\rceil)\le B^2e^{-m_0\tau/a_j}\to0$ for every $\tau>0$.
\end{enumerate}
\end{proposition}
\begin{proof}
Iterating Lemma~\ref{fg-lem:effmass}(i) gives $C_j(n)\le e^{-n\,m(\beta_j)}B^2$. Hence $m(\beta_j)\le n_j^{-1}\log(B^2/C_j(n_j))$. Divide by $a_j$ and use $a_jn_j\to\tau$; this proves (a). Part (b) is the same inequality read in the other direction.
\end{proof}

Part (b) is the fixed-lattice-rate obstruction. Part (a) shows that positive-time survival of a uniformly bounded probe imposes an \emph{upper} bound on the physical cutoff gaps. Neither statement supplies a positive lower bound on those cutoff gaps.

\subsection{A gapped limit need not have uniformly gapped cutoffs}

\begin{proposition}[Disappearing low-energy directions]\label{fg-prop:vanishinglow}
There are nonnegative cutoff Hamiltonians, probe families total at every cutoff, uniformly bounded probe norms, positive-time survival, and sparse-time low-energy exclusion for every $b<1$, whose limiting generated semigroup has gap one although the physical cutoff gaps tend to zero.
\end{proposition}
\begin{proof}
On the two-dimensional excitation space with orthonormal basis $e_1,e_2$, let
\[
H_j=\operatorname{diag}(j^{-1},1),\qquad
\varepsilon_j=e^{-j^2},\qquad
v_j^+=e_2+\varepsilon_j e_1,\quad
v_j^-=e_2-\varepsilon_j e_1.
\]
The two probes span the cutoff excitation space for every $j$ and have squared norms $1+\varepsilon_j^2\le2$. Their diagonal physical-time kernels are
\[
K_j^\pm(t)=e^{-t}+e^{-2j^2-t/j};
\]
the mixed kernel is $e^{-t}-e^{-2j^2-t/j}$. All kernels converge, including at time zero, to the rank-one kernel $e^{-t}$. The positive-kernel quotient therefore identifies the two limiting probes and gives the one-dimensional semigroup $e^{-t}$, whose excitation gap is one. Positive-time survival holds with $K_j^\pm(t)\ge e^{-t}$.

Take $T_j=j$. For every $0\le b<1$,
\[
e^{bT_j}K_j^\pm(T_j)
=e^{-(1-b)j}+e^{-2j^2+bj-1}\longrightarrow0.
\]
Thus the sparse-time exclusion hypothesis is satisfied for every fixed probe. Nevertheless the bottom of the cutoff excitation spectrum is $1/j$.

To express this in lattice units, choose $a_j=1/j$ and set $T_j^{\rm lat}=e^{-a_jH_j}$. Its lattice gap is $m_j=a_j/j=1/j^2$, so $m_j/a_j=1/j\to0$. Adjoining a separate vacuum eigenvector of energy zero changes none of these statements. The example is an abstract spectral construction, not a claim about actual Yang--Mills measures. It disproves the proposed general inference from a gapped surviving limit to a uniform positive lower bound on the complete cutoff spectra. \qedhere
\end{proof}

The degenerating Gram direction makes the mechanism explicit: $v_j^+-v_j^-=2\varepsilon_j e_1$ has vanishing norm and disappears in the limit. Totality at each individual cutoff does not supply uniform observability of every low-energy direction. A claim equating a gap in the limiting generated sector with a lower bound on all cutoff global gaps needs an additional uniform spectral-coverage hypothesis. The finite-probe conditioning factors in the spectral certificate theorem record precisely this type of issue.

\subsection{The asymptotic-scaling form}

The perturbative scaling convention discussed earlier \cite{BergClarke} uses the pure $SU(N)$ two-loop coefficients $b_0=\tfrac{11N}{3(4\pi)^2}$ and $b_1=\tfrac{34N^2}{3(4\pi)^4}$. For the Wilson action, $\beta=2N/g_0^2$, perturbative renormalization relates spacing and bare coupling by
\begin{equation}\label{fg-eq:AF}
a\Lambda_L=e^{-1/(2b_0g_0^2)}\,(b_0g_0^2)^{-b_1/(2b_0^2)}\bigl(1+O(g_0^2)\bigr),
\end{equation}
with $\Lambda_L$ the lattice $\Lambda$-parameter. Here $b_1/(2b_0^2)=51/121$ for every $N$ and $1/(2b_0g_0^2)=12\pi^2\beta/(11N^2)$. For $N=2$ this gives $a\Lambda_L=f(\beta)\,(1+O(1/\beta))$ with
\begin{equation}\label{fg-eq:f}
f(\beta)=\exp\Bigl(-\frac{3\pi^2\beta}{11}\Bigr)\Bigl(\frac{6\pi^2\beta}{11}\Bigr)^{51/121}.
\end{equation}
Equation~\eqref{fg-eq:AF} is a perturbative asymptotic relation with a specified lattice scale convention. A nonperturbative construction using it requires an independently justified scale-setting trajectory; it is neither an exact equality nor an unconditional estimate established here. One may instead formulate the required bounds directly in the independently defined physical spacing $a_j$.

\begin{target}[A stronger uniform-cutoff goal in a calibrated $SU(2)$ trajectory]\label{fg-tgt}
There are $\beta_1<\infty$ and $0<c\le C<\infty$ such that
\[
c\,f(\beta)\;\le\;m(\beta)\;\le\;C\,f(\beta)\qquad\text{for all }\beta\ge\beta_1 .
\]
\end{target}

If a trajectory independently satisfies $a\Lambda_L\asymp f(\beta)$, this lower inequality would give a positive uniform lower bound on the full cutoff physical gaps. The upper inequality is compatible with, and along a surviving trajectory follows from, Proposition~\ref{fg-prop:twosided} once its bounded-norm and positive-time survival assumptions hold. Quantification over every sufficiently large $\beta$ is a stronger research target than a subsequence construction. It is not proved here and is not a necessary condition for every method of obtaining a gapped continuum limit, as Proposition~\ref{fg-prop:vanishinglow} demonstrates.

An appropriate direct sufficient target along a chosen regulator sequence is
\[
\liminf_{j\to\infty}\frac{m_j}{a_j}>0,
\]
combined with a genuine continuum reconstruction, correct physical-time normalization, and a nonzero finite-energy surviving state. Alternatively, the sparse-time criteria can exclude low energy only in the surviving generated sector, with totality in the intended continuum excitation space. Neither route is completed by the fixed-spacing comparison theorem.

\paragraph{Scale setting and nontriviality.}
Mass-based scale setting is legitimate when positivity of the mass and a controlled nontrivial continuum construction have been independently established. Defining $a_j=m_j/\Delta$ by itself, however, only normalizes a number; it constructs no continuum field theory. The substantive requirements are convergence of the renormalized observable hierarchy, nontrivial surviving states, the appropriate spacetime axioms, and the required identification with Yang--Mills dynamics. Any asserted asymptotic calibration must be proved or stated explicitly as a further hypothesis. The definition of physical units and proof of physical content are separate steps.

\hypertarget{research_revision-physical-scale-calibration-flowed-observables-and-proof-dependencies}{%
\section{Physical scale, time smoothing and observable identification}\label{research_revision-physical-scale-calibration-flowed-observables-and-proof-dependencies}}

Physical scale calibration, smoothing and the choice of observable algebra impose separate requirements on a continuum construction. This section specifies those requirements and proves a reflection-positivity obstruction for time heat smoothing.

\hypertarget{research_revision-coupling-conventions-the-112-comparison-is-correct}{%
\subsection{Coupling normalization for a functional-inequality comparison}\label{research_revision-coupling-conventions-the-112-comparison-is-correct}}

Use the manuscript's Wilson convention

\[
 d\mu_{\beta_W}\propto \exp\!\left\{\frac{\beta_W}{N}
 \sum_{p}\operatorname{Re}\operatorname{Tr}U_p\right\}\,dU,
\]

with each geometric plaquette counted once. Shen--Zhu--Zhu use \(S=N\beta_{SZZ}\sum_p\operatorname{Re}\operatorname{Tr}U_p\), their equation (1.2). Thus

\[
\beta_W=N^2\beta_{SZZ}.
\]

Their SU(N) hypothesis is \(|\beta_{SZZ}|<1/[16(d-1)]\). For SU(2) in four dimensions this becomes exactly \(\beta_W<1/12\). Their result includes volume-uniform functional inequalities, uniqueness and exponential correlation decay. The sufficient total-variation range of Theorem IH.3 exceeds this displayed SU(2) condition. The theorems control different families of estimates, so this numerical comparison does not order all their conclusions. \cite{SZZ}

\hypertarget{research_revision-the-scale-relation-is-asymptotic-not-an-exact-theorem}{%
\subsection{The scale relation is asymptotic, not an exact theorem}\label{research_revision-the-scale-relation-is-asymptotic-not-an-exact-theorem}}

For $SU(2)$, $\beta_W=4/g_0^2$. The two-loop scale factor is defined in \eqref{fg-eq:f}. Its relation to a physical spacing is a perturbative asymptotic statement with a specified scale scheme \cite{BergClarke}.

For a rigorous continuum proof, one needs an actual trajectory \(a\mapsto\beta_W(a)\), a scale-setting definition, and proved estimates along that trajectory. A convenient exact target is

\[
\liminf_{a\downarrow0}\frac{m_{\rm lat}(a)}{a}\ge m_*>0.
\]

Together with a genuine nontrivial continuum reconstruction and uniform bounds for a total family of properly normalized physical states, this is sufficient for gap transfer. It is not necessary to prove a mass-gap statement at every bare coupling. The prefactor and onset of the correlation bound must also have the needed uniformity after renormalization. A perturbative formula for the running coupling alone does not supply these estimates.

\hypertarget{research_revision-the-u1z2-exclusion-rule-must-be-narrowed}{%
\subsection{Model-specific tests using Abelian and finite groups}\label{research_revision-the-u1z2-exclusion-rule-must-be-narrowed}}

Applicability to an Abelian or finite group is not, by itself, a contradiction. Strong-coupling decay may correctly apply to both groups. Moreover, exponential decay of gauge-invariant correlations at weak coupling is proved for finite gauge groups, including Z2, by Adhikari--Cao; Forsström treats finite Abelian groups directly. \cite{AdhikariCao,Forsstrom}

The useful obstruction is narrower: an argument claiming a weak-coupling continuum gap in a model with a proved massless phase has to identify an assumption that fails there. Fröhlich--Spencer's four-dimensional Abelian massless-phase theorem is a relevant check, with its exact action, observables and hypotheses preserved. Its large-N cyclic-group statement must not be replaced by a statement about Z2. \cite{FS}

A theorem using quaternion/S3 geometry is already specific to SU(2), even if some general machinery also applies to Abelian groups. Applicability to another group is not itself an error.

\hypertarget{research_revision-gradient-flow-is-useful-but-does-not-erase-reconstruction-obligations}{%
\subsection{Time smoothing and reflection positivity}\label{research_revision-gradient-flow-is-useful-but-does-not-erase-reconstruction-obligations}}

Lüscher's 2010 work motivates positive-flow-time gauge observables. Lüscher--Weisz prove their finiteness to all orders in perturbation theory once the underlying four-dimensional theory is renormalized. This does not by itself give a nonperturbative continuum construction or a uniform physical mass gap. \cite{WilsonFlow,FlowPert}

Four-dimensional gradient flow smooths in the Euclidean time direction as well as space. Consequently, a flowed observable nominally labelled by a point in positive time need not be measurable with respect to the original positive-time field algebra. The original reflection-positivity proof cannot simply be reused for the flowed algebra. The following direct counterexample makes that obstruction precise; the calculation is explicit.

\hypertarget{research_revision-proposition-rv.1-time-heat-smoothing-need-not-preserve-reflection-positivity}{%
\subsubsection{Proposition RV.1 (time heat smoothing need not preserve reflection positivity)}\label{research_revision-proposition-rv.1-time-heat-smoothing-need-not-preserve-reflection-positivity}}

Let \(\omega>0\), \(\tau>0\), and define the stationary covariance

\[
K_\tau(t)=\frac1{2\pi}\int_{\mathbb R}
\frac{e^{ipt}e^{-2\tau(p^2+\omega^2)}}{p^2+\omega^2}\,dp.
\]

The centered Gaussian process with covariance \(K_\tau\) is not reflection positive with respect to the positive-time algebra generated by its own time-labelled field. In contrast, the unsmoothed covariance \(K_0(t)=e^{-\omega|t|}/(2\omega)\) is reflection positive.

\textbf{Proof.} The Gaussian multiplier makes every differentiated integral absolutely convergent. Therefore \(K_\tau\) is a real even smooth function, and

\[
K_\tau(0)>0,\qquad K_\tau'(0)=0,\qquad
K_\tau''(0)=-\frac1{2\pi}\int_{\mathbb R}
\frac{p^2 e^{-2\tau(p^2+\omega^2)}}{p^2+\omega^2}\,dp<0.
\]

For small positive \(h\), Taylor expansion gives

\[
K_\tau(0)K_\tau(2h)-K_\tau(h)^2
=K_\tau(0)K_\tau''(0)h^2+O(h^4)<0.
\]

For each such \(h\), continuity implies that, for a sufficiently small \(\epsilon>0\),

\[
\det\begin{pmatrix}
K_\tau(2\epsilon)&K_\tau(2\epsilon+h)\\
K_\tau(2\epsilon+h)&K_\tau(2\epsilon+2h)
\end{pmatrix}<0.
\]

This is the reflection Gram matrix of the two time evaluations at \(\epsilon\) and \(\epsilon+h\). A negative determinant gives a vector of coefficients with negative reflection quadratic form, violating positivity. If one works only with smooth compactly supported test functions, approximate the two positive-time point evaluations by smooth nonnegative mollifiers with supports remaining in positive time. Continuity of the covariance preserves the negative quadratic form for sufficiently narrow mollifiers. For \(\tau=0\), the corresponding kernel is \(e^{-\omega s}e^{-\omega t}/(2\omega)\) for \(s,t>0\), a positive rank-one kernel. This proves both assertions. \(\square\)

The same calculation applies to each fixed spatial Fourier mode of a heat-smoothed free massive scalar field. It does not show that flowed observables are useless, nor that Yang--Mills gradient flow always violates a particular chosen physical construction. It shows that preservation of a positive-time algebra is an additional issue. One can instead retain the original physical Hilbert space, establish suitable approximation/localization theorems, or analyze an appropriate spatial-only smoothing scheme; each route needs its own proof.

\hypertarget{research_revision-gaussian-fields-and-non-gaussian-composite-observables}{%
\subsection{Gaussian fields and non-Gaussian composite observables}\label{research_revision-gaussian-fields-and-non-gaussian-composite-observables}}

A continuum Yang--Mills identification requires the appropriate local observables and dynamics, together with Euclidean symmetry, locality, clustering and nontriviality. Lattice symmetry alone does not establish these limiting properties.

A blanket demand that ``some observable have a nonzero fourth cumulant'' is too weak to certify interactions. For a standard normal variable \(X\), the composite observable \(Y=X^2-1\) already has fourth cumulant 48. Indeed,

\[
\log\mathbb E e^{tY}=-t-\tfrac12\log(1-2t)
=\sum_{n\ge2}\frac{2^{n-1}}n t^n,
\]

so \(\kappa_n(Y)=2^{n-1}(n-1)!\) for \(n\ge2\). Thus non-Gaussian gauge-invariant composites can occur even over a Gaussian underlying field. A claim of nontrivial interacting Yang--Mills behavior must specify the fields, algebra and dynamical property being distinguished.

\hypertarget{research_revision-what-an-exact-bootstrap-certificate-must-establish}{%
\subsection{What an exact bootstrap certificate must establish}\label{research_revision-what-an-exact-bootstrap-certificate-must-establish}}

Kazakov--Zheng derive linear closed SU(2) loop equations and bootstrap plaquette bounds using semidefinite optimization. Their paper identifies MOSEK and discusses numerical precision limitations. These finite-dimensional constraints motivate exact certification of the resulting bounds. \cite{KZFinite}

A rigorous certificate must include the exact loop identities and all normalization conventions, a mathematically valid feasible relaxation containing the actual Gibbs expectations, exact dual coefficients, positive-semidefinite slack certificates, and a verified weak-duality calculation. An interval eigenvalue bound is useful only after propagating every residual into an explicit objective correction. A list of floating-point optima is not a certificate. Finite-loop plaquette bounds alone also do not establish an infinite-distance mass gap.

The continuum application therefore requires renormalized correlation bounds and reconstruction assumptions in addition to any fixed-spacing comparison estimate.

\clearpage
\part{Confinement, matter, and controlled comparison models}
\section{Matter representations and screening}
\label{ef-sec:screening}
The comparison models in this part have different field content, geometry,
or dimension from four-dimensional pure Wilson theory. Their exact
calculations identify sufficient estimates for the specified models and
the additional matching assumptions needed for a microscopic application.
Published semiclassical mechanisms and exactly solvable gauge theories
retain their original attribution.

\subsection{Why the fermion representation matters}
\begin{proposition}[Centre-charge obstruction to screening]
For $SU(N)$, an external source in a representation of nonzero $N$-ality cannot form a singlet by tensoring with any finite number of adjoint particles.
\end{proposition}
\begin{proof}
Let $z=e^{2\pi i/N}I$ be the central generator. On a representation of $N$-ality $k$, $z$ acts as $e^{2\pi i k/N}$. It acts as $1$ on every adjoint factor. Therefore it acts as $e^{2\pi i k/N}\ne1$ on the entire tensor product. On a singlet it would act as $1$, so no singlet subrepresentation can occur.
\end{proof}
For $SU(2)$ this is the familiar fact that tensoring an odd number of half-integer spins with integer spins cannot produce spin zero. This is a kinematic obstruction to screening; it does not itself prove a positive string tension.

Fundamental dynamical fermions change the conclusion. If finite-energy static-light colour singlets exist, a separated pair supplies a variational state with
\begin{equation}
E_0(R)\le 2M_{\mathrm{static-light}}+o(1),\qquad R\to\infty,
\end{equation}
after a common static self-energy subtraction. A potential $E_0(R)\sim\sigma R$ with $\sigma>0$ cannot then persist to arbitrarily large $R$. This argument assumes the stated singlets and their asymptotic separation; it is not a construction of them. String breaking is compatible with the absence of isolated observable quarks.

\subsection{Kinematics is not dynamics: the two-dimensional lesson }
The centre-charge proposition is sometimes read as ``adjoint matter cannot screen fundamental probes''. That reading is false as a dynamical statement. Gross, Klebanov, Matytsin and Smilga showed that in two-dimensional $SU(N)$ gauge theory with \emph{massless} adjoint Majorana fermions a fundamental probe is screened: the string tension of the fundamental Wilson loop vanishes, even though no finite number of adjoints can neutralise its $N$-ality \cite{ef:gkms}. For a small adjoint mass $m$ confinement returns with a string tension proportional to $m$ times the fermion condensate. The mechanism is that the string tension is a statement about the vacuum energy in the presence of a background electric field, and with massless adjoints a chiral rotation can remove that background; the asymptotic state carries the centre charge non-locally rather than in a finite bound state. The lesson for the four-dimensional problem is that a proof of confinement must control energies, not representation theory. Whether the massless four-dimensional adjoint theory confines fundamental probes in the sense of an area law is not settled by any result used in this analysis.

\section{Symmetries and the mixed anomaly of adjoint QCD }
\label{ef-sec:anomaly}
The gauge contribution to the mixed anomaly constrains possible infrared phases; the constraints do not determine a mass gap by themselves.

Consider $SU(N)$ Yang-Mills with $n_f$ massless adjoint Weyl fermions $\lambda^a_{I}$, $I=1,\dots,n_f$. Because the matter is adjoint, the $\Z_N$ centre acts trivially on all fields and the theory has a $\Z_N^{(1)}$ one-form (centre) symmetry \cite{ef:gkks}. The classical $U(1)$ acting on all fermions by a common phase is broken by the anomaly to the discrete subgroup under which the measure in a unit-charge instanton background is invariant; an adjoint Weyl fermion has $2N$ zero modes in such a background, so the surviving symmetry is $\Z_{2Nn_f}$, containing fermion number $(-1)^F$ as its order-two element. The faithful symmetry group also involves identifications with flavour transformations; the argument below uses only this common axial transformation and the centre background.

\begin{proposition}[Mixed centre--chiral anomaly phase]
Work on a spin four-manifold with vanishing gravitational index contribution, for example a flat four-torus, and couple the theory to an admissible background $\Z_N^{(1)}$ gauge field. Assume the resulting $PSU(N)$ gauge bundle has topological charge $Q\in\tfrac1N\Z$ and adjoint Weyl index $2NQ$. Under the generator of $\Z_{2Nn_f}$ the fermion measure changes by the phase $e^{-2\pi iQ}$. For $Q=k/N$ with $N\nmid k$ this phase is nontrivial.
\end{proposition}
\begin{proof}
Under the stated geometric assumptions, the gauge part of the index theorem gives each adjoint Weyl operator index $n_L-n_R=2NQ$ by the index theorem (the adjoint Dynkin index is $2N$ in the normalisation where the fundamental has index $1$). The generator multiplies every fermion by $e^{2\pi i/(2Nn_f)}$, so the zero-mode measure (whose transformation depends only on the index) acquires $e^{-2\pi i\,n_f\,2NQ/(2Nn_f)}=e^{-2\pi iQ}$. For integer $Q$ this is $1$, so this axial subgroup is preserved in the absence of a centre background; for $Q=k/N$ it equals $e^{-2\pi ik/N}$.
\end{proof}
A phase that cannot be removed by a local counterterm is a mixed 't~Hooft anomaly between $\Z_N^{(1)}$ and $\Z_{2Nn_f}$ \cite{ef:gkks,ef:shimizu,ef:anberpoppitz2018}. The cited anomaly-matching arguments exclude a trivially gapped infrared phase that preserves both symmetries. Symmetry breaking, gapless excitations, or a nontrivial topological sector can instead match the anomaly. The following small-circle effective potential realises one symmetry-breaking possibility. For $SU(2)$ with $n_f=2$, the symmetry is $\Z_8$ and the small-circle analysis of Section~\ref{ef-sec:dual} realises its generator on the dual photon as
\begin{equation}
\sigma\longmapsto\sigma+\pi ,
\label{ef-eq:chiralshift}
\end{equation}
because the monopole vertex carries $e^{i\sigma}$ times $2n_f=4$ fermion zero modes, and $e^{2\pi i\cdot4/8}=-1$ must be compensated. The bion potential $\zeta(1-\cos2\sigma)$ is invariant under \eqref{ef-eq:chiralshift}; its two minima $\sigma=0,\pi$ are exchanged. Hence $\Z_8\to\Z_4$ with two vacua, in agreement with the anomaly. The $\pi$-walls of Sections~\ref{ef-sec:wall}--\ref{ef-sec:probes} are the domain walls between these vacua, and the deconfinement of fundamental probes on them (Theorem~\ref{ef-thm:probes}) is the effective-model realisation of discrete anomaly inflow on the wall \cite{ef:aps,ef:ksu,ef:coxpoppitzwong}.

\section{The Seiberg-Witten route: a worked local calculation}
\label{ef-sec:sw}
This comparison uses the magnetic effective superpotential near a monopole singularity in softly deformed $\mathcal N=2$ $SU(2)$ theory \cite{ef:sw}. The original vector multiplet contains an adjoint scalar and two Weyl fermions. It is not pure bosonic Yang-Mills with arbitrary extra quarks.

Take a holomorphic function $u(a)$ near $a=0$, with $u'(0)=c\ne0$, and
\begin{equation}
W(a,M,\widetilde M)=\sqrt2\,aM\widetilde M+\mu u(a),\qquad \mu\ne0.
\label{ef-eq:sw}
\end{equation}
Here $a=a_D$ is a magnetic special coordinate, and $M,\widetilde M$ have opposite magnetic charges. Assume a positive nonsingular local kinetic metric, and use a local normalization with coefficient $g_D^{-2}$ for the $a$ and gauge kinetic terms. The bosonic potential is
\begin{align}
V={}&2|a|^2\bigl(|M|^2+|\widetilde M|^2\bigr)
 +g_D^2\bigl|\sqrt2M\widetilde M+\mu u'(a)\bigr|^2\notag\\
&+\frac{g_D^2}{2}\bigl(|M|^2-|\widetilde M|^2\bigr)^2.
\label{ef-eq:swpotential}
\end{align}

\begin{proposition}[Isolated local Higgs vacuum]
In a neighbourhood where $u'(a)\ne0$, the zero-potential solution is, modulo the magnetic $U(1)$,
\begin{equation}
a=0,\qquad M\widetilde M=-\frac{\mu c}{\sqrt2},\qquad
|M|=|\widetilde M|=v,\quad v^2=\frac{|\mu c|}{\sqrt2}.
\end{equation}
Its quadratic potential has no zero directions after quotienting by gauge transformations.
\end{proposition}
\begin{proof}
Each term in \eqref{ef-eq:swpotential} is nonnegative. If both charged fields vanished, the middle term would be nonzero. At least one is nonzero, so the first term forces $a=0$. The last two terms give the product and equal magnitudes. Their phases are fixed up to the gauge action.

Choose constant field phases so that the local vacuum is represented by equal positive magnitudes $v$. The linearized F equations, with harmless phase factors absorbed, are
\[
\delta F_M=\sqrt2v\,\delta a,\quad
\delta F_{\widetilde M}=\sqrt2v\,\delta a,\quad
\delta F_a=\sqrt2v(\delta M+\delta\widetilde M)+\mu u''(0)\delta a.
\]
The linearized D equation is $2v\operatorname{Re}(\delta M-\delta\widetilde M)=0$.
Vanishing of all these equations gives $\delta a=0$, $\delta\widetilde M=-\delta M$ and $\operatorname{Re}\delta M=0$. This last direction is precisely the gauge orbit. A sum of positive squares is therefore positive definite on the finite-dimensional physical tangent space.
\end{proof}
With gauge kinetic term $-F_D^2/(4g_D^2)$ and canonical charged-scalar kinetic terms, $A_{D,c}=A_D/g_D$ gives
\begin{equation}
m_{A_D}^2=4g_D^2v^2>0.
\end{equation}

\subsection{The confining string is a critical vortex in the truncated model }
Set $a\equiv0$ in \eqref{ef-eq:swpotential} and restrict to the charge-conjugation-symmetric slice $\widetilde M=-e^{i\arg(\mu c)}\,\overline M$, which is consistent with the $D$-term. Writing $\phi=\sqrt2M$, a direct substitution gives
\begin{equation}
|D M|^2+|D\widetilde M|^2=|D\phi|^2,\qquad
V=\frac{g_D^2}{2}\bigl(|\phi|^2-\phi_0^2\bigr)^2,\qquad \phi_0^2=\sqrt2\,|\mu c|=2v^2 .
\label{ef-eq:ano}
\end{equation}
This is the abelian Higgs model $\mathcal E=\int\dd^2x\,[B^2/(2e^2)+|D\phi|^2+(\lambda/2)(|\phi|^2-\phi_0^2)^2]$ with $e^2=g_D^2$ and $\lambda=g_D^2$: the quartic and gauge couplings coincide, which is the critical (Bogomolny) point separating type I from type II.

\begin{proposition}[BPS string tension in the truncation]
In the truncated model \eqref{ef-eq:ano}, the energy per unit length of a static vortex of winding $n\ge0$ satisfies $\mathcal E\ge2\pi n\phi_0^2$, with equality on solutions of the first-order equations $(D_1+iD_2)\phi=0$, $B=e^2(\phi_0^2-|\phi|^2)$. Such solutions exist for every nonnegative integer $n$; negative winding uses the conjugate first-order equations and $|n|$ in the tension. Hence, in the normalisation of \eqref{ef-eq:sw},
\begin{equation}
T_n^{\mathrm{trunc}}=2\pi n\phi_0^2=2\sqrt2\,\pi\,n\,|\mu c| ,
\end{equation}
and $T_n^{\mathrm{trunc}}$ is an upper bound for the minimal energy per length in the sector of the full local model \eqref{ef-eq:swpotential} in which $M$ has winding $n$ (and $\widetilde M$ winding $-n$).
\end{proposition}
\begin{proof}
With $[D_1,D_2]=-iB$ one has $|D_1\phi|^2+|D_2\phi|^2=|(D_1+iD_2)\phi|^2+B|\phi|^2+\partial_i J_i$ for a current $J$ that integrates to zero for finite-energy fields, and $B^2/(2e^2)+(e^2/2)(|\phi|^2-\phi_0^2)^2=\frac1{2e^2}\bigl(B-e^2(\phi_0^2-|\phi|^2)\bigr)^2+B(\phi_0^2-|\phi|^2)$. Adding the two identities, $\mathcal E=\int[\,|(D_1+iD_2)\phi|^2+\frac1{2e^2}(B-e^2(\phi_0^2-|\phi|^2))^2\,]+\phi_0^2\int B\ge\phi_0^2\cdot2\pi n$, since the flux of a winding-$n$ configuration is $2\pi n$. Existence of solutions of the first-order system for every $n\ge0$ is Taubes' theorem \cite{ef:jaffetaubes}. The upper-bound statement holds because the truncated configurations, extended by $a\equiv0$, are admissible for the full local model and the full potential agrees with the truncated one on them.
\end{proof}
The supporting numerical calculation integrates the $n=1$ equations and evaluates the unreduced abelian-Higgs energy, obtaining agreement with $2\pi\phi_0^2$ to relative error below $10^{-7}$. This is a numerical check of the truncation, not a certified solution of the full magnetic model.

Two physical comments, both attributed. First, the $a$ field is sourced at order $\mu u''(0)$ in the vortex core, so the truncation is not exact; Vainshtein and Yung showed that the next-to-leading corrections in $\mu$ make the strings slightly type I, so that $k$ strings attract and the BPS equality is lifted \cite{ef:vy}. Second, the abelian low-energy theory has strings labelled by $n\in\Z$, whereas the full $SU(2)$ theory distinguishes only the $N$-ality $n\bmod2$: the $n=2$ string must be able to break by $W$-pair creation, which the abelian description cannot see. This was noted by Douglas and Shenker \cite{ef:ds} and is a general limitation of abelianised descriptions of confinement, including the one in Sections~\ref{ef-sec:dual}--\ref{ef-sec:probes}, where the analogue is the two-wall structure of the fundamental interface.

This is a local effective Higgs calculation. The condensate notation is a gauge-fixed description. The vacuum algebra does not prove the exact quantum spectrum or uniform control as $|\mu|$ becomes large. Removing all superpartners also removes the protected structure used to determine \eqref{ef-eq:sw}; the calculation cannot simply be continued to pure Yang-Mills.

\section{A small spatial circle and its centre-symmetric minimum}
\label{ef-sec:circle}
Now take $n_f$ massless adjoint Weyl flavours, periodic around $S^1_L$. Periodic spatial boundary conditions are essential; this is not the thermal ensemble. For a diagonal $SU(N)$ holonomy
\[
\Omega=\operatorname{diag}(e^{i\phi_1},\dots,e^{i\phi_N}),\qquad \textstyle\sum_j\phi_j\in2\pi\Z,
\]
the standard one-loop determinant gives, up to a holonomy-independent constant,
\begin{equation}
V_1(\Omega)=\frac{2(n_f-1)}{\pi^2L^4}
 \sum_{n=1}^{\infty}\frac{|\tr_F\Omega^n|^2}{n^4}.
\label{ef-eq:gpy}
\end{equation}
For $SU(2)$, $\Omega=\operatorname{diag}(e^{i\phi},e^{-i\phi})$ and $|\tr\Omega^n|^2=4\cos^2n\phi$, so $V_1=\frac{4(n_f-1)}{\pi^2L^4}f(2\phi)+\mathrm{constant}$ with $f$ as below. This determinant is an imported physical approximation \cite{conf:unsal,ef:gpy}; the following minimisations are exact for it.

\begin{lemma}[Fourier polynomial]
On $0\le x\le2\pi$,
\begin{equation}
f(x)=\sum_{n\ge1}\frac{\cos nx}{n^4}
=\frac{\pi^4}{90}-\frac{\pi^2x^2}{12}
 +\frac{\pi x^3}{12}-\frac{x^4}{48}
 =-\frac{\pi^4}{3}\,B_4\!\Bigl(\frac{x}{2\pi}\Bigr),
\end{equation}
where $B_4(t)=t^4-2t^3+t^2-\frac1{30}$ is the fourth Bernoulli polynomial.
\end{lemma}
\begin{proof}
For $0<r<1$, $\sum r^n e^{inx}/n=-\log(1-re^{ix})$. Taking imaginary parts and $r\uparrow1$ yields the sawtooth series $(\pi-x)/2$ on $(0,2\pi)$. Integration gives
$\sum_{n\ge1}\cos nx/n^2=C_2-\pi x/2+x^2/4$.
Its average over a period is zero, so $C_2=\pi^2/6$. Integrate twice more, using $f'(0)=0$ and zero average of $f$, to obtain the displayed polynomial. These integrated series converge absolutely and uniformly; endpoint values follow by continuity. The Bernoulli form is a direct expansion.
\end{proof}

\begin{proposition}[Minimum and barrier at one loop, $SU(2)$]
For $n_f>1$, the minimum of \eqref{ef-eq:gpy} in $[0,\pi]$ is $\phi=\pi/2$. Moreover
\begin{equation}
V_1(0)-V_1(\pi/2)=\frac{(n_f-1)\pi^2}{12L^4},\qquad
V_1''(\pi/2)=\frac{4(n_f-1)}{3L^4}.
\end{equation}
\end{proposition}
\begin{proof}
Differentiate the polynomial:
$f'(x)=-x(x-\pi)(x-2\pi)/12$.
It is negative on $(0,\pi)$ and positive on $(\pi,2\pi)$. Also $f(0)-f(\pi)=\pi^4/48$ and $f''(\pi)=\pi^2/12$. Substitute in \eqref{ef-eq:gpy}.
\end{proof}
At this minimum $\tr_F\Omega=0$. The off-diagonal gauge modes have lowest momentum $m_W=\pi/L$. The condition $L\Lambda\ll1$ makes the coupling at this scale weak. The $n_f=1$ case has a vanishing one-loop potential and needs a separate analysis \cite{ef:poppitzunsal}; it is not covered by this proposition. For $n_f=0$ the sign reverses and the centre-symmetric point is not the minimum.

\subsection{All $SU(N)$ at once: a universal-optimality argument }
Write $z_j=e^{i\phi_j}$ and $p_n=\tr\Omega^n=\sum_jz_j^n$. We consider the whole class of potentials
\begin{equation}
V(\Omega)=\sum_{n\ge1}a_n|p_n|^2,\qquad a_n=\int_{(0,1)}q^n\,\dd\nu(q),
\label{ef-eq:cmclass}
\end{equation}
with $\nu$ a nonzero positive Borel measure on $(0,1)$ such that $\int q(1-q)^{-1}\dd\nu(q)<\infty$. By Tonelli this is exactly $\sum_{n\ge1}a_n<\infty$, and ensures uniform convergence of $V$. The measure may have infinite mass near zero; the shifted sequence $(a_{n+1})_{n\ge0}$ is a Hausdorff moment sequence for the finite measure $q\,\dd\nu(q)$. The one-loop coefficients belong to the class: $1/n^4=\frac16\int_0^\infty t^3e^{-nt}\dd t$, i.e. $\dd\nu=\frac16(\log1/q)^3\dd q/q$.

\begin{theorem}[Centre-symmetric minimum with uniqueness]
\label{ef-thm:sun}
For $V$ as in \eqref{ef-eq:cmclass} and every $\Omega\in SU(N)$,
\begin{equation}
V(\Omega)\ \ge\ V(\Omega_c)=N^2\sum_{m\ge1}a_{mN},
\end{equation}
with equality if and only if the eigenvalues of $\Omega$ form a regular $N$-gon, i.e. $\Omega$ is conjugate to $e^{i\alpha}\operatorname{diag}(1,\omega,\dots,\omega^{N-1})$, $\omega=e^{2\pi i/N}$, where $e^{iN\alpha}=(-1)^{N-1}$ (the corresponding centre rotations give the same unordered eigenvalue set). For $a_n=1/n^4$, $V(\Omega_c)=\pi^4/(90N^2)$ and
\begin{equation}
V_1(\mathbf 1)-V_1(\Omega_c)=\frac{(n_f-1)\pi^2}{45L^4}\Bigl(N^2-\frac1{N^2}\Bigr).
\end{equation}
\end{theorem}
\begin{proof}
For fixed $q\in(0,1)$ put $E_q(z)=\sum_{n\ge1}q^n|p_n|^2=\sum_{j,k}P_q(\phi_j-\phi_k)$ with the Poisson-type kernel
\[
P_q(x)=\sum_{n\ge1}q^n\cos nx=-\frac12+\frac{1-q^2}{2(1-2q\cos x+q^2)} .
\]
In terms of the squared chordal distance $s=|z_j-z_k|^2=2-2\cos x\in[0,4]$,
\[
P_q=-\frac12+f_q(s),\qquad f_q(s)=\frac{(1-q^2)/2}{(1-q)^2+qs},\qquad
(-1)^kf_q^{(k)}(s)=\frac{k!\,q^k(1-q^2)/2}{((1-q)^2+qs)^{k+1}}>0 ,
\]
so $f_q$ is strictly completely monotone on $(0,4]$. Hence $E_q(z)=NP_q(0)-\tfrac12N(N-1)+2\sum_{j<k}f_q(|z_j-z_k|^2)$ is, up to a constant, the $f_q$-energy of $N$ points on the circle. The regular $N$-gon is a sharp configuration in $S^1$ (it has $\lfloor N/2\rfloor$ distinct inner products and is a spherical $(N-1)$-design), so by Theorem~1.2 of Cohn and Kumar \cite{ef:cohnkumar} it minimises the $f_q$-energy among all $N$-point configurations, and because $f_q$ is strictly completely monotone every minimiser is an isometric image of it. The cited linear-programming proof also excludes a tuple with repeated points. Indeed, write $a_q(t)=f_q(2-2t)$, whose derivatives of every positive order are strictly positive. Its Hermite interpolation polynomial $h$ at the distinct off-diagonal inner products of the regular polygon has $h\le a_q$ on $[-1,1]$ and $h(1)<a_q(1)$. For odd $N$ the interpolation nodes are doubled; for even $N$ the node $-1$ is single and the other nodes doubled. The interpolation remainder is positive at $t=1$. The positive-definite polynomial bound applies to labelled tuples as well as distinct sets, and each coincident off-diagonal pair contributes the strictly positive discrepancy $a_q(1)-h(1)$. Thus equality cannot occur with repeated eigenvalues. Consequently $E_q(z)\ge E_q(z^{\rm gon})$ for every $q$, with equality only for a regular polygon. Tonelli and the summability assumption permit integration against $\nu\ge0$, giving $V(\Omega)\ge V(\Omega_c)$.

For the $N$-gon, $p_n=\sum_j e^{in\alpha}\omega^{jn}=0$ unless $N\mid n$, in which case $|p_n|=N$; hence $E_q(z^{\rm gon})=N^2\sum_mq^{mN}$ and $V(\Omega_c)=N^2\sum_ma_{mN}$. If $V(\Omega)=V(\Omega_c)$ then $E_q(z)=E_q(z^{\rm gon})$ for $\nu$-almost every $q$, hence for at least one $q$, and the uniqueness clause identifies $z$ with a regular $N$-gon. The determinant condition $\prod_jz_j=e^{iN\alpha}\omega^{N(N-1)/2}=e^{iN\alpha}(-1)^{N-1}=1$ fixes $\alpha$ modulo $2\pi/N$. The displayed values follow from $\sum_m(mN)^{-4}=\pi^4/(90N^4)$ and $|p_n(\mathbf 1)|^2=N^2$.
\end{proof}

\begin{remark}
When the support of $\nu$ has an accumulation point in $(0,1)$, uniqueness can also be obtained from the inequality alone: equality for infinitely many $q$ accumulating in $(0,1)$ forces the analytic functions $\sum_nq^n|p_n|^2$ and $N^2\sum_mq^{mN}$ to coincide, so $p_1=\dots=p_{N-1}=0$; by Newton's identities the elementary symmetric functions $e_1,\dots,e_{N-1}$ of the $z_j$ vanish, so the $z_j$ are the roots of $z^N=(-1)^{N+1}e_N$ with $|e_N|=1$, a regular $N$-gon.
\end{remark}

For $N=2$ and $a_n=n^{-4}$ the theorem recovers the proposition above. For $N=3$ an elementary proof exists and is instructive because it shows why the Fourier argument alone does not suffice (the $N\mid n$ harmonics are \emph{maximised} at the centre-symmetric point).

\begin{proposition}[$SU(3)$ by hand]
Let $g_1,g_2,g_3\ge0$ be the three gaps between consecutive eigenvalue angles, $g_1+g_2+g_3=2\pi$. Then $V=\sum_{j,k}f(\phi_j-\phi_k)=3f(0)+2[f(g_1)+f(g_2)+f(g_3)]$, and the minimum over the simplex is attained only at $g_1=g_2=g_3=2\pi/3$.
\end{proposition}
\begin{proof}
Because $f$ is even and $2\pi$-periodic, $f(\phi_1-\phi_3)=f(g_1+g_2)=f(2\pi-g_3)=f(g_3)$. On the simplex $S(g)=\sum f(g_i)$ is a polynomial. At an interior critical point, Lagrange's condition reads $f'(g_1)=f'(g_2)=f'(g_3)=\lambda$, i.e. all three gaps are roots of the cubic $x^3-3\pi x^2+2\pi^2x+12\lambda=0$. Three distinct roots would sum to $3\pi\ne2\pi$. If two gaps equal $a$ and the third is $b=2\pi-2a$, the remaining root of the cubic is $c=3\pi-a-b=\pi+a$, and Vieta's second relation $ab+ac+bc=2\pi^2$ reduces exactly to $3a(\pi-a)=0$ (verified in rational arithmetic in \texttt{verify\_sun\_holonomy.py}), giving the boundary configurations $(0,0,2\pi)$ and $(\pi,\pi,0)$. Hence the only interior critical point is $g_1=g_2=g_3$, where the restricted Hessian is $f''(2\pi/3)\sum\delta g_i^2$ with $f''(2\pi/3)=\pi^2/18>0$. On the boundary ($g_3=0$, two coincident eigenvalues) $S=2f(g_1)+f(0)\ge2f(\pi)+f(0)=-\pi^4/120$, while $S(2\pi/3,2\pi/3,2\pi/3)=3f(2\pi/3)=-\frac{13}{810}\pi^4<-\pi^4/120$. 
\end{proof}

\begin{theorem}[Hessian spectrum at the centre-symmetric point]
\label{ef-thm:hessian}
Let $a_n$ be real with $\sum n^2|a_n|<\infty$, and parametrise $\phi_j=\phi_j^c+\delta_j$ with $\sum_j\delta_j=0$. In the Euclidean metric on the traceless $\delta$ the Hessian of $V=\sum a_n|p_n|^2$ at $\Omega_c$ has eigenvectors $\cos(2\pi kj/N)$, $\sin(2\pi kj/N)$ and eigenvalues
\begin{equation}
\lambda_k=N\,(c_k+c_{N-k}),\qquad c_k=\sum_{n\equiv k\ (N)}n^2a_n-\sum_{N\mid n}n^2a_n,\qquad k=1,\dots,N-1 .
\end{equation}
For $a_n=1/n^4$,
\begin{equation}
\boxed{\ \lambda_k=\frac{\pi^2}{N}\Bigl[\csc^2\frac{\pi k}{N}-\frac13\Bigr]>0\ },\qquad
\min_k\lambda_k=\begin{cases}\dfrac{2\pi^2}{3N}&N\text{ even},\\[6pt]
\dfrac{\pi^2}{N}\Bigl[\sec^2\dfrac{\pi}{2N}-\dfrac13\Bigr]&N\text{ odd}.\end{cases}
\end{equation}
\end{theorem}
\begin{proof}
Let $\hat\delta_k=\sum_j\omega^{jk}\delta_j$ be the discrete Fourier transform, $\hat\delta_0=0$, and expand $p_n(\phi^c+\delta)=\sum_je^{in\phi^c_j}(1+in\delta_j-\tfrac12n^2\delta_j^2+\dots)$. For $N\nmid n$ the zeroth-order term vanishes and $|p_n|^2=n^2|\hat\delta_{n\bmod N}|^2+o(\|\delta\|^2)$. For $N\mid n$, $e^{in\phi^c_j}$ is $j$-independent, so $|p_n|^2=N^2-Nn^2\sum_j\delta_j^2+o(\|\delta\|^2)=N^2-n^2\sum_{k}|\hat\delta_k|^2+o(\|\delta\|^2)$ by Parseval. The summability assumption gives uniform convergence of the second derivatives, so the summed potential is $C^2$ and has a quadratic Taylor expansion with $o(\|\delta\|^2)$ remainder. Collecting, $V=V(\Omega_c)+\sum_{k=1}^{N-1}c_k|\hat\delta_k|^2+o(\|\delta\|^2)$. For real $\delta$, $|\hat\delta_k|=|\hat\delta_{N-k}|$. When $k\ne N/2$, a normalised real mode $\delta_j=\sqrt{2/N}\cos(2\pi kj/N)$ has $|\hat\delta_k|^2+|\hat\delta_{N-k}|^2=N$. Since the quadratic term is $\frac12\delta^{\!\top}\!H\delta$, the eigenvalue is $\lambda_k=2\cdot\frac N2(c_k+c_{N-k})$. For even $N$ and $k=N/2$, the unit mode is $\delta_j=(-1)^j/\sqrt N$, so $|\hat\delta_{N/2}|^2=N$ and the same formula gives $\lambda_{N/2}=2Nc_{N/2}$. For $a_n=n^{-4}$, $\sum_{n\equiv k}n^{-2}=N^{-2}\psi'(k/N)$ and $\sum_{N\mid n}n^{-2}=N^{-2}\psi'(1)$ with $\psi'$ the trigamma function, so $\lambda_k=N^{-1}[\psi'(k/N)+\psi'(1-k/N)-2\psi'(1)]$; the reflection formula $\psi'(x)+\psi'(1-x)=\pi^2\csc^2\pi x$ and $\psi'(1)=\pi^2/6$ give the boxed form. Positivity follows from $\csc^2\ge1>\frac13$.
\end{proof}
For $N=2$ the single eigenvalue is $\pi^2/3$ in the orthonormal coordinate $(\delta,-\delta)/\sqrt2$, equivalently $V''(\pi/2)=2\pi^2/3$ in the coordinate $\phi$, reproducing $V_1''(\pi/2)=4(n_f-1)/(3L^4)$ after multiplication by $2(n_f-1)/(\pi^2L^4)$. Finite-difference checks for $N\le8$ agree with this formula. A random-restart search provides an additional numerical diagnostic; the analytic minimisation theorem supplies the global conclusion.

\begin{corollary}[Holonomy fluctuation masses]
Use the four-dimensional normalisation $\frac1{4g^2}F^a_{\mu\nu}F^{a\mu\nu}$ and $A_4=\operatorname{diag}(\phi_j)/L$. Then the three-dimensional kinetic term of the Cartan holonomy is $(g^2L)^{-1}\sum_j(\partial\phi_j)^2$, and the one-loop potential curvature masses, with tree-level kinetic normalisation, of its $N-1$ fluctuation modes at the centre-symmetric minimum of \eqref{ef-eq:gpy} are
\begin{equation}
m_k^2=\frac{g^2(n_f-1)}{NL^2}\Bigl[\csc^2\frac{\pi k}{N}-\frac13\Bigr],\qquad k=1,\dots,N-1 .
\end{equation}
For $SU(2)$, $m^2=g^2(n_f-1)/(3L^2)$: parametrically lighter than $m_W=\pi/L$ and heavier than the nonperturbatively small dual-photon mass.
\end{corollary}
\begin{proof}
Integrating the $A_4$ kinetic term over the circle gives the stated three-dimensional coefficient; the potential is $L\,V_1$. The mass matrix is $(g^2L/2)\cdot L\cdot\mathrm{Hess}V_1=(g^2L^2/2)(2(n_f-1)/(\pi^2L^4))\lambda_k$.
\end{proof}
These are one-loop masses with tree-level kinetic normalisation; the next corrections are not computed here.

\subsection{Where the one-loop centre-symmetric minimum ends }
With a common mass $m$ for the adjoint fermions (periodic boundary conditions) the standard one-loop potential becomes
\begin{equation}
V_1(\Omega;m)=\frac{2}{\pi^2L^4}\sum_{n\ge1}\frac{|\tr\Omega^n|^2}{n^4}\bigl[-1+n_f\,h(nmL)\bigr],\qquad h(x)=\tfrac12x^2K_2(x),
\label{ef-eq:massive}
\end{equation}
$h(0)=1$, $h$ decreasing to zero. The coefficients are no longer of the form \eqref{ef-eq:cmclass} (they become negative for large $n$), so Theorem~\ref{ef-thm:sun} does not apply, and indeed the centre-symmetric minimum is lost at a finite mass. For $SU(2)$ the curvature at $\phi=\pi/2$ is proportional to $\sum_n(-1)^{n+1}[n_fh(nmL)-1]/n^2$. A 400-term numerical sum gives a local-stability threshold near $mL=2.280$ for $n_f=2$ ($2.972$ and $3.417$ for $n_f=3,4$). Searching 2001 equally spaced angles indicates that the global minimum moves earlier, near $mL=2.00$ for $n_f=2$, consistent with a first-order transition of this one-loop potential. These are numerical estimates with finite series and angular discretisation, not certified transition values. They illustrate why fermion decoupling at fixed small $L$ requires an additional analysis of the holonomy potential.

\begin{figure}[htbp]
\centering
\includegraphics[width=\linewidth]{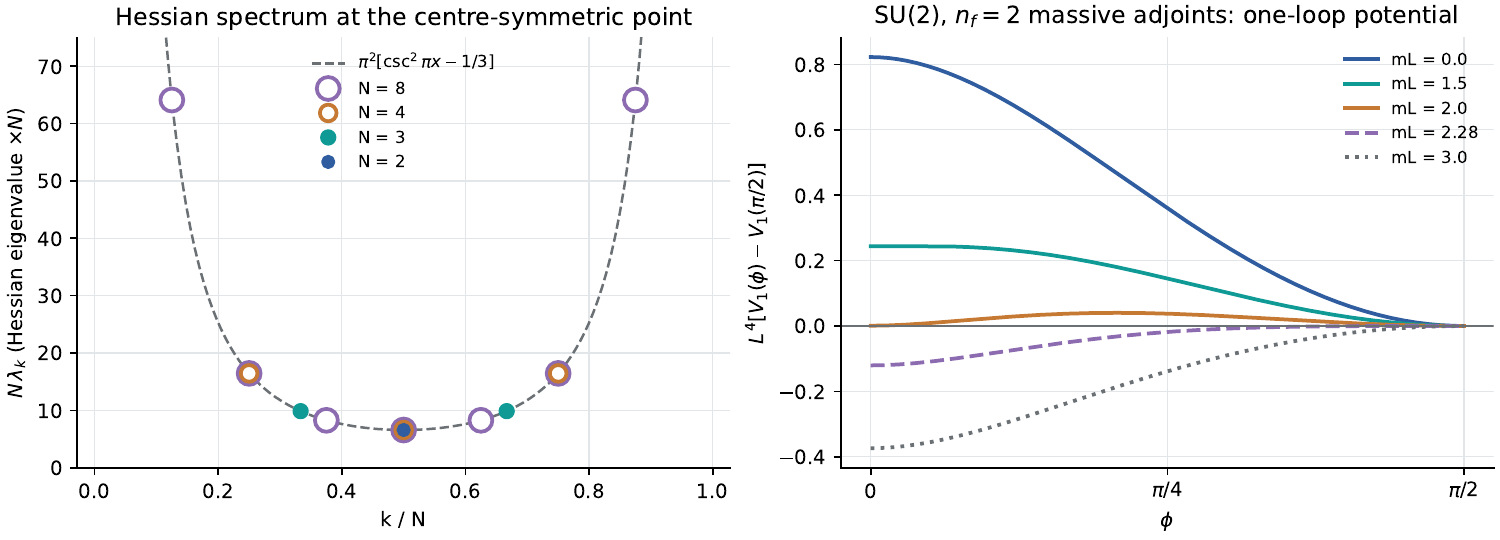}
\caption{Left: Hessian eigenvalues of the one-loop holonomy potential at the centre-symmetric point for several $N$ (dots, from finite differences) against the closed form of Theorem~\ref{ef-thm:hessian}. Right: the $SU(2)$, $n_f=2$ one-loop potential \eqref{ef-eq:massive} for increasing adjoint mass; the centre-symmetric point $\phi=\pi/2$ stops being the global minimum near $mL\approx2.0$ and the local minimum at $mL\approx2.28$.}
\label{ef-fig:sun}
\end{figure}

\section{Fermion exchange and an exactly evaluated pair integral}
In the semiclassical description, a fundamental monopole has action $S_0=4\pi^2/g^2$ and two zero modes per adjoint Weyl flavour. Its vertex contains fermion fields. Pairing a monopole with the appropriate anti-Kaluza-Klein monopole permits their contraction, leaving magnetic charge $2$ and topological charge $0$. This is the magnetic-bion mechanism \cite{conf:unsal}.

We now isolate the actual mathematical calculation. A three-dimensional massless fermion propagator scales as $r^{-2}$. Contracting $2n_f$ propagators contributes $r^{-4n_f}$. Let $A>0$ denote the Coulomb-repulsion length. Omitting overall determinants and core corrections, define
\begin{equation}
I_{n_f}(A)=4\pi\int_0^\infty r^{2-4n_f}e^{-A/r}\dd r.
\label{ef-eq:bion}
\end{equation}
The powers carry units; the omitted normalization supplies the corresponding units in a physical fugacity.

\begin{theorem}[Leading pair integral]
For real $n_f>3/4$,
\begin{equation}
I_{n_f}(A)=4\pi A^{3-4n_f}\Gamma(4n_f-3).
\end{equation}
For $n_f=2$, $I_2(A)=96\pi/A^5$. The normalized radial law has mode $A/6$ and mean $A/4$.
\end{theorem}
\begin{proof}
With $t=A/r$, the integral becomes $4\pi A^{3-4n_f}\int_0^\infty t^{4n_f-4}e^{-t}\dd t$. Its convergence condition and value follow immediately. The radial density has logarithmic derivative $(2-4n_f)/r+A/r^2$, so its mode is $A/(4n_f-2)$. When $n_f>1$, the ratio of its first moment to its normalization is $A\Gamma(4n_f-4)/\Gamma(4n_f-3)=A/(4n_f-4)$.
\end{proof}

\subsection{An explicit core-tail estimate}
The fraction of the \emph{model integral} coming from $0<r<r_0$ is
$\Gamma(4n_f-3,A/r_0)/\Gamma(4n_f-3)$.
For $n_f=2$ and $A/r_0=50$, this is $e^{-50}\sum_{j=0}^4 50^j/j!<6\cdot10^{-17}$. The support code certifies the inequality with rational arithmetic, using $e^{50}>\sum_{j=0}^{150}50^j/j!$.

This tiny number does \emph{not} bound the error of replacing the microscopic monopole core by the model. The true core weight would require its own comparison estimate. The calculation establishes integrability, a typical separation and an exact benchmark, not the complete bion determinant or a convergent cluster expansion.

\begin{center}
\includegraphics[width=.86\linewidth]{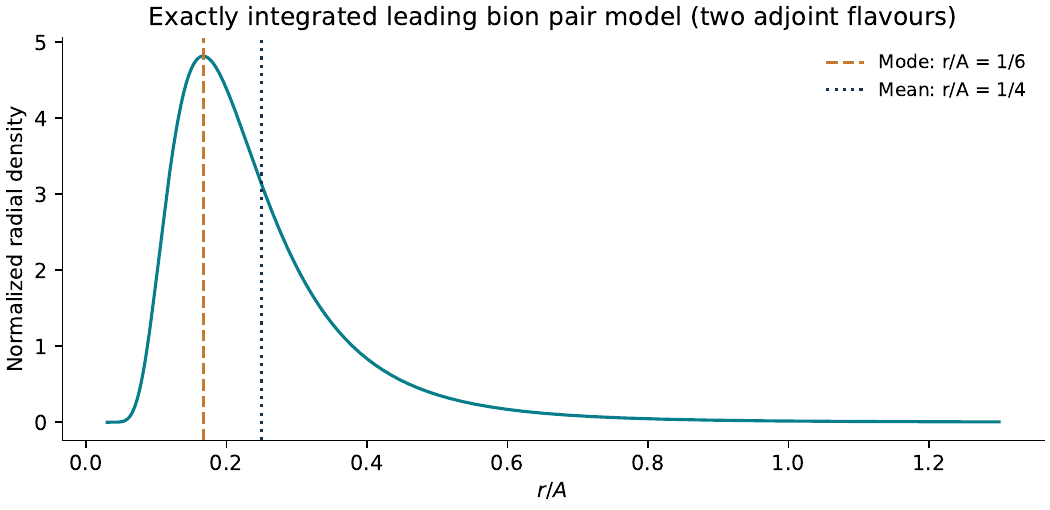}
\end{center}

\section{Dual normalization, the scalar action and a mass parameter}
\label{ef-sec:dual}
Use a Cartan gauge field normalized so the fundamental electric charge is $q=1/2$ and a basic monopole has flux $4\pi$. Put $g_3^2=g^2/L$. With $B_i=\epsilon_{ijk}F_{jk}/2$, the Maxwell action is $\int B^2/(2g_3^2)$. Imposing the Bianchi identity with a $2\pi$-periodic field $\sigma$ introduces $i\sigma\,\partial\cdot B/(4\pi)$.

Gaussian integration over $B$ gives a kinetic coefficient
\begin{equation}
S_{\mathrm{kin}}=\int\frac K2(\partial\sigma)^2\dd^3x,
\qquad K=\frac{g_3^2}{16\pi^2}>0.
\end{equation}
Equivalently, complete the square in $B$; the remaining term is $g_3^2(\partial\sigma)^2/(32\pi^2)$. Charge-$2$ objects contribute phases $e^{\pm2i\sigma}$. A positive leading fugacity therefore motivates the normalized action
\begin{equation}
S_0[\sigma]=\int_{\R^3}\left\{\frac K2(\partial\sigma)^2+
\zeta(1-\cos2\sigma)\right\}\dd^3x,
\qquad K,\zeta>0.
\label{ef-eq:sg}
\end{equation}
$\zeta$ denotes the entire coefficient of $1-\cos2\sigma$, absorbing the convention-dependent factor of two from the pair of charges. We do not assign it a numerical microscopic value. The shift $\sigma\to\sigma+\pi$ of \eqref{ef-eq:chiralshift} is a symmetry of \eqref{ef-eq:sg}; this is the dual-photon realisation of the unbroken-to-broken chiral $\Z_8\to\Z_4$ of Section~\ref{ef-sec:anomaly}.

\subsection{Quadratic theory}
Expanding about $\sigma=0$ gives $V''(0)=4\zeta$. With canonical field $\varphi=\sqrt K\sigma$,
\begin{equation}
m_0^2=\frac{4\zeta}{K},\qquad
\langle\sigma(x)\sigma(0)\rangle_{\mathrm{Gaussian}}
=\frac{e^{-m_0|x|}}{4\pi K|x|},\quad x\ne0.
\label{ef-eq:mass}
\end{equation}
To check the propagator, $( -\nabla^2+m_0^2)e^{-m_0r}/(4\pi r)=0$ for $r>0$; the integral of the outward gradient at a small sphere fixes its unit delta-function source. Division by $K$ gives the normalization.

The dimensions are $[K]=\mathrm{mass}$, $[\zeta]=\mathrm{mass}^3$, $[m_0]=\mathrm{mass}$, and $[T]=\mathrm{mass}^2$. Equation \eqref{ef-eq:mass} proves an exact Gaussian decay statement. It is not a nonperturbative spectral theorem for the interacting microscopic theory. The Weyl identification and the physical observable algebra must also be respected when interpreting a Cartan field as an observable.

\subsection{The Wilson charge fixes the wall count}
The insertion $\exp(iq\int_\Sigma B\cdot\dd S)$ shifts the dual gradient by a sheet of strength $4\pi q$. A fundamental probe thus imposes total monodromy $2\pi$. Adjacent minima of \eqref{ef-eq:sg} are separated by only $\pi$. Consequently a fundamental-charge interface requires two elementary walls. This normalization agrees with the double-wall structure studied in \cite{ef:aps}; a single $\pi$ wall is not the fundamental string. Section~\ref{ef-sec:probes} turns this counting into a theorem about the energy functional, including the complementary statement that the excess energy on a wall stays bounded with probe separation.

\section{Exact wall tension, profile and linear stability}
\label{ef-sec:wall}
For a planar profile, define the energy per area
\begin{equation}
E[\sigma]=\int_{-\infty}^{\infty}
\left\{\frac K2(\sigma')^2+\zeta(1-\cos2\sigma)\right\}\dd z.
\end{equation}

\begin{theorem}[Single-wall solution]
Among locally absolutely continuous profiles with $\sigma(-\infty)=0$ and $\sigma(+\infty)=\pi$, the minimum is
$T_\pi=4\sqrt{K\zeta}$.
It is attained by the translates
$\sigma_*(z)=2\arctan e^{m_0(z-z_0)}$, $m_0=2\sqrt{\zeta/K}$.
\end{theorem}
\begin{proof}
The pointwise inequality
$\frac K2(\sigma')^2+V(\sigma)\ge\sqrt{2KV(\sigma)}\,|\sigma'|$
follows by completing a square. Every connecting profile traverses $[0,\pi]$, so its energy is at least
\[
\int_0^\pi\sqrt{2K\zeta(1-\cos2s)}\dd s
=2\sqrt{K\zeta}\int_0^\pi\sin s\dd s=4\sqrt{K\zeta}.
\]
Equality requires $\sigma'=m_0\sin\sigma$ on this interval. Integration gives $\log\tan(\sigma/2)=m_0(z-z_0)$, with the required limits and finite energy.
\end{proof}

\begin{proposition}[The two-wall infimum]
For endpoints $0$ and $2\pi$, the infimum is $T_F^{(0)}=8\sqrt{K\zeta}$. It is not attained by a single finite-separation static profile in this one-component leading model.
\end{proposition}
\begin{proof}
The same integral lower bound over $[0,2\pi]$ is twice $T_\pi$. Two truncated single walls, separated by an increasingly long interval near $\pi$, approach that energy, proving the matching upper bound. Saturation would require $|\sigma'|=m_0|\sin\sigma|$. A solution starting below $\pi$ reaches it only at infinite $z$, since $\int\dd\sigma/|\sin\sigma|$ diverges there. It cannot then reach $2\pi$ at the other infinity as one saturating profile.
\end{proof}
For a finite Wilson loop, edge geometry and interactions determine the wall separation; the proposition concerns the asymptotic planar variational problem.

\begin{proposition}[No negative linear wall modes]
After canonical normalization, the one-dimensional fluctuation operator is
$\mathcal L=-\partial_z^2+m_0^2\{1-2\sech^2[m_0(z-z_0)]\}$.
It has a translational zero mode, no negative eigenvalues, and no other bound eigenvalues below the continuum threshold $m_0^2$.
\end{proposition}
\begin{proof}
Set $Q=\partial_z+m_0\tanh[m_0(z-z_0)]$. Then $\mathcal L=Q^\dagger Q\ge0$ and $QQ^\dagger=-\partial_z^2+m_0^2$. The zero mode solves $Q\psi=0$, giving $\psi\propto\sech[m_0(z-z_0)]$. Any positive bound eigenvalue below $m_0^2$ would be mapped by $Q$ to a nonzero square-integrable eigenfunction of the constant-coefficient partner, which has none. The decaying potential has essential spectrum $[m_0^2,\infty)$.
\end{proof}
\begin{remark}[Threshold resonance ]
The bounded solution $\psi=\tanh[m_0(z-z_0)]$ of $\mathcal L\psi=m_0^2\psi$ is a zero-energy resonance exactly at the continuum threshold (the reflectionless P\"oschl--Teller well). Under a generic perturbation of the potential or kinetic function this half-bound state either dissolves into the continuum or becomes a genuine bound state. The numerical calculations in Section~\ref{ef-sec:stability} indicate a wall-localised mode just below threshold for the specified example.
\end{remark}

\section{A uniform higher-harmonic stability theorem, and the linear response of the tension}
\label{ef-sec:harmonic}
This is a self-contained sufficient criterion for the \emph{static effective functional}. It is useful because the potential is allowed to contain infinitely many harmonics.

Let
\begin{equation}
V(s)=\zeta(1-\cos2s)+\sum_{n=2}^{\infty}c_n(1-\cos2ns),\qquad
\varepsilon=\frac1\zeta\sum_{n=2}^{\infty}n^2|c_n|<1.
\label{ef-eq:harmonic}
\end{equation}
with real coefficients $c_n$. Let the kinetic coefficient $Z(s)$ be positive, continuous and $\pi$-periodic, with
\begin{equation}
(1-\eta)K\le Z(s)\le(1+\eta)K,\qquad 0\le\eta<1.
\end{equation}
Use $E=\int\{Z(\sigma)(\sigma')^2/2+V(\sigma)\}\dd z$.

\begin{theorem}[Quantified wall and curvature bounds]
\label{ef-thm:harmonic}
Under these assumptions the only minima of $V$ are $\pi\Z$, and
\begin{align}
(1-\varepsilon)\zeta(1-\cos2s)&\le V(s)
\le(1+\varepsilon)\zeta(1-\cos2s),\label{ef-eq:pointbound}\\
T_\pi&=\int_0^\pi\sqrt{2Z(s)V(s)}\dd s,\label{ef-eq:Tint}\\
4\sqrt{K\zeta}\sqrt{(1-\eta)(1-\varepsilon)}
&\le T_\pi\le
4\sqrt{K\zeta}\sqrt{(1+\eta)(1+\varepsilon)},\label{ef-eq:Tbound}\\
\frac{1-\varepsilon}{1+\eta}m_0^2
&\le m_{\mathrm{curv}}^2:=\frac{V''(0)}{Z(0)}
\le\frac{1+\varepsilon}{1-\eta}m_0^2.\label{ef-eq:mbound}
\end{align}
The two-period fundamental-interface infimum is $2T_\pi$.
\end{theorem}
\begin{proof}
The identity $e^{2ins}-1=(e^{2is}-1)\sum_{j=0}^{n-1}e^{2ijs}$ gives $|\sin ns|\le n|\sin s|$. Hence
$0\le1-\cos2ns\le n^2(1-\cos2s)$.
The weighted absolute sum proves \eqref{ef-eq:pointbound}, uniform convergence, and strict positivity between neighbouring minima. It also permits two termwise derivatives, giving
$V''(0)=4\zeta+4\sum_{n\ge2}n^2c_n$.
This and the kinetic bounds prove \eqref{ef-eq:mbound}.

Completing the square with $Z(s)$ instead of $K$ proves the lower bound \eqref{ef-eq:Tint}. It is attained by the increasing solution defined implicitly by
$z-z_0=\int_{\pi/2}^{\sigma(z)}\sqrt{Z(s)/(2V(s))}\dd s$.
The pointwise bounds make $V(s)$ comparable to $s^2$ and $(\pi-s)^2$ near the endpoints, so this integral diverges to both infinities. It defines an admissible wall of precisely the integral energy. Equation \eqref{ef-eq:Tbound} follows by applying the two pointwise coefficient bounds under the integral. Periodicity and separated-wall approximants give $2T_\pi$ for the next period.
\end{proof}

An actual microscopic application must prove a bound on \emph{all} omitted harmonics in the weighted norm $\sum n^2|c_n|$, bound the kinetic correction, and control higher derivatives, additional fields and quantum fluctuations. A small unweighted potential error is insufficient: rapidly oscillating harmonics can have large curvature.

\subsection{Linear response and a quadrature-free enclosure }
The bound \eqref{ef-eq:Tbound} is crude because it replaces the perturbation by its worst-case envelope. The tension functional is in fact \emph{concave} in the data $(Z,V)$, and this yields a sharp one-sided bound plus an explicit remainder. Write $Z(s)=K(1+z(s))$ and $V(s)=\zeta(1-\cos2s)(1+v(s))$, so that $|z|\le\eta$ and, by \eqref{ef-eq:pointbound}, $|v|\le\varepsilon$ with
\begin{equation}
v(s)=\frac1\zeta\sum_{n\ge2}c_n\frac{\sin^2ns}{\sin^2s}=\frac1\zeta\sum_{n\ge2}c_nU_{n-1}(\cos s)^2 ,
\end{equation}
$U_{n-1}$ the Chebyshev polynomial of the second kind. Let $x=z+v+zv$, so $2ZV=4K\zeta\sin^2s\,(1+x)$ and $|x|\le\rho:=\eta+\varepsilon+\eta\varepsilon$.

\begin{theorem}[Concavity enclosure and harmonic-number response]
\label{ef-thm:concave}
Assume \eqref{ef-eq:harmonic}, $Z$ as above, and $\rho<1$. Define
\begin{equation}
\delta_1=\sqrt{K\zeta}\int_0^\pi\sin s\,\bigl(z+v+zv\bigr)(s)\,\dd s .
\end{equation}
Then
\begin{equation}
4\sqrt{K\zeta}+\delta_1-\frac{\sqrt{K\zeta}\,\rho^2}{2(1-\rho)^{3/2}}\ \le\ T_\pi\ \le\ 4\sqrt{K\zeta}+\delta_1 .
\label{ef-eq:concave}
\end{equation}
The potential part of $\delta_1$ is given exactly by odd harmonic numbers,
\begin{equation}
\sqrt{K\zeta}\int_0^\pi\sin s\,v(s)\,\dd s=2\sqrt{\frac K\zeta}\sum_{n\ge2}c_n\,h_n,\qquad
h_n=\sum_{k=0}^{n-1}\frac1{2k+1}=H_{2n}-\tfrac12H_n ,
\label{ef-eq:hn}
\end{equation}
so that $\left.\partial T_\pi/\partial c_n\right|_{c=0,Z=K}=2\sqrt{K/\zeta}\,h_n$ and $\left.\partial T_\pi/\partial z[w]\right|_{c=0,z=0}=\sqrt{K\zeta}\int_0^\pi\sin s\,w(s)\dd s$.
\end{theorem}
\begin{proof}
By \eqref{ef-eq:Tint}, $T_\pi=2\sqrt{K\zeta}\int_0^\pi\sin s\sqrt{1+x(s)}\,\dd s$. For $|x|\le\rho<1$ Taylor's theorem with Lagrange remainder gives
\[
1+\frac x2-\frac{x^2}{8(1-\rho)^{3/2}}\ \le\ \sqrt{1+x}\ \le\ 1+\frac x2 ,
\]
the upper inequality being concavity of the square root. Multiply by $2\sqrt{K\zeta}\sin s$, integrate, and use $\int_0^\pi\sin s\,\dd s=2$ and $\int_0^\pi\sin s\,x^2\dd s\le2\rho^2$ to obtain \eqref{ef-eq:concave}. For \eqref{ef-eq:hn}, the identity $\sum_{k=0}^{n-1}\sin(2k+1)s=\sin^2(ns)/\sin s$ (sum a geometric series in $e^{2is}$) gives $\int_0^\pi\sin^2(ns)/\sin s\,\dd s=\sum_{k<n}2/(2k+1)$. The bound is applied to the full relative perturbation $x=z+v+zv$ of the product $2ZV$, which is why the cross term $zv$ appears in $\delta_1$ rather than in the remainder; the geometric-mean inequality $\sqrt{(1+z)(1+v)}\le1+(z+v)/2$ gives another valid upper bound. Neither upper bound uniformly dominates the other, since the sign of the integrated cross term can change.
\end{proof}

\begin{remark}[Second order]
The next Taylor term is explicit: with $\delta_2=\frac14\sqrt{K\zeta}\int_0^\pi\sin s\,x^2\dd s$,
\begin{equation}
\Bigl|T_\pi-\bigl(4\sqrt{K\zeta}+\delta_1-\delta_2\bigr)\Bigr|\le\frac{\sqrt{K\zeta}\,\rho^3}{4(1-\rho)^{5/2}} .
\end{equation}
When $z$ is a polynomial in $\cos s$ and only finitely many harmonics occur, both $\delta_1/\sqrt{K\zeta}$ and $\delta_2/\sqrt{K\zeta}$ are polynomial integrals against $\sin s\,\dd s$. Rational polynomial coefficients then give exact rational values without numerical quadrature. This additional algebraic assumption is not needed for the general enclosure.
\end{remark}

\begin{remark}[Distinct harmonic sensitivity of tension and curvature]
The first derivative of the tension at the unperturbed model has coefficients $h_n\sim\tfrac12\log n$, while $V''(0)$ has coefficients proportional to $n^2$. This is a local linear-response statement. A nonlinear tension estimate under only a logarithmically weighted error requires additional control of the vacuum locations and nonnegativity of the potential; that control does not follow from the derivative formula. The weighted $n^2$ hypothesis of Theorem~\ref{ef-thm:harmonic} supplies the stated global positivity and curvature bounds.
\end{remark}

\section{A reproducible certified example}
\label{ef-sec:example}
\begin{figure}[htbp]
\centering
\includegraphics[width=.9\linewidth]{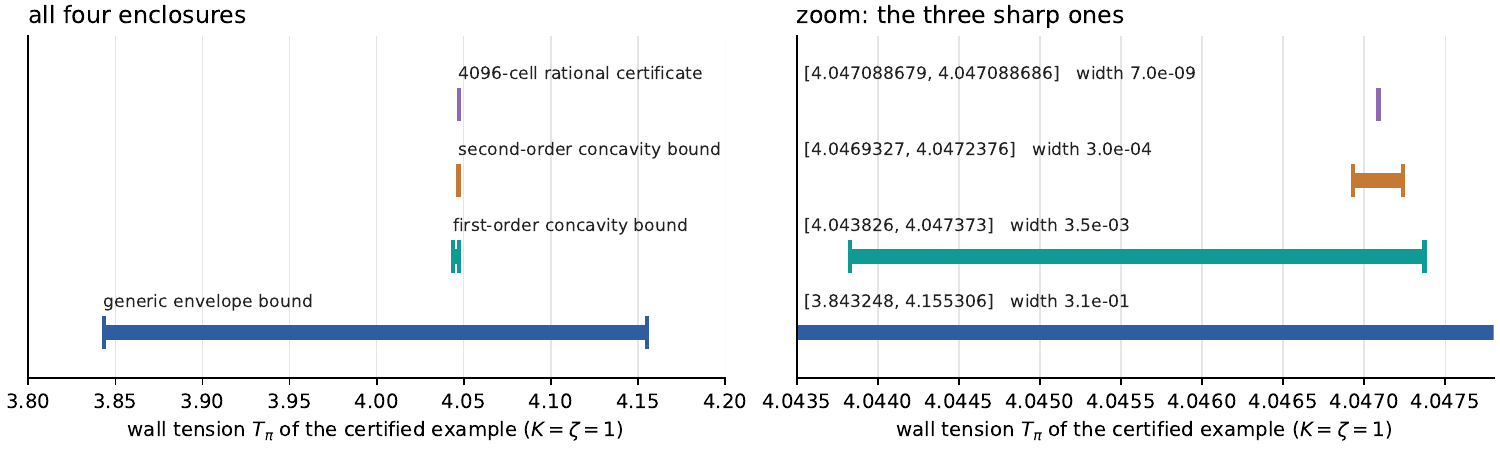}
\caption{Nested enclosures of the illustrative wall tension: the generic envelope bound \eqref{ef-eq:Tbound}, the first- and second-order concavity enclosures of Theorem~\ref{ef-thm:concave}, and the 4096-cell rational certificate. Widths differ by eight orders of magnitude; all are proofs for the specified functional.}
\label{ef-fig:enclosures}
\end{figure}

Use normalized parameters $K=\zeta=1$ and choose
\begin{align}
Z(s)&=1+\frac1{50}\sin^2s,\\
V(s)&=(1-\cos2s)+\frac1{100}(1-\cos4s)
                  -\frac1{500}(1-\cos6s).
\end{align}
These are illustrative specified coefficients, not fitted or derived Yang-Mills parameters. Exactly,
\begin{equation}
\varepsilon=\frac{29}{500},\quad \eta=\frac1{50},\quad \rho=\frac{1979}{25000},\quad
m_{\mathrm{curv}}^2=\frac{511}{125}=4.088.
\end{equation}
The general theorem gives $3.84324862<T_\pi<4.15530505$. Theorem~\ref{ef-thm:concave} gives, with $\delta_1=15544/328125\approx0.0473721$ (kinetic part $\frac{2}{75}$, potential part $2(\frac43\cdot\frac1{100}-\frac{23}{15}\cdot\frac1{500})=\frac{77}{3750}$, cross term $\frac{113}{656250}$, all exact rational values),
\begin{equation}
\begin{aligned}
4.043826&<T_\pi<4.047373 &&\text{(first order)},\\
4.0469327&<T_\pi<4.0472376 &&\text{(second order)}.
\end{aligned}
\end{equation}
and the sharper exact arithmetic certificate gives
\begin{equation}
\boxed{4.047088679<T_\pi<4.047088686.}
\end{equation}
For the corresponding two-wall infimum, double both endpoints. The leading unperturbed value was $4$ for one wall and $8$ for two. The three analytic enclosures are nested and contain the rational certificate; their widths are $0.31$, $3.5\cdot10^{-3}$, $3.0\cdot10^{-4}$ and $5.8\cdot10^{-9}$ respectively ($7\cdot10^{-9}$ for the last after outward decimal rounding; Figure~\ref{ef-fig:enclosures}).

\subsection{How the enclosure is proved}
Write $u=\cos s$ and $w=1-u^2$. Straight algebra gives
\begin{align}
T_\pi&=4\int_0^1\sqrt{Q(u)}\dd u,\\
Q(u)&=\left(1+\frac{w}{50}\right)
 \left(\frac{511}{500}+\frac{w}{125}-\frac{4w^2}{125}\right).
\end{align}
The pointwise theorem gives $Q\ge471/500>9/10$. Let $B_1$ and $B_2$ be the sums of absolute coefficients of $Q'$ and $Q''$. On $[0,1]$ they bound the respective derivatives. Since $\sqrt Q\ge9/10$, a valid bound for $f=\sqrt Q$ is
$|f''|\le M:=B_2/(2\cdot\frac9{10})+B_1^2/(4(\frac9{10})^3)$.
For $N=4096$ equal cells, the composite midpoint error for $4\int_0^1f$ is at most $M/(6N^2)$. All sample values of $Q$ are rational. Integer square roots provide outward bounds for $\sqrt Q$ on a $2^{-90}$ grid. Combining these bounds with the analytic quadrature remainder proves the displayed enclosure without floating-point decisions.

\begin{center}
\includegraphics[width=\linewidth]{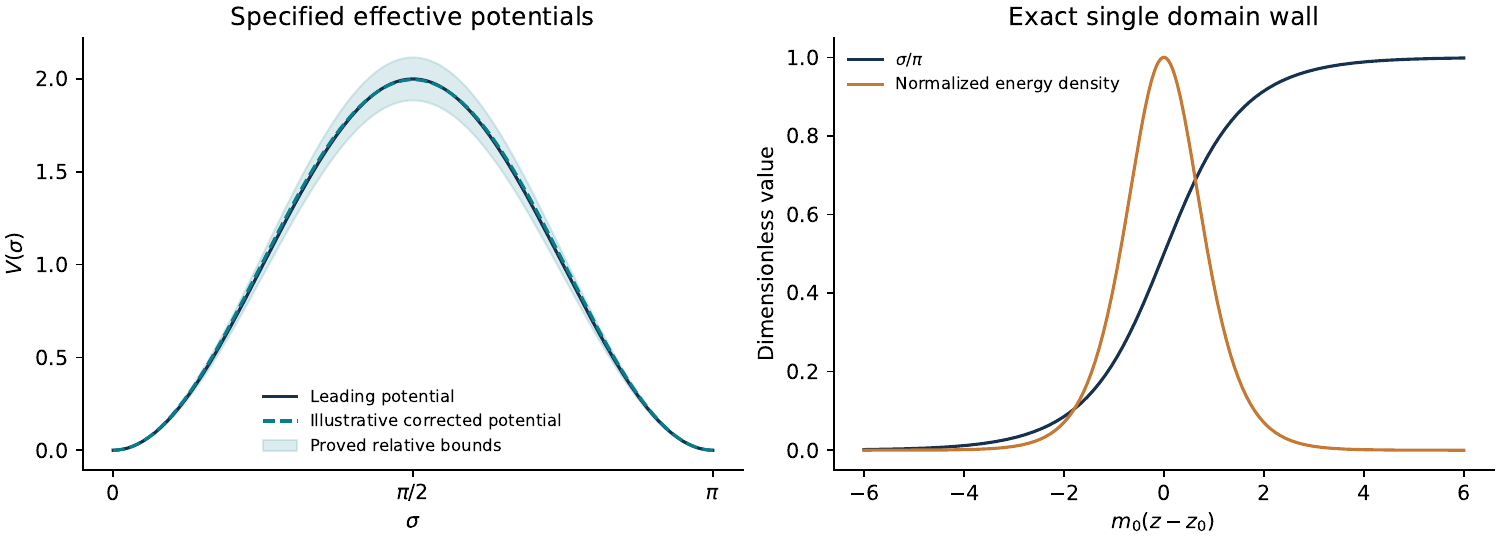}
\end{center}
\vspace{-4pt}
{\small The plots show the explicitly defined model. They are illustrations, not numerical evidence for microscopic four-dimensional confinement.}

\section{Linear stability of the perturbed wall }
\label{ef-sec:stability}
Theorem~\ref{ef-thm:harmonic} produced a wall for every admissible $(Z,V)$ as the minimiser of a one-dimensional problem. Minimisers have nonnegative second variation, but the structure of that second variation is what controls the fluctuation spectrum (and, in the quantum theory, the one-loop correction to the tension). The Bogomolny factorisation of the leading model extends to the whole class.

Assume in this section that $Z\in C^2$ and $V\in C^2$ (the latter follows from $\sum n^2|c_n|<\infty$), and put
\begin{equation}
W(s)=\sqrt{\frac{2V(s)}{Z(s)}},\qquad \sigma_*'=W(\sigma_*),\qquad w(z)=W'(\sigma_*(z)),
\end{equation}
so that $w(z)\to\pm m_{\rm curv}$ as $z\to\mp\infty$ by $\pi$-periodicity of $Z$ and $V$.

\begin{theorem}[Factorised second variation]
\label{ef-thm:stability}
For every real $\psi\in H^1(\R)$, the quadratic term in $E[\sigma_*+t\psi]$ is $t^2\mathcal Q[\psi]$, where
\begin{equation}
\mathcal Q[\psi]=\frac12\int_{\R}Z(\sigma_*)\bigl(\psi'-w\psi\bigr)^2\dd z\ \ge0 ,
\label{ef-eq:Q}
\end{equation}
with equality if and only if $\psi\in\R\,\sigma_*'$. Equivalently the fluctuation operator
\begin{equation}
\begin{aligned}
\mathcal L&=Z(\sigma_*)^{-1}\bigl[-\partial_zZ(\sigma_*)\partial_z+U\bigr],\\
U&=\bigl(Z(\sigma_*)w\bigr)'+Z(\sigma_*)w^2\\
 &=V''(\sigma_*)-\tfrac12Z''(\sigma_*)\sigma_*'^2-Z'(\sigma_*)\sigma_*''.
\end{aligned}
\end{equation}
defined by the closed form $2\mathcal Q$ on $L^2(\R,Z(\sigma_*)\dd z)$, is self-adjoint and nonnegative, has the simple eigenvalue $0$ with eigenfunction $\sigma_*'$, and essential spectrum $[m_{\rm curv}^2,\infty)$. In the leading model it is the P\"oschl--Teller operator of Section~\ref{ef-sec:wall}.
\end{theorem}
\begin{proof}
Expanding $E[\sigma_*+\psi]$ to second order and using the Euler--Lagrange equation for the first-order terms, the quadratic form is
$\mathcal Q[\psi]=\frac12\int[Z\psi'^2+2Z'\sigma_*'\psi\psi'+(\tfrac12Z''\sigma_*'^2+V'')\psi^2]$ with $Z,Z',Z'',V''$ evaluated at $\sigma_*(z)$. Integrating the middle term by parts ($2\psi\psi'=(\psi^2)'$) turns it into $-(Z'\sigma_*')'\psi^2=-(Z''\sigma_*'^2+Z'\sigma_*'')\psi^2$, which gives the second expression for $U$. On the other hand expanding \eqref{ef-eq:Q} and integrating $-2Zw\psi\psi'$ by parts gives $\frac12\int[Z\psi'^2+((Zw)'+Zw^2)\psi^2]$. The two expressions agree because, differentiating $ZW^2=2V$ twice in $s$,
\[
V''=\tfrac12Z''W^2+2Z'WW'+ZW'^2+ZWW'' ,
\]
and substituting $\sigma_*'=W$, $\sigma_*''=WW'$, $(Zw)'=W(Z'W'+ZW'')$ one finds $(Zw)'+Zw^2=V''-\tfrac12Z''W^2-Z'WW'$. Boundary terms vanish for $\psi\in H^1$ since all coefficients are bounded ($\sigma_*$ takes values in $[0,\pi]$). Nonnegativity is manifest from \eqref{ef-eq:Q}; $\mathcal Q[\psi]=0$ forces $\psi'=w\psi$ a.e., whose solutions are $\psi=C\exp\!\int w=C\exp\!\int W'(\sigma_*)\sigma_*'/W(\sigma_*)\dd z=C\,W(\sigma_*)=C\sigma_*'$, which is square integrable because $\sigma_*'$ decays exponentially: from \eqref{ef-eq:pointbound}, $W(s)\ge\mu|\sin s|$ with $\mu=m_0\sqrt{(1-\varepsilon)/(1+\eta)}$, so $\sigma_*'\ge\mu\sin\sigma_*$ and $\sigma_*(z)\le2\arctan e^{\mu(z-z_1)}$ for $z\to-\infty$ (similarly at $+\infty$). For a precise essential-spectrum argument, put $p(z)=\sqrt{Z(\sigma_*(z))}$. Its first two derivatives tend to zero exponentially. The unitary map $\psi\mapsto p\psi$ transforms the operator into
\[
-\partial_z^2+\frac{U}{Z}+\frac{p''}{p}.
\]
Continuity of $V''$ and the endpoint limits give $U/Z\to V''(0)/Z(0)=m_{\rm curv}^2$ at both ends. An exponential rate for this last convergence is unnecessary and does not follow from $C^2$ regularity alone. The transformed potential is bounded and differs from $m_{\rm curv}^2$ by a function tending to zero, hence is a relatively compact perturbation of $-\partial_z^2+m_{\rm curv}^2$. Weyl's theorem gives the claimed essential spectrum. The form is closed because $Z$ is uniformly positive and bounded, $w$ is bounded, and its form norm is equivalent to the $H^1$ norm after adding the $L^2$ norm.
\end{proof}

\begin{remark}[Bound states]
Unlike the leading model, the partner operator $QQ^\dagger$ is not of constant coefficients when $Z$ or $V$ is perturbed, so the theorem does not exclude further discrete eigenvalues in $(0,m_{\rm curv}^2)$. Such eigenvalues are internal (shape) modes of the wall. For the certified example of Section~\ref{ef-sec:example} two numerical methods produce a candidate eigenvalue
\begin{equation}
\lambda_1=4.08380\qquad(\text{threshold }m_{\rm curv}^2=4.088,\ \text{binding }m_{\rm curv}^2-\lambda_1=0.0042),
\end{equation}
from a generalised finite-difference eigenproblem on boxes up to $|z|\le384$ and a shooting computation of the Jost mismatch, agreeing to $2\cdot10^{-5}$. These calculations do not include interval bounds on discretisation, tails, or the number of eigenvalues.  Its wavefunction extends over $\sim1/\sqrt{m^2_{\rm curv}-\lambda_1}\approx15$ units, which is why small boxes miss it. The leading model has a threshold resonance. The numerical result is consistent with its binding under this perturbation. In the linearised $2+1$-dimensional effective theory such a mode would propagate along the wall with mass $\sqrt{\lambda_1}$ below the bulk linear-fluctuation threshold. The translational branch remains gapless along the wall, and no full interacting quantum spectral gap follows.
\end{remark}

\begin{figure}[htbp]
\centering
\includegraphics[width=\linewidth]{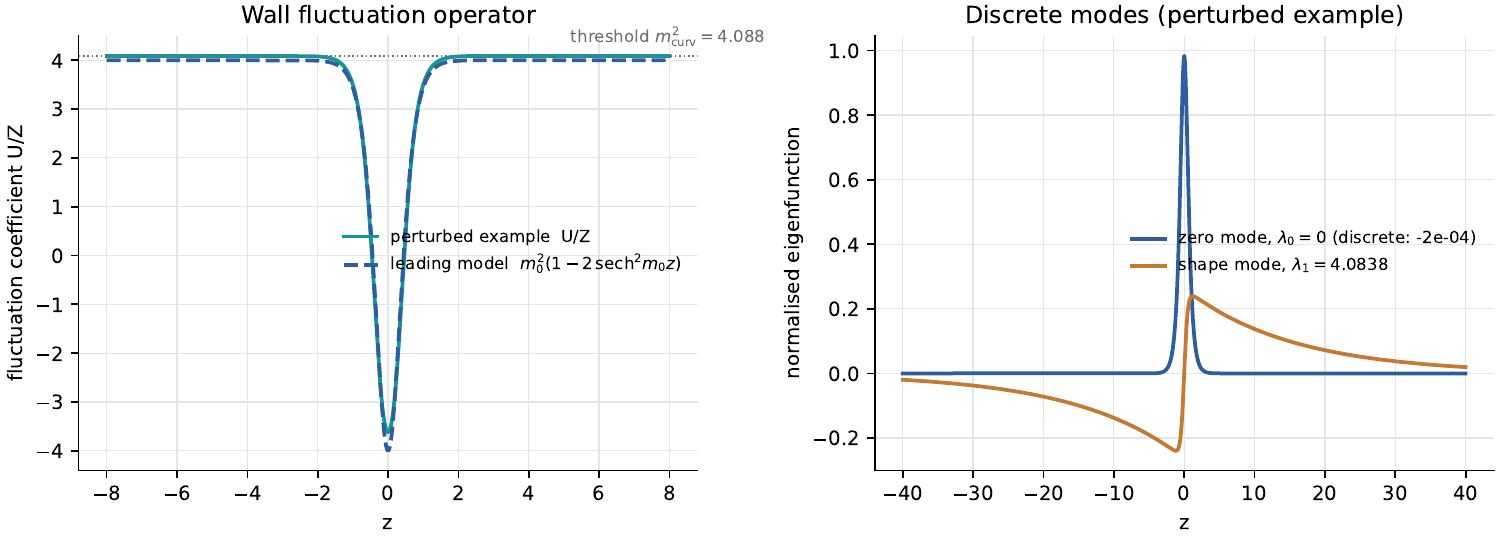}
\caption{Left: the coefficient $U/Z$ of the weighted fluctuation operator for the certified example compared with the P\"oschl--Teller well of the leading model. Right: the translation zero mode and the near-threshold shape mode of the perturbed wall (normalised), with the continuum threshold indicated.}
\label{ef-fig:spectrum}
\end{figure}

\section{Confinement in the bulk, deconfinement on the wall }
\label{ef-sec:probes}
Section~\ref{ef-sec:dual} identified a fundamental probe with a $2\pi$ monodromy of the dual field. We now prove, for the whole admissible class of Theorem~\ref{ef-thm:harmonic}, that a probe--antiprobe pair in the bulk costs a double wall of tension $2T_\pi$ per unit length, while the same pair placed on a $\pi$-wall costs an energy bounded uniformly in the separation. Both statements are about the classical effective functional; the physical picture is that of Anber, Poppitz and Sulejmanpasic \cite{ef:aps} and of discrete anomaly inflow \cite{ef:ksu,ef:coxpoppitzwong}.

\subsection*{Setting}
Work in the plane transverse to two parallel probe lines (energies are per unit length of the probes). Let $S_R=(-R,R)\times\R$ be a strip, let the probes sit at $P_\pm=(\pm d/2,0)$ with $d>4$, $R>d/2+2$, and let $\Gamma_d=[-d/2,d/2]\times\{0\}$ be the segment joining them. Fix a core radius $\varrho\in(0,\tfrac14)$ and remove the discs $D_\pm$ of radius $\varrho$ around $P_\pm$. An \emph{admissible field} is a function $\sigma\in H^1_{\rm loc}(S_R\setminus(\Gamma_d\cup D_+\cup D_-))$ which is continuous up to $\Gamma_d$ from both sides with
\begin{equation}
\sigma(x,0^+)-\sigma(x,0^-)=2\pi\qquad(|x|<d/2),
\label{ef-eq:jump}
\end{equation}
and finite energy $E[\sigma]=\int\{Z(\sigma)|\nabla\sigma|^2/2+V(\sigma)\}\dd x\dd y$, the integral taken off $\Gamma_d$ and the cores. By $2\pi$-periodicity of $Z$ and $V$ the energy density is unaffected by \eqref{ef-eq:jump}; the segment is a branch cut that may be moved by adding $2\pi$ on one side, so \eqref{ef-eq:jump} is the statement that $\sigma$ is a continuous multivalued function with monodromy $\pm2\pi$ around $P_\pm$. Two sectors are distinguished by the behaviour as $|y|\to\infty$: we require that $\sigma(x,y)$ tends, uniformly in $x$, to a constant $v_-\in\pi\Z$ as $y\to-\infty$ and to a constant $v_+\in\pi\Z$ as $y\to+\infty$ (after a global shift, $v_-=0$), with
\begin{description}[leftmargin=22pt]
\item[bulk:] $v_+\in2\pi\Z$;
\item[wall:] $v_+\in\pi+2\pi\Z$.
\end{description}
The integer in $v_+$ is not fixed in advance; the infimum below is over all admissible fields of the sector.
Denote the infima of $E$ over the two sectors by $E_{\rm bulk}(d;R)$ and $E_{\rm wall}(d;R)$; with no probes ($d=0$, no cut) the corresponding values are $0$ and $2RT_\pi$, the latter attained by $\sigma_*(y)$.

\begin{theorem}[Two walls in the bulk, none on the wall]
\label{ef-thm:probes}
Let $(Z,V)$ satisfy the hypotheses of Theorem~\ref{ef-thm:harmonic} with $Z\in C^1$, and let $T_\pi=\int_0^\pi\sqrt{2ZV}$. Then:
\begin{enumerate}[label=(\alph*)]
\item (bulk, lower bound) for $R\ge d$, $E_{\rm bulk}(d;R)\ge2T_\pi\,(d-2\varrho)$;
\item (bulk, upper bound) for every $s\ge0$, $E_{\rm bulk}(d;R)\le e(s)\,d+C_{\rm B}(\varrho,s)$, where $e(s)=E_{1D}[\rho(\cdot+s)+\rho(\cdot-s)]$ is the one-dimensional energy of two walls at separation $2s$, $\rho(y)=\sigma_*(-y)-\pi$ being the wall from $0$ to $-\pi$, and $C_{\rm B}$ does not depend on $d$ or $R$; since $e(s)\to2T_\pi$ as $s\to\infty$,
\[
\lim_{d\to\infty}\ \frac{E_{\rm bulk}(d;R)}{d}=2T_\pi\qquad\text{uniformly in }R\ge d ;
\]
\item (wall) $-4\varrho T_\pi\le E_{\rm wall}(d;R)-2RT_\pi\le C_{\rm W}(\varrho)$ with $C_{\rm W}$ independent of $d$ and $R$.
\end{enumerate}
In particular the probe--antiprobe potential $\mathcal V_{\rm bulk}(d)=E_{\rm bulk}(d;R)-E_{\rm bulk}(0;R)$ ($R\ge d$) grows linearly with asymptotic slope $2T_\pi$, while $\mathcal V_{\rm wall}(d)=E_{\rm wall}(d;R)-E_{\rm wall}(0;R)\in[-4\varrho T_\pi,C_{\rm W}]$ is bounded uniformly in $d$.
\end{theorem}
\begin{proof}
\emph{Lower bounds.} Fix $x$ with $|x\mp d/2|\ge\varrho$ and consider the vertical line $\{x\}\times\R$. Along it, define the continuous lift $\ell(y)=\sigma(x,y)-2\pi\,\mathbf 1_{y>0}\mathbf 1_{|x|<d/2}$; by \eqref{ef-eq:jump} it is continuous, and the one-dimensional energy of the line equals $\int\{Z(\ell)\ell'^2/2+V(\ell)\}\dd y$ because $Z,V$ are $2\pi$-periodic and $\partial_y\sigma=\ell'$ off the cut. Since $Z|\nabla\sigma|^2\ge Z(\partial_y\sigma)^2$, $E[\sigma]\ge\int\dd x\,E_{1D}[\ell(x,\cdot)]$. The pointwise inequality of Theorem~\ref{ef-thm:harmonic} gives $E_{1D}[\ell]\ge\left|\int_{\ell(-\infty)}^{\ell(+\infty)}\sqrt{2ZV}\right|=|\ell(+\infty)-\ell(-\infty)|\,T_\pi/\pi$, since $\sqrt{2ZV}$ is $\pi$-periodic and the endpoint values lie in $\pi\Z$ (in both sectors the limits are vacua). By definition of the sectors the limits $v_\pm$ are the same for all $x$; normalise $v_-=0$ and write $v_+=\pi+2\pi k$ (wall) or $2\pi k$ (bulk). For almost every $x$ the restriction of $\sigma$ to the vertical line is absolutely continuous off the cut, so the lift is well defined. In the wall sector $\ell(+\infty)=\pi+2\pi k$ for lines not crossing the cut and $\pi+2\pi(k-1)$ for lines crossing it; both are odd multiples of $\pi$, so every line contributes at least $T_\pi$, and the lines avoiding the cores have total measure $2R-4\varrho$; this is (c)'s lower bound. In the bulk sector $\ell(+\infty)=2\pi k$ or $2\pi(k-1)$ respectively, giving $E\ge2T_\pi[\,|k|(2R-d-2\varrho)+|k-1|(d-2\varrho)\,]$; for $R\ge d$ the minimum over $k\in\Z$ is at $k=0$, which is (a).

\emph{Upper bounds.} Let $\rho(y)=\sigma_*(-y)-\pi$; it is a wall from $0$ (at $y=-\infty$) to $-\pi$ (at $y=+\infty$) and has the same one-dimensional energy $T_\pi$ as $\sigma_*$, because $Z$ and $V$ are $\pi$-periodic and the energy density is invariant under $y\to-y$. Let $M(y)=\rho(y)+2\pi\mathbf 1_{y>0}-\sigma_*(y)$, i.e. $M(y)=\sigma_*(-y)+\pi-\sigma_*(y)$ for $y>0$ and $\sigma_*(-y)-\pi-\sigma_*(y)$ for $y<0$; it decays exponentially at $\pm\infty$ by the pointwise bounds and the first-order wall equation and $M(0^+)-M(0^-)=2\pi$. Let $\lambda\in C^\infty(\R)$ increase from $0$ on $(-\infty,-1]$ to $1$ on $[1,\infty)$, let $\chi(r)$ be a smooth cutoff equal to $1$ for $r\le1$ and $0$ for $r\ge2$, and let $\vartheta\in(-\pi,\pi]$ be the polar angle. Define the \emph{junction function}
\[
\Lambda(x,y)=\chi(r)\,\frac{1+\cos\vartheta}{2}+(1-\chi(r))\,\lambda(x),
\]
smooth on $\R^2\setminus\{0\}$, equal to $\lambda(x)$ for $r\ge2$, with $\Lambda(x,0)=1$ for $x>0$ and $\Lambda(x,0)=0$ for $x<0$. For the wall sector set
\[
\sigma_{\rm W}(x,y)=\sigma_*(y)+\bigl[\Lambda(x+\tfrac d2,y)-\Lambda(x-\tfrac d2,y)\bigr]M(y).
\]
Its jump across $y=0$ is $2\pi[\Lambda(x+\frac d2,0)-\Lambda(x-\frac d2,0)]=2\pi\mathbf 1_{|x|<d/2}$, so \eqref{ef-eq:jump} holds exactly on $\Gamma_d$ and nowhere else; $\sigma_{\rm W}=\sigma_*(y)$ for $|x|>d/2+2$ and $\sigma_{\rm W}=\sigma_*(-y)\pm\pi$ (a reversed $\pi$-wall of the same tension, by $\pi$-periodicity and spatial reflection invariance of the one-dimensional energy) for $|x|<d/2-2$; these regions contribute exactly $T_\pi$ per unit $x$. In the two transition regions $|x\mp d/2|\le2$, $|\nabla\sigma_{\rm W}|^2\le C(|\nabla\Lambda|^2M^2+|M'|^2+\sigma_*'^2)$ and $V(\sigma_{\rm W})-V(\text{wall})=O(|M|)$; since $|M|,|M'|\le Ce^{-\mu|y|}$, $|\nabla\Lambda|\le C/r$ for $r\le2$, and $|\nabla\Lambda|\le C$ for $r\ge2$, the excess energy of each transition region is bounded by $C\int_{\varrho}^{2}\dd r/r+C$, finite and independent of $d,R$. This is (c)'s upper bound, with $C_{\rm W}=O(\log1/\varrho)$ the logarithmic self-energy of the probe lines. For the bulk sector use the double-wall profile $M_s(y)=\rho(y+s)+\rho(y-s)+2\pi\mathbf 1_{y>0}$, whose continuous lift descends from $0$ to $-2\pi$ through the intermediate vacuum $-\pi$; it also decays exponentially at both ends and jumps by $2\pi$. Set $\sigma_{\rm B}=[\Lambda(x+\frac d2,y)-\Lambda(x-\frac d2,y)]M_s(y)$. Between the probes it is the one-dimensional double wall of energy $e(s)$ per unit length, outside it vanishes, and the same estimate bounds the transition regions by $C_{\rm B}(\varrho,s)$. Finally $e(s)\to2T_\pi$ exponentially, by the exponential approach of $\sigma_*$ to its limits; together with (a) this gives the limit in (b).
\end{proof}

\begin{remark}
The constants $C_{\rm B},C_{\rm W}$ contain the logarithmically divergent Coulomb self-energy of the probe lines (in the dual picture, the $1/r$ gradient of a vortex), which may be subtracted using a common fixed-core convention when comparing configurations with the same probes. The displayed bounds retain their stated $\varrho$ dependence. In the one-component leading model the two walls between bulk probes repel, so the minimiser pushes them apart as far as the geometry allows; the excess $E_{\rm bulk}(d;R)-2T_\pi d$ therefore grows slowly with $d$ (Figure~\ref{ef-fig:probes}), the ``double-string'' of \cite{ef:aps}. On the wall the two half-walls $0\to\pi$ and $0\to-\pi$ are exactly degenerate, which is the effective-model form of the statement that the probe's $2\pi$ monodromy is absorbed by the $\Z_2$ shift \eqref{ef-eq:chiralshift} across the wall.
\end{remark}

\subsection*{Lattice illustration }
\begin{figure}[htbp]
\centering
\includegraphics[width=\linewidth]{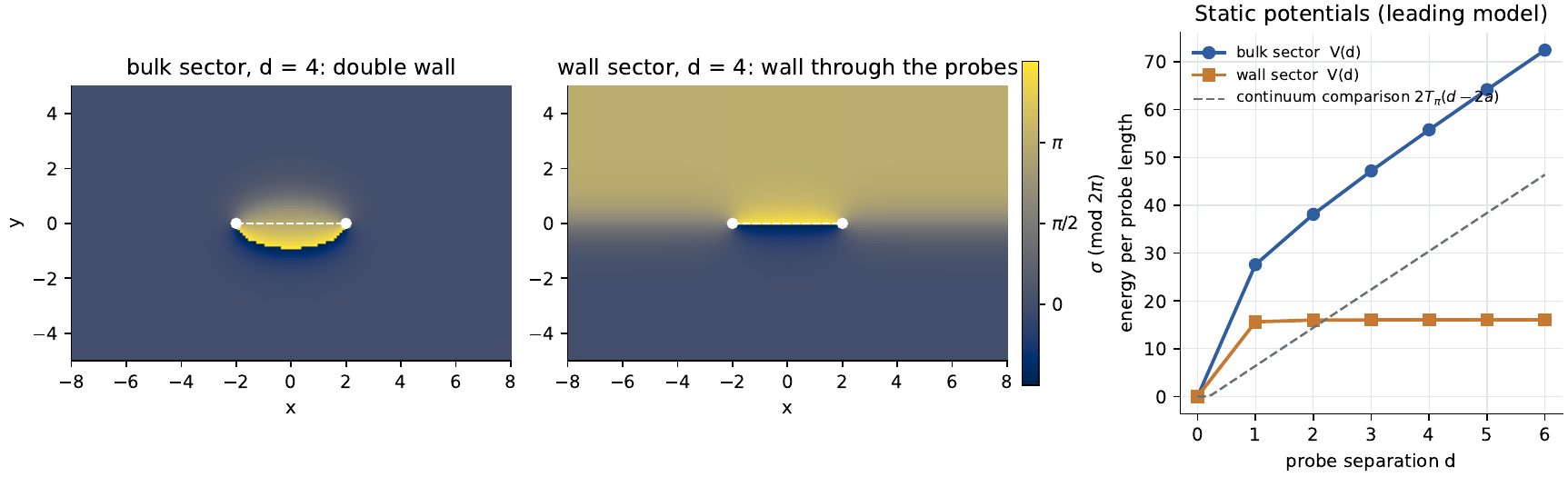}
\caption{Numerically relaxed lattice configurations of the leading model with two fundamental probes at separation $d=4$ (left: bulk sector, a double wall of total monodromy $2\pi$ joins the probes; middle: wall sector, the $\pi$-wall passes through both probes and changes from $0\to\pi$ to $0\to-\pi$ between them at no asymptotic cost), and the two static potentials (right) with the continuum comparison line $2T_\pi(d-2a)$. The continuum theorem is not a bound on this discretisation without an additional comparison estimate.}
\label{ef-fig:probes}
\end{figure}

A numerical local minimisation uses the lattice energy $\sum_{\rm links}\frac K2(\Delta\sigma-2\pi\eta)^2+a^2\sum_{\rm sites}V(\sigma)$ of the leading model ($K=\zeta=1$, $a=0.1$, strip $16\times10$) with the $2\pi$ shift $\eta$ on the vertical links crossing $\Gamma_d$. The results (Figure~\ref{ef-fig:probes}) show $\mathcal V_{\rm wall}(d)$ saturating at $16.03$ (two probe self-energies; increments fall below $10^{-3}$ beyond $d=5$) and $\mathcal V_{\rm bulk}(d)$ growing with fitted slope $8.41$ over $d\in[3,6]$, approaching $2T_\pi=8$ from above as the double wall widens. The finite-grid results are qualitatively consistent with the continuum theorem. No certified discretisation or global-minimum error bound is supplied for this illustration.

\section{Exact confinement calculations in two dimensions}
\label{ef-sec:2d}
For a fully exact gauge-theory comparison, consider pure two-dimensional Yang-Mills on the plane in its heat-kernel formulation \cite{Migdal1975,ef:witten2d,ef:nguyen,ef:gross}. There are no added fermions in this subsection. These results are established mathematics and physics, reproduced here as benchmarks.

Let $G=SU(N)$ with normalized Haar measure and Casimir convention $C_2(F)=(N^2-1)/(2N)$. The heat kernel of $\frac12\Delta_G$ is
\begin{equation}
K_t(U)=\sum_{\lambda}d_\lambda e^{-tC_2(\lambda)/2}
\chi_\lambda(U),\qquad t>0.
\end{equation}
It is positive, has total integral one and satisfies $K_s*K_t=K_{s+t}$. These are standard compact-group heat-equation facts; the sum is smooth for positive time.

\begin{theorem}[Planar simple-loop area law]
Give each planar face of area $A_f$ heat-kernel weight $K_{g_2^2A_f}$. For a simple loop $C$ of enclosed area $A$, the normalized Wilson character in an irreducible representation $R$ satisfies
$\langle W_R(C)\rangle=e^{-g_2^2C_2(R)A/2}$.
This expression is unchanged under any subdivision of the enclosed faces.
\end{theorem}
\begin{proof}
Apply the Haar-preserving tree reduction of Section~\ref{gauge_foundations-spanning-tree-reduction-with-its-measure-identity} on a planar disk. The independent-face argument of Section~\ref{gauge_foundations-rectangular-wilson-loops-and-central-convolution} gives the stated heat-kernel laws. Gluing two faces convolves their densities; the heat semigroup therefore replaces their two times by the sum. Repetition leaves the boundary holonomy with density $K_{g_2^2A}$. The same conclusion follows directly by integrating internal edges with character orthogonality.
Consequently, with $t=g_2^2A$,
$\langle W_R(C)\rangle
=\frac1{d_R}\int_G\chi_R(U)K_t(U)\dd U
=\frac1{d_R}d_{\overline R}e^{-tC_2(\overline R)/2}
=e^{-tC_2(R)/2}$.
The second equality uses $\int\chi_R\chi_\lambda=\delta_{\lambda,\overline R}$, and $R$ and its conjugate have the same dimension and Casimir. Since convolution adds face areas, subdivision does not change the answer. The heat-kernel continuum construction preserves this observable.
\end{proof}
For a rectangle of spatial width $R$ and Euclidean time $T$,
$V_R^{\mathrm{static}}(R)=-\lim_{T\to\infty}\frac1T\log\langle W_R(R\times T)\rangle
=\frac{g_2^2C_2(R)}2R$.
For a fundamental source, $\sigma_{SU(2)}=3g_2^2/8$ and $\sigma_{SU(3)}=2g_2^2/3$.

\subsection{Nested loops and the two screening channels }
\begin{proposition}[Nested two-loop correlator]
Let $C_1\subset C_2$ be nested simple loops with the same orientation, enclosing areas $A_1<A_2$, in representations $R_1,R_2$. Base their holonomies using a connecting path. Reversing one orientation replaces its representation by its dual. Then
\begin{equation}
\bigl\langle W_{R_1}(C_1)W_{R_2}(C_2)\bigr\rangle
=\frac{e^{-g_2^2C_2(R_2)(A_2-A_1)/2}}{d_{R_1}d_{R_2}}\sum_{\lambda}N^{\lambda}_{R_1R_2}\,d_\lambda\,e^{-g_2^2C_2(\lambda)A_1/2},
\end{equation}
where $N^\lambda_{R_1R_2}$ are the Clebsch--Gordan multiplicities of $R_1\otimes R_2$. For $SU(2)$ and two fundamental loops,
\begin{equation}
\langle W_F(C_1)W_F(C_2)\rangle=\tfrac14e^{-3g_2^2(A_2-A_1)/8}\bigl[1+3e^{-g_2^2A_1}\bigr].
\end{equation}
\end{proposition}
\begin{proof}
Let $U_1$ be the holonomy of $C_1$ (density $K_{t_1}$, $t_1=g_2^2A_1$) and $V$ that of the annular face (density $K_{t_2}$, $t_2=g_2^2(A_2-A_1)$), so the holonomy of $C_2$ is $U_1V$. The $V$-integral is a character convolution: $\int\chi_{R_2}(U_1V)K_{t_2}(V)\dd V=e^{-t_2C_2(R_2)/2}\chi_{R_2}(U_1)$ by orthogonality. The remaining integral $\int\chi_{R_1}\chi_{R_2}K_{t_1}$ is evaluated by decomposing the product of characters into irreducibles. For $SU(2)$, $F\otimes F=0\oplus1$ with $C_2(1)=2$ and $d=1,3$.
\end{proof}
The formula is an application of the Migdal--Rusakov rules \cite{Migdal1975,ef:rusakov,ef:witten2d}. Direct $SU(2)$ Haar quadrature checks its convolution step numerically. The annulus carries the representation $R_2$; the tensor-product channels occur in the inner area $A_1$. For two fundamental $SU(2)$ loops their algebraic weights are $1/4$ and $3/4$, associated with the singlet and adjoint channels. As $A_1$ grows the adjoint contribution is exponentially suppressed relative to the singlet term. These coefficients are character-decomposition weights, not probabilities of dynamical screening events. A single-loop area law and a singlet component in a two-loop correlator are distinct statements.

\subsection{With fermions: the Gross--Klebanov--Matytsin--Smilga dichotomy}
Adding dynamical adjoint fermions destroys the simple independent-face calculation, and the result changes qualitatively \cite{ef:gkms}: for massless adjoint Majorana fermions the fundamental string tension vanishes (screening), while for small mass $m$ it is proportional to $m\langle\bar\psi\psi\rangle$. Together with the proposition above, two-dimensional gauge theory supplies exact instances of three phenomena that the four-dimensional discussion must distinguish---an area law for fundamental probes, pair screening in the singlet channel, and dynamical screening by massless adjoint matter---none of which is decided by representation theory alone. Two-dimensional pure Yang-Mills has no propagating local gluons; its representation structure and continuum problem differ substantially from four dimensions. None of these exact statements transfers to four dimensions without proof.

\section{A finite Hamiltonian benchmark for the effective scalar theory}
\label{ef-sec:quantum}
The static functional also motivates a quantum scalar model. This section
specifies a finite regularisation, proves its twist equivalence and a
time-discretisation bound, and records numerical spectra. A representation
cutoff makes the Hamiltonian a finite matrix; the bounds below concern
that matrix and do not estimate the omitted rotor representations or
identify it with the microscopic Yang--Mills Hamiltonian.

\subsection{Hamiltonian and momentum compression}
The formal continuum Hamiltonian associated with \eqref{ef-eq:sg} is
\[
H=\int\dd^2x\left[\frac{\Pi^2}{2K}+
\frac K2(\nabla\sigma)^2+V(\sigma)\right].
\]
For a compact field, the globally defined site operators are $p_x$ and
$U_x=e^{i\sigma_x}$, with $p_x|p\rangle=p|p\rangle$,
$U_x|p\rangle=|p+1\rangle$ and $[p_x,U_x]=U_x$ on their natural
domain. There is no globally canonical bounded angle operator on the
circle. A compact square-lattice regularisation is
\begin{equation}
aH=\sum_x\left[\frac{p_x^2}{2\kappa}
 +\kappa\sum_\mu\bigl(1-\cos(\sigma_{x+\mu}-\sigma_x-\theta_{x\mu})\bigr)
 +\lambda(1-\cos2\sigma_x)
 +\sum_{n\ge2}\gamma_n(1-\cos2n\sigma_x)\right],
\label{ef-eq:rotor}
\end{equation}
where $\kappa=Ka$, $\lambda=\zeta a^3$, and $\gamma_n=c_na^3$ in
$2+1$ dimensions. The analogous chain uses $\kappa=K$,
$\lambda=\zeta a^2$, and $\gamma_n=c_na^2$.
Cosines in this formula are bounded multiplication operators on the
untruncated product of circles. Hard momentum compression means
$H_P=\Pi_PH\Pi_P$ on the product of site spaces
$\operatorname{span}\{|p\rangle:-P\le p\le P\}$, and is the
regularisation used in the numerical calculation.

A distinct noncompact discretisation uses
$\kappa(\sigma_{x+\mu}-\sigma_x)^2/2$ with site space $L^2(\R)$;
field-grid approximations to such scalar models are discussed in
\cite{ef:jlp}. Compact cosine gradients allow lattice vortices, whose core
energies are regularisation dependent. Matching their weights to
microscopic charged particles requires additional input.

In the slowly varying classical approximation, \eqref{ef-eq:rotor} gives
$m_0a=2\sqrt{\lambda/\kappa}$ and
$T_\pi a^2\simeq4\sqrt{\kappa\lambda}$ in two spatial dimensions
($M_\pi a\simeq4\sqrt{\kappa\lambda}$ on a chain). The tension formula
is a continuum-gradient approximation, not an exact identity for a
finite cosine-gradient lattice. The usual semiclassical requirements
$\kappa\gg1$ and $m_0a\ll1$ suggest $1/K\ll a\ll1/m_0$ in two
spatial dimensions. This is a possible resolution window for an
effective model; it is not a cutoff-removal construction at fixed $K$.

\subsection{Twists and sector-energy differences}
\begin{proposition}[Twist equivalence on a chain]
Let $H_\theta$ be the compact Hamiltonian \eqref{ef-eq:rotor} on a periodic
chain, with every $\theta_\ell\in\{0,\pi\}$. Its spectrum depends on
the link twists only through
$\Theta=\sum_\ell\theta_\ell\pmod{2\pi}$. The global operator
$C=\prod_x(-1)^{p_x}$ commutes with $H_\theta$, and reflection
$|p_x\rangle\mapsto|-p_x\rangle$ changes $\Theta$ to $-\Theta$.
These statements hold both before and after hard momentum compression.
\end{proposition}
\begin{proof}
For a site set $S$, $U_S=\exp(i\pi\sum_{x\in S}p_x)$ is unitary,
commutes with $\Pi_P$ and the kinetic terms, and conjugates $U_x$ to
$-U_x$ precisely for $x\in S$. Thus the on-site even harmonics are
unchanged, while a link twist changes by $\pi$ precisely when that
link crosses the boundary of $S$. Successive choices of intervals move
all twists to one link; their sum modulo $2\pi$ is the remaining
invariant. Taking $S$ to be the whole chain proves the $C$ symmetry.
Momentum reflection replaces every $U_x$ by $U_x^\dagger$ and changes
each twist to its negative. Since both unitaries commute with the hard
cutoff, the same conjugation identities hold for the compressed matrices.
\end{proof}
On a cylinder the corresponding classification requires a flat
$\Z_2$ link field, with zero twist around each contractible plaquette;
local plaquette fluxes are additional invariants for unrestricted twists.

For a periodic chain define the finite-size sector difference
$M_P=E_{0,P}^{(\pi)}-E_{0,P}^{(0)}$. For a flat twist on a cylinder
with physical extent $L_\parallel$ parallel to the interface, define
$\tau_P=(E_{0,P}^{(\pi)}-E_{0,P}^{(0)})/L_\parallel$.
Identifying a limit of these quantities with an infinite-volume kink
mass or wall tension requires control of size and momentum cutoff.
The twist changes signs of link terms without changing the Hilbert-space
dimension. The minimum register size at cutoff $P$ is
$N_s\lceil\log_2(2P+1)\rceil$ qubits, before auxiliary registers and
error correction; this dimension count is not a gate-complexity estimate.

\subsection{A support-counting time-discretisation bound}
In this subsection $H=A+B$ denotes the dimensionless compressed matrix
$aH_P$, with $A=\sum_yA_y$, $A_y=p_y^2/(2\kappa)$, and $B$ the
sum of the compressed on-site and nearest-neighbour potential terms.
Although these terms were multiplication operators before compression,
overlapping terms need not commute after hard compression. The following
estimate uses only their supports and operator norms.
For example, at $P=1$ the compressed multipliers $C_j=\Pi_P\cos(j\sigma)\Pi_P$
obey
\[
[C_1,C_2]=\frac14
\begin{pmatrix}0&-1&0\\1&0&1\\0&-1&0\end{pmatrix}\ne0
\]
in the momentum basis $(-1,0,1)$.

\begin{theorem}[Symmetric-product bound for the compressed rotor]
\label{ef-thm:trotter}
Suppose the lattice has $N_s$ sites and degree at most $2D$,
$\kappa>0$, $\lambda\ge0$, and $\sum_{n\ge2}|\gamma_n|<\infty$.
Set
\[
\alpha=\frac{P^2}{2\kappa},\qquad
b=2\lambda+4D\kappa+2\sum_{n\ge2}|\gamma_n|,
\qquad
S_2(\tau)=e^{-iB\tau/2}e^{-iA\tau}e^{-iB\tau/2}.
\]
For $t\ge0$ and every integer $r\ge1$,
\begin{equation}
\left\|S_2(t/r)^r-e^{-itH}\right\|
\le \frac{t^3}{r^2}N_s(2D+1)
\left[\frac{4\alpha b^2}{24}
      +\frac{4\alpha^2b}{12}\right].
\label{ef-eq:trotter-bound}
\end{equation}
\end{theorem}
\begin{proof}
The integral commutator estimate for the symmetric product
\cite{ef:childs} gives
\[
\|S_2(\tau)-e^{-i\tau(A+B)}\|
\le\frac{\tau^3}{24}\|[B,[B,A]]\|
  +\frac{\tau^3}{12}\|[A,[A,B]]\|.
\]
The first coefficient belongs to the outer operator $B$. This is a
norm estimate from integral remainders, not just a formal
Baker--Campbell--Hausdorff expansion. Telescoping the product of $r$
unitaries multiplies the one-step bound by at most $r$.

Let $B_y$ be the sum of potential terms whose supports contain $y$.
Compression does not increase their norms, so $\|B_y\|\le b$ and
$\|A_y\|\le\alpha$. Then
$[A,B]=\sum_yX_y$, where $X_y=[A_y,B_y]$ has norm at most $2\alpha b$
and support in $S_y=\{y\}\cup\operatorname{nbrs}(y)$,
with $|S_y|\le2D+1$. Therefore
\[
\|[A,[A,B]]\|
\le\sum_y2\left\|\sum_{x\in S_y}A_x\right\|\|X_y\|
\le4N_s(2D+1)\alpha^2b.
\]
For the other nested commutator, only potential terms meeting $S_y$
can contribute to $[B,X_y]$. The sum of their norms is at most
$\sum_{x\in S_y}b\le(2D+1)b$; counting a link twice only increases
this upper bound. Hence
\[
\|[B,[B,A]]\|
\le\sum_y2(2D+1)b\|X_y\|
\le4N_s(2D+1)\alpha b^2.
\]
Substitution and $\tau=t/r$ prove \eqref{ef-eq:trotter-bound}.
\end{proof}
This bound is extensive in the number of sites and controls only the
finite matrix's time discretisation. It does not control the hard
momentum cutoff, ground-state preparation, or finite-volume effects.

\subsection{Finite-matrix numerical spectra}
The calculation uses $N=5$ sites, $m_0a=0.8$,
$\kappa\in\{1.5,2,3,4,6\}$,
$\lambda=\kappa(m_0a)^2/4$, and
$P=\lceil3.5\sqrt{0.65\kappa}\rceil+1$.
Sparse numerical diagonalisation of the compressed matrices gives the
values in Table~\ref{ef-tab:rotor}. Comparing cutoffs $P$ and $P-1$
changes the sector difference by less than $0.5\%$ in these examples.
This is an adjacent-cutoff diagnostic, not a bound on the infinite
representation tail. Matrix eigenvalues and residuals are evaluated in
floating point; the term diagonalisation does not imply an exact
arithmetic spectral certificate.

The computed sector difference is roughly $10$--$41\%$ below the
classical lattice minimum. The finite-lattice one-loop estimate
$M_{\rm cl}+\tfrac12\sum(\omega^{\rm kink}-\omega^{\rm vac})$
is within about $2\%$ of the one-loop correction for these five
parameter choices. That agreement is a finite-model numerical result.
Near-degenerate even and odd eigenstates give very small reported
splittings at larger $\kappa$; values at the floating-point error floor
do not resolve the tunnelling rate or prove spontaneous symmetry breaking.

\begin{table}[htbp]
\centering\scriptsize
\caption{Numerical diagonalisation (ED) of the hard-compressed compact
rotor chain, $N=5$, $m_0a=0.8$. Energies are in units $1/a$.
The remainder column is $(M_{\rm ED}-M_{\rm 1\text{-}loop})/
(M_{\rm 1\text{-}loop}-M_{\rm cl})$. The continuum DHN formula
uses the bare comparison mass and is not expected to agree at this
lattice spacing. Very small parity splittings are unresolved numerical
values rather than certified positive gaps.}
\label{ef-tab:rotor}
\begin{tabular}{@{}rrrrrrrrr@{}}
\toprule
$\kappa$ & $P$ & $M_{\rm ED}$ & $M_{\rm 1\text{-}loop}$ & $M_{\rm cl}$ & DHN (cont.) & rem. & meson gap & $\Z_2$ split\\
\midrule
1.5 & 5 & 1.4676 & 1.4547 & 2.4671 & 2.145 & -1.3\% & 0.588 & 3.2e-04 \\
2.0 & 5 & 2.2973 & 2.2771 & 3.2895 & 2.945 & -2.0\% & 0.643 & 7.5e-06 \\
3.0 & 6 & 3.9356 & 3.9219 & 4.9343 & 4.545 & -1.4\% & 0.699 & 3.4e-09 \\
4.0 & 7 & 5.5710 & 5.5666 & 6.5790 & 6.145 & -0.4\% & 0.726 & unresolved \\
6.0 & 8 & 8.8443 & 8.8562 & 9.8686 & 9.345 & +1.2\% & 0.755 & unresolved \\
\bottomrule
\end{tabular}

\end{table}

The continuum sine-Gordon one-loop formula is
$M=8m/\beta_{\rm SG}^2-m/\pi$ in its stated renormalisation
convention \cite{ef:dhn,ef:rajaraman}. Here $\beta_{\rm SG}=2/\sqrt K$.
For comparison with the cosine-gradient chain, define
\[
R_{\rm lat}=\frac12\sum_k
(\omega_k^{\rm kink}-\omega_k^{\rm vac})
 +M_{\rm cl}\left[\langle\sigma^2\rangle+
              \tfrac14\langle(\Delta\sigma)^2\rangle\right]
 +\frac{m_0a}{\pi},
\]
where
\[
\langle\sigma^2\rangle=\frac1N\sum_k\frac1{2\kappa\omega_k},
\quad
\langle(\Delta\sigma)^2\rangle=\frac1N\sum_k
\frac{2-2\cos k}{2\kappa\omega_k},
\quad \omega_k^2=4\sin^2(k/2)+(m_0a)^2.
\]
The two displayed expectation values are harmonic-vacuum contractions
of the on-site and compact-gradient terms. Their use in the comparison
retains its one-loop status. In the continuum momentum integral,
$\langle(\Delta\sigma)^2\rangle\to2/(\pi\kappa)$ as $m_0a\to0$;
this follows by cancelling one factor of $|\sin(k/2)|$ and integrating
over the Brillouin zone. If also $M_{\rm cl}/(2\kappa m_0a)\to1$,
the corresponding gradient term approaches $m_0a/\pi$.
The computed residual ratios in Table~\ref{ef-tab:dhn} decrease for the
three reported lattices. This finite trend does not prove a remainder
rate or a continuum limit.

\begin{table}[htbp]
\centering\scriptsize
\caption{One-loop comparison on long compact-rotor chains at $\kappa=10$.
The residual is the explicitly defined $R_{\rm lat}$, in units $m_0a$;
the table gives numerical values without a uniform remainder certificate.}
\label{ef-tab:dhn}
\begin{tabular}{@{}rrrrrr@{}}
\toprule
$N$ & $m_0a$ & $\frac12\sum(\omega^{\rm kink}-\omega^{\rm vac})$ & $-M_{\rm cl}[\langle\sigma^2\rangle+\frac14\langle(\Delta\sigma)^2\rangle]-m/\pi$ & residual$/m$ & gradient term $/(m/\pi)$\\
\midrule
200 & 0.4 & -0.62983 & -0.62388 & -0.0149 & 0.944 \\
400 & 0.2 & -0.36123 & -0.36023 & -0.0050 & 0.982 \\
800 & 0.1 & -0.20300 & -0.20285 & -0.0015 & 0.995 \\
\bottomrule
\end{tabular}

\end{table}

\begin{table}[htbp]
\centering\small
\caption{Symmetric-product error for the $N=3$, $P=2$ compressed
instance ($\kappa=3$, $\lambda=1/2$, $t=2$). Operator norms and errors
are evaluated numerically. The extensive bound uses the rigorous
support-counting constants of Theorem~\ref{ef-thm:trotter}.}
\label{ef-tab:trotter}
\begin{tabular}{@{}rrrr@{}}
\toprule
steps $r$ & computed error & commutator estimate & extensive bound (Thm.) \\
\midrule
4 & 9.453e-02 & 4.746e-01 & 9.317e+01 \\
8 & 1.940e-02 & 1.186e-01 & 2.329e+01 \\
16 & 4.670e-03 & 2.966e-02 & 5.823e+00 \\
32 & 1.157e-03 & 7.415e-03 & 1.456e+00 \\
\bottomrule
\end{tabular}
\par\smallskip\footnotesize $\|[B,[B,A]]\|=17.792$, $\|[A,[A,B]]\|=2.493$ (numerical operator norms on the $5^3$-dimensional space).

\end{table}

\begin{figure}[htbp]
\centering
\includegraphics[width=\linewidth]{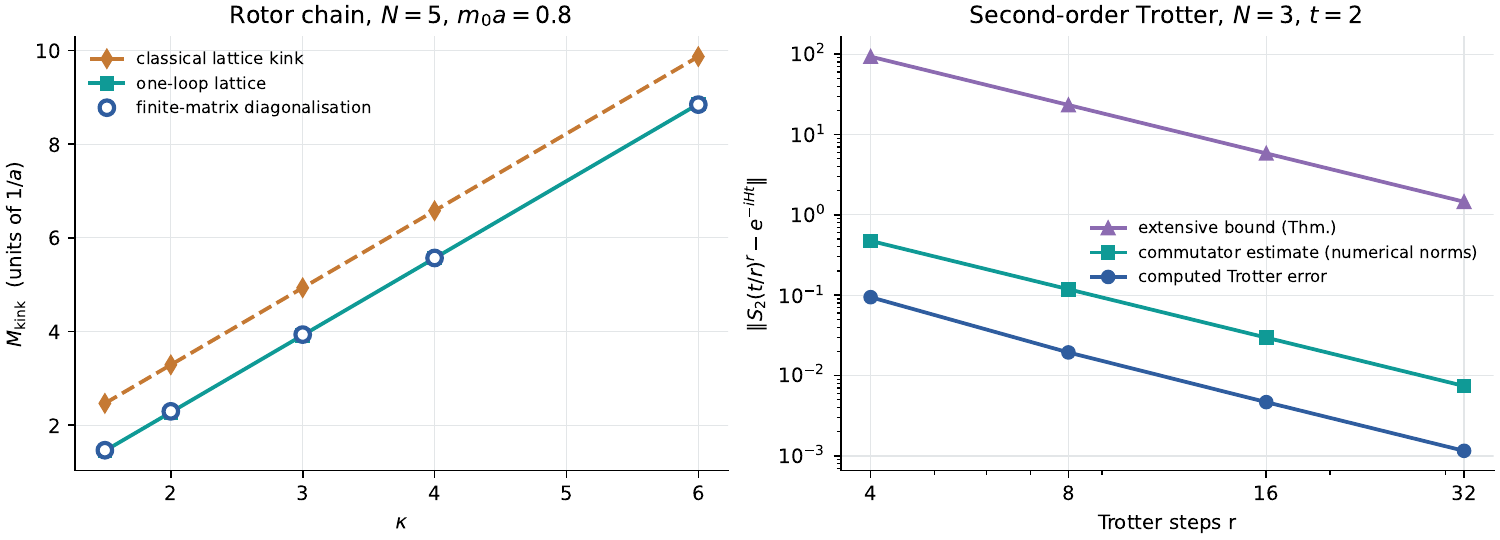}
\caption{Left: finite-matrix, one-loop, and classical kink-sector
energies of the compact rotor chain. Right: numerical symmetric-product
error, the estimate from computed nested-commutator norms, and the
analytic extensive bound. These are benchmarks of the stated scalar
regularisation.}
\label{ef-fig:quantum}
\end{figure}

\section{Microscopic matching and the limits of the effective descriptions}
\label{ef-sec:gap}
\subsection{The weak-coupling window is a real restriction}
For $SU(2)$ with $n_f$ adjoint Weyl flavours, write the one-loop coefficient as
\[
\mu\frac{\dd g}{\dd\mu}=-\frac{\beta_0}{16\pi^2}g^3+\cdots,
\qquad \beta_0=\frac{22-4n_f}{3}.
\]
Integrating gives $8\pi^2/g^2(1/L)=\beta_0\log(1/(\Lambda L))$ at this order. Hence
$e^{-S_0}=(\Lambda L)^{(11-2n_f)/3}$.
With $K$ proportional to $g^2/L$ and a leading bion coefficient of order $L^{-3}e^{-2S_0}$ times coupling powers, the power-law parts are
\begin{align}
\frac{m_0}{\Lambda}&\sim(\Lambda L)^{(8-2n_f)/3}
 \times\text{coupling/logarithmic factors},\\
\frac{T_F^{(0)}}{\Lambda^2}&\sim(\Lambda L)^{(5-2n_f)/3}
 \times\text{coupling/logarithmic factors}.
\end{align}
For $n_f=2$ these exponents are $4/3$ and $1/3$. They are asymptotic bookkeeping, not predictions at $L\Lambda\gtrsim1$. The missing prefactors and remainders have not been computed here. For an integer number of adjoint Weyl flavours, the one-loop asymptotic-freedom condition is $n_f\le5$. The small-circle analysis does not determine the phase of the corresponding uncompactified theory.

\subsection{Confining gauge dynamics need not gap all fermions}
The small-circle theory retains neutral fermions. At weak coupling, the monopole vertex for $n_f=2$ is a four-fermion operator; in three dimensions its free-field scaling dimension is $4>3$. The gauge-sector scalar mass calculation by itself therefore supplies no mass for every fermionic or gauge-invariant composite channel. More generally, a full gap requires all physical low-energy states to be controlled, not just one bosonic Hessian. The anomaly permits several infrared realisations. The specified small-circle potential has degenerate vacua, but neither that feature nor the anomaly supplies a gap in every channel.

\subsection{The decoupling and decompactification steps are unproved}
Increasing $L$ to obtain $\R^4$ leaves the small-circle expansion. Positive masses at each finite parameter do not imply a uniformly positive limit: $m(L)=e^{-L}$ is a counterexample to that inference.

Decoupling the adjoint fermions while keeping a small circle is also nontrivial, and the numerical one-loop calculation in Section~\ref{ef-sec:circle} illustrates the obstruction: the centre-symmetric point ceases to be the global minimum of \eqref{ef-eq:massive} at $mL\approx2.00$ and ceases to be a local minimum at $mL=2.280$ ($n_f=2$). At the heavy-fermion endpoint the pure-gauge sign of \eqref{ef-eq:gpy} is recovered,
$V_1(\pi/2)-V_1(0)=\pi^2/(12L^4)>0$.
Thus the centre-symmetric mechanism used here cannot simply be assumed to survive the fixed-small-$L$ heavy-fermion limit. A joint path in mass, circumference and possible deformations needs an actual phase and error analysis. Adiabatic continuity is a proposal to prove, not a substitute for such estimates; the quantum-simulation formulation of Section~\ref{ef-sec:quantum} is one way to test it numerically, not to prove it.

The fixed-coupling pure-gauge contraction certificates do not supply the weighted Fourier, fermionic or continuum bounds required here. Integrating fermions changes the effective measure, so the pure-gauge comparison constants require a new derivation.

\subsection{Uniform matching estimates needed for a microscopic application}
The next useful target for this route is to derive the effective action from a defined microscopic regularization and prove, for its actual coefficients,
\begin{equation}
\sup_{a,\,V_{\rm box}}
\frac{1}{\zeta}\sum_{n\ge2}n^2|c_n|\le\varepsilon_*<1,
\qquad
\sup_{a,\,V_{\rm box},\,s}|Z(s)/K-1|\le\eta_*<1,
\label{ef-eq:target}
\end{equation}
on a stated small-circle parameter region. Here $a$ is the ultraviolet regulator and $V_{\rm box}$ the physical spatial volume. One must also bound the omitted derivative terms, nonlocal interactions and additional fields. Equation \eqref{ef-eq:target} is a sufficient target, not a claim that the microscopic coefficients meet it. The local response in Theorem~\ref{ef-thm:concave} has logarithmic harmonic weights; a nonlinear replacement of the weighted $n^2$ hypothesis requires additional nonnegativity and fixed-vacuum assumptions.

A microscopic gap estimate must act on the full reconstructed physical vacuum complement. The dense-subspace spectral passage in Section~\ref{gap_transfer-exponential-decay-on-a-total-set-implies-a-hamiltonian-gap} applies once the corresponding physical-time estimates and the nontrivial continuum construction have been supplied.

Uniform cutoff estimates must survive reconstruction, infinite volume, the change of geometry and removal of added matter, while producing the intended nontrivial theory. Exponential decay for a finite list of observables does not establish the dense-subspace premise. An exact wall tension alone does not supply it either.

\section{Corrected variational bounds for periodic effective potentials}\label{sec:confinement_variational}

\subsection{Weak harmonic control and vacuum stability}

The weighted harmonic stability theorem applies under
\[
U(s)=\zeta(1-\cos 2s)+\sum_{n\ge2}c_n(1-\cos 2ns),\qquad
\zeta^{-1}\sum_{n\ge2}n^2|c_n|<1.
\]
Replacing this condition by the logarithmic norm alone requires an additional hypothesis preserving the ground states.

\begin{proposition}[Arbitrarily small logarithmic norm can destroy the old vacuum]\label{conf:weak_norm_counterexample}
Fix $\zeta>0$. For an integer $M\ge2$, set $c_M=-2\zeta/M^2$ and all other coefficients to zero. Then
\[
\sum_{n\ge2}|c_n|\log n=\frac{2\zeta\log M}{M^2}\longrightarrow0,
\qquad U''(0)=-4\zeta<0.
\]
For the unrecentered constant-kinetic energy with endpoints $0$ and $\pi$, the infimum is $-\infty$.
\end{proposition}
\begin{proof}
Twice differentiating the finite cosine sum gives $U''(0)=4\zeta+4M^2c_M=-4\zeta$. There is also an explicit negative point. At $s_M=1/(2M)$,
\[
\frac{U(s_M)}\zeta=(1-\cos(1/M))-\frac2{M^2}(1-\cos1)
\le \frac1{2M^2}-\frac{11}{12M^2}=-\frac5{12M^2}.
\]
Here $1-\cos x\le x^2/2$, and the alternating Taylor bound gives $1-\cos1\ge1/2-1/24=11/24$. Connect $0$ to $s_M$ over a fixed interval, remain at $s_M$ for a length $D$, then connect to $\pi$. The two transitions have a fixed finite energy, while the plateau contributes $DU(s_M)\to-\infty$.
\end{proof}
This does not contradict the weighted theorem: its relative $n^2$ norm is $2$, outside that theorem's admissible region. It refutes using the logarithmic norm as an independent replacement.

\textbf{A valid logarithmic bound with the missing hypothesis.} If $U\ge0$, $U(0)=U(\pi)=0$, the kinetic coefficient is constant $K>0$, and $S_h=\sum_{n\ge2}|c_n|h_n<\infty$, where
\[
h_n=\sum_{j=0}^{n-1}\frac1{2j+1},
\]
then the variational period cost $\tau(U)$ satisfies
\[
\left|\tau(U)-4\sqrt{K\zeta}\right|\le4\sqrt{K/\zeta}\,S_h.
\]
Indeed, $|\sqrt U-\sqrt{U_0}|\le|U-U_0|/\sqrt{U_0}$ in the interior of the period, with $U_0=2\zeta\sin^2s$, and
\[
\int_0^\pi\frac{\sin^2(ns)}{\sin s}\,\dd s=2h_n.
\]
Integrating and summing absolute values proves the bound. This preserves a legitimate weak-norm statement without silently assuming nonnegativity follows from that norm.

\subsection{Ground-state recentering and the exact winding infimum}

We can avoid controlling curvature if the goal is only a positive \emph{classical winding energy}. Subtract the actual ground-state energy instead of keeping an unstable reference vacuum.

\begin{theorem}[Exact winding infimum]\label{conf:winding}
Let $U,Z:\R\to\R$ be continuous and $\pi$-periodic, with $Z\ge K_->0$. Set
\[
m=\min_{s\in[0,\pi]}U(s),\qquad Q(s)=U(s)-m\ge0,\qquad
\tau=\int_0^\pi\sqrt{2Z(s)Q(s)}\,\dd s.
\]
Choose any minimizer $a$ of $U$. For every integer $k$,
\[
\inf_{\substack{\sigma(-\infty)=a\\\sigma(+\infty)=a+k\pi}}
\int_\R\left[\frac{Z(\sigma)}2|\sigma'|^2+Q(\sigma)\right]\dd x
=|k|\tau,
\]
where profiles are locally $H^1$ with the stated limits and finite energy. No curvature or isolated-vacuum assumption is needed.
\end{theorem}
\begin{proof}
Pointwise square completion gives
\[
\frac Z2|\sigma'|^2+Q\ge\sqrt{2ZQ}\,|\sigma'|.
\]
Let $G'(s)=\sqrt{2Z(s)Q(s)}$. The integrated right side is at least $|G(a+k\pi)-G(a)|=|k|\tau$.

For the upper bound with $k=1$, traverse $[a,a+\pi]$ according to
\[
\sigma'=\sqrt{2(Q(\sigma)+\epsilon)/Z(\sigma)},\qquad\epsilon>0.
\]
Continuity and periodic boundedness imply finite traversal time. Extend the profile constantly at both endpoints; the resulting derivative jumps are permitted in $H^1$. Its energy is
\[
\int_a^{a+\pi}\sqrt{Z(s)/2}\,
\frac{2Q(s)+\epsilon}{\sqrt{Q(s)+\epsilon}}\,\dd s
\longrightarrow\tau.
\]
For $0<\epsilon\le1$, the integrand is dominated by $\sqrt{2Z(Q+1)}$, so dominated convergence applies. Concatenating $|k|$ traversals and reversing orientation when necessary proves the result.
\end{proof}

\begin{proposition}[Unweighted harmonic control is enough after recentering]\label{conf:harmonic_recentered}
For $U(s)=2\zeta\sin^2s+\sum_{n\ge2}c_n(1-\cos2ns)$, suppose
$S=\sum_{n\ge2}|c_n|<\zeta$. Then
\[
\boxed{\displaystyle
\tau\ge\sqrt{2K_-(\zeta-S)}\,\arccos(S/\zeta)>0.}
\]
\end{proposition}
\begin{proof}
The series converges uniformly; $U(0)=0$ gives $m\le0$. Thus
$Q(s)\ge\pos{2\zeta\sin^2s-2S}$. On the interval where
$\sin^2s\ge(1+S/\zeta)/2$, $Q\ge\zeta-S$. This interval has length $\arccos(S/\zeta)$. Integrate the lower bound for $\sqrt{2ZQ}$ there.
\end{proof}
Extra minima may prevent attainment or split one period into several elementary walls. The statement is about the \emph{infimum for one winding}, not a unique kink or a quantum mass gap.

\subsection{The asymptotic classical probe energy}

Use the following strip and fixed probe cores: let $\mathcal S_R=(-R,R)\times\R$, put probes at $(\pm d/2,0)$, remove radius-$\rho$ discs with $0<\rho<1/4$, and cut along the segment joining them. Assume $d>4$ and $R\ge d$. A real lift of the compact field has jump $2\pi$ across the cut and no other jump. Its limits as $y\to\pm\infty$ are $a$ and $a+2\pi k$, uniformly in $x$, for one common integer $k$.

\begin{theorem}[Bulk probe confinement in the recentered scalar model]\label{conf:bulk}
Under Theorem~\ref{conf:winding}'s hypotheses, let $E_{\mathrm{bulk}}(d;R)$ be the infimum of
\[
\int_{\mathcal S_R\setminus\{\text{cut and cores}\}}
\left[\frac{Z(\sigma)}2|\nabla\sigma|^2+Q(\sigma)\right]\dd x\dd y
\]
over these fields and all integers $k$. Then
\[
E_{\mathrm{bulk}}(d;R)\ge2\tau(d-2\rho),\qquad
\lim_{d\to\infty}\frac{E_{\mathrm{bulk}}(d;R)}d=2\tau
\]
uniformly for $R\ge d$. In particular $S<\zeta$ in Proposition~\ref{conf:harmonic_recentered} gives a strictly positive classical asymptotic slope.
\end{theorem}
\begin{proof}
On each vertical line avoiding the cores, remove the $2\pi$ jump to obtain a continuous lift. Outside the cut its endpoints differ by $2\pi k$; inside they differ by $2\pi(k-1)$. Dropping the nonnegative horizontal gradient energy and applying Theorem~\ref{conf:winding} gives
\[
E\ge2\tau\{ |k|(2R-d-2\rho)+|k-1|(d-2\rho)\}.
\]
For $R\ge d$, minimizing over $k\in\Z$ yields the stated lower bound.

For the upper bound, choose a decreasing finite-interval profile $q(y)$ from $a$ to $a-2\pi$ with energy at most $2\tau+\epsilon$, as in Theorem~\ref{conf:winding}, and extend it constantly. Place its traversal across $y=0$. Then
\[
f(y)=q(y)-a+2\pi\mathbf1_{y>0}
\]
has compact support and jump $2\pi$. Let $\lambda$ be a smooth step from $0$ to $1$ on $[-1,1]$, and $\chi(r)$ equal $1$ for $r\le1$ and $0$ for $r\ge2$. For $r=\sqrt{x^2+y^2}>0$, set
\[
J(x,y)=\chi(r)\frac{1+x/r}{2}+(1-\chi(r))\lambda(x).
\]
It obeys $J(x,0)=\mathbf1_{x>0}$ and $|\nabla J|\le C/r+C$. Use
\[
\sigma(x,y)=a+\bigl[J(x+d/2,y)-J(x-d/2,y)\bigr]f(y).
\]
The jump is exactly $2\pi$ between the cores and zero elsewhere. Away from the two fixed-width transition regions the field is either the constant $a$ or the chosen double-period profile. The transition regions have finite vertical extent. There, $Z,Q,f$ are bounded, $f'$ is square integrable, and the $1/r$ gradient has finite squared integral after removing the fixed cores. Consequently
\[
E_{\mathrm{bulk}}(d;R)\le(2\tau+\epsilon)d+C_{\epsilon,\rho},
\]
where $C_{\epsilon,\rho}$ is independent of $d,R$. Divide by $d$, take the uniform upper limit, and then let $\epsilon\downarrow0$.
\end{proof}

\section{Heavy-matter estimates and the electric comparison}\label{sec:confinement_heavy_matter}

\subsection{Boundary-uniform mixing at strong coupling}

\label{conf:mixing_section}

The next estimates concern regulated quantum gauge measures. The first applies to the full pure Wilson measure with arbitrary fixed boundary links.

On a finite four-dimensional hypercubic box, take link variables in $SU(N)$, arbitrary fixed boundary links, and
\[
\dd\mu_\beta=\mathcal Z^{-1}
\exp\left\{\beta\sum_p\frac1N\operatorname{Re}\operatorname{Tr}U_p\right\}
\prod_e\dd U_e,\qquad \beta\ge0.
\]
Let $\mathcal G$ be the graph whose vertices are links, adjacent when they share a plaquette. Let $c_{ef}$ be the maximal total-variation change of the conditional law at $e$ when only link $f$ is changed.

\begin{theorem}[Explicit boundary-uniform lattice mixing]\label{conf:mixing}
Put $q=18\beta$. If $0<q<1$, then for bounded real observables $F,G$ supported on link sets $A,B$,
\[
|\Cov_{\mu_\beta}(F,G)|
\le\frac{2\|F\|_\infty\|G\|_\infty |A|}{1-q}
q^{d_{\mathcal G}(A,B)}.
\]
The constants are independent of volume and boundary links. At $\beta=0$, disjoint link sets are independent.
\end{theorem}
\begin{proof}
If two single-link conditional densities differ by a log tilt $h$ with oscillation at most $D$, their total variation is at most $\tanh(D/4)$. One way to see this is to scale $e^h$ to $[1,e^D]$, bound the absolute deviation from its mean by the endpoint chord, and maximize over that mean; the maximum is $(e^{D/2}-1)/(e^{D/2}+1)$.

For each plaquette shared by $e,f$, changing $f$ changes the conditional log weight at $e$ by a function of oscillation at most $4\beta$. Hence
\[
c_{ef}\le\tanh(\beta n_{ef})\le\beta n_{ef},\qquad
\sum_f c_{ef}\le\beta\sum_f n_{ef}\le18\beta=q.
\]
Here $n_{ef}$ counts shared plaquettes: a link belongs to at most six plaquettes, each containing three other links.

The Dobrushin comparison argument \cite{RVH} bounds the change of the expectation of $F$ under two assignments on $B$ by
$2\|F\|_\infty\sum_{e\in A,f\in B}D_{ef}$, where
$D=\sum_{n\ge0}C^n$ and $C=(c_{ef})$. This follows by coupling single-link updates: disagreement probabilities satisfy the comparison inequality $p\le Cp+b$, solved by the convergent series $D$. Conditioning on $B$ and multiplying by $G$ gives the same bound for the covariance with an extra $\|G\|_\infty$. Since $(C^n)_{ef}=0$ for $n<d_{\mathcal G}(A,B)$ and every row sum of $C^n$ is at most $q^n$, the displayed estimate follows.
\end{proof}

For the fixed-size local supports in Chatterjee's hypothesis, graph distance controls spatial distance and this supplies boundary-uniform exponential decay. His theorem therefore supplies a Wilson area law in this strong-coupling region \cite{conf:chatterjee}. This is an application of established machinery, not a new all-coupling confinement theorem.

\note{\textbf{Continuum obstruction.} The asymptotically free Wilson trajectory has $\beta(a)=2N/g_0(a)^2\to\infty$. It cannot remain inside $\beta<1/18$. The estimate is volume-uniform, but it does not control the continuum trajectory. Blocking weak-coupling data into this region with controlled defect errors is still unproved here.}

\subsection{An absolutely convergent heavy-fermion action}

This calculation does not assume confinement or small gauge coupling. It applies uniformly to every link configuration, including adjoint links. Work on an even four-dimensional torus with each extent at least four, or an open hypercubic box. Use Hermitian Euclidean gamma matrices and unitary gauge-covariant shifts $T_\mu$ (partial shifts for an open box). Define
\[
D_\kappa=I-\kappa H,\qquad
H=\sum_{\mu=1}^4\bigl[(1-\gamma_\mu)T_\mu+
(1+\gamma_\mu)T_\mu^\dagger\bigr].
\]
Let $d_{\mathrm{rep}}$ be the representation dimension, $r=4d_{\mathrm{rep}}$, and $n_D\ge1$ the integer number of degenerate Dirac flavours. The scalar normalization factored out of $D_\kappa$ cancels in normalized gauge expectations. For adjoint $SU(2)$, $r=12$; one Dirac flavour has the field content of two Weyl flavours.

\begin{theorem}[Absolutely convergent local fermion action]\label{conf:fermion_action}
For $0\le\kappa<1/8$, $\det D_\kappa>0$ and
\[
S_f(U):=-n_D\log\det D_\kappa
=n_D\sum_{n\ge4}\frac{\kappa^n}{n}\operatorname{Re}\operatorname{Tr}H^n.
\]
Expand each trace into rooted oriented closed walks $\gamma$. Let $\Phi_\gamma$ be its real contribution, let $|\gamma|$ be its length, and include a vertex in its support whenever the walk visits it. For $\alpha\ge0$ and $t=16\kappa e^\alpha<1$,
\[
\boxed{\|\Phi\|_\alpha:=
\sup_x\sum_{\gamma\ni x}e^{\alpha|\gamma|}\|\Phi_\gamma\|_\infty
\le n_Dr\,\frac{t^4}{1-t^2}.}
\]
The bound is independent of the volume, gauge coupling, and gauge configuration.
\end{theorem}
\begin{proof}
Writing $(1\pm\gamma_\mu)=2P_{\mu,\pm}$, each directional pair is block diagonal in the orthogonal spin projections and has norm at most $2$. Hence $\|H\|\le8$, so the matrix logarithm series converges for $8\kappa<1$. Gamma-five hermiticity makes the determinant real; invertibility and continuity from its value $1$ at $\kappa=0$ make it positive.

There are no length-one closed walks. Immediate reversal has spin factor $(1-\gamma_\mu)(1+\gamma_\mu)=0$, removing length two. Bipartiteness removes odd lengths. Thus the first possible term has length four. The series is the usual hopping expansion \cite{conf:hopping}; the following absolute estimate needs no numerical truncation.

For length $n$, a path's spin-gauge product has norm at most $2^n$, so its trace has absolute value at most $r2^n$. At most $8^n$ directed paths start at a fixed vertex. A rooted closed walk containing a specified $x$ can be rerooted at $x$ with at most $n$ choices for its original root, giving the upper count $n8^n$. This cancels the $1/n$ in the logarithm. Summing the resulting bound $n_Dr t^n$ over even $n\ge4$ proves the claim.
\end{proof}
Constant loop contributions may be retained in this norm; they cancel from normalized observables. Length-four winding walks can occur when a circumference is four sites, so the leading term is not identified solely with plaquettes in that case.

\subsection{Matter and winding tails}

The preceding local estimate permits termwise differentiation inside a strict convergence region. If $\kappa>0$ and $t=16\kappa e^\alpha<1$, then
\[
\boxed{\|\partial_\kappa\Phi\|_\alpha
\le\frac{n_Dr}{\kappa}\,
\frac{4t^4-2t^6}{(1-t^2)^2}.}
\]
Indeed, differentiating a length-$n$ term introduces $n/\kappa$, and
$\sum_{n\ge4,\ n\text{ even}}nt^n=(4t^4-2t^6)/(1-t^2)^2$.
Thus the total local-action change as $\kappa\downarrow0$ is integrable and tends to zero, uniformly in volume and in all gauge configurations.

\begin{proposition}[Fermionic winding contributions]\label{conf:winding_tail}
If a periodic direction has even extent $N_L=L/a\ge4$, the part of the fermion action made of walks winding nontrivially around that direction obeys
\[
\boxed{\|\Phi_{\mathrm{wind},L}\|_\alpha
\le n_Dr\,\frac{t^{N_L}}{1-t^2}.}
\]
\end{proposition}
\begin{proof}
A walk with nonzero winding has at least $N_L$ nearest-neighbour steps. It is a subset of the walks counted in Theorem~\ref{conf:fermion_action}. Sum the same absolute bound over even $n\ge N_L$.
\end{proof}
For $t\le t_*<1$, this yields exponential suppression of the determinant's winding terms as $L/a\to\infty$. This is an actual decompactification estimate for this part of the integrated fermion action. It is not an estimate for the full gauge measure's response to changing the circumference.

\section*{Where the uniformity stops}
With the usual bare Wilson mass convention,
\[
\kappa=\frac1{2am_0+8}.
\]
At $\alpha=0$, our conservative absolute local bound requires $am_0>4$. It covers a mass above the ultraviolet cutoff. A fixed \emph{renormalized} mass in the continuum instead requires critical-mass tuning; the usual asymptotically free Wilson trajectory approaches $\kappa=1/8$, not a compact subinterval of $\kappa<1/16$.

Consequently, this proves a uniform heavy-cutoff tail, not uniform physical-mass decoupling after the continuum limit. The numerical threshold is sufficient and not asserted to be optimal. Improving it somewhat would not by itself establish control up to the critical continuum trajectory.

\textbf{Why a small local action is not the missing loop estimate.} Differentiating a normalized Wilson expectation gives the ratio
\[
\Gamma(C)=\sum_\gamma
\frac{\Cov(W(C),\partial_s\Phi_\gamma)}{\langle W(C)\rangle}.
\]
A bound on $\|\partial_s\Phi\|_\alpha$ controls the local numerator inputs. It does not cancel the exponentially small Wilson denominator. That cancellation is precisely the additional defect estimate still needed.

\subsection{Volume-uniform decoupling of local expectations}

The local fermion-action estimate can be combined with the pure Wilson contraction. This gives more than a bound on the action: it controls quantum Euclidean expectations of bounded local observables, uniformly in the total volume.

\begin{theorem}[Heavy-matter decoupling for local observables]\label{conf:local_decoupling}
Let $\mu_0$ be the pure Wilson measure at $18\beta<1$, and let
\[
\dd\mu_\kappa=\frac{e^{-S_f}}{\mu_0(e^{-S_f})}\dd\mu_0
\]
use the positive Wilson determinant from Theorem~\ref{conf:fermion_action}, with $t=16\kappa<1$. Put
\[
L_\kappa=n_Dr\frac{t^4}{1-t^2}.
\]
For any bounded real observable $F$ supported on a finite set $A$ of links,
\[
\boxed{
|\mu_\kappa(F)-\mu_0(F)|
\le\frac{\|F\|_\infty|A|}{1-18\beta}\,L_\kappa.}
\]
The bound is uniform in the box, boundary links, and allowed periodic circumferences. No independent contraction hypothesis for $\mu_\kappa$ is needed.
\end{theorem}
\begin{proof}
At a fixed exterior configuration, the two conditional laws for link $e$ differ by the log tilt
\[
h_e=-\sum_{\gamma\ni e}\Phi_\gamma.
\]
The sum includes all rooted-loop occurrences, not only loops rooted at $e$. Every walk containing $e$ visits a fixed endpoint of $e$, so Theorem~\ref{conf:fermion_action} gives
\[
\operatorname{osc}_e h_e\le2L_\kappa,
\qquad b_e\le\tanh(L_\kappa/2)\le L_\kappa/2.
\]
Apply the comparison inequality with the \emph{pure} measure as the contracting reference. Its matrix $D=(I-C)^{-1}$ has row sums at most $(1-18\beta)^{-1}$. Since the coordinate oscillation of $F$ is at most $2\|F\|_\infty$ on $A$ and zero elsewhere,
\[
|\mu_\kappa(F)-\mu_0(F)|
\le\sum_{e\in A,f}2\|F\|_\infty D_{ef}b_f
\le\frac{\|F\|_\infty|A|L_\kappa}{1-18\beta}.
\]
This establishes the stated estimate without any comparison of extensive partition functions.
\end{proof}

\textbf{What is now established.} For a fixed finite edge support, the heavy-fermion limit of these local expectations is uniform in total volume and circumference, within the stated strong-coupling and heavy-cutoff regimes. In particular, no growing total-volume factor occurs in this comparison.

\textbf{What is not established.} A loop of fixed physical perimeter generally uses $|A|\sim P/a$ links. Moreover, this is an absolute expectation error. Dividing it by an exponentially small Wilson expectation does not give a bounded relative error. The estimate therefore supplies neither the all-area quantum transport premise nor control along $\beta(a)\to\infty$.

\subsection{A restricted joint decoupling path}

There is also a direct finite-volume comparison that holds at \emph{every} gauge coupling. It gives a genuine simultaneous cutoff/mass result on a very fast mass trajectory, but not on all trajectories.

Let $N_s$ be the number of lattice vertices, $d_{\mathrm{mat}}=rN_s$, and $u=8\kappa<1$. The operator-norm expansion from Theorem~\ref{conf:fermion_action} gives
\[
|S_f(U)|\le n_Dd_{\mathrm{mat}}\sum_{n\ge4}\frac{u^n}{n}
\le\delta:=\frac{n_DrN_s u^4}{4(1-u)}.
\]
Let $\mu_\kappa$ be the normalized gauge measure with this determinant and $\mu_0$ the same pure gauge measure. Since $e^{-\delta}\le e^{-S_f}\le e^\delta$,
\[
e^{-2\delta}\le\frac{\dd\mu_\kappa}{\dd\mu_0}\le e^{2\delta},
\qquad
|\langle O\rangle_\kappa-\langle O\rangle_0|
\le\|O\|_\infty(e^{2\delta}-1).
\]
This proof uses only normalization of the two measures; it does not assume that the pure gauge continuum limit exists.

\begin{proposition}[A fast-decoupling joint path]\label{conf:fast_path}
Along any family of lattices and gauge couplings for which
\[
u\le u_*<1,\qquad N_su^4\longrightarrow0,
\]
expectations under the two measures differ by a quantity tending to zero, uniformly over all observables with $\|O\|_\infty\le1$. In particular, if a pure-gauge sequence of such bounded observables has a limit, the determinant-modified sequence has the same limit.
\end{proposition}
\begin{proof}
The hypotheses give $\delta\to0$ in the preceding explicit bound.
\end{proof}

For fixed physical four-volume $V_4$, $N_s$ is of order $V_4/a^4$. When $am_0\to\infty$, $u$ is of order $1/(am_0)$. A sufficient condition is therefore
\[
\frac{V_4}{a^8m_0^4}\longrightarrow0.
\]
In fixed reference units, $m_0=a^{-p}$ with $p>2$ is one example. If the physical volume or circumference grows simultaneously, its contribution to $N_s$ must also be included in $N_su^4\to0$.

\note{\textbf{Precise scope.} This removes fermions much faster than the cutoff scale on a selected path. It proves equality of limits for bounded regulated observables if one limit already exists. It neither constructs the gauge continuum theory nor proves order-independent physical decoupling. In particular, perimeter-renormalized Wilson loops need not remain uniformly bounded as regulated observables, and absolute convergence does not preserve their large-area exponent.}

\subsection{The exactly solvable electric Hamiltonian}

An independent route starts from the electric part of the $SU(2)$ lattice Hamiltonian. Unlike the earlier static scalar functional, this is a quantum operator on a gauge-invariant Hilbert space.

\begin{theorem}[Exact electric-limit string tension and gap]\label{conf:electric}
On a finite open cubic spatial lattice with spacing $a$, impose Gauss law at every vertex, including its boundary, and let
\[
H_E=\frac{g^2}{2a}\sum_e E_e^2.
\]
In the sector with two fundamental static sources separated by graph distance $n$, the ground energy relative to the vacuum is $3g^2n/(8a)$. Thus for collinear separation $R=na$,
\[
V_E(R)=\sigma_E R,\qquad \sigma_E=\frac{3g^2}{8a^2}.
\]
If the box contains a plaquette, the source-free physical gap is
\[
\Delta_E=\frac{3g^2}{2a}=4a\sigma_E.
\]
Both statements are independent of the box size once it contains the required path and plaquette.
\end{theorem}
\begin{proof}
Use the spin-network basis, on which $E_e^2$ has eigenvalue $j_e(j_e+1)$. The central element $-I$ in each Gauss constraint implies that the subgraph of half-integer-spin edges has odd degree precisely at the source vertices. Such a subgraph contains a path connecting those vertices. At least $n$ edges therefore have Casimir at least $3/4$. A shortest path carrying spin $1/2$, with all other edges trivial and the invariant source and internal intertwiners, attains that energy. The vacuum with all spins zero has energy zero.

Without sources, a nonempty subgraph of nonzero-spin edges has no degree-one vertex: a single nontrivial representation cannot contain a singlet. A finite graph of minimum nonzero degree at least two has a cycle. On the cubic lattice every cycle has at least four edges, each with Casimir at least $3/4$. A spin-$1/2$ plaquette loop attains the lower bound $4(3/4)g^2/(2a)$.
\end{proof}

The exact identity $\Delta_E=4a\sigma_E$ is a sharp obstruction: finite positive string tension as $a\to0$ forces the electric-model gap to zero. A positive lower bound on the gap instead forces its string tension to diverge. This solved limit cannot supply both finite continuum string tension and a positive gap.

\textbf{Adding a positive magnetic term does not repair the proof.} Ground energies in both the vacuum and source sectors shift. For $M>s>0$, take
\[
H_v=\operatorname{diag}(0,M),\quad H_q=\operatorname{diag}(s,M),
\quad B=\operatorname{diag}(M,0)\ge0.
\]
The ground-energy difference starts at $s$. Adding the same positive $B$ makes both ground energies equal $M$, so the difference becomes zero. Thus $H_E+H_M\ge H_E$ does not imply a lower bound on the vacuum-subtracted static potential. This matrix example refutes that inference; it is not a proposed Yang--Mills model.

\section{Transport of quantum Wilson-loop bounds}\label{sec:confinement_transport}

\subsection{An exact normalized insertion identity}

A minimum classical energy is not a quantum free energy. To obtain a Wilson area law one needs the normalized quantum expectation, including fluctuations and the vacuum subtraction. The following identity makes the missing quantity explicit.

Work first at a finite regulator and volume. On a common configuration space, let
\[
\dd\mu_s=\mathcal Z_s^{-1}e^{-S_s}\dd\nu,\qquad
w_s(C)=\langle W(C)\rangle_s>0,\qquad F_s(C)=-\log w_s(C).
\]
Assume differentiation under the integral is justified, the loop observable is $s$-independent, and $s\mapsto F_s(C)$ is locally absolutely continuous. Direct differentiation gives, almost everywhere,
\[
\partial_s w_s(C)=-\Cov_s(W(C),\partial_sS_s),
\qquad
\boxed{\partial_sF_s(C)=\Gamma_s(C):=
\frac{\Cov_s(W(C),\partial_sS_s)}{\langle W(C)\rangle_s}.}
\]
The covariance is the bilinear connected expectation. If $S_s=\sum_x\mathcal L_{s,x}$, then $\Gamma_s$ is the sum of loop-normalized connected action-density insertions. Writing it as a Wilson-tilted expectation is algebraically permissible, but that tilt is generally \emph{not a probability measure}: the Wilson observable is not pointwise positive.

\begin{theorem}[Conditional transport of an area bound]\label{conf:area_transport}
Suppose, in fixed physical units,
\[
F_0(C)\ge\sigma_0 A(C)-b_0P(C)-c_0,\qquad\sigma_0>0,
\]
and, for $0\le s\le s_*$,
\[
\Gamma_s(C)\ge-\epsilon(s)A(C)-b(s)P(C)-c(s).
\]
Assume the coefficient functions are integrable and the estimates are uniform in the relevant loops, regulator and volume, with
\[
\sigma_*:=\sigma_0-\int_0^{s_*}\epsilon(s)\dd s>0.
\]
Then the endpoint satisfies
\[
F_{s_*}(C)\ge\sigma_*A(C)
-\left(b_0+\int_0^{s_*}b(s)\dd s\right)P(C)
-c_0-\int_0^{s_*}c(s)\dd s.
\]
\end{theorem}
\begin{proof}
Integrate $\partial_sF_s=\Gamma_s$ and add the initial inequality. Passing to any endpoint or regulator limit requires convergence of the quantities in the inequality and finite limiting coefficients; it is not furnished by the finite-system identity itself.
\end{proof}

\note{\textbf{What remains unproved.} The effective-model analysis establishes neither the starting \emph{quantum} bound nor the loop-normalized covariance estimate. Only the downward change in $F$ needs control. A bound on its absolute derivative would be sufficient but stronger than necessary. For an unbounded decoupling/decompactification path, the loss must be integrable over the entire path.}

Changing the geometry or integrating out matter requires a valid common-variable formulation and matched loop observables. If the observable itself depends on $s$, its derivative contributes an additional term. These matters cannot be omitted by merely naming a path.

\subsection{Mixed differences and area growth}

\label{conf:creutz_section}

For a fixed physical increment $h>0$, define the rectangular mixed difference
\[
\begin{split}
\chi_{s,h}(R,T)={}&F_s(R+h,T+h)+F_s(R,T)\\
&-F_s(R+h,T)-F_s(R,T+h).
\end{split}
\]
Equivalently, it is the negative logarithm of the corresponding four-loop ratio. If $F=\sigma RT+\alpha R+\beta T+c$, then $\chi_h=\sigma h^2$. Thus additive perimeter and equal rectangular-corner counterterms cancel exactly. This does not itself prove that the continuum loop ratios exist.

The interpolation identity implies
\[
\partial_s\chi_{s,h}=\Delta_R^h\Delta_T^h\Gamma_s.
\]
An explicit sufficient microscopic target is therefore
\[
\boxed{\Delta_R^h\Delta_T^h\Gamma_s(R,T)\ge-\epsilon(s)h^2,
\qquad\int_0^{s_*}\epsilon(s)\dd s<\sigma_0,}
\]
together with the initial \emph{quantum} inequality $\chi_{0,h}\ge\sigma_0h^2$. Integration gives $\chi_{s_*,h}\ge\sigma_*h^2>0$.

\begin{proposition}[From the mixed-difference bound to area growth]\label{conf:creutz}
Suppose $\chi_h(R,T)\ge\sigma_*h^2$ on the grid $R=R_0+ih$, $T=T_0+jh$, with $i,j\ge0$. Then
\[
\begin{split}
F(R,T)\ge{}&\sigma_*(R-R_0)(T-T_0)\\
&+F(R,T_0)+F(R_0,T)-F(R_0,T_0).
\end{split}
\]
If the last line has a lower bound $-B(R+T)-D$, this is an area lower bound with only linear and constant corrections.
\end{proposition}
\begin{proof}
For $R=R_0+nh$, $T=T_0+mh$, double telescoping gives the exact identity
\[
\begin{split}
F(R,T)={}&F(R,T_0)+F(R_0,T)-F(R_0,T_0)\\
&+\sum_{i=0}^{n-1}\sum_{j=0}^{m-1}
\chi_h(R_0+ih,T_0+jh).
\end{split}
\]
The sum has $nm$ terms; use $nmh^2=(R-R_0)(T-T_0)$.
\end{proof}
For positive bare normalized loop expectations at a fixed regulator, $w\le1$ gives $F\ge0$, making the boundary term harmless there. After continuum renormalization, the stated boundary lower bound and convergence still need proof. A grid theorem is enough to establish the large-grid area coefficient; extending it to every rectangle needs regularity or an additional comparison.

\textbf{An alternative estimate.} For locally absolutely continuous $\chi_{s,h}$, if $\partial_s\chi_{s,h}\ge-g(s)\chi_{s,h}$ almost everywhere and $\int g<\infty$, multiplication by $\exp(\int_0^s g)$ yields
$\chi_{s_*,h}\ge\chi_{0,h}\exp(-\int g)>0$. This may be a useful formulation for an RG argument, but no such Yang--Mills estimate is established here.

\subsection{Obstructions to absolute-error transport}

\label{conf:transport_obstructions}

\textbf{Absolute accuracy does not preserve string tension.} Fix $0<\delta<1$ and consider the positive functions
\[
w_0(A)=e^{-\sigma A},\qquad
w_\delta(A)=(1-\delta)e^{-\sigma A}+\delta.
\]
For every $A\ge0$, $|w_\delta(A)-w_0(A)|\le\delta$. Nevertheless,
\[
\lim_{A\to\infty}-\frac1A\log w_\delta(A)=0.
\]
This is a counterexample to an inference from absolute errors, not a claim that these functions construct a Yang--Mills measure. Uniformly small errors in loop values can dominate an exponentially small signal. The useful control is relative, logarithmic, or through mixed differences.

\textbf{A positive limit cannot be inferred from positivity at finite size.} The smooth function $\sigma(L)=\sigma_0e^{-L/L_0}$ is positive for every finite $L$ and has zero limit. A proposal asserting no finite-$L$ phase transition still needs a quantitative lower bound at $L=\infty$.

\textbf{A bulk action estimate is not automatically a defect estimate.} Bounding the difference of two actions by a small constant times the total volume does not control a loop free energy after the infinite-volume limit. One must retain cancellation between the vacuum normalization and the inserted expectation. The quantity $\Gamma_s$ defined above does exactly that algebraically; bounding it is the hard step.

\textbf{Classical tension does not control quantum entropy.} Theorem~\ref{conf:winding}, Proposition~\ref{conf:harmonic_recentered}, and Theorem~\ref{conf:bulk} concern a classical energy functional. A path integral also sums over fluctuations and interfaces. A lower bound on a minimizing classical configuration is not a lower bound on the logarithm of the normalized quantum integral. Treating these quantities as identical would assume the required quantum estimate.

\textbf{The small-circle expansion cannot be extended by substitution.} In a centre-symmetric $SU(N)$ background,
\[
m_W=\frac{2\pi}{NL},\qquad NL\Lambda\ll1
\]
is the semiclassical scale window. At fixed $N$ and physical $\Lambda$, decompactification exits it. Increasing $N$ also worsens this particular scale separation. Centre-stabilized large-$N$ volume independence does not, by itself, preserve the abelian calculation across this loss of control \cite{conf:uy}.

\textbf{Matter content matters.} Removing adjoint fermions changes the holonomy potential and the quantum measure. The effective-model analysis itself shows that its one-loop centre-symmetric minimum is lost in the fixed-small-circle heavy-fermion limit. Fundamental dynamical matter introduces possible string breaking, so an unscreened fundamental-probe area law is not the universal definition of confinement for every theory with fermions.

These obstructions explain why solving a more accurate scalar wall, or simulating that same scalar theory on more sites, does not close the stated four-dimensional target.

\subsection{Boundary-uniform covariance as an external confinement criterion}

Chatterjee's lattice result \cite{conf:chatterjee}, Definition 2.3 and Theorem 2.4, gives a useful alternative: exponential decay of bounded local covariances, uniformly over the specified cubes and \emph{all} boundary conditions, implies rectangular Wilson area laws for irreducible representations nontrivial on the centre. Schematically, its premise has the form
\[
|\Cov_{B,\delta}(f,g)|\le K_1e^{-K_2\operatorname{dist}(f,g)},
\]
with $|f|,|g|\le1$, the support convention specified in that paper, and constants independent of the cube and boundary data. A vacuum mass estimate does not establish this stronger premise.

Even a positive lattice area coefficient $c(\beta)$ at each finite coupling does not suffice for a physical continuum claim. Along a prescribed scaling trajectory one needs, at minimum, quantitative control compatible with
\[
\liminf_{a\downarrow0}\frac{c(\beta(a))}{a^2}>0,
\]
plus renormalized loop convergence and control of the lower-order terms. The physical area is $a^2$ times the lattice plaquette count. This requirement is not provided by the effective-model analysis.

\subsection{Geometry of the surface response}

The relevant four-dimensional geometry can be controlled exactly. Let
$\Sigma=[0,R]\times[0,T]\times\{0,0\}$ with $R,T\in a\mathbb N$, use $\ell^1$ distance, and define $b_a(\mu)=a\coth(\mu a/2)$.

\begin{lemma}[Surface-tube sum]\label{conf:surface}
For $\mu>0$,
\[
a^4\sum_{x\in a\Z^4}e^{-\mu\operatorname{dist}_1(x,\Sigma)}
=b_a^2RT+b_a^3(R+T)+b_a^4,
\qquad b_a\le a+2/\mu.
\]
\end{lemma}
\begin{proof}
With $z=e^{-\mu a}$, the one-dimensional sum for an interval is
\[
a\sum_{j\in\Z}e^{-\mu\operatorname{dist}(aj,[0,R])}
=a\left(R/a+1+\frac{2z}{1-z}\right)=R+b_a.
\]
For a singleton it is $b_a$. The four sums factor, giving $b_a^2(R+b_a)(T+b_a)$. Finally $e^x-1\ge x$ gives $b_a=a+2a/(e^{\mu a}-1)\le a+2/\mu$.
\end{proof}
Finite boxes give a smaller sum. On periodic lattices with a nonwrapping rectangle, the same upper bound follows by bounding the maximum over periodic images by their sum and unfolding the fundamental domain.

Take real $W(C)$ and real action densities, with $\langle W(C)\rangle_s>0$; for example use the normalized real-trace Wilson observable. Suppose a renormalized action-density decomposition is justified,
$\partial_sS_s=a^4\sum_x O_s(x)$, and define
\[
g_s(C;x)=\frac{\Cov_s(W(C),O_s(x))}{\langle W(C)\rangle_s}.
\]
For $Q(s)\ge0$ and $\mu(s)>0$, consider the \emph{additional microscopic hypothesis}
\[
\pos{-g_s(C;x)}\le Q(s)e^{-\mu(s)\operatorname{dist}_1(x,\Sigma_C)},
\tag{H}
\]
uniformly in cutoff, volume, mass, circumference and loop size, then the interpolation identity and the lemma prove
\[
\partial_sF_s(R,T)\ge
-Q(s)b_a^2RT-Q(s)b_a^3(R+T)-Q(s)b_a^4.
\]
For $a\le a_0$, a sufficient area-loss budget is
\[
\int Q(s)\bigl(a_0+2/\mu(s)\bigr)^2\dd s<\sigma_0,
\]
with the corresponding third and fourth powers integrable for the lower-order terms. In a justified continuum passage the geometric area coefficient approaches $4\int Q/\mu^2$.

\note{\textbf{Hypothesis (H) is not proved here.} Ordinary clustering bounds an absolute covariance; division by an exponentially small loop expectation can destroy uniformity. The determinant estimates do not supply this cancellation. Contact terms and perimeter/corner counterterms must also be controlled. A mixed finite difference has alternating signs, so one cannot take a mixed difference of this one-sided bound and claim a Creutz bound.}

\subsection{A sufficient criterion for simultaneous limits}

The configuration space changes with lattice spacing. Discrete scale comparisons avoid a formal derivative with respect to that spacing. Choose compatible physical lattices and parameters
\[
a_j=a_0 2^{-j},\qquad m_k=m_*2^k,\qquad L_\ell=L_*2^\ell,
\]
and matched, renormalized physical rectangles with fixed increment $h$. Write
\[X_{j,k,\ell}=\chi_{j,k,\ell,h}(R,T)/h^2.\]

\begin{theorem}[Conditional simultaneous-limit theorem]\label{conf:joint_limits}
Suppose, uniformly in the other indices,
\[
\begin{aligned}
|X_{j+1,k,\ell}-X_{j,k,\ell}|&\le A_j,\\
|X_{j,k+1,\ell}-X_{j,k,\ell}|&\le B_k,\\
|X_{j,k,\ell+1}-X_{j,k,\ell}|&\le C_\ell,
\end{aligned}
\qquad
\sum_j A_j+\sum_kB_k+\sum_\ell C_\ell<\infty.
\]
Then there is a joint limit, independent of limit order and cofinal path, and
\[
|X_{j,k,\ell}-X_\infty|
\le\sum_{i\ge j}A_i+\sum_{i\ge k}B_i+\sum_{i\ge\ell}C_i.
\]
If the bounds are also uniform in the loop family, convergence is uniform on that family.
\end{theorem}
\begin{proof}
Compare any two triples by changing one coordinate at a time. Telescoping bounds their difference by the three sequence tails beginning at the smaller indices. Those tails tend to zero, so the family is jointly Cauchy. Completeness of $\R$ proves existence; sending one triple to infinity proves the error bound. The same argument covers iterated limits.
\end{proof}

For preservation of positivity alone, one-sided increments with total negative budget $E<\sigma_0$ suffice: starting from $X_{0,0,0}\ge\sigma_0$ gives $X_{j,k,\ell}\ge\sigma_0-E$. Such one-sided bounds do \emph{not} prove existence or order independence. A fourth summable sequence handles box size, on compatible domains that contain the loops. Identification with the intended continuum quantum theory and control of the boundary terms in Section~\ref{conf:creutz_section} remain necessary.

\textbf{A precise failure of exchanging limits.} The smooth positive scalar family
\[
X(a,m)=\sigma_0\frac{am}{1+am}
\]
has
\[
\lim_{m\to\infty}\lim_{a\to0}X=0,
\qquad
\lim_{a\to0}\lim_{m\to\infty}X=\sigma_0,
\]
while $am=1$ gives $\sigma_0/2$. This refutes exchanging limits on the basis of positivity or pointwise estimates alone. It is a counterexample to an inference rule, not a Yang--Mills construction. The microscopic sequences $A_j,B_k,C_\ell$ satisfying this theorem have not been established here.

\subsection{Twisted partition functions and sector normalization}

The Tomboulis--Yaffe route provides an independent microscopic comparison. For a reflection-positive Wilson $SU(N)$ theory on a suitable torus, let $Z^{[k]}$ be its partition function with the standard centre twist $k$ on a plaquette stack winding around the $d-2$ transverse periodic directions, so its dual surface is closed and wraps the torus. Set
\[
q_\Lambda=1-\frac1N\sum_{k=0}^{N-1}\frac{Z^{[k]}}Z.
\]
For a normalized rectangular Wilson loop of nonzero $N$-ality, Kanazawa's generalization gives \cite{conf:ty}
\[
|\langle W(C)\rangle|\le
2q_\Lambda^{A_{\mathrm{lat}}(C)/(N_\mu N_\nu)}.
\]
The stated reflection construction uses compatible doubling sizes $N_\mu=2^p n_\mu$, $N_\nu=2^q n_\nu$ for the rectangle's lattice side lengths. Thus a sufficient defect estimate would be
\[
-\log q_\Lambda\ge\sigma_*a^2N_\mu N_\nu
\]
uniformly along appropriate thermodynamic and continuum sequences. A single finite-box value $q_\Lambda<1$ does not establish this rate. The inequality controls bare lattice loops; renormalized continuum observables still require their own limiting argument.

It is not legitimate to assume that a local RG interpolation can be continued over every required scale. Ito and Seiler give the elementary counterexample $\Psi(\lambda,t)=e^{-t}-1+2\lambda$: its locally regular solution $t=-\log(1-2\lambda)$ cannot reach $\lambda=1$ \cite{conf:itoseiler}. Uniform continuation is a proof obligation, not a consequence of the local implicit-function theorem.

\subsection{The physical continuum scale}

The weak-coupling trajectory imposes a different quantitative scale. For the normalized Wilson action, $\beta_{\mathrm W}=2N/g_0^2\to\infty$ as $a\to0$. The universal perturbative coefficients are \cite{conf:beta}
\[
\mu\frac{\dd g}{\dd\mu}=-b_0g^3-b_1g^5+O(g^7),\qquad
b_0=\frac{11N}{48\pi^2},\quad b_1=\frac{34N^2}{3(16\pi^2)^2}.
\]
Integrating this ultraviolet expansion gives the usual asymptotic scaling formula
\[
a\Lambda=e^{-1/(2b_0g_0^2)}(b_0g_0^2)^{-51/121}[1+O(g_0^2)].
\]
This formula identifies the candidate physical scale. It does not establish the infrared dynamics or the existence of the continuum theory. In particular, \emph{if} $\sigma_{\rm phys}\to s_*\Lambda^2$ with $0<s_*<\infty$, the dimensionless lattice tension must satisfy
\[
\sigma_{\rm lat}=a^2\sigma_{\rm phys}\sim s_*\Phi(g_0),\qquad
\boxed{\Phi(g)=e^{-1/(b_0g^2)}(b_0g^2)^{-102/121}.}
\]
The ratio $b_1/b_0^2=102/121$ is independent of $N$. Above active matter thresholds, the pure-theory coefficients cannot be substituted without threshold matching.

\begin{proposition}[Absolute perturbative accuracy does not resolve the target]\label{conf:flatness}
For each finite real $p$, $g^p/\Phi(g)\to\infty$ as $g\downarrow0$. Extending $\Phi$ by zero at $g=0$ gives a smooth function with every derivative zero there. Consequently the functions $s\Phi(g)$, $s\ge0$, have identical formal power series, although their ratio to $\Phi(g)$ is $s$.
\end{proposition}
\begin{proof}
The logarithm of the first ratio is $1/(b_0g^2)$ plus a constant and a multiple of $\log g$; the positive inverse square dominates. Each derivative of $\Phi$ is $\Phi$ times a finite sum of inverse powers of $g$, hence tends to zero by the same argument. Induction proves smooth extension and flatness.
\end{proof}

Thus an absolute error budget $Cg^p$ alone cannot certify a positive coefficient of order $\Phi(g)$. This does not exclude a favorable one-sided estimate, exact cancellation, or a nonperturbative bound. Nor is it an impossibility theorem for renormalization methods.

For a concrete illustration, the positive functions
\[
w_s(g;n,m)=\exp[-2g^2(n+m)-s\Phi(g)nm]
\]
have identical perturbative expansions at every fixed lattice rectangle. Their unit-increment Creutz differences equal $s\Phi(g)$ exactly. When $a^2nm\to A$ and $\Phi(g)\sim(a\Lambda)^2$, the area term approaches $s\Lambda^2A$. These functions illustrate nonuniformity in loop size; they are not asserted to arise from Yang--Mills measures.

\textbf{Correct dimensions for additive blocking errors.} Let $a_j=b^ja_0$, $a_J=\ell$, and suppose matched loop free energies obey
\[
F_j(bn,bm)\ge F_{j+1}(n,m)-\varepsilon_j nm-p_j(n+m)-c_j.
\]
For matched asymptotic area coefficients this implies
\[
\tau_j\ge b^{-2}(\tau_{j+1}-\varepsilon_j),\qquad
\sigma_{\rm UV}\ge\frac{\tau_J}{\ell^2}
-\sum_{j=0}^{J-1}\frac{\varepsilon_j}{a_{j+1}^2}.
\]
The comparison concerns aligned rectangles; existence and identification of the corresponding asymptotic rates are additional hypotheses. The physical weights $a_{j+1}^{-2}$ cannot be dropped. Summability of unscaled lattice errors does not suffice.

\subsection{A sufficient relative ultraviolet estimate}

A relative estimate avoids subtracting an error larger than the nonperturbative signal. The following statement keeps the loop-size and perimeter requirements explicit, so a positive bound is not lost by exchanging the large-loop and continuum limits.

\begin{theorem}[Conditional relative transport at fixed physical rectangles]\label{conf:relative_transport}
For each regulator depth $d$, let $F_{d,k}(R,T)$, $0\le k\le d$, be real matched loop free energies on a common physical rectangle family. Think of refinement spacings $a_k=\ell b^{-k}$. Write $A=RT$, $P=2(R+T)$. Assume, uniformly in $d$, all other regulator parameters, and the rectangle family,
\[
F_{d,0}\ge\sigma_0 A-B_0P-C_0,\qquad \sigma_0>0,
\]
and, for $k<d$,
\[
\boxed{F_{d,k+1}\ge(1-\eta_k)F_{d,k}-u_kP-v_k.}\tag{R}
\]
Suppose $B_0,C_0,u_k,v_k\ge0$, $0\le\eta_k\le\rho<1$, and
\[E:=\sum_k\eta_k<\infty,\quad U:=\sum_ku_k<\infty,\quad V:=\sum_kv_k<\infty.\]
Then, uniformly in $d$ and those parameters,
\[
F_{d,d}(R,T)\ge
\underbrace{\sigma_0e^{-E/(1-\rho)}}_{\sigma_*>0}A
-(B_0+U)P-(C_0+V).
\]
Any pointwise limit of these matched free energies on this family inherits the bound.
\end{theorem}
\begin{proof}
Iterate (R). The coefficient of the initial area term is $\prod_{k<d}(1-\eta_k)$. Every negative perimeter or constant term is multiplied by a product in $[0,1]$, so replacing those products by one only weakens the lower bound. Finally,
\[
\log(1-x)=-\int_0^x\frac{\dd t}{1-t}\ge-\frac{x}{1-\rho}
\quad(0\le x\le\rho)
\]
gives the stated product bound. Passing to an existing pointwise limit proves the last assertion.
\end{proof}

\textbf{Why an $O(g^4)$ relative loss is enough in the ultraviolet tail.}
If the \emph{actual} running couplings and defect losses satisfy
\[
g_k^2\le\frac1{\alpha+\gamma k},\qquad
\eta_k\le Cg_k^4,\qquad \alpha,\gamma>0,
\]
then comparison with a decreasing integral gives
\[
E\le C\sum_{k\ge0}(\alpha+\gamma k)^{-2}
\le C\left(\alpha^{-2}+\frac1{\alpha\gamma}\right).
\]
This proves a positive lower product, provided the uniform $\rho<1$ condition also holds. A finite crossover preceding the ultraviolet tail needs its own uniformly positive product. An $O(g_k^2)$ upper bound alone does not guarantee this: the admissible example $\eta_k=c/(\alpha+\gamma k)$, $0<c<\alpha$, has a product tending to zero.

\note{\textbf{The microscopic hypotheses remain unproved.} Neither the positive physical-scale seed nor (R) follows from a perturbative beta function. The matched perimeter and corner budgets also require proof; divergent bare-loop counterterms cannot be ignored. This theorem preserves a bound in limits that exist. It does not construct those limits, prove their order independence, or establish matter-threshold and circumference uniformity.}

\subsection{Exact blocking and the full action}

For an integer block factor $b$, assign each coarse link the ordered product of the fine links in its straight chain:
\[
V_e=B(U)_e=\prod_{f\in\operatorname{chain}(e)}U_f.
\]
The chains use disjoint fine edges. Product Haar measure therefore pushes forward to product coarse Haar measure. For a finite regulated action, define its exact unnormalized coarse density by
\[
\rho_B(V)=\mathbb E_{\rm Haar}[e^{-S(U)}\mid B(U)=V],
\qquad S_B=-\log\rho_B.
\]
Conditional expectation immediately gives $\int\rho_B\,\dd V=Z$. If $C$ follows coarse chains, ordered multiplication gives the pointwise identity
\[
W_{\rm fine}(C;U)=W_{\rm coarse}(C;B(U)),
\]
so its expectation is preserved exactly. The action $S_B$ generally contains many interactions; it is not asserted to be another Wilson plaquette action. For nonaligned observables the exact coarse insertion is instead $\mathbb E_\mu[W\mid B]$.

Reflection positivity is also preserved if $B$ intertwines the reflections and pulls the coarse positive-half algebra into the fine positive-half algebra: for such $G$,
\[
(B_*\mu)(\overline{G\circ\theta_c}G)
=\mu(\overline{(G\circ B)\circ\theta}(G\circ B))\ge0.
\]
This identity gives no contraction or area coefficient for $S_B$.

For a centre twist $S^{[k]}=S+\Delta S_k$, put $h_k=e^{-\Delta S_k}$. The exact coarse insertion and twisted density satisfy
\[
h_{k,B}=\mathbb E_\mu[h_k\mid B],\qquad
\rho_B^{[k]}=\rho_B h_{k,B},\qquad
\frac{Z^{[k]}}Z=\langle h_{k,B}\rangle_{B_*\mu}.
\]
One must retain sector normalizations. For instance, actions $S$ and $S+c$ have identical normalized measures but partition functions differing by $e^{-c}$.

\textbf{A common bulk residual does not cancel automatically.}
Let approximate weights on two states be $(1,\epsilon)$ in one sector and $(\epsilon,1)$ in another. Their partition ratio equals one. Multiply \emph{both} by $e^{-R}$, with $R=(0,M)$. The new ratio is
\[
\frac{\epsilon+e^{-M}}{1+\epsilon e^{-M}}\sim e^{-M}
\quad\text{when }\epsilon=e^{-2M},\ M\to\infty.
\]
Thus identical action residuals can alter a defect ratio arbitrarily. In general, if $S_0=\widetilde S_0+R$, $S_d=\widetilde S_d+R+\delta R$, positivity gives
\[
\left|\log\frac{Z_d/Z_0}{\widetilde Z_d/\widetilde Z_0}\right|
\le\operatorname{osc}(R)+\|\delta R\|_\infty.
\]
Indeed, each residual expectation lies between the exponential of its negative maximum and negative minimum. Subtraction leaves the displayed range bound. Its first term can scale with total volume.

Even relative accuracy for each near-unity $Z^{[k]}/Z$ need not control their small difference $q_\Lambda$. In $SU(2)$, $r_1=1-2e^{-\sigma A}$ gives $q=e^{-\sigma A}$, whereas replacing $r_1$ by $1-\delta$ gives $q'=\delta/2$. For sufficiently large $A$ these are positive ratios with a small absolute change but different area exponents. This is an inference counterexample, not a construction of gauge partition functions.

Ba\l aban's ultraviolet-stability theorem controls bulk effective densities while the effective couplings remain small; it does not supply the required loop or flux-sector estimate \cite{conf:balaban}. Exact blocking removes approximation errors only if the full action is retained. Proving a uniform physical-scale area bound for that exact action is still the missing dynamical step.

\subsection{Failure of single-link total-variation contraction at weak coupling}

The restriction in Section~\ref{conf:mixing_section} is not merely a loose numerical constant. In pure $SU(2)$ Wilson theory the actual total-variation influence coefficients violate the standard one-link Dobrushin criterion for sufficiently large $\beta$.

\begin{proposition}[An explicit $SU(2)$ obstruction]\label{conf:weak_influence}
Use density $\exp[(\beta/2)\sum_p\operatorname{ReTr}U_p]$ relative to product Haar measure. Let $e$ be an interior edge whose adjacent plaquettes have free remaining edges, and define
\[
c_{ef}=\sup_{\omega=\omega'\ {\rm off}\ f}
\|\mu_e(\cdot\mid\omega)-\mu_e(\cdot\mid\omega')\|_{\rm TV}.
\]
Set $p_{\rm cap}=1/3-\sqrt3/(4\pi)$. Then
\[
\sum_f c_{ef}\ge6\max\{0,1-p_{\rm cap}^{-1}e^{-\beta/2}\}.
\]
In particular, this row sum exceeds one whenever
\[
\boxed{\beta>2\log\frac6{5p_{\rm cap}}
=3.6290215626\ldots .}
\]
\end{proposition}
\begin{proof}
The conditional density of $U_e$ is proportional to
$\exp[(\beta/2)\operatorname{ReTr}(U_eK)]$, where $K$ is the sum of six oriented staples. Each staple contains a distinct opposite parallel edge. Set the transverse edges to identity; prescribing those six opposite edges then prescribes the staples independently, using an adjoint when orientation requires it.

Choose one of these opposite edges $f$. Assign the other five staples the matrices
\[
D_j=\operatorname{diag}(e^{2\pi ij/5},e^{-2\pi ij/5}),
\qquad j=0,\ldots,4.
\]
Both diagonal entries sum to zero. Vary only $f$ so the remaining staple is $I$ or $-I$. The conditional laws are therefore exactly
\[
p_\pm(U)=Z_\beta^{-1}e^{\pm\beta x}\,\dd U,
\qquad x=\tfrac12\operatorname{Tr}U.
\]
Their normalizations agree by $U\mapsto-U$. Haar measure gives $x$ the density $(2/\pi)\sqrt{1-x^2}$ on $[-1,1]$, hence
\[
\Pr_{\rm Haar}(x\ge1/2)=p_{\rm cap},\qquad
Z_\beta\ge p_{\rm cap}e^{\beta/2}.
\]
The unnormalized $p_+$ mass on $x\le0$ is at most $1/2$. Since $p_+\ge p_-$ exactly on $x\ge0$,
\[
\|p_+-p_-\|_{\rm TV}
=1-2\Pr_{p_+}(x\le0)
\ge1-p_{\rm cap}^{-1}e^{-\beta/2}.
\]
Apply the construction separately to each of the six opposite edges. Their influence suprema may use different exterior assignments. Summing proves the claim. These assignments are allowed without gauge fixing, and continuity also permits approximation by neighborhoods of positive Haar measure.
\end{proof}

The result applies sufficiently far inside a nondegenerate box with arbitrary fixed outer boundary data. It rules out extending the earlier \emph{single-link total-variation row-sum method} to the asymptotically free trajectory merely by improving constants. It does not rule out block methods, other metrics, or confinement itself.

\note{\textbf{Status after this physical-continuum continuation.} A sharper sufficient relative-error criterion and an exact obstruction to the earlier contraction method have been proved. No positive physical-scale quantum seed and no microscopic estimate (R) have been established along the required trajectory. Consequently, four-dimensional continuum confinement remains unproved by this work.}

\section{An explicit reference-loop trajectory and exact blocking}\label{sec:trajectory}
\paragraph{Conventions.} $SU(2)$ traces are real, so $\ReTr=\Tr$ on single group elements. For $X\in SU(2)$ we write $X=x_0+i\,\vec x\cdot\vec\sigma$ with $|x|=1$. Every sum $K$ of $SU(2)$ matrices is a real multiple of an $SU(2)$ matrix, and $|K|:=\sqrt{\tfrac12\Tr KK^\dagger}$. Haar measure is normalized. Rectangles are planar, lie along lattice axes, and do not wind around the torus; on a torus of side $N$ their lattice-unit sides satisfy $R,T\in\{1,\dots,N-1\}$, and our results are stated for sides at most $N/2$. We write
\[
w(R,T)=\langle W_C\rangle,\qquad f(R,T)=-\log w(R,T),\qquad w(R,0):=1 .
\]

\subsection{The theory and the trajectory}
\subsubsection{Lattices, boxes and boundary conditions}
Fix a physical length $\ell>0$ and an even integer $M\ge4$. At level $k\ge0$ the lattice spacing is $a_k=\ell2^{-k}$. The reference box is the periodic torus
\[
\Lambda_{k}=\bigl(a_k\Z/L\Z\bigr)^4,\qquad L=M\ell,\qquad N_k=M2^k\ \text{sites per direction}.
\]
Every link carries normalized Haar measure, and
\[
\dd\mu_{k,\beta}(U)=Z^{-1}e^{-S_\beta(U)}\prod_e\dd U_e,\qquad S_\beta(U)=\beta\sum_p\Bigl(1-\tfrac12\Tr U_p\Bigr).
\]
For uniformity statements, an \emph{enclosing box} is any periodic torus $\Lambda_k'=(a_k\Z/L'\Z)^4$ with $L'=M'\ell$, where $M'\ge M$ is even. The same $\beta_k$, defined in the reference box, is used in every enclosing box. A physical rectangle with sides $R=n\ell$, $T=m\ell$ lies on every lattice. At level $k$ it has $n2^k\times m2^k$ lattice sides.

\subsubsection{Positivity, concavity and the two coupling limits}
Reflection positivity (RP) of the Wilson action under reflections in hyperplanes through sites (\emph{site reflections}) and hyperplanes bisecting links (\emph{link reflections}) on an even torus is classical \cite{OSLattice,traj:seilerbook}. Let $\theta$ be such a reflection perpendicular to the direction $\nu$, and let $\mathcal A_+$ be the complex functions of the links in the closed upper half (for link reflections, the links with both endpoints in the upper half). Set $(\theta F)(U)=\overline{F(\theta U)}$. Then
\begin{equation}\label{traj:eq:rp}
\langle F\,\theta F\rangle\ge0,\qquad |\langle F\,\theta G\rangle|^2\le\langle F\,\theta F\rangle\,\langle G\,\theta G\rangle\qquad(F,G\in\mathcal A_+).
\end{equation}
For link reflections, gauge-invariant expectations are unchanged if the crossing $\nu$-links are set to $1$ by a gauge transformation on one slice adjacent to each of the two fixed hyperplanes of the torus. We use this normalization whenever a loop crosses a link-reflection plane.

\begin{lemma}[Strict positivity and log-convexity]\label{traj:lem:pos}
Let $N\ge4$ be even and $\beta>0$. For $1\le R,T\le N/2$:
\begin{enumerate}[label=(\alph*)]
\item for nonnegative integers $a,b$ with $a+b=T$ and $\max(2a,2b)<N$, $w(R,T)^2\le w(R,2a)\,w(R,2b)$; for $a,b\ge1$ with $a+b=T+1$ and $\max(2a,2b)\le N$, $w(R,T)^2\le w(R,2a-1)\,w(R,2b-1)$;
\item $w(R,T)^2\le w(R,T-1)\,w(R,T+1)$ for $1\le T\le N/2-1$, and similarly in $R$;
\item $w(R,T)>0$.
\end{enumerate}
\end{lemma}
\begin{proof}
(a) Place a site reflection plane so that $C$ has $a$ steps in the $\nu$-direction above it and $b$ below. The plane meets $C$ in two sites $c_1,c_2$. Let $P_+$ be the part of $C$ above the plane, from $c_1$ to $c_2$. Let $P_-'$ be the reflection into the upper half of the part below the plane, also oriented from $c_1$ to $c_2$. With $F_{ij}=M(P_+)_{ij}$ and $G_{ij}=M(P'_-)_{ij}$ (holonomy matrix entries),
\[
W_C=\tfrac12\Tr\bigl(M(P_+)\,M(\theta P'_-)^{-1}\bigr)=\tfrac12\sum_{ij}F_{ij}\,\theta G_{ij}.
\]
Moreover $\tfrac12\sum_{ij}F_{ij}\,\theta F_{ij}$ is the Wilson loop of $P_+\cup\theta P_+$, a rectangle of height $2a$, and similarly for $G$. Apply \eqref{traj:eq:rp} to each pair $(i,j)$ and then Cauchy--Schwarz to the sum over $(i,j)$. For a link reflection, set the two crossing $\nu$-links of $C$ to $1$. A part containing $a$ vertical steps, the crossing step included, then doubles to height $2a-1$, and $a+b=T+1$. If $b=0$, the lower part lies in the plane and its doubled loop is trivial, with value $1$. The strict inequality $2a<N$ in the site case keeps $P_+\cup\theta P_+$ from reaching the second fixed plane, where it would close around the torus.

(b) For odd $T$, use (a) with site splitting $a=(T+1)/2$, $b=(T-1)/2$. For even $T$, use link splitting $a=T/2+1$, $b=T/2$.

(c) By translation and hypercubic symmetry, all plaquettes have the same expectation $\langle P\rangle_\beta$. Moreover,
\[
\frac{\dd}{\dd\beta}\langle P\rangle_\beta=\frac{\Var(\sum_pP_p)}{\#\{p\}}>0,
\qquad P_p:=\tfrac12\Tr U_p,
\]
and $\langle P\rangle_0=0$; hence $w(1,1)>0$. Every factor on the right in (a) is $\ge0$ by RP. Positivity of the left side therefore forces both factors to be strictly positive. If $w(R,T)>0$, the two parts of (a) give $w(R,s)>0$ for every $s\le\min(2T,N/2)$. Starting from $w(1,1)>0$ and iterating gives $w(1,s)>0$ for all $s\le N/2$. Repeating the argument in the other direction gives the claim.
\end{proof}

Part (b) says that $T\mapsto f(R,T)$ is \emph{concave} on $\{0,\dots,N/2\}$, with $f(R,0)=0$. This is the familiar RP statement behind Seiler's linear upper bound on the potential \cite{traj:seiler78} and Bachas' concavity theorem \cite{traj:bachas}. It is used again in the asymptotic-coefficient reduction.

\begin{lemma}[The two coupling limits]\label{traj:lem:limits}
Fix $N\ge4$ even and $1\le R,T\le N/2$.
\begin{enumerate}[label=(\alph*)]
\item As $\beta\downarrow0$, $w(R,T)=(\beta/4)^{RT}\bigl(1+O(\beta)\bigr)$.
\item For $\beta\ge1$, uniformly in $N$,
\[
1-\langle P\rangle_\beta\le\frac{2(\log\beta+6.4)}\beta,\qquad
1-w(R,T)\le(RT)^2\bigl(1-\langle P\rangle_\beta\bigr).
\]
In particular $w(R,T)\to1$ as $\beta\to\infty$.
\end{enumerate}
\end{lemma}
\begin{proof}
(a) Expand $e^{-S_\beta}=e^{-\beta n_p}\prod_p\sum_m(\beta/2)^m(\Tr U_p)^m/m!$. Because $-I\in SU(2)$, a Haar integral of a product of matrix entries of one link vanishes unless that link occurs an even number of times. Therefore every multiset of plaquettes contributing to $\langle W_C\rangle$ reduces mod $2$ to a $\Z_2$ 2-chain $\sigma$ with $\partial\sigma=C$. Project $\sigma$ onto the $(\mu,\nu)$ coordinate torus of $C$. The projection is a $\Z_2$ 2-chain on a two-dimensional torus with boundary $C$, so it is either the rectangle or its complement. Since $RT\le N^2/4<N^2-RT$, $\sigma$ contains at least $RT$ plaquettes. At order $\beta^{RT}$ the multiset must therefore be a set of $RT$ $(\mu\nu)$-plaquettes projecting bijectively onto the rectangle. If two neighbouring plaquettes had different transverse positions, a link strictly inside the rectangle would occur once in $\partial\sigma$. Hence $\sigma$ is the planar rectangle itself. Gluing with $\int\Tr(AU)\Tr(U^\dagger B)\dd U=\tfrac12\Tr(AB)$ and $\int\Tr(AUBU^\dagger)\dd U=\tfrac12\Tr A\Tr B$ gives the factor $(\tfrac12)^{RT-1}$. Here $2RT-R-T$ interior links contribute factors $\tfrac12$, and $(R-1)(T-1)$ interior vertices contribute factors $2$. The last integral is $\tfrac12\int(\Tr X)^2\dd X=\tfrac12$. The coefficient is therefore $(\beta/2)^{RT}(\tfrac12)^{RT}=(\beta/4)^{RT}$, while $Ze^{\beta n_p}=1+O(\beta^2)$.

(b) Let $E(\beta)=\langle\sum_p(1-P_p)\rangle_\beta=-\frac{\dd}{\dd\beta}\log Z(\beta)$, where $Z(\beta)=\int e^{-S_\beta}\le1$. Since $E$ is nonincreasing,
\[
\frac\beta2E(\beta)\le\int_{\beta/2}^\beta E=\log Z(\beta/2)-\log Z(\beta)\le-\log Z(\beta).
\]
On the set where every link satisfies $\|U_e-1\|_F\le\delta$, we have $1-P_p=\tfrac14\|U_p-1\|_F^2\le4\delta^2$. The Haar measure of $\{\|U-1\|_F\le\delta\}$ is at least $c_H\delta^3$, with $c_H=4/(3\sqrt2\pi^3)$; this uses the class density $\tfrac2\pi\sin^2\vartheta$ together with $\sin\vartheta\ge2\vartheta/\pi$. With $\delta^2=1/\beta$, $n_\ell/n_p=2/3$ and $\log(1/c_H)<3.5$, we obtain $-\log Z\le n_p\{4+\log\beta+\tfrac23\cdot3.5\}$. This is the first bound. For the second bound, the lattice non-abelian Stokes formula writes $U_C$ as an ordered product of conjugates of the $RT$ enclosed plaquette variables. Unitarity gives $\|U_C-1\|_F\le\sum_p\|U_p-1\|_F$. Cauchy--Schwarz then gives $1-W_C\le RT\sum_{p\subset C}(1-P_p)$; take expectations.
\end{proof}

\subsubsection{The renormalization condition and the definition of $\beta_k$}
At level $k$, put $n=2^k$ and define, on the reference box,
\begin{equation}\label{traj:eq:creutz}
\chi_k(\beta):=-\log\frac{w(2n,2n)\,w(n,n)}{w(2n,n)^2}.
\end{equation}
This is the physical Creutz ratio \cite{traj:creutz} of the loops with sides $\ell\times\ell$, $2\ell\times\ell$ and $2\ell\times2\ell$. At $k=0$ these are the lattice $1\times1$, $2\times1$ and $2\times2$ loops. The perimeter and the four right-angle corners cancel identically.

\begin{theorem}[The trajectory is well defined]\label{traj:thm:traj}
For every $c_*>0$ and $k\ge0$, $\chi_k$ is real-analytic on $(0,\infty)$,
\[
\chi_k(\beta)=-4^k\log(\beta/4)+O(\beta)\quad(\beta\downarrow0),\qquad
\chi_k(\beta)\to0\quad(\beta\to\infty).
\]
Hence $\mathcal Z_k:=\{\beta>0:\chi_k(\beta)=c_*\}$ is nonempty and compact, and
\[
\boxed{\beta_k:=\min\mathcal Z_k}
\]
is well defined.
\end{theorem}
\begin{proof}
All three loop expectations are strictly positive (Lemma~\ref{traj:lem:pos}) and are ratios of entire functions of $\beta$, so $\chi_k$ is real-analytic. The exponent in Lemma~\ref{traj:lem:limits}(a) is $4n^2+n^2-2\cdot2n^2=n^2$. Lemma~\ref{traj:lem:limits}(b) gives $\chi_k\to0$. The intermediate value theorem gives a root. $\mathcal Z_k$ is closed in $(0,\infty)$, bounded away from $0$ because $\chi_k\to\infty$, and bounded because $\chi_k\to0<c_*$.
\end{proof}

\begin{definition}[Reference-loop trajectory]\label{traj:def:traj}
The trajectory is the tuple
\[
\begin{aligned}
&a_k=\ell2^{-k},\qquad \beta_k=\min\{\beta>0:\chi_k(\beta)=c_*\},\\
&\text{box: reference torus }(\Z/M2^k\Z)^4\text{ (side }L=M\ell\text{) and enclosing tori }M'\ge M\text{ even},\\
&\text{boundary conditions: periodic for all links},\qquad \text{renormalization condition: }\chi_k(\beta_k)=c_* .
\end{aligned}
\]
The perimeter and corner convention of the normalized-ratio construction is fixed by the same three loops. With $F_k(R,T)=-\log w_k(R,T)$ in physical units and $P=2(R+T)$, set
\begin{equation}\label{traj:eq:lambda}
\lambda_k:=\frac{F_k(2\ell,\ell)-F_k(\ell,\ell)-c_*}{2\ell},\qquad
\kappa_k:=F_k(\ell,\ell)-c_*-4\ell\lambda_k,\qquad
\widehat F_k:=F_k-\lambda_kP-\kappa_k.
\end{equation}
Then $\widehat F_k(C)=c_*A(C)/\ell^2$ exactly on the three reference loops at every $k$.
\end{definition}
The last identity follows by substituting \eqref{traj:eq:creutz} into \eqref{traj:eq:lambda}. Any other convention differing by $k$-independent constants changes $\widehat F$ only by a fixed finite renormalization.

\begin{proposition}[The first scale is in the strong-coupling region]\label{traj:prop:start}
Let $0<\beta_*<\sqrt2/3$ and
\[
C(\beta_*):=-5\log\Bigl[\frac{\beta_*}4\Bigl(1-\frac{9\beta_*^2}2\Bigr)\Bigr].
\]
If $c_*>C(\beta_*)$, then $\mathcal Z_0\subset(0,\beta_*)$, for every even $M\ge4$. For $\beta_*=1/20$, $C(\beta_*)=21.967\ldots$.
\end{proposition}
\begin{proof}
By Lemma~\ref{traj:lem:pos}(a), $w(2,1)\ge w(1,1)^2$ and $w(2,2)\ge w(2,1)^2$. Since $w(2,1)\le1$,
\[
\chi_0(\beta)\le-\log w(2,2)-\log w(1,1)\le-5\log\langle P\rangle_\beta .
\]
Since $\langle P\rangle_\beta$ is increasing, for $\beta\ge\beta_*$ this is at most $-5\log\langle P\rangle_{\beta_*}\le C(\beta_*)<c_*$, by Lemma~\ref{traj:lem:plaq} below. Hence no root lies in $[\beta_*,\infty)$.
\end{proof}

\begin{remark}[What is not known about the trajectory]\label{traj:rem:traj}
\begin{enumerate}[label=(\roman*)]
\item \emph{Uniqueness.} Monotonicity of $\chi_k$ in $\beta$ is not known: Griffiths-type inequalities are not available for $SU(2)$. The minimal-root rule removes the ambiguity, but it is a convention.
\item \emph{Escape to weak coupling.} We cannot prove $\beta_k\to\infty$. If $\sup_k\beta_k=B<\infty$, the lattice Creutz ratio at size $2^k$ would stay equal to $c_*$ at couplings in $(0,B]$. Excluding this requires a lower bound on large lattice Creutz ratios at fixed coupling, which is a confinement statement in its own right. If the strong-coupling cluster expansion is granted, $\beta_k$ eventually leaves the strong-coupling region. Nothing proved here controls it afterwards.
\item \emph{The ultraviolet budget presupposes asymptotic freedom along the trajectory.} The budget $\eta_k\le Cg_k^4$ with $g_k^2=4/\beta_k\le1/(\alpha+\gamma k)$ requires $\beta_k\ge4(\alpha+\gamma k)$. This is a nonperturbative statement about Definition~\ref{traj:def:traj} and is not established. Perturbatively, the identification is $\Delta(1/g^2)=2b_0\log2$ per halving of $a$, with $b_0=11/(24\pi^2)$. This gives $\Delta\beta\to8b_0\log2=0.2575\ldots$ and $\gamma=0.0644$; three-loop data are in \cite{conf:beta}. It identifies the expected ultraviolet scaling and plays no role in any proof.
\end{enumerate}
\end{remark}

\subsection{One exact link integration}
Let $e=(x,\mu)$ be a link and $S_1,\dots,S_6$ its oriented staples. Thus the six plaquettes containing $e$ have holonomies $U_eS_i$ up to cyclic order, and
\[
S=S_{\rm rest}-\frac\beta2\ReTr(U_eK),\qquad K=\sum_{i=1}^6S_i .
\]

\begin{proposition}[One-link integration]\label{traj:prop:link}
Write $K=rV$ with $r=|K|$ and $V\in SU(2)$ when $r>0$. Then
\[
\int_{SU(2)}e^{(\beta/2)\ReTr(UK)}\dd U=J(\beta r),\qquad J(z)=\frac2\pi\int_0^\pi e^{z\cos\vartheta}\sin^2\vartheta\,\dd\vartheta=\frac{2I_1(z)}z,\quad J(0)=1,
\]
so integrating out $U_e$ gives exactly
\[
\boxed{S_{\rm eff}=S_{\rm rest}-\log\frac{2I_1(\beta r)}{\beta r}.}
\]
The conditional law of $U_e$ given the other links has mean $\E[U_e\mid\text{rest}]=\dfrac{I_2(\beta r)}{I_1(\beta r)}\,V^\dagger$ (and $0$ if $r=0$). Moreover
\[
r^2=\tfrac12\sum_{i,j}\Tr(S_iS_j^\dagger)=6+\sum_{i<j}\Tr(S_iS_j^\dagger)=6+2\sum_{i<j}W_{ij},
\]
where $W_{ij}$ is the normalized Wilson loop of the six-link loop $S_iS_j^\dagger$. Three of these fifteen loops are planar $1\times2$ rectangles, from opposite staples in one plane. The other twelve are bent loops around two plaquettes in orthogonal planes sharing $e$.
\end{proposition}
\begin{proof}
A real combination of unit quaternions is a quaternion, so $K=rV$. Right invariance $U\mapsto UV^\dagger$ turns the integrand into the class function $e^{(\beta r/2)\Tr U}$. Weyl's formula, with class density $\tfrac2\pi\sin^2\vartheta$ and $\Tr U=2\cos\vartheta$, gives the first integral. The integral representation of $I_1$ \cite[10.32.2]{DLMF} gives $J=2I_1/z$. For the mean, $U'=UV$ has a conjugation-invariant law, so $\E U'$ commutes with $SU(2)$ and is a scalar $c$. Here $c=\E\tfrac12\Tr U'=J'(z)/J(z)$, and $J'=2I_2/z$ because $(z^{-1}I_1)'=z^{-1}I_2$ \cite[10.29.4]{DLMF}. The identity for $r^2$ uses $\Tr(S_jS_i^\dagger)=\overline{\Tr(S_iS_j^\dagger)}=\Tr(S_iS_j^\dagger)$, which is real for $SU(2)$. The loop $S_iS_j^\dagger$ runs from $x+\mu$ through staple $i$ to $x$, then back through staple $j$.
\end{proof}

\begin{lemma}[The generated pair interactions have nonnegative expectation]\label{traj:lem:pairs}
On an even periodic box, $\langle W_{ij}\rangle\ge0$ for all fifteen pairs. Hence $\langle r^2\rangle\ge6$ for every $\beta\ge0$.
\end{lemma}
\begin{proof}
Each six-link loop is mapped to itself by the link reflection that bisects $e$. That reflection exchanges $x$ and $x+\mu$, fixing both staple planes. After the two crossing $\mu$-links are set to $1$, $W_{ij}=\tfrac12\Tr(Y\,(\theta Y)^\dagger)=\tfrac12\sum_{ab}Y_{ab}\,\theta Y_{ab}$. Here $Y$ is the two-link path on the $x+\mu$ side. Apply \eqref{traj:eq:rp}.
\end{proof}

\begin{lemma}[An exact plaquette identity and a volume-uniform lower bound]\label{traj:lem:plaq}
On an even periodic box, for every link $e$,
\[
6\langle P\rangle_\beta=\Bigl\langle r_e\,\frac{I_2(\beta r_e)}{I_1(\beta r_e)}\Bigr\rangle_\beta,
\qquad\text{and}\qquad
\langle P\rangle_\beta\ge\frac\beta4\Bigl(1-\frac{9\beta^2}2\Bigr)\quad(0\le\beta\le\sqrt2/3).
\]
\end{lemma}
\begin{proof}
$\sum_{p\ni e}P_p=\tfrac12\ReTr(U_eK)$, and Proposition~\ref{traj:prop:link} gives $\E[\tfrac12\ReTr(U_eK)\mid\text{rest}]=r\,I_2(\beta r)/I_1(\beta r)$. By symmetry the left side has expectation $6\langle P\rangle$. For the bound, write $t=z^2/4$ and $a_m=1/(m!(m+1)!)$. Then
\[
\frac{I_2(z)}{I_1(z)}=\frac z4\cdot\frac{\sum_ma_m\frac2{m+2}t^m}{\sum_ma_mt^m}.
\]
The difference of the two series is at most $\sum_{m\ge1}a_mt^m\le\tfrac t2\sum_ma_mt^m$, since $a_{m+1}\le a_m/2$. Hence $I_2/I_1\ge\tfrac z4(1-z^2/8)$. Since $0\le r\le6$,
\[
6\langle P\rangle\ge\tfrac\beta4\bigl(1-\tfrac{36\beta^2}8\bigr)\langle r^2\rangle\ge\tfrac{6\beta}4\bigl(1-\tfrac{9\beta^2}2\bigr),
\]
using Lemma~\ref{traj:lem:pairs}.
\end{proof}
The bound is sharp at leading order, since $\langle P\rangle=\beta/4+O(\beta^2)$, and it is uniform in the box. It is what makes Proposition~\ref{traj:prop:start} explicit. The identity $6\langle P\rangle=\langle rI_2/I_1\rangle$ was also checked against heat-bath data: $3.6150(14)$ versus $3.6152(14)$ on $8^4$ at $\beta=2.3$ (the numerical evidence in Section~\ref{sec:empirical}).

\paragraph{Small-$z$ and large-$z$ forms are not uniform.} The exact series are
\[
J=1+\frac{z^2}8+\frac{z^4}{192}+\frac{z^6}{9216}+\cdots,\qquad
-\log J=-\frac{z^2}8+\frac{z^4}{384}-\frac{z^6}{9216}+\frac{z^8}{184320}-\cdots .
\]
Thus $-\log J(\beta r)=-\tfrac34\beta^2-\tfrac{\beta^2}4\sum_{i<j}W_{ij}+O(\beta^4r^4)$. The fourth-order term contains products $W_{ij}W_{kl}$ of loops. Along the trajectory, however, $z=\beta r$ ranges over $[0,6\beta]$. The quartic truncation gives $-1.333$, $+2.667$ and $+103.7$ at $z=4,8,15$, against the exact values $-1.585$, $-4.605$ and $-10.686$. Conversely, the large-$z$ form $-\log J\approx-z+\tfrac32\log z+\text{const}$ fails near $r=0$. Such configurations are allowed: the eighteen links of the six staples are distinct, so the staples are independent and $S_2=-S_1$, $S_4=-S_3$, $S_6=-S_5$ gives $K=0$. Only the exact Bessel form is uniform.

\begin{proposition}[Midpoint integration of a split coarse link]\label{traj:prop:mid}
Let a coarse link $e$ be split as $U_{e,1}=Q$, $U_{e,2}=Q^{-1}V_e$. Let $K_1,K_2$ be the staple sums of the two fine links. At fixed $V_e$ and fixed other links,
\[
\int\dd Q\;e^{\frac\beta2\ReTr(QK_1)+\frac\beta2\ReTr(Q^{-1}V_eK_2)}=J(\beta\tilde r),\qquad
\tilde r=\bigl|K_1+K_2^\dagger V_e^\dagger\bigr|,
\]
a sum of twelve $SU(2)$ matrices, and $\tilde r^2=12+\sum_{i<j}\Tr(X_iX_j^\dagger)$. The cross terms $\Tr(S^{(1)}_iV_eS^{(2)}_j)$ are loops through the coarse link.
\end{proposition}
\begin{proof}
No plaquette contains both halves, and $\ReTr(Q^{-1}X)=\ReTr(QX^\dagger)$. Then apply Proposition~\ref{traj:prop:link}.
\end{proof}
Different midpoints $Q_e,Q_{e'}$ meeting at a coarse site share fine plaquettes. For example, the fine plaquette at $x$ in the $(\mu,\nu)$ plane contains $U_{e,1}$ and $U_{e',1}$. The complete blocking integral therefore does \emph{not} factorize into Bessel functions. Only one variable at a time can be integrated in closed form, after which the action is no longer of Wilson type.

\subsection{Exact blocking and the defect $D_k$}
Identify the coarse torus $\Lambda_k$ with the even sites of $\Lambda_{k+1}$. For a coarse link $e$, let $V_e=U_{e,1}U_{e,2}$, which defines $B:U\mapsto V$. The \emph{fibre variables} are the midpoints $Q_e=U_{e,1}$ and all fine links that do not lie on coarse links. The change of variables $(U_1,U_2)\mapsto(Q,V)=(U_1,U_1U_2)$ preserves Haar$\times$Haar. Hence $B_*(\text{Haar})=\text{Haar}$, and given $B(U)=V$ the fibre variables are independent and Haar distributed. Define
\[
e^{-S^B_{k+1}(V)}=\E_{\rm Haar}\bigl[e^{-S_{\beta_{k+1}}(U)}\bigm|B(U)=V\bigr],\qquad D_k=S^B_{k+1}-S_{\beta_k}.
\]

\begin{proposition}[Exact representation and structure of $D_k$]\label{traj:prop:Dk}
Let $\widetilde D_k(U):=S_{\beta_{k+1}}(U)-S_{\beta_k}(B(U))$ on the fine torus. Then:
\begin{enumerate}[label=(\alph*)]
\item $\displaystyle D_k(V)=-\log\E_{\rm Haar}\bigl[e^{-\widetilde D_k(U)}\bigm|B(U)=V\bigr]$ exactly;
\item $\widetilde D_k$ is an explicit sum of bounded local terms,
\[
\widetilde D_k=\sum_{P}\Psi_P+\sum_{p\in\mathcal I}\beta_{k+1}\bigl(1-\tfrac12\Tr U_p\bigr),\qquad
\Psi_P=\beta_{k+1}\sum_{p\subset P}\bigl(1-\tfrac12\Tr U_p\bigr)-\beta_k\bigl(1-\tfrac12\Tr V_P\bigr),
\]
where $P$ runs over coarse plaquettes, $V_P$ is the eight-link fine loop around $P$, and $\mathcal I$ is the set of fine plaquettes contained in no coarse plaquette. Per coarse cell there are $96$ fine plaquettes, $24$ in coarse plaquettes and $72$ in $\mathcal I$;
\item $S^B_{k+1}$ is invariant under coarse gauge transformations and under the symmetries of the coarse torus, and $B_*\mu_{k+1}$ is reflection positive for coarse site and link reflections;
\item $0\le S^B_{k+1}\le2\beta_{k+1}n_p^{\rm fine}$. Thus $D_k$ is bounded, but only by a quantity proportional to the volume;
\item $\langle W_C\rangle$ under $e^{-S^B_{k+1}}$ equals $w_{k+1}(C)$ for every loop following coarse links.
\end{enumerate}
\end{proposition}
\begin{proof}
(a) $S_{\beta_k}\circ B$ is constant on each fibre. (b) A fine plaquette lies in a coarse plaquette exactly when its two transverse coordinates are even. (c) $B$ commutes with gauge transformations at even sites, and the odd-site gauge freedom is integrated out by Haar invariance. Coarse site planes are fine site planes at even coordinates. Coarse link planes are fine site planes at odd coordinates. In both cases $F\in\mathcal A^{\rm coarse}_+$ implies $F\circ B\in\mathcal A^{\rm fine}_+$ and $\theta(F\circ B)=(\theta F)\circ B$. Hence $\langle F\theta F\rangle_{B_*\mu}=\langle(F\circ B)\theta(F\circ B)\rangle_\mu\ge0$. (d) is immediate from $0\le S\le2\beta n_p$. (e) The holonomies agree exactly.
\end{proof}

\begin{remark}[Why this is not yet a controlled expansion]
The representation (a) is exact, but it does not provide an interaction range bounded independently of the box for $S^B_{k+1}$. The first midpoint integration already produces $-\log J(\beta\tilde r)$ of twelve-staple sums (Proposition~\ref{traj:prop:mid}). Iterating can produce interactions on larger connected sets; their decay must be controlled separately. A convergent polymer expansion of $-\log\E[e^{-\widetilde D_k}\mid B]$ can be sought in a sufficiently small-coupling domain. The definition of the trajectory does not prove either that it stays in such a domain or that it escapes to weak coupling (Remark~\ref{traj:rem:traj}). On the weak-coupling side, Ba\l aban's renormalization-group programme provides localized effective-action and coupling-renormalization estimates in the small-field regime of four-dimensional lattice gauge theory \cite{Balaban4,conf:balaban}. It uses its own averaging blocking and gauge fixing, and it does not address the large-loop ratio required in the scale comparison. Its conclusions are not assumed to control the loop ratios below. Replacing $D_k$ by a shift of one plaquette coupling is therefore an uncontrolled approximation in four dimensions. The next proposition shows exactly what such a replacement does in the one case where everything is computable.
\end{remark}

\subsubsection{An exactly solvable control case: two dimensions}
In two dimensions with free boundary conditions, after maximal-tree gauge fixing, the plaquette holonomies are independent with density $\propto e^{(\beta/2)\Tr U}$. The character coefficients are
\[
\int e^{(\beta/2)\Tr U}\chi_j(U)\,\dd U=(2j+1)c_j(\beta),\qquad c_j=\frac{2I_{2j+1}(\beta)}\beta,\qquad x_j:=\frac{c_j}{c_0}=\frac{I_{2j+1}}{I_1}.
\]
A $2\times2$ blocked plaquette is a product of four independent fine plaquettes, up to conjugation. Hence $\E[D^j(U_p)]=x_j\mathbf 1$ for each fine plaquette and the blocked density has coefficients $x_j(\beta_{k+1})^4$.

\begin{proposition}[Two-dimensional exact blocking]\label{traj:prop:2d}
In two dimensions, $w_k(C)=x_{1/2}(\beta_k)^{A(C)/a_k^2}$. The condition $\chi_k=c_*$ reads $x_{1/2}(\beta_k)=e^{-c_*4^{-k}}$. Since $x_{1/2}(\beta)$ is the single-plaquette expectation of $\tfrac12\Tr U$, its derivative is the corresponding variance; hence it is strictly increasing and the condition has a unique solution, and $\beta_k\sim\tfrac32 4^k/c_*$. Along it, for every fundamental rectangle,
\[
\mathcal R_k(C)\equiv1,\qquad\text{so (A) holds with }\eta_k=u_k=v_k=0 .
\]
The exact defect is local but not of Wilson form:
\[
D_k(V)=\sum_P\Bigl[-\log\sum_j(2j+1)y_j\chi_j(V_P)-\beta_k\bigl(1-\tfrac12\Tr V_P\bigr)\Bigr]+\text{const},\qquad y_j=x_j(\beta_{k+1})^4 .
\]
For $c_*=0.9$ and $k=1$, the bracket varies by about $12$ between $\vartheta=0$ and $\vartheta=3.1$; its values are recorded in the numerical evidence in Section~\ref{sec:empirical}.
The Wilson-form replacement is exact in the fundamental sector by construction. In the adjoint sector its error per coarse plaquette is $\delta_k=\log x_1(\beta_k)-4\log x_1(\beta_{k+1})$, with the exact limit $\delta_k\beta_k^3\to25/16$, proved below.
\end{proposition}
\begin{proof}
Independence gives $w=x_{1/2}^{\#\text{plaquettes}}$. With $n=2^k$, $\chi_k=-n^2\log x_{1/2}(\beta_k)$. Therefore $x_{1/2}(\beta_{k+1})^4=x_{1/2}(\beta_k)$, which is exactly the statement that the blocked fundamental coefficient equals the Wilson one. Consequently $w_{k+1}(C)=w_k(C)$. The large-$\beta$ form follows from $\log x_{1/2}=-\tfrac3{2\beta}+O(\beta^{-2})$.
\end{proof}

\begin{proposition}[Exact leading mismatch for every fixed representation]
\label{traj:prop:representation-mismatch}
Let $j\in\{0,\tfrac12,1,\ldots\}$ be fixed and $C_j=j(j+1)$. Set
$x_j(\beta)=I_{2j+1}(\beta)/I_1(\beta)$ and choose $\beta^+$ by
$x_{1/2}(\beta^+)^4=x_{1/2}(\beta)$. Then, as $\beta\to\infty$,
\[
\log x_j(\beta)-4\log x_j(\beta^+)
=\frac{5C_j(4C_j-3)}{32\beta^3}+O_j(\beta^{-4}).
\]
In particular the adjoint coefficient is exactly $25/16$. For $j=0$ and
$j=1/2$ the mismatch vanishes identically. The estimate is for each fixed
representation; no uniform assertion as $j\to\infty$ is made.
\end{proposition}
\begin{proof}
Write $z=1/\beta$. The fixed-order large-argument Bessel expansion
\cite[\S10.40(i)]{DLMF} has coefficients
\[
 I_\nu(\beta)=\frac{e^\beta}{\sqrt{2\pi\beta}}
 \left\{\sum_{r=0}^{3}
 \frac{(-1)^r\prod_{q=1}^{r}(4\nu^2-(2q-1)^2)}{r!\,8^r}\,z^r
 +O_\nu(z^4)\right\}.
\]
Taking the logarithm and subtracting the $\nu=1$ expression gives
\[
 \log x_j(\beta)=-2C_jz-C_jz^2+
 \frac{C_j(8C_j-9)}{12}z^3+O_j(z^4).
\]
For the fundamental representation this is
$-3z/2-3z^2/4-3z^3/16+O(z^4)$.
The leading derivative is nonzero, so asymptotic inversion of the matching
relation (or substitution followed by the mean-value theorem) yields
\[
 z^+:=(\beta^+)^{-1}=\frac z4+\frac{3z^2}{32}
 +\frac{3z^3}{512}+O(z^4).
\]
Substitute this into the general expression. The linear and quadratic
terms cancel. The remaining cubic coefficient is
\[
 \frac{15}{16}\frac{C_j(8C_j-9)}{12}
 +\frac{3C_j}{64}+\frac{3C_j}{16}
 =\frac{5C_j(4C_j-3)}{32}.
\]
The remainders stay $O_j(z^4)$ under this composition. The exact
zero statements follow from $x_0=1$ and the defining matching condition.
\end{proof}
The accompanying rational calculation checks the coefficient algebra for
$j=0,\tfrac12,\ldots,6$. The proof applies to every fixed $j$. This
refines the numerical limit in the two-dimensional benchmark; it does not
supply a four-dimensional blocking estimate.

Table~\ref{traj:tab:2d} lists the computed trajectory for $c_*=0.9$. the numerical evidence in Section~\ref{sec:empirical} reports a Monte Carlo confirmation of the fourth-power law at $\beta=2.6$: the blocked plaquette is $0.0729(8)$, versus $x^4=0.0733$. Two lessons transfer, and one does not. First, a Wilson-form replacement can be exact for the renormalized observable while wrong elsewhere. For an adjoint loop of fixed physical area the accumulated error $A4^k\delta_k/\ell^2=O(4^{-2k})$ is summable. Second, positivity of every quantity is automatic here. What does \emph{not} transfer: the two-dimensional trajectory has $\beta_k\propto4^k$ because $g^2$ has mass dimension two. It gives no information about the logarithmic four-dimensional budget.

\begin{table}[h]\centering\small
\begin{tabular}{@{}rrrrr@{}}
\toprule
$k$ & $\beta_k$ & $\beta_{k+1}$ & adjoint mismatch $\delta_k$ & $\delta_k\beta_k^3$\\\midrule
0 & 1.8415 & 7.1383 & $2.260\times10^{-1}$ & 1.412\\
1 & 7.1383 & 27.1615 & $6.406\times10^{-3}$ & 2.330\\
2 & 27.1615 & 107.1655 & $8.728\times10^{-5}$ & 1.749\\
3 & 107.1655 & 427.1664 & $1.307\times10^{-6}$ & 1.608\\
4 & 427.1664 & 1707.1666 & $2.019\times10^{-8}$ & 1.574\\
5 & 1707.1666 & 6827.1666 & $3.146\times10^{-10}$ & 1.565\\
6 & 6827.1666 & 27307.1667 & $4.912\times10^{-12}$ & 1.563\\\bottomrule
\end{tabular}
\caption{Exact two-dimensional trajectory for $c_*=0.9$ and the error of the one-coupling replacement in the adjoint sector. The fundamental sector is matched exactly at every step.}\label{traj:tab:2d}
\end{table}

\subsection{The quantum ratio}
Because $e^{-S^B_{k+1}}=e^{-S_{\beta_k}-D_k}$ on coarse variables, Proposition~\ref{traj:prop:Dk}(e) gives the exact reweighting
\[
w_{k+1}(C)=\frac{\langle W_Ce^{-D_k}\rangle_k}{\langle e^{-D_k}\rangle_k},\qquad
\mathcal R_k(C)=\frac{\langle W_Ce^{-D_k}\rangle_k}{\langle W_C\rangle_k\langle e^{-D_k}\rangle_k},\qquad
F_{k+1}(C)-F_k(C)=-\log\mathcal R_k(C).
\]
\textbf{Positivity at both endpoints.} A physical rectangle with sides $R,T\le L/2$ is a lattice rectangle at levels $k$ and $k+1$, with sides at most half the torus. Lemma~\ref{traj:lem:pos} therefore gives $w_k(C)>0$ and $w_{k+1}(C)>0$ for every $\beta>0$, in every enclosing box. All logarithms above are defined.

\textbf{Renormalized ratio.} With the convention \eqref{traj:eq:lambda},
\[
\widehat{\mathcal R}_k(C)=e^{(\lambda_{k+1}-\lambda_k)P(C)+\kappa_{k+1}-\kappa_k}\mathcal R_k(C),\qquad
\widehat F_{k+1}-\widehat F_k=-\log\widehat{\mathcal R}_k,
\]
and $\widehat{\mathcal R}_k(C)=1$ exactly on the three reference loops in the reference torus used to determine the parameters. This matching does not automatically persist in larger enclosing tori with those parameters fixed. Condition (A) requires control beyond these three reference-box equalities, including its stated uniformity over loops and enclosing volumes.

\textbf{The two open inputs, as used below.} With nonnegative constants $\eta_k\le\rho<1$ and $\sum_k(\eta_k+u_k+v_k)<\infty$, uniform over all sufficiently large rectangles and all enclosing boxes,
\begin{align*}
\text{(A)}\quad \log\widehat{\mathcal R}_k(C)
 &\le\eta_k\widehat F_k(C)+u_kP(C)+v_k,\\
\text{equivalently}\quad \widehat F_{k+1}(C)
 &\ge(1-\eta_k)\widehat F_k(C)-u_kP(C)-v_k.
\end{align*}
\[
\text{(B)}\quad \widehat F_0(C)\ge\sigma_0A(C)-B_0P(C)-C_0,\qquad\sigma_0>0 .
\]
The multiplicative form of (A) matters for the asymptotic-coefficient reduction. An additive bound $\widehat F_{k+1}\ge\widehat F_k-\eta_kA-\cdots$ would instead give $\Sigma_{k+1}\ge\Sigma_k-\eta_k$, and it would need $\sum\eta_k<\Sigma_0$.

\subsection{The scale-comparison ratio}
\subsubsection{An interpolation with a local defect along which positivity is proved}
The linear coarse path $S_k+sD_k$ has two drawbacks. Its defect $D_k$ is only implicitly defined, and positivity of $\langle W_C\rangle_{k,s}$ for $0<s<1$ does not follow from endpoint positivity. We know of no proof that $e^{-sD_k}$ preserves reflection positivity. The following lifted path removes both problems.

\begin{theorem}[Reflection-positive lifted interpolation]\label{traj:thm:lift}
On the fine torus $\Lambda_{k+1}$ define, for $s\in[0,1]$,
\[
\dd\nu_s(U)\propto\exp\bigl\{-(1-s)S_{\beta_k}(B(U))-s\,S_{\beta_{k+1}}(U)\bigr\}\prod_e\dd U_e .
\]
\begin{enumerate}[label=(\alph*)]
\item $B_*\nu_0=\mu_{k,\beta_k}$ and $\nu_1=\mu_{k+1,\beta_{k+1}}$. Hence for every loop $C$ following coarse links, $\nu_0(W_C)=w_k(C)$ and $\nu_1(W_C)=w_{k+1}(C)$.
\item $\nu_s$ is reflection positive for every reflection in a fine site hyperplane, at even and at odd coordinates.
\item For every $s\in[0,1]$ and every coarse-aligned rectangle with coarse sides at most $N_k/2$, $\nu_s(W_C)>0$.
\item $F_{k,s}(C):=-\log\nu_s(W_C)$ is real-analytic on $[0,1]$, and
\[
\frac{\dd F_{k,s}(C)}{\dd s}=\Gamma_s(C):=\sum_X\Bigl[\frac{\nu_s(W_C\Phi_X)}{\nu_s(W_C)}-\nu_s(\Phi_X)\Bigr],
\qquad
-\log\mathcal R_k(C)=\int_0^1\Gamma_s(C)\,\dd s,
\]
where $\Phi_X$ runs over the local terms $\Psi_P$ and $\beta_{k+1}(1-\tfrac12\Tr U_p)$, $p\in\mathcal I$, of Proposition~\ref{traj:prop:Dk}(b), each bounded by $2\beta_{k+1}$ or $8\beta_{k+1}+2\beta_k$.
\item The coarse marginal of $\nu_s$ has unnormalized density $e^{-S_{\beta_k}}\E[e^{-s\widetilde D_k}\mid B]\le e^{-S_{\beta_k}-sD_k}$, with equality at $s=0,1$. The lifted path and the linear coarse path share their endpoints.
\end{enumerate}
\end{theorem}
\begin{proof}
(a) Fine Haar measure pushes forward to coarse Haar measure, and the $s=0$ weight depends on $U$ only through $B(U)$. At $s=1$ the coarse loop is a fine loop.

(b) The factor $e^{-sS_{\beta_{k+1}}}$ is a Wilson weight, reflection positive for all site reflections \cite{OSLattice}. For $e^{-(1-s)S_{\beta_k}\circ B}=\prod_P\exp\{\lambda(\tfrac12\Tr V_P-1)\}$, $\lambda=(1-s)\beta_k$, consider first a plane at an even coordinate. Every coarse plaquette lies in one closed half, so the weight factorizes as $e^{-A_\Pi}\,e^{-A_+}\,\theta e^{-A_+}$. For a plane at an odd coordinate $h$, coarse plaquettes either lie in one closed half or are bisected. A bisected $P$ satisfies $\tfrac12\Tr V_P=\tfrac12\Tr(Y_P(\theta Y_P)^\dagger)$, where $Y_P$ is its four-link half on the upper side, joining the two points where $P$ meets the plane. The character expansion
\[
e^{\frac\lambda2\Tr(Y(\theta Y)^\dagger)}=\sum_jd_j(\lambda)\chi_j\bigl(Y(\theta Y)^\dagger\bigr)=\sum_jd_j(\lambda)\sum_{m,n}D^j_{mn}(Y)\,\theta\bigl(D^j_{mn}(Y)\bigr),
\]
with $d_j=(2j+1)\cdot2I_{2j+1}(\lambda)/\lambda\ge0$, writes each crossing factor as $\sum_\alpha a_\alpha G_\alpha\,\theta G_\alpha$ with $a_\alpha\ge0$ and $G_\alpha\in\mathcal A_+$. Finite products of densities of the form $e^{-A_\Pi}\sum_\alpha a_\alpha(G_\alpha e^{-A_+})\,\theta(G_\alpha e^{-A_+})$ have the same form. For such a density $\rho$, conditioning on the links of the plane gives
\[
\int F\,\theta F\,\rho=\sum_\alpha a_\alpha\int\dd U_\Pi\,e^{-A_\Pi}\Bigl|\int\dd U_+\,F\,G_\alpha e^{-A_+}\Bigr|^2\ge0 .
\]

(c) First the coarse plaquette. Let $\nu_{\lambda,s}\propto\exp\{-\lambda\sum_P(1-\tfrac12\Tr V_P)-s\beta_{k+1}\sum_p(1-\tfrac12\Tr U_p)\}$. It is invariant under coarse translations and the hypercubic group fixing a coarse site, so all coarse plaquettes have a common expectation $f(\lambda)$. Then $f'(\lambda)=\Var(\sum_P\tfrac12\Tr V_P)/n_P>0$. If $s>0$, then $f(0)=w_{2\times2}$ for the fine Wilson measure at $s\beta_{k+1}$, which is $>0$ by Lemma~\ref{traj:lem:pos}. If $s=0$, then $f(0)=0$ and $f$ is strictly increasing. Either way $\nu_s(V_P)=f((1-s)\beta_k)>0$. Next, propagate. A coarse rectangle of fine height $2T'$ occupies $[0,2T']$. Reflecting in the fine site plane at height $h\in\{0,\dots,2T'\}$ doubles its two parts into rectangles occupying $[2h-2T',2T']$ and $[0,2h]$. Both have even corners, so they are coarse rectangles of coarse heights $2T'-h$ and $h$. By (b) and the argument of Lemma~\ref{traj:lem:pos}(a,c), positivity of coarse height $T'$ implies positivity of every coarse height $\le2T'$. The same holds in the other direction.

(d) Finite-dimensional compact integrals of smooth functions; $\frac{\dd}{\dd s}S_s=\widetilde D_k$ and $\nu_s(W_C)>0$.

(e) The unnormalized coarse marginal density is $e^{-S_{\beta_k}}\E[e^{-s\widetilde D_k}\mid B]$, and Jensen's inequality for $x\mapsto x^s$ gives the bound by $e^{-S_{\beta_k}}(\E[e^{-\widetilde D_k}\mid B])^s$.
\end{proof}

\note{\textbf{What Theorem~\ref{traj:thm:lift} settles.} The quantity in (A) is now an integral of an explicit local derivative over a path of reflection-positive measures, with strictly positive loop expectations throughout. This removes the positivity obstruction raised in the scale comparison, for this path. It does \emph{not} bound $\Gamma_s$. Each $\Phi_X$ is of order $\beta$ and there are of order $|\Lambda|$ of them. Without cancellation, the trivial bound $|\Gamma_s|\le(1+\nu_s(|W_C|)/\nu_s(W_C))\sum_X\|\Phi_X\|_\infty$ is a bulk estimate with an additional signed-tilt amplification factor.}

\subsubsection{The Wilson-tilted functional is far from a probability measure}
\begin{lemma}[Total variation of the tilt]\label{traj:lem:tv}
Let $\mu$ be the Wilson measure at any level, or $\nu_s$. Let $C$ be a rectangle as in Lemma~\ref{traj:lem:pos} (respectively Theorem~\ref{traj:thm:lift}(c)). Then
\[
\int|W_C|\,\dd\mu\ \ge\ \int W_C^2\,\dd\mu\ =\ \tfrac14\bigl(1+\mu(\chi_1(U_C))\bigr)\ \ge\ \tfrac14,
\qquad\text{hence}\qquad
\frac{\|W_C\,\mu\|_{\rm TV}}{\mu(W_C)}\ \ge\ \frac{e^{F(C)}}4 .
\]
\end{lemma}
\begin{proof}
$|W_C|\le1$ gives $|W_C|\ge W_C^2$, and $\chi_{1/2}^2=\chi_0+\chi_1$. The adjoint loop satisfies $\chi_1(U_C)=\sum_{ab}D^1(M(P_+))_{ab}\,\theta D^1(M(P_+))_{ab}$ for the reflection in the symmetry plane of $C$, as in Lemma~\ref{traj:lem:pos}. Hence $\mu(\chi_1(U_C))\ge0$ by \eqref{traj:eq:rp}.
\end{proof}
Thus the signed measure $W_C\mu/\mu(W_C)$, whose expectations enter $\Gamma_s$, has total variation at least $e^{F(C)}/4$. Under an area law this is exponentially large in the area. Any estimate of $\Gamma_s$ that treats the tilt as a bounded perturbation of a probability measure loses the entire signal. Heat-bath data agree: on $8^4$ at $\beta=2.3$ the ratio $\langle|W|\rangle/\langle W\rangle$ for the $4\times4$ loop is $85.8$, against the bound $50.6$ (the numerical evidence in Section~\ref{sec:empirical}).

\begin{proposition}[Insertions in a reflection hyperplane]\label{traj:prop:plane}
Let $\mu$ be reflection positive for the site reflection in a hyperplane $\Pi$ perpendicular to the $T$-direction of $C$. Let $C$ have $a$ steps above $\Pi$ and $b$ below, with $a+b=T$. For every real $O$ depending only on links in $\Pi$,
\[
|\mu(W_CO)|\le\|O\|_\infty\sqrt{w(R,2a)\,w(R,2b)}\le\|O\|_\infty\,w(R,T)\,e^{f(R,T)-\frac12f(R,2T)},
\]
the last inequality when $2T\le N/2$. If $\Pi$ is the midplane of $C$ ($a=b$), then $|\mu(W_CO)|\le\|O\|_\infty\,\mu(W_C)$.
\end{proposition}
\begin{proof}
Assume $\|O\|_\infty\le1$ and put $\rho_\pm=1\pm O\ge0$. These are $\theta$-invariant and supported in $\Pi$. Hence $\sqrt{\rho_\pm}F\in\mathcal A_+$ whenever $F\in\mathcal A_+$, and $\langle F,G\rangle_\pm:=\mu(F\rho_\pm\,\theta G)$ are positive semidefinite Hermitian forms. With $F_{ij},G_{ij}$ as in Lemma~\ref{traj:lem:pos}, $W_CO=\tfrac14\sum_{ij}\bigl(F_{ij}\rho_+\theta G_{ij}-F_{ij}\rho_-\theta G_{ij}\bigr)$. Cauchy--Schwarz in each form and then in the sum over $(i,j,\pm)$ gives
\[
|\mu(W_CO)|\le\tfrac14\Bigl(\sum_{ij,\pm}\langle F_{ij},F_{ij}\rangle_\pm\Bigr)^{1/2}\Bigl(\sum_{ij,\pm}\langle G_{ij},G_{ij}\rangle_\pm\Bigr)^{1/2}=\sqrt{w(R,2a)\,w(R,2b)} ,
\]
because $\sum_\pm\langle\cdot,\cdot\rangle_\pm=2\mu(\cdot\,\theta\,\cdot)$ and $\sum_{ij}\mu(F_{ij}\theta F_{ij})=2w(R,2a)$. For the last inequality, $f(R,\cdot)$ is concave with $f(R,0)=0$. For the Wilson measure this is Lemma~\ref{traj:lem:pos}(b). For $\nu_s$, fine site reflections of both parities give $w(T')^2\le w(h)\,w(2T'-h)$ in coarse units, hence concavity in coarse units. Therefore $f(R,2a)+f(R,2b)\ge\frac aTf(R,2T)+\frac bTf(R,2T)=f(R,2T)$.
\end{proof}
So on the midplane of the loop the tilt acts as a contraction, exactly as a probability measure would. Off the midplane, the loss is controlled by the concavity defect $f(T)-\tfrac12f(2T)$. For an area law with perimeter, this defect is of order the perimeter term on the $R$-sides, so it is not uniform. Most terms of $\Gamma_s$ are supported off every symmetry plane of $C$. Even on the midplane, summing the individual bounds over the whole hyperplane produces a three-dimensional volume factor unless the tilted correlations decay away from $C$. That decay, a \emph{relative clustering} estimate for $\nu_s$, is not proved here.

\subsubsection{Testing proposed simplifications against the ratio}
\begin{enumerate}[label=(T\arabic*)]
\item \emph{Small-$\beta r$ expansion of the one-link action.} Not uniform along the trajectory: at $z=15$ the quartic truncation is off by $114$ (the one-link formula).
\item \emph{Large-$\beta r$ (saddle) expansion.} Not uniform: $r=0$ is attained. Discarding the small-$r$ region would require either its probability to be small \emph{relative to} $e^{-F(C)}$, or cancellation between the two terms of $\Gamma_s$. Neither is available.
\item \emph{Replacing $D_k$ by a one-coupling shift.} In two dimensions this is exact for fundamental loops (Proposition~\ref{traj:prop:2d}). In four dimensions the replacement error $\delta D=\sum_X\delta\Phi_X$ contributes $\sum_X[\nu(W_C\delta\Phi_X)/\nu(W_C)-\nu(\delta\Phi_X)]$, and the only general bound is $(1+\nu(|W_C|)/\nu(W_C))\sum_X\|\delta\Phi_X\|_\infty$, proportional to the volume.
\item \emph{Bounding $\Gamma_s$ by the size of the defect.} For smooth fields, $\sum_{p\subset\text{cell}}(1-\tfrac12\Tr U_p)$ at spacing $a_{k+1}$ and $\sum_P(1-\tfrac12\Tr V_P)$ at spacing $a_k$ agree to leading order, so the classical part of $\widetilde D_k$ is $(\beta_{k+1}-\beta_k)$ times the fine action density. This is a heuristic statement. Relative to the action, this is of order $(\beta_{k+1}-\beta_k)/\beta_{k+1}\approx2b_0\log2\,g_k^2$. Since $\sum_kg_k^2$ diverges logarithmically when $g_k^2\sim1/(\alpha+\gamma k)$, no bound proportional to the size of $\widetilde D_k$ can meet the budget $\eta_k\le Cg_k^4$. A proof of (A) must use the cancellation enforced by the renormalization condition, which removes this leading term for the reference loops by construction.
\item \emph{Absolute instead of relative errors.} Already refuted in Section~\ref{conf:transport_obstructions}. Lemma~\ref{traj:lem:tv} quantifies the loss as $e^{F}/4$.
\item \emph{Inferring large-loop behaviour from finite loops.} Theorem~\ref{traj:thm:mono} below shows that finite-loop data give only \emph{upper} bounds on the asymptotic area coefficient. No finite-size check can establish (A) or (B).
\end{enumerate}

\note{\textbf{Status of (A).} No inequality of type (A) is proved here for any $k$. What would suffice is a one-sided, area-proportional bound on $\int_0^1\Gamma_s\,\dd s$ for large loops. By the asymptotic-coefficient reduction, it is enough to have it for the asymptotic area coefficient only. In the crossover region ($\beta$ between the strong-coupling radius and the onset of asymptotic scaling), no rigorous technique for such a bound is known to us. On the weak-coupling side it would require relative clustering of the reflection-positive family $\nu_s$, with constants uniform in the loop size.}

\subsection{A reduction of (A) and (B) to one number per level}
\begin{theorem}[RP monotonicity; classical]\label{traj:thm:mono}
Let $N\ge4$ be even and $\beta>0$, and let $1\le R,T\le N/2$ (lattice units).
\begin{enumerate}[label=(\alph*)]
\item $T\mapsto f(R,T)$ is concave on $\{0,\dots,N/2\}$ with $f(R,0)=0$, and likewise in $R$.
\item $\varphi(R,T):=f(R,T)/(RT)$ is nonincreasing in $R$ and in $T$.
\item With the half-box value $\sigma_N:=f(N/2,N/2)/(N/2)^2$,
\[
f(R,T)\ge\sigma_N\,RT\qquad\text{for every rectangle in the box}.
\]
No perimeter or constant term appears.
\item Let $w_\infty$ be any local limit of the periodic-box loop expectations along a subsequence $N_j\to\infty$. Then (a)--(b) hold for $f_\infty$ on $\{0,1,2,\dots\}$. The limits
\[
\Sigma:=\lim_{R,T\to\infty}\frac{f_\infty(R,T)}{RT}=\inf_{R,T}\frac{f_\infty(R,T)}{RT},\qquad V(R):=\lim_{T\to\infty}\frac{f_\infty(R,T)}T
\]
exist. $V$ is concave and nondecreasing, and for all $R\ge R'\ge0$, $T\ge1$,
\[
f_\infty(R,T)\ge\Sigma\,RT,\qquad f_\infty(R,T)\ge V(R)\,T,\qquad V(R)-V(R')\ge\Sigma\,(R-R').
\]
\end{enumerate}
\end{theorem}
\begin{proof}
(a) is Lemma~\ref{traj:lem:pos}(b). (b) A concave function with $f(0)=0$ has $f(T)/T$ nonincreasing; apply this in each variable. (c) The minimum of $\varphi$ over $[1,N/2]^2$ is at the corner. (d) The limiting loop expectations are strictly positive; the asymptotic area coefficient is allowed to be zero. Concavity gives $f_N(R,T)\le RT\,f_N(1,1)$, so $w_N(R,T)\ge\langle P\rangle_{N,\beta}^{RT}$. Since $\langle P\rangle$ increases in $\beta$, Lemma~\ref{traj:lem:plaq} at $\beta'=\min(\beta,0.3)$ gives $\langle P\rangle_{N,\beta}\ge\tfrac{\beta'}4(1-\tfrac92\beta'^2)>0$ uniformly in $N$. Hence $f_\infty$ is finite. The finitely many inequalities used in (a) pass to local limits. Monotone bounded sequences converge, and the double infimum of a function nonincreasing in each variable equals both iterated limits. $V$ is a pointwise limit of concave functions of $R$, hence concave, with $V(0)=0$. Its increments $s_R=V(R+1)-V(R)$ are nonincreasing with limit $s_\infty$. Then $V(R)/R\to s_\infty$, so $s_\infty=\lim_RV(R)/R=\Sigma\ge0$. Hence $s_R\ge\Sigma$, which gives monotonicity and the last inequality.
\end{proof}
These are the RP consequences behind Seiler's linear upper bound \cite{traj:seiler78} and Bachas' concavity of the static potential \cite{traj:bachas}. Only the packaging in (c) and its use below are specific to this note. Heat-bath data on $8^4$ at $\beta=2.3$ satisfy all tested instances: for example $w(4,4)=0.00494(18)\ge w(4,2)^2=0.00198(3)$.

\paragraph{The asymptotic coefficient along the trajectory.} Choose one subsequence of enclosing boxes along which all loop expectations at all levels converge; a diagonal argument over countably many loops makes this possible. Let $\Sigma_k$ be the limit $\Sigma$ of Theorem~\ref{traj:thm:mono}(d) at level $k$, converted to physical units (divide by $a_k^2$). Equivalently, $\Sigma_k=\sigma_{\rm lat}(\beta_k)/a_k^2$, where $\sigma_{\rm lat}$ is the asymptotic lattice area coefficient for that limit.

\begin{corollary}[Consequences for asymptotic area coefficients]\label{traj:cor:reduce}
\begin{enumerate}[label=(\alph*)]
\item If (A) and (B) hold with constants uniform in the enclosing boxes, then
\[
\Sigma_{k+1}\ge(1-\eta_k)\,\Sigma_k\quad(\mathrm{A}'),\qquad \Sigma_0\ge\sigma_0>0\quad(\mathrm{B}').
\]
\item Consequences of the reduced statements: (A$'$) and (B$'$) with $0\le\eta_k\le\rho<1$ and $\sum\eta_k<\infty$ imply
\[
\Sigma_K\ge\sigma_*:=\Sigma_0\exp\Bigl[-\frac{\sum_k\eta_k}{1-\rho}\Bigr]>0\qquad\text{for all }K .
\]
Consequently, for every level $K$, every rectangle $C$ and the chosen infinite-volume limit,
\[
\boxed{F_{K,\infty}(C)\ge\sigma_*\,A(C),\qquad V_K(R)-V_K(R')\ge\sigma_*(R-R')\quad(R\ge R').}
\]
\item The finite-box form also holds. If $\sigma_{K,M'}:=F_{K,M'}(C_{\rm half})/A(C_{\rm half})$ for the half-box square, then $F_{K,M'}(C)\ge\sigma_{K,M'}A(C)$ for every rectangle with sides at most $L'/2$.
\end{enumerate}
\end{corollary}
\begin{proof}
(a) Pass to the infinite-volume limit in (A) for each fixed physical loop. Divide by $A(C)$, and let $R,T\to\infty$. Then $P/A\to0$ and $1/A\to0$, and $\widehat F_k/A\to\Sigma_k$ because the subtractions are of lower order. The same applies to (B). (b) Iterate and use $\log(1-\eta)\ge-\eta/(1-\rho)$. Then apply Theorem~\ref{traj:thm:mono}(d) at level $K$; differences of $V_K$ are unchanged by the perimeter subtraction $2\lambda_K$. (c) is Theorem~\ref{traj:thm:mono}(c).
\end{proof}

\note{\textbf{Consequences for the programme.} (i) For a bare regulated area bound, the asymptotic coefficient $\Sigma_k$ suffices; a finite box admits the half-box lower bound. These reduced statements do not imply the original finite-loop comparison (A), uniform half-box estimates, or convergence of renormalized continuum loops. Perimeter and corner control remains necessary for that continuum passage. (ii) The conclusion is \emph{stronger} than the target of the starting bound in one respect. The bare bound $F_{K,\infty}\ge\sigma_*A$ has no perimeter or constant term, and the potential bound needs no additive constant $B_*$. (iii) In dimensionless form, (A$'$) and (B$'$) say that $\rho_k:=\Sigma_k\ell^2/c_*$, the asymptotic area coefficient in units of the scale-$\ell$ Creutz ratio, is bounded away from $0$ along the trajectory. This is the most economical form of ``confinement survives the continuum limit'' in this framework. It is \emph{not} proved here. (iv) Finite data only bound $\Sigma_k$ from \emph{above}, since $\Sigma_k\le\varphi(R,T)/a_k^2$ for every finite loop. A lower bound on $\Sigma_k$ is genuinely asymptotic.}

\subsection{The starting bound}
\textbf{Strong-coupling input (SC).} There are $\beta_{\rm sc}\in(0,\sqrt2/3)$ and $\sigma_{\rm sc}>0$ with the following property. For every $\beta\in(0,\beta_{\rm sc}]$, the periodic-box Wilson measures converge locally as $N\to\infty$. The limit satisfies $w_\infty(R,T)\le C_\beta\,e^{-\sigma_{\rm sc}RT+b_\beta(R+T)}$ in lattice units.

This is the classical strong-coupling area law, obtained by the convergent cluster (character) expansion \cite{OSLattice,traj:seilerbook}. An alternative route for $18\beta<1$ combines the boundary-uniform exponential decay of Theorem~\ref{conf:mixing} (via Dobrushin uniqueness, which also gives the local convergence), with Chatterjee's theorem \cite{conf:chatterjee}. That theorem asserts the area law for every infinite-volume Gibbs measure. We have \emph{not} re-derived the uniformity of $\sigma_{\rm sc}$ in $\beta\in(0,\beta_{\rm sc}]$ from either source; this is the one external input of this section.

\begin{proposition}[Starting bound on the trajectory]\label{traj:prop:B}
Assume (SC) and choose $c_*>C(\beta_{\rm sc})$, as in Proposition~\ref{traj:prop:start}. Then $\beta_0\in(0,\beta_{\rm sc})$, and for every rectangle $C$, in the infinite-volume limit,
\[
F_{0,\infty}(C)\ge\frac{\sigma_{\rm sc}}{\ell^2}A(C),\qquad
\widehat F_{0,\infty}(C)\ge\frac{\sigma_{\rm sc}}{\ell^2}A(C)-|\lambda_0|P(C)-|\kappa_0| .
\]
In particular (B$'$) holds with $\Sigma_0\ge\sigma_{\rm sc}/\ell^2$, and (B) holds in infinite volume with $\sigma_0=\sigma_{\rm sc}/\ell^2$, $B_0=|\lambda_0|$, $C_0=|\kappa_0|$.
\end{proposition}
\begin{proof}
Proposition~\ref{traj:prop:start} places $\beta_0$ below $\beta_{\rm sc}$. (SC) gives $f_\infty(R,T)/(RT)\ge\sigma_{\rm sc}-o(1)$, so $\Sigma\ge\sigma_{\rm sc}$. Theorem~\ref{traj:thm:mono}(d) then gives $f_\infty\ge\sigma_{\rm sc}RT$ for \emph{every} rectangle; convert with $a_0=\ell$.
\end{proof}

\paragraph{Scope of the starting estimate.}
The reference-loop trajectory uses the Wilson action itself, and
Proposition~\ref{traj:prop:start} places its first point in the stated
strong-coupling region. Assumption (SC), including its uniformity in
$\beta$, is explicit. The resulting area bound concerns the infinite-volume
limit. Local convergence transfers a comparison for each fixed loop to
sufficiently large boxes, with a tolerance and a box size that may depend
on that loop. It does not yield a positive uniform half-box estimate when
the loop grows with the box. Corollary~\ref{traj:cor:reduce}(c) leaves its
half-box coefficient undetermined.

\paragraph{Choice of the reference Creutz ratio.}
The sufficient numerical choice $c_*>21.97$ fixes a coarse reference scale
through the stated finite-loop equation. If an independently justified
area-dominated regime gave $\chi(\ell,\ell)\simeq\sigma\ell^2$, it would
suggest $\ell\simeq4.7\,\sigma^{-1/2}$ at this threshold. That interpretation
is heuristic here. Neither departure from strong coupling nor the number
of blocking steps needed to reach a weak-coupling regime is proved by the
matching construction. Estimate (A$'$) remains unproved along the selected
sequence, irrespective of which coupling regime it visits.

\subsection{Limits}
\textbf{What survives.} If (A$'$) and (B$'$) were proved, Corollary~\ref{traj:cor:reduce}(b) would hold at every level, for the chosen infinite-volume limit. Both conclusions survive any further limit that exists. If, along a subsequence, potential differences $V_k(R)-V_k(R')$ converge as $k\to\infty$, the limit obeys
\[
V(R)-V(R')\ge\sigma_*(R-R'),
\]
which is the requested $V(R)\ge\sigma_*R-2B_*$ with the additive reference fixed by $V(R')$. By Theorem~\ref{traj:thm:mono}, the limit is also concave and nondecreasing, so $V'(R)\ge\sigma_*$ wherever it is differentiable.

\textbf{What is not established.}
\begin{enumerate}[label=(\roman*)]
\item \emph{Existence of the continuum limit.} A Cauchy bound $|\widehat F_{k+1}(C)-\widehat F_k(C)|\le M_C(k+1)^{-1-\delta}$ is not proved. Note that a Cauchy bound with loop-dependent $M_C$ would not imply (A$'$). Confinement needs $M_C$ controlled by $\eta A(C)$, which is a statement of the same type as (A).
\item \emph{Infinite volume.} A unique, subsequence-independent infinite-volume state at fixed $k$ is not established here outside the stated strong-coupling regime. Elsewhere compactness gives subsequential local limits, which may depend on the subsequence.
\item \emph{Identification.} Even if all limits existed, identifying them with a nontrivial continuum Yang--Mills theory requires the axioms (reflection positivity survives; Euclidean invariance and nontriviality do not follow from anything here).
\item \emph{Mass gap.} An area law does not by itself give a spectral gap; this is a separate problem.
\end{enumerate}

\paragraph{Endpoint of the construction.}
The finite-level matching, exact blocking representation and lifted
reflection-positive interpolation are explicit. They identify the
normalized action insertion whose control would propagate an area bound.
The reduced estimate (A$'$), together with the starting bound, is a
sufficient criterion for a positive bare asymptotic area coefficient at
every level. It has not been established along the four-dimensional
trajectory. Renormalized loop convergence, physical spectral control and
identification of a nontrivial continuum theory remain additional tasks.

\clearpage
\part{Numerical evidence and conclusions}
\begingroup
\def\E{\mathbb E}\def\Cov{\operatorname{Cov}}\def\Var{\operatorname{Var}}
\section{Lattice blocking experiments and effective-action diagnostics}\label{emp:early}
The retained simulation records contain two complementary experiments:
matching blocked Wilson observables across lattice sizes, and estimating
coefficients of a finite effective-action ansatz. Their role is to test
possible coarse-graining constructions. Their fitted masses and action
coefficients are statistical estimates; they do not certify a uniform
physical spectral bound.

\subsection{Simulation and blocking definitions}
The ensembles use the four-dimensional Wilson weight
$\exp(\beta\sum_p\operatorname{Tr}U_p/2)$, with
$\beta=4/g_0^2$. Links are unit quaternions. The supplied simulation codes
use heat-bath updates, with over-relaxation in the larger runs
\cite{traj:creutz,emp:KennedyPendleton1985}. Lattice sizes extend to
$32^4$ for blocking and $22^3\times40$ for the scalar-channel measurements.

For each coarse link, let $W$ be its two-link straight transporter and
let $Q_1,\ldots,Q_6$ be the four-link paths that move one fine step in a
transverse direction, two steps in the coarse-link direction, and back
one transverse step. Define the normalized staple map
\[
 \mathcal B_c(U)_\mu(X)=
 \frac{W_\mu(X)+c\sum_{k=1}^6Q_{k,\mu}(X)}
 {|W_\mu(X)+c\sum_{k=1}^6Q_{k,\mu}(X)|}.
\]
The paths have common endpoints, so the map is gauge covariant wherever
the denominator is nonzero. This is a Swendsen-type blocking construction
\cite{emp:Swendsen1979}. The exceptional zero set is excluded when
constructing numerical configurations; no smooth positive blocked
Gibbs density follows merely from this definition.

For two-lattice matching, configurations on $L^4$ and $(L/2)^4$ are
blocked until their final lattice sizes agree. Matching one blocked
plaquette defines an observable-dependent effective coupling and a
step $\Delta\beta$. Agreement of this one observable does not prove
equality of the blocked measures or a universal renormalization trajectory.
The more extensively thermalized $32^4/16^4$ records at $\beta_0=2.4$
give $\Delta\beta=0.2412(5)$ at matching depth one and $0.243(2)$ at
depth two. The supplied two-loop comparison is approximately $0.275$.
An initial $16^4/8^4$ survey gave a depth-dependent spread near $0.21$;
the larger-lattice values replace that survey as the more informative
estimate. Neither set controls the asymptotic weak-coupling trajectory.

\subsection{Exact identities behind action estimation}
For a smooth function on $SU(2)^E$, let $L_{l,a}$ be differentiation by
left multiplication of link $l$ by $e^{t\tau_a}$, where
$\tau_a=i\sigma_a$, and set
$\nabla_l=(L_{l,1},L_{l,2},L_{l,3})$,
$\Delta_l=\sum_aL_{l,a}^2$.
\begin{lemma}[Schwinger--Dyson identity]\label{emp:sd}
If $S,F\in C^2(SU(2)^E)$ and $d\nu=Z^{-1}e^S dU$, then
\[
 \E_\nu[\Delta_lF+\nabla_lS\cdot\nabla_lF]=0.
\]
If a loop traverses $l$ once, with holonomy $h=U_lX$ and
$F=f(h_0)$, then
\[
 L_{l,a}F=-f'(h_0)h_a,\qquad
 \Delta_lF=(1-h_0^2)f''(h_0)-3h_0f'(h_0).
\]
In particular $\Delta_l\chi_j=-4j(j+1)\chi_j$.
\end{lemma}
\begin{proof}
Left invariance of Haar measure gives $\int L_{l,a}G\,dU=0$.
Use $G=e^SL_{l,a}F$, divide by $Z$ and sum over $a$.
For the loop formula, quaternion multiplication gives
$L_{l,a}h_0=-h_a$ and $L_{l,a}h_a=h_0$. Apply the chain rule and
$\sum_ah_a^2=1-h_0^2$. The character equation follows from the
$SU(2)$ character differential equation, or directly for each polynomial
character used by the estimator.
\end{proof}

\begin{proposition}[The omitted-action residual]\label{emp:sdresidual}
For $S=\sum_i c_iO_i+R$, define
\[
 G_{ij}=|E|^{-1}\sum_l\E[\nabla_lO_i\cdot\nabla_lO_j],\quad
 b_j=-|E|^{-1}\sum_l\E[\Delta_lO_j],\quad
 r_j=|E|^{-1}\sum_l\E[\nabla_lR\cdot\nabla_lO_j].
\]
Then $Gc=b-r$ and $G\succeq0$. If $G$ is invertible, the estimator that
omits $R$ differs from the true retained coefficient vector by
\[
 G^{-1}b-c=G^{-1}r.
\]
\end{proposition}
\begin{proof}
Insert the action expansion in Lemma~\ref{emp:sd} and sum over links.
The matrix $G$ is the Gram matrix of the vector fields $\nabla O_i$.
Its kernel consists exactly of linear combinations with zero gradient
almost everywhere. The last identity is matrix algebra.
\end{proof}
The six-operator ansatz uses fundamental, adjoint and spin-$3/2$
plaquettes, rectangles, chairs and twisted six-link loops.
For a blocked measure, applying Proposition~\ref{emp:sdresidual} requires
a positive $C^2$ density or an appropriate weak integration-by-parts
formulation. The supplied calculation assumes that regularity and does
not prove it. Stability under changing the ansatz is a diagnostic; it does
not bound the residual $r$ or all omitted interactions.

\subsection{Haar blocking: an exact distinction between smoothing and decimation}
\begin{lemma}[Straight decimation preserves product Haar]\label{emp:haar}
For independent Haar-distributed fine links on an even periodic lattice,
the decimated links $V_\mu(X)=U_\mu(2X)U_\mu(2X+e_\mu)$ are independent
Haar variables.
\end{lemma}
\begin{proof}
The coarse links use disjoint pairs of fine variables. Each product is Haar
by left invariance, and disjoint pairs are independent.
\end{proof}

\begin{theorem}[Leading Haar plaquette created by staple blocking]\label{emp:haarstaple}
On the infinite lattice, or an even periodic lattice of side at least six,
let the fine links be independent Haar variables. The plaquette expectation
after one normalized staple blocking satisfies
\[
 P_c:=\E\left[\tfrac12\operatorname{Tr}(\mathcal B_c(U))_p\right]
       =\frac98c^2+O(c^3)\qquad(c\longrightarrow0).
\]
The remainder is uniform over fine configurations in any fixed
$|c|\le c_0<1/6$. Consequently product Haar is not fixed by this smoothing
map for all sufficiently small positive $c$.
\end{theorem}
\begin{proof}
For a coarse link, write $S=W+cQ$, $Q=\sum_{k=1}^6Q_k$, and
$t=\langle W,Q\rangle$. Since $|S|\ge1-6|c|$, normalized blocking is
uniformly analytic near zero, with
\[
 \frac{S}{|S|}=W+c(Q-tW)
 +c^2\left[\left(\tfrac32t^2-\tfrac12|Q|^2\right)W-tQ\right]+O(c^3).
\]
Independent Haar measure is invariant under multiplying any one fine link
by the central element $-1$. A monomial with an odd occurrence of that
fine link therefore has zero expectation. This parity condition, rather
than unrestricted word cancellation, is sufficient for the following
enumeration.

Label the four coarse sides by their positive orientations:
$(0,e_\mu)$, $(2e_\mu,e_\nu)$,
$(2e_\nu,e_\mu)$, $(0,e_\nu)$, with the last two reversed in the
plaquette. Each straight path has two edges; its six bent alternatives
have steps $(\pm e_\rho,e_\mu,e_\mu,\mp e_\rho)$ with $\rho\ne\mu$,
and likewise for a $\nu$ side. Every first-order summand has the same
fine-link incidence parity as its selected bent path, because the
projection term inserts two extra copies of the straight path.
A second-order correction on one side has the parity of that straight
path and two selected bent paths. The 24 first-order possibilities and
144 single-side second-order possibilities all have an odd fine-link
incidence. Of the 216 pairs of first-order corrections on distinct sides,
only two have every incidence even: the two opposite sides both bend
towards the plaquette interior. These counts follow by listing the stated
steps; the exact integer enumeration and word reductions are supplied
in \path{code/conditioning/block_path_certificate.py}.

For either surviving pair, let $Q_1,Q_3$ be the inward bent paths and
$W_2,W_4$ the unchanged sides. Their product
$Q_1W_2Q_3^*W_4^*$ reduces to the identity. The two replaced straight
transporters $W_1,W_3$ are independent Haar variables independent of these
four paths. The Haar second moment is
$\E[\langle W,Q\rangle W]=Q/4$, so integrating each first-order
correction $Q_i-\langle W_i,Q_i\rangle W_i$ replaces it by $3Q_i/4$.
Each surviving pair contributes $(3/4)^2=9/16$. The two contributions
sum to $9/8$. The uniform Taylor remainder proves the asymptotic statement.
\end{proof}
Observables of the blocked Haar field with disjoint fine-link dependency
sets are exactly independent, even though its plaquette mean is positive.
A plaquette mean and a worst-case sufficient influence criterion therefore
measure different properties.

\begin{table}[htbp]\centering\small
\caption{Retained Haar-input blocking diagnostics. Parentheses give the
reported statistical uncertainty. The last column evaluates a sufficient
criterion on a six-term fitted action; it is not a certified influence
bound for the complete blocked measure.}\label{emp:haartable}
\begin{tabular}{@{}rrrr@{}}\toprule
Staple parameter $c$ & $P_c$ & Fitted fundamental coupling & Fitted criterion\\
\midrule
$0.1$ & $0.0109(3)$ & $0.0439(12)$ & $2.2$\\
$0.25$ & $0.0485(2)$ & $0.2005(11)$ & $47$\\
$0.5$ & $0.06692(14)$ & $0.2770(6)$ & $75$\\
$1$ & $0.03272(12)$ & $0.1287(5)$ & $15$\\
$2$ & $0.00996(16)$ & $0.0395(7)$ & $3.8$\\
$4$ & $0.00268(17)$ & $0.0107(8)$ & $0.63$\\
$8$ & $0.00075(18)$ & $0.0030(8)$ & $0.45$\\
\bottomrule\end{tabular}
\end{table}

\subsection{Measured action flow and the proposed coefficient envelope}
At $\beta_0=2.4$, the $32^4$ hot-start records give blocked plaquettes
$0.58507(11)$, $0.3859(4)$, $0.1420(12)$ and $0.067(5)$ at levels
one through four. Cold-start controls are compatible at the recorded
precision. The level-one fitted action has fundamental coefficient
$3.787(4)$, rectangle coefficient $-0.3300(4)$, chair coefficient
$-0.1033(4)$ and twisted-loop coefficient $0.0916(6)$.
Thus fitting a Wilson coefficient alone discards sizeable compensating
interactions.

For the three fitted plaquette characters at this level, the normalized
coefficients are approximately $(0.6463,0.3291,0.1373)$. The comparator
Migdal--Kadanoff step started at $\beta_0+0.25=2.65$ gives
$(0.5526,0.2072,0.0532)$. The spin-$3/2$ ratio is approximately $2.58$.
A direct quadrature of the fitted single-plaquette weight freshly
reproduces $0.137339$ for the last coefficient. This supplies numerical
evidence against that particular coefficientwise envelope for this
blocking and ansatz. Because the full omitted-action residual is unknown,
it is not an exact counterexample for the complete effective action.

The parameter scan $c=1/2,1,2,4,8$ changes these comparisons.
For $c\ge2$ the first-step fitted plaquette ratios fall below one in the
recorded runs, while the omitted-loop diagnostic remains too large for
the proposed small-remainder criterion. Approaching the blocked-Haar
plaquette value at deeper levels does not establish proximity of complete
measures, and cannot by itself supply the conditional locality hypotheses
of Section~\ref{cond:coarse}.

\subsection{Scalar-channel mass measurements}
The initial simulation series used variational smeared-plaquette operators
at three couplings. Its effective masses at time separation $1\to2$ were
$1.274(214)$ on $8^3\times16$ at $\beta=2.25$,
$0.895(108)$ on $12^3\times16$ at $\beta=2.4$, and
$0.816(121)$ on $16^3\times16$ at $\beta=2.475$.
These are the supplied simulation estimates. The comparison values
$1.390(30)$, $0.953(19)$ and $0.754(10)$, respectively, are published
measurements of Lucini and Teper \cite{emp:LuciniTeper2001} and retain that
attribution.

The second series measures both scalar correlations and smeared Wilson
loops, fitting a static potential to estimate the string tension.
Its recorded results are collected in Table~\ref{emp:earlymasses}.
\begin{table}[htbp]\centering\small
\caption{Retained scalar-channel and string-tension estimates. The mass
column is the supplied early-time exponential fit, not a certified
spectral bound. At $\beta=2.6$ the source reports no plateau in the fitted
range, so no ground-state determination is assigned to that point.}
\label{emp:earlymasses}
\begin{tabular}{@{}ccrrrr@{}}\toprule
$\beta$ & Lattice & Measurements & Fitted $am$ & $a\sqrt\sigma$ & $m/\sqrt\sigma$\\\midrule
$2.3$ & $10^3\times20$ & 1950 & $1.5(3)$ & $0.371(5)$ & $4.0(8)$\\
$2.4$ & $12^3\times24$ & 1450 & $0.91(6)$ & $0.2701(18)$ & $3.4(2)$\\
$2.5$ & $16^3\times32$ & 550 & $0.72(5)$ & $0.189(2)$ & $3.8(3)$\\
$2.6$ & $22^3\times40$ & 250 & $0.64(4)$ & $0.140(2)$ & unresolved\\
\bottomrule\end{tabular}
\end{table}

A three-point linear fit in $a^2\sigma$, excluding the unresolved point,
gives the supplied intercept $3.82\pm0.45$; a constant fit gives
$3.58\pm0.17$. These are descriptive extrapolations of the supplied
estimates. They do not establish continuum convergence. The analysis
uses a fixed variational vector across jackknife samples and an early
logarithmic exponential fit; the code also takes absolute values before
logging projected correlations. Those choices require further study of
excited-state contamination, correlation signs, autocorrelation, fit
covariance and finite temporal extent before these quoted uncertainties
can be treated as a complete error budget. In particular a noisy
finite-periodic-time fit is not a rigorous upper bound merely because
vacuum infinite-time correlators have positive spectral measures.

The preserved raw observations and analysis programs are in
\path{code/conditioning/early/lattice_i} and
\path{code/conditioning/early/lattice_ii}. Their two diagnostic checkers
were rerun: all ten checks in each passed their stated numerical
comparisons. Those reruns re-evaluate stored observations and formulas;
they do not generate new independent ensembles or certify the statistical
errors. The exact small-staple incidence certificate was recomputed
separately and supports Theorem~\ref{emp:haarstaple}.
\endgroup

\section{Numerical evidence and its relation to the spectral claims}
\label{sec:empirical}

The numerical material has three distinct roles. Published glueball
measurements provide evidence about physical excitations. Recorded flow
measurements permit an exact arithmetic check of a finite empirical covariance.
The coupled-replica experiment concerns an update algorithm. None of these
roles is interchangeable with a lower bound on the full vacuum-complement
spectrum.

\subsection{An attributed reanalysis of published scalar masses}

Athenodorou and Teper measured the spectrum and string tensions of pure
$SU(N)$ Wilson gauge theories and extrapolated them to the continuum.
Their $SU(2)$ result for the lightest scalar is
\cite[Tables 23 and 34]{emp:AthenodorouTeper2021}
\[
  \frac{m_{0^{++}}}{\sqrt\sigma}=3.781(23).
\]
The measurements and the published continuum result are theirs. The
calculation here is a secondary fit to the six printed $SU(2)$ data rows
in their Table 1, reproduced below so that the input is explicit.

\begin{table}[htbp]
\centering\small
\begin{tabular}{ccccc}
\toprule
$\beta$ & $L_s^3\times L_t$ & $a\sqrt\sigma$ & $am_{0^{++}}$ & $L_s a\sqrt\sigma$\\
\midrule
2.2986 & $12^3\times16$ & $0.36778(69)$ & $1.224(16)$ & 4.41336\\
2.3714 & $14^3\times16$ & $0.29023(50)$ & $1.025(12)$ & 4.06322\\
2.427  & $20^3\times16$ & $0.24013(41)$ & $0.8469(76)$ & 4.80260\\
2.509  & $22^3\times20$ & $0.18011(22)$ & $0.6563(56)$ & 3.96242\\
2.60   & $30^4$ & $0.13283(30)$ & $0.5001(41)$ & 3.98490\\
2.70   & $40^4$ & $0.09737(23)$ & $0.3652(35)$ & 3.89480\\
\bottomrule
\end{tabular}
\caption{Published measurements from Athenodorou and Teper
\cite[Table 1]{emp:AthenodorouTeper2021}; the final column is calculated
from the printed values. Parentheses denote the authors' statistical errors.}
\label{emp:tab:mass_input}
\end{table}

Put $s=a\sqrt\sigma$, $x=s^2$ and $y=(am)/s$. The leading-cutoff model is
$y=R_0+cx$. Weighted least squares gives
$\widehat\theta=(X^TWX)^{-1}X^TWy$, with parameter covariance
$(X^TWX)^{-1}$, without rescaling by the residual chi-square. The
mass-only calculation uses $\operatorname{se}(y)=\operatorname{se}(am)/s$.
A sensitivity calculation propagates both printed marginal errors with
zero covariance; a further orthogonal-distance fit uses
$am=R_0s+cs^3$ in the original $(s,am)$ variables.

\begin{table}[htbp]
\centering\small
\begin{tabular}{lccc}
\toprule
Fit and retained rows & $R_0$ & Standard error & $\chi^2$/d.o.f.\\
\midrule
Mass errors only, six rows & 3.781592 & 0.022733 & 6.910233/4\\
Both marginal errors, six rows & 3.780692 & 0.023307 & 6.619031/4\\
Orthogonal distance, six rows & 3.780688 & 0.023246 & 6.659460/4\\
Both marginal errors, omit coarsest & 3.789563 & 0.027912 & 6.285331/3\\
Both marginal errors, omit two coarsest & 3.825300 & 0.032427 & 1.597646/2\\
\bottomrule
\end{tabular}
\caption{Freshly reproduced secondary fits. Errors are model-dependent
statistical fit errors, not certified total uncertainties. The orthogonal-distance
row uses the unscaled parameter covariance.}
\label{emp:tab:mass_fit}
\end{table}

The primary six-point ratio fit agrees with the published continuum
number. Its movement under removal of coarse points is a useful sensitivity
diagnostic. The original joint covariance and resampling data are not
available in the printed table, so independence is an approximation;
the fits do not bound higher cutoff terms, finite-volume effects, or
operator-overlap systematics. Bennett and collaborators reported
$m_{0^{++}}/\sqrt\sigma=3.841(84)$ for $Sp(2)\simeq SU(2)$, an
independent published calculation consistent with this scale
\cite[Table IV]{emp:Bennett2021}.

\begin{figure}[htbp]
\centering
\includegraphics[width=\textwidth]{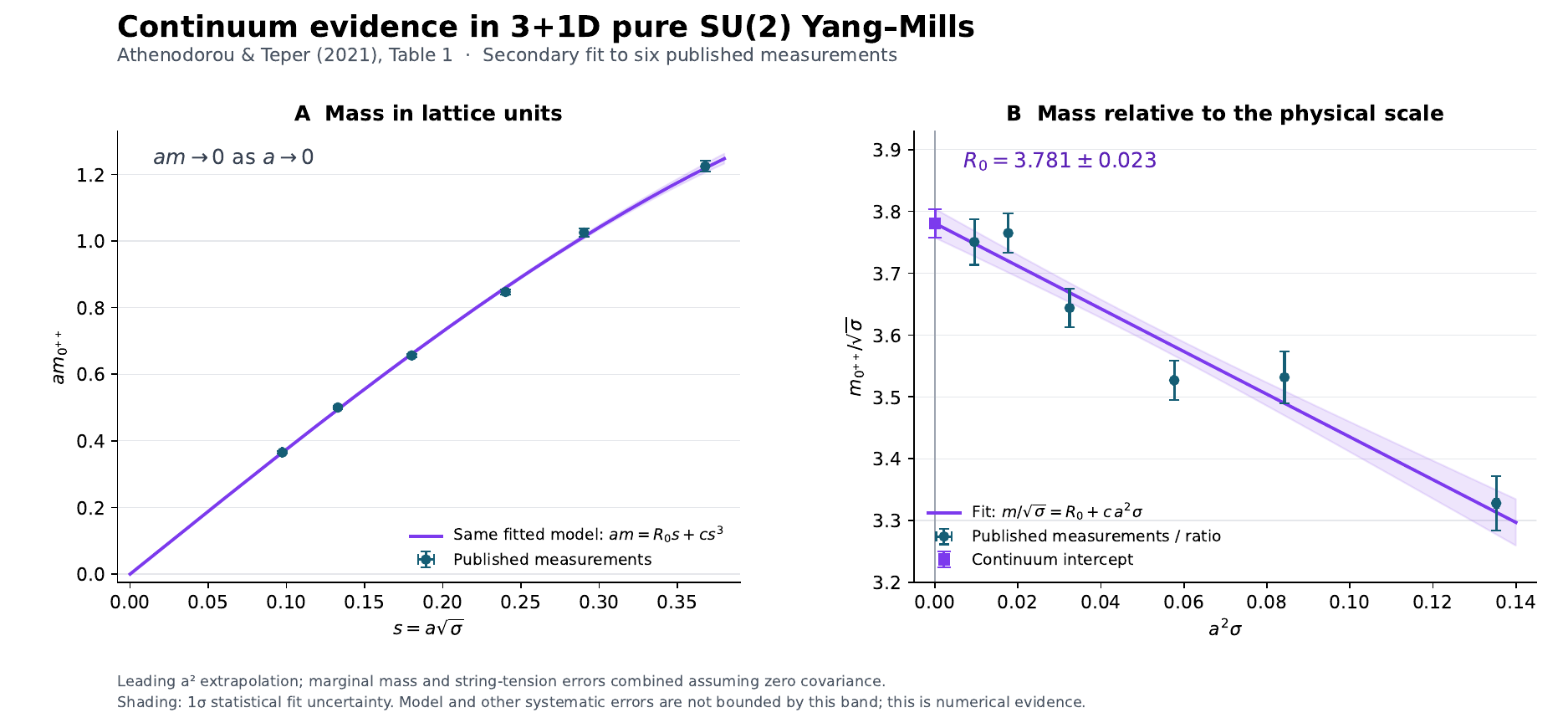}
\caption{Secondary fit to the measurements in Table~\ref{emp:tab:mass_input}.
The shaded band is the propagated one-standard-error fit band under the
stated covariance approximation. A nonzero dimensionless mass ratio can
coexist with $am\to0$; the plot is numerical evidence, not a spectral
lower-bound certificate.}
\end{figure}

A positive scalar energy obtained with a chosen operator basis does not
exclude an unseen lower state with small overlap. The continuum spectral
criterion in the mathematical sections concerns a dense family in the
entire physical vacuum complement. It is therefore stronger than fitting
one channel. The ratio to $\sqrt\sigma$ also keeps the overall scale
explicit: no numerical value in GeV is assigned to pure $SU(2)$ without
an additional scale convention.

\subsection{Consistency checks for exact link and reflection identities}

The supplied Wilson-trajectory calculation includes heat-bath records
on $8^4$ at $\beta=2.3$ and $6^4$ at $\beta=1.0$.
These are supporting statistical checks of exact identities. On the
$8^4$ ensemble, the recorded plaquette mean is $0.60250(24)$, while the
two sides of the link identity are
\[
6\langle P\rangle=3.6150(14),\qquad
\left\langle r\,I_2(\beta r)/I_1(\beta r)\right\rangle=3.6152(14).
\]
For the $4\times4$ loop, $w(4,4)=0.00494(18)$ and
$w(4,2)^2=0.00198(3)$, consistent with the reflection inequality.
The recorded tilted total variation is $85.8$, compared with its
lower bound $e^{F}/4\simeq50.6$. These ensembles and errors are
retained from the research record; the full trajectories were not
regenerated during this consolidation. The exact link integrals,
coefficient identities and two-dimensional blocking calculation
were recomputed. Numerical consistency is not used as a step in
any of the reflection-positivity proofs.

\subsection{Connected fluctuations in published SU(2) flow measurements}
\label{sec:published-flow-covariance}

We recompute a connected covariance matrix from the public
Monte Carlo data of Bennett, Hong, Lee, Lin, Lucini, Piai and
Vadacchino.\cite{bridge:flowpaper,bridge:flowdata}
The configurations and original measurements remain their work. No new Monte
Carlo simulation or continuum extrapolation is performed here.

The selected ensemble is pure SU(2), $\beta=2.60$, on a periodic $24^4$
lattice. Its complete log has $3999$ configurations, each with $251$ flow
measurements, in agreement with the published ensemble table. Section IV of
the paper describes the stored ensembles as thermalized. The released analysis
workflow numbers records from one and applies $N_{\rm therm}=100$ by retaining
ordinal labels $100,\ldots,3999$.\cite{bridge:flowcode}
We follow that rule, leaving $N=3900$ records; the compact extraction preserves
all $3999$ original records so the selection is auditable.

For each retained configuration $n$, define the three dimensionless recorded
observables
\[
 X_{n,i}=t_i^2 E_{\rm cl}^{(n)}(t_i),
 \qquad (t_1,t_2,t_3)/a^2=(1.9,3.8,7.6),
\]
using the log's \texttt{t2*Esym} column. Here $E_{\rm cl}$ is the clover
discretization of the volume-averaged flowed energy density. These are auxiliary
Wilson flow times, not physical Euclidean time separations.
We compute
\[
 \bar X_i=\frac1N\sum_{n=1}^N X_{n,i},\qquad
 C_{ij}=\frac1{N-1}\sum_{n=1}^N
       (X_{n,i}-\bar X_i)(X_{n,j}-\bar X_j).
\]

\begin{center}
\begin{tabular}{rcc}
\hline
$t_i/a^2$ & Empirical mean $\bar X_i$ & Empirical variance $C_{ii}$\\
\hline
$1.9$ & $0.06477695179$ & $5.965196521\times10^{-6}$\\
$3.8$ & $0.10604599949$ & $6.287743472\times10^{-5}$\\
$7.6$ & $0.20848698517$ & $6.922014520\times10^{-4}$\\
\hline
\end{tabular}
\end{center}

The full connected matrix, rounded here for readability, is
\[
 10^6 C\simeq
 \begin{pmatrix}
 5.965196521&18.187950407&52.258244553\\
 18.187950407&62.877434721&197.019388977\\
 52.258244553&197.019388977&692.201451979
 \end{pmatrix}.
\]
The pair correlations are approximately $0.93912624$, $0.81325396$ and
$0.94437680$ for pairs $(1,2)$, $(1,3)$ and $(2,3)$ respectively. Thus the
observables are strongly correlated, while the recorded covariance has three
nondegenerate directions.

\paragraph{Exact finite-sample arithmetic.}
Every printed decimal in the selected columns is parsed as a rational number.
The checker reconstructs $C=LDL^T$ exactly, with positive pivots
\[
 10^6\operatorname{diag}D\simeq
 (5.965196521,\;7.422172046,\;43.067790289).
\]
The full fractions are supplied in the certificate. For the empirical
correlation matrix $R$, $\operatorname{tr}R=3$ and
$\det R=\det C/(C_{11}C_{22}C_{33})$.
Consequently
\[
 \lambda_{\min}(R)\geq\frac{4\det R}{9}
 =0.00326417750635\ldots>0,
\]
since the product of its two largest eigenvalues is at most $9/4$.
This is an exact arithmetic statement about the finite recorded sample.

\paragraph{Provenance and scope.}
Only the complete $58{,}754{,}203$-byte compressed log entry was retrieved from
the public archive. Its decompressed length, SHA256, ZIP CRC32, complete stream,
configuration identities and measurement counts were checked. The retrieval
script can recreate the compact table directly from the public archive; the
arithmetic checker runs offline on that table.
No confidence interval or rigorous lower bound for the population covariance is
asserted. Serial dependence, flow-integration error, finite volume and cutoff
bias remain properties of the underlying data. The calculation gives
attributable evidence for joint fluctuations of smooth gauge-invariant
observables at this lattice spacing. It supplies neither local conditional-block
angles nor a physical transfer-gap bound, and does not construct a continuum
limit.

\subsection{A reported winding-channel comparison}
\label{sec:reported-winding}

The finite-box certificate in Section~\ref{sec:finite-transfer} is accompanied
in the supplied research record by a preliminary Monte Carlo report at
$\beta=177/500$ on $1^3\times N_t$ periodic lattices. That report describes
four hot/cold chains with $1{,}000{,}000$ retained measurements at $N_t=16$
and two chains with $300{,}000$ retained measurements at $N_t=24$.
These are reported measurement counts, not independent-sample counts.
The raw six-chain records and their analysis executable were not included
in the supplied archives. Accordingly, the following summary preserves
the reported comparison but is not a reproduced numerical result of this
version and contributes no input to the mathematical certificate.

The gauge-invariant observable is the spatial winding loop
$W_i(t)=\tfrac12\Tr U_i(t)$, with correlation
\[
 C(d)=\frac1{3N_t}\sum_{i=1}^3\sum_t
       \langle W_i(t)W_i(t+d)\rangle.
\]
On this spatial torus the loop is closed and center odd. The report fits
$A(e^{-xd}+e^{-x(N_t-d)})$ at $d=1,2,3$, with blocked covariance and
$1{,}000$ block-bootstrap refits. Its quoted results are
\begin{center}
\begin{tabular}{@{}ccc@{}}
\toprule
$N_t$ & Reported $x$ and statistical error & Reported $95\%$ interval\\
\midrule
16 & $2.29485\pm0.01384$ & $[2.26836,2.32148]$\\
24 & $2.29529\pm0.02055$ & $[2.25546,2.33656]$\\
\bottomrule
\end{tabular}
\end{center}
Both quoted percentile intervals contain the independently certified
infinite-time transfer interval $[2.27102703117506,2.27225894375269]$.
This numerical containment can be checked directly; validating the
sampling errors, thermalization and fit procedure requires the missing
original records. A finite-$N_t$ fit also has possible excited-state and
thermal contributions. The channel is center-charged electric flux and
does not determine the infinite-volume scalar-glueball mass.

\section{Coupled-replica Monte Carlo: an algorithmic comparison}
\label{sec:replica}

This section analyses the coupled-update experiment recorded with the research data. Its evolution parameter is the number of update sweeps, not physical
Euclidean time. No synchronization rate, coupling time, or failure to
synchronize is substituted for a physical transfer-operator gap.

\subsection{A coupling with the correct Wilson heat-bath marginals}

The simulated measure is four-dimensional pure $SU(2)$ Wilson theory,
\[
 d\mu(U)\ \propto\
 \exp\!\left[\beta\sum_p\tfrac12\operatorname{ReTr}U_p\right]
 \prod_e dU_e,
\]
on periodic $L^4$ lattices with $L=2,3,4$. Every plaquette contains four
distinct links at these extents. The usual conditional action is therefore
linear in a chosen link $U$:
\[
 d\mu(U\mid\text{other links})\ \propto\
 e^{\beta\operatorname{ReTr}(US)/2}\,dU,
 \qquad \|S\|\le6,
\]
where $S$ is the sum of its six staples and the norm is the Euclidean norm
of its quaternion coordinates. The source uses
$U=q_0I-i\boldsymbol q\cdot\boldsymbol\sigma$.

For replicas $A$ and $B$, hold their respective staples $S_A,S_B$ fixed
during a link update. Generate a common sequence of independent Haar
matrices $Q_j$ and uniforms $u_j$. Replica $r\in\{A,B\}$ accepts the first
candidate satisfying
\[
 u_j\le
 \exp\!\left\{\beta\left[
 \tfrac12\operatorname{ReTr}(Q_jS_r)-\|S_r\|
 \right]\right\}.
\]
If one replica accepts first, it retains that candidate while proposals
continue for the other. Each marginal is consequently ordinary rejection
sampling from its own exact conditional distribution. Unequal rejection
counts do not change either marginal: each replica uses its first accepted
candidate in the common independent sequence. The next link starts with
fresh randomness. Links are updated in the fixed site-major, direction-minor
order recorded in the source. The product of these conditional kernels
preserves the Wilson measure for each replica.

This coupling uses common coordinate frames for the two configurations.
The joint coupling rule is not invariant under independently changing the
replicas' gauge frames. A gauge-invariant observable does not remove that
dependence of the coupling itself. This distinction is relevant when
interpreting coalescence statistics. Coupling-scheme dependence of damage spreading is established in other Markov-chain models \cite{emp:Bernard2010}.

\subsection{An exact algorithmic coalescence bound}

\begin{proposition}[Finite-box algorithmic coalescence]\label{emp:coalescence}
For the ideal exact-arithmetic shared-candidate coupling just defined,
let $E=4L^4$ be the number of links and let $\tau$ be the first sweep at
which the two configurations agree on every link. For any initial pair,
any finite $\beta\ge0$ and any integer $n\ge1$,
\[
 \boxed{\displaystyle
 \Pr(\tau>n)\le
 \left(1-e^{-12\beta E}\right)^n,
 \qquad \mathbb E\tau\le e^{12\beta E}.}
\]
At $\beta=0$, the configurations agree after one full sweep. Equality is
absorbing under this coupling.

\end{proposition}\begin{proof}
For either replica and every Haar candidate,
$\tfrac12\operatorname{ReTr}(QS_r)\ge-\|S_r\|$ and $\|S_r\|\le6$.
Thus its acceptance probability is at least $e^{-12\beta}$. If the first
uniform in a link update is at most $e^{-12\beta}$, both replicas accept
the same first candidate. Conditional on the entire past, that event has
probability $e^{-12\beta}$. Requiring it at every one of the $E$ updates
in a sweep has probability at least $e^{-12\beta E}$. Previously updated
links are not updated again during that sweep, so this event leaves every
link equal at its end. Once all links agree, the staples and all subsequent
acceptance decisions agree, proving absorption. Iterating the conditional
failure bound proves the tail inequality, and summing the tail proves the
expectation bound. At $\beta=0$ every first candidate is accepted.
\end{proof}

This bound is very conservative and deteriorates exponentially with the
number of links. It proves eventual coalescence at every finite volume and
coupling, so finite-horizon noncoalescence cannot be a literal finite-box
transition to infinite coupling time. The proposition concerns the Monte
Carlo coupling and gives no physical transfer-gap bound.

\subsection{Frozen experiment and observables}

A retained exploratory pilot established feasibility and selected a grid.
Before production, the full design, seeds, source hashes, endpoints, and
statistical analysis were frozen in \nolinkurl{production_design.json}.
Production comprised $1{,}512$ independently seeded pairs:
\begin{itemize}
\item Main calculation: 32 hot/cold pairs at each of 10 couplings
$\beta=0$, $0.025$, $0.05$, $0.1$, $0.12$, $0.14$, $0.16$, $0.18$, $0.2$, $0.354$, on each of the
three volumes, with a 512-sweep horizon (960 pairs).
\item Small-perturbation calculation: 24 pairs at each of
$\beta=0.05,0.12,0.16,0.2,0.354$ on each volume (360 pairs).
One configuration was warmed for 512 ordinary heat-bath sweeps; its copy
was perturbed by an independent random-axis quaternion rotation of angle
$0.02$ on every link. The initial mean link distance is approximately
$1-\cos(0.02)=2.0\times10^{-4}$. The warmup is a specified initialization,
not a proof of exact equilibrium.
\item Higher-coupling diagnostics: eight pairs for each initialization,
volume, and $\beta=0.7,1.5,2.5$, with a separately labeled 128-sweep horizon
(144 pairs).
\item Independent-noise controls: eight hot/cold pairs on $3^4$ at each of
$\beta=0$, $0.05$, $0.16$, $0.354$, $1.5$, $2.5$, with 128 sweeps (48 pairs).
\end{itemize}

Two discrepancies were measured after each sweep:
\[
 D_U=\frac1{2E}\sum_e|q_e^A-q_e^B|^2,
 \qquad
 D_P=\left[\frac1{N_p}\sum_p
 (p_p^A-p_p^B)^2\right]^{1/2},
 \quad p_p^r=\tfrac12\operatorname{ReTr}U_p^r,
 \quad N_p=6L^4.
\]
$D_U$ compares the link coordinates in a common frame; $D_P$ is invariant
under separate gauge transformations of the two configurations. Coalescence
means exact componentwise equality of every link, with no distance tolerance.
Shared pairs stopped at this absorbing state. Subsequent damage is
mathematically zero, but no additional simulated data rows were invented.
Independent controls ran their full horizon.

The calculation produced $309{,}623$ actual recorded rows, including initial
configurations. Rows within a pair are correlated; independent pair seeds
are the statistical units. The initial production run took 94.5 seconds
of wall time with two workers; the sum of individually recorded process
elapsed times was 181.4 seconds. The latter is not a CPU-time measurement.

\subsection{Results and statistical interpretation}

The main endpoint is the number of pairs coalesced by 512 sweeps:
\begin{center}
\begin{tabular}{c c c c}
\hline
$\beta$ & $2^4$ & $3^4$ & $4^4$\\
\hline
0 & 32/32 & 32/32 & 32/32\\
0.025 & 32/32 & 32/32 & 32/32\\
0.05 & 32/32 & 32/32 & 32/32\\
0.10 & 32/32 & 32/32 & 32/32\\
0.12 & 32/32 & 32/32 & 0/32\\
0.14 & 32/32 & 7/32 & 0/32\\
0.16 & 32/32 & 0/32 & 0/32\\
0.18 & 32/32 & 0/32 & 0/32\\
0.20 & 32/32 & 0/32 & 0/32\\
0.354 & 0/32 & 0/32 & 0/32\\
\hline
\end{tabular}
\end{center}
Under the independent-pair binomial model, the pointwise 95\% exact
Clopper--Pearson intervals are $[0.8911,1]$ for 32/32,
$[0,0.1089]$ for 0/32, and $[0.0928,0.3997]$ for 7/32.
These are not simultaneous confidence bounds over the grid.

At $\beta=0.05$, the median coalescence sweeps were 3, 4 and 5 on the
three volumes. At $\beta=0.10$, they were 6, 16 and 25. The pronounced
volume dependence near $\beta=0.12$--$0.14$ is an empirical finite-horizon
effect of this coupling; no critical coupling or thermodynamic phase
transition is fitted.

At $\beta=0.354$, the final gauge-invariant discrepancies for hot/cold
pairs were
\[
 D_P(512)=
 \begin{cases}
 0.62780\pm0.00979,&2^4,\\
 0.64446\pm0.00415,&3^4,\\
 0.63995\pm0.00207,&4^4,
 \end{cases}
\]
with standard errors across the 32 independent pairs.
For the independently updated $3^4$ control at that coupling,
$D_P(128)=0.70914\pm0.00781$ across eight pairs. The difference between
these measurements describes the chosen joint dynamics; it is not a mass.

Small perturbations had a different probability of healing. At
$\beta=0.354$, the numbers coalesced by 512 sweeps were 14/24, 1/24 and
0/24 for $L=2,3,4$. At $\beta=0.12$, they were 24/24, 24/24 and 9/24.
Every hot/cold pair in the separately scoped higher-coupling calculation
remained distinct through 128 sweeps. Full counts and uncertainty intervals,
including the perturbed higher-coupling pairs, are retained in
\texttt{replica\_analysis.json} and \texttt{replica\_summary.csv}.

\begin{center}
\includegraphics[width=\textwidth]{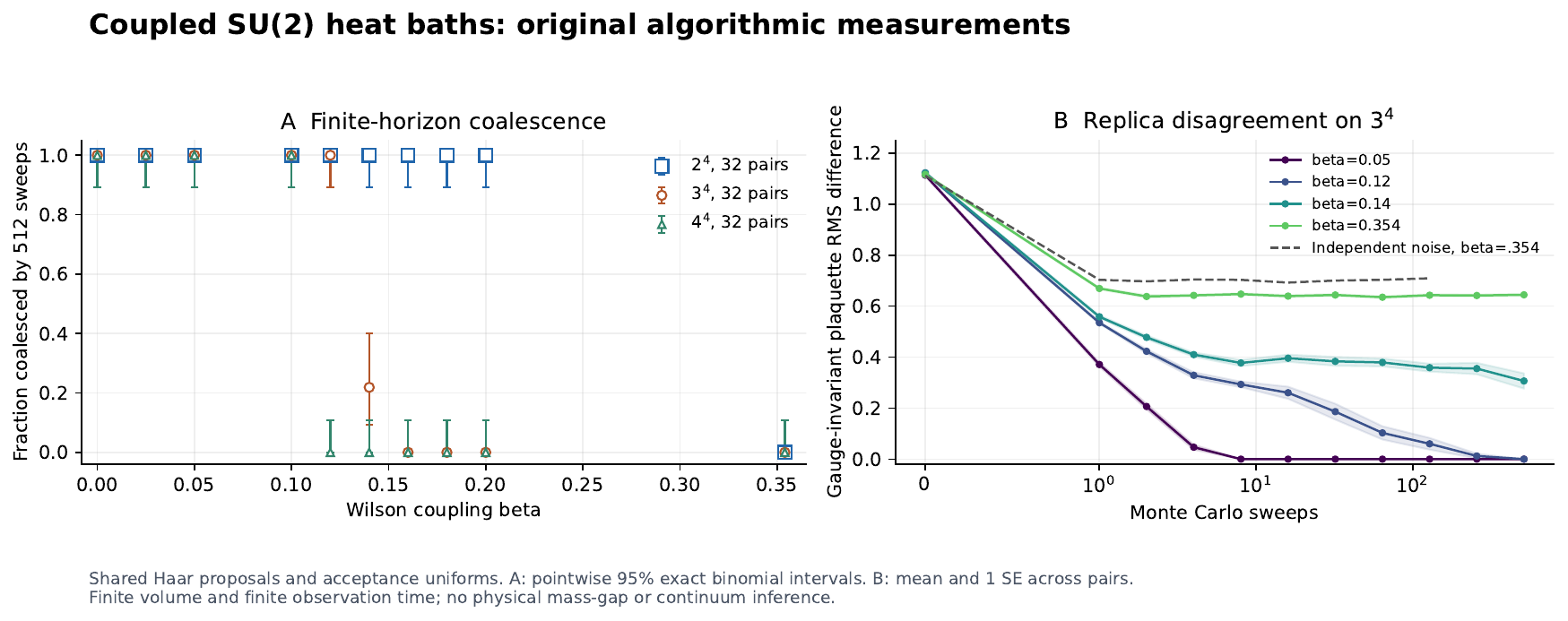}
\end{center}

\subsection{Numerical validation and reproducibility}

Validation covered the new coupling as well as its inherited geometry.
The maximum observed discrepancy between staple-predicted and full-action
changes on $L=2,3,4$ was $4.91\times10^{-14}$. Tests confirmed exact
one-sweep coalescence at $\beta=0$, and absorption of identical pairs
across the tested couplings through $\beta=2.5$. Separate gauge
transformations changed $D_P$ by at most $2.23\times10^{-16}$ in the test.
Four sets of $150{,}000$ coupled conditional draws checked eight marginal
projected means against the exact Bessel expectation $I_2(\kappa)/I_1(\kappa)$;
all deviations were below 1.64 estimated standard errors. The maximum link
norm-squared error over production was $8.89\times10^{-16}$.

The frozen source, design and all 1,512 trajectory/metadata pairs have been
checked against their recorded hashes. Twenty-eight affected trajectory files
were recovered by rerunning their original seeds with the unchanged source;
each recovered trajectory matched its original completion hash. No outcome
was excluded. The final analysis uses the complete recorded data. The V5
check revalidates the complete data and recomputes the summary; it is not
an independent regeneration of every stochastic trajectory.

The source, fixed design, raw measurements, original run hashes, validation,
recovery evidence and analysis are supplied. These Monte Carlo results are
finite-precision statistical measurements. Their uncertainty statements
have a different meaning from the deterministic error certificates for
the physical transfer operator.

\section{Conclusions and the next mathematical target}
\label{sec:conclusion}

The established endpoint is a collection of physical lattice results and
controlled passage criteria. The finite-box bounds concern the full
gauge-invariant vacuum complement, and their representation tails are
included. The strong-coupling theory has estimates that survive spatial
growth at fixed lattice coupling. At arbitrary fixed spatial cutoff and
positive Hamiltonian coupling, subsequential thermodynamic dynamics and
nonzero local correlations are constructed. These statements resolve
different regulator-level questions and retain their distinct uniformities.

The concentration and fidelity results explain why spatial refinement
cannot be handled by an unexamined fixed electric basis or a global
comparison with Haar measure. The improved local estimate supplies a
volume-independent bound on a fixed conditional sector, but the retained
sector and its coupling to the discarded sector still matter. A complete
lower bound for the interacting vacuum complement must control them
collectively. The finite-dimensional Schur certificates show how this can
be done at fixed parameters, including all omitted modes; they do not
provide a uniform complexity or gap estimate along cutoff removal.

The conditioning calculations have a separate conclusion. The stated
finite functional is bounded throughout the certified field domain,
including its analytic remainders. The corresponding extended physical
coupling range is conditional on the imported block-comparison
hypotheses. Exact sensitivity and ceiling calculations also delimit that
method. A successful interval computation cannot by itself validate a
missing comparison between the functional and the physical transfer
spectrum.

The explicit reference-loop trajectory proves that the renormalization
equation has a selected solution at every finite level. Exact blocking preserves the expectations of aligned loops. The lifted
reflection-positive interpolation connects the two endpoint measures and
makes their generally nonzero normalized insertion explicit. The
construction does not prove that its couplings escape to the weak-coupling
regime, that its renormalized observables converge, or that a positive
asymptotic area coefficient persists. The corrected scalar wall and
heavy-matter results remain useful comparisons within their own models.

The next substantial advance would establish three controls on one and
the same sequence:
\begin{enumerate}
\item Renormalized gauge-invariant states and local observables with
nonzero limiting norms and compatible physical-time correlations as
$a_s\to0$, together with the local field properties required to identify
the continuum theory.
\item A collective estimate on the actual interacting vacuum measure,
or a stronger full spectral inequality, whose physical lower bound
stays above a fixed $M>0$. The estimate must include all sectors and
all omitted representations relevant to the limiting physical space.
\item Uniformity as physical spatial volume grows, including control
of vacuum subtraction, density of the limiting probe family, and the
passage from finite-volume correlations to the infinite-volume theory.
\end{enumerate}

The continuum transfer theorems prove what follows once these controls
are available. They do not establish the controls themselves. The
published lattice spectrum gives substantial numerical motivation, and
the certified benchmarks supply concrete tests for proposed methods.
The work therefore ends with an explicit, reproducible boundary between
the proved lattice theory and the remaining continuum construction.
It does not determine an exact continuum mass or claim that boundary
has been crossed.

\section*{Declarations}
\addcontentsline{toc}{section}{Declarations}

\paragraph{Authorship and funding.}
Aleksandar Mihajlovic is the sole author. The author reports that this
research was undertaken independently and received no external financial
support, grant funding or third-party sponsorship. The institutional
affiliation does not imply institutional funding or endorsement.

\paragraph{Attribution and data availability.}
Published measurements, external theorems and third-party software are
credited to their original authors. Original calculations and reanalyses
are distinguished from those inputs. The accompanying research package
contains the source manuscript, executable calculations, numerical records
and mathematical certificates needed to reproduce the stated finite
calculations. No external review or endorsement is implied by their
inclusion. The original measurement data remain attributable to the
publications and releases cited in the text.

\paragraph{Competing interests.}
The author reports no third-party financial support or commercial
sponsorship. No additional competing interests have been declared in the
research materials supplied for this version.

\paragraph{Computer-assisted preparation and disclosure.}
Generative AI tools, including ChatGPT/Codex and Claude, were used for
organization, formatting and preparation of the supplied material and,
as recorded in its source files, for drafting some mathematical
derivations, explanatory text, code and verification workflows.
The tools are not authors and their responses are not independent
mathematical validation. Mathematical claims are supported by the
written proofs and the explicitly scoped reproducible calculations;
responsibility for the submitted manuscript and its conclusions remains
with the sole author. This disclosure does not attribute published
measurements or external results to the author or to these tools.

\clearpage
\appendix
\part*{Mathematical calculation appendices}
\addcontentsline{toc}{part}{Mathematical calculation appendices}
\hypertarget{calculation_certificates-exact-calculation-records-and-reproducible-certificates}{%
\section{Exact calculation records and reproducible certificates}\label{calculation_certificates-exact-calculation-records-and-reproducible-certificates}}

\hypertarget{calculation_certificates-what-the-calculation-records-establish}{%
\subsection{What the calculation records establish}\label{calculation_certificates-what-the-calculation-records-establish}}

The preceding arguments use inequalities quantified over all admissible volumes, separations, or cutoff indices. This section records finite calculations of quantities whose mathematical meaning has already been established analytically. Its purpose is to test normalization, verify implementations, and make the worked examples reproducible. The calculations are not estimates of an unknown four-dimensional mass gap.

Three objects must be distinguished throughout. The first is the four-dimensional Wilson measure. The second is the exactly solvable one-plaquette integral for \(SU(2)\). The third is a two-dimensional, free-boundary rectangular lattice whose gauge fixing makes its plaquette variables independent. The last two objects supply the numerical tables in this section. Their exact integrals are useful controls for calculations in the first model, but their laws cannot be substituted for its interacting plaquette distribution.

For the fundamental representation of \(SU(2)\), write \[
w(U)=\tfrac12\operatorname{Tr}U=\cos\theta,
\qquad 0\le\theta\le\pi.
\] The normalized Haar class measure is \((2/\pi)\sin^2\theta\,d\theta\). The one-plaquette partition function and the normalized spin-\(j\) character expectation are \[
z(\beta)=\frac{2I_1(\beta)}{\beta},\qquad
a_j(\beta)=\frac{I_{2j+1}(\beta)}{I_1(\beta)},\qquad \beta>0.
\] The exact-model section proves these formulas from Haar integration. The present section proves an elementary enclosure algorithm for their values. It uses neither asymptotic expansions nor floating-point special-function libraries in the certificate arithmetic.

The elementary character identity \[
\frac{\chi_1(U)}3=\frac{4w(U)^2-1}{3}
\] gives the one-plaquette variance \[
v(\beta)=\frac{1+3a_1(\beta)}4-a_{1/2}(\beta)^2.
\] On a planar rectangular lattice with free boundary, a fundamental Wilson loop enclosing \(A\) plaquettes has expectation \(a_{1/2}(\beta)^A\). The integer \(A\) is the number of plaquettes, not a four-dimensional physical area in unspecified units.

\hypertarget{calculation_certificates-a-positive-series-enclosure-theorem}{%
\subsection{A positive-series enclosure theorem}\label{calculation_certificates-a-positive-series-enclosure-theorem}}

For an integer \(n\ge0\) and a positive rational number \(b\), define \[
t_k(n,b)=\frac{(b/2)^{2k+n}}{k!(k+n)!},\qquad
S_K(n,b)=\sum_{k=0}^K t_k(n,b).
\] Every term and every partial sum is rational. The series for the modified Bessel function is \[
I_n(b)=\sum_{k=0}^{\infty}t_k(n,b).
\] Only nonnegative terms occur. Consequently truncation gives a lower bound with no cancellation error. A complete upper bound needs a bound on the infinite remainder.

\textbf{Theorem A.1 (rational enclosure of a positive Bessel series).} Put \[
r_K=\frac{b^2}{4(K+2)(K+n+2)}.
\] Whenever \(r_K<1\), \[
S_K(n,b)\le I_n(b)
\le S_K(n,b)+\frac{t_{K+1}(n,b)}{1-r_K}.
\tag{A1}
\] For every \(b>0\) and every prescribed relative tolerance \(\eta>0\), some finite \(K\) satisfies \(r_K<1\) and \[
\frac{t_{K+1}(n,b)}{1-r_K}\le\eta S_K(n,b).
\tag{A2}
\]

\textbf{Proof.} Direct division gives \[
\frac{t_{k+1}}{t_k}
=\frac{b^2}{4(k+1)(k+n+1)}.
\] For successive terms after \(t_{K+1}\), this ratio is at most \(r_K\). Therefore \[
\sum_{k=K+1}^{\infty}t_k
\le t_{K+1}\sum_{\ell=0}^{\infty}r_K^\ell
=\frac{t_{K+1}}{1-r_K}.
\] This proves (A1), including the entire infinite tail. As \(K\to\infty\), the ratios tend to zero. Beyond a finite index they are at most \(1/2\), so the terms decrease geometrically and the displayed remainder tends to zero. Meanwhile \(S_K\ge t_0>0\). This proves the eventual relative inequality (A2). No numerical convergence test is substituted for this argument. \(\square\)

The generator chooses \(K\) adaptively for each pair \((n,b)\) and requires \(\eta=10^{-45}\). Its stopping decision compares rational numbers exactly. The tolerance describes the Bessel-series enclosure; subsequent ratios and variance expressions have their own propagated interval widths. Those widths are retained explicitly rather than being assumed equal to \(10^{-45}\).

The distinction between \(r_K\) and \(t_{K+1}/t_K\) matters. The tail starts at \(t_{K+1}\), so its later ratios begin with \(t_{K+2}/t_{K+1}\). Using an earlier, larger ratio would remain valid once below one but would be less efficient. Using a later, smaller ratio without accounting for the first omitted term would not justify the same geometric bound.

For rational \(b\), factorial denominators can be large, but that is a computational cost rather than a mathematical uncertainty. Python's integer and rational arithmetic retains the numerator and denominator exactly. The mathematical proof of (A1) is independent of the language used to implement those integer operations.

\hypertarget{calculation_certificates-propagating-the-enclosures-through-observable-formulas}{%
\subsection{Propagating the enclosures through observable formulas}\label{calculation_certificates-propagating-the-enclosures-through-observable-formulas}}

Suppose \(0<L_n\le I_n(b)\le U_n\). Positivity of \(I_1\) gives \[
\frac{L_n}{U_1}\le\frac{I_n(b)}{I_1(b)}\le\frac{U_n}{L_1}.
\tag{A3}
\] For the partition function, \[
\frac{2L_1}{b}\le z(b)\le\frac{2U_1}{b}.
\] Let \(m_-\le a_{1/2}\le m_+\) and \(s_-\le a_1\le s_+\). The bounds used below satisfy \(0<m_-\le m_+<1\). Monotonicity then gives \[
\frac{1+3s_-}{4}-m_+^2
\le v(b)\le
\frac{1+3s_+}{4}-m_-^2.
\tag{A4}
\] No independence between the uncertain ratios is assumed. In fact, they share the same denominator. Interval evaluation remains valid because it encloses every allowed simultaneous value, though ignoring that dependence may enlarge the interval. This conservative propagation is preferable to displaying a narrow decimal interval whose error has not been proved.

For a nonnegative integer area \(A\), \[
m_-^A\le a_{1/2}(b)^A\le m_+^A.
\tag{A5}
\] The records use \(A=4\) and \(A=16\). Very small positive values can have an outward-rounded lower decimal endpoint equal to zero. That does not assert that the exact expectation is zero. The exact rational endpoint remains positive in the machine-readable certificate.

The ratio \(I_n/I_1\) and the plaquette variance have different qualitative behavior. As \(b\downarrow0\), the fundamental mean tends to zero, while the variance tends to \(1/4\). As \(b\to\infty\), the mean tends to one and the variance tends to zero. These statements follow from the exact integral and its asymptotics, not from the finite parameter grid. The grid is a consistency check on those independently derived limits.

\hypertarget{calculation_certificates-decimal-output-with-a-proved-direction-of-rounding}{%
\subsection{Decimal output with a proved direction of rounding}\label{calculation_certificates-decimal-output-with-a-proved-direction-of-rounding}}

For a rational number \(x=p/q\), \(q>0\), define its \(d\)-place lower and upper decimal representatives by \[
\underline{x}_d=10^{-d}\left\lfloor10^d p/q\right\rfloor,
\qquad
\overline{x}_d=10^{-d}\left\lceil10^d p/q\right\rceil.
\] Then \(\underline{x}_d\le x\le\overline{x}_d\). The implementation evaluates the two integer divisions directly. It does not convert the rational number to a binary float and then attempt to infer which way that float rounded.

\textbf{Proposition A.2 (displayed enclosures).} If \(L\le x\le U\), then \[
\underline L_d\le x\le\overline U_d.
\] The width introduced by decimal display is less than \(2\times10^{-d}\) in addition to the rational interval width.

\textbf{Proof.} The enclosing inequalities follow by transitivity. Each endpoint moves outward by an amount in \([0,10^{-d})\). Adding the two changes proves the width statement. \(\square\)

The CSV file displays 14 decimal places. The manuscript tables use fewer places to remain readable. All full rational numerators and denominators are supplied separately. A reader can reproduce the decimal tables from those fractions without evaluating a Bessel function.

\hypertarget{calculation_certificates-parameter-grid-and-worked-examples}{%
\subsection{Parameter grid and worked examples}\label{calculation_certificates-parameter-grid-and-worked-examples}}

The calibration grid is \[
\left\{\frac1{36},\frac1{18},\frac1{10},\frac14,\frac12,
1,2,4,8,16,32\right\}.
\] The first two values connect the examples to the four-dimensional strong-coupling comparison constants. The remaining values illustrate the exact one-plaquette law beyond that comparison regime. No implication is drawn that the four-dimensional model has the same one-plaquette marginal at those values.

At \(b=1\), for example, the outward-rounded certified intervals include \[
1.13031820798497\le z(1)\le1.13031820798498,
\] \[
0.24019372387008\le a_{1/2}(1)\le0.24019372387009,
\] and \[
0.22172580340314\le v(1)\le0.22172580340315.
\] The variance is not the square of the mean and is not the variance under Haar measure. It is computed under the tilted one-plaquette law with density \(e^{b w}/z(b)\).

At \(b=1/36\), the mean is enclosed by \[
0.00694422118989\le a_{1/2}(1/36)\le0.00694422118990.
\] Its leading small-\(b\) approximation \(b/4=1/144\) is larger, consistent with the first correction \(-b^3/96\). This is a useful normalization test: replacing the Wilson weight by \(e^{2b w}\) would change the leading coefficient and invalidate comparison of the tables with the displayed action.

At \(b=32\), the partition function is large while the normalized mean remains below one. Calculating the ratio by separately evaluating two large floating-point Bessel values can be numerically delicate at still larger \(b\). The rational series method used here does not overflow, although its integer operands grow. A scaled special-function implementation could be more efficient; efficiency is not needed for the small reproducibility grid.

% Generated from exact rational certificates; directed decimal rounding.
\noindent The tables display outward-rounded bounds with nine decimal places in the normalized mantissa. Their full rational endpoints are included in the calculation records.
\begin{longtable}{rcc}
\caption{Certified interval for $z(\beta)$, the single-plaquette partition function.}\\
\toprule $\beta$ & Lower endpoint & Upper endpoint \\
\midrule\endfirsthead
\toprule $\beta$ & Lower endpoint & Upper endpoint \\
\midrule\endhead
\bottomrule\endfoot
$1/36$ & $1.000096453$ & $1.000096454$ \\
$1/18$ & $1.000385852$ & $1.000385853$ \\
$1/10$ & $1.001250520$ & $1.001250521$ \\
$1/4$ & $1.007832871$ & $1.007832872$ \\
$1/2$ & $1.031577221$ & $1.031577222$ \\
$1$ & $1.130318207$ & $1.130318208$ \\
$2$ & $1.590636854$ & $1.590636855$ \\
$4$ & $4.879732576$ & $4.879732577$ \\
$8$ & $9.996828419\times10^{1}$ & $9.996828420\times10^{1}$ \\
$16$ & $1.081324294\times10^{5}$ & $1.081324295\times10^{5}$ \\
$32$ & $3.439278444\times10^{11}$ & $3.439278445\times10^{11}$ \\
\end{longtable}
\begin{longtable}{rcc}
\caption{Certified interval for $a_{1/2}(\beta)$, the fundamental character mean.}\\
\toprule $\beta$ & Lower endpoint & Upper endpoint \\
\midrule\endfirsthead
\toprule $\beta$ & Lower endpoint & Upper endpoint \\
\midrule\endhead
\bottomrule\endfoot
$1/36$ & $6.944221189\times10^{-3}$ & $6.944221190\times10^{-3}$ \\
$1/18$ & $1.388710311\times10^{-2}$ & $1.388710312\times10^{-2}$ \\
$1/10$ & $2.498958983\times10^{-2}$ & $2.498958984\times10^{-2}$ \\
$1/4$ & $6.233787272\times10^{-2}$ & $6.233787273\times10^{-2}$ \\
$1/2$ & $1.237179282\times10^{-1}$ & $1.237179283\times10^{-1}$ \\
$1$ & $2.401937238\times10^{-1}$ & $2.401937239\times10^{-1}$ \\
$2$ & $4.331274267\times10^{-1}$ & $4.331274268\times10^{-1}$ \\
$4$ & $6.580472673\times10^{-1}$ & $6.580472674\times10^{-1}$ \\
$8$ & $8.192494103\times10^{-1}$ & $8.192494104\times10^{-1}$ \\
$16$ & $9.078148458\times10^{-1}$ & $9.078148459\times10^{-1}$ \\
$32$ & $9.535031514\times10^{-1}$ & $9.535031515\times10^{-1}$ \\
\end{longtable}
\begin{longtable}{rcc}
\caption{Certified interval for $a_1(\beta)$, the spin-one character mean.}\\
\toprule $\beta$ & Lower endpoint & Upper endpoint \\
\midrule\endfirsthead
\toprule $\beta$ & Lower endpoint & Upper endpoint \\
\midrule\endhead
\bottomrule\endfoot
$1/36$ & $3.214865538\times10^{-5}$ & $3.214865539\times10^{-5}$ \\
$1/18$ & $1.285760208\times10^{-4}$ & $1.285760209\times10^{-4}$ \\
$1/10$ & $4.164064234\times10^{-4}$ & $4.164064235\times10^{-4}$ \\
$1/4$ & $2.594036347\times10^{-3}$ & $2.594036348\times10^{-3}$ \\
$1/2$ & $1.025657377\times10^{-2}$ & $1.025657378\times10^{-2}$ \\
$1$ & $3.922510451\times10^{-2}$ & $3.922510452\times10^{-2}$ \\
$2$ & $1.337451465\times10^{-1}$ & $1.337451466\times10^{-1}$ \\
$4$ & $3.419527326\times10^{-1}$ & $3.419527327\times10^{-1}$ \\
$8$ & $5.903752948\times10^{-1}$ & $5.903752949\times10^{-1}$ \\
$16$ & $7.730462885\times10^{-1}$ & $7.730462886\times10^{-1}$ \\
$32$ & $8.808121060\times10^{-1}$ & $8.808121061\times10^{-1}$ \\
\end{longtable}
\begin{longtable}{rcc}
\caption{Certified interval for $\operatorname{Var}(w)$, the single-plaquette variance.}\\
\toprule $\beta$ & Lower endpoint & Upper endpoint \\
\midrule\endfirsthead
\toprule $\beta$ & Lower endpoint & Upper endpoint \\
\midrule\endhead
\bottomrule\endfoot
$1/36$ & $2.499758892\times10^{-1}$ & $2.499758893\times10^{-1}$ \\
$1/18$ & $2.499035803\times10^{-1}$ & $2.499035804\times10^{-1}$ \\
$1/10$ & $2.496878252\times10^{-1}$ & $2.496878253\times10^{-1}$ \\
$1/4$ & $2.480595168\times10^{-1}$ & $2.480595169\times10^{-1}$ \\
$1/2$ & $2.423863045\times10^{-1}$ & $2.423863046\times10^{-1}$ \\
$1$ & $2.217258034\times10^{-1}$ & $2.217258035\times10^{-1}$ \\
$2$ & $1.627094921\times10^{-1}$ & $1.627094922\times10^{-1}$ \\
$4$ & $7.343834340\times10^{-2}$ & $7.343834341\times10^{-2}$ \\
$8$ & $2.161187474\times10^{-2}$ & $2.161187475\times10^{-2}$ \\
$16$ & $5.656921995\times10^{-3}$ & $5.656921996\times10^{-3}$ \\
$32$ & $1.440819720\times10^{-3}$ & $1.440819721\times10^{-3}$ \\
\end{longtable}
\begin{longtable}{rcc}
\caption{Certified interval for $a_{1/2}(\beta)^4$, the planar area-four loop.}\\
\toprule $\beta$ & Lower endpoint & Upper endpoint \\
\midrule\endfirsthead
\toprule $\beta$ & Lower endpoint & Upper endpoint \\
\midrule\endhead
\bottomrule\endfoot
$1/36$ & $2.325381338\times10^{-9}$ & $2.325381339\times10^{-9}$ \\
$1/18$ & $3.719175227\times10^{-8}$ & $3.719175228\times10^{-8}$ \\
$1/10$ & $3.899747712\times10^{-7}$ & $3.899747713\times10^{-7}$ \\
$1/4$ & $1.510107664\times10^{-5}$ & $1.510107665\times10^{-5}$ \\
$1/2$ & $2.342774863\times10^{-4}$ & $2.342774864\times10^{-4}$ \\
$1$ & $3.328485132\times10^{-3}$ & $3.328485133\times10^{-3}$ \\
$2$ & $3.519352279\times10^{-2}$ & $3.519352280\times10^{-2}$ \\
$4$ & $1.875116951\times10^{-1}$ & $1.875116952\times10^{-1}$ \\
$8$ & $4.504686270\times10^{-1}$ & $4.504686271\times10^{-1}$ \\
$16$ & $6.791866215\times10^{-1}$ & $6.791866216\times10^{-1}$ \\
$32$ & $8.265869246\times10^{-1}$ & $8.265869247\times10^{-1}$ \\
\end{longtable}
\begin{longtable}{rcc}
\caption{Certified interval for $a_{1/2}(\beta)^{16}$, the planar area-sixteen loop.}\\
\toprule $\beta$ & Lower endpoint & Upper endpoint \\
\midrule\endfirsthead
\toprule $\beta$ & Lower endpoint & Upper endpoint \\
\midrule\endhead
\bottomrule\endfoot
$1/36$ & $2.923995710\times10^{-35}$ & $2.923995711\times10^{-35}$ \\
$1/18$ & $1.913315377\times10^{-30}$ & $1.913315378\times10^{-30}$ \\
$1/10$ & $2.312842440\times10^{-26}$ & $2.312842441\times10^{-26}$ \\
$1/4$ & $5.200338902\times10^{-20}$ & $5.200338903\times10^{-20}$ \\
$1/2$ & $3.012466474\times10^{-15}$ & $3.012466475\times10^{-15}$ \\
$1$ & $1.227401035\times10^{-10}$ & $1.227401036\times10^{-10}$ \\
$2$ & $1.534090440\times10^{-6}$ & $1.534090441\times10^{-6}$ \\
$4$ & $1.236270311\times10^{-3}$ & $1.236270312\times10^{-3}$ \\
$8$ & $4.117733158\times10^{-2}$ & $4.117733159\times10^{-2}$ \\
$16$ & $2.127925851\times10^{-1}$ & $2.127925852\times10^{-1}$ \\
$32$ & $4.668250200\times10^{-1}$ & $4.668250201\times10^{-1}$ \\
\end{longtable}

\begin{figure}[htbp]
\centering
\includegraphics[width=\textwidth]{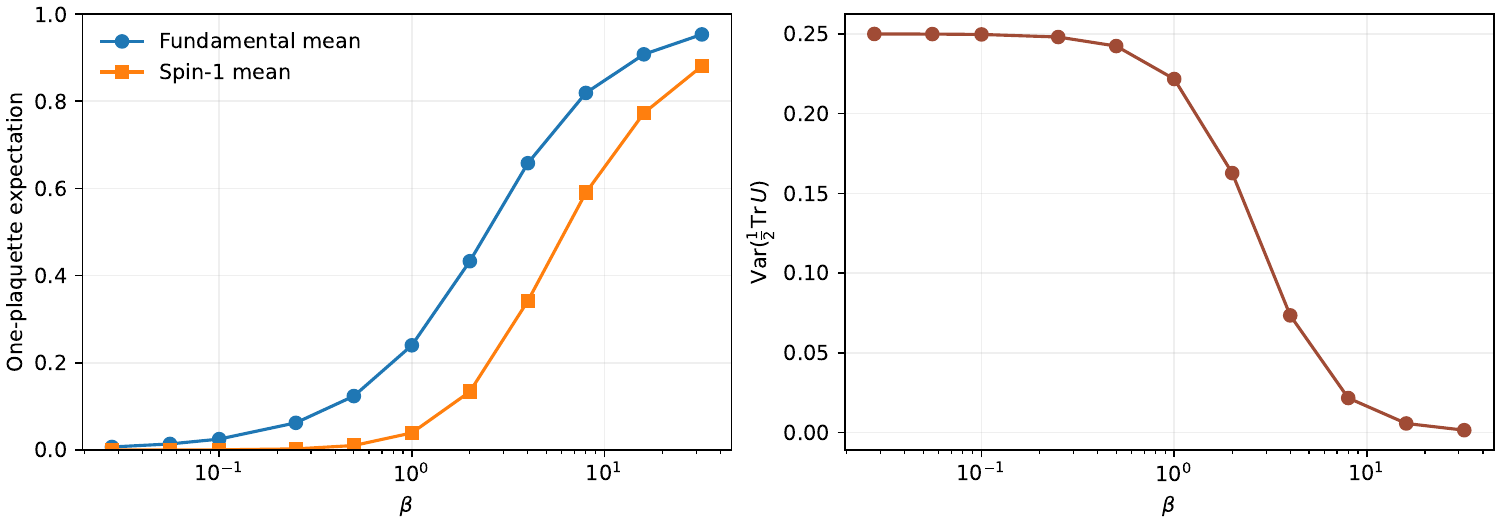}
\caption{Exactly solvable $SU(2)$ one-plaquette quantities at the certified parameter values. Lines join computed points for readability. The figures are not four-dimensional simulation results. The numerical certificates are rational; plotting coordinates are floating-point renderings of their midpoints.}
\end{figure}

\hypertarget{calculation_certificates-certifying-the-sharpened-comparison-threshold}{%
\subsection{Certifying the sharpened comparison threshold}\label{calculation_certificates-certifying-the-sharpened-comparison-threshold}}

The locality section improves the coarse influence coefficient \(\beta\) to \(\tanh\beta\) on the ordinary hypercubic lattice. In four dimensions its sufficient contraction condition is \[
18\tanh\beta<1,
\qquad\text{equivalently}\qquad
\beta<\operatorname{arctanh}(1/18).
\] The threshold is a sufficient condition for this comparison proof. It is not a determination of a phase transition or a physical critical coupling.

For \(0<x<1\), \[
\operatorname{arctanh}x=\sum_{k=0}^{\infty}\frac{x^{2k+1}}{2k+1}.
\] This identity follows by integrating the geometric series for \((1-x^2)^{-1}\) on a compact subinterval of \((-1,1)\), where uniform convergence justifies termwise integration. Positivity provides a particularly simple exact certificate.

\textbf{Proposition A.3 (threshold enclosure).} Put \[
T_K(x)=\sum_{k=0}^K\frac{x^{2k+1}}{2k+1}.
\] Then \[
T_K(x)\le\operatorname{arctanh}x
\le T_K(x)+\frac{x^{2K+3}}{(2K+3)(1-x^2)}.
\tag{A6}
\]

\textbf{Proof.} In every omitted term, the denominator is at least \(2K+3\). Hence \[
\sum_{k=K+1}^{\infty}\frac{x^{2k+1}}{2k+1}
\le\frac{x^{2K+3}}{2K+3}\sum_{j=0}^{\infty}x^{2j}.
\] Summing the geometric series proves (A6). \(\square\)

The certificate uses \(x=1/18\) and \(K=18\). It also proves directly that the improved threshold exceeds \(1/18\), because all terms after the first are positive. The improvement is a refinement of the displayed sufficient estimate. 

\hypertarget{calculation_certificates-an-analytic-scaling-control-with-no-sampled-field-interpretation}{%
\subsection{An analytic scaling control with no sampled-field interpretation}\label{calculation_certificates-an-analytic-scaling-control-with-no-sampled-field-interpretation}}

One figure plots the explicit comparison functions \[
B_p(a)=a^{-p}e^{-1/a},\qquad p\in\{0,2,4\}.
\] Their logarithms are \(-p\log a-1/a\). The graph is not evidence for a limit: the limit follows immediately from the following elementary lemma.

\textbf{Lemma A.4 (exponentials dominate polynomial normalization).} For every \(c>0\) and real \(p,N\), \[
\lim_{a\downarrow0}a^{-p-N}e^{-c/a}=0.
\]

\textbf{Proof.} Set \(x=c/a\). Up to a fixed positive constant, the expression is \(x^{p+N}e^{-x}\). Choose a nonnegative integer \(k>\max\{p+N,0\}\). The exponential series gives \(e^{x/2}\ge (x/2)^k/k!\) for \(x>0\). Consequently \[
x^{p+N}e^{-x}
\le2^k k!\,x^{p+N-k}e^{-x/2}\longrightarrow0.
\] The final factors are bounded and converge to zero; in particular, \(p+N-k<0\). \(\square\)

This elementary result enters the fixed strong-coupling obstruction because the interaction-graph distance grows at least proportionally to physical separation divided by \(a\). The theorem concerns observables obeying its support-size and norm hypotheses. A graph of \(B_p\) cannot replace those hypotheses or establish that arbitrary renormalized field observables satisfy them.

\begin{figure}[htbp]
\centering
\includegraphics[width=.82\textwidth]{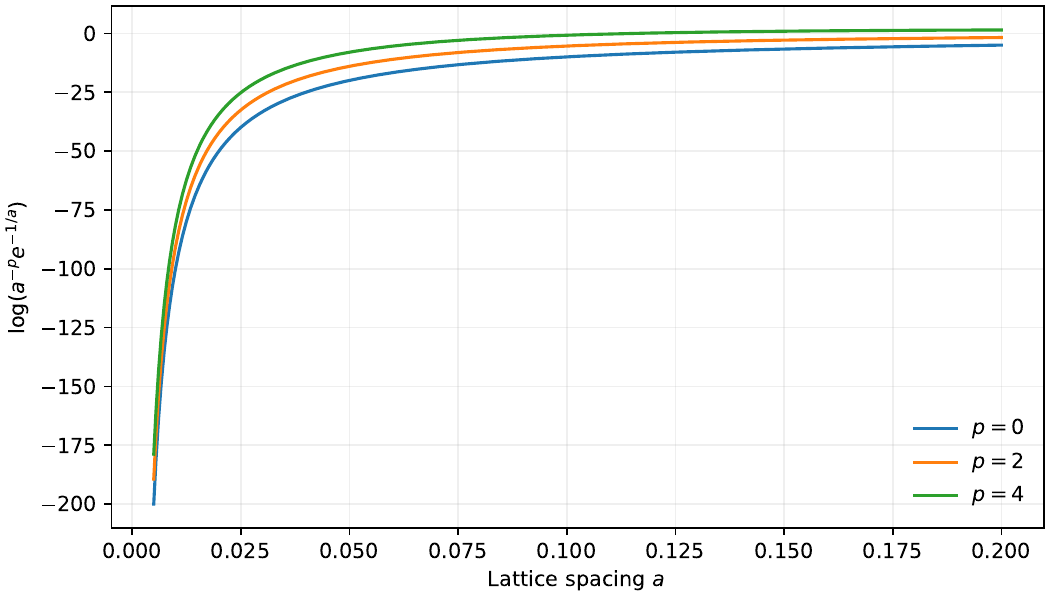}
\caption{Logarithms of three explicit scaling bounds. The common exponential term dominates every fixed polynomial normalization as $a$ tends to zero. This is a visualization of Lemma A.4, not a numerical estimate of a Yang--Mills correlator.}
\end{figure}

The exact records contain 33 Bessel enclosures and 66 derived intervals on the stated eleven-coupling grid, together with the comparison-threshold interval. All are quantities of the specified solvable calibration models. The accompanying rational data permit direct checking of the tail and interval inequalities above.

\hypertarget{calculation_certificates-how-the-calculations-constrain-a-proposed-proof}{%
\subsection{How the calculations constrain a proposed proof}\label{calculation_certificates-how-the-calculations-constrain-a-proposed-proof}}

A proposed argument must preserve normalization across its stages. If a character expansion uses \(\operatorname{Tr}U\) while a comparison estimate uses \(N^{-1}\operatorname{ReTr}U\), then the coupling parameter changes by a factor of \(N\). The series calculations make such errors visible. At small \(\beta\), the fundamental one-plaquette mean begins at \(\beta/4\) in the convention used here; this coefficient is an exact local check.

A proposed argument must also preserve the model's dependence structure. The planar expectation \(a_{1/2}^A\) uses independent plaquette variables after a legitimate two-dimensional gauge change. The one-cube calculation in the exact-model section gives a direct obstruction to extending that independence into higher dimensions. Replacing a four-dimensional loop expectation by the planar formula would remove the very curvature constraints that need analysis.

Finally, a proposed argument must preserve the units of its rate. A number obtained from \(-\log a_{1/2}(\beta)\) is a dimensionless decay parameter for a specified solvable observable. Calling it a four-dimensional particle mass requires a physical time variable, a transfer semigroup, a nonzero state, and the correct factor of the lattice spacing. The spectral and reconstruction sections state those requirements as mathematical assumptions and prove the deductions from them.

These controls are useful because they have exact answers. They permit a candidate calculation to fail quickly and specifically: a wrong small-coupling coefficient is a normalization error; an unjustified product of plaquette means is a dependence error; a missing factor of \(1/a\) is a units error. Passing these checks does not prove the remaining continuum estimate, but failing one identifies a repairable mathematical defect before it propagates through the manuscript.

\section{The 96 plaquette classes of the perfect forest}\label{fg-app:table}

For each parity class $x\bmod2$ (first column) the second column gives $e_{\mu(x)}$ from \eqref{fg-eq:rule}. The remaining columns give, for the plaquette $P(x;a,b)$, the unique link of $\Fs$ it contains. The symbol $\circ_k$ is the link leaving the base corner $x$ in direction $k$. The symbol $i_j$ is the link leaving the corner $x+e_i$ in direction $j$. Each entry can be checked by looking up $\mu$ at the indicated corner. The program confirms both that the entry is correct and that it is unique.

\begin{center}\small
\begin{tabular}{@{}cc|cccccc@{}}\toprule
$x\bmod2$ & $e_{\mu(x)}$ & $(1,2)$ & $(1,3)$ & $(1,4)$ & $(2,3)$ & $(2,4)$ & $(3,4)$\\\midrule
0000 & $e_{1}$ & $\circ_{1}$ & $\circ_{1}$ & $\circ_{1}$ & $2_{3}$ & $4_{2}$ & $3_{4}$ \\
0001 & $e_{2}$ & $\circ_{2}$ & $1_{3}$ & $4_{1}$ & $\circ_{2}$ & $\circ_{2}$ & $3_{4}$ \\
0010 & $e_{4}$ & $1_{2}$ & $3_{1}$ & $\circ_{4}$ & $2_{3}$ & $\circ_{4}$ & $\circ_{4}$ \\
0011 & $e_{4}$ & $2_{1}$ & $1_{3}$ & $\circ_{4}$ & $3_{2}$ & $\circ_{4}$ & $\circ_{4}$ \\
0100 & $e_{3}$ & $2_{1}$ & $\circ_{3}$ & $1_{4}$ & $\circ_{3}$ & $4_{2}$ & $\circ_{3}$ \\
0101 & $e_{2}$ & $\circ_{2}$ & $3_{1}$ & $1_{4}$ & $\circ_{2}$ & $\circ_{2}$ & $4_{3}$ \\
0110 & $e_{3}$ & $1_{2}$ & $\circ_{3}$ & $4_{1}$ & $\circ_{3}$ & $2_{4}$ & $\circ_{3}$ \\
0111 & $e_{1}$ & $\circ_{1}$ & $\circ_{1}$ & $\circ_{1}$ & $3_{2}$ & $2_{4}$ & $4_{3}$ \\
1000 & $e_{1}$ & $\circ_{1}$ & $\circ_{1}$ & $\circ_{1}$ & $3_{2}$ & $2_{4}$ & $4_{3}$ \\
1001 & $e_{3}$ & $1_{2}$ & $\circ_{3}$ & $4_{1}$ & $\circ_{3}$ & $2_{4}$ & $\circ_{3}$ \\
1010 & $e_{2}$ & $\circ_{2}$ & $3_{1}$ & $1_{4}$ & $\circ_{2}$ & $\circ_{2}$ & $4_{3}$ \\
1011 & $e_{3}$ & $2_{1}$ & $\circ_{3}$ & $1_{4}$ & $\circ_{3}$ & $4_{2}$ & $\circ_{3}$ \\
1100 & $e_{4}$ & $2_{1}$ & $1_{3}$ & $\circ_{4}$ & $3_{2}$ & $\circ_{4}$ & $\circ_{4}$ \\
1101 & $e_{4}$ & $1_{2}$ & $3_{1}$ & $\circ_{4}$ & $2_{3}$ & $\circ_{4}$ & $\circ_{4}$ \\
1110 & $e_{2}$ & $\circ_{2}$ & $1_{3}$ & $4_{1}$ & $\circ_{2}$ & $\circ_{2}$ & $3_{4}$ \\
1111 & $e_{1}$ & $\circ_{1}$ & $\circ_{1}$ & $\circ_{1}$ & $2_{3}$ & $4_{2}$ & $3_{4}$ \\
\bottomrule
\end{tabular}
\captionof{table}{The perfect forest $\Fs$: every plaquette class contains exactly one forest link.}\label{fg-tab:forest}
\end{center}

\clearpage
\phantomsection

\end{document}